\pdfoutput=1

\documentclass[envcountchap,graybox,sectrefs,vecphys]{svmono}

\usepackage{type1cm}
\usepackage{newtxtext}
\usepackage{newtxmath}

\usepackage{chapterbib}		 
\usepackage{cite}			 
\usepackage[type={CC},modifier={by-nc},version={4.0}]{doclicense}
\usepackage[bottom]{footmisc}
\usepackage{graphicx}        

\usepackage{amsfonts}   
\usepackage{bbold}		
\usepackage{feynmf}		
\usepackage[pdftex,colorlinks,linkcolor=purple,citecolor=red,urlcolor=blue]{hyperref}
\usepackage{mathrsfs}	
\usepackage{mathtools}	
\usepackage{slashed}	
\usepackage{xparse}		

\allowdisplaybreaks

\let\a=\alpha
\let\b=\beta
\let\c=\chi
\let\d=\delta
\let\eps=\epsilon
\let\g=\gamma
\let\k=\kappa
\let\l=\lambda
\let\m=\mu
\let\n=\nu
\let\o=\omega
\let\p=\phi
\let\s=\sigma
\let\S=\Sigma
\let\t=\theta
\let\x=\xi

\let\ve=\varepsilon
\let\vp=\varphi
\let\vr=\varrho

\let\w=\wedge

\newcommand{\iA}{\mathfrak{A}}
\newcommand{\C}{\mathbb{C}}
\newcommand{\N}{\mathbb{N}}
\newcommand{\R}{\mathbb{R}}
\newcommand{\Z}{\mathbb{Z}}

\newcommand{\trgr}{\{e\}}
\newcommand{\un}{\mathbb{1}}
\newcommand{\df}{F}
\newcommand{\dfg}{K}
\newcommand{\DF}{\mathbb{F}}
\newcommand{\dx}{X}
\newcommand{\DX}{\mathbb{X}}
\newcommand{\dA}{\tilde F}
\newcommand{\dAg}{\tilde K}
\newcommand{\dm}{\vr}
\newcommand{\Aa}{\mathcal{A}}
\newcommand{\Da}{\mathcal{D}}
\newcommand{\Ha}{\mathscr{H}}
\newcommand{\hf}{\psi}
\newcommand{\La}{\mathscr{L}}
\newcommand{\Fa}{\mathscr{F}}
\newcommand{\M}{\mathcal{M}}
\newcommand{\MN}{\mathcal{N}}
\newcommand{\mf}{\chi}
\renewcommand{\O}{\mathcal{O}}
\newcommand{\nf}{n_\mathrm{f}}
\newcommand{\bigO}{\mathcal{O}}
\newcommand{\transfo}{P}
\newcommand{\proj}{\mathcal{P}}
\newcommand{\Raut}{\mathcal{R}}
\newcommand{\rep}{\mathcal{R}}
\newcommand{\pau}{\tau}
\newcommand{\T}{\mathcal{T}}
\newcommand{\upix}{\mathfrak{U}}
\newcommand{\U}{\mathcal{U}}
\newcommand{\MM}{\mathcal{M}}
\newcommand{\mc}{\omega}
\newcommand{\mcu}{\omega_\parallel}
\newcommand{\mcb}{\omega_\perp}
\newcommand{\mcp}{\omega_P}
\newcommand{\MC}{\Omega}
\newcommand{\MCU}{\Omega_\parallel}
\newcommand{\MCB}{\Omega_\perp}
\newcommand{\con}{\Omega}
\newcommand{\LC}{\hat\Gamma}
\newcommand{\kil}{\xi}
\newcommand{\XiEFT}{\Xi}

\newcommand{\abs}[1]{\left\lvert{#1}\right\rvert}
\newcommand{\adj}[1]{\overline{#1}}
\newcommand{\fdeg}[2]{#1^{(#2)}}
\newcommand{\fr}[1]{\smash{\underline{\smash{#1}}}}
\newcommand{\gr}[1]{\mathrm{#1}}
\newcommand{\he}[1]{{#1}^{\dagger}}
\newcommand{\inner}[2]{\langle{#1},{#2}\rangle}
\newcommand{\lie}[1]{\mathfrak{#1}}
\newcommand{\Od}[2]{\D{#1}/\D{#2}}
\newcommand{\OD}[2]{\frac{\D{#1}}{\D{#2}}}
\newcommand{\Pd}[2]{\de{#1}/\de{#2}}
\newcommand{\PD}[2]{\frac{\de{#1}}{\de{#2}}}
\newcommand{\skal}[2]{\vec{#1}\cdot\vec{#2}}
\DeclareDocumentCommand\tran{ o m }{\IfNoValueTF{#1}{\mathcal T_{#2}}{\mathcal T_{{#1}\to{#2}}}}
\newcommand{\tri}[1]{\mathfrak{#1}}
\newcommand{\at}[2]{\left.{#1}\right\rvert_{#2}}
\newcommand{\vev}[1]{\langle{#1}\rangle}
\newcommand{\bra}[1]{\left\langle{#1}\right\rvert}
\newcommand{\ket}[1]{\left\lvert{#1}\right\rangle}
\newcommand{\braket}[2]{\left\langle{#1}\middle\vert{#2}\right\rangle}
\newcommand{\amplitude}[3]{\left\langle{#1}\middle\vert{#2}\middle\vert{#3}\right\rangle}

\newcommand{\uiota}{\allmodesymb{\greeksym}{i}}

\let\cd=\nabla
\let\de=\partial
\DeclareMathOperator{\diag}{diag}
\DeclareMathOperator{\divg}{div}
\DeclareMathOperator{\ho}{\star}
\DeclareMathOperator{\id}{id}
\DeclareMathOperator{\IM}{im}
\newcommand{\ix}[1]{\uiota_{#1}}
\newcommand{\ld}[1]{\mathcal{L}_{#1}}
\DeclareMathOperator{\KER}{ker}
\DeclareMathOperator{\rank}{rank}
\DeclareMathOperator*{\Res}{Res}
\DeclareMathOperator{\sech}{sech}
\DeclareMathOperator{\sgn}{sgn}
\DeclareMathOperator{\tr}{tr}
\DeclareMathOperator{\vol}{vol}

\newcounter{exno}[chapter]
\renewcommand{\theexno}{\thechapter.\arabic{exno}}
\newcommand{\refex}[1]{Example~\ref{#1}}
\newenvironment{illustration}{\refstepcounter{exno}\begin{tips}{Example~\theexno}}{\end{tips}}
\newenvironment{watchout}{\begin{svgraybox}}{\end{svgraybox}}

\begin{document}

\author{Tom\'a\v{s} Brauner}
\title{Effective Field Theory for Spontaneously Broken Symmetry}
\maketitle


\frontmatter
\begin{dedication}
To Laura and Julie
\end{dedication}

\preface


Symmetry has accompanied our attempts to understand the inner workings of nature ever since the birth of science. The powerful concept of \emph{spontaneous symmetry breaking} (SSB) makes it possible to reconcile the complex experimental phenomenology with the simplicity of underlying natural laws. It is therefore a necessary ingredient of virtually any textbook on quantum field theory. However, a typical textbook-level treatment offers a mere simplified version of the state of the art of the subject in its early days of the 1960s. This hardly does justice to the richness and beauty of SSB. Moreover, the subject has undergone development in the last two decades that has substantially sharpened our understanding of SSB in quantum many-body systems.

This book is an attempt to fill the gap in the literature, and to provide an up-to-date survey of the landscape of SSB from the perspective of \emph{effective field theory} (EFT). In writing the book, I was aiming at learners of all seniority levels who are familiar with the basic notions of quantum field theory. The style of the text and the mathematical background required will probably be a best match for graduate students of theoretical physics. However, the introductory parts of the book might be accessible even to advanced undergraduates. Those will benefit from studying first Part~\ref{part:prologue} of the book, which introduces the concepts of SSB and the EFT description thereof at an elementary level. More experienced readers might proceed directly to Part~\ref{part:foundations}. This offers an overview of the physics of SSB at a standard textbook level. However, it already contains numerous details that are hard to find in a student-friendly form elsewhere. The core techniques of EFT for SSB are developed in Parts~\ref{part:internalSSB} and~\ref{part:spacetimeSSB}. These provide a comprehensive account of the subject, from the pioneering works of the 1960s to some recent developments that are otherwise only found scattered in the research literature.

The most important recent results included are, as of writing this book, about a decade old. This applies in particular to the completed classification of \emph{Nambu--Goldstone} (NG) \emph{bosons} (Chap.~\ref{chap:NGbosons}), the EFT formalism for spontaneously broken internal symmetries in quantum many-body systems (Chap.~\ref{chap:effLagrangian}), and some applications of EFT to systems with spontaneously broken spacetime symmetry (Chaps.~\ref{chap:spacetimequantum} and~\ref{chap:spacetimeclassical}). However, some of the material included also reflects new works that appeared while I was already writing the book. This is the case especially for the discussion of scattering of NG bosons in Chap.~\ref{chap:scattering}. Finally, Chap.~\ref{chap:cosetspacetime} puts forward a novel formalism for implementation of spontaneously broken spacetime symmetry that, to the best of my knowledge, has not appeared in this form before. This significantly influences the narrative surrounding the EFT for broken spacetime symmetries in Chaps.~\ref{chap:spacetimequantum} and~\ref{chap:spacetimeclassical}. My hope is that the change is for the better.

The one thing I have learned while writing this book is that one's understanding of any topic is a never-ending process of evolution. I would therefore like to extend my gratitude to those who helped me along the way. First and foremost, I am indebted to Angelo Esposito, Yoshimasa Hidaka and Ji\v{r}\'{\i} Novotn\'{y}, who have read critically parts of the manuscript and given me constructive feedback, helping me improve the text on multiple fronts. My personal view of the subject of this book and of effective field theory in general has been shaped by interactions with numerous mentors and collaborators throughout the years, including Jens O.~Andersen, Christoph P.~Hofmann, Ji\v{r}\'{\i} Ho\v{s}ek, Martin Mojahed, Sergej Moroz, Hitoshi Murayama, Ji\v{r}\'{\i} Novotn\'{y}, Riccardo Penco, Aleksi Vuorinen, Haruki Watanabe and Naoki Yamamoto. Last but not least, I would like to thank Stefan Theisen for encouraging me to write this book.


\vspace{\baselineskip}
\begin{flushright}\noindent
Stavanger, Norway\hfill Tom\'a\v{s} Brauner\\
September 2023\hfill{\phantom{Brauner}}\\
\end{flushright}
\tableofcontents
\extrachap{Notation and Conventions}


\section*{List of Acronyms}

Acronyms that appear in the book and their use in established collocations:
\begin{description}[CCWZ]
\item[ChPT]{Chiral perturbation theory}
\item[CS]{Chern--Simons (Lagrangian, theory)}
\item[DBI]{Dirac--Born--Infeld (Lagrangian, theory)}
\item[DM]{Dzyaloshinskii--Moriya (interaction, term)}
\item[EFT]{Effective field theory}
\item[EM]{Energy--momentum (tensor)}
\item[EoM]{Equation of motion}
\item[GW]{Goldstone--Wilczek (current)}
\item[IHC] Inverse Higgs constraint
\item[LC]{Levi-Civita (connection, symbol, tensor)}
\item[LO]{Leading order}
\item[MC]{Maurer--Cartan (equation, form)}
\item[NG]{Nambu--Goldstone (boson, field, mode)}
\item[NLO]{Next-to-leading order}
\item[QCD]{Quantum chromodynamics}
\item[SSB]{Spontaneous symmetry breaking}
\item[VEV]{Vacuum expectation value}
\item[VPD]{Volume-preserving diffeomorphism}
\item[WZ]{Wess--Zumino (action, Lagrangian, term)}
\end{description}
For the reader's convenience, each of these acronyms is also defined in the text the first time it appears in a particular chapter.


\section*{Mathematical Conventions}

This book aims at audiences from diverse areas of physics, which makes it virtually impossible to maintain coherent notation without alienating some of the readers. I therefore try to adopt conventions that are as close as possible to practice, at the cost of being somewhat dependent on the context.

Thus, a ``relativistic'' notation is used to discuss physical systems whose dynamics is Lorentz-invariant. I assume a flat Minkowski spacetime unless (exceptionally) stated otherwise. Much of the content of the book is valid for spacetimes of any dimension $D\geq3$; see Sect.~\ref{sec:nogo} for a justification of this constraint. The timelike Minkowski metric,
\begin{equation*}
g_{\m\n}\equiv\diag(+1,-1,-1,\dotsc)\;,
\end{equation*}
is implicitly assumed when needed. Spacetime vectors (or, in $D=4$ dimensions, four-vectors) are denoted using italics: $p,q,\dotsc$. An inner product of spacetime vectors, implied by the Minkowski metric, is indicated with a dot,
\begin{equation*}
p\cdot q\equiv g_{\m\n}p^\m q^\n=p^\m q_\m=p_\m q^\m\;.
\end{equation*}
Here the Einstein summation convention is applied to any pair of repeated indices, one covariant and one contravariant. The Levi-Civita tensor $\smash{\ve_{\m\n\l\dotsb}}$ in Minkowski spacetime is normalized so that $\smash{\ve^{012\dotsb}=1}$. Finally, I consistently use the ``natural units'' of high-energy physics where both the speed of light $c$ and the reduced Planck constant $\hbar$ are set to one. Thus, the energy and momentum of a relativistic particle are identified respectively with the frequency and wave vector of the corresponding quantum-mechanical plane wave. Moreover, the Minkowski square of the energy--momentum of the particle is $p^2\equiv p\cdot p=m^2$, $m$ being the rest mass of the particle.

On the contrary, in systems lacking manifest Lorentz invariance, typical for applications to condensed-matter physics, I adopt a ``nonrelativistic'' notation. Here, spatial vectors are denoted using boldface italics: $\vec a,\vec b,\dotsc$. The Einstein summation convention is still used, but indices are no longer raised or lowered with the Minkowski metric. Instead, upper and lower indices are used interchangeably just for visual clarity and are treated as equivalent. Formally, this amounts to using the flat Euclidean metric, $g_{rs}\equiv\d_{rs}$. The dot product of two spatial vectors can then be written in a number of equivalent forms such as
\begin{equation*}
\skal ab=\d_{rs}a^rb^s=a_rb_r=a^rb^r=a^rb_r=a_rb^r\;.
\end{equation*}
The symbol $\vec\nabla$ is used for the gradient operator, that is a spatial-vector differential operator with components $\de_r\equiv\Pd{}{x^r}$. Wherever needed explicitly, the number of spatial dimensions is denoted as $d\equiv D-1$.


\mainmatter
\begin{partbacktext}
\part{Prologue}
\label{part:prologue}
\end{partbacktext}
\chapter{Introduction}
\label{chap:intro}

\abstract*{This chapter introduces in layman terms the two central concepts of the book: effective field theory and spontaneous symmetry breaking. The purpose is to set up the basic framework for the book and to underline its broad relevance. The introduction is followed by an outline of the contents of the book. This informs the reader of the hierarchical structure of the material and of the dependencies among the various parts. A list of basic references for further reading is appended.}


\section{What Is Effective Field Theory?}
\label{sec:whatisEFT}

The laws of nature have a hierarchical structure, stratified by the resolution scale of our observations. When Isaac Newton published in 1687 his gravitational law explaining Kepler's laws of planetary motion, he did not need to know the details of the planets' inner structure. His theory gives an accurate description of the dynamics of the solar system while treating the planets as point-like objects.

The phenomenological understanding of the basic material properties of solids, liquids and gases culminated in the mid-19\textsuperscript{th} century through the work of Clausius, Gibbs, Kelvin, Maxwell and others. Much about these different phases of matter and the transitions between them can be, and was, learned without insight into the molecular structure of matter. It was only about a half-century later that quantum mechanics provided an adequate framework for the description of atoms and molecules.

Quantum mechanics itself is extremely successful in explaining the chemical properties of different substances and mechanical properties of solids, liquids and gases. In fact, virtually all natural phenomena observed at macroscopic scales boil down either to gravity or to Maxwell's electromagnetism combined with quantum mechanics. Up to rare exceptions, for instance radioactivity, one does not need to understand the structure of atomic nuclei. Nuclear physics itself only advanced later, in the 1930s. And again, much about nuclear structure and reactions was understood  before the discovery in the 1960s that individual nucleons are composed of yet smaller particles, the quarks.

The smallest currently known constituents of matter are quarks, bound in atomic nuclei, and leptons, of which the electron is but one example. How do we know that the chain of successive divisions into smaller and smaller constituents ends here? We do not. Perhaps, one day, even smaller building blocks of nature will be discovered. Nevertheless, we do have a successful theory that is able to account for experimental observations at the current resolution frontier: the Standard Model of particle physics. This theory does not depend on the details of as yet undiscovered microscopic physics. Our ignorance of such microscopic structure currently out of reach is subsumed into the values of the input parameters of the Standard Model.

This is the essence of \emph{effective field theory} (EFT). Going back to where we started, the information about the detailed structure of planets, unknown to Newton, could be subsumed into the values of the planets' masses. Likewise, the quantum-mechanical nature of molecular interactions only affects macroscopic thermodynamics through material constants such as specific heats or elastic moduli. All that quantum mechanics in turn needs to know about atomic nuclei are their masses and electric charges. Finally, the quark structure of nucleons only manifests itself through binding forces between nucleons that hold the nuclei together. These nuclear forces were studied thoroughly already in the mid-1930s by Fermi, Yukawa and others, three decades before the existence of quarks was conceived by Gell-Mann and Zweig.

Generally speaking, EFT is a framework for a quantitative description of nature at certain level of resolution. Whatever physics might exist at shorter, unresolved length scales can be taken into account through the values of the parameters of the EFT. In the absence of further information about the microscopic physics, these parameters have to be determined by experiment. This line of reasoning has implicitly accompanied the evolution of physics from its early stages. In the second half of the 20\textsuperscript{th} century, it was however developed into a full-fledged quantitative formalism with tremendous impact, from concrete physical applications to the philosophy of science. One of the founders of modern EFT, Steven Weinberg, offers a nice account of the early history of EFT in~\cite{Weinberg2021a}. The array of successful applications of the EFT program has grown so long in the last decades that I cannot even list them all here. Instead, I point the reader to the literature. References to introductory-level expositions of EFT with applications are given below in Sect.~\ref{subsec:furtherreading}.


\section{Broken Symmetry Zoo}
\label{sec:SSBZOO}

Most macroscopic phenomena observed in nature are dominated by the collective dynamics of the fundamental constituents of matter. In hindsight, this is the reason why 19\textsuperscript{th}-century physics was so successful in describing nature. Consider, for instance, a single-component fluid in thermodynamic equilibrium. All the thermodynamic properties of the fluid are determined by the values of two independent observables such as temperature and pressure. If we now disturb the fluid locally, the interactions among the fluid molecules will cause the perturbation to propagate. This propagation manifests macroscopically as a sound wave, possibly accompanied by a background flow. Such small perturbations from thermodynamic equilibrium are described by hydrodynamics, which is an early example of an EFT of the modern type. The microscopic input required here is the equation of state of the fluid along with a set of transport coefficients. Other than that, hydrodynamics is based only on a set of local conservation laws including energy, momentum  and whatever other conserved charge the fluid might carry. We know from Noether's theorem that local conservation laws are a consequence of continuous symmetries. One can therefore say that the dynamics of fluids near equilibrium is largely dictated by symmetry. That is typical of \emph{spontaneous symmetry breaking} (SSB).

\begin{figure}[t]
\sidecaption[t]
\includegraphics[width=2.9in]{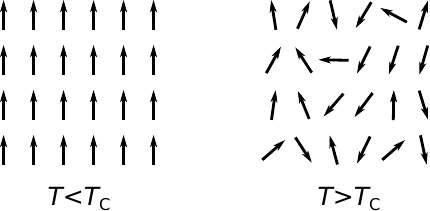}
\caption{Schematic picture of the ordered state of a ferromagnet below the Curie temperature $T_\mathrm{C}$ and of the disordered state above the Curie temperature $T_\mathrm{C}$}
\label{fig:ferromagnet}
\end{figure}

Let me illustrate this on another example. Ferromagnets are materials which, below certain temperature called the Curie temperature, exhibit spontaneous alignment of magnetic moments of their atoms. A solid ferromagnet can be pictured as a lattice, each of whose nodes carries a single spin degree of freedom, see Fig.~\ref{fig:ferromagnet}. The mutual interactions between the spins may be perfectly isotropic, that is invariant under spatial rotations. Yet, the ferromagnetic state below the Curie temperature obviously possesses a preferred direction of alignment of the spins. This direction is a priori arbitrary and may be set for instance by boundary conditions, not by the internal dynamics of the ferromagnet itself. The emergence of an equilibrium state that breaks the intrinsic symmetry of the system is a hallmark of SSB.

A local perturbation of the ferromagnetic state will be propagated by spin--spin interactions and manifest itself macroscopically as a spin wave. The existence of wave-like excitations which dominate the physics at long distances is another general feature of SSB; these are called \emph{Nambu--Goldstone} (NG) \emph{bosons}. Just like for sound waves in fluids, the dynamics of ferromagnetic spin waves is controlled by symmetry. A quantitatively accurate EFT of spin waves can be constructed solely based on the underlying spacetime translation invariance, spatial rotation invariance, and the fundamental commutation relations of angular momentum (spin).

The above two examples hint at a particular type of EFT that governs the long-distance physics of systems with SSB. The relevant degrees of freedom of this EFT are the NG bosons whereas the details of the EFT are dictated by symmetry. It should therefore not come as a surprise that many of the techniques developed in this book rely on the theory of Lie groups, and to a lesser extent on differential geometry. The precise mathematical structure of the EFT depends on the kind of symmetry in question. It is common to use the term \emph{internal symmetry} for symmetries that act directly on the dynamical degrees of freedom of a given physical system. SSB of an internal symmetry lies for instance behind the phenomena of (anti)ferromagnetism or superfluidity. On the other hand, a \emph{spacetime symmetry} is a geometric property of space and time. It affects the dynamical degrees of freedom indirectly as fields living in the physical spacetime. SSB of spacetime symmetries is more subtle yet ubiquitous and relevant for \emph{all} phases of matter. This is the reason why the form of hydrodynamics is largely fixed by the local conservation laws of energy and momentum. A different pattern of SSB gives rise to the elasticity theory of solids. There are also examples of phenomena that exhibit a combination of SSB of internal and spacetime symmetry. These include for instance vortex lattices in rotating superfluids or type-II superconductors, anisotropic Cooper pairing in $p$-wave superconductors, or the helical order in chiral magnets.


\section{Structure of This Book}
\label{sec:structureofbook}

The subject of this book is of central importance to quantum field and many-body theory. As such, it connects several branches of physics including, but not limited to, high-energy physics, condensed-matter physics, astrophysics and cosmology. With the explicit aim to cater to these different communities, I have tried my best to make the text relevant and comprehensible to audiences with different backgrounds. It is up to the reader to judge to what extent this effort has been successful.

\begin{watchout}%
The book aims primarily at graduate students of theoretical physics regardless of their concrete specialization. The text should in principle be accessible to anyone who has taken a first course on quantum field theory. With this target group in mind, I have interspersed the text with two didactic elements. On the one hand, there are numbered examples, graphically separated from the main text, that mostly serve to illustrate newly introduced theoretical concepts or arguments. Occasional more advanced examples mention interesting applications that would disrupt the line of discussion if included in the text. On the other hand, gray blocks such as this one point to important facts or subtleties that might otherwise be missed. They are meant to encourage the reader to take a critical look at the presented material.
\end{watchout}

The mathematics background required to benefit from this book likewise corresponds to that of a typical student of theoretical physics. It consists mostly of linear algebra, advanced calculus including complex calculus and the calculus of variations, and the theory of partial differential equations. The only piece of background I assume that might go beyond the curriculum of some graduate programs is certain familiarity with group theory and its applications to physics. The level and extent as covered by~\cite{Georgi1999a} is fully sufficient; Chaps.~14 and 15 of~\cite{Stone2009a} are a good start that will get the reader very far. Some more advanced parts of the book rely on basic background in differential geometry. Everything that is needed in this regard (and presumably more) is covered in a self-contained manner in Appendix~\ref{app:diffgeom}.

In order to make the book potentially useful also for more experienced researchers, the content is composed of several layers. In the following Chaps.~\ref{chap:ourfirstmodel} and~\ref{chap:firstmodelgeneralizations}, I introduce the concepts of SSB, NG bosons and the EFT description thereof through the case study of a simple toy model. This is intended for an uninitiated reader. Anybody else can proceed directly to Part~\ref{part:foundations}. Chapters~\ref{chap:SSB} and~\ref{chap:NGbosons} of this part give a thorough introduction to the physics of SSB. This serves as a supplement to advanced courses on field theory and may thus be suitable for graduate students or junior researchers. An expert reader can skip this part as well.

Parts~\ref{part:internalSSB} and~\ref{part:spacetimeSSB} constitute the core of this book that will bring the reader close to the research frontier. Part~\ref{part:internalSSB} is dedicated to spontaneously broken internal symmetries. Here the EFT methodology is by now well-settled, and is developed in detail in Chaps.~\ref{chap:CCWZ} and~\ref{chap:effLagrangian}. Chapters~\ref{chap:internalexamples} and~\ref{chap:scattering} work out some concrete applications of the general formalism. In contrast to internal symmetries, spontaneously broken spacetime symmetries remain a rather active area of research with a number of open questions left to answer. The choice and organization of material in Part~\ref{part:spacetimeSSB} is therefore necessarily more subjective. I however try to draw on the analogy with broken internal symmetries as much as possible. This largely fixes the content of Chap.~\ref{chap:cosetspacetime}, which details the necessary modifications to the techniques of Chap.~\ref{chap:CCWZ}. The following Chaps.~\ref{chap:spacetimequantum} and~\ref{chap:spacetimeclassical} then develop the EFT for spontaneously broken spacetime symmetry using numerous examples, roughly sorted by increasing complexity.

In spite of the rather large extent of the book, I was forced to make compromises regarding the choice of material. Sadly, some very exciting aspects of SSB had to be left out altogether. Covering them at the same level of detail as in the rest of the book would have required a substantial amount of additional background. Some of these topics are mentioned at least briefly in Part~\ref{part:epilogue}.

Let me conclude this introduction with a remark on the chosen style of the book. In writing the text, I have not tried to give a review of all existing research literature on the subject. My intention was instead to create a unified narrative that would give the reader the tools necessary to critically assess existing results and approaches, and to launch a research program of their own. In line with this philosophy, I have deliberately adopted a very restrictive policy regarding bibliography. The purpose of references in this book is primarily to aid the reader. Generally, I thus only include references to point the reader to specified supplementary information, or occasionally to fill in a gap in the argumentation. Only when I explicitly borrow an idea or example from a single identifiable source, do I credit this with a citation. I therefore apologize to all those who might be missing their name in the reference list.


\subsection{Further Reading}
\label{subsec:furtherreading}

This book revolves around the application of EFT to physical systems with SSB. Most of the graduate-level physics background that is needed to follow the advanced parts of the book is covered in Part~\ref{part:foundations}. This however does not mean that the text is comprehensive. Here is a (biased) list of suggestions for readers who wish to learn more details than what this book can cover.

A monograph fully devoted to EFT was missing on the market for years. This gap has now been filled by the new opus by Burgess~\cite{Burgess2021a}. A reader seeking a more concise introduction to EFT will find a variety of excellent resources with slightly different balance of topics included. Thus, for instance, \cite{Kaplan2005a} and~\cite{Manohar2018a} cover a spectrum of applications to particle phenomenology and nuclear physics. Reference~\cite{Gripaios2015a} includes a discussion of physics beyond the Standard Model and of the application of EFT to fluid dynamics, which is hard to find in a student-friendly form elsewhere. Cohen's lectures~\cite{Cohen2019a} are a comprehensive source on EFT with a very useful list of further references including annotations. There are also great texts that mostly focus on EFT for NG bosons. This is true especially of~\cite{Scherer2012a,Pich2018a} and to some extent of~\cite{Penco2020a}.

SSB is rarely a subject of a dedicated monograph. An exception is the classic by Strocchi~\cite{Strocchi2021}, which will satisfy a mathematically oriented reader desiring a rigorous treatment of SSB. The comprehensive early review~\cite{Guralnik1968a} remains a valuable source of insight, including details that seem to have been forgotten by practitioners. The lecture notes~\cite{Beekman2019a} give a modern introduction to SSB that covers recent developments. They omit a discussion of EFT for NG bosons though, and are thus nicely complemented by~\cite{Naegels2021a} which focuses largely on the latter. Finally, the specific topic of classification of NG bosons is addressed in detail by~\cite{Watanabe2020a}.


\bibliographystyle{spphys}
\bibliography{references}
\chapter{Our First Model}
\label{chap:ourfirstmodel}

\begin{fmffile}{feynman02}



\newcommand{\fvertHHH}{\parbox{10mm}{
\begin{fmfgraph}(10,10)
\fmfleftn{l}{1}
\fmfrightn{r}{2}
\fmf{plain}{l1,v}
\fmf{plain}{r1,v,r2}
\fmfdot{v}
\end{fmfgraph}}}

\newcommand{\fvertHpp}{\parbox{10mm}{
\begin{fmfgraph}(10,10)
\fmfleftn{l}{1}
\fmfrightn{r}{2}
\fmf{plain}{l1,v}
\fmf{dashes}{r1,v}
\fmf{dashes}{r2,v}
\fmfdot{v}
\end{fmfgraph}}}

\newcommand{\fvertHff}{\parbox{10mm}{
\begin{fmfgraph}(10,10)
\fmfset{arrow_len}{3mm}
\fmfleftn{l}{1}
\fmfrightn{r}{2}
\fmf{plain}{l1,v}
\fmf{fermion}{r1,v,r2}
\fmfdot{v}
\end{fmfgraph}}}

\newcommand{\fvertpff}{\parbox{10mm}{
\begin{fmfgraph}(10,10)
\fmfset{arrow_len}{3mm}
\fmfleftn{l}{1}
\fmfrightn{r}{2}
\fmf{dashes}{l1,v}
\fmf{fermion}{r1,v,r2}
\fmfdot{v}
\end{fmfgraph}}}

\newcommand{\fvertHHHH}{\parbox{10mm}{
\begin{fmfgraph}(10,10)
\fmfleftn{l}{2}
\fmfrightn{r}{2}
\fmf{plain}{l1,v,l2}
\fmf{plain}{r1,v,r2}
\fmfdot{v}
\end{fmfgraph}}}

\newcommand{\fvertHHpp}{\parbox{10mm}{
\begin{fmfgraph}(10,10)
\fmfleftn{l}{2}
\fmfrightn{r}{2}
\fmf{plain}{l1,v,l2}
\fmf{dashes}{r1,v}
\fmf{dashes}{r2,v}
\fmfdot{v}
\end{fmfgraph}}}

\newcommand{\fvertpppp}{\parbox{10mm}{
\begin{fmfgraph}(10,10)
\fmfleftn{l}{2}
\fmfrightn{r}{2}
\fmf{dashes}{l1,v}
\fmf{dashes}{l2,v}
\fmf{dashes}{r1,v}
\fmf{dashes}{r2,v}
\fmfdot{v}
\end{fmfgraph}}}

\newcommand{\fvertppff}{\parbox{10mm}{
\begin{fmfgraph}(10,10)
\fmfset{arrow_len}{3mm}
\fmfleftn{l}{2}
\fmfrightn{r}{2}
\fmf{dashes}{l1,v,l2}
\fmf{fermion}{r1,v,r2}
\fmfdot{v}
\end{fmfgraph}}}

\newcommand{\fscatHpHp}{\parbox{15mm}{
\begin{fmfgraph}(15,15)
\fmfleftn{l}{2}
\fmfrightn{r}{2}
\fmf{dashes}{l1,v}
\fmf{dashes}{r1,v}
\fmf{plain}{l2,v,r2}
\fmfdot{v}
\end{fmfgraph}}}

\newcommand{\fscatHpHps}{\parbox{20mm}{
\begin{fmfgraph}(20,15)
\fmfleftn{l}{2}
\fmfrightn{r}{2}
\fmf{dashes}{l1,vl,vr}
\fmf{dashes}{r1,vr}
\fmf{plain}{l2,vl}
\fmf{plain}{r2,vr}
\fmfdot{vl,vr}
\end{fmfgraph}}}

\newcommand{\fscatHpHpt}{\parbox{15mm}{
\begin{fmfgraph}(15,15)
\fmfleftn{l}{2}
\fmfrightn{r}{2}
\fmf{plain}{vu,vd}
\fmf{plain}{l2,vu,r2}
\fmf{dashes}{l1,vd}
\fmf{dashes}{r1,vd}
\fmfdot{vu,vd}
\end{fmfgraph}}}

\newcommand{\fscatHpHpu}{\parbox{15mm}{
\begin{fmfgraph}(15,15)
\fmfleftn{l}{2}
\fmfrightn{r}{2}
\fmf{dashes}{vu,vd}
\fmf{plain}{l2,vu}
\fmf{phantom}{vu,r2}
\fmf{dashes}{l1,vd}
\fmf{phantom}{vd,r1}
\fmffreeze
\fmf{dashes}{r1,vu}
\fmf{plain}{vd,r2}
\fmfdot{vu,vd}
\end{fmfgraph}}}

\newcommand{\fscatpppp}{\parbox{15mm}{
\begin{fmfgraph}(15,15)
\fmfleftn{l}{2}
\fmfrightn{r}{2}
\fmf{dashes}{l1,v}
\fmf{dashes}{r1,v}
\fmf{dashes}{l2,v}
\fmf{dashes}{r2,v}
\fmfdot{v}
\end{fmfgraph}}}

\newcommand{\fscatpppps}{\parbox{20mm}{
\begin{fmfgraph}(20,15)
\fmfleftn{l}{2}
\fmfrightn{r}{2}
\fmf{dashes}{l1,vl}
\fmf{dashes}{l2,vl}
\fmf{dashes}{r1,vr}
\fmf{dashes}{r2,vr}
\fmf{plain}{vl,vr}
\fmfdot{vl,vr}
\end{fmfgraph}}}

\newcommand{\fscatppppt}{\parbox{15mm}{
\begin{fmfgraph}(15,15)
\fmfleftn{l}{2}
\fmfrightn{r}{2}
\fmf{dashes}{l1,vd}
\fmf{dashes}{r1,vd}
\fmf{dashes}{l2,vu}
\fmf{dashes}{r2,vu}
\fmf{plain}{vu,vd}
\fmfdot{vu,vd}
\end{fmfgraph}}}

\newcommand{\fscatppppu}{\parbox{15mm}{
\begin{fmfgraph}(15,15)
\fmfleftn{l}{2}
\fmfrightn{r}{2}
\fmf{plain}{vu,vd}
\fmf{dashes}{l2,vu}
\fmf{phantom}{vu,r2}
\fmf{dashes}{l1,vd}
\fmf{phantom}{vd,r1}
\fmffreeze
\fmf{dashes}{r1,vu}
\fmf{dashes}{r2,vd}
\fmfdot{vu,vd}
\end{fmfgraph}}}

\newcommand{\fscatfpfpa}{\parbox{20mm}{
\begin{fmfgraph}(20,15)
\fmfset{arrow_len}{3mm}
\fmfleftn{l}{2}
\fmfrightn{r}{2}
\fmf{fermion}{l2,vl,vr,r2}
\fmf{dashes}{l1,vl}
\fmf{dashes}{r1,vr}
\fmfdot{vl,vr}
\end{fmfgraph}}}

\newcommand{\fscatfpfpb}{\parbox{20mm}{
\begin{fmfgraph}(20,15)
\fmfset{arrow_len}{3mm}
\fmfleftn{l}{2}
\fmfrightn{r}{2}
\fmf{fermion}{l2,vl,vr,r2}
\fmf{phantom}{l1,vl}
\fmf{phantom}{vr,r1}
\fmffreeze
\fmf{dashes}{l1,vr}
\fmf{dashes}{r1,vl}
\fmfdot{vl,vr}
\end{fmfgraph}}}

\newcommand{\fscatfpfpc}{\parbox{15mm}{
\begin{fmfgraph}(15,15)
\fmfset{arrow_len}{3mm}
\fmfleftn{l}{2}
\fmfrightn{r}{2}
\fmf{fermion}{l2,vu,r2}
\fmf{dashes}{l1,vd}
\fmf{dashes}{r1,vd}
\fmf{plain}{vu,vd}
\fmfdot{vu,vd}
\end{fmfgraph}}}

\newcommand{\fscatfpfpd}{\parbox{15mm}{
\begin{fmfgraph}(15,15)
\fmfset{arrow_len}{3mm}
\fmfleftn{l}{2}
\fmfrightn{r}{2}
\fmf{fermion}{l2,v,r2}
\fmf{dashes}{l1,v,r1}
\fmfdot{v}
\end{fmfgraph}}}


\abstract*{The primary target of the book are graduate students; the minimum background required is to have taken a first course on quantum field theory. This chapter aims at readers who have already taken such a course but have not heard about spontaneous symmetry breaking yet. Using a simple toy model, the text explains what spontaneous symmetry breaking is and why it implies the presence of massless particles in the spectrum. Furthermore, it is shown that these Nambu--Goldstone bosons tend to interact weakly at low energies. This simple setting also serves as a primer on some more advanced aspects of field theory. These include the invariance of the $S$-matrix under field redefinitions, and the construction of effective field theory for Nambu--Goldstone bosons.}


The purpose of this chapter is to introduce the uninitiated reader to the key concepts underlying the book. I assume that the reader has taken a first course on quantum field theory, but has not necessarily heard about \emph{spontaneous symmetry breaking} (SSB). The field theory tools we are going to need include identification of a continuous symmetry and the corresponding conserved current via Noether's theorem, extraction of interaction vertices from the Lagrangian and their representation by Feynman diagrams, and their use in a perturbative calculation of scattering amplitudes.

I will start straight away by introducing a simple toy model for a complex scalar field $\phi$ and a Dirac field $\Psi$, defined by the Lagrangian density
\begin{equation}
\begin{split}
\La={}&\de_\m\p^*\de^\m\p+m^2\p^*\p-\l(\p^*\p)^2+(\eps^*\p+\eps\p^*)\\
&+\adj\Psi\I\slashed\de\Psi-g(\adj\Psi_\mathrm{L}\Psi_\mathrm{R}\p+\adj\Psi_\mathrm{R}\Psi_\mathrm{L}\p^*)\;.
\end{split}
\label{toylag}
\end{equation}
Here $m$, $\l$ and $g$ are parameters that are assumed to be real and positive but otherwise arbitrary. The parameter $\eps$ is complex and should be sufficiently small in a sense made more precise below. Finally, the usual slash notation indicates contraction of a Lorentz vector with Dirac $\g$-matrices, thus $\slashed{\de}\equiv\g^\m\de_\m$. I am not going to explain the motivation for the choice of this particular model. Let us rather take this as an invitation to explore with open mind its interesting physical properties.

\begin{watchout}%
I have chosen the relativistic notation just to make the mathematical analysis of the model as simple as possible. It is not essential for qualitative understanding of the results; much of the calculation could be repeated with nonrelativistic (Schr\"odinger) fields. This would of course require some minor modifications of the Lagrangian~\eqref{toylag}. First, the kinetic terms for $\p$ and $\Psi$ would have to be changed. Second, the chiral components $\Psi_\mathrm{L,R}$ of the Dirac field $\Psi$ would have to be replaced with two species of nonrelativistic spin-$1/2$ fermions.
\end{watchout}

Apart from Poincar\'e symmetry, guaranteed by the manifestly relativistic setup, the model~\eqref{toylag} has two natural symmetries,
\begin{equation}
\begin{aligned}
\p&\to\p\;,\quad&\Psi&\to\exp(\I\eps_\mathrm{V})\Psi\qquad&&\text{(exact)}\;,\\
\p&\to\exp(-2\I\eps_\mathrm{A})\p\;,\quad&\Psi&\to\exp(\I\eps_\mathrm{A}\g_5)\Psi\qquad&&\text{(approximate)}\;.
\end{aligned}
\label{vectoraxial}
\end{equation}
In the context of relativistic theories with fermionic degrees of freedom, the transformations with the parameters $\eps_\mathrm{V,A}$ are known respectively as \emph{vector} and \emph{axial}. The axial transformation only becomes a true symmetry of the Lagrangian~\eqref{toylag} in the limit $\eps\to0$. For small nonzero $\eps$, it is therefore sensible to think of it as an approximate symmetry. By means of Noether's theorem, the vector and axial symmetries imply the existence of the following currents and conservation laws,
\begin{equation}
\begin{aligned}
J^\m_\mathrm{V}&=\adj\Psi\g^\m\Psi\;,\qquad&J^\m_\mathrm{A}&=\adj\Psi\g^\m\g_5\Psi-2\I(\p^*\de^\m\p-\de^\m\p^*\p)\;,\\
\de_\m J^\m_\mathrm{V}&=0\;,\qquad&\de_\m J^\m_\mathrm{A}&=2\I(\eps^*\p-\eps\p^*)\;.
\end{aligned}
\label{toyconservation}
\end{equation}
The expression for $\de_\m J^\m_\mathrm{A}$ confirms the observation that the axial symmetry becomes exact in the limit $\eps\to0$.

In the rest of this chapter, we will analyze the physical properties of our toy model. To that end, we will utilize the basic workflow of perturbative quantum field theory, rooted in the theory of oscillations of mechanical systems~\cite{Goldstein2013a}:
\begin{itemize}
\item Find the ground state (Sect.~\ref{sec:firstmodelSSB}). This is the mandatory first step for any quantum system and is carried out by minimizing the (classical) Hamiltonian of the model.
\item Identify the spectrum of excitations above the ground state (Sect.~\ref{sec:firstmodelNGboson}). This is based on the part of the Lagrangian, bilinear in the fluctuations of the fields $\p,\Psi$ around the ground state.
\item Work out the physical consequences of interactions of the excitations (Sect.~\ref{sec:firstmodelNGboson}). This follows from the part of the Lagrangian of higher orders in the fluctuations.
\end{itemize}
Finally, in Sect.~\ref{sec:firstmodelEFT} we will see how the most distinguishing features of the model can be captured by a low-energy \emph{effective field theory} (EFT).


\section{Spontaneous Symmetry Breaking}
\label{sec:firstmodelSSB}

As the first step, we would like to find the (classical) ground state of the model~\eqref{toylag}. To that end, we need the classical Hamiltonian density,
\begin{equation}
\Ha=\de_0\p^*\de_0\p+\vec\nabla\p^*\cdot\vec\nabla\p+V(\p,\p^*)-\adj\Psi\I\skal\g\nabla\Psi+g(\adj\Psi_\mathrm{L}\Psi_\mathrm{R}\p+\adj\Psi_\mathrm{R}\Psi_\mathrm{L}\p^*)\;,
\label{toyham}
\end{equation}
where the scalar potential $V(\p,\p^*)$ is given by
\begin{equation}
V(\p,\p^*)\equiv-m^2\p^*\p+\l(\p^*\p)^2-(\eps^*\p+\eps\p^*)\;.
\label{toypotential}
\end{equation}
In the ground state of any quantum system, the \emph{vacuum expectation value} (VEV) of a fermionic field must be zero. We can therefore focus on the scalar sector of the Hamiltonian~\eqref{toyham}. The part thereof containing derivatives of $\p,\p^*$ is positive-semidefinite. Hence, the lowest-energy state will be such that the VEV of $\p$, $\vev\p$, is a coordinate-independent constant. The value of this constant is determined by minimizing the potential $V(\p,\p^*)$. It is rather obvious that the ground state is nontrivial in that $\vev\p\neq0$. First, for any nonzero $\eps$, the first partial derivatives of $V(\p,\p^*)$ at $\p=\p^*=0$ are nonzero as well. Second, even in the limit $\eps\to0$, the Hessian matrix of $V(\p,\p^*)$ at $\p=\p^*=0$ is negative-definite. In the jargon of high-energy physics, this is a consequence of the ``mass term'' $m^2\p^*\p$ in~\eqref{toylag} having a ``wrong sign.''

What is the ground state then? The VEV $\vev\p$ must be a solution to the condition $0=\Pd{V}{\p^*}=-m^2\p+2\l\p(\p^*\p)-\eps$. It must therefore have the same complex phase as $\eps$, possibly up to an overall minus sign. Let us introduce the notation
\begin{equation}
\eps=\frac{w}{\sqrt2}\E^{\I\t}\;,\qquad
\vev\p=\frac{v}{\sqrt2}\E^{\I\t}\;,
\end{equation}
where $\t$ is the common complex phase and $v,w$ are real. The value of $v$ characterizing the scalar VEV is then related to the parameter $w$ by
\begin{equation}
v(\l v^2-m^2)=w\;.
\label{vw}
\end{equation}
Being a cubic equation, \eqref{vw} has up to three real solutions for $v$. In the limit $w\to0$, these are $v=0,\pm m/\sqrt\l$. Note that for $\eps=0$, the potential $V(\p,\p^*)$ only depends on $\p^*\p$ and thus is insensitive to the phase $\t$. There is then a continuum of states of lowest energy, having $v=\pm m/\sqrt\l$ and arbitrary $\t$. This can be traced to the fact that for $\eps=0$, the axial transformation in~\eqref{vectoraxial}, which changes the phase of $\p$, is an exact symmetry of the Lagrangian~\eqref{toylag}. For any field configuration $\p(x)$, the field $\E^{\I\t}\phi(x)$ with constant $\t$ therefore has the same energy as $\p(x)$ itself. The existence of degenerate ground states is a hallmark of SSB. The defining property of SSB, in the context of our toy model, is that the VEV $\vev\p$ is not invariant under the axial transformation~\eqref{vectoraxial}. The ground state has a lower symmetry than the Lagrangian.

The reader might be concerned that having multiple degenerate ground states could lead to subtleties. (It does.) In any case, this was not anticipated by our algorithmic workflow outlined below~\eqref{toyconservation}. A simple workaround is to keep nonzero $\eps$, which ensures the existence of a unique state of lowest energy. For positive $w$, this is the solution of~\eqref{vw} with the highest value of $v$. Using the closed formula for the solutions of a cubic equation would be impractical. It is however easy to find an approximate expression for $v$ as long as $w$ is small enough. One just needs to rewrite~\eqref{vw} in a form suitable for iteration,
\begin{equation}
v=\frac{m}{\sqrt\l}\sqrt{1+\frac{w}{m^2v}}=\frac{m}{\sqrt\l}\left(1+\frac{\sqrt\l w}{2m^3}+\dotsb\right)=\frac{m}{\sqrt\l}+\frac{w}{2m^2}+\dotsb\;.
\label{vexpansion}
\end{equation}
We end up with an infinite series with the expansion parameter $\sqrt\l w/m^3$. This gives us a precise mathematical condition for $w$ being ``small enough,'' $w\ll{m^3}/{\sqrt\l}$. The condition guarantees that the value of $v$ in the ground state differs very little from $m/\sqrt\l$. In other words, $\eps$ now acts merely as a perturbation that enables us to pin down a unique ground state, without modifying this state appreciably. In the following, I will implicitly assume that the condition $w\ll{m^3}/{\sqrt\l}$ is satisfied.


\section{Nambu--Goldstone Boson and Its Interactions}
\label{sec:firstmodelNGboson}

The next step is to identify the spectrum of excitations above our ground state, and to work out the consequences of their interactions. To that end, we need to find field variables with vanishing VEV that make the bilinear part of the Lagrangian diagonal. Such variables correspond to the \emph{normal modes}, well-known from the classical theory of small oscillations~\cite{Goldstein2013a}. Upon expansion in powers of the new fields, the bilinear part of the Lagrangian determines the spectrum of normal modes, and the higher-order parts their interactions.

The change of variables to the normal modes can be thought of as a choice of parameterization of $\p$. (The Dirac field $\Psi$ already has a vanishing VEV, so we do not have to do anything about it.) There is no a priori preferred choice of parameterization. Below, I will illustrate two different options, one of which is intuitively natural, whereas the other is more physical and practically convenient.


\subsection{Linear Parameterization}
\label{subsec:firstmodellinear}

One obvious possibility how to parameterize $\phi$ is to shift it by its VEV, and represent the resulting complex field in terms of its real and imaginary parts. This is still not completely unambiguous, but a smart choice is for instance\footnote{The somewhat unusual notation is required for compatibility with the later discussion in Chap.~\ref{chap:CCWZ}.}
\begin{equation}
\p(x)=\frac{\E^{\I\t}}{\sqrt2}[v+\mf(x)+\I\pi(x)]\;.
\label{toyparamlinear}
\end{equation}
Taking out an overall factor $\E^{\I\t}$ ensures that the phase $\t$ drops out of the Lagrangian. Inserting this parameterization into~\eqref{toylag}, we get a Lagrangian in terms of $\mf,\pi,\Psi$ that can be organized by the dependence on the scalar fields,
\begin{equation}
\La=\La_\mathrm{vac}+\La_\mathrm{bilin}+\La_\mathrm{int}+\La_\Psi\;.
\label{toylagexpansion}
\end{equation}
The piece independent of the fields, $\La_\mathrm{vac}=(3/4)\l v^4-(1/2)m^2v^2$, determines up to a minus sign the energy density of the ground state. There are no terms linear in $\mf,\pi$; this is ensured by their definition~\eqref{toyparamlinear} through removing from $\p$ its VEV. The bilinear part of the Lagrangian is of greatest interest to us,
\begin{equation}
\La_\mathrm{bilin}=\frac12(\de_\m\mf)^2-\frac12m_\mf^2\mf^2+\frac12(\de_\m\pi)^2-\frac12m_\pi^2\pi^2\;,
\end{equation}
where the mass parameters are given by
\begin{equation}
m_\mf^2=2m^2+\frac{3w}v\;,\qquad
m_\pi^2=\frac wv\;.
\label{vwmass}
\end{equation}
An obvious corollary is that in the limit of exact axial symmetry ($\eps=0$), the field $\pi$ becomes massless. This is our first \emph{Nambu--Goldstone} (NG) \emph{boson}. It is easy to see the origin of the massless mode in the spectrum. With a canonically normalized kinetic term for $\p$ in the Lagrangian, the mass spectrum is determined by the eigenvalues of the Hessian matrix of the potential $V(\p,\p^*)$ in the ground state. But I have already argued that for any $\p(x)$, changing its overall phase will give a configuration of the same energy. Hence for any $\p\neq0$, there is a direction in the field space in which the potential does not change. This guarantees that the Hessian has an eigenvector with zero eigenvalue. The existence of the NG boson is therefore a direct consequence of the axial symmetry and its spontaneous breakdown, independent of the specific Lagrangian. This is the essence of \emph{Goldstone's theorem}~\cite{Goldstone1961a,Goldstone1962a}.

Equation~\eqref{vwmass} likewise tells us that in the limit of exact axial symmetry, that is $\eps=0$, the mass of $\mf$ approaches $m\sqrt2$. The massive counterpart of the NG boson is usually referred to as the \emph{Higgs boson} (or \emph{Higgs mode}). The interaction part of the scalar Lagrangian in~\eqref{toylagexpansion} consists of cubic and quartic terms,
\begin{equation}
\La_\mathrm{int}=-\l v\mf(\mf^2+\pi^2)-\frac\l4(\mf^2+\pi^2)^2\;.
\label{toyintlinear}
\end{equation}
Finally, the fermion part of the Lagrangian reads
\begin{equation}
\La_\Psi=\adj\hf(\I\slashed\de-m_\hf)\hf-\frac g{\sqrt2}\adj\hf(\mf+\I\g_5\pi)\hf\;,
\label{toyfermilinear}
\end{equation}
where I have redefined the Dirac field by $\Psi=\exp\left[-(\I/2)\t\g_5\right]\hf$ in order to bring the kinetic term into the standard Dirac form. We can see that the originally massless fermion has acquired mass, $m_\hf=gv/\sqrt2$. This is an example of generation of fermion mass by spontaneous breaking of chiral symmetry, which plays an important role in the Standard Model of particle physics.


\subsection{Scattering of Nambu--Goldstone Bosons}
\label{subsec:firstmodelscattering}

Now that we know the mass spectrum of our model, let us see how the different particles interact with each other. We are going to need Feynman rules for the interaction vertices, encoded in the Lagrangians~\eqref{toyintlinear} and~\eqref{toyfermilinear},
\begin{align}
\notag
\fvertHHH&=-6\I\l v\;,\quad
&\fvertHpp&=-2\I\l v\;,\quad
&\fvertHff&=-\frac{\I g}{\sqrt2}\;,\quad
&\fvertpff&=\frac{g\g_5}{\sqrt2}\;,\\
\fvertHHHH&=-6\I\l\;,\quad
&\fvertHHpp&=-2\I\l\;,\quad
&\fvertpppp&=-6\I\l\;.
\label{toyFruleslinear}
\end{align}
The undirected solid and dashed lines represent respectively $\mf$ and $\pi$. The oriented solid lines stand for the fermion $\hf$. These Feynman rules for the interaction vertices must be augmented with standard relativistic propagators for $\mf,\pi,\hf$ with respective masses $m_\mf,m_\pi,m_\hf$.

\begin{figure}[t]
\sidecaption[t]
\includegraphics[width=2.0in]{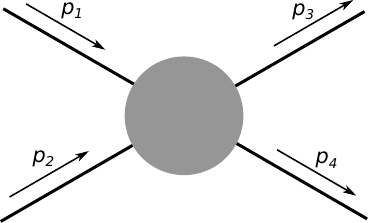}
\caption{Kinematics of the four-particle scattering processes discussed in this chapter. The arrows indicate flow of energy--momentum. In other words, the initial state includes two particles with momenta $\vec p_1,\vec p_2$, whereas the final state two particles with momenta $\vec p_3,\vec p_4$}
\label{fig:scattering}
\end{figure}

I will work out a few sample scattering processes including the NG boson $\pi$. All of these processes will have the simple four-particle kinematics displayed in Fig.~\ref{fig:scattering}. In all cases, I will use the same notation $p_1,p_2$ for the energy--momentum of the incoming particles and $p_3,p_4$ for the energy--momentum of the outgoing particles. For the sake of brevity, I will also use the Lorentz-invariant \emph{Mandelstam variables} $s\equiv(p_1+p_2)^2$, $t\equiv(p_1-p_3)^2$ and $u\equiv(p_1-p_4)^2$. As a consequence of energy and momentum conservation, the Mandelstam variables satisfy the constraint
$s+t+u=m_1^2+m_2^2+m_3^2+m_4^2$, where $m_1,m_2,m_3,m_4$ are the masses of the particles participating in the scattering process.

Let us first inspect $\pi\pi\to\pi\pi$ scattering. The invariant amplitude for this process is given in terms of Feynman diagrams by
\begin{equation}
-\I\Aa_{\pi\pi\to\pi\pi}=\fscatpppp+\fscatpppps+\fscatppppt+\fscatppppu\;.
\label{toyApipipipigraphs}
\end{equation}
Using the Feynman rules~\eqref{toyFruleslinear}, this evaluates to
\begin{equation}
\Aa_{\pi\pi\to\pi\pi}=6\l+4\l^2 v^2\biggl(\frac1{s-m_\mf^2}+\frac1{t-m_\mf^2}+\frac1{u-m_\mf^2}\biggr)\;.
\label{toyApipipipiaux}
\end{equation}
Of particular interest is the behavior of this amplitude at low energies. Given that the NG boson $\pi$ becomes massless in the limit $\eps\to0$, we can certainly find a kinematical regime such that $s,t,u\ll m_\mf^2$ for sufficiently small $w$. It then makes sense to expand the amplitude in powers of the Mandelstam variables. Upon a bit of manipulation using~\eqref{vw} and \eqref{vwmass}, one finds
\begin{equation}
\Aa_{\pi\pi\to\pi\pi}=-\frac{4\l w}{m_\mf^4v}\left(m^2-\frac{w}{2v}\right)-\frac{4\l^2v^2}{m_\mf^6}(s^2+t^2+u^2)+\bigO(s^3,t^3,u^3)\;.
\label{toyApipipipi}
\end{equation}
Interestingly, the leading term vanishes in the limit $\eps\to0$. But that is not all. In the same limit, we find that $s=2p_1\cdot p_2=2p_3\cdot p_4$, $t=-2p_1\cdot p_3=-2p_2\cdot p_4$ and $u=-2p_1\cdot p_4=-2p_2\cdot p_3$. If we now take the additional limit in which the momentum of any of the four particles goes to zero, then all the Mandelstam variables, and thus the amplitude $\Aa_{\pi\pi\to\pi\pi}$, will vanish. This property is concealed in~\eqref{toyApipipipiaux}, where all the four individual contributions coming from the diagrams in~\eqref{toyApipipipigraphs} have a nonzero limit. The algebraic cancellation leading to the eventual vanishing of the amplitude requires an interplay of the cubic and quartic interactions in the Lagrangian. This seems to be too much of a coincidence. In fact, it is our first example of the \emph{Adler zero} principle: scattering amplitudes of a NG boson vanish in the limit that its momentum goes to zero. In plain terms, low-energy NG bosons interact weakly.

I have just made a much stronger statement than what the single amplitude worked out so far would seem to justify. We need more examples to check whether the Adler zero principle actually holds. To keep things simple, I will from now on take the limit of exact axial symmetry, $\eps=0$, in which the NG boson $\pi$ is exactly massless. The next example we will look at is $\mf\pi\to\mf\pi$ scattering, with the amplitude
\begin{equation}
-\I\Aa_{\mf\pi\to\mf\pi}=\fscatHpHp+\fscatHpHps+\fscatHpHpt+\fscatHpHpu\;.
\label{toyApsipipsipigraphs}
\end{equation}
Here a straightforward application of the Feynman rules gives
\begin{equation}
\Aa_{\mf\pi\to\mf\pi}=2\l+4\l^2v^2\biggl(\frac1s+\frac3{t-m_\mf^2}+\frac1u\biggr)\;.
\label{toyApsipipsipi}
\end{equation}
Suppose that we again take the momentum of one of the NG bosons, $p_2$ or $p_4$, to zero. In this limit, $\smash{s,u\to m_\mf^2=2m^2}$ whereas $t\to0$. Using finally the fact that $v=m/\sqrt\l$, it is easy to see that the whole amplitude $\Aa_{\mf\pi\to\mf\pi}$ vanishes.

We are starting to see a pattern. We have now checked two different scattering processes, which together involve all the scalar interaction vertices in~\eqref{toyFruleslinear} except for the $\mf^4$ one. The Adler zero principle seems to hold, but it emerges out of a nontrivial interplay of different interaction terms and a cancellation among different Feynman diagrams.

As the last example, consider the scattering of a NG boson off a fermion, $\hf\pi\to\hf\pi$. The invariant amplitude for this process is
\begin{equation}
-\I\Aa_{\hf\pi\to\hf\pi}=\fscatfpfpa+\fscatfpfpb+\fscatfpfpc\;.
\label{toyAfpifpigraphs}
\end{equation}
The Feynman rules~\eqref{toyFruleslinear} now give us
\begin{equation}
\begin{split}
\Aa_{\hf\pi\to\hf\pi}={}&\frac{g^2}2\adj u(p_3)\biggl(\frac1{\slashed p_1+\slashed p_2+m_\hf}+\frac1{\slashed p_3-\slashed p_2+m_\hf}\biggr)u(p_1)\\
&+\frac{\sqrt2\l gv}{t-m_\mf^2}\adj u(p_3)u(p_1)\;,
\end{split}
\label{toyAfpifpiaux}
\end{equation}
where $u(p_1)$ and $u(p_3)$ are the usual plane-wave Dirac spinors with polarization indices suppressed for brevity. In order to further simplify the expression for the amplitude, we invert the Dirac propagators and use the Dirac equation for the Dirac spinors, $(\slashed p_1-m_\hf)u(p_1)=\adj u(p_3)(\slashed p_3-m_\hf)=0$. Using finally the fact that $\smash{s-m_\hf^2=2p_1\cdot p_2}$ and $\smash{u-m_\hf^2=-2p_2\cdot p_3}$, the amplitude~\eqref{toyAfpifpiaux} can be cast as
\begin{equation}
\Aa_{\hf\pi\to\hf\pi}=\frac{g^2}4\adj u(p_3)\biggl(\frac{\slashed p_2}{p_1\cdot p_2}+\frac{\slashed p_2}{p_2\cdot p_3}\biggr)u(p_1)+\frac{\sqrt2\l gv}{t-m_\mf^2}\adj u(p_3)u(p_1)\;.
\label{toyAfpifpi}
\end{equation}
This form is suitable for checking the limit $p_2\to0$. Should one prefer to take the momentum of the other NG boson, $p_4$, to zero, it is possible to replace the expression in the large parentheses with $\slashed p_4/(p_3\cdot p_4)+\slashed p_4/(p_1\cdot p_4)$. Either way, the amplitude does not go to zero in the limit of vanishing momentum of the NG boson, in contrast to the previously discussed scalar amplitudes. The problem lies in the first term in~\eqref{toyAfpifpi}, which does not even have a well-defined limit when $p_2$ or $p_4$ is taken to zero. That is because the fermion propagators become singular in this limit. I will only be able to offer a proper discussion of this issue in Chap.~\ref{chap:scattering}. For the moment, it suffices to keep in mind that the Adler zero principle has exceptions.


\subsection{Nonlinear Parameterization}
\label{subsec:firstmodelnonlinear}

The whole preceding discussion of scattering amplitudes in our toy model was based on the linear parameterization~\eqref{toyparamlinear}. However, as already pointed out, this is not the only choice. Let us try something else. In elementary complex calculus, one learns about two different representations of complex numbers: the linear (``Cartesian'') and exponential (``polar''). I will thus replace~\eqref{toyparamlinear} with
\begin{equation}
\p(x)=\frac{\E^{\I[\t+\pi(x)/v]}}{\sqrt2}[v+\mf(x)]\;.
\label{toyparamexp}
\end{equation}
With this parameterization, it is natural to think of $\mf$ as the fluctuation of the magnitude of $\p$, and of $\pi$ as the fluctuation of its phase.

We can now follow the same steps as in Sect.~\ref{subsec:firstmodellinear}. Using the partition~\eqref{toylagexpansion} of the Lagrangian, we find that the constant part $\La_\mathrm{vac}$ and the bilinear part $\La_\mathrm{bilin}$ remain unchanged. This follows from the fact that we are expanding around the same ground state as before, and that the parameterizations~\eqref{toyparamlinear} and~\eqref{toyparamexp} agree to first order in $\mf,\pi$. The scalar interaction Lagrangian is now different though,\footnote{Remember that I have already set $\eps=0$. Otherwise, the interaction Lagrangian would include additional terms proportional to $w$. This is in contrast to~\eqref{toyintlinear}, which is correct for any $\eps\in\C$.}
\begin{equation}
\La_\mathrm{int}=-\l v\mf^3-\frac\l4\mf^4+\biggl(\frac\mf v+\frac{\mf^2}{2v^2}\biggr)(\de_\m\pi)^2\;.
\label{toyintexp}
\end{equation}
Likewise, the fermionic part of the Lagrangian~\eqref{toyfermilinear} turns into
\begin{equation}
\La_\Psi=\adj\hf(\I\slashed\de-m_\hf)\hf-\frac g{\sqrt2}\mf\adj\hf\hf+\frac{\de_\m\pi}{2v}\adj\hf\g^\m\g_5\hf\;,
\label{toyfermiexp}
\end{equation}
where I have redefined the fermion field as $\Psi=\exp[-(\I/2)(\t+\pi/v)\g_5]\hf$.

Note that the NG field $\pi$ now enters the Lagrangian only through its derivatives. This remarkable feature is a direct consequence of the exponential parameterization~\eqref{toyparamexp}. Namely, since $\pi(x)$ appears therein together with the constant phase $\t$, any constant field $\pi$ must drop out of the Lagrangian as a consequence of the axial symmetry. Another way to look at this is that the axial transformation~\eqref{vectoraxial} acts on the fields $\mf,\pi,\hf$ via
\begin{equation}
\mf\to\mf\;,\qquad
\pi\to\pi-2v\eps_\mathrm{A}\;,\qquad
\hf\to\hf\;.
\label{axialtransfo}
\end{equation}
While $\mf$ and $\hf$ are left intact by the axial transformation, $\pi$ is shifted by a constant. Hence, there is no other way to construct a Lagrangian invariant under the axial symmetry than to forbid operators containing $\pi$ without any derivatives. This is typical for NG bosons, and it gives us insight into the origin of the Adler zero principle. If the NG field only interacts via its derivatives, it should not be surprising that its interactions are suppressed at low energies.

Let us check how this works explicitly. We will need the Feynman rules for the interaction vertices contained in~\eqref{toyintexp} and~\eqref{toyfermiexp},
\begin{equation}
\begin{alignedat}{3}
\fvertHHH&=-6\I\l v\;,\qquad
&\fvertHpp&=-\frac{2\I}vp\cdot q\;,\qquad
&\fvertHff&=-\frac{\I g}{\sqrt2}\;,\\
\fvertpff&=\frac{1}{2v}\slashed p\g_5\;,\qquad
&\fvertHHHH&=-6\I\l\;,\qquad
&\fvertHHpp&=-\frac{2\I}{v^2}p\cdot q\;.
\end{alignedat}
\label{toyFrulesexp}
\end{equation}
The notation is such that $p,q$ are energy--momenta carried by the NG legs, oriented towards the vertex. A comparison with~\eqref{toyFruleslinear} shows that, unsurprisingly, the interaction vertices involving only $\mf$ or $\hf$ remain unchanged. The vertices of $\pi$ now depend on energy--momentum though.

Since there is no $\pi^4$ vertex, the $\pi\pi\to\pi\pi$ amplitude is now given by the last three diagrams in~\eqref{toyApipipipigraphs}. A brief calculation leads to
\begin{equation}
\Aa_{\pi\pi\to\pi\pi}=\frac1{v^2}\biggl(\frac{s^2}{s-m_\mf^2}+\frac{t^2}{t-m_\mf^2}+\frac{u^2}{u-m_\mf^2}\biggr)\;,
\label{toyscataux}
\end{equation}
which is easily seen to be equivalent to our previous result~\eqref{toyApipipipiaux} in the limit $\eps\to0$. Note, however, that in the exponential parameterization~\eqref{toyparamexp}, the Adler zero property is manifest. Each of the three contributions to~\eqref{toyscataux} vanishes separately when the momentum of any of the four particles is taken to zero. There is no need for cancellation between different Feynman diagrams. This is a direct consequence of the fact that the NG field $\pi$ is now derivatively coupled. An important lesson of this exercise is that while the scattering amplitude can be calculated in any parameterization of $\phi$, not all parameterizations are equal. Indeed, some make the physical properties of the amplitude manifest.

\begin{watchout}%
I have glossed over the fact that the scattering amplitude $\Aa_{\pi\pi\to\pi\pi}$, and in fact the whole $S$-matrix, is independent of the choice of parameterization of $\phi$ around its VEV. This may seem intuitively obvious. After all, it is common lore that physical predictions of a theory should not depend on arbitrary choices such as a reference frame or a coordinate system. Yet, the mathematical proof of such \emph{reparameterization invariance} of the $S$-matrix is nontrivial~\cite{Chisholm1961a,Kamefuchi1961a}. Moreover, the independence on the choice of parameterization really only applies to physical observables. Quantities that are in principle unobservable, such as off-shell Green's functions of fields, may differ in different parameterizations. Finally, verifying the reparameterization invariance of the $S$-matrix explicitly beyond the tree-level (classical) approximation may require taking carefully account of the Jacobian of the functional integral measure~\cite{Honerkamp1971}.
\end{watchout}

The next amplitude to check is that for the $\mf\pi\to\mf\pi$ process. In this case, the same four diagrams as in~\eqref{toyApsipipsipigraphs} contribute, although the individual graphs of course take different values in the two parameterizations~\eqref{toyparamlinear} and~\eqref{toyparamexp}. A straightforward application of the Feynman rules~\eqref{toyFrulesexp} leads to
\begin{equation}
\Aa_{\mf\pi\to\mf\pi}=\frac1{v^2}\biggl[t+\frac{(s-m_\mf^2)^2}s+\frac{3m_\mf^2t}{t-m_\mf^2}+\frac{(u-m_\mf^2)^2}u\biggr]\;,
\label{toyscatexp}
\end{equation}
which is equivalent to the previous result~\eqref{toyApsipipsipi}. In its present form, the amplitude however vanishes manifestly, diagram by diagram, in the limit where the momentum of one of the NG bosons is taken to zero.

The last amplitude we are interested in is for $\hf\pi\to\hf\pi$. Again the same diagrams as in~\eqref{toyAfpifpigraphs} contribute, but again they take individually different values than before due to the different Feynman rules~\eqref{toyFrulesexp}. The result can be written in a number of different forms; one that is not too far from a direct application of Feynman rules is
\begin{equation}
\begin{split}
\Aa_{\hf\pi\to\hf\pi}={}&\frac1{8v^2}\adj u(p_3)\biggl[\frac{\slashed p_4(\slashed p_1-m_\hf)\slashed p_2}{p_1\cdot p_2}-\frac{\slashed p_2(\slashed p_1-m_\hf)\slashed p_4}{p_2\cdot p_3}\biggr]u(p_1)\\
&+\frac{g}{\sqrt2\,v}\frac{t}{t-m_\mf^2}\adj u(p_3)u(p_1)\;.
\end{split}
\end{equation}
This time it takes more effort to prove equivalence with our previous result~\eqref{toyAfpifpi}. I will spare the reader of the details, making use of the Dirac equation and the properties of the Dirac $\g$-matrices. What matters is that the contribution of every single diagram now carries a factor of $p_2$ and $p_4$ in the numerator, coming from the derivative couplings of the NG field $\pi$. Yet, this is not sufficient to make the amplitude vanish if either $p_2$ or $p_4$ goes to zero.


\section{Low-Energy Effective Field Theory}
\label{sec:firstmodelEFT}

The spectrum of our model~\eqref{toylag} contains the massless NG boson $\pi$, the Higgs mode $\mf$ with mass $m_\mf=m\sqrt2$, and the Dirac fermion $\hf$ with mass $m_\hf=gv/\sqrt2=mg/\sqrt{2\l}$. At energies well below the mass scales of $m_\mf$ and $m_\hf$, these massive modes will not be excited. As a consequence, the physics of the toy model will reduce to that of the NG boson $\pi$. Yet, in both parameterizations~\eqref{toyparamlinear} and~\eqref{toyparamexp}, the Higgs field is needed even at low energies since it mediates interactions between NG bosons. In fact, in the exponential parameterization~\eqref{toyparamexp}, the Lagrangian does not contain any direct self-interaction of $\pi$ at all. This sounds like an overkill. We should not need any other fields just to describe self-interactions of the NG bosons. In the spirit of EFT, introduced in Chap.~\ref{chap:intro}, we should be able to describe the low-energy physics of NG bosons using the $\pi$ field alone.

Such a low-energy EFT should respect the symmetries of the underlying theory defined by~\eqref{toylag}. With the transformation rules~\eqref{axialtransfo}, this does not place any constraints on $\mf,\hf$, which can thus be safely dropped. The shift transformation of $\pi$, on the other hand, forbids operators without derivatives on $\pi$. It follows that the effective Lagrangian for the NG boson must be described by some, a priori unknown, function of derivatives of $\pi$.

We also observed above that derivatives in the interaction vertices suppress their contribution to the $S$-matrix at low energies. More derivatives imply stronger suppression. This suggests an organization principle, whereby operators (with the same number of $\pi$ fields) are hierarchically ordered according to the number of derivatives they contain. The dominant contributions to the $S$-matrix will come from operators with the fewest derivatives possible. Since each field $\pi$ must still carry at least one derivative, the EFT will be dominated by interaction operators with \emph{exactly} one derivative per each $\pi$. Lorentz invariance then constrains the effective Lagrangian to
\begin{equation}
\La_\mathrm{eff}=\frac12(\de_\m\pi)^2+\sum_{n=2}^\infty c_{2n}[(\de_\m\pi)^2]^n+\dotsb\;,
\label{toyEFT}
\end{equation}
where the ellipsis stands for operators with more than one derivative per $\pi$.

The as yet undetermined couplings $c_{2n}$ govern the low-energy properties of scattering amplitudes of NG bosons. It remains to find out what these couplings are. In a concrete physical system, this could be done by performing a set of scattering experiments. It is however also possible to fix the values of $c_{2n}$ theoretically from the underlying model~\eqref{toylag}. I will outline two strategies for doing so.


\subsection{Matching}
\label{subsec:firstmodelmatching}

One possibility is to evaluate a set of scattering amplitudes, or other observables, in both the underlying model~\eqref{toylag} and its low-energy EFT~\eqref{toyEFT}. Comparing the predictions allows one to fix $c_{2n}$ in terms of the parameters $m,\l,g$. In the jargon of EFT, this is called \emph{matching}.

Let us see how it works in practice on the example of the $\pi\pi\to\pi\pi$ scattering. Within the EFT, this is described by the $c_4$ coupling. There is a single Feynman diagram, namely the first diagram in~\eqref{toyApipipipigraphs}. The corresponding invariant amplitude reads, in terms of the Mandelstam variables, $\Aa_{\pi\pi\to\pi\pi}=-2c_4(s^2+t^2+u^2)$. Note how the dependence on the particle momenta exactly copies the leading contribution to the previously calculated amplitude~\eqref{toyApipipipi} in the limit of exact axial symmetry. Upon comparing the coefficients of the kinematical invariant $s^2+t^2+u^2$, we find
\begin{equation}
c_4=\frac{\l}{4m^4}\;.
\label{c4matching}
\end{equation}

The same procedure could in principle be followed to determine $c_6$, $c_8$ and so on. That would however be a tedious task, since the number of Feynman diagrams contributing to an $n$-particle process grows rapidly with $n$. Sometimes, an alternative approach is feasible whereby the EFT is deduced from the underlying theory directly at the level of the Lagrangian. I will now demonstrate how to do this in the case of our toy model.

\begin{watchout}%
A cautious reader might be wondering why I have made no mention of Feynman diagrams containing loops in all the discussion above. Within the toy model~\eqref{toylag}, restricting to tree level amounts to considering the leading contribution to scattering amplitudes in a power expansion in the small couplings $\l,g$. The result~\eqref{c4matching} should therefore likewise be interpreted as the leading contribution to $c_4$, induced by the presence of the heavy modes $\mf$ and $\hf$. See Sect.~2.3 of~\cite{Burgess2021a} for more details.
\end{watchout}


\subsection{Eliminating the Heavy Modes}
\label{subsec:firstmodelintegratingout}

In the exponential parameterization~\eqref{toyparamexp}, all tree-level scattering amplitudes of NG bosons in the model~\eqref{toylag} arise from Feynman diagrams that include virtual $\mf$ modes in the propagators. Recall now that quantum field theory at tree level is equivalent to its classical limit. Including the effect of virtual $\mf$ quanta is then equivalent to solving the classical \emph{equation of motion} (EoM) for $\mf$ and inserting the result back to the Lagrangian. In this way, one can obtain an EFT for $\pi$ alone that takes into account all the interactions of the original model~\eqref{toylag}.

Let us see this program through to its end. We start by dropping the fermion field and putting the scalar interactions~\eqref{toyintexp} together with the kinetic term into a complete scalar Lagrangian,
\begin{equation}
\La=\frac12(\de_\m\mf)^2-m^2\mf^2-\l v\mf^3-\frac\l4\mf^4+\frac12\left(1+\frac\mf v\right)^2(\de_\m\pi)^2\;.
\label{toylagscalar}
\end{equation}
The parameters $m,\l,v$ are related by $v=m/\sqrt\l$. The EoM for $\mf$, which we would like to solve for $\mf$ in terms of $\pi$, reads
\begin{equation}
\Box\mf+2m^2\mf+3\l v\mf^2+\l\mf^3-\frac1v\left(1+\frac\mf v\right)(\de_\m\pi)^2=0\;.
\label{toyEoM}
\end{equation}
It is of course not possible to solve this equation in a closed form, we can however still extract useful information from it. At low energies, we expect $\mf$ to be very small. (The probability to excite virtual $\mf$ quanta far away from their mass shell is tiny.) In the first approximation, the terms in~\eqref{toyEoM} quadratic and cubic in $\mf$ can therefore be neglected. We can likewise drop the $\Box\mf$ term, suppressed by two derivatives. Equation~\eqref{toyEoM} then becomes a linear algebraic equation for $\mf$ with the solution $\mf\approx(\de_\m\pi)^2/(2m^2v)$. In order to go beyond this approximation, we rewrite~\eqref{toyEoM} in a form suitable for iteration,
\begin{equation}
\mf=\frac1v\left[2m^2+\Box-\frac1{v^2}(\de_\m\pi)^2+3\l v\mf+\l\mf^2\right]^{-1}(\de_\m\pi)^2\;.
\label{toyEoMiterative}
\end{equation}
All terms but $2m^2$ in the square brackets are small due to containing either derivatives or extra factors of $\mf$. This makes it possible to iteratively expand~\eqref{toyEoMiterative} in inverse powers of $m^2$ up to any desired order in $\pi$ and its derivatives. Up to and including the first subleading contribution to $\mf$, we thus get
\begin{equation}
\mf=\frac{(\de_\m\pi)^2}{2m^2v}-\frac{[(\de_\m\pi)^2]^2}{8m^4v^3}+\dotsb\;.
\label{psiiterate}
\end{equation}
When inserted back into~\eqref{toylagscalar}, this is enough to generate the couplings $c_4$ and $c_6$ in the Lagrangian~\eqref{toyEFT}. When the dust settles, we find $c_4=\l/(4m^4)$ and $c_6=0$. The result for $c_4$ agrees with our previous calculation relying on direct matching of scattering amplitudes. It may however come as a surprise that $c_6$ vanishes.

To get further insight, we multiply the EoM~\eqref{toyEoM} by $\mf/2$ and add to the Lagrangian~\eqref{toylagscalar}, giving
\begin{equation}
\La\simeq\frac{\l v}2\mf^3+\frac\l4\mf^4+\frac12\left(1+\frac\mf v\right)(\de_\m\pi)^2\;.
\label{toylagaux}
\end{equation}
The symbol $\simeq$ indicates that I have dropped a surface term. The obtained Lagrangian can be further manipulated by rewriting the EoM~\eqref{toyEoM} as
\begin{equation}
\l\mf(\mf+2v)=\frac{(\de_\m\pi)^2}{v^2}-\frac{\Box\mf}{\mf+v}\;.
\end{equation}
Using this twice in sequence, it is possible to remove from~\eqref{toylagaux} all terms that only contain $\mf$ without any derivatives. One thus arrives at an equivalent Lagrangian,
\begin{equation}
\La\simeq\frac12(\de_\m\pi)^2+\frac\l{4m^4}[(\de_\m\pi)^2]^2-\biggl[\frac{(\de_\m\pi)^2}{m^2}+\mf^2\biggr]\frac{\Box\mf}{4(\mf+v)}\;.
\label{toylagaux2}
\end{equation}
The first two terms here are just the kinetic term for $\pi$ and the $c_4$ coupling. The last term, upon expansion in $\pi$ using~\eqref{toyEoMiterative}, contains only operators with more than one derivative per $\pi$. This allows one to make the strong conclusion that \emph{all} the couplings in the effective Lagrangian~\eqref{toyEFT} except for $c_4$ vanish.


\section{Moral Lessons}
\label{sec:firstmodelmorals1}

In this introductory chapter, we have analyzed a simple toy model, introducing along the way some concepts that will play a key role throughout the rest of the book. Let me briefly summarize what we have found, and draw a few morals.

\runinhead{Lesson \#1} Some physical systems have multiple degenerate ground states. Barring accidental degeneracy, this happens typically when the system possesses a symmetry under which the ground states are not invariant. This is the defining property of SSB.

\runinhead{Lesson \#2} When the spontaneously broken symmetry is continuous, the spectrum of the system contains a massless particle: the NG boson. This is Goldstone's theorem.

\runinhead{Lesson \#3} The scattering amplitude for a process involving a NG boson vanishes in the limit where its momentum is taken to zero. That is, NG bosons interact weakly at low energy. This is the Adler zero principle. The principle may be violated in case taking the NG boson momentum to zero brings some of the virtual particles in the process on the mass shell.

\runinhead{Lesson \#4} It is convenient to choose a nonlinear field parameterization in which a constant NG field can be eliminated by a symmetry transformation. In such a parameterization, the NG field is derivatively coupled, making its physical properties of zero mass and weak interactions at low energy manifest.

\runinhead{Lesson \#5} The nonlinear field parameterization allows one to construct a low-energy EFT in terms of the NG field alone. This captures the physics of NG bosons at energies well below the mass scale of other, massive particles present in the system.\\[-1ex]


\end{fmffile}

\bibliographystyle{spphys}
\bibliography{references}
\chapter{Generalizations of the Model}
\label{chap:firstmodelgeneralizations}

\abstract*{This chapter further develops the concepts introduced in the previous chapter at a level suitable for a reader without prior knowledge of spontaneous symmetry breaking. The material covered includes physical systems with multiple Nambu--Goldstone bosons and systems lacking Lorentz invariance. The focus is on understanding the connection between the pattern of symmetry breaking, the number of Nambu--Goldstone bosons, and their dispersion relations. It is shown how this connection can be conveniently addressed using a low-energy effective field theory. Altogether, this chapter makes the reader ready for the general, state-of-the-art discussion of spontaneous symmetry breaking, Nambu--Goldstone bosons and their effective field theory description, which follows later in the book.}


In the previous chapter, I worked out in detail a model of a single complex scalar field, in which a continuous symmetry is spontaneously broken. The corresponding axial transformations span the Abelian group $\gr{U}(1)$. Due to its simplicity, such a model is suitable for a first encounter with \emph{spontaneous symmetry breaking} (SSB). However, it falls short of exhibiting the full spectrum of possible realizations of SSB. In this chapter, I will therefore generalize the model in two aspects. First, I will allow for multiple scalar fields carrying a nontrivial representation of a possibly non-Abelian symmetry group. This will take us to the level of standard expositions of SSB in textbooks on quantum field theory, oriented towards high-energy physics. Second, I will pay off my debt to readers with other backgrounds and show how the story changes if one gives up relativistic (Lorentz) invariance. This is particularly relevant for applications to condensed-matter physics, but also concerns dense relativistic matter that one deals with, for instance, in astrophysics and cosmology.

The primary goal of this chapter is to introduce the reader to the intricate interplay between the pattern of SSB and the spectrum of the associated \emph{Nambu--Goldstone} (NG) \emph{bosons}. This goes a long way towards a broad qualitative understanding of the behavior of physical systems with SSB at low energies. Namely, as explained in the previous chapter, NG bosons tend to interact weakly. In the absence of other massless particles, the low-temperature thermodynamics of the system will then be accurately described in terms of a free gas of NG bosons. The latter is completely characterized by the number of NG bosons in the spectrum and their dispersion relations.

Within the landscape of relativistic field theory, a NG boson is always a massless particle whose dispersion relation is fixed by Lorentz invariance. Section~\ref{sec:firstmodelnonAbelian} illustrates how, in this case, the number of NG bosons can be determined solely from group theory. As soon as we give up Lorentz invariance, however, interesting things start to happen. Importantly, the NG state in the spectrum requires more data to specify than mere rest mass. Section~\ref{sec:firstmodelNR} shows that the asymptotic behavior of the dispersion relation of NG bosons at low momentum is intimately related to their number. Within this chapter, I will not discuss interactions of NG bosons any further, since there are no qualitative changes compared to the toy model of Chap.~\ref{chap:ourfirstmodel}.


\section{Relativistic Models with Non-Abelian Symmetry}
\label{sec:firstmodelnonAbelian}

Suppose we are given a set of real scalar fields, $\p^i$. Similarly to the previous chapter, I will introduce at once a class of toy model Lagrangians with a fixed kinetic term and a generic potential,
\begin{equation}
\La=\frac12\d_{ij}\de_\m\p^i\de^\m\p^j-V(\p)\;.
\label{toylagmulti}
\end{equation}
Here $V(\p)$ is an a priori arbitrary function of $\p^i$, only restricted by the requirements that it is bounded from below and its Taylor expansion around $\p^i=0$ starts at the quadratic order. Since any complex scalar field can always be represented by two real ones, the class of Lagrangians~\eqref{toylagmulti} includes our original toy model~\eqref{toylag}. I have however dropped the fermion sector of~\eqref{toylag}. I have also set to zero the terms proportional to $\eps$ and $\eps^*$, as they only serve to select a unique ground state. With the experience gathered in Chap.~\ref{chap:ourfirstmodel}, the reader should by now be comfortable with degenerate ground states.

The symmetry of the theory defined by~\eqref{toylagmulti} will play a key role throughout this chapter. I will assume that the fields $\p^i$ carry a linear representation of some Lie group $G$. Invariance of the whole Lagrangian~\eqref{toylagmulti} under transformations from $G$ requires that the kinetic term and the potential are invariant separately. Let the total number of scalar fields be $n$. Then the kinetic term alone is invariant under continuous rotations of $\p^i$, forming the orthogonal group $\gr{SO}(n)$. We therefore expect that $G$ is a subgroup of $\gr{SO}(n)$, depending on the concrete choice of potential. With this in mind, I will focus in the following on the symmetry of the potential $V(\p)$ itself.

Suppose that the fields $\p^i$ transform in some (real) representation $\rep$ of $G$. The invariance of the potential $V(\p)$ under $G$ is expressed by the condition
\begin{equation}
V(\rep(g)\p)=V(\p)\quad\text{for any }g\in G\;.
\label{Vinvariance}
\end{equation}
For continuous symmetries, it is usually more convenient to work with infinitesimal transformations. Denoting the generators of $G$ as $Q_{A,B,\dotsc}$, the condition~\eqref{Vinvariance} then boils down to
\begin{equation}
\PD{V(\p)}{\p^i}\rep(Q_A)^i_{\phantom ij}\p^j=0\quad\text{for any }Q_A\in\lie g\;,
\label{Vinvinfty}
\end{equation}
where $\lie g$ is the Lie algebra of $G$. This will be the starting point for the analysis of the spectrum of NG bosons below.

\begin{illustration}%
It is straightforward to construct a potential that is invariant under the complete group of rotations, $\gr{SO}(n)$. Indeed, any function of the quadratic invariant, $\Phi^2\equiv\d_{ij}\p^i\p^j$, will do. For instance, a generic $\gr{SO}(n)$-invariant quartic polynomial potential will take the form $V(\p)=-m^2\Phi^2+\l\Phi^4$, which generalizes~\eqref{toypotential} from $n=2$ to any $n$. The fields $\p^i$ transform under the vector representation of $\gr{SO}(n)$, in which $\rep(g)$ are $n\times n$ orthogonal matrices. The generators of $\gr{SO}(n)$ are labeled by a pair of vector indices; up to overall normalization, $\rep(Q_{kl})^i_{\phantom ij}$ equals $\d^i_k\d_{jl}-\d^i_l\d_{jk}$.
\end{illustration}


\subsection{Spectrum of Nambu--Goldstone Bosons}
\label{subsec:firstmodelnonAbelianspectrum}

Let us choose a potential $V(\p)$ so that there is a ground state in which some of the fields $\p^i$ have a nonzero \emph{vacuum expectation value} (VEV), $\vev{\p^i}$. The elements of $G$ that leave the VEVs unchanged form a group $H$, referred to as the \emph{unbroken subgroup} of $G$,
\begin{equation}
H\equiv\{h\in G\,\vert\,\rep(h)^i_{\phantom ij}\vev{\p^j}=\vev{\p^i}\}\;.
\label{toydefunbroken}
\end{equation}
The generators of $H$ will be labeled as $Q_{\a,\b,\dotsc}$. As an immediate consequence of~\eqref{toydefunbroken}, such unbroken generators satisfy
\begin{equation}
\rep(Q_\a)^i_{\phantom ij}\vev{\p^j}=0\quad\text{for any }Q_\a\in\lie h\;,
\label{unbrokensubalgebra}
\end{equation}
where $\lie h$ is the Lie algebra of $H$. In many physical systems, $H$ is a proper subgroup of $G$. That is, there are transformations from $G$ that do change some of the VEVs $\vev{\p^i}$, hence also the ground state itself. Representing a symmetry of the Lagrangian~\eqref{toylagmulti} but not of the ground state, these are said to be spontaneously broken.

By definition of symmetry, acting with an element of $G$ on any field configuration gives a (possibly different) field configuration with the same energy. Hence, as already observed in the previous chapter, the existence of symmetry transformations that do not leave the ground state invariant implies the existence of degenerate ground states. We also previously found that this is closely related to the presence of massless particles in the spectrum.

Let us see how the NG bosons emerge in the present general setting. We start by choosing a basis $\{Q_\a,Q_a\}$ of the Lie algebra $\lie g$ such that $\{Q_\a\}$ is a basis of $\lie h$ and $Q_a\notin\lie h$, that is $\rep(Q_a)^i_{\phantom ij}\vev{\p^j}\neq0$. Next, we take a derivative of~\eqref{Vinvinfty} with respect to $\p$. Upon some relabeling of indices, this becomes
\begin{equation}
\frac{\de^2V(\p)}{\de\p^i\de\p^j}\rep(Q_A)^j_{\phantom jk}\p^k+\PD{V(\p)}{\p^j}\rep(Q_A)^j_{\phantom ji}=0\;.
\end{equation}
Finally, we evaluate this condition in the chosen ground state and use the fact that by definition, the first derivatives of $V(\p)$ in the ground state vanish,
\begin{equation}
\at{\frac{\de^2V(\p)}{\de\p^i\de\p^j}}{\p=\vev\p}\rep(Q_A)^j_{\phantom jk}\vev{\p^k}=0\;.
\label{hessian}
\end{equation}
This does not say anything about the unbroken generators $Q_\a$, for which the left-hand side identically vanishes. However, for the broken generators $Q_a$, this implies that $\smash{\rep(Q_a)^i_{\phantom ij}\vev{\p^j}}$ is an eigenvector of the Hessian matrix of $V(\p)$ in the ground state with zero eigenvalue (\emph{zero mode}). Moreover, the set of eigenvectors $\rep(Q_a)^i_{\phantom ij}\vev{\p^j}$ with all the different choices of $Q_a$ is linearly independent. To see why, suppose that there are coefficients $c^a$ such that $c^a\rep(Q_a)^i_{\phantom ij}\vev{\p^j}=0$. Then by~\eqref{unbrokensubalgebra}, $c^aQ_a\in\lie h$. But we have assumed that the set $\{Q_\a,Q_a\}$ is linearly independent, hence $c_a=0$.

This proves that for every broken symmetry generator $Q_a$, the Hessian matrix of $V(\p)$ in the ground state has a corresponding independent zero mode. Now recall that with a canonically normalized kinetic term as in~\eqref{toylagmulti}, the Hessian of $V(\p)$ represents the mass matrix of the theory. That is, its eigenvalues give the squared masses of elementary one-particle excitations in the spectrum. Thus, there is one NG boson in the spectrum for each broken symmetry generator. The total number of NG bosons in the toy model~\eqref{toylagmulti} is $\dim G-\dim H$.

\begin{watchout}%
What I have just presented is (a simplified version of) one of the original proofs of Goldstone's theorem~\cite{Goldstone1962a}. It is worth stressing that the number of zero modes of the Hessian matrix of the potential may be larger than $\dim G-\dim H$. First, I have not proven that the Hessian cannot have zero modes entirely unrelated to symmetry. (It can.) Second, it is possible that the symmetry group of the potential $V(\p)$ alone is larger than that of the kinetic term, and thus of the whole Lagrangian~\eqref{toylagmulti}. In both cases, a classical analysis as worked out in this and the preceding chapter may yield ``fake'' NG bosons. These are scalar fields that have no classical mass term in the Lagrangian, but acquire mass solely due to quantum corrections~\cite{Weinberg1972a,Coleman1973b}. As I will explain in Chap.~\ref{chap:NGbosons}, the vanishing of the mass of true NG bosons is guaranteed by SSB. This is an exact result that must also hold within any approximation that respects the symmetry of the given theory.

Let me stress that everything said so far relies on the kinetic term in the Lagrangian being a positive-definite quadratic form in the time derivatives of $\p^i$. Finding the mass spectrum of a scalar theory such as~\eqref{toylagmulti} is then equivalent to finding the normal modes using the theory of small oscillations of classical mechanical systems~\cite{Goldstein2013a}. Section~\ref{sec:firstmodelNR} revolves largely around the subtleties that arise when this requirement is not satisfied, which may happen once we depart from the landscape of Lorentz-invariant field theory.
\end{watchout}

\begin{illustration}%
\label{ex:SO(n)vectormodel}%
Consider a model of an $n$-plet of scalar fields $\p^i$ whose Lagrangian is invariant under the group $G\simeq\gr{SO}(n)$. Any ground state in which at least one of the VEVs $\vev{\p^i}$ is nonzero will be invariant under the subgroup $H\simeq\gr{SO}(n-1)$ of transformations that leave $\vev{\p^i}$ fixed. In other words, $H$ consists of field rotations in the $n-1$ directions orthogonal to $\vev{\p^i}$. By the general argument developed above, we expect the spectrum of such a model to contain $\dim\gr{SO}(n)-\dim\gr{SO}(n-1)=n-1$ NG bosons. Note that the special case of $n=2$ correctly recovers the single NG boson we found in Chap.~\ref{chap:ourfirstmodel}.
\end{illustration}


\subsection{Low-Energy Effective Field Theory}
\label{subsec:firstmodelnonAbelianEFT}

On general grounds, we expect the dynamics of NG bosons to be captured by a low-energy \emph{effective field theory} (EFT). In Sect.~\ref{subsec:firstmodelintegratingout}, I showed how to derive such an EFT explicitly from the underlying Lagrangian. Here we do not know the precise form of the potential $V(\p)$ or its symmetry. It is however still possible to gain useful insight by following, if only schematically, the same argument as in Sect.~\ref{subsec:firstmodelintegratingout}.

The first step towards the EFT for NG bosons is a suitable choice of parameterization of the fields $\p^i$. The exponential parameterization~\eqref{toyparamexp} for a single complex field can be generalized to the set of real fields $\p^i$ as
\begin{equation}
\phi^i(x)=U^i_{\phantom ij}(\pi(x))\bigl[\vev{\p^j}+\mf^j(x)\bigr]\;.
\label{toyparamexpgen}
\end{equation}
Here $U(\pi)$ is a matrix taking values in the representation $\rep$ of $G$. It encodes a set of NG fields $\pi^a$, one for each broken generator $Q_a$. One can for instance choose the generators as real antisymmetric matrices and imagine that $U(\pi)$ is the orthogonal matrix $\exp[\I\pi^a\rep(Q_a)]$. But a precise form of $U(\pi)$ is not important. All that I will need is that when expanded in powers of $\pi^a$, the leading term is $U(0)=\un$ and the linear term is proportional to $\pi^a\rep(Q_a)$. Finally, $\mf^i$ encodes a set of \emph{Higgs modes} whose masses are expected to be nonzero.

\begin{watchout}%
In spite of the suggestive notation, the different components of $\mf^i$ are not linearly independent. By mere counting of degrees of freedom, the number of independent Higgs fields should be $n-\dim G+\dim H$. What exactly these independent linear combinations of $\mf^i$ are, can be identified using group theory. One just has to decompose the space of $\p^i$ into irreducible representations of $H$ and drop the representations corresponding to the NG bosons.
\end{watchout}

The advantage of the parameterization~\eqref{toyparamexpgen} is that the matrix $U(\pi)$ drops out of the potential $V(\p)$, because the latter is by construction $G$-invariant, cf.~\eqref{Vinvariance}. The fact that $U(\pi)$ depends implicitly on spacetime coordinates through the fields $\pi^a(x)$ does not need to bother us since the potential does not contain any derivatives. In this parameterization, the Lagrangian~\eqref{toylagmulti} will therefore have a similar structure to that in~\eqref{toylagscalar}. The NG fields $\pi^a$ will only enter through the kinetic term, that is together with derivatives. The nonderivative potential, on the other hand, will only depend on the Higgs fields $\mf^i$. Using the shorthand notation $\vec\p,\vec\mf$ for the vectors $\p^i,\mf^i$, the Lagrangian reads explicitly
\begin{equation}
\begin{split}
\La={}&\frac12\de_\m\vec\mf\cdot\de^\m\vec\mf+\de_\m\vec\mf\cdot U^T\de^\m U\cdot(\vev{\vec\p}+\vec\mf)\\
&+\frac12(\vev{\vec\p}+\vec\mf)\cdot\de_\m U^T\de^\m U\cdot(\vev{\vec\p}+\vec\mf)-V(\mf)\;.
\end{split}
\label{toylagmultiaux}
\end{equation}
We would now like, at least in principle, to eliminate $\mf^i$ by using its \emph{equation of motion} (EoM). It is important to make sure that the expansion of $\mf^i$ in powers of $\pi^a$ generalizing~\eqref{psiiterate} starts with a term with (at least) two derivatives and two powers of $\pi^a$. This in turn requires that the Lagrangian~\eqref{toylagmultiaux} does not contain any mixing term, linear in both $\p^i$ and $\mf^i$. The vanishing of such mixing terms can be achieved by choosing the vector $\vec\mf$ to be orthogonal to all $\rep(Q_a)\cdot\vev{\vec\p}$.

The rest of the argument is simple. Inserting the solution for $\mf^i$ in the Lagrangian~\eqref{toylagmultiaux} gives terms that contain at least four derivatives. The effective Lagrangian for the NG fields will however be dominated by operators with only two derivatives, since a higher number of derivatives means stronger suppression at lower energies. Such two-derivative terms are obtained by simply setting $\mf^i\to0$ in~\eqref{toylagmultiaux},
\begin{equation}
\La_\mathrm{eff}=\frac12\vev{\vec\p}\cdot\de_\m U^T\de^\m U\cdot\vev{\vec\p}+\dotsb\;,
\label{toyEFTlag}
\end{equation}
where the ellipsis stands for contributions with more than two derivatives. Remarkably, this leading contribution to the EFT arises solely from the kinetic term in~\eqref{toylagmulti}. The concrete choice of potential $V(\p)$ does not matter. All we need to know are the VEVs $\vev{\p^i}$ and the ensuing \emph{symmetry-breaking pattern} $G\to H$. These together fix the matrices $U(\pi)$ up to a choice of parameterization in terms of the NG fields $\pi^a$.

\begin{illustration}%
\label{ex:NLSMfirst}%
Let me illustrate this on the case of spontaneous breaking $\gr{SO}(n)\to\gr{SO}(n-1)$ by an $n$-vector field $\vec\p$, introduced in \refex{ex:SO(n)vectormodel}. Here $U(\pi)$ takes values in the defining (vector) representation of $\gr{SO}(n)$. Thus, $U(\pi)\cdot\vev{\vec\p}$ is a vector of the same length as $\vev{\vec\p}$. By a suitable choice of units, we can make this length whatever we want. It is conventional to set $U(\pi)\cdot\vev{\vec\p}\equiv v\vec n(\pi)$, where $\vec n$ is a unit vector field and $v$ is a dimensionful constant. Then the effective Lagrangian~\eqref{toyEFTlag} acquires the form
\begin{equation}
\La_\mathrm{eff}=\frac{v^2}2\de_\m\vec n\cdot\de^\mu\vec n+\dotsb\;,
\label{toylagNLSM}
\end{equation}
which is known as the \emph{nonlinear sigma model}. Note that in spite of the appearance, this is not a noninteracting field theory. The unit vector field $\vec n$ takes values from the unit $(n-1)$-sphere, $S^{n-1}$. The NG fields $\pi^a$ can be thought of as $n-1$ independent coordinates on the sphere. Once a choice of coordinates is made and the Lagrangian~\eqref{toylagNLSM} is expanded in powers of $\pi^a$, it is going to contain operators with two derivatives and an arbitrarily high number of NG fields.
\end{illustration}

This example underlines an important distinction between the broader class of theories~\eqref{toylagmulti} and the special case analyzed in Chap.~\ref{chap:ourfirstmodel}, where the symmetry group $G\simeq\gr{U}(1)$ was Abelian. In the latter case, the leading contribution to the effective Lagrangian contains operators with exactly one derivative on each NG field $\pi$. The part of the Lagrangian with two derivatives is then just the kinetic term, and any interactions necessarily come with four or more derivatives. Once $G$ is allowed to be non-Abelian, the two-derivative part of the Lagrangian~\eqref{toyEFTlag} contains interactions bringing together an arbitrarily high number of NG bosons.

Equation~\eqref{toyEFTlag} gives us the first hint that the form of the EFT for NG bosons is  controlled by symmetry, regardless of the details of the underlying model. The Lagrangian can be cast solely in terms of the NG fields $\pi^a$,
\begin{equation}
\begin{split}
\La_\mathrm{eff}&=\frac12g_{ab}(\pi)\de_\mu\pi^a\de^\mu\pi^b+\dotsb\;,\\ g_{ab}(\pi)&\equiv\vev{\vec\p}\cdot\PD{U(\pi)^T}{\pi^a}\PD{U(\pi)}{\pi^b}\cdot\vev{\vec\p}\;.
\end{split}
\label{toygab}
\end{equation}
I will show in Chap.~\ref{chap:effLagrangian} that the matrix function $g_{ab}(\pi)$ can be interpreted in terms of the geometry of the Lie groups $G$ and $H$. The dependence of $g_{ab}(\pi)$ on the NG fields is fixed by this geometry. All that is left of the concrete model leading to the EFT~\eqref{toygab} is the numerical value of the constant matrix $g_{ab}(0)$.


\section{Going Nonrelativistic}
\label{sec:firstmodelNR}

In all the examples worked out in the previous and this chapter so far, I restricted myself to relativistic, Lorentz-invariant field theories. This choice was based on technical simplicity. The notation using Lorentz indices is more compact, and the calculations of scattering amplitudes in terms of the relativistic Mandelstam variables are more transparent. There are good reasons to step out of the box, though. First, the great majority of natural phenomena occurs at energies well below the scale at which relativity starts to matter. In the spirit of EFT, such phenomena should therefore have an accurate description in terms of nonrelativistic field theory. Second, as we shall see in this section, relaxing the requirement of Lorentz invariance leads to rich phenomenology. The examples developed below will prepare the reader for the more thorough discussion of SSB that comes in Part~\ref{part:foundations} of the book.

Operationally, what I will do is to modify the previously discussed toy models in a way that respects invariance under spatial rotations, and spatial and temporal translations. While this does not place any restrictions on the potential $V(\p)$, the kinetic term may now assume a more general form. The Lagrangian may then contain independent terms with either two spatial derivatives, or one or two temporal derivatives. The very possibility of adding terms with a single time derivative constitutes the main qualitative difference compared to relativistic field theory.


\subsection{Single Schr\"odinger Field}
\label{subsec:firstmodelNRSchrodinger}

In order to copy as closely as possible the previous discussion of relativistic field theory, let us start with the following model,
\begin{equation}
\La=2\I M\p^*\de_0\p-\vec\nabla\p^*\cdot\vec\nabla\p+m^2\p^*\p-\l(\p^*\p)^2\;.
\label{toylagSchrodinger}
\end{equation}
Here $\p$ is a complex scalar field and $\de_0$ stands for a time derivative. This Lagrangian is almost identical to the scalar sector of~\eqref{toylag}, except for the temporal part of the kinetic term. For $m=\l=0$, the corresponding EoM is the Schr\"odinger equation for a free particle of mass $M$. It is therefore natural to refer to the field $\p$ as a ``Schr\"odinger field.'' 

The analysis of the model~\eqref{toylagSchrodinger} now follows the same steps as in Chap.~\ref{chap:ourfirstmodel}. The classical Hamiltonian density reads
\begin{equation}
\Ha=\vec\nabla\p^*\cdot\vec\nabla\p-m^2\p^*\p+\l(\p^*\p)^2\;.
\end{equation}
The lowest energy is achieved by a coordinate-independent state such that
\begin{equation}
\vev\p=\frac v{\sqrt2}\E^{\I\t}\;,\qquad
v\equiv\frac m{\sqrt\l}\;,
\label{toySchrvacuum}
\end{equation}
where $\t$ is an arbitrary phase, reflecting the ground state degeneracy. Making use of the exponential parameterization~\eqref{toyparamexp} brings the Lagrangian to the form
\begin{equation}
\La=\La_\mathrm{vac}+\La_\mathrm{bilin}+\La_\mathrm{int}\;.
\end{equation}
The vacuum piece equals $\La_\mathrm{vac}=(1/2)m^2v^2-(\l/4)v^4=m^4/(4\l)$. The bilinear and interaction parts of the Lagrangian read, respectively,
\begin{equation}
\begin{split}
\La_\mathrm{bilin}&=-2M\mf\de_0\pi-\frac12(\vec\nabla\mf)^2-m^2\mf^2-\frac12(\vec\nabla\pi)^2\;,\\
\La_\mathrm{int}&=-\l v\mf^3-\frac\l4\mf^4-\frac Mv\mf^2\de_0\pi-\left(\frac\mf v+\frac{\mf^2}{2v^2}\right)(\vec\nabla\pi)^2\;,
\end{split}
\label{toylagSchrodingerbilin}
\end{equation}
where I have discarded some contributions that amount to a total time derivative.

Just like for the relativistic model~\eqref{toylag}, the NG field $\pi$ enters the Lagrangian only with derivatives. This is an immediate consequence of the choice of parameterization~\eqref{toyparamexp}. Here is where the similarity ends, though. It would be a mistake to conclude that $\pi$ represents a NG boson and $\mf$ describes a massive, Higgs mode. The two fields are mixed by the bilinear term $-2M\mf\de_0\pi$, which makes $\mf,\pi$ canonically conjugated to each other. We therefore expect that the spectrum of the model~\eqref{toylagSchrodinger} contains just one type of excitation, which should then be a NG boson.

In order to describe this excitation more accurately, let us rewrite the bilinear part of the Lagrangian in a matrix form,
\begin{equation}
\setlength\arraycolsep{0.5ex}
\La_\mathrm{bilin}\simeq\frac12
\bigl(\begin{matrix}
\mf & \pi
\end{matrix}\bigr)
\raisebox{-2ex}{$\begin{pmatrix}
\vec\nabla^2-2m^2 & -2M\de_0\\
2M\de_0 & \vec\nabla^2
\end{pmatrix}
\begin{pmatrix}
\mf\\
\pi
\end{pmatrix}$}\;.
\label{lagtoybilinSchrodinger}
\end{equation}
The symbol $\simeq$ again indicates that I have dropped a surface term. What is entirely new compared to relativistic field theory is that the bilinear Lagrangian cannot be diagonalized by any local field redefinition. This is just a minor nuisance when it comes to finding the spectrum of one-particle states. It is however a major complication if one wants to study scattering of the particles. Calculating a cross-section for a scattering process now requires a careful mapping of one-particle states to fields in the Lagrangian, and a revision of rules for phase space integration. Some discussion of these issues can be found for instance in~\cite{Brauner2006b,Gongyo2016a,Brauner2018a}.

In order to find the dispersion relation of the NG mode, we Fourier-transform and set the determinant of the matrix in~\eqref{lagtoybilinSchrodinger} to zero. This gives the energy $E$ of the NG boson as a function of its momentum $\vec p$,
\begin{equation}
E(\vec p)=\frac{\abs{\vec p}}{2M}\sqrt{\vec p^2+2m^2}\;.
\label{toySchrdisprel}
\end{equation}
Note that in the limit $m\to0$, this recovers the conventional dispersion relation of a nonrelativistic particle, $E(\vec p)=\vec p^2/(2M)$.

\begin{watchout}%
Once the dispersion relation is no longer fixed by Lorentz invariance and the rest mass, it makes little sense to refer to a NG boson as a massless particle. Instead, it is common to say that the NG boson is a \emph{gapless} mode or excitation. This is however a bit of a misnomer. The gap in the excitation spectrum is defined as $\min_{\vec p}E(\vec p)$. On the other hand, the characteristic property of NG bosons is, as I will explain in detail in Chap.~\ref{chap:NGbosons}, that
\begin{equation}
\lim_{\vec p\to\vec0}E(\vec p)=0\;.
\end{equation}
Since the excitation energy above a stable ground state is by definition positive, a NG boson is always gapless. However, a gapless excitation may not necessarily be a NG boson. There are physical systems where the dispersion relation of a one-particle excitation develops a local minimum due to dynamical effects. This is the case for instance for the ``roton'' mode in superfluid helium~\cite{Leggett2006a}. I am however not aware of any example where the energy of the excitation at the local minimum could be tuned to zero without breaking some symmetry.
\end{watchout}

One might hope that it is possible to get rid of the annoying mixing between the $\mf,\pi$ fields by eliminating the Higgs field $\mf$ as in Sect.~\ref{subsec:firstmodelintegratingout}. Let us see how the derivation of the low-energy EFT for $\pi$ alone goes. We start by writing the bilinear and interaction parts of the Lagrangian~\eqref{toylagSchrodingerbilin} together as
\begin{equation}
\La\simeq-\frac12(\vec\nabla\mf)^2-m^2\mf^2-\l v\mf^3-\frac\l4\mf^4-\frac12\left(1+\frac\mf v\right)^2\XiEFT\;,
\label{toylagintegrateoutSchr}
\end{equation}
up to terms that are a total time derivative. I have introduced a shorthand notation,
\begin{equation}
\XiEFT\equiv2Mv\de_0\pi+(\vec\nabla\pi)^2\;.
\label{toyXidef}
\end{equation}
Note that the Lagrangian~\eqref{toylagintegrateoutSchr} is nearly identical to~\eqref{toylagscalar} except for the replacement $(\de_\m\mf)^2\to-(\vec\nabla\mf)^2$ and a different identification of $\XiEFT$; in~\eqref{toylagscalar} it is simply $-(\de_\m\pi)^2$. We can therefore follow the same steps as in Sect.~\ref{subsec:firstmodelintegratingout} to eliminate $\mf$. All we have to do is to take the results thereof and substitute $(\de_\m\pi)^2\to-\XiEFT$ and $\Box\to-\vec\nabla^2$ wherever appropriate. Thus, an iterative solution of the EoM for $\mf$ is obtained at once from~\eqref{toyEoMiterative},
\begin{equation}
\begin{split}
\mf&=-\frac1v\left(2m^2-\vec\nabla^2+\frac\XiEFT{v^2}+3\l v\mf+\l\mf^2\right)^{-1}\XiEFT\\
&=-\left(1-\frac{\vec\nabla^2}{2m^2}\right)^{-1}\frac\XiEFT{2m^2v}+\bigO(\XiEFT^2)\;.
\end{split}
\end{equation}
By using this in~\eqref{toylagaux2}, we then readily determine the first two terms of the effective Lagrangian expanded in powers of $\XiEFT$,
\begin{equation}
\La_\mathrm{eff}\simeq-\frac\XiEFT2+\frac\XiEFT{4m^2v^2}\left(1-\frac{\vec\nabla^2}{2m^2}\right)^{-1}\XiEFT+\bigO(\XiEFT^3)\;.
\label{lagtoyXi}
\end{equation}

While restricted to the lowest orders in $\XiEFT$ and hence $\pi$, this Lagrangian still contains terms with an arbitrarily high number of derivatives. The price for having eliminated the kinetic mixing between $\mf$ and $\pi$ therefore is an introduction of nonlocal operators. This is necessary to get the dispersion relation of the NG boson right. Indeed, by substituting from~\eqref{toyXidef}, we find
\begin{equation}
\La_\mathrm{eff}\simeq2M^2\de_0\pi\frac{\de_0}{2m^2-\vec\nabla^2}\pi-\frac12(\vec\nabla\pi)^2+\bigO(\pi^3)\;.
\end{equation}
Fourier transformation to the energy--momentum space then recovers our previous result~\eqref{toySchrdisprel}.


\subsection{Multiple Nambu--Goldstone Fields}
\label{subsec:firstmodelNRnonAbelian}

We would of course like to understand also what happens in models that contain multiple nonrelativistic scalar fields. Are there any further surprises awaiting us? Instead of developing a general picture akin to the relativistic framework of Sect.~\ref{sec:firstmodelnonAbelian}, I will focus here on some instructive examples. A more complete analysis will follow in Chap.~\ref{chap:effLagrangian} using the powerful machinery of EFT.

Let us first contemplate what we expect to find. Consider a model with multiple (real) scalar fields $\p^i$ and a generic potential $V(\p)$ as in~\eqref{toylagmulti}. The argument of Sect.~\ref{subsec:firstmodelnonAbelianspectrum} then goes through without change. The conclusion that there is a well-defined injective mapping from broken symmetry generators to zero modes of the Hessian matrix of $V(\p)$ in the ground state remains. One can say that the number of linearly independent NG \emph{fields} still equals $\dim G-\dim H$, in one-to-one correspondence with the broken generators. However, with a generalized, nonrelativistic kinetic term, we are no longer guaranteed the existence of an independent NG \emph{mode} for each such NG field. In presence of terms in the Lagrangian with a single time derivative, some of the NG fields may be canonically conjugated to each other. We then expect one NG mode in the spectrum to be associated with a pair of NG fields.

Let me conclude this chapter and the whole Part~\ref{part:prologue} of the book with a couple of illustrative examples.

\begin{illustration}%
Consider the limit $m=\l=0$ of~\eqref{toylagSchrodinger}, already briefly mentioned previously,
\begin{equation}
\La=2\I M\p^*\de_0\p-\vec\nabla\p^*\cdot\vec\nabla\p\;.
\label{lagSchrodinger}
\end{equation}
Here the potential is vanishing, hence the real and imaginary parts of $\p$ are both trivially zero modes. There is a single one-particle excitation in the spectrum with dispersion relation $E(\vec p)=\vec p^2/(2M)$. This is a theory of a noninteracting Schr\"odinger field, which seems to be rather uninspiring. Things get more interesting, though, once we look at the theory from the point of view of symmetry.

There is a $\gr{U}(1)\simeq\gr{SO}(2)$ symmetry under which the field transforms as $\p\to\E^{\I\eps}\p$. The natural choice of ground state with $\vev{\p}=0$ leaves this symmetry unbroken. However, the free Schr\"odinger theory~\eqref{lagSchrodinger} has another symmetry, namely $\p\to\p+\eps_1+\I\eps_2$, where both $\eps_{1,2}$ are real parameters. This shifts the Lagrangian by a total time derivative and thus leaves the action invariant. Such a shift symmetry is necessarily spontaneously broken no matter how the VEV $\vev\p$ is chosen. At the end of the day, we have two spontaneously broken generators but only one NG mode with a quadratic dispersion relation. In some aspects, the free Schr\"odinger theory~\eqref{lagSchrodinger} constitutes a rather nontrivial realization of a nonrelativistic NG boson. A detailed discussion of SSB in the quantized free Schr\"odinger theory is offered in Sect.~\ref{sec:SSBsubtleties}.
\end{illustration}

For an example of an interacting field theory, replace the single complex field $\p$ in~\eqref{toylagSchrodinger} with a two-component complex field (doublet) $\Phi$,
\begin{equation}
\La=2\I M\he\Phi\de_0\Phi-\vec\nabla\he\Phi\cdot\vec\nabla\Phi+m^2\he\Phi\Phi-\l(\he\Phi\Phi)^2\;.
\label{toylagSU(2)}
\end{equation}
Here the dagger indicates Hermitian conjugation. This Lagrangian possesses a $\gr{U}(2)$ symmetry. It has an Abelian subgroup, $\gr{U}(1)$, under which the doublet $\Phi$ changes its overall phase, $\Phi\to\E^{\I\eps}\Phi$. What is new is the non-Abelian $\gr{SU}(2)$ subgroup of $\gr{U}(2)$ under which $\Phi\to\rep(g)\Phi$, where $g\in\gr{SU}(2)$ and $\rep$ stands for the fundamental representation thereof.

The analysis of the model~\eqref{toylagSU(2)} proceeds as before. The classical Hamiltonian is minimized by any constant field $\Phi$ such that $\vev{\he\Phi\Phi}=v^2/2\equiv m^2/(2\l)$. Geometrically, the set of all degenerate ground states corresponds to a 3-sphere, $S^3$. This is easy to see if one thinks of $\Phi$ as a four-component real vector whose length is fixed by minimization of the Hamiltonian. While any of the different degenerate ground states is equally good, the conventional choice that simplifies notation is\footnote{Any other choice of ground state can be obtained from our $\vev\Phi$ by multiplication with a constant $\gr{U}(2)$ matrix. This generalizes the arbitrary phase $\E^{\I\t}$ in~\eqref{toySchrvacuum}.}
\begin{equation}
\vev\Phi=\frac v{\sqrt2}
\begin{pmatrix}
0 \\ 1
\end{pmatrix}\;.
\label{toySU(2)vacuum}
\end{equation}
Out of all the generators of $\gr{U}(2)$, the only one (up to normalization) that leaves this VEV unchanged is $\un+\pau_3$, where $\pau_3$ is the third Pauli matrix. The remaining three linearly independent generators are spontaneously broken. We therefore expect three of the four degrees of freedom contained in $\Phi$ to represent NG fields.

The low-energy spectrum of the model~\eqref{toylagSU(2)} is most easily addressed within an EFT for the three NG fields. To that end, we need to choose a parameterization for $\Phi$. Inspired by~\eqref{toyparamexpgen}, let us set
\begin{equation}
\Phi(x)=\frac1{\sqrt2}U(\pi(x))
\begin{pmatrix}
0 \\ v+\mf(x)
\end{pmatrix}\;,
\end{equation}
where $U(\pi)$ is a unitary matrix encoding the three NG degrees of freedom and $\mf$ is a (real) Higgs field. A straightforward manipulation then leads to a Lagrangian of the type~\eqref{toylagintegrateoutSchr}, except that we now have to set
\begin{equation}
\XiEFT=-2\I Mv^2(\he U\de_0U)_{22}+v^2(\vec\nabla\he U\cdot\vec\nabla U)_{22}\;.
\end{equation}
The leading contributions to the effective Lagrangian are again given by~\eqref{lagtoyXi}. To complete the analysis, we parameterize $U(\pi)$ in terms of a triplet of NG fields, $\vec\pi(x)$,
\begin{equation}
U(\pi)=\exp\left(\frac\I v\skal\pau\pi\right)\;.
\end{equation}
Up to second order in the NG fields, which is needed to pin down the bilinear part of the effective Lagrangian, one finds
\begin{equation}
\XiEFT=-2Mv\de_0\pi^3-2M(\vec\pi\times\de_0\vec\pi)^3+\d_{ab}\vec\nabla\pi^a\cdot\vec\nabla\pi^b+\bigO(\pi^3)\;.
\end{equation}
Upon using this in~\eqref{lagtoyXi}, one thus obtains
\begin{equation}
\begin{split}
\La\simeq{}&2M^2\de_0\pi^3\frac{\de_0}{2m^2-\vec\nabla^2}\pi^3-\frac12(\vec\nabla\pi^3)^2\\
&+M(\pi^1\de_0\pi^2-\pi^2\de_0\pi^1)-\frac12(\vec\nabla\pi^1)^2-\frac12(\vec\nabla\pi^2)^2+\bigO(\pi^3)\;.
\end{split}
\label{toynonrelaux}
\end{equation}

The dispersion relation of $\pi^3$ is the same as for the model~\eqref{toylagSchrodinger} with a single complex field, namely~\eqref{toySchrdisprel}. The $\pi^1,\pi^2$ sector is however very different. The second line of~\eqref{toynonrelaux} is just the free Schr\"odinger theory in disguise; $\pi^1,\pi^2$ are the real and imaginary parts of the complex Schr\"odinger field. The corresponding dispersion relation is $E(\vec p)=\vec p^2/(2M)$. It is interesting to note that the bilinear Lagrangian in the $\pi^1,\pi^2$ sector comes entirely from the term in the effective Lagrangian~\eqref{lagtoyXi} linear in $\XiEFT$. It is therefore insensitive to the choice of potential in the underlying Lagrangian~\eqref{toylagSU(2)}.

The above discussion has an immediate generalization to a class of models described by the same Lagrangian~\eqref{toylagSU(2)}, in which $\Phi$ is an $n$-component complex field. The symmetry is then $G\simeq\gr{U}(n)$. The ground state can still be chosen as any constant field such that $\vev{\he\Phi\Phi}=v^2/2$. It is convenient to pick $\vev\Phi$ so that, analogously to~\eqref{toySU(2)vacuum}, only its the $n$-th component is nonzero and equals $v/\sqrt2$. The symmetry group $G$ is then spontaneously broken down to the subgroup $H\simeq\gr{U}(n-1)$, acting on the first $n-1$ components of $\Phi$. Accordingly, there are $\dim\gr{U}(n)-\dim\gr{U}(n-1)=2n-1$ broken symmetry generators as well as NG fields. The analysis of the excitation spectrum also proceeds following the same steps, except for the necessary complications due to the presence of unitary $n\times n$ matrices. At the end of the day, one finds one NG boson with dispersion relation~\eqref{toySchrdisprel}. The remaining $2n-2$ NG fields pair up into $n-1$ modes with the Schr\"odinger-like dispersion relation $E(\vec p)=\vec p^2/(2M)$.


\section{Moral Lessons}
\label{sec:firstmodelmorals}

In this chapter, I have refined the discussion of toy models for SSB in two ways. First, I have allowed for the possibility of multiple scalar fields and multiple broken symmetries. I have however remained within the territory of \emph{internal} symmetries, that is symmetries only acting on the fields, independently of spacetime coordinates. Second, we explored the consequences of giving up Lorentz invariance, in particular by adding to the Lagrangian terms with a single time derivative. With the newly gathered experience, let me briefly revisit the four lessons drawn in Sect.~\ref{sec:firstmodelmorals1}.

\runinhead{Lesson \#1} Nothing to change. The statement made in Sect.~\ref{sec:firstmodelmorals1} is generally valid.

\runinhead{Lesson \#2} Suppose that the given system possesses a continuous symmetry group $G$ which is spontaneously broken to its subgroup $H$. If the action of the system is Lorentz-invariant, its spectrum contains $\dim G-\dim H$ massless particles: the NG bosons. If the system is not Lorentz-invariant, NG bosons are characterized as one-particle excitations with dispersion relation $E(\vec p)$ such that $\lim_{\vec p\to\vec0}E(\vec p)=0$. The low-energy dynamics of the system can still be captured in terms of $\dim G-\dim H$ NG \emph{fields}. However, the actual number of NG modes in the spectrum may be lower than $\dim G-\dim H$. This happens when two NG fields are canonically conjugated by a term with a single time derivative, leading to a single NG mode. The dispersion relation of such a NG mode is typically quadratic in momentum.

\runinhead{Lesson \#3} Nothing to change. The statement made in Sect.~\ref{sec:firstmodelmorals1} is generally valid.

\runinhead{Lesson \#4} It is convenient to choose a field parameterization in which a set of constant NG fields can be eliminated by a symmetry transformation. In such a parameterization, every operator in the Lagrangian containing NG fields carries at least one derivative. If the symmetry group $G$ is non-Abelian, it is however not necessary that every NG field carries a derivative. The nonlinear constraints imposed by the broken symmetry nevertheless ensure that the Adler zero property of scattering amplitudes is preserved, modulo the exceptions alluded to in Lesson \#3.

\runinhead{Lesson \#5} Nothing to change. The statement made in Sect.~\ref{sec:firstmodelmorals1} is generally valid.


\bibliographystyle{spphys}
\bibliography{references}
\begin{partbacktext}
\part{Foundations}
\label{part:foundations}
\end{partbacktext}
\chapter{Symmetry and Conservation Laws}
\label{chap:symmetry}

\abstract*{This chapter introduces the general concept of symmetry in both classical and quantum physics. The part devoted to classical physics stresses the importance of conservation laws and the role of Noether's theorem for their understanding. A fairly general discussion of Noether's theorem is offered, including ambiguities in the definition of the Noether current, possible existence of generalized, tensor conservation laws, and the inverse of Noether's theorem. The Hamiltonian formalism is then used to motivate the transition from classical to quantum physics, and the definition of symmetry in the latter. In anticipation of the discussion of spontaneous symmetry breaking in the following chapter, the text here emphasizes the difference between implementing symmetry in terms of transformations on the algebra of physical observables, and in terms of unitary mappings on the Hilbert space of states.}


Anybody who has taken a first course on field theory has been exposed to the correspondence between symmetry and conservation laws via Noether's theorem. Indeed, I already assumed basic familiarity with this correspondence in Part~\ref{part:prologue}. The semiclassical analysis of the toy models therein would have hardly been possible without the concepts of symmetry of a Lagrangian and Noether current.

In this chapter, I will however restart the discussion of symmetries largely from scratch. With the rich spectrum of modern applications that emerged in the last decades, it appears appropriate to start by carefully defining the basic notions. I will not attempt to give the most general definition of symmetry. In particular, most of this chapter is phrased in the language of classical field theory, tailored to the needs of Parts~\ref{part:internalSSB} and~\ref{part:spacetimeSSB} of the book. Some remarks on a further generalization of the concept of symmetry developed here are postponed to the concluding  Sect.~\ref{sec:generalizedsymmetries}. Nevertheless, within the conservative exposition offered here, I will devote more space to stressing exceptions rather than to repeating ad nauseam familiar concepts.


\section{What Is Symmetry?}
\label{sec:whatissym}

In very general terms, any definition of symmetry must include two ingredients. The first of these is an object that we wish to declare to be symmetric. The second is the operation, or transformation, that should constitute the desired symmetry. In classical physics, one usually defines symmetry by its action on a set of local fields. I will use the notation $\psi^i$ for a set of fields treated as functions $\psi^i:x^\m\to\psi^i(x)$. Here $x^\m$ denotes collectively a set of coordinates, which may for the time being include space, time, or both. The case where $x^\m$ includes only time $t$ corresponds to mechanics with $\psi^i(t)$ as the dynamical variables. In mathematics, it is common to refer to $x^\m$ and $\psi^i$ respectively as the \emph{independent} and \emph{dependent variables}. These (in)dependent variables may take values from some linear space or from a more general mathematical structure such as a manifold. In the latter case, $x^\m$ and $\psi^i$ are identified with the corresponding local coordinates in the sense of Appendix~\ref{appsec:manifolds}.


\subsection{Symmetry Transformations}
\label{subsec:symtransfo}

Let us initially focus on the second ingredient of symmetry, that is the operation. This book revolves largely around the consequences of continuous symmetries, for which it is sufficient to consider infinitesimal transformations. We shall deal with the following generic class of simultaneous transformations of the fields and coordinates,
\begin{equation}
\begin{split}
\udelta\psi^i(x)&\equiv\psi'^i(x')-\psi^i(x)=\eps\df^i[\psi,x](x)\;,\\
\udelta x^\m&\equiv x'^\m-x^\m=\eps\dx^\m[\psi,x](x)\;.
\end{split}
\label{generalizedsym}
\end{equation}
Here $\eps$ is an infinitesimal parameter of the transformation. The square bracket notation indicates that $\df^i$ and $\dx^\m$ are local functions of the fields and (a finite number of) their derivatives, possibly depending explicitly on the coordinates. Transformations of the type~\eqref{generalizedsym} are known in mathematical literature as \emph{generalized local transformations}. The terminology is historical. Namely, the class~\eqref{generalizedsym} generalizes so-called \emph{point transformations}, first studied by Sophus Lie in the 1860s. For those, $\df^i$ and $\dx^\m$ are only allowed to depend on fields and coordinates, not not on field derivatives.

\begin{watchout}%
Indicating that the fields and coordinates should be transformed simultaneously is, while conventional, a red herring. The coordinates are independent variables that can be chosen at will. Any nonsingular transformation of coordinates can be undone by a change of variables. The content of~\eqref{generalizedsym} can therefore be equivalently encoded in a transformation of the fields alone,
\begin{equation}
\psi'^i(x)-\psi^i(x)=\eps\df^i[\psi,x](x)-\eps\dx^\m[\psi,x](x)\de_\m\psi^i(x)\;.
\label{evolutionary}
\end{equation}
This is the \emph{evolutionary form} of the transformation~\eqref{generalizedsym}. More generally, the representation~\eqref{generalizedsym} of the transformation is ambiguous with respect to the redefinition $\dx^\m\to\dx^\m+\tilde\dx^\m$ and $\smash{\df^i\to\df^i+\tilde\dx^\m\de_\m\psi^i}$, where $\tilde\dx^\m[\psi,x]$ is any local function of the coordinates, fields and their derivatives. The evolutionary form can thus be viewed as fixing the ambiguity by setting $\dx^\m=0$.
\end{watchout}

The class of transformations~\eqref{generalizedsym} is so broad that it is often convenient to consider special cases. Throughout this book, I will use the term \emph{internal symmetry} for a point transformation with $\dx^\m=0$ and a coordinate-independent $\df^i$, that is
\begin{equation}
\udelta\psi^i(x)=\eps\df^i(\psi(x))\;,\quad
\udelta x^\m=0\qquad\text{(internal symmetry)}\;.
\label{definternal}
\end{equation}
A point transformation for which $\dx^\m$ is nonzero but only depends on the spacetime coordinates will be referred to as a \emph{spacetime symmetry},
\begin{equation}
\udelta\psi^i(x)=\eps\df^i(\psi(x),x)\;,\quad
\udelta x^\m=\eps\dx^\m(x)\neq0\qquad\text{(spacetime symmetry)}\;.
\label{defspacetime}
\end{equation}
These definitions of internal and spacetime symmetries define the agenda for Parts~\ref{part:internalSSB} and~\ref{part:spacetimeSSB} of the book. The reader should however be warned that these are not standard definitions aligned with published literature. Of all the reasons, this is because a precise definition of internal and spacetime symmetries is rarely found.

\begin{figure}[t]
\sidecaption[t]
\includegraphics[width=2.9in]{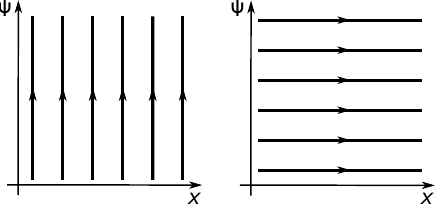}
\caption{Schematic illustration of internal (left panel) and spacetime (right panel) symmetries. The oriented curves indicate the flow in the space of fields and coordinates (see Appendix~\ref{appsec:manifoldmaps}), generated respectively by the infinitesimal transformations~\eqref{definternal} and~\eqref{defspacetime}. To stress the difference between the two cases, the right panel shows a ``purely spacetime'' transformation, for which $\df^i=0$}
\label{fig:pointsymmetryflow}
\end{figure}

\begin{illustration}%
A spacetime translation in the direction of coordinate $x^\n$ is most naturally implemented by setting $\udelta_\n\psi^i(x)=0$ and $\udelta_\n x^\m=\eps\d^\m_\n$. According to~\eqref{evolutionary}, this can be equivalently encoded as $\udelta_\n\psi^i(x)=-\eps\de_\n\psi^i(x)$ without any change of coordinates. Translations are an example of a ``purely spacetime'' transformation, illustrated in the right panel of Fig.~\ref{fig:pointsymmetryflow}. As the figure shows, one can intuitively think of internal and spacetime symmetries as point transformations, generating flows along different subspaces of the space of fields and coordinates.
\end{illustration}

\subsection{Object of Symmetry}
\label{subsec:symtobject}

I will assume throughout the book that the set of all symmetry transformations, generated by the infinitesimal motions~\eqref{generalizedsym}, constitutes a group. Historically, the mathematical field of Lie groups arose from the study of symmetries of \emph{differential equations}. This is now a mature subject with a large body of literature devoted to it. An interested reader is referred to~\cite{Olver1986a,Bluman2002a,Bluman2010a} for more details. Very briefly, the major applications of symmetry methods to differential equations include:
\begin{itemize}
\item Finding new solutions of a given differential equation from already known ones.
\item Finding solutions respecting the symmetry of the differential equation.
\item Classifying differential equations with given symmetry.
\end{itemize}
In physics, it is much more common to define symmetry by invariance of the \emph{action} of a system under some transformation. It is worth stressing that the two notions of symmetry are not equivalent. On the one hand, there are differential equations that do not originate from any variational principle, yet may have nontrivial symmetries. On the other hand, differential equations that do descend from an action functional may have a larger symmetry group than the action itself.

The chief advantage of defining symmetry through an action functional is that this gives us a direct link to conservation laws. The link is supplied by the celebrated Noether theorem, which is the subject of most of Sect.~\ref{sec:SSBLagrangian}. It is in principle also possible to define conservation laws directly on the level of a differential equation. This allows one to deduce a restricted correspondence between symmetries and conservation laws that generalizes the Noether theorem~\cite{Bluman2010a}. However, as far as I know, such a generalized notion of symmetry is of limited use in physics, and I will therefore not pursue it any further.


\section{Lagrangian Approach to Symmetry}
\label{sec:SSBLagrangian}

The correspondence between continuous symmetries and conservation laws has a fascinating history. A nice overview of the developments following Noether's groundbreaking discovery, including an account of the contributions of various authors to the subject, can be found in~\cite{Kosmann-Schwarzbach2011a}. Noether's theorem is covered to some extent in virtually any textbook on classical mechanics or (classical or quantum) field theory. Unfortunately, this is frequently done under unnecessary restrictions on the Lagrangian or on the type of symmetry transformation. Here I will present a fairly general version of Noether's theorem following a trick, which to the best of my knowledge goes back to Gell-Mann and L\'evy~\cite{Gell-Mann1960a}.


\subsection{Noether's Theorem}
\label{subsec:noether}

I consider a class of field theories defined by a Lagrangian density $\La$, which is a local function of a set of fields $\psi^i$ and their derivatives,
\begin{equation}
S=\int\D^D\!x\,\La[\psi,x](x)\;.
\label{action}
\end{equation}
Here $D$ is the dimension of the space of independent variables, which may still include space, time or both. The Lagrangian density may depend explicitly on the coordinates, and there is no restriction on the order of field derivatives it contains. Suppose that the action $S$ is invariant under the infinitesimal transformation~\eqref{generalizedsym}.\footnote{When evaluating the variation of the action under~\eqref{generalizedsym}, it is allowed to drop whatever boundary terms might appear. Thus, the ``invariance of the action'' should be more accurately interpreted as invariance of the Lagrangian density up to the divergence of a vector field.} Then there is a local vector field $J^\m[\psi,x]$, called \emph{Noether current}, which is divergence-less,
\begin{equation}
\de_\m J^\m[\psi,x]=0\qquad
\text{(on-shell)}\;.
\label{noether}
\end{equation}
The qualifier \emph{on-shell} reminds us that the conservation law~\eqref{noether} only holds for fields satisfying the Euler--Lagrange \emph{equation of motion} (EoM).

To prove this claim, we evaluate the variation of the action $\udelta S$ under a modification of~\eqref{generalizedsym} where the parameter $\eps(x)$ is allowed to depend on the coordinates. With the assumption of invariance under the original transformation~\eqref{generalizedsym}, $\udelta S$ may only depend on derivatives of $\eps(x)$. Using integration by parts, it can then always be brought to the form
\begin{equation}
\udelta S=\int\D^D\!x\,J^\m[\psi,x](x)\de_\m\eps(x)\qquad
\text{(off-shell)}\;.
\label{noethertrick}
\end{equation}
Here the qualifier \emph{off-shell} indicates that I have not yet imposed the EoM on the fields. We can see that the Noether current can be extracted as the coefficient of $\de_\mu\eps$ in the variation of the action. Once the EoM is applied, $\udelta S$ must vanish by the definition of the variational principle. This guarantees the on-shell conservation law~\eqref{noether}.

\begin{illustration}%
\label{ex:shiftscalar}%
For a simple example, consider a free massless relativistic scalar field $\p$ with the Lagrangian density $\La[\p]=(1/2)(\de_\m\p)^2$. This is obviously invariant under a shift of the field, $\p\to\p+\eps$. Making the shift coordinate-dependent, we find the variation of the action $\smash{\udelta S=\int\D^D\!x\,\de^\m\p(x)\de_\m\eps(x)}$. Comparison with~\eqref{noethertrick} tells us that the Noether current is $J^\m[\p]=\de^\m\p$. Conservation of this current is equivalent to the EoM for $\p$, which is the massless Klein--Gordon equation. As an aside, such an equivalence holds for any theory of a real scalar $\p$, invariant under the shift $\p\to\p+\eps$. This follows immediately from the variation of the action under the localized transformation $\p(x)\to\p(x)+\eps(x)$,
\begin{equation}
\udelta S=\int\D^D\!x\,\frac{\udelta S}{\udelta\p(x)}\eps(x)=-\int\D^D\!x\,\de_\m J^\m[\p,x](x)\eps(x)\;,
\end{equation}
which implies that $\de_\m J^\m[\p,x](x)=-\udelta S/\udelta\p(x)$ off-shell.
\end{illustration}

The statement of Noether's theorem is well-known. Instead of collecting numerous examples, I will thus focus on a few comments and illustrations that go somewhat off the beaten track. To start with, one should keep in mind that the Noether current is not uniquely determined by the assumed symmetry. The variation~\eqref{noethertrick} only allows us to extract $J^\m[\psi,x]$ up to addition of a vector field whose divergence vanishes off-shell. Such a modification of the current does not affect the conservation law~\eqref{noether}.

There is another, somewhat more subtle ambiguity in the Noether current, related to the definition of the transformation used to produce~\eqref{noethertrick}. Namely, it would be tempting to simply take~\eqref{generalizedsym} and make it local by replacing $\eps\to\eps(x)$. But there is a more general possibility that the localized transformation also depends on the derivatives of $\eps(x)$. All that is required is that for constant $\eps$, the transformation reduces to~\eqref{generalizedsym}. To see how this ambiguity affects the Noether current, consider a local transformation of the form
\begin{equation}
\udelta\psi^i(x)=\eps(x)\df^i[\psi,x](x)+\de_\m\eps(x)\dfg^{i\m}[\psi,x](x)\;,
\label{noethersymde}
\end{equation}
where $\dfg^{i\m}[\psi,x]$ is an arbitrary local function of the fields and their derivatives. To keep things simple, I have used the evolutionary form of the transformation where the coordinates $x^\m$ do not change. Applying~\eqref{noethertrick}, one finds that the new term in the transformation rule for $\psi^i$ shifts the Noether current by a term linear in $\dfg^{i\m}[\psi,x]$,
\begin{equation}
J^\m[\psi,x](x)=J^\m[\psi,x](x)\Bigr\rvert_{\dfg=0}+\dfg^{i\m}[\psi,x](x)\frac{\udelta S}{\udelta\psi^i(x)}\;.
\label{noetherimproved}
\end{equation}
The argument is easily generalized to transformations depending on arbitrarily high derivatives of $\eps(x)$. All the ensuing corrections to the current are proportional to $\udelta S/\udelta\psi^i$, which defines the EoM for the fields. The moral is that the ambiguity in the definition of the localized symmetry transformation leads to new contributions to the Noether current that vanish on-shell~\cite{Brauner2020a}. Neither this ambiguity does therefore affect the conservation law~\eqref{noether}.

\begin{illustration}%
\label{ex:canonicalEM}%
By the naive replacement $\eps\to\eps(x)$, we can localize the action of spacetime translation in the $x^\n$-direction as $\udelta_\n\psi^i(x)=-\eps(x)\de_\n\psi^i(x)$. Assuming that the Lagrangian density does not depend explicitly on the coordinates and on higher than first derivatives of $\psi^i$,~\eqref{noethertrick} gives us the set of currents
\begin{equation}
T^\m_{\phantom\m\n}=\d^\m_\n\La-\PD{\La}{(\de_\m\psi^i)}\de_\n\psi^i\;.
\label{EMcanonical}
\end{equation}
This is the familiar \emph{canonical energy--momentum} (EM) \emph{tensor}. (The unusual overall sign is a consequence of the conventions used here.) This EM tensor is known to possess some undesired features that have inspired various ``improvements'' with a history as long as Noether's theorem itself. For an illustration, consider a relativistic theory of a scalar field $\p$ and a vector field $A_\m$, defined by
\begin{equation}
\La[\p,A]=A^\m\de_\m\p-\frac12A^\m A_\m\;.
\label{AphiLag}
\end{equation}
The canonical EM tensor of this theory is $T^{\m\n}=g^{\m\n}\La-A^\m\de^\n\p$; I have raised the second index with the flat Minkowski metric $g^{\m\n}$. This EM tensor is notably not symmetric, which is a general trait shared by theories that contain nonscalar fields. This is quite troubling, if only for the fact that~\eqref{AphiLag} is actually the free massless scalar theory in disguise. Indeed, using the EoM for the vector field, $A_\m=\de_\m\p$, turns the Lagrangian into $\La[\p]=(1/2)(\de_\m\p)^2$.

One way to solve this problem is to note that the naive local translation, $\udelta_\n A_\m(x)=-\eps(x)\de_\n A_\m(x)$, is not compatible with the EoM for $A_\m$. Under a local translation, $A_\m$ should transform as a covariant vector field. Let us therefore try
\begin{equation}
\udelta_\n\p(x)=-\eps(x)\de_\n\p(x)\;,\qquad
\udelta_\n A_\m(x)=-\eps(x)\de_\n A_\m(x)-A_\n(x)\de_\m\eps(x)\;.
\label{Aphitransfo}
\end{equation}
This is a generalized local transformation of the type~\eqref{noethersymde}, which leads to the correspondingly modified EM tensor, $\tilde T^{\m\nu}=g^{\m\n}\La+A^\m A^\n-(A^\m\de^\n\p+A^\n\de^\m\p)$. This is symmetric off-shell, which is ultimately because the transformation~\eqref{Aphitransfo} has a well-defined geometric meaning~\cite{Brauner2020a}. In the language of Appendix~\ref{appsec:manifoldmaps},~\eqref{Aphitransfo} represents the Lie derivative of $\p$ and $A_\m$ along the vector field $-\eps(x)\de_\n$.
\end{illustration}

The above example shows that the ambiguity in the definition of the local transformation used to produce the Noether current is not just a nuisance. It may be used as a tool to construct ``improved'' Noether currents with desired properties. Quite recently, this idea was exploited to systematically construct EM tensors tailored to Lorentz, scale and conformal symmetry~\cite{Kourkoulou2022a}. For further background and references on Noether's theorem and improvement of Noether currents, see for instance~\cite{Gieres2022}.


\subsection{Tensor Conservation Laws}
\label{subsec:noethertensor}

Before closing the discussion of Noether's theorem, let me stress that~\eqref{noether}, while generic, is not the only form the ensuing conservation law may take. It may happen that the variation of the action~\eqref{noethertrick}, or its parts, only depends on higher than first derivatives of $\eps(x)$. This leads to conservation laws with likewise higher than first derivatives of a generalized tensor current.

\begin{illustration}
Consider the theory of a free real \emph{Lifshitz scalar} field $\p$~\cite{Griffin2013a},
\begin{equation}
\La[\p]=\frac12(\de_0\p)^2-\frac12(\de_r\de_s\p)^2\;.
\end{equation}
This Lagrangian is, just like that of the massless relativistic scalar theory, invariant under the shift $\p\to\p+\eps$. Making the replacement $\eps\to\eps(x)$ leads to the variation
\begin{equation}
\udelta S=\int\D^D\!x\,\bigl\{J^0[\p](x)\de_0\eps(x)-J^{rs}[\p](x)\de_r\de_s\eps(x)\bigr\}\;,
\end{equation}
where $J^0[\p]\equiv\de_0\p$ and $J^{rs}[\p]\equiv\de_r\de_s\p$. The corresponding on-shell conservation law takes the form
\begin{equation}
\de_0J^0[\p]+\de_r\de_sJ^{rs}[\p]=0\;.
\label{dipole}
\end{equation}
This is, unsurprisingly, equivalent to the EoM for $\p$. 
\end{illustration}

From now on I will always assume that the coordinates $x^\m$ include both space and time. We can then integrate the local conservation law~\eqref{noether} over \emph{space}. Assuming asymptotic behavior of the fields at infinity such that the boundary term produced by integrating $\de_r J^r[\psi,x]$ vanishes, we conclude that the integral charge
\begin{equation}
Q\equiv\int\D^d\!\vec x\,J^0[\psi,x](x)
\end{equation}
is time-independent. Curiously, a local conservation law of the type~\eqref{dipole} is stronger. Namely, apart from conservation of $Q$, it also implies time-independence of
\begin{equation}
Q^r\equiv\int\D^d\!\vec x\,x^rJ^0[\psi,x](x)\;.
\end{equation}
This looks like the dipole moment of the charge distribution defined by the density $J^0[\psi,x]$. Conservation laws of the dipole type~\eqref{dipole} have recently attracted considerable attention due to their relevance for so-called \emph{fracton} phases of matter. The reader is referred to~\cite{Seiberg2020a,Grosvenor2022a} for more details about this intriguing subject.


\section{Symmetry and Conservation Laws in~Hamiltonian~Formalism}
\label{sec:SSBHamiltonian}

The Lagrangian formalism provides a natural framework for the discussion of symmetries and conservation laws, yet it also offers numerous other benefits. For applications to \emph{effective field theory} (EFT), it is particularly important that the formalism works without change for Lagrangians depending on higher field derivatives. Also, it makes the transition from classical to quantum theory straightforward within the path-integral approach to quantization. However, we will see in Chap.~\ref{chap:SSB} that the most striking manifestations of \emph{spontaneous symmetry breaking} (SSB) include peculiar properties of the quantum ground state and the spectrum of excitations above it. Such features are not easily addressed using path integrals; here it is more natural to use the operator approach to quantization.

With this in mind, I will now devote some space to the Hamiltonian formalism, which is the classical counterpart of the operator language of quantum field theory. This comes at the cost of having to restrict the discussion to theories whose action functional depends only on the first time derivatives of fields. In return, we gain the reverse of Noether's theorem, allowing us to extract the corresponding symmetry from a given conservation law. What follows is a brief survey of the \emph{symplectic} formulation of the Hamiltonian formalism, adapted to local field theory. The background developed here will prove useful in Chap.~\ref{chap:effLagrangian}. For a further generalization of the approach outlined below, see for instance~\cite{Jose1998a}. A reader interested in modern developments of this approach is advised to consult~\cite{Gieres2021}.


\subsection{Symplectic Formulation of Hamiltonian Dynamics}
\label{subsec:symplectic}

The starting point is a manifold called the \emph{target space} of the given theory. This manifold carries a geometric structure fixed by a (locally defined) 1-form $\o$ called the \emph{symplectic potential}. I will use the notation $\o\equiv\o_i(\x)\D\x^i$ where $\x^i$ is a set of (local) coordinates on the target space. The \emph{phase space} of the theory consists of all time-independent fields taking values in the target space. With some abuse of notation, I will denote such fields as $\x^i(\vec x)$. The Hamiltonian of the theory is a local functional on the phase space, $H=\int\D^d\!\vec x\,\Ha[\xi,\vec x](\vec x)$. As indicated by the square bracket notation, the Hamiltonian density $\Ha$ is a local function of the fields and their spatial derivatives, possibly also depending explicitly on spatial coordinates.

The action is now a functional of trajectories on the phase space, which I with some further abuse of notation denote as $\x^i(\vec x,t)\equiv\x^i(x)$. It is fixed by the choice of symplectic potential and Hamiltonian,
\begin{equation}
S=\int\D^D\!x\,\bigl\{\o_i(\x(x))\dot\x^i(x)-\Ha[\x,\vec x](x)\bigr\}\;.
\label{sympaction}
\end{equation}
The EoM for the variational principle based on~\eqref{sympaction} takes the form
\begin{equation}
\Omega_{ij}(\x(x))\dot\x^j(x)=\frac{\udelta H}{\udelta\x^i(x)}\;,\qquad
\Omega_{ij}(\x)\equiv\PD{\o_j(\x)}{\x^i}-\PD{\o_i(\x)}{\x^j}\;.
\label{sympHameq}
\end{equation}
Here $\udelta/\udelta\x^i(x)$ indicates taking the variational derivative of a functional on the phase space with respect to $\x^i(\vec x)$, and substituting the trajectory $\x^i(x)$ in the result. The antisymmetric matrix $\Omega_{ij}(\x)$ collects the components of the \emph{symplectic 2-form}, $\Omega\equiv(1/2)\Omega_{ij}(\x)\D\x^i\w\D\x^j$. This is an object of central importance for the Hamiltonian approach to mechanics and field theory. In the language of Appendix~\ref{appsec:cohomology}, it is a closed 2-form, for it is related to the symplectic potential via $\Omega=\D\o$. Moreover, the matrix $\Omega_{ij}(\x)$ is assumed to be nonsingular so that~\eqref{sympHameq} constitutes a complete set of evolution equations for the fields $\x^i$. One can then rewrite~\eqref{sympHameq} as
\begin{equation}
\dot\x^i(x)=\Omega^{ij}(\x(x))\frac{\udelta H}{\udelta\x^j(x)}\;,
\label{sympevolution}
\end{equation}
where $\Omega^{ij}(\x)$ is the matrix inverse of $\Omega_{ij}(\x)$.

\begin{illustration}%
Consider a Lagrangian field theory of $n$ scalar fields $\p^i$, taking values in $\R^n$. The target space of the Hamiltonian description of this theory is $\R^n\times\R^n$, spanned on the fields $\p^i$ and their conjugate momenta $\pi_i$. The Lagrangian and Hamiltonian densities are related by the Legendre transform, $\La=\pi_i\dot\p^i-\Ha$. Matching this to~\eqref{sympaction} allows one to identify the symplectic potential, $\o(\p,\pi)=\pi_i\D\p^i$. The symplectic 2-form in turn becomes $\Omega(\p,\pi)=\D\pi_i\w\D\p^i$, and the EoM~\eqref{sympHameq} reduces to the familiar form
\begin{equation}
\dot\p^i(x)=\frac{\udelta H}{\udelta\pi_i(x)}\;,\qquad
\dot\pi_i(x)=-\frac{\udelta H}{\udelta\p^i(x)}\;.
\end{equation}
In general, coordinates $\p^i,\pi_i$ on the target space in which the symplectic 2-form acquires the simple form $\Omega(\p,\pi)=\D\pi_i\w\D\p^i$ are called \emph{Darboux coordinates}. By the \emph{Darboux theorem}, such coordinates exist at least locally on any manifold endowed with a symplectic 2-form (see for instance Sect.~43 of~\cite{Arnold1989a}). The global existence of Darboux coordinates on the target space is however not guaranteed. In fact, it is ruled out whenever the symplectic 2-form is closed but not exact. \refex{ex:sympspin} below provides a nontrivial illustration of this possibility. The geometry of manifolds carrying a symplectic structure is therefore locally identical to that of $\R^n\times\R^n$, but may be globally nontrivial.
\end{illustration}

With the symplectic 2-form at hand, one can define the \emph{Poisson bracket} of any two local functionals $F,G$ on the phase space,
\begin{equation}
\{F,G\}\equiv\int\D^d\!\vec x\,\Omega^{ij}(\x(\vec x))\frac{\udelta F}{\udelta\x^i(\vec x)}\frac{\udelta G}{\udelta\x^j(\vec x)}\;.
\label{poisson}
\end{equation}
A special case is the fundamental Poisson bracket of the phase space coordinates,
\begin{equation}
\{\x^i(\vec x),\x^j(\vec y)\}=\Omega^{ij}(\x(\vec x))\d^d(\vec x-\vec y)\;.
\label{poissonfund}
\end{equation}
In local Darboux coordinates, \eqref{poisson} reduces to the textbook definition of Poisson bracket in terms of derivatives with respect to canonical coordinates $\smash{\psi^i(\vec x)}$ and momenta $\smash{\pi_i(\vec x)}$. Likewise, \eqref{poissonfund} generalizes the fundamental brackets $\{\psi^i(\vec x),\psi^j(\vec y)\}=\{\pi_i(\vec x),\pi_j(\vec y)\}=0$ and $\{\psi^i(\vec x),\pi_j(\vec y)\}=\d^i_j\d^d(\vec x-\vec y)$ in a way independent of the choice of coordinates on the target space. It is easy to check that~\eqref{poisson} has the following properties. First, $\{F,G\}$ is obviously linear in both arguments and antisymmetric. Moreover, it satisfies the Leibniz (product) rule. Finally, it satisfies the Jacobi identity; this takes more effort to prove and requires using the closedness of the symplectic 2-form.

The Poisson bracket offers a very compact expression for time evolution. For any local functional $F$ on the phase space that does not depend explicitly on time,
\begin{equation}
\dot F=\int\D^d\!\vec x\,\frac{\udelta F}{\udelta\x^i(x)}\dot\x^i(x)=\{F,H\}\;,
\label{dotF}
\end{equation}
where I used~\eqref{sympevolution}. The latter is itself equivalent to $\dot\x^i=\{\x^i,H\}$. In this form, it is natural to think of time evolution as a flow on the phase space, generated by $H$.

This completes the preparation required to address the main objective of this survey of the Hamiltonian formalism: the connection of symmetries and conservation laws. First, everything said in Sect.~\ref{sec:SSBLagrangian} about Noether's theorem remains valid in the present Hamiltonian setting. For the sake of identifying the Noether current, there is no difference between the action functionals~\eqref{action} and~\eqref{sympaction}. We can therefore conclude at once that a continuous symmetry of the action~\eqref{sympaction} implies the existence of a current $J^\m[\x,\vec x]$ conserved on-shell. This current can be identified by localizing the symmetry transformation and using~\eqref{noethertrick}.

\begin{illustration}%
The Hamiltonian density of a free massless relativistic scalar field $\p$ is $\Ha[\p,\pi]=(1/2)\pi^2+(1/2)(\vec\nabla\p)^2$. The symplectic potential takes the Darboux form, $\o(\p,\pi)=\pi\D\p$. The action functional
\begin{equation}
S=\int\D^D\!x\,\Bigl\{\pi(x)\dot\p(x)-\frac12\pi(x)^2-\frac12[\vec\nabla\p(x)]^2\Bigr\}
\end{equation}
is invariant under the constant shift $\p\to\p+\eps$ with $\pi$ kept unchanged. Making the shift coordinate-dependent, the action varies by $\udelta S=\int\D^D\!x\,[\pi(x)\de_0\eps(x)-\vec\nabla\p(x)\cdot\vec\nabla\eps(x)]$. This leads to the identification of the temporal and spatial components of the Noether current as $J^0[\p,\pi]=\pi$, $J^r[\p,\pi]=-\de_r\p=+\de^r\p$. Upon using the EoM for $\p$, $\dot\p=\pi$, the current is seen to coincide with that derived in \refex{ex:shiftscalar}, $J^\m[\p]=\de^\m\p$.
\end{illustration}

What is new in the Hamiltonian framework is that we can now reverse Noether's theorem and reconstruct the symmetry from a given conservation law. By~\eqref{dotF}, a local functional $F$ that does not depend explicitly on time defines an integral of motion (conserved charge) if and only if $\{F,H\}=0$. Moreover, if $F$ and $G$ are both conserved, then so is $\{F,G\}$ thanks to the Jacobi identity. Thus, all conserved charges of the theory furnish a Lie algebra with respect to the Poisson bracket. Finally, every conserved charge generates a flow on the phase space, defined by the infinitesimal transformation
\begin{equation}
\udelta\x^i(\vec x)=\eps\{\x^i(\vec x),F\}=\eps\Omega^{ij}(\x(\vec x))\frac{\udelta F}{\udelta\x^j(\vec x)}\;.
\label{sympflow}
\end{equation}
We thus have an explicit realization of a geometric transformation, associated with a given conserved charge. It remains to demonstrate that this is in fact a symmetry of the action~\eqref{sympaction}. To that end, a short calculation using the definition of the symplectic 2-form shows that under the transformation~\eqref{sympflow}, the action varies by
\begin{equation}
\udelta S=-\eps\int\D^D\!x\,\frac{\udelta F}{\udelta\x^i(x)}\dot\x^i(x)-\eps\int\D t\,\{H,F\}(t)\;.
\end{equation}
The first term vanishes upon integration, being a total time derivative. The second term vanishes by the assumption that $F$ is conserved. This completes the proof that every conserved charge (that does not depend explicitly on time) generates a symmetry of the action.

\newpage

\begin{illustration}%
\label{ex:sympspin}%
The low-energy physics of ferromagnets can be encoded in a continuous field theory whose basic degree of freedom is the local spin (magnetization) density. In the first approximation, fluctuations of the ferromagnetic equilibrium correspond to local changes in spin orientation, while the magnitude $M$ of spin density remains constant. The target space is thus equivalent to the sphere $S^2$. The local spin variable $n^i(\vec x)$ is a unit three-vector field that satisfies the fundamental Poisson bracket
\begin{equation}
\{n^i(\vec x),n^j(\vec y)\}=\frac1M\ve^{ij}_{\phantom{ij}k}n^k(\vec x)\d^d(\vec x-\vec y)\;,
\label{spinbracket}
\end{equation}
reflecting the Lie algebra of angular momentum. This is a prominent example of a system with a phase space whose symplectic structure is largely fixed by symmetry alone. Other examples can be found in~\cite{Dzyaloshinskii1980}.

For any Hamiltonian that is a local functional of $n^i$, the ensuing EoM is $\dot n^i=\{n^i,H\}$. Using the chain rule, this can be put into a neat vector form\footnote{This is a slight abuse of notation. I am using boldface italics both for spatial vectors living in $\R^d$ with arbitrary $d$ and for the spin vector living on $S^2\subset\R^3$.}
\begin{equation}
\dot{\vec n}(x)=\frac1M\frac{\udelta H}{\udelta\vec n(x)}\times\vec n(x)\;,
\label{LLaux}
\end{equation}
known as the \emph{Landau--Lifshitz equation}. The concrete form of the Hamiltonian depends on the physical system. For isotropic ferromagnets, the simplest choice is quadratic in gradients of $\vec n(\vec x)$, as suggested by the Landau theory of phase transitions,
\begin{equation}
H=\int\D^d\!\vec x\,\frac{\vr_\mathrm{s}}2\delta_{ij}\vec\nabla n^i(\vec x)\cdot\vec\nabla n^j(\vec x)\;.
\label{LLHam}
\end{equation}
Here $\vr_\mathrm{s}$ is the so-called spin stiffness. The EoM then reduces to
\begin{equation}
\dot{\vec n}(x)=\frac{\vr_\mathrm{s}}M\vec n(x)\times\vec\nabla^2\vec n(x)\;.
\label{LLequation}
\end{equation}

Matching~\eqref{LLaux} to the general EoM~\eqref{sympHameq} allows one to extract the symplectic 2-form on $S^2$, corresponding to the Poisson bracket~\eqref{spinbracket},
\begin{equation}
\Omega=-\frac M2\ve_{ijk}n^i\D n^j\w\D n^k\;.
\label{spinsymp2form}
\end{equation}
This is, up to normalization, the area 2-form on $S^2$, see \refex{ex:embeddingmetric} in Appendix~\ref{appsec:manifoldmaps}. The area 2-form is closed but not exact. Hence the sphere $S^2$ is an example of a symplectic manifold on which globally well-defined Darboux coordinates do not exist. This fact has deep consequences for the topological properties of ferromagnets. As to symmetry, the symplectic 2-form~\eqref{spinsymp2form} and the Hamiltonian~\eqref{LLHam} are both manifestly invariant under $\gr{SO}(3)$ rotations of the spin variable $n^i$. This symmetry corresponds to the conservation of total spin, $\vec S=M\int\D^d\!\vec x\,\vec n(\vec x)$.

Ferromagnets are fascinating materials that exhibit many of the nontrivial features of SSB. This example is just a taster; in Sect.~\ref{sec:spinwaves}, I will serve the reader a much more thorough discussion of the low-energy physics of spin systems.
\end{illustration}


\subsection{Symmetry in Quantum Physics}
\label{subsec:SSBquantum}

Our discussion of symmetries in field theory has been strictly classical so far. The implementation of symmetries in quantum field theory comes with numerous subtleties. Some of these are related to SSB and I will return to them in the next chapter. Here I will therefore just briefly outline the transition from classical to quantum physics. As is well known, this transition is streamlined in the Hamiltonian formalism. Local functionals on the phase space $F,G,\dotsc$ are replaced with operators $\hat F,\hat G,\dotsc$ on the Hilbert space of physical states. The commutator of these operators is then obtained from the Poisson bracket of their classical counterparts, roughly speaking, by the replacement $\{F,G\}\to-\I[\hat F,\hat G]$. In the following, I will drop the hat on operators, but otherwise closely follow the Hamiltonian representation of symmetries outlined above. For the sake of simplicity, I will only consider conserved charges that do not explicitly depend on time.

With this qualification, a Hermitian operator $Q$ represents a conserved charge of a quantum system if and only if it commutes with the Hamiltonian, $[Q,H]=0$. By analogy with~\eqref{sympflow}, the conserved charge generates a flow on the algebra of observables. For a given Hermitian operator $A$, the  shift induced by $Q$ is
\begin{equation}
\udelta A=-\I\eps[A,Q]\;.
\label{operatorflow}
\end{equation}
This can be extended to a transformation with a finite parameter $\eps$. One can think of such a transformation as a map $A\to A(\eps)$ where $A(\eps)$ satisfies the ``flow equation''
\begin{equation}
\OD{A(\eps)}{\eps}=-\I[A(\eps),Q]=\I[Q,A(\eps)]\;.
\label{Aflow}
\end{equation}
This has the formal solution
\begin{equation}
A(\eps)=\E^{\I\eps Q}A(0)\E^{-\I\eps Q}\;.
\label{Aeps}
\end{equation}
The duality between the Schr\"odinger and Heisenberg pictures of quantum mechanics suggests an interpretation of~\eqref{Aeps} in terms of a formal unitary operator $U(\eps)\equiv\E^{-\I\eps Q}$, acting on the Hilbert space of states. Instead of transforming operators via~\eqref{Aeps}, we could then equivalently transform physical states by $U(\eps)$. In either case, the fact that $Q$ is a conserved charge is reflected by the invariance of the Hamiltonian under the transformation generated by $Q$, $H(\eps)=H$ or $[U(\eps),H]=0$.

\begin{watchout}%
As I will demonstrate in Sect.~\ref{sec:SSBsubtleties}, when the symmetry in question is spontaneously broken, the operator $U(\eps)$ may in fact not exist. This is one of the quirks of SSB. The symmetry may not be realized by unitary operators on the Hilbert space. As we will see, this feature is related to the nontrivial structure of the Hilbert space in presence of spontaneously broken symmetries. Even then, it makes sense to ask how the symmetry affects results of measurements. The transformation of physical observables under the symmetry remains well-defined and is still expressed by the flow equation~\eqref{Aflow}.
\end{watchout}

The starting point of the discussion in this chapter was that a definition of symmetry requires a transformation and an object. All the objects we have worked with so far---actions, Hamiltonians, and EoM---capture the dynamics of the entire physical system. It is however no less interesting and useful to study the symmetries of a particular \emph{state} of the system, whether classical or quantum. This brings us to the realm of SSB, which will be addressed in detail in the next chapter.


\bibliographystyle{spphys}
\bibliography{references}
\chapter{Spontaneous Symmetry Breaking}
\label{chap:SSB}

\abstract*{The roots of the modern understanding of symmetries in physics can be traced to the work of Sophus Lie on transformations of differential equations in the 19\textsuperscript{th} century. In this chapter, the history of the subject is used to emphasize the distinction between the symmetry of the dynamics of a physical system and the symmetry of its specific state. The central concept of spontaneous symmetry breaking is then introduced as the discrepancy between the two notions of symmetry. It is shown how the presence of spontaneous symmetry breaking can be detected through the sensitivity of measurements on the physical state to external perturbations. This naturally leads to the concepts of the order parameter and the vacuum manifold. The rest of the chapter is devoted to highlighting some subtle features of spontaneous symmetry breaking. These include the properties of the ground state and the tower of low-lying excited states in a finite volume, and the absence of well-defined unitary operators representing broken symmetry in the thermodynamic limit.}


In this chapter, I will finally spell out a formal definition of \emph{spontaneous symmetry breaking} (SSB). While the previous chapter was largely phrased in terms of classical field theory, I will now switch entirely to the quantum-theoretic language. This is more than justified by Sect.~\ref{sec:SSBsubtleties}, which gives the reader a flavor of the many subtleties, associated with SSB in quantum systems. Some further details on the physical aspects of SSB can be found for instance in the recent lecture notes~\cite{Beekman2019a}. A mathematically oriented reader will find the classic book~\cite{Strocchi2021} a unique source of additional information.

To motivate what comes below, let me briefly return to the roots of Lie group theory in the study of symmetries of differential equations. Suppose that we have a set of (partial) differential equations invariant under some group $G$ of transformations. A given solution to the equations can fall into one of two classes:
\begin{description}[(ii)]
\item[(i)] The solution is not invariant under $G$. This is the generic case, in which one can use the symmetry to obtain new solutions by the action of transformations from $G$ on the solution already known.
\item[(ii)] The solution is invariant under $G$. Such solutions are special, and are usually easier to find than a general solution to the same equations. The reason for this is that assuming a priori $G$-invariance of the solution reduces the number of independent variables of the differential equations.
\end{description}
The situation in quantum physics maps closely to what one does in the context of the theory of differential equations. The major difference is that one tends to focus on the eigenstates of the quantum Hamiltonian as the primary tool to investigate physical observables. For this reason, it is common to start with the ground state, or at least a metastable equilibrium state, and study the excitations above it. This is the quantum counterpart of the theory of small oscillations in classical mechanics~\cite{Goldstein2013a}. We will see that quantum physics offers natural analogs to both of the above-mentioned cases (i) and (ii) one encounters in differential equations. In both cases, symmetry has profound consequences for the structure of the excitation spectrum. For most of this chapter, I will focus on the symmetry properties of the equilibrium state. I will return to the excitation spectrum in Sect.~\ref{sec:SSBsubtleties}, and then again in detail in Chap.~\ref{chap:NGbosons}.


\section{Physical State and Its Symmetry}
\label{sec:SSBstate}

A proper discussion of SSB requires a number of new concepts. In order to make the narrative as natural as possible, I will augment it with some simple examples from condensed-matter and statistical physics. This will highlight the close connection between SSB and phase transitions. To parallel the language of statistical mechanics, I will represent the state of a quantum system with a \emph{density operator} (or \emph{density matrix}) $\dm$. Readers who need to refresh their memory of the density operator formalism will find an undergraduate-level introduction for instance in~\cite{Ballentine1998a,Basdevant2002a,Balian2007a}.

We already know that symmetries of quantum systems are represented, if only formally, by unitary operators. An operator $U$ is said to constitute a symmetry of the (pure or mixed) state $\dm$ if the density operators $\dm$ and $U\dm\he U$ are indistinguishable by any measurement. The latter means equal probabilities for any specific outcome of any measurement, and by extension equal averages of all observables. Insofar as our description of the state of the system is free of redundancies, this implies that the density operators must be equal,
\begin{equation}
U\dm\he U=\dm\qquad
\text{(symmetry of a state)}\;.
\label{symdef}
\end{equation}
Suppose that $\dm$ is a pure state, that is $\dm=\ket\psi\bra\psi$ for some normalized ket-vector $\ket\psi$ in the Hilbert space. Then a unitary operator $U$ represents a symmetry of $\dm$ if and only if $\ket\psi$ is an eigenstate of $U$. One side of the equivalence is obvious: if $\ket\psi$ is an eigenstate of $U$, then $\smash{U\dm\he U=\dm}$. To prove the opposite implication, write $\abs{\bra\psi U\ket\psi}^2$ as $\smash{\bra\psi U\ket\psi\bra\psi\he U\ket\psi=\bra\psi U\dm\he U\ket\psi=1}$. Thus, the states $\ket\psi$ and $U\ket\psi$ are both normalized to unity and have a unit overlap, which is only possible if they are equal up to a phase. It follows that $\ket\psi$ is an eigenstate of $U$.

In the following, I will frequently use a shorthand notation for the average of an observable $A$ in the state $\dm$,
\begin{equation}
\vev{A}_\dm\equiv\tr(\dm A)\;.
\end{equation}
It is a simple corollary of~\eqref{symdef} that a symmetry transformation of observables, $A\to\he UAU$, does not affect their average in symmetric states. To see this, write $\smash{\vev{\he UAU}_\dm}$ as $\smash{\tr(\dm\he UAU)=\tr(U\dm\he UA)=\tr(\dm A)}$, which is just $\vev{A}_\dm$. This conclusion holds for discrete and continuous symmetries alike.


\subsection{Broken and Unbroken Symmetry}
\label{subsec:SSBbrokenunbroken}

Suppose now that the Hamiltonian of a system is invariant under a set of transformations that span a group $G$. These are represented, even if just formally, by a set of unitary operators $U(g)$, $g\in G$. A randomly chosen state $\dm$ will not be symmetric under all of these. Those transformations that are symmetries of $\dm$ will span a subgroup of $G$ called the \emph{unbroken subgroup} of $\dm$, denoted by
\begin{equation}
H_\dm\equiv\{g\in G\,\vert\,U(g)\dm\he{U(g)}=\dm\}\;.
\end{equation}
All other elements of $G$ are said to be \emph{spontaneously broken} in the state $\dm$. This is our formal definition of SSB.

\begin{illustration}%
\label{ex:spin12particle}%
Pure spin states of a spin-$1/2$ particle comprise a two-dimensional Hilbert space isomorphic to $\C^2$. This Hilbert space carries a representation of the spin group, $G\simeq\gr{SU}(2)$, generated by the spin operator $\vec S$. Consider the projection of spin to a direction, defined by the unit vector $\vec n=(\sin\t\cos\vp,\sin\t\sin\vp,\cos\t)$, 
\begin{equation}
\skal nS=\frac\hbar2\skal n\pau=\frac\hbar2
\begin{pmatrix}
\cos\t & \E^{-\I\vp}\sin\t\\
\E^{\I\vp}\sin\vp & -\cos\t
\end{pmatrix}\;.
\end{equation}
Here $\t,\vp$ are the standard spherical angles and $\vec\pau$ is the vector of Pauli matrices. The projected spin operator has an eigenvector $\ket{\vec n,+}$ with eigenvalue $+\hbar/2$,
\begin{equation}
\skal nS\ket{\vec n,+}=\frac\hbar2\ket{\vec n,+}\;,\qquad
\ket{\vec n,+}=\begin{pmatrix}
\cos\t/2\\
\E^{\I\vp}\sin\t/2
\end{pmatrix}\;.
\end{equation}
Any nonzero vector in $\C^2$ can, up to an overall factor, be cast in this form with a suitable choice of $\t,\vp$. Hence, any nonzero vector in this Hilbert space is an eigenstate of a projection of the spin operator to some direction. We conclude that for any pure state of a spin-$1/2$ particle, the unbroken subgroup is isomorphic to $\gr{U}(1)$, and consists of spin rotations generated by $\skal nS$.
\end{illustration}

How do we check whether a given symmetry is spontaneously broken in a given state $\dm$? For the density operators $\dm$ and $U\dm\he U$ to be different, there must be at least one observable $A$ whose average in the two states differs. This observable must itself not be invariant under the symmetry, as $\he UAU=A$ would imply
\begin{equation}
\vev{A}_{U\dm\he U}=\tr(U\dm\he UA)=\tr(\dm\he UAU)=\tr(\dm A)=\vev{A}_\dm\;.
\end{equation}
In order to detect SSB, one must therefore use an observable that itself breaks the symmetry. This fairly simple observation forms a conceptual foundation of the approach to SSB developed in Sect.~\ref{sec:SSBperturbation}.

One observable that is by definition invariant under all the operators $U(g)$ from the symmetry group is the Hamiltonian itself. Hence, for those $g\in G$ that are spontaneously broken, the state $U(g)\dm\he{U(g)}$ differs from $\dm$ but has the same average energy. This presents a serious conundrum. Suppose that $\dm$ is a purported ground state of a quantum system. From quantum mechanics, we are used to thinking about the ground state as a unique vector in the Hilbert space. But now $U(g)\dm\he{U(g)}$ for any $g\notin H_\dm$ is a different state of the same energy. How do we know which of the different states having the same energy to choose as \emph{the} ground state?

We noticed already in Chap.~\ref{chap:ourfirstmodel} that SSB implies the existence of degenerate ground states. This property is so universal that it earned the privilege of being my first ``moral lesson'' in Sect.~\ref{sec:firstmodelmorals1}. But now our problem is even worse. In the classical analysis of Chap.~\ref{chap:ourfirstmodel}, the various candidate ground states could be uniquely labeled by the expectation value of a scalar field. In the present fully quantum setting, we have not only the set of density operators $U(g)\dm\he{U(g)}$ with $g\in G$. We can even make linear combinations of these states. In the desire to identify a unique ground state, we might wish to take a ``democratic average'' of all the would-be ground states, connected by transformations from $G$. This is an idea that certainly deserves at least consideration, even if it will eventually turn out to be physically incorrect. However, the related mathematical procedure of averaging over different elements of the symmetry group will actually prove quite useful.


\subsection{Symmetrization by Group Averaging}
\label{subsec:SSBlemma}

When performing an average over the symmetry group, I will repeatedly make use of a simple statement that I will formulate as a standalone lemma. Let $G$ be a finite group or a compact Lie group, and $U(g)$ be its unitary representation on a finite-dimensional vector space. Then
\begin{equation}
\proj\equiv\frac1{|G|}\sum_{g\in G}U(g)
\label{lemma}
\end{equation}
is a projector to the subspace of vectors invariant under $G$. In particular, if the representation of $G$ by $U(g)$ does not contain in its decomposition any singlet (that is, a trivial one-dimensional representation of $G$), then the action of $\proj$ on any vector gives zero. The proof of the lemma is very simple and I will therefore leave it up to the reader. Let me just add a comment on the sum over group elements. Equation~\eqref{lemma} as it stands is only valid for finite groups; $|G|$ then denotes the number of group elements. For compact Lie groups, the sum has to be replaced by group integration with an invariant measure on $G$; $|G|$ then stands for the volume of $G$ with respect to this measure. An interested reader will find more details in Chap.~3 of~\cite{Barut1977a}. In the following, I will pragmatically use a sum over $g\in G$ as in~\eqref{lemma} without repeating this disclaimer.

Let us now get back to density operators. The action of the group $G$ on these is defined by $g:\dm\to U(g)\dm\he{U(g)}$. This suggests that we consider the following statistical average,
\begin{equation}
\bar\dm\equiv\frac1{|G|}\sum_{g\in G}U(g)\dm\he{U(g)}\;.
\label{rhosymmetrized}
\end{equation}
This average still satisfies the requirements on a density operator: it is Hermitian, positive-semidefinite and has a unit trace. Moreover, it is manifestly invariant under all transformations from $G$. It therefore represents a (mixed) physical state of the system which is perfectly symmetric under the whole group $G$, yet has the same average energy as $\dm$.

\begin{watchout}%
In case the original state is pure, $\dm=\ket\psi\bra\psi$, we have the alternative choice to symmetrize the ket-vector $\ket\psi$ rather than the density operator,
\begin{equation}
\ket\psi\to\proj\ket\psi=\frac1{|G|}\sum_{g\in G}U(g)\ket\psi\;.
\label{psiaverage}
\end{equation}
This averaged ket-vector, if well-defined, represents a pure state of the same energy as $\ket\psi$ that is symmetric under the whole group $G$. The catch is that the ket-vector may be ill-defined. By our symmetrization lemma, this happens whenever $\ket\psi$ belongs to a representation of $G$ that does not contain any singlet. Recall, for instance, \refex{ex:spin12particle}. There, the whole Hilbert space $\C^2$ constitutes an irreducible representation of the spin group $G\simeq\gr{SU}(2)$. The naive average~\eqref{psiaverage} therefore gives zero no matter which pure state we start from. On the other hand, the average density operator~\eqref{rhosymmetrized} must describe \emph{some} physical state of the spin-$1/2$ particle. By Schur's lemma, $\bar\dm$ must be proportional to the identity operator on $\C^2$. The requirement that $\tr\bar\dm=1$ then fixes the normalization so that $\bar\dm=(1/2)\id$, for any choice of the initial state $\dm$. This mixed state describes an unpolarized distribution on the Hilbert space.
\end{watchout}

It seems that we managed to have our cake and eat it. By the simple averaging procedure, we can make any state $G$-invariant without changing its energy. Have we done away with the whole notion of SSB? To see that the group averaging is not so innocuous, we need to work out the consequences. Consider a set of observables $A^i$ such that under the action of the operators $U(g)$, they transform by some (real and orthogonal) matrix representation $\rep$ of $G$,
\begin{equation}
\he{U(g)}A^iU(g)=\rep(g)^i_{\phantom ij}A^j\;,\qquad
g\in G\;.
\label{UdAU}
\end{equation}
Suppose that all these observables and their nontrivial linear combinations break (part of) the symmetry under $G$. In other words, the representation $\rep$ does not contain any singlet of $G$ in its decomposition. It then follows as a simple consequence of the definition of the density operator~\eqref{rhosymmetrized} and the symmetrization lemma that
\begin{equation}
\begin{split}
\vev{A^i}_{\bar\dm}&=\tr(\bar\dm A^i)=\frac1{|G|}\sum_{g\in G}\tr\bigl[U(g)\dm\he{U(g)}A^i\bigr]\\
&=\frac1{|G|}\sum_{g\in G}\tr\bigl[\dm\he{U(g)}A^iU(g)\bigr]=\frac1{|G|}\sum_{g\in G}\rep(g)^i_{\phantom ij}\vev{A^j}_{\dm}=0\;.
\label{noOPatT0}
\end{split}
\end{equation}
We could have anticipated this: in the $G$-invariant state $\bar\dm$, the average of any observable breaking the symmetry vanishes.

Why should this be bad? Any quantum system possesses a distinguished state invariant under its symmetry group $G$: the thermodynamic equilibrium. This is represented by the canonical density operator
\begin{equation}
\dm_\b\equiv\frac1Z\E^{-\b H}\;,
\end{equation}
where $\b$ is the inverse temperature and $Z\equiv\tr\exp(-\b H)$ the canonical partition function. The claim therefore is that in thermodynamic equilibrium, any observable breaking the symmetry of the system must average to zero. There cannot be any SSB in thermodynamic equilibrium. This would be very pretty, were it not in a blatant contradiction with empirical evidence.

\begin{figure}[t]
\sidecaption[t]
\includegraphics[width=2.9in]{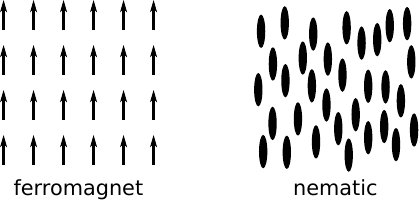}
\caption{Comparison of the symmetry-breaking states of ferromagnets and nematic liquid crystals. The symmetry of the former corresponds to that of a regular lattice of oriented, mutually aligned spins. Nematics, on the other hand, can be represented by mutually aligned undirected rods with random positions}
\label{fig:orderparameter}
\end{figure}

\begin{illustration}%
Solid ferromagnets can be visualized in terms of a lattice of mutually aligned spins (see the left panel of Fig.~\ref{fig:orderparameter}). The ferromagnetic state obviously breaks rotational symmetry. Moreover, the presence of the crystal lattice also breaks spatial translations. Both rotations and translations are fundamental symmetries that are expected to be preserved by the microscopic dynamics of any material. Hence ferromagnetism certainly is a macroscopic manifestation of SSB.

In nematic liquid crystals (right panel of Fig.~\ref{fig:orderparameter}), long organic molecules are aligned along the same direction. However, the positions of the molecules are random as in ordinary liquids.  As far as symmetry is concerned, the molecules can be represented by (undirected) rigid rods or ellipsoids with randomized positions. Hence the nematic state breaks rotational symmetry but not spatial translations. See Sect.~2.7 of~\cite{Chaikin1995a} for basic phenomenology of liquid crystals.
\end{illustration}

One of the greatest accomplishments of physics is the discovery that the fundamental laws of nature possess a high degree of symmetry. How is it then possible at all that we observe so many manifestly asymmetric macroscopic states of matter? There seem to be two logical possibilities to avoid a contradiction:
\begin{itemize}
\item The observed asymmetric states are not in thermodynamic equilibrium, and thus may be at most metastable.
\item The asymmetry of the observed states is a result of perturbations breaking the symmetry of the microscopic interactions.
\end{itemize}
Both of these possibilities are relevant. In order to understand their role, we have to dive deeper into the thermodynamics of physical systems with SSB. In the next section, I will focus solely on the canonical density operator $\dm_\b$ as the state of greatest importance for macroscopic observations.


\section{Effect of External Perturbations}
\label{sec:SSBperturbation}

In Sect.~\ref{subsec:SSBlemma}, I introduced the set of operators $A^i$ as mere observables. Let us now see what happens if we allow them to actually affect the dynamics of the system. To that end, I couple each $A^i$ to an external field $\l_i$ and perturb the Hamiltonian of the system by a term linear in both,
\begin{equation}
H(\l)\equiv H-\l_iA^i\;.
\end{equation}
In the case of a spin-$1/2$ particle discussed in \refex{ex:spin12particle}, $A^i$ can be chosen for instance as the components of the spin operator. The perturbations $\l_i$ can then be interpreted as an external magnetic field. In general, one can think of $\l_i$ as a set of ``chemical potentials'' associated with the observables $A^i$. In presence of the perturbations, the canonical density operator is deformed to
\begin{equation}
\dm_{\b,\l}=\frac1{Z(\l)}\E^{-\b H(\l)}\;,
\end{equation}
and the modified partition function reads $\smash{Z(\l)=\tr\E^{-\b H(\l)}=\tr\exp[-\b(H-\l_iA^i)]}$. The symmetry of the Hamiltonian $H$ under the group $G$ is reflected in the invariance of the partition function under the transformation $\smash{\l_i\xrightarrow{g}\l_j\rep(g)^j_{\phantom ji}}$ of its variables. This follows from a simple manipulation,
\begin{equation}
\begin{split}
Z(\l)&=\tr\bigl[\he{U(g)}\E^{-\b H(\l)}U(g)\bigr]=\tr\exp\bigl[-\b\he{U(g)}H(\l)U(g)\bigr]\\
&=\tr\exp\bigl\{-\b\bigl[H-\l_i\rep(g)^i_{\phantom ij}A^j\bigr]\bigr\}=Z(\l\rep(g))\;.
\end{split}
\end{equation}
It is convenient to trade the partition function for the free energy $F(\l)$ by $Z(\l)\equiv\smash{\E^{-\b F(\l)}}$. The free energy inherits the symmetry of the partition function under $G$-trans\-for\-ma\-tions of $\l_i$. Moreover, it makes it easy to evaluate the statistical average of $A^i$ in presence of the perturbations,
\begin{equation}
\vev{A^i}_{\dm_{\b,\l}}=\frac1{Z(\l)}\tr\bigl\{A^i\exp\bigl[-\b(H-\l_jA^j)\bigr]\bigr\}=-\PD{F(\l)}{\l_i}\;.
\label{vevAlambda}
\end{equation}

One of the distinguishing features of ferromagnets is that once magnetized by an external magnetic field, the \emph{macroscopic} magnetization persists even after the external field is turned off. With this in mind, we will be particularly interested in the behavior of $\vev{A^i}_{\dm_{\b,\l}}$ when the volume $V$ of the system is large and simultaneously $\l_i$ is small. In other words, we want to take the double limit $V\to\infty$ and $\l_i\to0$. This is where interesting things start to happen.


\subsection{Taking the Thermodynamic Limit}
\label{subsec:SSBTDlimit}

In a finite volume and for any nonzero temperature, the partition function, and thus the free energy, is analytic in $\l_i$. When all the $\l_i$ are small, one can therefore perform a Taylor expansion,
\begin{equation}
F(\l)=F(0)+\at{\PD{F(\l)}{\l_i}}{\l=0}\l_i+\bigO(\l^2)\;.
\end{equation}
But the linear term in the expansion is forbidden by the $G$-invariance of the free energy combined with the symmetrization lemma. Thus, in a finite volume and at nonzero temperature, one has a well-defined limit
\begin{equation}
\lim_{\l\to0}\vev{A^i}_{\dm_{\b,\l}}=0\;.
\label{limAil0}
\end{equation}
This should not be particularly surprising.

How does the behavior of free energy as a function of $\l_i$ change when we take the limit of infinite volume or zero temperature? Some common mathematical features of these two limits may be discussed jointly. It is however the large-volume (thermodynamic) limit that is responsible for the existence of macroscopically distinct phases of matter. Only in special cases may the zero-temperature limit lead to nontrivial physics even in a finite volume; I will give an explicit example below. The general discussion of the effects of external perturbations will however be phrased in terms of the limit $V\to\infty$. The temperature will be assumed to be fixed and nonzero unless explicitly stated otherwise.

\begin{figure}[t]
\sidecaption[t]
\includegraphics[width=2.9in]{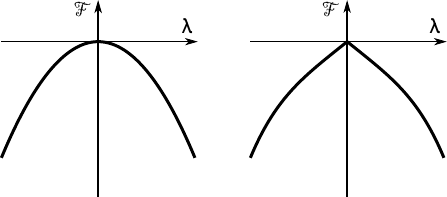}
\caption{Schematic illustration of scenarios in which the free energy density $\Fa$ is (left panel) and is not (right panel) analytic at $\l_i=0$. In the latter case, the observables $A^i$ may retain a nonzero average even in the limit $\l\to0$. Its value may however depend on the way the limit is taken}
\label{fig:freeenergy}
\end{figure}

It is convenient to trade the free energy for its density, $\Fa(\l)\equiv{F(\l)}/V$, which is expected to be well-defined for any volume, including the limit $V\to\infty$. It is sensible to assume that even in this limit, the free energy density remains continuous as a function of $\l_i$. However, the analyticity may be lost. With regard to the behavior of the average of $A^i$ around the point $\l_i=0$, there are three possible scenarios:
\begin{itemize}
\item The free energy density remains analytic in the limit $V\to\infty$. This automatically guarantees the validity of~\eqref{limAil0}. There is no SSB once the external field $\l_i$ is turned off.
\item The free energy density becomes nonanalytic in the limit $V\to\infty$, yet~\eqref{limAil0} remains valid. This is an interesting borderline case that is relevant for the physics of low-dimensional systems.
\item The free energy density becomes nonanalytic in the limit $V\to\infty$. The average $\vev{A^i}_{\dm_{\b,\l}}$ remains nonzero even if the limit $\l_i\to0$ is subsequently taken. This is the paradigm of SSB, illustrated by the right panel of Fig.~\ref{fig:freeenergy}.
\end{itemize}

\begin{watchout}%
An alert reader might wonder why I focused on possible nonanalytic behavior of $\Fa$ at $\l_i=0$, yet tacitly assumed that analyticity survives the thermodynamic limit for nonzero $\l_i$. This is intimately related to the degeneracy of the ground (equilibrium) state. What I really assumed was that adding the external fields $\l_i$ breaks the symmetry of the system sufficiently so as to make the ground state unique. I will further elaborate on this assumption in Sect.~\ref{subsec:SSBOPvacmanifold}. The point is that the free energy density typically becomes nonanalytic as a consequence of a competition of degenerate ground states with different values of $\vev{A^i}_{\dm_{\b,\l}}$. This is something one expects only at the $G$-invariant point $\l_i=0$.
\end{watchout}

It is instructive to work out at least one concrete example to verify our prediction how SSB may emerge when a particular limit of the equilibrium state is taken. Ideally, this example should be exactly solvable in order to leave no doubt about the validity of the conclusions. It is therefore time to meet the \emph{Ising model}, the most well-known and best-understood model of ferromagnetism in statistical physics. The price for all its benefits is that the limit in which the free energy density becomes nonanalytic is realized by taking the temperature to zero; the $V\to\infty$ limit itself is nearly trivial. I will largely follow the treatment in Sect.~2.2 of~\cite{Mussardo2010}.

\begin{illustration}%
\label{ex:isingmodel}%
Consider a one-dimensional chain of $N$ spin variables $\s_i$, allowed to take values $\pm1$. The Hamiltonian is given by
\begin{equation}
H=-J\sum_{i=1}^N(\s_i\s_{i+1}-1)-B\sum_{i=1}^N\s_i\;,
\label{isingham}
\end{equation}
where $\s_{N+1}$ is identified with $\s_1$ to ensure a periodic boundary condition. The (positive) coupling $J$ represents interaction of nearest-neighbor spins, and $B$ an external magnetic field. In the absence of the external field, the Hamiltonian~\eqref{isingham} has a $G\simeq\Z_2$ symmetry under the spin flip $\s_i\to-\s_i$.

The partition function of the model is easily calculable thanks to the fact that the exponential of a sum equals a product of exponentials. Thus, the contribution of a spin configuration $\{\s_i\}_{i=1}^N$ to the partition function equals $\prod_{i=1}^NV(\s_i,\s_{i+1})$, where
\begin{equation}
V(\s,\s')\equiv\exp\left[\b J(\s\s'-1)+\frac{\b B}2(\s+\s')\right]\;.
\end{equation}
The latter can be thought of as a matrix element of the so-called \emph{transfer matrix},
\begin{equation}
V=\begin{pmatrix}
\E^{\b B} & \E^{-2\b J}\\
\E^{-2\b J} & \E^{-\b B}
\end{pmatrix}\;.
\end{equation}
With the periodic boundary condition on the spins, the partition function of the Ising model is now $Z=\tr V^N=\l_+^N+\l_-^N$, where $\smash{\l_\pm=\cosh\b B\pm\sqrt{\smash{\sinh^2\b B}+\E^{-4\b J}}}$ are the eigenvalues of the transfer matrix. The average spin at each site of the chain is obtained by applying~\eqref{vevAlambda} to $\Fa(B)\equiv F(B)/N$,
\begin{equation}
\vev{\s_i}_{\dm_{\b,B}}=\frac{\sinh\b B}{\sqrt{\sinh^2\b B+\E^{-4\b J}}}\frac{\l_+^N-\l_-^N}{\l_+^N+\l_-^N}\xrightarrow{N\to\infty}\frac{\sinh\b B}{\sqrt{\sinh^2\b B+\E^{-4\b J}}}\;.
\label{isingOP}
\end{equation}
The total number of spins $N$ plays the role of volume, and sending it to infinity amounts to the thermodynamic limit.

It is obvious from~\eqref{isingOP} that the Ising model does not feature SSB at any nonzero temperature. Indeed, whether $N$ is finite or not, the average spin necessarily vanishes in the limit $B\to0$ whenever the temperature is nonzero. On the other hand, taking the temperature to zero gives
\begin{equation}
\lim_{\b\to\infty}\vev{\s_i}_{\dm_{\b,B}}=\sgn B\;,
\label{sgnB}
\end{equation}
again regardless of whether $N$ is finite or not. Taking subsequently a (one-sided) limit of vanishing magnetic field will render the average spin nonzero, with a sign aligned with that of the magnetic field. This can be traced to the properties of the canonical density operator, which in the zero-temperature limit becomes a projector to the subspace of states with lowest energy. For the Ising Hamiltonian~\eqref{isingham},
\begin{equation}
\begin{split}
B>0:&\qquad
\lim_{\b\to\infty}\dm_{\b,B}=\ket{++\dotsb}\bra{++\dotsb}\;,\\
B=0:&\qquad
\lim_{\b\to\infty}\dm_{\b,B}=\smash{\frac12}\bigl(\ket{++\dotsb}\bra{++\dotsb}+\ket{--\dotsb}\bra{--\dotsb}\bigr)\;,\\
B<0:&\qquad
\lim_{\b\to\infty}\dm_{\b,B}=\ket{--\dotsb}\bra{--\dotsb}\;,
\end{split}
\label{isinggroundstate}
\end{equation}
where $\ket{++\dotsb}$ and $\ket{--\dotsb}$ denote respectively the normalized states with all spins being positive and negative. Note that the $\Z_2$-symmetric state on the second line of~\eqref{isinggroundstate} is unstable: an infinitesimally weak external field will project it onto one of the two fully polarized states.
\end{illustration}

The example demonstrates that taking an appropriate (infinite-volume or zero-temperature) limit in the presence of a perturbation may lead to an equilibrium state violating the symmetry of the system. The system then remains in an asymmetric state even if the perturbation is subsequently removed. It is very important that the limits of infinite volume and vanishing external field are performed in this order. Indeed, taking $\l_i\to0$ first would delete the average of any symmetry-breaking observable in line with~\eqref{limAil0}. Colloquially, one often speaks of the ``noncommutativity of limits,''
\begin{equation}
\lim_{\l\to0}\lim_{V\to\infty}\vev{A^i}_{\dm_{\b,\l}}\neq\lim_{V\to\infty}\lim_{\l\to0}\vev{A^i}_{\dm_{\b,\l}}\;,
\end{equation} 
as a smoking gun of SSB.

This concludes our excursion to thermodynamics. We have seen why asymmetric macroscopic states of matter can---and why symmetric macroscopic states may not---be stable. Identifying a (meta)stable equilibrium is a necessary first step if we ever want to understand the spectrum of a quantum system. Our next goal is therefore to establish a general operational procedure for finding a stable ground state.


\subsection{Order Parameter and the Vacuum Manifold}
\label{subsec:SSBOPvacmanifold}

In the Ising model, it was enough to distinguish the two possible asymmetric ground states by the sign of the external field. This is eventually because the model possesses a discrete $\Z_2$ symmetry. In systems with continuous symmetry we expect a continuum of degenerate equilibrium states. An already familiar example is the isotropic ferromagnet, where the spontaneous magnetization can take an arbitrary direction in space. Thus, we must be careful when taking the limit of vanishing external fields.

Suppose that we are given a specific set of fields, $\hat\l_i$. I will define the limit of vanishing perturbations by changing the magnitude but keeping the ``direction'' of these fields. This can be implemented precisely by using a positive scaling parameter $\eps$. We can thus construct a macroscopic thermal state $\smash{\dm^\infty_{\b,\hat\l}}$ by first taking the thermodynamic limit and then removing the external fields via
\begin{equation}
\dm^\infty_{\b,\hat\l}\equiv\lim_{\eps\to0^+}\lim_{V\to\infty}\dm_{\b,\eps\hat\l}\;.
\label{equilibrium}
\end{equation}
In presence of SSB, the average
\begin{equation}
\vev{A^i}_{\dm^\infty_{\b,\hat\l}}=\lim_{\eps\to0^+}\lim_{V\to\infty}\vev{A^i}_{\dm_{\b,\eps\hat\l}}
\label{OPdef}
\end{equation}
is expected to be nonzero. This is an important proxy of SSB, known as the \emph{order parameter}. It does exactly what its name suggests: parameterize the appearance of order in the thermodynamic state of matter described by the density operator $\smash{\dm^\infty_{\b,\hat\l}}$.

Is the state~\eqref{equilibrium} stable? This question is closely related to the choice of observables $A^i$. Namely, the set $A^i$ should be ``complete'' in the sense that the values of the order parameters $\smash{\vev{A^i}_{\dm^\infty_{\b,\hat\l}}}$ determine the thermal state $\smash{\dm^\infty_{\b,\hat\l}}$ uniquely. In the limit of zero temperature, this means that $\smash{\dm^\infty_{\b,\hat\l}}$ should be a pure state, analogously to the $B\neq0$ states in~\eqref{isinggroundstate} for the Ising model.

\begin{illustration}%
In the Ising model, the role of the order parameter is played by the average spin~$\vev{\s_i}$. According to~\eqref{sgnB}, in the zero-temperature limit in the presence of a nonzero magnetic field, $\vev{\s_i}$ acquires one of two possible values. These uniquely specify the corresponding symmetry-breaking pure ground states via~\eqref{isinggroundstate}.

In an isotropic ferromagnet, the order parameter can be taken as the average of the operator of total spin (magnetization) $\vec S$. Specifying the vector $\vev{\vec S}$ is then sufficient to determine the stable macroscopic equilibrium uniquely. This is however not necessarily the only possible choice of order parameter. Instead of the three components of the vector $S_i$, we might try to consider for instance the traceless symmetric tensor $T_{ij}\equiv S_iS_j+S_jS_i-(2/3)\d_{ij}\vec S^2$. This set of operators transforms under a nontrivial irreducible representation of the spin group $G\simeq\gr{SU}(2)$, hence it satisfies the requirements we have imposed on the operators $A^i$. The average $\vev{T_{ij}}$, however, does not specify a unique equilibrium state. Namely, this order parameter cannot distinguish between the two states that only differ by the overall sign of $\vev{\vec S}$.
\end{illustration}

The order parameter is usually much easier to determine, or at least estimate, than the actual equilibrium state, $\smash{\dm^\infty_{\b,\hat\l}}$. Its real value lies in the fact that it carries the same information about the symmetry of the equilibrium state as the density operator $\smash{\dm^\infty_{\b,\hat\l}}$ itself. To see this, just note that
\begin{equation}
\vev{A^i}_{U(g)\dm^\infty_{\b,\hat\l}\he{U(g)}}=\tr\bigl[U(g)\dm^\infty_{\b,\hat\l}\he{U(g)}A^i\bigr]=\rep(g)^i_{\phantom ij}\vev{A^j}_{\dm^\infty_{\b,\hat\l}}\;.
\end{equation}
This shows that for any $g\in G$ that is a symmetry of $\smash{\dm^\infty_{\b,\hat\l}}$, the matrix $\rep(g)$ leaves the order parameter unchanged. But the opposite is true as well. For any $g\in G$ such that $\smash{\rep(g)^i_{\phantom ij}\vev{A^j}_{\dm^\infty_{\b,\hat\l}}=\vev{A^i}_{\dm^\infty_{\b,\hat\l}}}$, the assumed completeness of the set of operators $A^i$ guarantees that the states $\smash{\dm^\infty_{\b,\hat\l}}$ and $\smash{U(g)\dm^\infty_{\b,\hat\l}\he{U(g)}}$ are the same. The pattern of SSB in a macroscopic equilibrium state is therefore completely determined by the associated order parameter(s).

Finally, let me briefly return to the correspondence between the order parameter and the pure ground state (\emph{vacuum}) obtained by taking the zero-temperature limit of~\eqref{equilibrium}. It is in principle possible that the physical system possesses two completely unrelated ground states of equal energy. (This happens for instance at a first-order phase transition.) Barring such accidental degeneracy, however, one expects the possible candidate vacua to be mutually related by (broken) symmetry transformations. It is therefore sufficient to know just one of the vacuum states; it should be possible to reconstruct all the others by the action of the operators $U(g)$, $g\in G$. We will see in Sect.~\ref{sec:SSBsubtleties} that this procedure is subtle in the limit of infinite volume. What one can however do safely is to trade the chosen vacuum state $\dm$ for the corresponding order parameter $\vev{A^i}_\dm$. All other possible values of the order parameter can then be generated by acting on $\vev{A^i}_\dm$ with the matrices $\rep(g)$.

In case $G$ is a Lie group, the possible vacuum values of the order parameter span a manifold of dimension $\dim G-\dim H_\dm$; see Appendix~\ref{appsec:manifolds} for the mathematical background. This is known as the \emph{vacuum manifold}. The central tenet of this book is that the physics of systems with spontaneously broken continuous symmetry can be captured by a low-energy \emph{effective field theory} (EFT). The form of this EFT is largely fixed by the geometry of the vacuum manifold. The latter can in turn be understood in terms of $G$ and its unbroken subgroup $H_\dm$. To develop the EFT framework based on the pattern of SSB is the goal of Parts~\ref{part:internalSSB} and~\ref{part:spacetimeSSB} of the book.

\begin{illustration}%
\label{ex:QCDfirsttime}%
\emph{Quantum chromodynamics} (QCD) is a gauge theory of the strong nuclear interaction. Strongly-interacting matter is represented by the quark field $\Psi^\a_i$. This is a Dirac spinor where $\a$ is an index of the fundamental representation of the $\gr{SU}(3)$ gauge group of QCD. The index $i$ represents quark flavor and belongs to the fundamental representation of $\gr{SU}(\nf)$. In applications, the number of relevant (light) quark flavors $\nf$ is usually two or three.

In the limit of vanishing quark masses, QCD possesses a $G\simeq\gr{SU}(\nf)_\mathrm{L}\times\gr{SU}(\nf)_\mathrm{R}$ symmetry, consisting of independent flavor transformations of the left- and right-handed components of $\Psi^\a_i$. In the ground state of QCD, this symmetry is spontaneously broken by the complex order parameter $\smash{\vev{\S_{ij}}\equiv\vev{\d_{\a\b}\adj\Psi^\a_{i\mathrm{L}}\Psi^\b_{j\mathrm{R}}}}$. Note that $\smash{\tr(\S+\he\S)}$ is the usual mass term for a Dirac spinor, here with equal masses for all quark flavors. We can therefore think of ``switching on'' small equal quark masses as an analog of the external field $\l_i$ that selects a particular vacuum state. In this vacuum state, $\vev{\S_{ij}}=\s\d_{ij}$, where the constant $\s$ is an intrinsic scale of QCD. The unbroken, ``vector'' subgroup of $G$, $H\simeq\gr{SU}(\nf)_\mathrm{V}$, consists of identical unitary transformations of left- and right-handed quarks. Broken symmetry transformations convert this vacuum to other vacuum states, in which $\vev{\S_{ij}}/\s$ is unitary. The vacuum manifold is therefore diffeomorphic to the Lie group $\gr{SU}(\nf)$.
\end{illustration}


\subsection{Intermediate Summary}
\label{subsec:SSBmidsummary}

We have taken quite a conceptual tour to arrive at the notions of order parameter and vacuum manifold, essential for understanding the physics of SSB. Let us therefore make a little break and review what we have done in this chapter so far.

I started Sect.~\ref{sec:SSBstate} with a basic definition of a state of a quantum system in terms of a density operator, and of its symmetry. Section~\ref{subsec:SSBbrokenunbroken} makes the distinction between broken and unbroken symmetries, and highlights energetic degeneracy of different states as a universal feature of SSB. A naive way out of the puzzle this poses might be to consider only states respecting the full symmetry $G$ of the system. In Sect.~\ref{subsec:SSBlemma}, I motivated the consideration of such $G$-symmetric states by recalling the thermal (canonical) density operator. The latter however renders the average of any observable breaking the symmetry zero. This is in a striking contradiction with the existence of asymmetric, macroscopically stable states of matter in nature. A careful consideration of the effects of small external perturbations in Sect.~\ref{subsec:SSBTDlimit} shows that these may drive the system towards a symmetry-breaking state. In fact, in the large-volume limit, even an infinitesimally small perturbation may be sufficient to destabilize the $G$-invariant thermal state.

For most applications, it is mandatory to build a physical description of a given system upon a stable ground state. Such vacuum states are constructed in Sect.~\ref{subsec:SSBOPvacmanifold} by taking the thermodynamic limit in presence of a perturbation, before the latter may be switched off. The different, degenerate vacuum states can now be labeled uniquely by different values of an order parameter, corresponding to the average of a suitably chosen set of observables. The choice of the order parameter itself is not unique. Depending on this choice, the same vacuum state may be represented by a point in different order parameter spaces. However, the set of all degenerate vacuum states spans a vacuum manifold, whose geometric structure is determined solely by the symmetry group $G$ and its unbroken subgroup.

This completes the background needed to understand the following chapters. When talking about a vacuum or ground state, I will always implicitly have in mind a pure state, obtained by the limiting procedure in~\eqref{equilibrium}. Also, I will from now on always assume vanishing thermodynamic temperature. Nonzero temperature was only used in this chapter as a tool to avoid singularities in the partition function.

The story of SSB, however, does not end here. A closer look reveals a number of intriguing aspects that I have skipped so far in the desire to provide a minimal self-contained introduction to SSB. While not of direct relevance for the rest of the book, it would be a pity to omit these aspects altogether. I therefore mention some of them at least briefly in the next section, in the form of a case study.


\section{Some Subtle Features of Spontaneous Symmetry Breaking}
\label{sec:SSBsubtleties}

In the previous section, the thermodynamic limit played a crucial role in selecting a unique symmetry-breaking ground state. This begs the question of what one can expect in volumes that are large yet finite. After all, real macroscopic systems in nature certainly are finite, albeit possibly large from the point of view of microscopic physics. To what extent are then symmetry-breaking states stabilized by an external perturbation? Besides, how does the finite-volume quantum ground state look in the absence of perturbations?

These are some of the questions I will address here. I am not aware of any formal framework that would allow one to tackle these questions in full generality. In fact, the answer to some of them may depend on the specific physical system. Instead of trying to be general, I will therefore work out in detail one concrete example, just to illustrate what is at stake. Some rigorous results on the spectrum of quantum systems with SSB in a finite volume can be found in~\cite{Koma1994} or in Part I of~\cite{Tasaki2020}.


\subsection{Free Schr\"odinger Field in Finite Volume}
\label{subsec:NRfield}

Following Sect.~3 of~\cite{Brauner2010a}, consider the theory of a free Schr\"odinger field $\psi$, defined by the Lagrangian density
\begin{equation}
\La=\I\he\psi\de_0\psi-\frac1{2m}\vec\nabla\he\psi\cdot\vec\nabla\psi\;,
\label{SchrLag}
\end{equation}
where $m$ is a mass parameter. I will quantize this theory in a $d$-dimensional box of size $L$, with volume $V=L^d$, endowed with a periodic boundary condition. The boundary condition admits only discrete values of momentum $\vec p$, given component-wise by
\begin{equation}
p_r=\frac{2\pi n_r}L\;,\qquad
n_r\in\Z\;.
\label{Schrmomenta}
\end{equation}
The field operator can be Fourier-expanded in terms of plane-wave solutions to the Schr\"odinger equation. In the Schr\"odinger picture,
\begin{equation}
\psi(\vec x)=\frac1{\sqrt V}\sum_{\vec p}a_{\vec p}\E^{\I\skal px}\;,
\label{SchrfieldfiniteV}
\end{equation}
where $a_{\vec p}$ is the annihilation operator for a single-particle state of momentum $\vec p$. This is normalized so that together with the creation operator $\smash{\he a_{\vec p}}$, it satisfies $[a_{\vec p},\he a_{\vec q}]=\d_{\vec p\vec q}$. The normalization of the plane-wave expansion by the factor $1/\sqrt V$ then guarantees that the canonical coordinate $\psi(\vec x)$ and its conjugate momentum $\smash{\I\he\psi(\vec x)}$ obey the canonical commutation relation $[\psi(\vec x),\I\he\psi(\vec y)]=\I\d^d(\vec x-\vec y)$.

Within second quantization, all these operators act on a Hilbert space, built using the Fock construction. The starting point is the Fock vacuum $\ket0$, defined by the condition $a_{\vec p}\ket0=0$ for all allowed values of $\vec p$. An orthogonal basis of the space is then obtained by acting on $\ket0$ with a finite number of creation operators in all possible ways. The Schr\"odinger Hamiltonian has a simple expression in terms of the annihilation and creation operators,
\begin{equation}
H=\frac1{2m}\int\D^d\!\vec x\,\vec\nabla\he\psi(\vec x)\cdot\vec\nabla\psi(\vec x)=\sum_{\vec p}\frac{\vec p^2}{2m}\he a_{\vec p}a_{\vec p}\;.
\label{SchrHam}
\end{equation}
This is a positive-semidefinite operator. The lowest (zero) energy is reached by any state $\ket\Omega$ in the Hilbert space that satisfies
\begin{equation}
a_{\vec p}\ket\Omega=0\quad\text{for all }\vec p\neq\vec0\qquad\text{(ground state)}\;.
\label{SchOmega}
\end{equation}
There are infinitely many such states in the Hilbert space, in particular any state obtained from $\ket0$ using the zero-momentum creation operator $\he a_{\vec0}$. Hence, the subspace of states of zero energy is identical to the Fock space of a linear harmonic oscillator.

The infinite degeneracy of the ground state hints at SSB. To see what symmetry might be spontaneously broken here, let us think what conserved charges there are in the first place. There turn out to be infinitely many of them. One can take for instance any operator of the form
\begin{equation}
Q_f\equiv\sum_{\vec p}f(\vec p)\he a_{\vec p}a_{\vec p}=\int\D^d\!\vec x\,\he\psi(\vec x)f(-\I\vec\nabla)\psi(\vec x)\;,
\label{SchrQf}
\end{equation}
where $f$ is assumed to be a real analytic function. This class of operators includes the Hamiltonian itself, with $f(\vec p)=\vec p^2/(2m)$, the momentum operator $\vec P=\sum_{\vec p}\vec p\he a_{\vec p}a_{\vec p}$, and the operator of particle number $\smash{Q_1=\sum_{\vec p}\he a_{\vec p}a_{\vec p}}$. From~\eqref{SchrQf} it is obvious that all the operators $Q_f$ mutually commute with each other, hence all are conserved. By~\eqref{operatorflow}, they induce a local transformation of $\psi(\vec x)$ with parameter $\eps$, $\udelta\psi(\vec x)=-\I\eps f(-\I\vec\nabla)\psi(\vec x)$. In addition to the class~\eqref{SchrQf}, any Hermitian operator constructed out of $a_{\vec0}$ and $\he a_{\vec0}$ commutes with the Hamiltonian, and thus represents a conserved charge as well. Not all of these, however, descend from a local Noether current. Two notable exceptions are the two possible Hermitian operators linear in $a_{\vec0}$ and $\smash{\he a_{\vec0}}$. I will list them alongside the symmetry transformations they generate:
\begin{equation}
\begin{alignedat}{2}
Q_\mathrm{R}&\equiv\I\int\D^d\!\vec x\,[\psi(\vec x)-\he\psi(\vec x)]=\I\sqrt V(a_{\vec0}-\he a_{\vec0})\;,\qquad&&\psi\to\psi-\eps\;,\\
Q_\mathrm{I}&\equiv\int\D^d\!\vec x\,[\psi(\vec x)+\he\psi(\vec x)]=\sqrt V(a_{\vec0}+\he a_{\vec0})\;,\qquad&&\psi\to\psi-\I\eps\;.
\end{alignedat}
\label{SchrQRQI}
\end{equation}
With all the symmetry transformations at hand, the associated Noether currents can be extracted from the Lagrangian~\eqref{SchrLag} using the technique outlined in Sect.~\ref{subsec:noether}.

The Lie algebra of our conserved charges is defined by the commutators
\begin{equation}
[Q_f,Q_\mathrm{R}]=-\I f(\vec0)Q_\mathrm{I}\;,\qquad
[Q_f,Q_\mathrm{I}]=+\I f(\vec0)Q_\mathrm{R}\;,\qquad
[Q_\mathrm{R},Q_\mathrm{I}]=2\I V\;.
\label{Schrsymalgebra}
\end{equation}
These make it natural to split the space of charges $Q_f$ into the one-dimensional subspace of charges proportional to $Q_1$, and a subspace of charges with $f(\vec0)=0$. The latter commute with $Q_\mathrm{R},Q_\mathrm{I}$ and moreover annihilate any ground state $\ket\Omega$ as defined by~\eqref{SchOmega}. We can therefore drop them and focus on the set $Q_1,Q_\mathrm{R},Q_\mathrm{I}$. By~\eqref{Schrsymalgebra}, this set furnishes a central extension of the Lie algebra $\lie{iso}(2)$, corresponding to Euclidean transformations in the complex plane of $\psi$. The generator $Q_1$ induces (phase) rotations, whereas $Q_\mathrm{R},Q_\mathrm{I}$ generate respectively translations of the real and imaginary parts of $\psi$.

The Fock vacuum $\ket0$ is annihilated by $Q_1$, but not by $Q_\mathrm{R},Q_\mathrm{I}$. This suggests that alternative vacuum states may be obtained from $\ket0$ by transformations generated by $Q_\mathrm{R},Q_\mathrm{I}$. We can pick any $z\equiv z_1+\I z_2\in\C$ and define a new vacuum state by
\begin{equation}
\ket z\equiv\exp\bigl[\I(z_1Q_\mathrm{R}+z_2Q_\mathrm{I})\bigr]\ket0=\exp\bigl[\sqrt V(z\he a_{\vec0}-z^*a_{\vec0})\bigr]\ket0\;.
\label{Schrcoherent}
\end{equation}
This is a coherent state in the Fock space of zero-energy states. Indeed, it is easily seen to be an eigenstate of the annihilation operator, $a_{\vec0}\ket z=\sqrt Vz\ket z$. In fact, $\ket z$ is simultaneously an eigenstate of the field operator at any point in space, $\psi(\vec x)\ket z=z\ket z$. We thus end up with a continuum of vacuum states, labeled by a complex parameter $z$. The order parameter distinguishing these states from each other is provided by the expectation value of $\psi(\vec x)$. The vacuum manifold is the complex plane $\C$. For any choice of $z$, the state $\ket z$ possesses an unbroken $\gr{SO}(2)$ symmetry, corresponding to rotations of the complex plane around the point $z$. Its generator is $\prescript{z}{}Q_1\equiv Q_1+z_2Q_\mathrm{R}-z_1Q_\mathrm{I}$. The state $\ket z$ is an eigenstate of this operator, $\prescript{z}{}Q_1\ket z=-V\abs{z}^2\ket z$.

Interestingly, it is not possible to select the state $\ket z$ by adding to the Hamiltonian a perturbation linear in $\psi$ as I did in Sect.~\ref{sec:SSBperturbation}. That would make the Hamiltonian unbounded from below. This is the price to pay for the vacuum manifold not being compact. An alternative is to add to the Hamiltonian a chemical potential $\m$ for the conserved charge $\prescript{z}{}Q_1$, so that it becomes $H_\m\equiv H-\m\prescript{z}{}Q_1$. The expectation value of such a perturbed Hamiltonian in the state $\ket{z'}$ with any $z'\in\C$ is
\begin{equation}
\bra{z'}H_\m\ket{z'}=-V\mu\bigl[\abs{z-z'}^2-\abs{z}^2\bigr]\;.
\end{equation}
An infinitesimally small \emph{negative} $\m$ will then select $\ket z$ as the unique ground state with energy $V\m\abs{z}^2$.


\subsection{Pathologies of the Infinite-Volume Limit}
\label{subsec:NRfieldproblem}

The analysis of the free Schr\"odinger theory has been perfectly clean so far. Most of the properties of the quantized theory match their classical counterparts, in particular the existence of infinitely many conservation laws, and infinitely many degenerate ground states. What happens when we try to take the limit of infinite volume? Let us evaluate the overlap of two different coherent states of the type~\eqref{Schrcoherent},
\begin{equation}
\begin{split}
\abs{\braket{z'}{z}}^2&=\abs{\amplitude{z'}{U(z)}{0}}^2=\exp\bigl(-V\abs{z-z'}^2\bigr)\;,\\
U(z)&\equiv\exp\bigl[\I(z_1Q_\mathrm{R}+z_2Q_\mathrm{I})\bigr]\;.
\end{split}
\label{Schroverlap}
\end{equation}
We therefore find that
\begin{equation}
\lim_{V\to\infty}\abs{\braket{z'}{z}}^2=\lim_{V\to\infty}\abs{\amplitude{z'}{U(z)}{0}}^2=\d_{zz'}\;.
\end{equation}

In the limit $V\to\infty$, any two states $\ket z$ and $\ket{z'}$ with $z\neq z'$ become orthogonal. It is easy to generalize the above calculation to show that, in fact, any vector in the Fock space built above $\ket z$ must be orthogonal to $\ket{z'}$ in this limit.\footnote{Recall that the basis of the Fock space consists of states obtained by acting with a \emph{finite} number of creation operators on the vacuum.} In other words, the ket-vector $\ket{z'}$ does not lie in the Fock space built above $\ket z$ at all. Moreover, the matrix element of any observable constructed using a finite number of annihilation and creation operators between the $\ket{z}$ and $\ket{z'}$ states vanishes. Altogether, the states $\ket z$ with $z\in\C$ therefore give rise to a completely disconnected Fock space each. Transitions between Fock spaces with different vacuum states are forbidden. This is certainly not what we expected!

This singular behavior of the $V\to\infty$ limit is easy to understand. Just remember that $\ket z$ is an eigenstate of $a_{\vec0}$ with the eigenvalue $\sqrt Vz$. Thus, if we insist on keeping $z$ fixed, then scaling $V$ towards infinity makes the eigenvalue diverge as well. This is ultimately the reason why $\ket{z'}$ does not lie in the Fock space built above $\ket z$. We might, on the other hand, try to keep the eigenvalue of $a_{\vec0}$ fixed, which would require scaling $z$ as $1/\sqrt V$ as the volume grows. Roughly speaking, this means that in a large but finite volume $V$, transitions between Fock spaces with different $z$ are allowed, but only within a range $\abs{z-z'}\lesssim1/\sqrt V$.

There is another way of looking at this, emphasizing the properties of the operator $U(z)$ defined by~\eqref{Schroverlap}. Since the states $\ket0$ and $\ket{z'}$ do not lie in the same Fock space, the operator $U(z)$ becomes ill-defined in the limit $V\to\infty$. Spontaneously broken symmetry may not be realized by unitary operators on the Hilbert space of the physical system! That however does not mean that in presence of SSB, it makes no sense to speak of symmetry transformations that are spontaneously broken. The transformation of the field operator by $U(z)$,
\begin{equation}
\he{U(z)}\psi(\vec x)U(z)=\psi(\vec x)+z\;,
\end{equation}
is perfectly well-defined even in the limit $V\to\infty$. The same applies to any local operator built out of a finite number of $\psi$ and $\he\psi$. In each of the Fock spaces built above any $\ket z$, we thus have a well-defined notion of broken symmetry transformations in terms of averages of observables. We just have to be careful not to think of broken symmetry as an operator connecting different \emph{states} in the same space.

\begin{watchout}%
The distinction between the realizations of broken symmetries by operators on states and by transformations of observables is not a mere mathematical curiosity one can safely ignore. It has observable consequences. One of the earliest applications of group theory to physics was to classify energy levels of a quantum system in terms of representations of its symmetry group. In quantum mechanics, the so-called Wigner theorem guarantees that any symmetry is represented by a unitary or antiunitary operator on the Hilbert space (see Chap.~2 of~\cite{Weinberg1995a}). Energy levels are therefore classified by irreducible representations of the symmetry group $G$. In quantum field theory, on the other hand, only the unbroken subgroup $H$ is as a rule realized by unitary operators on the Hilbert space. Thus, energy levels form irreducible multiplets of $H$. SSB implies degeneracy of the ground state, accompanied by a reduced degeneracy of the excitation spectrum.

For example, the spectrum of hadrons is organized into (approximately degenerate) multiplets of the isospin group $\gr{SU}(2)$. We know from \refex{ex:QCDfirsttime} that this is the unbroken subgroup of QCD with two light quark flavors. There is an even higher degeneracy, corresponding to multiplets of $\gr{SU}(3)$. This is however less accurate due to the large splitting between the masses of the up, down and strange quarks. There are no multiplets of the  $\gr{SU}(\nf)_\mathrm{L}\times\gr{SU}(\nf)_\mathrm{R}$ symmetry of QCD. Such multiplets would be easily noticeable. For instance, the pseudoscalar mesons such as pions or kaons would have to appear in the spectrum together with particles of opposite parity. The lack of such particles is a directly observable manifestation of the nonexistence of unitary operators representing spontaneously broken symmetry.
\end{watchout}


\subsection{Uniqueness of the Finite-Volume Ground State}
\label{subsec:NRfieldtwist}

The above excursion into the subtleties of SSB was purposefully limited to a single, exactly solvable example. The conclusion that in infinite volume, spontaneously broken symmetry is not realized by unitary operators on the Hilbert space is, however, general~\cite{Guralnik1968a}. Likewise, it is generally true that different symmetry-breaking vacuum states belong to disconnected Hilbert spaces.

One property of the free Schr\"odinger theory~\eqref{SchrLag} that is not general is the existence of exactly degenerate ground states already in a finite volume. This is typical for systems where the order parameter is constructed using an operator that commutes with the Hamiltonian. In our case, this operator can be taken as the spatial average of the Schr\"odinger field,
\begin{equation}
\bar\psi\equiv\frac1V\int\D^d\!\vec x\,\psi(\vec x)=\frac1{\sqrt V}a_{\vec0}\;.
\end{equation}
Another system of this type is the ferromagnet, where the order parameter is supplied by a selected component of total spin. Historically, the very special properties of such systems led to a controversy (involving some famous individuals~\cite{Peierls1991a}) as to whether they should count as SSB at all or not.

An alternative is that in any finite volume, the system possesses a unique ground state, which is then necessarily $G$-invariant. Above this ground state, there is a tower of closely spaced energy levels, which all become exactly degenerate in the limit $V\to\infty$. A notable, much studied example of such a system are antiferromagnets. The realization of SSB is then quite nontrivial. Although a detailed discussion would be beyond the scope of this book, let me make at least a few remarks. The reader is encouraged to consult Sect.~19.1 of~\cite{Weinberg1996a} or Sect.~2 of~\cite{Beekman2019a} for further information.

If one tries to take the naive limit $V\to\infty$, the unique, finite-volume $G$-invariant ground state turns into a likewise $G$-invariant state in infinite volume. The latter is, however, unstable under infinitesimally small external perturbations. The instability has its precursor already in a finite volume. Namely, the magnitude of a perturbation needed to drive the system away from the $G$-invariant vacuum scales as the inverse of the system's volume. Such a perturbation may be supplied in particular by the measurement of any symmetry-breaking observable. This makes the system relax into one of the pure symmetry-breaking states constructed in Sect.~\ref{subsec:SSBOPvacmanifold}.

Another way to understand the distinction between the various would-be ground states is through correlations of spatially separated observables. For any two local observables $A(\vec x)$ and $B(\vec x)$, one may construct the spatial correlation function $\vev{A(\vec x)B(\vec y)}$; the angular brackets indicate expectation value in a chosen ground state. The so-called \emph{cluster decomposition principle} dictates that
\begin{equation}
\lim_{\abs{\vec x-\vec y}\to\infty}\bigl[\vev{A(\vec x)B(\vec y)}-\vev{A(\vec x)}\vev{B(\vec y)}\bigr]=0\;.
\label{clusterdecomposition}
\end{equation}
It encodes the fundamental requirement that the outcomes of distant measurements should not be correlated with each other. The naive $G$-invariant ground state obtained by taking the $V\to\infty$ limit violates this principle, and thus displays long-range entanglement. The entanglement is destroyed by any measurement of a local symmetry-breaking observable, which brings the system to a stable, symmetry-breaking ground state satisfying~\eqref{clusterdecomposition}.

To illustrate some of the above general observations, I will conclude the discussion of SSB by working out one more simple toy model. We return to the free Schr\"odinger theory, but twist the boundary condition to
\begin{equation}
\psi(\vec x+\vec e_rL)=\E^{\I\t_r}\psi(\vec x)\;.
\label{twistedbc}
\end{equation}
Here $\vec e_r$ is a unit vector along the $r$-th Cartesian coordinate axis, and $\t_r\in(-\pi,+\pi]$ is an arbitrarily chosen phase. I assume that at least one of $\t_r$ is nonzero, otherwise~\eqref{twistedbc} would be just the periodic boundary condition we used previously.

\begin{figure}[t]
\sidecaption[t]
\includegraphics[width=2.9in]{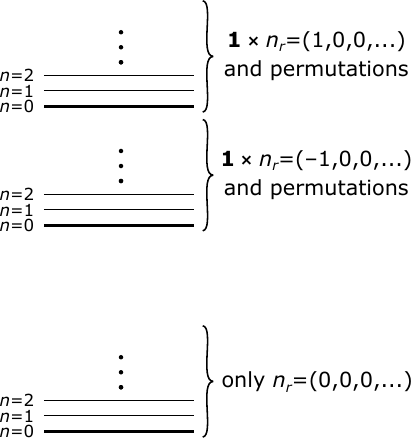}
\caption{Schematic visualization of the spectrum of the free Schr\"odinger theory with the twisted boundary condition~\eqref{twistedbc}, for small (positive) $\t_r$. Each of the indicated groups of energy levels consists of states with multiple quanta of momentum $\vec\t/L$; the number of such quanta is denoted as $n$. The lowest-lying multiplet includes no other particles. The higher multiplets include one or more particles where some of the $n_r$ in~\eqref{twistedmomenta} are nonzero. For simplicity, all the $\t_r$ are assumed to be equal; other\-wise the spectrum would feature additional structure}
\label{fig:Andersontower}
\end{figure}

The quantization of the theory proceeds as before, except for the allowed values of momenta, where~\eqref{Schrmomenta} is replaced with
\begin{equation}
p_r=\frac{\t_r}L+\frac{2\pi n_r}L\;,\qquad
n_r\in\Z\;.
\label{twistedmomenta}
\end{equation}
The Hamiltonian~\eqref{SchrHam} now has a unique ground state, namely the Fock vacuum $\ket0$. The lowest-lying excitation of this ground state corresponds to $n_r=0$ for all $r$, that is, $\vec p=\vec\t/L$. This is a one-particle state $\ket{\vec\t/L}\equiv\he a_{\vec\t/L}\ket0$ with excitation energy
\begin{equation}
E_{\vec\t/L}=\frac{\vec\t^2}{2mL^2}\;.
\end{equation}
As long as all the $\t_r$ are sufficiently small, the low end of the excitation spectrum will consist entirely of states with $n\in\N$ quanta of momentum $\vec\t/L$ and excitation energy $nE_{\vec\t/L}$. Only above these will one find states for which some of the $n_r$ are nonzero. See Fig.~\ref{fig:Andersontower} for a visualization of the spectrum of energy levels.

How is the Fock vacuum $\ket0$ affected by perturbations? Since the $\vec p=\vec0$ state is excluded by the boundary condition, we can now add a perturbation linear in the field without running into an instability. Let us therefore add to~\eqref{SchrLag} the term
\begin{equation}
\La_\eps[\psi,\vec x](x)\equiv\frac{\eps^*}{\sqrt V}\psi(x)\E^{-\I\skal\t x/L}+\frac\eps{\sqrt V}\he\psi(x)\E^{+\I\skal\t x/L}\;,
\end{equation}
where $\eps$ is a small constant parameter with the dimension of energy. The exponential, coordinate-dependent factors are added in order that $\La_\eps$ itself satisfies a periodic boundary condition. The corresponding correction to the Hamiltonian~\eqref{SchrHam} is
\begin{equation}
H_\eps=-\bigl(\eps^*a_{\vec\t/L}+\eps\he a_{\vec\t/L}\bigr)\;.
\end{equation}
This perturbation will change the energy eigenstates within the multiplets shown in Fig.~\ref{fig:Andersontower}. It will however not lead to any mixing between states from different multiplets, that is states with different $n_r$. Let us focus on the lowest-lying multiplet, which is where the ground state resides. Here the Hamiltonian is projected down to
\begin{equation}
\begin{split}
H_{\vec\t/L}&\equiv E_{\vec\t/L}\he a_{\vec\t/L}a_{\vec\t/L}-\bigl(\eps^*a_{\vec\t/L}+\eps\he a_{\vec\t/L}\bigr)\\
&=E_{\vec\t/L}\bigl[(\he a_{\vec\t/L}-z^*_\eps)(a_{\vec\t/L}-z_\eps)-\abs{z_\eps}^2\bigr]\;,
\end{split}
\end{equation}
where $z_\eps\equiv\eps/E_{\vec\t/L}$. There is still a unique ground state, but it is now the coherent state $\ket{z_\eps}$. The perturbation has lowered the energy compared to the Fock vacuum $\ket0$ to $-E_{\vec\t/L}\abs{z_\eps}^2=-\abs\eps^2/E_{\vec\t/L}$.

Let us see what we have found. The original vacuum $\ket0$ is nondegenerate. Its separation from the nearest excited states is however only of the order of $E_{\vec\t/L}\sim1/L^2$. In the limit of infinite volume, all the different multiplets of states shown in Fig.~\ref{fig:Andersontower} become degenerate. Accordingly, the Fock vacuum becomes unstable under small perturbations. For small fixed $\t_r$, one can always find $\eps$ that satisfies the bounds
\begin{equation}
1\ll\frac{\abs\eps}{E_{\vec\t/L}}\ll\frac{2\pi}{\abs{\vec\t}}\;.
\label{Schrtwistbound}
\end{equation}
The lower bound makes the change in energy caused by the perturbation much larger than the level spacing within a given multiplet. Thus, even a small perturbation can drive the ground state from $\ket0$ to a coherent state $\ket{z_\eps}$ with large $z_\eps$. The upper bound in~\eqref{Schrtwistbound} ensures that the shift of energy eigenvalues induced by the perturbation is much smaller than the separation of the different multiplets in the spectrum. 

\begin{watchout}%
The free Schr\"odinger theory with the twisted boundary condition~\eqref{twistedbc} serves as a good example of the instability of the finite-volume ground state with respect to small perturbations. Likewise, it provides a decent illustration of the structure of the spectrum of systems with SSB in a finite volume. One important feature that it does not illustrate is the $G$-invariance of the finite-volume ground state. This is because the symmetries~\eqref{SchrQRQI} of the free Schr\"odinger theory are broken by the boundary condition~\eqref{twistedbc} for any nonzero $\t_r$. In fact, insofar as the expectation value $\vev{\psi(\vec x)}$ is well-defined, shift symmetries of the type~\eqref{SchrQRQI} must always be spontaneously broken.
\end{watchout}


\bibliographystyle{spphys}
\bibliography{references}
\chapter{Nambu--Goldstone Bosons}
\label{chap:NGbosons}

\abstract*{The existence of gapless excitations, the Nambu--Goldstone bosons, is the most striking consequence of spontaneous symmetry breaking. This chapter starts with an intuitive explanation of the presence of such gapless modes in the spectrum. The insight thus developed is already sufficient to understand the intimate relationship between the number of Nambu--Goldstone modes, their dispersion relations, and the nature of broken symmetry. The argument is then put on a solid mathematical basis. First, the Goldstone theorem, which asserts the very existence of Nambu--Goldstone bosons, is proven. The text then continues with a discussion of redundancies between certain, in particular spacetime, symmetries. This leads to a prediction for the number of independent Nambu--Goldstone fields. Finally, a counting rule relating the numbers of independent fields and states in the spectrum is established. The chapter concludes with a brief overview of Nambu--Goldstone-like modes, which are also related to symmetry yet lack one or more of the attributes of genuine Nambu--Goldstone bosons.}


In Chaps.~\ref{chap:ourfirstmodel} and~\ref{chap:firstmodelgeneralizations}, we saw examples of how \emph{spontaneous symmetry breaking} (SSB) gives rise to massless particles in the spectrum. Our operational understanding of these \emph{Nambu--Goldstone} (NG) \emph{bosons} was based on a classical (tree-level) analysis of certain scalar field potentials. The purpose of this chapter is to develop a general understanding of NG bosons without the limitations of a particular model or approximation. The basic intuition is built in Sect.~\ref{sec:NGintuitive}. This explains both where NG bosons come from, and how their spectrum is related to the nature of broken symmetry. The subsequent sections put this intuition on a solid footing. In Sect.~\ref{sec:GoldstoneThm} I prove the so-called Goldstone theorem, which asserts the very existence of a NG boson as a consequence of SSB. Section~\ref{sec:NGclassification} then delves into the question how many NG bosons there are in a given system, and how their number relates to their dispersion relations. This generalizes the observations made in Chap.~\ref{chap:firstmodelgeneralizations}; see Sect.~\ref{sec:firstmodelmorals} for a quick summary. Altogether, this chapter completes the background needed in Parts~\ref{part:internalSSB} and~\ref{part:spacetimeSSB}, where the \emph{effective field theory} (EFT) formalism for SSB is developed in detail.


\section{Intuitive Picture}
\label{sec:NGintuitive}

The intuitive understanding of NG bosons relies heavily on the concept of order parameter. Consider for simplicity a system whose ground state is spatially uniform, and imagine that we disturb it by a weak, local perturbation. Locally, it is energetically favorable for the system to remain in one of the degenerate ground states. The order parameter will therefore respond to the perturbation by developing spatial variation while remaining on the vacuum manifold everywhere. The energy cost of creating such a spatially varying excited state must be proportional to \emph{gradients} of the order parameter. This is because changing the order parameter uniformly just amounts to a different choice of ground state. We can then imagine order parameter configurations that vary over progressively longer and longer length scales. In the limit that the scale of spatial variation goes to infinity, the gradient of the order parameter vanishes, and the energy cost must go to zero.

This is a completely general observation, valid regardless of the choice of field variables and dynamics behind SSB. The only assumption that is really necessary is that the spontaneously broken symmetry is continuous. This makes smooth spatial variation of the order parameter possible while keeping it on the vacuum manifold everywhere. We conclude that NG bosons are local, propagating fluctuations of the order parameter whose energy goes to zero in the infinite-wavelength limit.

The identification of NG bosons with fluctuations of the order parameter is not necessarily one-to-one. Eventually, we would like to know the precise spectrum of NG bosons; this among others governs the low-temperature thermodynamics of systems with SSB. What we really need in order to establish a dictionary between broken symmetries and NG bosons is to answer the following questions:
\begin{itemize}
\item How many different types of order parameter fluctuations (NG fields) are there in a given system?
\item What is the correspondence between the various order parameter fluctuations (NG fields) and NG modes in the spectrum?
\end{itemize}
I will deal with these two questions in the given order in the next two subsections.


\subsection{Redundancy of Order Parameter Fluctuations}
\label{subsec:NGredundancy}

I will build upon the intuitive picture of a NG boson in terms of a fluctuation of the order parameter that locally remains on the vacuum manifold. One can imagine such a fluctuation as being generated from the ground state by a broken symmetry transformation with a coordinate-dependent parameter. This leads immediately to the important observation that fluctuations induced by different broken symmetry generators may coincide.

\begin{illustration}%
\label{ex:scalarshifts}%
The Lagrangian of a free massless relativistic scalar field $\p$ is $\La[\p]=(1/2)(\de_\m\p)^2$. The action of this theory is invariant under the polynomial shift transformation $\p(x)\to\p(x)+\eps_1+\eps_{2\m}x^\m$, where $\eps_1,\eps_{2\m}$ are constant parameters. This symmetry is necessarily spontaneously broken, and the order parameter can be chosen as $\vev{\p(x)}$. One might expect to obtain $D+1$ different fluctuations of the order parameter by applying a polynomial shift with coordinate-dependent parameters $\eps_1(x)$, $\eps_{2\m}(x)$. However, shifting $\p(x)$ by $\smash{\eps_{2\m}(x)x^\m}$ is identical to shifting it by $\eps_1(x)$ if we set $\smash{\eps_1(x)\equiv\eps_{2\m}(x)x^\m}$. The set of $D+1$ broken generators associated with our polynomial shift therefore corresponds to a single independent fluctuation, induced by $\p(x)\to\p(x)+\eps_1(x)$. The only independent NG field in the theory is $\p$ itself.
\end{illustration}

Symmetries that become equivalent once their parameters are made coordinate-dependent are called \emph{redundant}. Such a redundancy is typical for spacetime symmetries. As \refex{ex:scalarshifts} shows, however, it is also possible for point symmetries that do not affect spacetime coordinates. Here is a less trivial example, relevant for any crystalline phase of matter.
\begin{illustration}%
\label{ex:crystal}%
In relativistic theories of scalar fields, the generators of spacetime rotations~$J^{\m\n}$ and translations $P^\m$ are known to satisfy the relation $J^{\m\n}=x^\m P^\n-x^\n P^\m$. A local rotation $\exp[(\I/2)\eps_{2\m\n}(x)J^{\m\n}]$ with antisymmetric matrix parameter $\eps_{2\m\n}$ is thus equivalent to a local translation $\exp[\I\eps_{1\m}(x)P^\m]$ with $\eps_{1\m}(x)\equiv-\eps_{2\m\n}(x)x^\n$. Imagine now a system where spacetime rotations and translations (or a subset thereof) are both spontaneously broken. For example, in crystalline solids, all the continuous spatial rotations and translations are spontaneously broken. The rotations are clearly redundant; the only independent NG fields are those associated with broken translations. These parameterize the vibrations of the crystal lattice.
\end{illustration}
The basic moral to remember is that the number of independent order parameter fluctuations may be lower than the number of broken symmetry generators. I will further refine this observation in Sect.~\ref{subsec:NGclassificationfluctuations}. The discussion of redundancy of local symmetry transformations will be of great importance for the development of EFT for broken spacetime symmetries in Part~\ref{part:spacetimeSSB}.


\subsection{Canonical Conjugation of Nambu--Goldstone Fields}
\label{subsec:NGcanonicalconjugation}

The number of NG modes in the spectrum may be lower than the number of independent NG fields in case some of the latter are canonically conjugated. We saw this in Sect.~\ref{sec:firstmodelNR}. To make statements independent of a specific model, we need to relate the possibility of canonical conjugation of NG fields directly to the broken symmetry.

Suppose we have two NG fields, $\pi^1$ and $\pi^2$, associated with two broken symmetry generators $Q_{1,2}$. We can assume without loss of generality that under the respective transformations generated by $Q_{1,2}$, these fields transform as
\begin{equation}
\pi^a\to\pi^a+\eps^b[\d^a_b+\bigO(\pi)]\;,\qquad a=1,2\;.
\label{NGfieldshift}
\end{equation}
Indeed, one can always choose the fields so that the ground state corresponds to $\pi^a=0$. That the leading, constant piece of the Taylor expansion of the transformation rule for $\pi^a$ around this ground state is nonzero, follows from the assumption of broken symmetry. The bases of NG fields and broken generators can then be aligned so that this constant piece is simply $\pi^a\to\pi^a+\eps^a$.

Suppose in addition that we can construct a low-energy EFT for the NG fields whose Lagrangian contains a term with a single time derivative,
\begin{equation}
\La_\mathrm{eff}=\vr_{12}\pi^1\de_0\pi^2+\dotsb\;.
\label{NGfieldshiftlag}
\end{equation}
The ellipsis stands for terms with more than one derivative or more than two NG fields. Recall from Sect.~\ref{subsec:noether} that the Noether currents corresponding to~\eqref{NGfieldshift} can be identified by making the parameters $\eps^a$ coordinate-dependent. From~\eqref{NGfieldshiftlag} alone, one can then extract the leading contributions to the Noether charge densities,
\begin{equation}
J^0_1[\pi]=-\vr_{12}\pi^2+\dotsb\;,\qquad
J^0_2[\pi]=+\vr_{12}\pi^1+\dotsb\;.
\label{NGfieldshiftcurrent}
\end{equation}
The ellipses now represent terms with derivatives or more than one NG field.

Here comes the key step: I will evaluate the transformation of $J^0_2[\pi]$ under the symmetry generated by $Q_1$ in two different ways. On the one hand, it is obvious that under~\eqref{NGfieldshift}, $J_2^0[\pi]\to J_2^0[\pi]+\vr_{12}\eps^1+\dotsb$. On the other hand, one may think of the currents as quantum operators and represent the same transformation as
\begin{equation}
\exp(\I\eps^1Q_1)J_2^0\exp(-\I\eps^1Q_1)=J_2^0+\I\eps^1[Q_1,J_2^0]+\bigO((\eps^1)^2)\;.
\end{equation}
The NG fields vanish in the ground state. Hence, upon taking the \emph{vacuum expectation value} (VEV), a comparison of our two little calculations leads to
\begin{equation}
\vr_{12}=\I\vev{[Q_1,J_2^0]}=-\I\vev{[Q_2,J_1^0]}\;,
\label{BWmatrixintuitive}
\end{equation}
where the second equality follows by running the same argument on $J_1^0[\pi]$ and $Q_2$.

This is a remarkable result that is as model-independent as it gets. In any theory that realizes the same symmetry-breaking pattern, one can construct the Noether currents via Noether's theorem. Evaluating the VEV of the commutator~\eqref{BWmatrixintuitive} then gives us a simple criterion for when to expect two NG fields to be canonically conjugated: whenever~\eqref{BWmatrixintuitive} is nonzero.

\begin{illustration}%
We already saw the relation between canonical conjugation and the charge commutator~\eqref{BWmatrixintuitive} at work in Sect.~\ref{subsec:NRfield}. Indeed, one can think of the free Schr\"odinger field $\psi$ therein as corresponding to two real NG fields. These are in turn associated with the invariance of the free Schr\"odinger theory under independent shifts of the real and imaginary parts of $\psi$. The commutator of the generators of these shifts, $Q_\mathrm{R}$ and $Q_\mathrm{I}$, turns out to be nonzero upon quantization. It is easy to check that the normalization of the commutator as shown in~\eqref{Schrsymalgebra} agrees with~\eqref{BWmatrixintuitive}. The factor of spatial volume $V$ in~\eqref{Schrsymalgebra} arises from the latter displaying a commutator of two charges rather than a charge and a charge density. 
\end{illustration}

Although noninteracting, the free Schr\"odinger theory is nontrivial in that it makes the VEV~\eqref{BWmatrixintuitive} arise from a central charge in the symmetry algebra. Let us have a look at one more example, which is mathematically simpler and physically familiar.

\newpage

\begin{illustration}%
In the ideal (that is spatially uniform and isotropic) approximation, both ferromagnets and antiferromagnets possess a symmetry $G\simeq\gr{SU}(2)$, generated by the operator $\vec S$ of total spin. In the ground state of both, this is spontaneously broken down to $H\simeq\gr{U}(1)$, consisting of spin rotations around the axis along which the spins are aligned. Suppose that one chooses the ground state oriented along the third direction in the spin space. Then $H$ is generated by $S_3$, whereas the two independent broken generators can be chosen as $S_{1,2}$.

The difference between ferro- and antiferromagnets is that the former feature a nonzero net magnetization. This amounts to a nonzero VEV $\vev{S_3}=-\I\vev{[S_1,S_2]}$. As a consequence, the two NG fields in a ferromagnet are canonically conjugated. The spectrum only contains a single NG mode: the ferromagnetic \emph{magnon}. Its dispersion relation at long wavelengths is known to be quadratic in momentum. In antiferromagnets, on the other hand, $\vev{\vec S}=\vec0$. As a consequence, the spectrum of antiferromagnets features two different magnon branches. Their dispersion relation is known to be linear in momentum in the long-wavelength limit.
\end{illustration}


\subsection{The Big Picture}
\label{subsec:NGbigpicture}

Let me summarize what we have learned so far. Spontaneous breaking of a continuous symmetry implies the existence of a gapless mode in the spectrum: the NG boson. This can be viewed as a propagating fluctuation of the order parameter. The number of such NG modes may be lower than the number of broken symmetry generators. There are two basic mechanisms how such a reduction may occur.

\begin{figure}[t]
\sidecaption[t]
\includegraphics[width=2.9in]{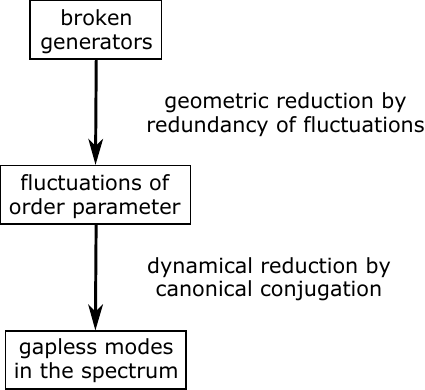}
\caption{Basic mechanisms that may cause the number of independent NG modes in the spectrum to be lower than the number of broken symmetry generators. The geometric reduction arises whenever fluctuations of the order parameter, induced by different generators, are indistinguishable from each other. The dynamical reduction is signaled by a nonzero VEV of the charge commutator~\eqref{BWmatrixintuitive}, or by a term in the effective Lagrangian for NG fields with a single time derivative}
\label{fig:bigpicture}
\end{figure}

First, different broken symmetry generators may induce fluctuations of the order parameter that are indistinguishable from each other. I will refer to this mechanism as ``geometric'' to keep in mind that it represents a restriction on the configuration space of off-shell NG fields. Second, there may not be a one-to-one correspondence between the independent NG fields and the independent NG modes in the spectrum. This happens whenever a pair of NG fields is canonically conjugated, which in turn requires a nonzero VEV of the charge commutator~\eqref{BWmatrixintuitive}. This mechanism I will refer to as ``dynamical'' since it restricts the space of on-shell fields, or eigenstates of the Hamiltonian. For the reader's convenience, an outline of the correspondence between broken symmetry, NG fields and NG modes is shown graphically in Fig.~\ref{fig:bigpicture}. This scheme is intuitively simple yet somewhat imprecise. I will return to the question of how many NG modes there are in a given system in Sect.~\ref{sec:NGclassification}.

\section{Goldstone Theorem}
\label{sec:GoldstoneThm}

The Goldstone theorem is one of the most profound exact results in quantum field theory that deserves a more careful justification than the intuitive but hand-waving argument of Sect.~\ref{sec:NGintuitive}. In the next two subsections, I will give two different proofs, following in spirit the original work of Goldstone, Salam and Weinberg~\cite{Goldstone1961a,Goldstone1962a}. Along the way, I will clarify the technical assumptions on which the theorem is based. A reader seeking a higher level of mathematical rigor than what I can offer here is advised to consult the early review~\cite{Guralnik1968a} or the book~\cite{Strocchi2021}.


\subsection{Operator Proof}
\label{subsec:GoldstoneOpProof}

I will start with a proof using the operator formalism of quantum field theory in the Heisenberg picture. Before proceeding to the details of the proof, however, I need to spend some time listing its technical assumptions.

First, I assume that in the limit of infinite spatial volume, the system possesses unbroken continuous (spacetime) translation invariance. This ensures the existence of a basis of the Hilbert space consisting of eigenstates of the energy--momentum operator $P^\m$. It is always possible, and I will implicitly do so, to define this operator so that the energy and momentum of the ground state are both zero. The assumption of continuous spatial translation invariance is, in fact, unnecessarily strong. The minimum requirement is unbroken discrete translation invariance, which is necessary to have a well-defined notion of quasiparticles. The proof then proceeds along the same steps at the cost of a somewhat cluttered notation~\cite{Watanabe2012a}.

Second, I assume the existence of a local operator $J^0(x)$, interpreted as the density of a conserved charge. The charge density should not depend explicitly on spacetime coordinates; this is needed to ensure the translation property
\begin{equation}
J^0(x)=\E^{\I P\cdot x}J^0(0)\E^{-\I P\cdot x}\;.
\label{currenttranslation}
\end{equation}
Now define the total charge contained in a finite spatial domain $\Omega$,
\begin{equation}
Q_\Omega(t)\equiv\int_\Omega\D^d\!\vec x\,J^0(\vec x,t)\;.
\label{QOmega}
\end{equation}
The last assumption we need is the existence of a local, time-independent operator $\Phi$ such that the VEV in the chosen ground state $\ket0$,
\begin{equation}
\bra0[Q_\Omega(t),\Phi]\ket0\;,
\label{orderparameter}
\end{equation}
is nonzero and time-independent in the limit $\Omega\to\infty$.

It is this last assumption that captures the essence of SSB. Think of $Q_\Omega$ as the Noether charge associated with a continuous symmetry. One would be tempted to say that in the limit $\Omega\to\infty$, $Q_\Omega$ itself becomes time-independent. But we saw in Sect.~\ref{sec:SSBsubtleties} that in the infinite-volume limit, the operators representing spontaneously broken symmetry may be ill-defined. The limit $\Omega\to\infty$ is only safe if performed on the commutator in~\eqref{orderparameter}, not on $Q_\Omega$ itself.

\begin{watchout}%
Usually, the time-independence of the integral charge is a consequence of a local conservation law, $\de_\m J^\m=0$. That is however strictly speaking not necessary. Taking this additional step requires the vanishing of the surface integral $\int_\Omega\D^d\!\vec x\,\de_rJ^r(\vec x,t)$ as $\Omega\to\infty$. This condition, or even its weaker form imposed only on the VEV $\bra0[\de_rJ^r(\vec x,t),\Phi]\ket0$, may be compromised in systems with long-range interactions~\cite{Guralnik1968a}. Here I avoid having to deal with this issue simply by assuming right away the time-independence of~\eqref{orderparameter}.
\end{watchout}

Think of $\bra0\Phi\ket0$ as an order parameter. Then~\eqref{orderparameter} is, up to a factor, its infinitesimal variation under the transformation $\Phi\to\exp(\I\eps Q_\Omega)\Phi\exp(-\I\eps Q_\Omega)$. The assumption that~\eqref{orderparameter} is nonzero guarantees that $\ket0$ cannot be an eigenstate of $Q_\Omega$. Hence the symmetry generated by $Q_\Omega$ must be spontaneously broken.

We are now ready to prove the existence of a NG mode in the system. I will initially assume that the system is enclosed in a finite volume $V$ with a periodic boundary condition. The Hilbert space then admits a basis of momentum eigenstates $\ket{n,\vec p}$, where $\vec p$ stands for a set of discrete momenta consistent with the boundary condition. In a finite volume, the basis states can be normalized as $\braket{m,\vec p}{n,\vec q}=\d_{mn}\d_{\vec p\vec q}$. The label $n$ indicates all other quantum numbers the states may possess, such as relative momenta in multiparticle states, or internal degrees of freedom.

Upon inserting the partition of unity in terms of $\ket{n,\vec p}$ and using the translation property~\eqref{currenttranslation}, the VEV~\eqref{orderparameter} can be rewritten as
\begin{equation}
\begin{split}
\bra0[Q_\Omega(t),\Phi]\ket0=\sum_{n,\vec p}\int_\Omega&\D^d\!\vec x\,\Bigl[\exp(-\I p_n\cdot x)\bra0J^0(0)\ket{n,\vec p}\bra{n,\vec p}\Phi\ket0\\
&-\exp(+\I p_n\cdot x)\bra0\Phi\ket{n,\vec p}\bra{n,\vec p}J^0(0)\ket0\Bigr]\;.
\end{split}
\label{Goldstoneproof1}
\end{equation}
Here $p_n$ is a shorthand notation for the energy--momentum of $\ket{n,\vec p}$. Now recall that SSB implies the existence of degenerate ground states. Even if one picks a unique vacuum $\ket0$, the other, alternative ground states do not disappear. They contribute to~\eqref{Goldstoneproof1} among the states $\ket{n,\vec0}$ with $\vec p=\vec0$. Is this something to worry about?

In the free Schr\"odinger theory, worked out in Sect.~\ref{subsec:NRfield}, $\bra0J^0(0)\ket{n,\vec0}$ scales as $1/\sqrt V$, cf.~\eqref{SchrQRQI}. This observation turns out to be general. With the periodic boundary condition, the integral charge $Q_V(t)$ is translationally invariant. A classic argument~\cite{Fabri1966a} then shows that the norm of $Q_V(t)\ket0$ scales as $\sqrt V$,
\begin{equation}
\bigl\|Q_V(t)\ket0\bigr\|^2=\int_V\D^d\!\vec x\,\bra0 J^0(\vec x,t)Q_V(t)\ket0=V\bra0J^0(0)Q_V(0)\ket0\;.
\end{equation}
This means, roughly, that the norm of $J^0(0)\ket0$ scales as $1/\sqrt V$, and so does therefore $\bra0J^0(0)\ket{n,\vec0}$. Altogether, the contribution of the alternative ground states to~\eqref{Goldstoneproof1} at fixed $\Omega$ and increasing $V$ is suppressed as $1/\sqrt V$. Should~\eqref{orderparameter} be nonzero in the infinite-volume limit, it must be dominated by excitations with nonzero momentum.

We can now switch to infinite volume. This requires changing the normalization of the basis states to $\braket{m,\vec p}{n,\vec q}=(2\pi)^d\d_{mn}\d^d(\vec p-\vec q)$ and replacing the sum over $\vec p$ in~\eqref{Goldstoneproof1} with an integral. Keeping $\Omega$ still finite, the integration over $\vec x$ then amounts to a Fourier transform of unity, $\smash{\int_\Omega\D^d\!\vec x\,\E^{\I\skal px}\equiv(2\pi)^d\d^d_\Omega(\vec p)}$. The notation underlines that $\d^d_\Omega(\vec p)$ is a finite-volume approximation of the Dirac $\d$-function. We then have
\begin{align}
\label{Goldstoneproof2}
\bra0[&Q_\Omega(t),\Phi]\ket0=\sum_n\int\D^d\!\vec p\,\bigl\{\exp[-\I E_n(\vec p)t]\d^d_\Omega(\vec p)\bra0 J^0(0)\ket{n,\vec p}\\
\notag
&\times\bra{n,\vec p}\Phi\ket0-\exp[+\I E_n(\vec p)t]\d^d_\Omega(-\vec p)\bra0\Phi\ket{n,\vec p}\bra{n,\vec p}J^0(0)\ket0\bigr\}\;,
\end{align}
where $E_n(\vec p)=p_n^0$ is the dispersion relation of the state $\ket{n,\vec p}$.

By our assumptions, \eqref{orderparameter} should be nonzero and time-independent in the limit $\Omega\to\infty$. For large $\Omega$, the function $\d^d_\Omega(\vec p)$ is smooth but sharply peaked around $\vec p=\vec0$. The time-independence thus requires that only states such that
\begin{equation}
\lim_{\vec p\to\vec0}E_n(\vec p)=0
\label{gapless}
\end{equation}
contribute to~\eqref{Goldstoneproof2}. The assumption that~\eqref{orderparameter} should be nonzero guarantees that such states exist. These must be one-particle states, for if the one-particle spectrum had a gap, then so would the multiparticle one. We have thus established the existence of one-particle states $\ket{n,\vec p}$ with the property~\eqref{gapless}, for which both $\bra0J^0(0)\ket{n,\vec p}$ and $\bra0\Phi\ket{n,\vec p}$ are nonzero. This concludes the proof of Goldstone's theorem.

There is a simple, intuitive way to understand why NG modes must satisfy~\eqref{gapless}. Suppose we know beforehand that there is a one-particle state $\ket{n,\vec p}$ created by $J^0$ \emph{and} that the latter descends from a local conservation law, $\de_\m J^\m=0$. Extending the translation property~\eqref{currenttranslation} to $J^\m$, we get at once $\bra0J^\m(x)\ket{n,\vec p}=\exp(-\I p_n\cdot x)\bra0J^\m(0)\ket{n,\vec p}$. Upon using current conservation and the fact that $\bra0J^0(0)\ket{n,\vec p}$ is nonzero, it then follows that $E_n(\vec p)={\vec p\cdot\bra0\vec J(0)\ket{n,\vec p}}/{\bra0J^0(0)\ket{n,\vec p}}$. This shows that the vanishing of the energy of NG modes in the long-wavelength limit is an immediate consequence of symmetry. The nontrivial part of Goldstone's theorem is that such NG states exist at all.


\subsection{Effective Action Proof}
\label{subsec:GoldstoneEffActionProof}

I will now present an entirely different proof of Goldstone's theorem which will prepare the ground for the discussion of redundancies among NG fields in the next section. In spirit, this proof is not based on current conservation but rather directly on the symmetry of the action. I will follow closely the appendix of~\cite{Watanabe2011a}.

The starting assumption is the existence of an action $\Gamma$ as a functional of a set of fields $\psi^i$. The notation indicates that $\Gamma$ is the quantum effective action, since we want to make exact statements about a quantum field theory. The argument below applies in principle equally well to the classical action $S$. However, the fields $\psi^i$ need not be the ``elementary'' fields that appear in $S$. They may be composite operators, and are chosen so that the VEV $\vev{\psi^i}$ is a suitable order parameter for SSB. The action is assumed to be invariant under the infinitesimal transformation
\begin{equation}
\udelta\psi^i(x)=\eps\df^i[\psi,x](x)\;.
\label{NGfieldtransfo}
\end{equation}
Unlike in the previous, operator proof of Goldstone's theorem, the symmetry transformation, and thus the Noether current, is allowed to depend explicitly on coordinates.

\begin{watchout}%
It is not guaranteed a priori that the quantum effective action $\Gamma$ inherits verbatim all the symmetries of the classical action $S$. It turns out that the symmetry transformations on the classical and quantum actions are the same at least if $\df^i[\psi,x]$ is linear in the fields; see Sect.~16.4 of~\cite{Weinberg1996a}. This is something I assumed already back in Sect.~\ref{sec:SSBperturbation} when I introduced the very concept of order parameter. The only restriction involved in the transition to the quantum effective action therefore is that the symmetry should preserve the functional integral measure. This is needed to exclude quantum anomalies, whose presence would invalidate Goldstone's theorem.
\end{watchout}

The invariance of the action is encoded in the condition
\begin{equation}
\int\D^D\!y\,\frac{\udelta\Gamma}{\udelta\psi^j(y)}\df^j[\psi,y](y)=0\;.
\end{equation}
Taking one more variational derivative and setting the fields equal to their VEVs, $\vev{\psi^i(x)}\equiv\vp^i(x)$, one gets
\begin{equation}
\int\D^D\!y\,\at{\frac{\udelta^2\Gamma}{\udelta\psi^i(x)\udelta\psi^j(y)}}{\psi=\vp}\df^j[\vp,y](y)=0\;.
\label{Goldstoneproof3}
\end{equation}
Now we need the assumption of continuous translation invariance of the action and of the ground state. The second variational derivative of $\Gamma$ can thus be traded for its Fourier transform, which is the inverse propagator of the theory, $\smash{G^{-1}_{ij}[\vp](p)}$. Multiplying~\eqref{Goldstoneproof3} with $\E^{\I p\cdot x}$ and integrating over $x$ then yields the condition
\begin{equation}
G^{-1}_{ij}[\vp](p)\int\D^D\!y\,\E^{\I p\cdot y}\df^j[\vp,y](y)=0\;.
\label{Goldstoneproof4}
\end{equation}
Hence, the propagator $G_{ij}[\vp](p)$ has a singularity at energy--momentum $p^\m$ whenever the integral in~\eqref{Goldstoneproof4} is nonzero. The spectral representation ensures that such singularities arise solely from states in the spectrum of the Hamiltonian.

By the assumption of broken symmetry, $\df^j[\vp,y]$, which represents the symmetry transformation of the order parameter, is nonzero. This alone implies that the integral in~\eqref{Goldstoneproof4} must be nonzero for \emph{some} $p^\m$. For coordinate-independent symmetries, the integral is proportional to $\d^D(p)$. Moreover, for known examples of coordinate-dependent symmetries, $\df^j[\vp,y]$ is polynomial in coordinates. In such cases, integration over $y$ leads to a linear combination of derivatives of $\d^D(p)$. One way or another, we conclude that SSB implies the existence of a mode whose energy vanishes in the limit of zero momentum, as expected.


\section{Classification and Counting}
\label{sec:NGclassification}

None of the above two proofs of the Goldstone theorem addresses the question how many NG bosons there are. To that end, we have to put in some extra effort. I will follow closely the scheme outlined in Sect.~\ref{subsec:NGbigpicture}. This means that as the first step, we need to know, given the symmetry-breaking pattern, how many independent NG \emph{fields}, or order parameter fluctuations, there are.


\subsection{Independent Order Parameter Fluctuations}
\label{subsec:NGclassificationfluctuations}

We add to~\eqref{NGfieldtransfo} an index $A$, distinguishing the action of different symmetry generators $Q_A$, $\udelta\psi^i(x)=\eps^A\df^i_A[\psi,x](x)$. Recall the view of NG fields as local fluctuations of the order parameter. The easiest (albeit not the only; cf.~Sect.~\ref{subsec:noether}) way to generate such fluctuations is by replacing $\eps^A\to\eps^A(x)$. Redundancies among order parameter fluctuations are then detected by the existence of nonzero functions $\eps^A(x)$ such that
\begin{equation}
\eps^A(x)\df^i_A[\vp,x](x)=0\;.
\label{lowmanohar}
\end{equation}
This relation between symmetry transformations implies an analogous dependence between the zero modes of the inverse propagator of the quantum theory. Namely, trading the coordinate inside $\eps^A(x)$ for a derivative with respect to the momentum variable $p^\m$, denoted as $\nabla_\m$, a combination of~\eqref{Goldstoneproof4} and~\eqref{lowmanohar} gives
\begin{equation}
\eps^A(-\I\nabla)\int\D^D\!y\,\E^{\I p\cdot y}\df^j_A[\vp,y](y)=0\;.
\end{equation}
The conclusion is that the number of independent NG fields equals the dimension of the symmetry group $G$ minus the number of linearly independent solutions of~\eqref{lowmanohar}. Note that the linear independence is meant in the functional sense: multiplying all the $\eps^A(x)$ by the same function $f(x)$ does not count as a new solution.

\begin{watchout}%
The rule for counting the number of independent NG fields does not have an established form in the literature. My formulation is close in spirit to~\cite{Low2002a} where, however, the functions $\eps^A(x)$ were only allowed to depend on coordinates corresponding to unbroken continuous translations. That seems too restrictive for instance in case of crystalline solids where only a discrete translation invariance survives in the ground state.

Another condition for redundancy similar to~\eqref{lowmanohar} was put forward in~\cite{Watanabe2013a}; see also the review~\cite{Watanabe2020a}. Their criterion is however phrased in terms of the vacuum ket-vector and charge density operators. As such, it also includes the redundancy of NG states under canonical conjugation, which I treat separately in Sect.~\ref{subsec:NGclassificationcounting}. Finally, in~\cite{Hayata2014a} a rule counting independent NG fields is formulated directly in terms of the zero modes of the inverse propagator.

In case of broken spacetime symmetries, the number of independent would-be NG fields identified with the help of~\eqref{lowmanohar} turns out to depend on the precise choice of order parameter(s). Remarkably, some of these fields may couple to gapped states in the spectrum. This subtlety is missed by the classical counting rule for NG fields based on~\eqref{lowmanohar}. It is best addressed having the full EFT machinery at hand. I will do so in Chap.~\ref{chap:spacetimequantum}.
\end{watchout}

As a basic sanity check, consider spontaneously broken internal symmetries in systems with unbroken translation invariance. In that case, $\df^i_A[\vp,x]$ does not depend on coordinates, either explicitly or implicitly through the order parameter. Also, it is an ordinary local function of $\vp^i$, hence the condition~\eqref{lowmanohar} reduces to $\eps^A(x)\df^i_A(\vp)=0$. The only nontrivial solutions for $\eps^A(x)$ are then those corresponding to generators of the unbroken subgroup $H$. The number of independent NG fields therefore equals $\dim G-\dim H$. For spontaneously broken internal symmetries, there is a one-to-one correspondence between NG fields and broken symmetry generators. This is one of the moral lessons we reached in Chap.~\ref{chap:firstmodelgeneralizations} by analyzing a set of specific models.

Oftentimes, the solutions to~\eqref{lowmanohar} descend from redundancy among local symmetry transformations acting on an arbitrary field configuration. This was certainly the case for \refex{ex:scalarshifts} and \refex{ex:crystal}. The condition~\eqref{lowmanohar} is however much weaker, as it only demands redundancy of local symmetry transformations of the ground state. Here is a nontrivial example of this latter type.

\begin{illustration}%
\label{ex:helimagnet}%
Helimagnets are magnetic materials in which the alignment axis of spins gradually changes with position; see~\cite{Kishine2015a} for basic phenomenology. The simplest type of helimagnetic state can be modeled by the unit-vector order parameter
\begin{equation}
\vev{\vec n(\vec x,t)}=(\cos kz,\sin kz,0)\;,
\label{helix}
\end{equation}
where $k$ is a positive constant. The local axis of alignment lies in the $xy$ plane but rotates around the $z$ axis, forming a helix with the pitch $2\pi/k$.

Suppose for simplicity that the material in which the helical order develops has a cubic crystal lattice. This is the case for instance for the inorganic compounds MnSi and FeGe; see Sect.~2 of~\cite{Togawa2016a} for a more comprehensive list of known helimagnetic materials. Then the low-energy physics of the material exhibits the following approximate continuous symmetries (in addition to time translations):
\begin{itemize}
\item Spatial translations (T): $\vec n'(\vec x',t)=\vec n(\vec x,t)$ with $\vec x'=\vec x+\vec\eps$, $\vec\eps\in\R^3$.
\item Space--spin rotations (R): $\vec n'(\vec x',t)=R\vec n(\vec x,t)$ with $\vec x'=R\vec x$, $R\in\gr{SO}(3)$.
\end{itemize}
Unlike in ordinary (anti)ferromagnets, spin rotations cannot be considered separately from spatial rotations as a consequence of the spin-orbit coupling. This induces the so-called Dzyaloshinskii--Moriya interaction, which is eventually responsible for the development of the helical order~\eqref{helix}.

In their evolutionary form, the above transformations correspond respectively to the functions $\smash{\prescript{\mathrm{T}}{}\df^i_r=-\de_rn^i}$ and $\smash{\prescript{\mathrm{R}}{}\df^i_r=-\ve_{rs}^{\phantom{rs}u}x^s\de_u n^i+\ve^i_{\phantom irj}n^j}$. Denoting the corresponding infinitesimal parameters as $\eps^r(\vec x,t)$ (translation in the $r$-th direction) and $\o^r(\vec x,t)$ (rotation around the $r$-th axis), \eqref{lowmanohar} turns into
\begin{equation}
\eps^r(\vec x,t)\de_r\vev{n^i(\vec x,t)}+\o^r(\vec x,t)\bigl[\ve_{rs}^{\phantom{rs}u}x^s\de_u\vev{n^i(\vec x,t)}-\ve^i_{\phantom irj}\vev{n^j(\vec x,t)}\bigr]=0\;.
\end{equation}
This boils down to the following constraints on the parameter functions,
\begin{equation}
\begin{split}
\o^1(\vec x,t)\sin kz&=\o^2(\vec x,t)\cos kz\;,\\
\o^3(\vec x,t)&=k[\eps^3(\vec x,t)+y\o^1(\vec x,t)-x\o^2(\vec x,t)]\;.
\end{split}
\label{helimagnet}
\end{equation}
Some of the solutions of~\eqref{helimagnet} are trivial. First of all, $\eps^1$ and $\eps^2$ can be chosen arbitrarily since they do not appear in~\eqref{helimagnet} at all. These represent unbroken translations in the $x,y$ directions. Furthermore, there is another unbroken symmetry, corresponding to simultaneous translation in the $z$-direction and  rotation around the $z$-axis. This extends to the solution of~\eqref{helimagnet} with $\o^3(\vec x,t)=k\eps^3(\vec x,t)$ and $\o^{1,2}(\vec x,t)=0$.

In addition to the three trivial solutions that identify unbroken symmetries, there is one solution of~\eqref{helimagnet} that represents a genuine redundancy of order parameter fluctuations. This can be thought of for instance as picking $\o^1$ arbitrarily and adjusting $\o^2$ and $\o^3$ so that~\eqref{helimagnet} is satisfied. Altogether, this makes four independent solutions of~\eqref{lowmanohar} in case of the order parameter~\eqref{helix}. With six different symmetry generators, three translations and three rotations, this leaves us with mere two independent NG fields. That is a result we could have guessed: the unit vector field $\vec n(\vec x,t)$ contains exactly two degrees of freedom.
\end{illustration}

This example demonstrates that while possibly tedious, finding the solutions to~\eqref{lowmanohar} as a set of linear equations for $\eps^A(x)$ is completely straightforward. Even though~\eqref{lowmanohar} does not give an explicit expression for the number of independent NG fields, it therefore offers a streamlined algorithmic procedure how to find them.


\subsection{Type-A and Type-B Nambu--Goldstone Bosons}
\label{subsec:NGclassificationcounting}

Now we finally get to the question how many NG bosons there are in the spectrum. I will start with the simpler case of spontaneously broken internal symmetry in systems with unbroken translation invariance. In this case, there are guaranteed to be $\dim G-\dim H$ NG fields, in a one-to-one correspondence to broken symmetry generators. Finding the number of NG modes requires just a mild generalization of the argument in Sect.~\ref{subsec:NGcanonicalconjugation}.

As the first step, one introduces the commutator matrix
\begin{equation}
\vr_{AB}\equiv\I\vev{[Q_A,J^0_B]}\;.
\label{BWmatrix}
\end{equation}
Translation invariance allows to rewrite the matrix as $\vr_{AB}=\I\lim_{V\to\infty}\vev{[Q_A,Q_B]}/V$. This makes it clear that $\vr_{AB}=0$ whenever $Q_A$ or $Q_B$ is unbroken. The nontrivial part of the commutator matrix therefore resides in its restriction to broken generators $Q_{a,b,\dotsc}$, that is $\vr_{ab}$. Moreover, $\vr_{AB}$ is real antisymmetric, and as such can be block-diagonalized by an orthogonal change of basis of symmetry generators. With the shorthand notation $r_\vr\equiv\rank\vr$, this implies that the part of the effective Lagrangian for NG fields $\pi^a$, linear in time derivatives, takes the form
\begin{equation}
\La_\mathrm{eff}=\vr_{12}\pi^1\de_0\pi^2+\dotsb+\vr_{r_\vr-1,r_\vr}\pi^{r_\vr-1}\de_0\pi^{r_\vr}+\dotsb\;.
\label{L01}
\end{equation}
The ellipsis indicates operators with more than one derivative or two NG fields.

This allows a unique categorization of NG fields. The $r_\vr$ NG fields $\pi^a$ with $a=1,\dotsc,r_\vr$ are canonically conjugated into $r_\vr/2$ pairs. Each such pair corresponds to one \emph{type-B NG boson}. Barring accidental cancellations in the spatial gradient part of the effective Lagrangian, the dispersion relation of type-B NG bosons is typically quadratic in momentum. The remaining NG fields $\pi^a$ with $a=r_\vr+1,\dotsc,\dim G-\dim H$ do not appear in any bilinear operator with a single time derivative. They correspond to one \emph{type-A NG boson} each. As a rule of thumb, free type-A bosons are described by bilinear operators in the Lagrangian with two temporal or two spatial derivatives. Hence, their dispersion relation is linear in momentum.

We have arrived at a simple counting rule for the respective numbers of type-A and type-B NG bosons~\cite{Watanabe2012b,Hidaka2013b},
\begin{equation}
n_\mathrm{A}=\dim G-\dim H-\rank\vr\;,\qquad
n_\mathrm{B}=\frac12\rank\vr\;.
\label{NGcounting}
\end{equation}
This characterizes the spectrum of NG bosons based solely on the pattern of symmetry breaking and a bit of additional information about the ground state, encoded in the matrix $\vr_{AB}$. The possible values of $\vr_{AB}$ are strongly restricted by the unbroken symmetry. To start with, in systems with unbroken Lorentz invariance, terms in the Lagrangian with a single time derivative are forbidden, hence $\vr_{AB}=0$. This means that $n_\mathrm{A}=\dim G-\dim H$ and $n_\mathrm{B}=0$. All relativistic NG bosons are massless particles with a strictly linear dispersion relation.

Let us now relax the requirement of Lorentz invariance, but keep the assumption that only internal symmetry is spontaneously broken and that the system possesses continuous translation invariance. If in addition the internal symmetry group $G$ is compact and semisimple, we can write
\begin{equation}
\vr_{AB}=\I\lim_{V\to\infty}\frac{\vev{[Q_A,Q_B]}}V=-f^C_{AB}\lim_{V\to\infty}\frac{\vev{Q_C}}V\;,
\label{BWmatrixaux}
\end{equation}
where $f^C_{AB}$ are the structure constants of $G$. This shows that the VEV of the conserved charges themselves can serve as an order parameter for SSB. The set of VEVs $\vev{Q_C}$ furnishes a vector in the adjoint representation of $G$. At the same time, it must remain unchanged under the action of $H$. As a consequence, the vector $\vev{Q_C}$ must correspond to a trivial one-dimensional representation (singlet) of $H$. Finding such singlets in the decomposition of the Lie algebra $\lie g$ of $G$ into irreducible multiplets of $H$ is a quick way to isolate candidate order parameters.

Focusing on singlets of $H$ in the decomposition of the adjoint representation of $G$ may not be restrictive enough in case $H$ is small or even trivial. We can however do much better than that. Namely, it is always possible to choose the ground state and basis of generators $Q_A$  such that all the charges that have a nonzero VEV mutually commute. A detailed proof of this statement is given in the appendix of~\cite{Watanabe2011a}. Thus, to identify conserved charge order parameters, it is sufficient to restrict to a Cartan subalgebra of $\lie g$ and find singlets of $H$ therein. By the usual root decomposition of Lie algebras, all the generators of $G$ are organized into pairs whose commutator lies in the Cartan subalgebra of $\lie g$. This maps type-B NG bosons uniquely to pairs of mutually conjugate roots of $\lie g$.

\begin{illustration}%
\label{ex:SU(n)}%
Consider the symmetry-breaking pattern $\gr{SU}(n)\to\gr{U}(n-1)$ with integer $n\geq2$. The unbroken subgroup $H\simeq\gr{U}(n-1)$ can be realized explicitly for instance as the set of unitary block-diagonal matrices with blocks of sizes $1$ and $n-1$, respectively. The Lie algebra $\lie g\simeq\lie{su}(n)$ contains a unique singlet of $H$, namely the generator of the $\gr{U}(1)$ factor of $H$, $Q\equiv\diag(-n+1,1,\dotsc,1)$. 

The $2n-2$ broken generators can be chosen as matrices with nonzero elements located in the first row and column. These span a complex $(n-1)$-plet of $\gr{SU}(n-1)\subset H$. Since all the broken generators reside in a single multiplet of $H$, we expect all the corresponding NG bosons to be either type-A or type-B. Which of these two options is realized depends on the VEV $\vev Q$. If $\vev Q=0$, we find that $n_\mathrm{A}=2n-2$ and $n_\mathrm{B}=0$. This happens in Lorentz-invariant theories, but also for instance in antiferromagnets where $n=2$. If, on the other hand, $\vev Q\neq0$ then the broken generators contribute pairwise to $\vr_{AB}$. As a result, $\rank\vr=2n-2$. According to~\eqref{NGcounting}, we then find $n_\mathrm{A}=0$ and $n_\mathrm{B}=n-1$. The special case of $n=2$ corresponds to ferromagnets.
\end{illustration}

\begin{figure}[t]
\sidecaption[t]
\includegraphics[width=2.9in]{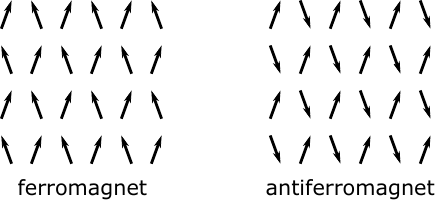}
\caption{Schematic illustration of spin order in canted (anti)ferromagnets. In ferromagnets, the individual spins are slightly tilted away from perfect alignment. Antiferromagnets, on the other hand, exhibit overall antialignment of spins, tilted so that a nonzero net magnetization arises. In both cases, the whole $\gr{SU}(2)$ group of spin rotations is spontaneously broken}
\label{fig:cantedferromagnet}
\end{figure}

\begin{illustration}%
For an example of a system where a type-B NG boson arises from the VEV of a broken symmetry generator, consider the symmetry-breaking pattern $\gr{SU}(2)\to\trgr$. Here the unbroken subgroup is trivial and any generator may develop a nonzero VEV. However, the rank of antisymmetric matrices is always even. The only possibilities are therefore $\rank\vr=0$ and $\rank\vr=2$. In the latter case, we find $n_\mathrm{A}=n_\mathrm{B}=1$. This is realized in so-called canted (anti)ferromagnets, schematically shown in Fig.~\ref{fig:cantedferromagnet}.
\end{illustration}

So far I have assumed that the internal symmetry group $G$ is compact and semisimple so that~\eqref{BWmatrixaux} holds. When this assumption is not satisfied, the commutator matrix~\eqref{BWmatrix} may receive contributions from central charges of the Lie algebra $\lie g$. This is the case for instance for the free Schr\"odinger field. Without going into detail, let me just mention that such central extensions are classified by the so-called second \emph{Lie algebra cohomology} of $\lie g$. See Chap.~6 of~\cite{Azcarraga1995} for further information.

How does the counting of NG bosons change for symmetries that are not internal, such as spacetime symmetries? In this case, there are no established general results in the literature. However, the simplicity of the argument given in Sect.~\ref{subsec:NGcanonicalconjugation} suggests that the reduction of independent NG modes based on the charge commutator~\eqref{BWmatrix} is robust. Suppose that we have already identified a set of independent NG fields $\pi^a$. Let us further assume that we can map these to a set of broken generators $Q_a$ that act on the fields by a shift such as~\eqref{NGfieldshift}. We can restrict the definition of the commutator matrix to the generators $Q_a$. The VEV in~\eqref{BWmatrix} may now in principle be coordinate-dependent, and it may no longer be appropriate to replace $J^0_B$ with the spatial average of $Q_B$. However, the rest of the argument will still go through. We will then get the same identification of type-B NG bosons with pairs of broken generators whose commutator has a nonzero VEV. Nontrivial examples of systems where the counting rule~\eqref{NGcounting} for type-B NG bosons was still found to be valid include the centrally extended algebra of spatial translations in an external magnetic field~\cite{Watanabe2012b}, and central extensions of commutators of spatial and internal symmetries in presence of topological defects such as a domain wall~\cite{Kobayashi2014b} or a line defect~\cite{Kobayashi2014a}.


\section{Nambu--Goldstone-Like Modes}
\label{sec:NGclassificationNGlike}

The defining property of a NG boson is its relation to symmetry and its spontaneous breaking. In the operator proof of Goldstone's theorem (Sect.~\ref{subsec:GoldstoneOpProof}), this manifests through the coupling of the NG state to the broken symmetry current, $\bra0J^0(0)\ket{n,\vec p}$. In the effective action proof (Sect.~\ref{subsec:GoldstoneEffActionProof}), the NG boson appears through a flat direction of the effective potential or action. Many physical systems possess excitations with the above properties that however lack the main attribute of a NG boson, that is being exactly gapless. Due to their similarity to genuine NG modes, I will call such excitations \emph{NG-like bosons}.

The existence of SSB-related modes that are not exactly gapless is of course only possible if some of the assumptions of Goldstone's theorem are violated. The usual suspect is the symmetry. There are very few \emph{exact} symmetries in nature; such usually only exist in our models, valid to certain, possibly high, accuracy. In this last section of the chapter, I will give a brief survey of different mechanisms how a NG-like boson may arise from a symmetry that is only approximate.


\subsection{Pseudo-Nambu--Goldstone Bosons}
\label{subsec:pseudoNGbosons}

Consider a theory whose Lagrangian can be split into a part invariant under a group $G$, and a small perturbation invariant only under some proper subgroup of $G$,
\begin{equation}
\La=\La_\mathrm{inv}+\eps\La_\mathrm{pert}\;.
\end{equation}
This is a setup that I used previously to isolate a unique ground state. The difference is that now I want to keep the parameter $\eps$ small but nonzero. This is called \emph{explicit} symmetry breaking.

Suppose that in the limit $\eps\to0$, the symmetry under $G$ is broken spontaneously. It is sensible to expect that by restoring small but nonzero $\eps$, the spectrum will evolve adiabatically. The NG mode predicted by Goldstone's theorem will then survive, but its mass (gap) will no longer be exactly zero. Such ``approximate'' NG bosons are usually called \emph{pseudo-NG bosons}.

Let us try to make an educated guess at how the gap of a pseudo-NG boson depends parametrically on $\eps$. Imagine that we have already derived a low-energy EFT for the (pseudo-)NG bosons in our system. The contributions of the perturbation $\La_\mathrm{pert}$ to the effective Lagrangian can be organized by powers of $\eps$. The leading contribution to the bilinear part of the effective Lagrangian typically comes from a nonderivative operator linear in $\eps$. I will justify this claim more carefully in Chap.~\ref{chap:effLagrangian}, where I actually construct the effective Lagrangian.

The rest of the argument depends on the type of the NG boson in the limit of vanishing perturbation. This type is still defined by the presence or absence of a term with a single time derivative, containing the given NG field. For type-A NG bosons, the bilinear Lagrangian contains two time derivatives which translate to squared energy in the Fourier space. For type-B NG bosons, on the other hand, a term with a single time derivative is present and dominates in the low-energy limit. This leads to the following parametric dependence of energy at zero momentum on $\eps$,
\begin{equation}
\begin{aligned}
\lim_{\vec p\to\vec0}E(\vec p)&\propto\sqrt\eps\quad&&\text{(``type-A'' pseudo-NG boson)}\;,\\
\lim_{\vec p\to\vec0}E(\vec p)&\propto\eps\quad&&\text{(``type-B'' pseudo-NG boson)}\;.
\end{aligned}
\label{pseudoNG}
\end{equation}

\begin{watchout}%
I have just used quotation marks for a reason. There is no established, unambiguous classification of pseudo-NG bosons. All I did was to rely on the continuity of the spectrum in the limit $\eps\to0$. Moreover, the parametric scaling in~\eqref{pseudoNG} is based on the assumption of a rather specific form of perturbation in the effective Lagrangian. It is imaginable that even for an originally type-A NG boson, $\La_\mathrm{pert}$ will induce an operator in the effective Lagrangian with a single time derivative and a coefficient proportional to some power of $\eps$. That may change the way the gap of the pseudo-NG mode scales with $\eps$. This is exactly what happens for the ``massive'' NG bosons, introduced in Sect.~\ref{subsec:massiveNGbosons}.
\end{watchout}

It is a simple application of the discussion in Sect.~\ref{subsec:NGclassificationcounting} to find how many pseudo-NG bosons one should expect in a given system. I will denote as $G_\eps$ and $H_\eps$ the symmetries of the action and of the ground state in presence of the perturbation. In analogy with~\eqref{BWmatrix}, I will define the commutator matrix $\vr_\eps$ by
\begin{equation}
\vr_{\eps,AB}\equiv\I\vev{[Q_A,J^0_B]}_\eps\;.
\end{equation}
The $\eps$ in the subscripts reminds us that the indices $A,B$ run over the generators of $G_\eps$, and that the VEV is taken in the perturbed vacuum. According to~\eqref{NGcounting}, there are altogether $\dim G-\dim H-(1/2)\rank\vr$ NG modes in the limit $\eps\to0$. By the same formula, $\dim G_\eps-\dim H_\eps-(1/2)\rank\vr_\eps$ genuine NG bosons survive upon switching on the perturbation. On the assumption of continuity of the spectrum, the number of pseudo-NG bosons therefore equals
\begin{equation}
n_\mathrm{pseudoNG}=(\dim G-\dim G_\eps)-(\dim H-\dim H_\eps)-\frac12(\rank\vr-\rank\vr_\eps)\;.
\label{pseudoNGcounting}
\end{equation}
We saw an example of a pseudo-NG boson already in Chap.~\ref{chap:ourfirstmodel}. In that case, $G\simeq\gr{U}(1)_\mathrm{V}\times\gr{U}(1)_\mathrm{A}$ and $H\simeq\gr{U}(1)_\mathrm{V}$, whereas $G_\eps\simeq H_\eps\simeq\gr{U}(1)_\mathrm{V}$. The counting rule~\eqref{pseudoNGcounting} is trivially satisfied and the prediction~\eqref{pseudoNG} for the scaling of the gap agrees with our explicit calculation of the mass, cf.~\eqref{vwmass}. Next, let us have a look at a couple of less trivial examples.

\begin{illustration}%
We already met \emph{quantum chromodynamics} (QCD) in Chap.~\ref{chap:SSB}, see \refex{ex:QCDfirsttime}. In this case, the quark mass plays the role of the perturbation $\eps$. With $\nf$ flavors of massless quarks, QCD possesses a $G\simeq\gr{SU}(\nf)_\mathrm{L}\times\gr{SU}(\nf)_\mathrm{R}$ symmetry, spontaneously broken down to the ``vector'' subgroup $H\simeq\gr{SU}(\nf)_\mathrm{V}$. In presence of the perturbation (with equal masses of all quarks), this reduces to $G_\eps\simeq H_\eps\simeq\gr{SU}(\nf)_\mathrm{V}$. The vacuum of QCD is Lorentz-invariant, so the matrices $\vr$ and $\vr_\eps$ are both identically zero. According to the counting rule~\eqref{pseudoNGcounting}, this leaves us with $\nf^2-1$ pseudo-NG bosons, which can be identified with a multiplet of pseudoscalar mesons. The fact that their mass scales with the square root of the quark mass, as predicted by~\eqref{pseudoNG}, has been verified in numerical lattice simulations of QCD.
\end{illustration}

\begin{illustration}%
\label{ex:antiferromagnetcounting}%
As discussed before, ideal (isotropic) antiferromagnets possess a $G\simeq\gr{SU}(2)$ spin symmetry, spontaneously broken down to $H\simeq\gr{U}(1)$. The low-energy EFT for antiferromagnetic spin waves (magnons) is a nonrelativistic version of the nonlinear sigma model, introduced in \refex{ex:NLSMfirst},
\begin{equation}
\La_\mathrm{eff}=\frac{\vr_\mathrm{s}}{2v^2}\bigl(\d_{ij}\de_0n^i\de_0n^j-v^2\d_{ij}\vec\nabla n^i\cdot\vec\nabla n^j\bigr)+\eps(n^3)^2\;.
\end{equation}
Here $\vr_\mathrm{s}$ is the so-called spin stiffness, $\vec n$ a unit-vector variable encoding the two NG fields, and $v$ the magnon phase velocity in the symmetric limit $\eps=0$.

For $\eps>0$, the perturbation proportional to $(n^3)^2$ describes a crystal anisotropy, corresponding to ``easy-axis'' antiferromagnets; see~\cite{Rezende2019} for a mild introduction to antiferromagnetic magnons. The perturbation breaks the continuous spin symmetry explicitly down to $G_\eps\simeq\gr{U}(1)$, consisting of rotations around the third spin axis. At the same time, it leaves only two possible ground states, $\vev{\vec n}_\eps=(0,0,\pm1)$, both of which preserve $H_\eps\simeq\gr{U}(1)$. By the counting formula~\eqref{pseudoNGcounting}, both magnons become pseudo-NG bosons. Their gap is proportional to $\sqrt\eps$ in accord with~\eqref{pseudoNG}. 
\end{illustration}


\subsection{Quasi-Nambu--Goldstone Bosons}
\label{subsec:quasiNGbosons}

The above general discussion of pseudo-NG bosons applies in principle to any NG-like mode, associated with an approximate symmetry. However, the strength of the perturbation, and hence the size of the gap in the spectrum, may depend on the concrete mechanism of explicit symmetry breaking. It is in particular possible to have systems with a well-defined classical limit whose symmetry is explicitly broken only by quantum corrections. The ensuing pseudo-NG modes have been dubbed \emph{quasi-NG bosons}~\cite{Uchino2010a}. I refer the reader to~\cite{Takahashi2015a,Nitta2015a} for a detailed discussion of the spectrum of quasi-NG modes.

To be more concrete, suppose we are given a Lorentz-invariant Lagrangian whose potential has a higher symmetry than the kinetic term. At the classical level, the spectrum of NG bosons is determined by the zero modes of the Hessian of the potential in the ground state. But some of these zero modes may correspond to symmetries of the potential that do not preserve the kinetic term. The corresponding fluctuations of the order parameter will receive a mass generated by quantum corrections.

The lifting of the mass of the would-be NG modes does not have to be induced by the kinetic term. What matters is that the classical ground state is determined entirely by a part of the Lagrangian that has a higher symmetry than the whole. The reduction of the symmetry may also be caused for instance by other contributions to the classical potential, or by a coupling to gauge fields.


\subsection{Massive Nambu--Goldstone Bosons}
\label{subsec:massiveNGbosons}

What the pseudo-NG bosons and their subclass, quasi-NG bosons, have in common is that their gap can rarely be determined from first principles. In general, it depends on all parameters a given theory might have. One says that the gap is not protected by symmetry, in contrast to genuine NG bosons whose gap is guaranteed to vanish by SSB. There is however a remarkable special class of pseudo-NG bosons whose gap can be calculated exactly using symmetry alone. Their existence was first pointed out about a decade ago~\cite{Nicolis2013a} and I will call them \emph{massive NG bosons} following~\cite{Watanabe2013b}.

Consider a system with a ``microscopic'' Hamiltonian $H$, possessing an internal symmetry group $G$. Pick one generator, $Q$, of $G$. In statistical physics, the many-body ground state of the system with a fixed average value of $Q$ is determined by minimizing the grandcanonical Hamiltonian $H_\m\equiv H-\mu Q$. Here $\m$ is the corresponding chemical potential, which is fixed by external conditions imposed on the system. If needed, one can always shift $H_\m$ by a constant so that the many-body ground state $\ket0_\m$ satisfies $H_\m\ket0_\m=0$. In this ground state, the symmetry under $G$ may be spontaneously broken. We are particularly interested in the fate of those generators of $G$ that are spontaneously broken but do not commute with $Q$.

As long as $G$ is compact, we can always choose $Q$ to be part of a Cartan subalgebra of the Lie algebra $\lie g$ of $G$. Those generators of $G$ that do not commute with $Q$ then organize themselves into pairs $Q_\pm$ such that $Q_-=\he Q_+$ and
\begin{equation}
[Q,Q_\pm]=\pm qQ_\pm\;.
\label{QQpm}
\end{equation}
The factor $q$ is determined by the corresponding root vector, and is thus fixed by symmetry. We can always choose the generators so that both $\m$ and $q$ are positive. I will now rerun the operator proof of Goldstone's theorem detailed in Sect.~\ref{subsec:GoldstoneOpProof} on the VEV $\prescript{}{\m}{\bra0}[Q_{+\Omega}(t),J^0_-(0)]\ket0_\m$.

Of course, \eqref{QQpm} is only formal since we know that generators of spontaneously broken symmetry are ill-defined as operators on the Hilbert space. I therefore assume that~\eqref{QQpm} also holds for the finite-volume charges $Q_{\pm\Omega}$, possibly up to a correction that vanishes in the limit $\Omega\to\infty$. This can be ensured for instance if the charge densities satisfy the local commutation relation $[Q,J^0_\pm(x)]=\pm qJ^0_\pm(x)$. One can then define a related pair of charges via ``time evolution'' with respect to $H_\m$,
\begin{equation}
\begin{split}
\prescript{\mu}{}Q_{\pm\Omega}(t)&\equiv\int_\Omega\D^d\!\vec x\,\exp(\I H_\m t-\I\skal Px)J^0_\pm(0)\exp(-\I H_\m t+\I\skal Px)\\
&=\E^{-\I\m Qt}Q_{\pm\Omega}(t)\E^{+\I\m Qt}=\E^{\mp\I\m qt}Q_{\pm\Omega}(t)\;.
\end{split}
\end{equation}
The point of this redefinition is that the spectrum of the many-body system consists of eigenstates of $H_\m$ rather than $H$. I will keep the notation $E_n(\vec p)$ for the eigenvalue of $H_\m$ (excitation energy) in the eigenstate $\ket{n,\vec p}$. Following the same steps as in Sect.~\ref{subsec:GoldstoneOpProof} then leads to an analog of~\eqref{Goldstoneproof2},
\begin{align}
\notag
&\prescript{}{\mu}{\bra0}[Q_{+\Omega}(t),J^0_-(0)]\ket0_\m=\E^{\I\m qt}\prescript{}{\mu}{\bra0}[\prescript{\mu}{}Q_{+\Omega}(t),J^0_-(0)]\ket0_\m\\
&=\E^{\I\m qt}\smash{\sum_n\int}\D^d\!\vec p\,\bigl\{\exp[-\I E_n(\vec p)t]\d^d_\Omega(\vec p)\abs{\prescript{}{\mu}{\bra0}J^0_+(0)\ket{n,\vec p}}^2\\
\notag
&\qquad\qquad\qquad\qquad-\exp[+\I E_n(\vec p)t]\d^d_\Omega(-\vec p)\abs{\prescript{}{\mu}{\bra0}J^0_-(0)\ket{n,\vec p}}^2\bigr\}\;.
\end{align}

With positive $\mu$ and $q$, the assumed time independence of VEVs of commutators of $Q_{+\Omega}(t)$ in the $\Omega\to\infty$ limit guarantees that $\lim_{\vec p\to\vec0}\prescript{}{\mu}{\bra0}J^0_-(0)\ket{n,\vec p}=0$ for any eigenstate $\ket{n,\vec p}$. Assuming that the VEV we started with is nonzero, on the other hand, ensures the existence of a state for which $\prescript{}{\mu}{\bra0}J^0_+(0)\ket{n,\vec p}\neq0$ and
\begin{equation}
\lim_{\vec p\to\vec0}E_n(\vec p)=\mu q\;.
\label{massiveNGmass}
\end{equation}
This is our massive NG boson. With some extra effort, one can also derive a counting rule for the number of massive NG modes in a system. To formulate such a rule, we adapt~\eqref{BWmatrix} for the present problem by introducing $\smash{\vr_{\m,AB}\equiv\I\vev{[Q_A,J^0_B]}_\m}$ and $\smash{\tilde\vr_{\m,AB}\equiv\I\vev{[\tilde Q_A,\tilde J^0_B]}_\m}$. The matrix $\tilde\vr$ is constructed using the generators of the subgroup of $G$ that commutes with $Q$. The number of massive NG modes then is
\begin{equation}
n_\mathrm{massiveNG}=\frac12(\rank\vr_\m-\rank\tilde\vr_\m)\;.
\label{massiveNGcounting}
\end{equation}
See~\cite{Watanabe2013b} for a detailed proof.

The counting rule~\eqref{massiveNGcounting} may resemble the last term in~\eqref{pseudoNGcounting} that counts pseudo-NG bosons. It may thus be worthwhile to stress the difference. Namely, the matrices $\vr$ and $\vr_\eps$ in~\eqref{pseudoNGcounting} are evaluated in different ground states, corresponding respectively to the absence and presence of the perturbation. On the contrary, both matrices $\vr_\m$ and $\tilde\vr_\m$ in~\eqref{massiveNGcounting} are evaluated in the same ground state, ``perturbed'' by the presence of the chemical potential. The difference is best illustrated on an example.

\begin{illustration}%
Let us have one more look at antiferromagnets, this time assuming perfect isotropy and spin symmetry, that is $G\simeq\gr{SU}(2)$ and $H\simeq\gr{U}(1)$. In an external magnetic field $\vec B$, the Hamiltonian of the antiferromagnet receives a contribution $-\m_\mathrm{m}\skal BS$, where $\m_\mathrm{m}$ is the magnetic moment and $\vec S$ the operator of total spin. The combination $\m_\mathrm{m}\abs{\vec B}$ plays the role of a chemical potential for the projection of $\vec S$ into the direction of $\vec B$. The latter can without loss of generality be chosen as, say, $S_3$.

Let us first treat the effect of the magnetic field as a perturbation in the sense of Sect.~\ref{subsec:pseudoNGbosons}. The field breaks the spin symmetry explicitly down to $G_\eps\simeq\gr{U}(1)$. This necessarily implies $\rank\vr_\eps=0$. Somewhat against the intuition, the antiferromagnet responds to the magnetic field by aligning its spins largely in a direction perpendicular to $\vec B$. This ground state breaks the residual exact symmetry so that $H_\eps\simeq\trgr$. At the same time, in the unperturbed antiferromagnetic ground state, there is no net magnetization so that $\rank\vr=0$. Equations~\eqref{BWmatrix} and~\eqref{pseudoNGcounting} then tell us that there is one genuine, type-A NG boson and one pseudo-NG boson. Unlike in the case of an easy-axis anisotropy, one of the two magnons therefore remains gapless.

The fact that the magnetic field acts as a chemical potential allows us to make a stronger statement about the spectrum. The magnetic field does induce net magnetization in the perturbed ground state, given by nonzero $\vev{S_3}_\m=-\I\vev{[S_1,S_2]}_\m$. This implies $\rank\vr_\m=2$. At the same time, $\rank\tilde\vr_\m=0$ simply because there are not enough symmetry generators left intact by the magnetic field. In accord with~\eqref{massiveNGcounting}, we then expect one massive NG mode in the spectrum. In other words, the previously predicted pseudo-NG magnon has a gap exactly fixed by~\eqref{massiveNGmass}. The algebraic factor $q$ equals one thanks to the commutator of $S_3$ with the ladder operators $S_\pm\equiv S_1\pm\I S_2$, $[S_3,S_\pm]=\pm S_\pm$. Hence the gap of the magnon equals $\m_\mathrm{m}\abs{\vec B}$.
\end{illustration}

An interested reader will find further examples of massive NG bosons in~\cite{Watanabe2013b}. Let me just stress that the exact expression~\eqref{massiveNGmass} for the gap is not the only special property of massive NG bosons. The peculiar perturbation by a chemical potential coupled to a conserved charge leaves us with a set of exact, albeit modified, conservation laws. These in turn impose exact constraints on the interactions of massive NG bosons. As a result, the Adler zero principle (cf.~Sect.~\ref{sec:firstmodelNGboson}) remains valid for massive NG modes, although it is generally violated for pseudo-NG bosons~\cite{Brauner2018a}.

As an aside, note that not every NG boson that acquires a gap due to a chemical potential is a massive NG boson~\cite{Nicolis2013b}. The simplest example of such an exception is provided by choosing $G\simeq\gr{SU}(2)$ and breaking it completely in the ground state. Provided none of the three generators develops a VEV, the spectrum contains three type-A NG bosons. Upon adding a chemical potential $\m$ for one of the generators, two of the NG modes receive a gap, proportional to $\m$. Only one of these is however a massive NG boson, as is easily checked using~\eqref{massiveNGcounting}.


\bibliographystyle{spphys}
\bibliography{references}
\begin{partbacktext}
\part{Spontaneously Broken Internal Symmetry}
\label{part:internalSSB}
\end{partbacktext}
\chapter{Nonlinear Realization of Symmetry}
\label{chap:CCWZ}

\abstract*{Spontaneously broken symmetries are not realized by unitary operators on the Hilbert space of states of the quantum system. Likewise, it is rarely convenient to represent them by linear transformations on fields, either classical or quantum. This chapter develops the technique of nonlinear realization of symmetry, which plays a key role in the construction of effective theories for spontaneously broken symmetries. The material covered includes both the conceptual foundations of the method and practical details of the so-called standard nonlinear realization. Most of the text is formulated in the language of field theory. The last section, however, offers a complementary geometric viewpoint. This provides intuitive insight into some of the previously discussed technical details. It also builds ground for some of the material of the subsequent chapters.}


In Parts~\ref{part:prologue} and~\ref{part:foundations} of the book, I reviewed the basic physics of~\emph{spontaneous symmetry breaking} (SSB) and the corresponding~\emph{Nambu--Goldstone} (NG) \emph{bosons}. In order to keep the discussion simple, I relied mostly on general but elementary field-theoretic arguments. Where explicit results were necessary or useful, I resorted to simple models. However, one of the main virtues of SSB is universality: the low-energy physics is largely independent of the microscopic, short-distance details. This agrees with the general spirit of \emph{effective field theory} (EFT); cf.~Sect.~\ref{sec:whatisEFT}. What SSB does for us is ensure a separation of resolution scales, which makes it possible to build a low-energy EFT solely using the NG degrees of freedom.

I will now set out on the main quest of this book: to develop an EFT framework for SSB that is based on symmetry alone. Thanks to the universality of SSB, the predictions of such a framework are guaranteed to match those of any microscopic theory with the same symmetry-breaking pattern. The construction of the EFT requires two main steps. The first step is to realize the action of the symmetry solely in terms of a set of NG fields. This problem is dealt with in the present chapter. The results will be used in the next chapter to construct effective actions for NG bosons.

The celebrated Wigner theorem of quantum mechanics states that any symmetry of a quantum system can be realized by a linear unitary or antilinear antiunitary operator on its Hilbert space (see Chap.~2 of~\cite{Weinberg1995a} for a detailed discussion). As a consequence, states in the spectrum of a quantum system can be organized into multiplets, corresponding to irreducible representations of its symmetry group. In quantum field theory, the concept of linear realization of symmetry is naturally lifted from states to fields. In a perturbative setting, one usually has a one-to-one correspondence between one-particle states and (elementary) fields. Moreover, the fields span a linear space. One then expects the symmetry to act on fields linearly via some representation just like it does on states in the Hilbert space. 

\begin{illustration}%
\label{ex:U(1)symmetry}%
A field theory of a single complex scalar field $\p$ may possess a $\gr{U}(1)$ symmetry under which the field transforms as $\phi\to\E^{\I\eps}\p$, where $\eps$ is a real parameter. In this case, the field $\p$ belongs to a complex one-dimensional representation of $\gr{U}(1)$. Similarly, the $\gr{SO}(n)$ ``linear sigma model'' includes an $n$-plet of real scalar fields $\p^i$, subject to symmetry transformations $\smash{\p^i\to R^i_{\phantom ij}\p^j}$, where $R\in\gr{SO}(n)$. In this case, the fields $\p^i$ belong to the vector representation of $\gr{SO}(n)$.
\end{illustration}

As we saw in Sect.~\ref{sec:SSBsubtleties}, however, a symmetry that is spontaneously broken is not necessarily realized by a set of unitary operators on the Hilbert space of states. Likewise, as illustrated in Chap.~\ref{chap:ourfirstmodel}, it may be convenient to use field variables that do not belong to a linear representation of the symmetry. This may even become a necessity in the low-energy EFT where only the NG bosons of the broken symmetry are present; we simply do not have enough degrees of freedom to fill complete multiplets of the symmetry group.

\begin{illustration}%
The complex scalar field of \refex{ex:U(1)symmetry} can be represented by its real and imaginary parts, $\p=\p^1+\I\p^2$. These span the vector representation of $\gr{SO}(2)\simeq\gr{U}(1)$. When the $\gr{U}(1)$ symmetry is spontaneously broken, it may however be more convenient to use the exponential parameterization of the field, $\p=\vr\E^{\I\t}$, in terms of its modulus $\vr$ and phase $\t$. In the low-energy EFT, the modulus field is integrated out and the only remaining degree of freedom is the NG field $\t$. The latter transforms under $\E^{\I\eps}\in\gr{U}(1)$ as $\t\to\t+\eps$, which is not a linear representation. This is not merely a matter of bad choice of parameterization; the EFT contains a single real field, yet the group $\gr{U}(1)$ does not have any nontrivial real one-dimensional representations.
\end{illustration}

With the above observations in mind, I will develop in this chapter a formalism for nonlinear realization of symmetries. Mathematically, this amounts to generalizing the concept of a linear representation of a symmetry group to that of an \emph{action} of the group. The space on which the group acts need not be linear itself; we can think of it as some manifold. Section~\ref{sec:CCWZgroupaction} introduces the necessary mathematical terminology. The main argument, leading to a classification of possible nonlinear realizations of symmetry, is presented in Sect.~\ref{sec:CCWZclassification}. A reader interested mainly in ready-made results of the formalism may want to proceed directly to Sect.~\ref{sec:CCWZstandard}; this collects a number of practically useful formulas that I will refer to in the following chapters. Finally, Sect.~\ref{sec:CCWZgeometry} offers an alternative geometric viewpoint which illuminates some of the mathematical structure used to realize a symmetry nonlinearly. While most of the chapter is phrased in a rather elementary language, this last section relies on some concepts of differential geometry in an extent covered in Appendix~\ref{app:diffgeom}.


\section{Group Action on Manifolds}
\label{sec:CCWZgroupaction}

To motivate the mathematical language that I need to introduce, suppose that we are given a theory of a set of fields, $\psi^i$, that possesses a symmetry group, $G$.\footnote{Many of the concepts introduced below can be applied without change to any, even finite, group. However, I will always have implicitly in mind a connected Lie group, or the component of a Lie group connected to the unit element.} We would like to understand how to realize the symmetry group in terms of a set of transformations of the fields,
\begin{equation}
T_g:\psi^i\to\psi'^i\equiv\DF^i(\psi,g)\;,\qquad
g\in G\;,
\label{pointtransfo}
\end{equation}
where the functions $\DF^i$ are assumed to be smooth in both of their arguments. The set of transformations $T_g$ is constrained by the requirement that it respects the group structure of $G$. Thus, the unit element $e\in G$ must be represented by the identity map $\id$, $\DF^i(\psi,e)=\psi^i$. Consistency with group multiplication requires that
\begin{equation}
\DF^i(\psi,g_1g_2)=\DF^i(\DF(\psi,g_2),g_1)\;,\qquad
g_1,g_2\in G\;.
\label{grouptimes}
\end{equation}
Finally, the transformation induced by the inverse of an element $g\in G$ has to satisfy
\begin{equation}
\DF^i(\DF(\psi,g),g^{-1})=\psi^i\;,\qquad
g\in G\;.
\label{groupinverse}
\end{equation}

In the terminology introduced in Chap.~\ref{chap:symmetry},~\eqref{pointtransfo} is an example of a \emph{point transformation}~\cite{Olver1986a,Bluman2002a}. The class of point transformations is clearly much broader than that of mere linear transformations, induced by a representation of $G$ on the fields. It is therefore worthwhile to recall that even point transformations do not exhaust all conceivable, and physically relevant, realizations of symmetry. First, the field transformation may in principle depend explicitly on the spacetime coordinates. This feature can be included under the umbrella of point symmetries by treating fields and coordinates on the same footing; this will become relevant later when we talk about spacetime symmetries. Moreover, it is perfectly possible that the transformation of the fields also depends on their derivatives; this was dubbed \emph{generalized local transformation} in Sect.~\ref{sec:whatissym}. Such generalized symmetries play a minor role in this book, yet we will see some concrete examples in Chap.~\ref{chap:scattering}.

The restriction to point symmetries of the type~\eqref{pointtransfo}, which I will make from now on unless explicitly stated otherwise, is a matter of practical convenience. Namely, it will allow us to disregard the fact that $\psi^i$ actually are fields, that is functions of spacetime coordinates. Instead, I will treat them as independent variables that the group $G$ acts upon. With this important technical assumption, we can now reformulate the problem of finding all possible realizations of a given symmetry group $G$ on the fields $\psi^i$ in geometric terms.

Consider a manifold $\M$ such that each point $x\in\M$ is uniquely specified by a set of values $\psi^i$. We can think of $\psi^i$ as a set of (possibly only locally defined) coordinates on $\M$. An \emph{action} of the group $G$ on $\M$ is a set of smooth invertible maps $T_g:\M\to\M$ that satisfy the group constraints
\begin{equation}
T_e=\id\;,\qquad
T_{g_1g_2}=T_{g_1}\circ T_{g_2}\;,\qquad
T_{g^{-1}}=(T_g)^{-1}\;,\qquad
g,g_1,g_2\in G\;.
\end{equation}
In somewhat more abstract terms, the action of the group on the manifold is defined by a homomorphism from $G$ to the group of diffeomorphisms on $\M$. 
\begin{watchout}%
It is important to distinguish actual symmetry transformations on $\M$ from a mere change of coordinates. The action of the group $G$ on $\M$ is defined geometrically by the maps $T_g$ without reference to a particular set of coordinates $\psi^i$. Depending on the choice of coordinates, the same map $T_g$ may correspond to different functions $\DF^i$ as defined by~\eqref{pointtransfo}. The freedom to choose coordinates on $\M$ mirrors the freedom to choose field variables in a given field theory. On the one hand, the independence of geometric properties of manifolds on the choice of local coordinates is a cornerstone of the language of differential geometry. On the other hand, it is an important result of quantum field theory that physical observables such as the $S$-matrix are invariant under (nearly) arbitrary field redefinitions~\cite{Chisholm1961a,Kamefuchi1961a}. See~\cite{Criado2019} for a recent pedagogical discussion of this issue.
\end{watchout}
Let us now introduce some further terminology. For a given action of the group $G$ on the manifold $\M$, the \emph{orbit} of a point $x\in\M$ is the set of all points on $\M$ that can be reached from $x$ by the action of some group element,
\begin{equation}
\O_x\equiv\{T_gx\,\vert\,g\in G\}\;.
\label{orbit}
\end{equation}
The relation $y\sim x$ if and only if there is a $g\in G$ such that $y=T_gx$ is an equivalence. Orbits of the group $G$ on the manifold $\M$ are the equivalence classes of this relation. As a consequence, $\M$ is a disjoint union of a (possibly infinite) set of orbits.

For a given point $x\in\M$, one defines its \emph{isotropy group} (also called the \emph{stabilizer} or the \emph{little group} of $x$) as the subgroup of $G$ that maps $x$ to itself,
\begin{equation}
H_x\equiv\{h\in G\,\vert\,T_hx=x\}\;.
\end{equation}
Two points lying on the same orbit of $G$ have isomorphic isotropy groups. Indeed, if $y=T_gx$, then for any $h\in H_x$ we have $T_{ghg^{-1}}y=T_gT_hT_{g^{-1}}y=T_gT_hx=T_gx=y$. Conversely, it is easy to check that $T_hy=y$ implies $g^{-1}hg\in H_x$. Thus, $H_x$ and $H_y$ are conjugate as subgroups,
\begin{equation}
H_{T_gx}=gH_xg^{-1}\;.
\label{isotropyconjugation}
\end{equation}

\begin{figure}[t]
\sidecaption[t]
\includegraphics[width=2.0in]{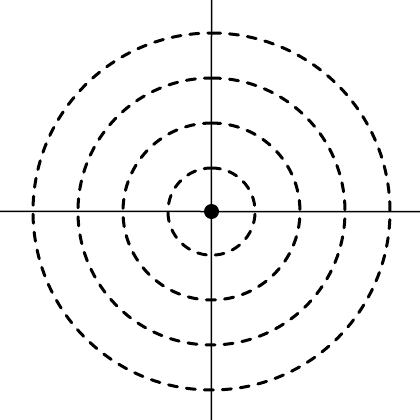}
\caption{Illustration of the action of the rotation group $\gr{SO}(2)$ on the Euclidean plane. Some orbits of $\gr{SO}(2)$ for which the isotropy group is trivial are displayed using \emph{dashed lines}. The \emph{black dot} indicates the single orbit whose isotropy group is the whole of $\gr{SO}(2)$}
\label{fig:SO(2)actioninplane}
\end{figure}

\begin{illustration}%
\label{ex:SO(2)actioninplane}%
The rotation group $\gr{SO}(2)$ acts on the Euclidean plane by rotations around the origin, see Fig.~\ref{fig:SO(2)actioninplane} for an illustration. All points of the plane away from the origin have a trivial isotropy group. The corresponding orbits of $\gr{SO}(2)$ are circles centered at the origin. The origin itself forms an orbit with the isotropy group $\gr{SO}(2)$.

This example has a straightforward generalization to the action of $\gr{SO}(n)$ on the Euclidean space $\R^n$. There, the origin has the isotropy group $\gr{SO}(n)$. All other points $x\in\R^n$ have the isotropy group $\gr{SO}(n-1)$, corresponding to $(n-1)$-dimensional rotations that leave the line connecting $x$ to the origin fixed. The corresponding orbits are $(n-1)$-dimensional spheres centered at the origin.
\end{illustration}

In somewhat loose terms, we can say that all points lying on the same orbit of $G$ on the manifold $\M$ have ``the same'' properties, since they can be related by a symmetry transformation. We thus expect manifolds consisting of a single orbit to have particularly simple geometric properties. In this case, the action of the group $G$ on the manifold $\M$ is called \emph{transitive}; any point on the manifold can be reached from any other point by the action of a suitable group element. A manifold equipped with a transitive action of a group is referred to as a \emph{homogeneous space}.

Homogeneous spaces will play a central role throughout the rest of this book. One of the special properties of a homogeneous space is that its structure is completely determined by the group $G$ and its subgroup $H$ that specifies the isotropy group of the homogeneous space.\footnote{Since the isotropy group is the same up to conjugation for all points on the same orbit, and hence all points of the homogeneous space, we can drop the subscript $x$.} To see why, let us introduce a relation between two elements of $G$: $g_2\sim g_1$ if and only if there is an $h\in H$ such that $g_2=g_1h$. This is an equivalence relation. As a consequence, the group $G$ is partitioned into a disjoint union of the corresponding equivalence classes. The equivalence class
\begin{equation}
gH\equiv\{gh\,\vert\,h\in H\}\;,\qquad
g\in G\;,
\end{equation}
is called the \emph{left coset} (or simply \emph{coset}) of $H$ in $G$. Incidentally, the cosets themselves can be viewed as orbits of a group action if we treat the group $G$ as the manifold on which the subgroup $H$ acts by multiplication from the right.

The quotient set $G/H$, called the \emph{coset space}, is the set of all cosets of $H$ in $G$. Despite the different appearance, this is just another mathematical realization of a homogeneous space. Indeed, consider a homogeneous space $\M$ along with a single point $x\in\M$ and its isotropy group $H$. It is easy to see that for any two elements $g_1,g_2\in G$, $T_{g_1}x=T_{g_2}x$ if and only if $g_1\sim g_2$, that is if $g_1$ and $g_2$ belong to the same coset. Hence there is a one-to-one correspondence between the elements of the coset space $G/H$ and the points of the manifold $\M$.

\begin{illustration}%
The Euclidean group $\gr{ISO}(n)$ consists of proper rotations and translations in $\R^n$ and their combinations. It obviously acts transitively on $\R^n$ for the simple reason that any point in $\R^n$ can be reached from any other point by a suitable translation. Each point in $\R^n$ has a its stabilizer a particular subgroup of $\gr{ISO}(n)$, consisting of $\gr{SO}(n)$ rotations around that point. Hence $\R^n$ equipped with the action of $\gr{ISO}(n)$ is a homogeneous space, equivalent to the coset space $\gr{ISO}(n)/\gr{SO}(n)$. This view of the Euclidean space in terms of its symmetry group follows the ``Erlangen program,'' put forward by Felix Klein in 1872.
\end{illustration}

\begin{illustration}%
\label{ex:SO(n)actionSn}%
The rotation group $\gr{SO}(n+1)$ acts naturally on the $n$-dimensional unit sphere $S^n$ if one thinks of the latter as embedded in $\R^{n+1}$. We can however think of the sphere $S^n$ in itself as our manifold $\M$. In that case, the action of $\gr{SO}(n+1)$ becomes transitive; any point on the sphere can be reached from any other point by a suitable rotation. The isotropy group of any point on the sphere is a particular $\gr{SO}(n)$ subgroup of $\gr{SO}(n+1)$. Hence, the sphere $S^n$ equipped with the action of $\gr{SO}(n+1)$ is a homogeneous space, equivalent to the coset space $\gr{SO}(n+1)/\gr{SO}(n)$.
\end{illustration}

Having set up the necessary mathematical background, I can now give a concise formulation of the main goal of this chapter. In order to understand how a given symmetry group $G$ can act on a given set of fields, we need to classify all possible actions of $G$ on a given manifold $\M$. This will be accomplished in the following section. A reader seeking further details on the mathematical background reviewed above is recommended to consult Chap.~13 of~\cite{Fecko2011a}.


\section{Classification of Nonlinear Realizations}
\label{sec:CCWZclassification}

At first sight, the task to classify all actions of an arbitrary group $G$ on an arbitrary manifold $\M$ seems hopeless. We would only know what to do in the case of a linear action on a vector space, where the problem boils down to the good old representation theory of Lie groups. How can we make use of this?

In this section, I will answer the above question, following the classic work of Coleman, Wess and Zumino~\cite{Coleman1969a}. The key step is to observe that the action of a group on a manifold can be \emph{partially} linearized by a suitable choice of coordinates on the manifold. Let us choose a fixed point $x_0\in\M$. We can always introduce a set of coordinates $\psi^i$ in the neighborhood of $x_0$ such that $x_0$ maps to the origin, $\psi^i=0$. Now any linear transformation leaves the origin intact. On the contrary, only the isotropy group $H_{x_0}$ keeps $x_0$ fixed; all the other elements of $G$ translate $x_0$ to some other point on the manifold. Thus, the best we can hope for is that we find a set of coordinates in which the action of $H_{x_0}$, not of the whole group $G$, becomes linear.

We will see that provided the isotropy group $H_{x_0}$ is compact, one can indeed construct a set of local coordinates on $\M$ in the neighborhood of $x_0$ in which the action of $H_{x_0}$ is linear. The good news is that this is in fact sufficient to classify all nonlinear realizations of the whole group $G$. The details of the argument are the subject of the following two subsections.

\begin{table}[t]
\caption{Correspondence between the mathematical terminology used in this chapter and the physics terminology introduced in Chap.~\ref{chap:SSB}}
\label{tab:SSBvsgeometry}
\begin{tabular}{p{4cm}p{3.8cm}}
\hline\noalign{\smallskip}
Mathematics terminology & Physics terminology\\
\noalign{\smallskip}\svhline\noalign{\smallskip}
``Origin'' $x_0\in\M$ & Vacuum state (order parameter)\\
Isotropy group $H_{x_0}$ & Unbroken subgroup $H$\\
Coset space $G/H_{x_0}$ & Vacuum manifold\\
\noalign{\smallskip}\hline\noalign{\smallskip}
\end{tabular}
\end{table}

While this whole chapter is intended to develop a mathematical formalism that will prove invaluable later, it might be helpful for the reader already now to keep in mind the corresponding concepts pertinent to SSB, introduced in Chap.~\ref{chap:SSB}. The fixed point $x_0\in\M$ corresponds to the selected vacuum state, or the associated order parameter. The subgroup $H_{x_0}$ that leaves $x_0$ intact is analogous to the subgroup of unbroken symmetries. Finally, the coset space $G/H_{x_0}$, which will turn out to span a submanifold of $\M$, is analogous to the vacuum manifold. For the reader's convenience, this correspondence is summarized in Table~\ref{tab:SSBvsgeometry}.


\subsection{Linearization of Group Action}
\label{subsec:CCWZlinearization}

To simplify the notation, I will from now on denote the functions $\DF^i$ defined in~\eqref{pointtransfo} directly as $\psi'^i$ whenever possible. Suppose that we have a set of local coordinates $\psi^i$ on $\M$ in which the chosen point $x_0$ maps to zero. In these coordinates, the isotropy group $H_{x_0}$ will be represented by some nonlinear functions of $\psi^i$ whose Taylor expansion in $\psi^i$ starts at the linear order,
\begin{equation}
\psi'^i(\psi,h)=D(h)^i_{\phantom ij}\psi^j+\bigO(\psi^2)\;,
\label{linearizationaux}
\end{equation}
where $h\in H_{x_0}$ and $D(h)^i_{\phantom ij}$ is a set of matrix coefficients. The conditions~\eqref{grouptimes} and~\eqref{groupinverse} require that the matrices $D(h)$ form a representation of $H_{x_0}$. We can then change the coordinates in the vicinity of $x_0$ to
\begin{equation}
\Psi^i(\psi)\equiv\int_{H_{x_0}}\D h\,D(h^{-1})^i_{\phantom ij}\psi'^j(\psi,h)\;,
\label{linearization}
\end{equation}
where $\D h$ is an invariant measure on $H_{x_0}$ normalized so that the total volume of $H_{x_0}$ is one.\footnote{A reader not familiar with group integration may find more information in Chap.~3 of~\cite{Barut1977a}. I will not dwell on details, since we shall not need group integration again in this book.} The normalization ensures that near the origin, $\Psi^i=\psi^i+\bigO(\psi^2)$. Hence, $\Psi^i$ are a well-defined set of coordinates in some neighborhood of the origin. In these new coordinates, the action of the isotropy group is linear. This can be seen upon a short manipulation,
\begin{align}
\notag
\Psi'^i&(\Psi,h')=\int_{H_{x_0}}\D h\,D(h^{-1})^i_{\phantom ij}\DF^j(\DF(\psi,h'),h)=\int_{H_{x_0}}\D h\,D(h^{-1})^i_{\phantom ij}\DF^j(\psi,hh')\\
&=D(h')^i_{\phantom ij}\int_{H_{x_0}}\D(hh')\,[D(hh')^{-1}]^j_{\phantom jk}\psi'^k(\psi,hh')=D(h')^i_{\phantom ij}\Psi^j\;,
\end{align}
for any $h'\in H_{x_0}$, where I used respectively the group composition law~\eqref{grouptimes}, the invariance of the integration measure, the representation property of the matrices $D(h)$, and the definition~\eqref{linearization} of the new coordinates. This concludes the proof of the statement that the action of the isotropy group $H_{x_0}$ can be linearized by a suitable choice of coordinates.
\begin{watchout}%
In defining the new coordinates~\eqref{linearization}, I tacitly assumed that $H_{x_0}$ has a finite volume so that the invariant measure on $H_{x_0}$ in fact \emph{can} be normalized to unity. This is a key step in the proof, which requires that the isotropy group $H_{x_0}$ be compact. The technique of nonlinear realization of symmetry developed in this chapter is often introduced straight away under the assumption that the whole group $G$ is compact. That is, however, not necessary.
\end{watchout}
Equation~\eqref{linearization} can serve as a useful practical tool to find explicitly the coordinates that linearize the action of the isotropy group. Let us have a look at a simple example, following~\cite{Joseph1970a}.

\begin{illustration}%
The action of $\gr{SO}(2)$ on the Euclidean plane, introduced in \refex{ex:SO(2)actioninplane}, can be recast in terms of an action of $\gr{U}(1)\simeq\gr{SO}(2)$ on the complex plane $\C$. Thus, under $\E^{\I\eps}\in\gr{U}(1)$, the complex coordinate $z$ transforms as $z\to z'\equiv\E^{\I\eps}z$. This action is linear. One can however tweak it by changing the coordinate $z$ to $w$ such that $z=f(w)$, where $f$ is a function analytic in the neighborhood of the origin of $\C$ such that $f(w)=w+\bigO(w^2)$. In the new coordinate $w$, the $\gr{U}(1)$ group acts via
\begin{equation}
w\to w'=f^{-1}(z')=f^{-1}(\E^{\I\eps}f(w))\;.
\end{equation}
In general, $w'$ will be a nonlinear function of $w$. The original coordinate $z$, in which the action of $\gr{U}(1)$ is linear, can be reconstructed using~\eqref{linearization}. In case of $\gr{U}(1)$, the group integration in~\eqref{linearization} amounts to averaging over the phase $\eps$ of the $\gr{U}(1)$ rotation. It can be written as an integral over a unit circle in the complex plane,
\begin{equation}
\frac1{2\pi}\int_0^{2\pi}\D\eps\,\E^{-\I\eps}f^{-1}(\E^{\I\eps}f(w))=-\frac\I{2\pi}\oint\frac{\D c}{c^2}f^{-1}(cf(w))\;,
\label{epsaverage}
\end{equation}
where I introduced a new complex integration variable $c\equiv\E^{\I\eps}$. Given the assumptions I made on $f$, the function $f^{-1}(cf(w))/c^2$ of the complex variable $c$ has a simple pole at the origin with the residue $f(w)$. It then follows at once from the residue theorem that the integral in~\eqref{epsaverage} evaluates to $f(w)=z$, as expected.

Note that the coordinate $z$ is not uniquely specified by the requirement that the action of $\gr{U}(1)$ is linear. We can for instance introduce a new variable $w$ via
\begin{equation}
w=zf(z\bar z)\;,
\end{equation}
where $f$ is a smooth real function such that near the origin, $f(z\bar z)=1+\bigO(z\bar z)$. Then $w$ is a well-defined coordinate in some neighborhood of the origin of $\C$, upon which $\E^{\I\eps}\in\gr{U}(1)$ acts linearly as $w\to\E^{\I\eps}w$.
\end{illustration}

The above example suggests that the local coordinates $\Psi^i$ in which the isotropy group $H_{x_0}$ acts linearly are generally ambiguous. Namely, under the action of $H_{x_0}$, the manifold $\M$ splits into a disjoint union of orbits. The action of $H_{x_0}$ will remain linear if we rescale the coordinates $\Psi^i$ by an arbitrary $H_{x_0}$-invariant function on $\M$, that is a function which takes a constant value on any orbit of $H_{x_0}$. The only constraint is that such a rescaling leads to a well-defined set of new coordinates. The following example provides a nontrivial illustration of this ambiguity.

\begin{illustration}%
\label{ex:linearcoordinates}%
Consider the action of $G\simeq\gr{SU}(2)\times\gr{SU}(2)$ on $\M\simeq\gr{SU}(2)$. As already hinted in \refex{ex:QCDfirsttime} and explained in detail in Sect.~\ref{sec:ChPT}, this is important for a low-energy EFT description of hadron physics. For a given $\U\in\M$ and a given element $(g_\mathrm{L},g_\mathrm{R})\in G$, the action is defined by
\begin{equation}
\U\to g_\mathrm{L}\U g_\mathrm{R}^{-1}\;.
\end{equation}
The isotropy group of $\U_0\equiv\un$ is the ``diagonal'' subgroup of $G$, $H_{\U_0}\simeq\gr{SU}(2)$, consisting of elements of the type $(g,g)$, that is $g_\mathrm{L}=g_\mathrm{R}=g$. It is easy to guess a triplet of coordinates $\psi^i$, parameterizing $\M$ in the vicinity of $\U_0$, on which $H_{\U_0}$ acts linearly. Some common choices are
\begin{equation}
\U=\E^{\I\skal\psi\pau}\;,\qquad
\U=\frac{\un+\frac\I2\skal\psi\pau}{\un-\frac\I2\skal\psi\pau}\;,\qquad
\U=\un\sqrt{1-\vec\psi^2}+\I\skal\psi\pau\;,
\label{chiralcoset}
\end{equation}
where $\vec\pau$ is the vector of Pauli matrices. All these choices coincide to linear order when expanded in powers of $\psi^i$, $\U=\un+\I\skal\psi\pau+\bigO(\psi^2)$. All of them are mutually connected by coordinate redefinitions of the type $\psi'^i=\psi^if(\vec\psi^2)$, where $f$ is a suitably chosen function. The isotropy group $H_{\U_0}$ acts on $\U$ in all parameterizations shown in~\eqref{chiralcoset} via rotations of $\vec\psi$. Hence $\vec\psi^2$ is invariant under the action of $H_{\U_0}$ and any function $f(\vec\psi^2)$ is constant on the orbits of $H_{\U_0}$, as expected.
\end{illustration}


\subsection{From Linear Representation to Nonlinear Realization}
\label{subsec:CCWZclassification}

The ambiguity in the choice of coordinates that linearize the action of $H_{x_0}$ around a chosen point $x_0\in\M$ can be used to complete the classification of group actions. First of all, note that the set $\{T_gx_0\,\vert\,g\in G\}$ defines a submanifold of $\M$. On this submanifold, $G$ acts transitively; it is thus equivalent to the coset space $G/H_{x_0}$. We can choose a set of coordinates $\pi^a$, $a=1,\dotsc,\dim G/H_{x_0}$, on it. Then we add another set of coordinates $\mf^\vr$, $\vr=1,\dotsc,\dim\M-\dim G/H_{x_0}$, so that $(\pi^a,\mf^\vr)$ together is a well-defined coordinate system on $\M$ in the vicinity of $x_0$. In this coordinate system, the coset space $G/H_{x_0}$ is embedded in $\M$ as the set of points $(\pi^a,0)$.

As the next step, we subject the coordinates $(\pi^a,\mf^\vr)$ to the linearization~\eqref{linearization}; with a slight abuse of notation, I will use the same symbols $(\pi^a,\mf^\vr)$ for the resulting new coordinates. Importantly, the condition $\mf^\vr=0$ is preserved by the procedure. We thus end up with a set of coordinates $(\pi^a,\mf^\vr)$ in which the isotropy group $H_{x_0}$ is represented by linear transformations, and moreover the subset $(\pi^a,0)$, parameterizing $G/H_{x_0}$, is invariant under the action of $G$. The latter implies that the representation of $H_{x_0}$ on the coordinates $(\pi^a,\mf^\vr)$ has an invariant subspace, that is, it is reducible. I now use once again the assumption that $H_{x_0}$ is compact. This ensures that the representation of $H_{x_0}$ is \emph{completely} reducible. The action of $H_{x_0}$ can then be brought to a block-diagonal form,
\begin{equation}
T_h(\pi,\mf)\equiv(\pi'^a,\mf'^\vr)=(D^{(\pi)}(h)^a_{\phantom ab}\pi^b,D^{(\mf)}(h)^\vr_{\phantom\vr\s}\mf^\s)\;,\qquad
h\in H_{x_0}\;,
\label{CCWZorthogonalization}
\end{equation}
by orthogonalization that leaves the subspace $(\pi^a,0)$ intact. The latter feature ensures that we can still use the coordinates $\pi^a$ to parameterize the submanifold $G/H_{x_0}$.

A given point $x\in G/H_{x_0}$ with coordinates $\pi^a$ can be reached from $x_0$ by the action of any element of $G$ that lies in the coset of $x$. To proceed, we need to choose a concrete representative element of the coset. In line with the notation common in theoretical physics, I will denote this coset representative as $U(\pi)$. The concrete choice of the representative can be made fairly arbitrarily. There are however some natural requirements that will make our life easier:
\begin{itemize}
\item $U(\pi)$ should be a smooth function of $\pi^a$ near the origin $x_0$.
\item The origin $x_0$ itself, i.e.~the trivial coset $eH_{x_0}$, should be represented by $U(0)=e$.
\item The choice of $U(\pi)$ should reflect the linearity of the action of $H_{x_0}$ in the coordinates $\pi^a$.
\end{itemize}
The first two constraints can always be satisfied. As to the third, the linearity of the action of $H_{x_0}$ requires that $T_hx=T_hT_{U(\pi)}x_0=T_{U(\pi')}x_0$ where $\pi'^a$ are linear in $\pi^a$. Given that $T_hx=T_hT_{U(\pi)}T_{h^{-1}}x_0=T_{hU(\pi)h^{-1}}x_0$, it is natural to pick $U(\pi)$ so that
\begin{equation}
U(\pi')=hU(\pi)h^{-1}\;.
\label{Upiprime}
\end{equation}
We need to make sure, however, that this can be done consistently.

Consider the Lie algebra $\lie g$ of $G$, which carries an adjoint action of $H_{x_0}$,
\begin{equation}
Q\to hQh^{-1}\;,\qquad
Q\in\lie g\;,\quad
h\in H_{x_0}\;.
\label{adjointactionofH}
\end{equation}
This defines a linear representation of $H_{x_0}$ which has an invariant subspace, namely the Lie algebra $\lie h$ of $H_{x_0}$. Thus, the representation is reducible. Suppose now that it is, in fact, completely reducible. This is certainly the case when $H_{x_0}$ is compact. More generally, coset spaces $G/H_{x_0}$ for which the Lie algebra $\lie g$ can be split as $\lie g\simeq\lie h\oplus\lie g/\lie h$ where both $\lie h$ and $\lie g/\lie h$ are invariant subspaces under the adjoint action of $H_{x_0}$~\eqref{adjointactionofH}, are called \emph{reductive}. For a reductive coset space, we can always choose $U(\pi)$ so that the linear action of $H_{x_0}$ on $G/H_{x_0}$ is realized by~\eqref{Upiprime}. We can for instance set $U(\pi)=\exp(\I\pi^aQ_a)$, where $Q_{a,b,\dotsc}$ is a basis of $\lie g/\lie h$. This is however by far not the only choice, as illustrated by \refex{ex:linearcoordinates}.

\begin{illustration}%
There are many examples of reductive coset spaces for which the isotropy group $H$ is not compact. One can for instance start with a compact $H$, for which the reductive property is guaranteed, and then switch to a related noncompact Lie group. For a concrete example, take as $G$ the Poincar\'e group of isometries of $D$-dimensional Minkowski spacetime and as $H$ its subgroup, the Lorentz group $\gr{SO}(d,1)$. The subspace $\lie g/\lie h$ can be spanned on the generators of spacetime translations, which carry the vector representation of the Lorentz group. This coset space is a cousin of the Euclidean space $\R^D\simeq\gr{ISO}(D)/\gr{SO}(D)$, for which $H\simeq\gr{SO}(D)$ is compact.
\end{illustration}

It is easy to promote the action~\eqref{Upiprime} of $H_{x_0}$ on $G/H_{x_0}$ to an action of the whole group $G$. Indeed, the action of any $g\in G$ on an element $x=T_{U(\pi)}x_0$ of $G/H_{x_0}$ is completely characterized by the product $gU(\pi)$. The latter can be, at least in the vicinity of the unit element, uniquely decomposed as $U(\pi')h(\pi,g)$, where $\pi'^a$ is defined by $T_gx=T_{U(\pi')}x_0$ and the factor $h(\pi,g)\in H_{x_0}$ ensures that the correct representative of the coset $T_gx$ is used.

In order to lift the action of $G$ from the submanifold $G/H_{x_0}$ to the whole of $\M$, we now make one last change of coordinates. Namely, we define new coordinates $\tilde\pi^a$ and $\tilde\mf^\vr$ by the requirement that the point $(\pi^a,\mf^\vr)\in\M$ can be expressed as
\begin{equation}
(\pi^a,\mf^\vr)=T_{U(\tilde\pi)}(0,\tilde\mf)\;.
\label{lastchangeofcoordinates}
\end{equation}
This notation can be intuitively thought of as defining a slicing of $\M$ by orbits of $G$, starting from the subset of points $(0,\tilde\mf^\vr)$. The submanifold $G/H_{x_0}$ is a special case corresponding to $\tilde\mf^\vr=0$. In order to see that $\tilde\pi^a,\tilde\mf^\vr$ in fact are well-defined coordinates on $\M$, note that~\eqref{lastchangeofcoordinates} defines uniquely $\pi^a,\mf^\vr$ for given $\tilde\pi^a,\tilde\mf^\vr$. It follows from the fact that $U(0)=e$ that if $\tilde\pi^a=0$, then $\pi^a=0$ and $\mf^\vr=\tilde\mf^\vr$. Likewise, it follows from the definition of coordinates on the submanifold $G/H_{x_0}$ that if $\tilde\mf^\vr=0$, then $\mf^\vr=0$ and $\pi^a=\tilde\pi^a$. Hence the Jacobian matrix $\Pd{(\pi,\mf)}{(\tilde\pi,\tilde\mf)}$ equals $\un$ at the origin, and~\eqref{lastchangeofcoordinates} defines a valid coordinate system in some neighborhood of $x_0$.

The linearity of the action of $H_{x_0}$ is preserved in the new coordinates since
\begin{equation}
T_hT_{U(\tilde\pi)}(0,\tilde\mf)=T_{hU(\tilde\pi)h^{-1}}T_h(0,\tilde\mf)=T_{hU(\tilde\pi)h^{-1}}(0,D^{(\mf)}(h)\tilde\mf)
\end{equation}
for any $h\in H_{x_0}$. It is now a matter of a short manipulation to see that the action of an arbitrary $g\in G$ on~\eqref{lastchangeofcoordinates} is already completely fixed,
\begin{equation}
\begin{split}
T_gT_{U(\tilde\pi)}(0,\tilde\mf)&=T_{gU(\tilde\pi)}(0,\tilde\mf)=T_{U(\tilde\pi')h(\tilde\pi,g)}(0,\tilde\mf)=T_{U(\tilde\pi')}T_{h(\tilde\pi,g)}(0,\tilde\mf)\\
&=T_{U(\tilde\pi')}(0,D^{(\mf)}(h(\tilde\pi,g))\tilde\mf)\;.
\end{split}
\label{endofline}
\end{equation} 

This is the end of the line. Equation~\eqref{endofline} shows that in some neighborhood of a chosen point $x_0\in\M$, the action of any Lie group $G$ can by a change of coordinates be brought to a ``standard form'' such that (dropping the tildes)
\begin{equation}
U(\pi)\to U(\pi')=gU(\pi)h(\pi,g)^{-1}\;,\hspace{0.5em}
\mf^\vr\to\mf'^\vr=D^{(\mf)}(h(\pi,g))^\vr_{\phantom\vr\s}\mf^\s\;,
\end{equation}
where $h(\pi,g)\in H_{x_0}$ and $D^{(\mf)}$ is a matrix representation of $H_{x_0}$. The only technical assumption that was required in the proof was that the isotropy group $H_{x_0}$ is compact. This is the main result of the chapter. We can now harvest the fruits of our labors.


\section{Standard Realization of Symmetry}
\label{sec:CCWZstandard}

Having in mind that the reader might have skipped the last, somewhat technical section, let me give here a brief but self-contained summary of its main result. Consider a manifold $\M$ equipped with an action of a group $G$, and let us choose a fixed point $x_0\in\M$. Suppose that the isotropy group $H_{x_0}$ of $x_0$ is compact. Then it is always possible to redefine coordinates in a neighborhood of $x_0$ so that the new, ``standard'' coordinates $(\pi^a,\mf^\vr)$ have the following properties. First, the point $x_0$ itself corresponds to the origin $(0,0)$. The subset $(\pi^a,0)$ spans a submanifold of $\M$, equivalent to the coset space $G/H_{x_0}$. Every point on the coset space can be uniquely characterized by a choice of a representative element $U(\pi)$ of the corresponding coset. The representative $U(\pi)$ can be chosen so that $U(0)=e$, and that the adjoint action of $h\in H_{x_0}$, $U(\pi)\to hU(\pi)h^{-1}$, defines a linear transformation of the coordinates $\pi^a$. The group $G$ acts on the coset space via left multiplication, which defines implicitly an element $h(\pi,g)$ of $H_{x_0}$ through
\begin{equation}
gU(\pi)=U(\pi'(\pi,g))h(\pi,g)\;,\qquad
g\in G\;.
\label{cosettransfo}
\end{equation}
The last two properties of $U(\pi)$ imply that $h(\pi,g)=g$ for any $g\in H_{x_0}$.

The action of an element $g\in G$ on the whole manifold $\M$ is now defined in terms of the standard coordinates $(\pi^a,\mf^\vr)$ as
\begin{equation}
\begin{split}
U(\pi)&\xrightarrow{g}U(\pi'(\pi,g))=gU(\pi)h(\pi,g)^{-1}\;,\\
\mf^\vr&\xrightarrow{g}\mf'^\vr(\mf,\pi,g)=D(h(\pi,g))^\vr_{\phantom\vr\s}\mf^\s\;,
\end{split}
\label{CCWZstandard}
\end{equation}
where the matrices $D(h)$ define some linear representation of $H_{x_0}$. Altogether, the action of $G$ is fully specified by the choice of coset representative $U(\pi)$, which fixes the first line of~\eqref{CCWZstandard}, and the choice of representation $D(h)$ of $H_{x_0}$, which fixes the second line thereof.
\begin{watchout}%
The above construction of the standard realization of group action goes through without change even if $H_{x_0}$ is noncompact provided the coset space $G/H_{x_0}$ is reductive. In that case, however, the line of argument in Sect.~\ref{sec:CCWZclassification} fails and it is no longer guaranteed that the standard realization is unique up to a coordinate redefinition. There may then be more mutually inequivalent nonlinear realizations of the group $G$ on the manifold $\M$, of which the standard realization is but one example.

Finally, one may try to follow the same steps of the construction of the standard realization even when the coset space $G/H_{x_0}$ is not reductive. Then, however, many of the simple features of the standard realization are lost. A concrete example of a nonreductive coset space is worked out in~\cite{Alonso2016a}.
\end{watchout}
In the standard nonlinear realization~\eqref{CCWZstandard}, the coordinates $\pi^a$ transform under $G$ on their own, independently of $\mf^\vr$. The transformation of the latter, on the other hand, is nonlinear in $\pi^a$ but linear in $\mf^\vr$ themselves. It is therefore possible to set the $\mf^\vr$s to zero consistently. This is not surprising. In the field theory language, the coordinates $\pi^a$ correspond to NG bosons and their universal presence therefore mirrors the Goldstone theorem as reviewed in Chap.~\ref{chap:NGbosons}. The remaining coordinates on $\M$, $\mf^\vr$, represent other degrees of freedom that are not of NG nature. In the jargon of EFT, they are usually called \emph{matter fields}. Among all the nonlinear realizations of the given symmetry group $G$ with the given subgroup $H$, there is therefore a ``minimal'' nonlinear realization, defined on the coset space $G/H$, which only includes the NG degrees of freedom. On a general manifold $\M$, additional matter fields may be present. Throughout this book, I will focus mostly on minimal nonlinear realizations due to their significance for low-energy EFT description of broken symmetries.

\begin{illustration}%
One can gain insight into the standard realization~\eqref{CCWZstandard} of the action of $G$ by looking at some special choices of the isotropy group. If there is a point $x_0\in\M$ such that $H_{x_0}\simeq G$, then in its vicinity, we do not have any coordinates $\pi^a$. The coset space $G/H_{x_0}$ consists of a single point that we can represent with $U=e$, which is consistent with the first line of~\eqref{CCWZstandard} if $h(g)=g$ for all $g\in G$. All coordinates in the neighborhood of $x_0$ are of the $\mf^\vr$ type. The second line of~\eqref{CCWZstandard} guarantees that the action of $G$ can be completely linearized, $\mf'^\vr(\mf,g)=D(g)^\vr_{\phantom\vr\s}\mf^\s$. This is the usual textbook realization of symmetry via a linear representation.

The opposite extreme, $H_{x_0}\simeq\trgr$, is more interesting. Here~\eqref{CCWZstandard} reduces to
\begin{equation}
U(\pi)\xrightarrow{g}gU(\pi)\;,\qquad
\mf^\vr\xrightarrow{g}\mf^\vr\;.
\label{trivialH}
\end{equation}
The coset space $G/H_{x_0}$ corresponds to the group manifold $G$ and carries an action of $G$ defined by simple left multiplication. Whatever other coordinates $\mf^\vr$ on $\M$ are present can always be chosen to be invariant under $G$. This is quite surprising. We are used to working with fields spanning linear multiplets of $G$; it is not obvious that the same physical content can be encoded in a set of fields that do not transform under $G$ at all. The resolution of this apparent paradox lies in the freedom to choose coordinates at will. Namely, if we start with a set of fields $\Psi^\vr$ transforming under $G$ as $\smash{\Psi^\vr\xrightarrow{g}D(g)^\vr_{\phantom\vr\s}\Psi^\s}$, we can make the redefinition $\smash{\mf^\vr\equiv D(U(\pi)^{-1})^\vr_{\phantom\vr\s}\Psi^\s}$. The new variables $\mf^\vr$ are invariant under $G$ in accord with~\eqref{trivialH}.
\end{illustration}


\subsection{Nonlinear Realization on Coset Spaces}
\label{subsec:CCWZcosetspaces}

Can we be more explicit about the way that the coordinates $(\pi^a,\mf^\vr)$ transform under the action of $G$? The standard realization~\eqref{CCWZstandard} of the group action requires the knowledge of the nonlinear functions $\pi'^a(\pi,g)$ and $h(\pi,g)$ . We would like to be able to compute these, at least for small transformations, that is for $g\in G$ infinitesimally close to unity.\footnote{I will only introduce the concept of a metric on a group, and more generally on a homogeneous space, in Sect.~\ref{sec:CCWZgeometry}. Statements about infinitesimal distance of group elements should therefore be interpreted within a faithful matrix representation of the group using some standard matrix norm. The same remark applies whenever a sum or difference of group elements is considered below.} To that end, it is sufficient to consider the action of $G$ on the coset space $G/H$; from now on will I drop the subscript $x_0$ on $H$ unless it is needed to explicitly distinguish the isotropy groups of different points on the coset space. Once the minimal realization of $G$ on $G/H$ is found, it can be extended to any other manifold by specifying the linear representation $D(h)$ of $H$ under which the additional coordinates $\mf^\vr$ transform.

The general algorithm for calculation of the desired functions $\pi'^a(\pi,g)$ and $h(\pi,g)$ is as follows. Take the first line of~\eqref{CCWZstandard}, multiply it with $U(\pi)^{-1}$, and subtract $U(\pi)^{-1}U(\pi)=e$. This gives the master relation
\begin{equation}
U(\pi)^{-1}\udelta U(\pi)=U(\pi)^{-1}gU(\pi)h(\pi,g)^{-1}-e\;,
\label{varmaster}
\end{equation}
where $\udelta U(\pi)\equiv U(\pi')-U(\pi)$. For $g\in G$ that is infinitesimally close to unity, both $U(\pi)^{-1}gU(\pi)$ and $h(\pi,g)$ are infinitesimally close to unity as well. By a systematic comparison of the left- and right-hand sides of~\eqref{varmaster}, one can then determine $\udelta\pi^a(\pi,g)\equiv\pi'^a(\pi,g)-\pi^a$ as well as $h(\pi,g)$.

To make further progress, we first have to establish some notation. In order to be able to discuss different symmetry transformations from $G$ separately, we choose a basis $Q_{A,B,\dotsc}$ of $\lie g$. A subset of these spans a basis of the Lie subalgebra $\lie h$ and will be denoted with Greek indices, $Q_{\a,\b,\dotsc}$. The rest of the generators spans the subspace $\lie g/\lie h$ and will be denoted as $Q_{a,b,\dotsc}$. The structure constants of the Lie algebra $\lie g$ will be called $\smash{f^C_{AB}}$ with a conventional factor of $\I$ in the commutation relations, that is, $\smash{[Q_A,Q_B]=\I f^C_{AB}Q_C}$. The structure constant does not have to be fully antisymmetric in its three indices even if $G$ is compact. It does have to be antisymmetric under the exchange of its lower two indices though. It also has to satisfy the Jacobi identity,
\begin{equation}
f^E_{AB}f^D_{EC}+f^E_{BC}f^D_{EA}+f^E_{CA}f^D_{EB}=0\;.
\label{jacobi}
\end{equation}
The basic commutator of the Lie algebra $\lie g$, $[Q_A,Q_B]=\I f^C_{AB}Q_C$, can be unfolded into three separate conditions on the subsets of generators $Q_{\a,\b,\dotsc}$ and $Q_{a,b,\dotsc}$,
\begin{equation}
\begin{split}
[Q_\a,Q_\b]&=\I \smash{f^\g_{\a\b}}Q_\g\;,\\
[Q_\a,Q_b]&=\I f^c_{\a b}Q_c\;,\\
[Q_a,Q_b]&=\I(f^\g_{ab}Q_\g+f^c_{ab}Q_c)\;.
\end{split}
\label{commutators}
\end{equation}
The first of these encodes the requirement that the generators $Q_{\a,\b,\dotsc}$ span a closed Lie algebra (that is $\smash{f^c_{\a\b}=0}$). The second of these likewise expresses the assumption that the coset space $G/H$ is reductive (that is $\smash{f^\g_{\a b}=0}$).

Next, we are going to need two simple statements from linear algebra, which I reproduce here for the sake of completeness. The first of these is usually known under the name \emph{Hadamard lemma},
\begin{equation}
\E^AB\E^{-A}=B+[A,B]+\frac1{2!}[A,[A,B]]+\frac1{3!}[A,[A,[A,B]]]+\dotsb\;,
\label{hadamard}
\end{equation}
where $A,B$ are arbitrary (square) matrices. The second identity, which to the best of my knowledge does not have an established name, reads
\begin{equation}
\E^{-A}\D\E^A=\int_0^1\D t\,\E^{-tA}(\D A)\E^{tA}\;.
\label{AdA}
\end{equation}
Here the symbol $\D$ acting on $A$ and $\E^A$ can be thought of as a differential, but also as a derivative with respect to whatever parameter $A$ might depend on. Once multiplied from the left with $\E^A$, we can think of~\eqref{AdA} as a continuous version of the Leibniz (product) rule, applied to $\E^A$. It follows from~\eqref{AdA} combined with the Hadamard lemma~\eqref{hadamard} that whenever $A$ is a function with values in a Lie algebra, $\E^{-A}\D\E^A$ will take values in the same Lie algebra.

Equation~\eqref{AdA} prepares the ground for the introduction of a concept of central importance for calculus on coset spaces: the \emph{Maurer--Cartan} (MC) \emph{form},
\begin{equation}
\mc(\pi)\equiv-\I U(\pi)^{-1}\D U(\pi)\;.
\label{MCformdef}
\end{equation}
For the time being, the reader may think of the $\D$ herein as an ordinary differential of a function. The true geometric significance of the MC form as a differential 1-form will become clear in Sect.~\ref{sec:CCWZgeometry}. Recall that any element of a Lie group sufficiently close to unity, in our case $U(\pi)$, can be obtained as the exponential of an element of the corresponding Lie algebra. Equation~\eqref{AdA} then implies that the MC form takes values in the Lie algebra $\lie g$. We can split it into pieces that belong to $\lie h$ and $\lie g/\lie h$, and represent each of these in terms of their components in a chosen basis of generators,
\begin{equation}
\begin{split}
\mc&\equiv\mcu+\mcb\;,\\
\mc&\equiv\mc^AQ_A\;,\qquad
\mcu\equiv\mc^\a Q_\a\;,\qquad
\mcb\equiv\mc^aQ_a\;.
\end{split}
\label{MCsplit}
\end{equation}
Finally, by writing $\D U(\pi)=[\Pd{U(\pi)}{\pi^a}]\D\pi^a$, one can introduce explicit components of the MC form in a chosen set of local coordinates $\pi^a$ on $G/H$, $\mc^A\equiv\mc^A_a\D\pi^a$.

It is instructive to check how the MC form is affected by the action of $G$ on the coset space. This follows directly from~\eqref{CCWZstandard}. It is a simple exercise to verify that for given $g\in G$,
\begin{equation}
\begin{split}
\mcu(\pi)&\xrightarrow{g}\mcu(\pi'(\pi,g))=h(\pi,g)\mcu(\pi)h(\pi,g)^{-1}-\I h(\pi,g)\D h(\pi,g)^{-1}\;,\\
\mcb(\pi)&\xrightarrow{g}\mcb(\pi'(\pi,g))=h(\pi,g)\mcb(\pi)h(\pi,g)^{-1}\;.
\end{split}
\label{MCtransfo}
\end{equation}
Note how~\eqref{AdA} guarantees that $\mcu(\pi')$ still takes values in the Lie algebra $\lie h$. While not of direct relevance right here and now, the transformation rules~\eqref{MCtransfo} will help us understand the geometric meaning of the MC form in Sect.~\ref{sec:CCWZgeometry}.

We still need a few last pieces of notation. Conjugation of elements of $\lie g$ by $U(\pi)$ will be abbreviated as
\begin{equation}
U(\pi)^{-1}Q_AU(\pi)\equiv\nu^B_A(\pi)Q_B\;,
\label{cosetconjugation}
\end{equation}
which defines a set of nonlinear functions $\nu^B_A(\pi)$ on the coset space. Finally, for the action of an element $g\in G$ infinitesimally close to unity, I will use the notation
\begin{equation}
\begin{split}
g&\approx e+\I\eps^AQ_A\;,\\
\udelta\pi^a(\pi,g)&\approx\eps^A\kil^a_A(\pi)\;,\qquad
h(\pi,g)\approx e+\I\eps^Ak^\a_A(\pi)Q_\a\;.
\end{split}
\label{killingdef}
\end{equation}
The $\approx$ symbol indicates that I have expanded all the quantities to linear order in the transformation parameters $\eps^A$, defined by the first line of~\eqref{killingdef}. The second line thereof introduces notation for the infinitesimal versions of the functions $\pi'^a(\pi,g)$ and $h(\pi,g)$. Of particular interest are the functions $\kil^a_A(\pi)$ that realize the motion induced on the coset space $G/H$ by the group $G$.

With all the notation at hand, we can now expand~\eqref{varmaster} to linear order in $\eps^A$ and compare coefficients of the various generators of $\lie g$ on the left- and right-hand sides. This leads to the identities
\begin{equation}
\begin{split}
\nu^\a_A(\pi)&=\kil^a_A(\pi)\mc^\a_a(\pi)+k^\a_A(\pi)\;,\\
\nu^a_A(\pi)&=\kil^b_A(\pi)\mc^a_b(\pi)\;.
\end{split}
\label{cosettransfoexact}
\end{equation}
These are still valid for any choice of the coset representative $U(\pi)$. Once it is fixed, the functions $\mc^A_a(\pi)$ are determined by~\eqref{MCformdef}. Likewise, the functions $\nu^B_A(\pi)$ are fixed by~\eqref{cosetconjugation}. The identities~\eqref{cosettransfoexact} then constitute a set of linear equations for $\kil^a_A(\pi)$ and $k^\a_A(\pi)$. At the origin, $\nu^B_A(0)=\d^B_A$ as a consequence of the fact that $U(0)=e$. The second line of~\eqref{cosettransfoexact} then implies that $\mc^a_b(0)$ is nonsingular. By continuity, it must remain nonsingular in some neighborhood of the origin. This guarantees that a solution of~\eqref{cosettransfoexact} for $\kil^a_A(\pi)$ and $k^\a_A(\pi)$ exists and it is unique.

I have now achieved the main goal of this subsection: to give an algorithm how, for a chosen set of coordinates $\pi^a$, to realize the action of the group $G$ on the coset space $G/H$. None of the nonlinear functions involved---$\mc^A_a(\pi)$, $\nu^B_A(\pi)$, $\kil^a_A(\pi)$ and $k^\a_A(\pi)$---can however in general be evaluated in a closed form. For practical applications, it is useful to have explicit expressions for these functions, even if just as a series expansion in a specific set of coordinates $\pi^a$. One popular choice of parameterization for which this can easily be done is
\begin{equation}
U(\pi)=\exp(\I\pi^aQ_a)\;.
\label{expparam}
\end{equation}
The Hadamard lemma~\eqref{hadamard} then tells us at once that
\begin{equation}
\nu^A_B(\pi)=\d^A_B-f^A_{Ba}\pi^a+\frac12f^C_{Ba}f^A_{Cb}\pi^a\pi^b+\bigO(\pi^3)\;.
\label{nu}
\end{equation}
Likewise, it follows quickly from~\eqref{AdA} that
\begin{equation}
\mc^A_a(\pi)=\d^A_a-\frac12f^A_{ab}\pi^b+\frac16f^B_{ab}f^A_{Bc}\pi^b\pi^c+\bigO(\pi^3)\;.
\label{omega}
\end{equation}
Finally, \eqref{cosettransfoexact} can be solved iteratively for the remaining pieces,
\begin{equation}
\begin{split}
\kil^a_A(\pi)={}&\d^a_A-\left(f^a_{Ab}-\frac12\delta^e_Af^a_{eb}\right)\pi^b\\
&+\frac12\left(f^\a_{Ab}f^a_{\a c}-\frac13\delta^e_Af^B_{eb}f^a_{Bc}+\frac12\delta^e_Af^d_{eb}f^a_{dc}\right)\pi^b\pi^c+\bigO(\pi^3)\;,\\
k^\a_A(\pi)={}&\d^\a_A-\left(f^\a_{Aa}-\frac12\delta^e_Af^\a_{ea}\right)\pi^a\\
&+\frac12\left(f^\b_{Aa}f^\a_{\b b}-\frac13\delta^e_Af^B_{ea}f^\a_{Bb}+\frac12\delta^e_Af^d_{ea}f^\a_{db}\right)\pi^a\pi^b+\bigO(\pi^3)\;.
\end{split}
\label{expparamresults}
\end{equation}
With these explicit expressions at hand, it is easy to illustrate some of the general properties of the standard nonlinear realization. For instance, the action of the isotropy group $H$ reduces to the linear adjoint transformation of the coordinates, $\kil^a_\a(\pi)=-f^a_{\a b}\pi^b$. Likewise, we find that $k^\a_\b(\pi)=\d^\a_\b$, which is an infinitesimal version of the relation $h(\pi,g)=g$ for any $g\in H$.

\begin{illustration}%
All the structure introduced above takes a particularly simple form when the group $G$ is Abelian. Then $\kil^a_A(\pi)=\d^a_A$: the group $G$ acts on the standard coordinates $\pi^a$ by a mere set of shifts. The MC form reduces to $\mc(\pi)=\D\pi^aQ_a$. This suggests a simple interpretation of $\mcb=\mc^a_b\D\pi^bQ_a$ in the general situation when $G$ is non-Abelian. Namely, $\mc^a$ supplies us with a generalized derivative (or differential) of $\pi^a$ which, as~\eqref{MCtransfo} shows, is covariant under the action of $G$.
\end{illustration}


\subsection{Symmetric Coset Spaces}
\label{subsec:CCWZsymmetriccoset}

Up to some technical assumptions, we have already accomplished a complete classification of possible nonlinear realizations of symmetry on manifolds. In the language of field theory, this amounts to finding all possible point symmetry transformations of a given set of fields under a given symmetry group $G$. The price we had to pay for the generality of this result was the rather complicated transformation rule~\eqref{CCWZstandard}. This involves nonlinear functions on the coset space $G/H$ that cannot be evaluated in a closed explicit form. There is, however, an important class of coset spaces for which we can do much better.

The coset space $G/H$ is called \emph{symmetric}, and the associated homogeneous space is called a \emph{symmetric space}, if the Lie algebra $\lie g$ admits an involutive automorphism under which $\lie h$ is even and $\lie g/\lie h$ is odd. In other words, we require that there is a linear map $\Raut:\lie g\to\lie g$ such that
\begin{equation}
\begin{aligned}
\Raut([Q_1,Q_2])&=[\Raut(Q_1),\Raut(Q_2)]\;,&\quad
Q_1,Q_2&\in\lie g\;,\\
\Raut(Q)&=Q\;,&\quad
Q&\in\lie h\;,\\
\Raut(Q)&=-Q\;,&\quad
Q&\in\lie g/\lie h\;.
\end{aligned}
\end{equation}
This property guarantees the vanishing of any structure constant with an odd number of $\lie g/\lie h$ indices. In particular, $f^\g_{\a b}=0$: any symmetric coset space is automatically reductive. In addition, $f^c_{ab}=0$, that is the last term in the commutation relations~\eqref{commutators} is missing.

\begin{illustration}%
The fundamental commutation relation of the $\gr{SO}(n)$ group reads
\begin{equation}
[J_{ij},J_{kl}]=\I(\d_{ik}J_{jl}+\d_{jl}J_{ik}-\d_{il}J_{jk}-\d_{jk}J_{il})\;,
\end{equation}
where $i,j,k,l=1,\dotsc,n$ and $J_{ij}$ is the antisymmetric tensor of angular momentum in $n$ spatial dimensions. The Lie algebra of $\gr{SO}(n)$ possesses an automorphism $\Raut$ under which $\Raut(J_{\a\b})=J_{\a\b}$ and $\Raut(J_{\a n})=-J_{\a n}$, where $\a,\b=1,\dotsc,n-1$. One can think of this automorphism geometrically as an inversion of the $n$-th coordinate axis. The coset space $\gr{SO}(n)/\gr{SO}(n-1)\simeq S^{n-1}$ is therefore symmetric.

In a similar vein, the Euclidean space $\R^n$ is a symmetric space. This can be seen most easily by recalling that $\R^n\simeq\gr{ISO}(n)/\gr{SO}(n)$. The desired automorphism $\Raut$ is the spatial inversion. Under this, all the generators of $\gr{SO}(n)$ (rotations) remain intact, whereas the remaining generators of $\gr{ISO}(n)/\gr{SO}(n)$ (translations) change sign.
\end{illustration}

\begin{illustration}%
\label{ex:chiralcoset}%
Consider the ``chiral'' coset spaces of the type $G_\mathrm{L}\times G_\mathrm{R}/G_\mathrm{V}$, where all the three groups $G_\mathrm{L},G_\mathrm{R},G_\mathrm{V}$ are isomorphic to the same Lie group $G$; this is a generalization of \refex{ex:linearcoordinates} where $G\simeq\gr{SU}(2)$. The chiral group $G_\mathrm{L}\times G_\mathrm{R}$ consists of elements $(g_\mathrm{L},g_\mathrm{R})$ where $g_\mathrm{L},g_\mathrm{R}\in G$. The ``vector'' isotropy group $G_\mathrm{V}$ consists of elements of the type $(g,g)$, that is $g_\mathrm{L}=g_\mathrm{R}=g$. The generators of the chiral group include two copies, $Q_{\mathrm{L},A}$ and $Q_{\mathrm{R},A}$, of the generators of $G$. The Lie algebra of the chiral group is defined in terms of the structure constants $f^C_{AB}$ of $G$ by
\begin{equation}
\begin{gathered}
[Q_{\mathrm L,A},Q_{\mathrm L,B}]=\I f^C_{AB}Q_{\mathrm L,C}\;,\qquad
[Q_{\mathrm R,A},Q_{\mathrm R,B}]=\I f^C_{AB}Q_{\mathrm R,C}\;,\\
[Q_{\mathrm L,A},Q_{\mathrm R,B}]=0\;.
\end{gathered}
\end{equation}
These commutation relations are invariant under the exchange of $Q_{\mathrm{L},A}$ and $Q_{\mathrm{R},A}$, which defines the desired automorphism,
\begin{equation}
\Raut(Q_{\mathrm{L},A})=Q_{\mathrm{R},A}\;,\qquad
\Raut(Q_{\mathrm{R},A})=Q_{\mathrm{L},A}\;.
\end{equation}
The generators of $G_\mathrm{V}$, equal to $Q_{\mathrm{L},A}+Q_{\mathrm{R},A}$ up to overall normalization, are even under this automorphism. The generators of the complementary space $\lie g/\lie h$, which are to be odd under $\Raut$, can be taken as $Q_{\mathrm{L},A}-Q_{\mathrm{R},A}$ up to an overall factor.
\end{illustration}

The automorphism $\Raut$ of the Lie algebra $\lie g$ can be lifted, at least locally near the unit element, to the Lie group $G$. It is then possible to choose the coset representative $U(\pi)$ so that $\Raut(U(\pi))=U(\pi)^{-1}$; one can use for instance the exponential parameterization~\eqref{expparam}. We now take the first line of~\eqref{CCWZstandard} and multiply it with the inverse of its image under $\Raut$. The result is a surprise: for symmetric coset spaces, there is a parameterization of $G/H$ in which the whole group $G$ is realized linearly,
\begin{equation}
\S(\pi)\equiv U(\pi)^2\;,\qquad
\S(\pi)\xrightarrow{g}\S(\pi'(\pi,g))=g\S(\pi)\Raut(g)^{-1}\;.
\label{Sigmadef}
\end{equation}
This is of such utility that whenever one deals with a symmetric coset space, one almost always uses the linearly transforming variable $\S$ instead of working directly with the coordinates $\pi^a$. Note, however, that one may need the coordinates $\pi^a$ if one wishes to extend the coset space $G/H$ to a larger manifold $\M$. This is because the transformation of $\mf^\vr$ in~\eqref{CCWZstandard} requires the functions $h(\pi,g)$ that depend on $\pi^a$.

With the automorphism $\Raut$ at hand, one may easily project out the $\mcu$ and $\mcb$ components of the MC form,
\begin{equation}
\mcu=\frac12[\mc+\Raut(\mc)]\;,\qquad
\mcb=\frac12[\mc-\Raut(\mc)]\;.
\end{equation}
The latter has a practically convenient expression in terms of $\S$,
\begin{equation}
\mcb=-\frac\I2U^{-1}\D\S U^{-1}=\frac\I2U\D\S^{-1}U\;.
\label{mcb_symmetric}
\end{equation}

\begin{illustration}%
\label{ex:chiralcosetSigma}%
Let us see how the variable $\S(\pi)$ is realized on the chiral coset spaces $G_\mathrm{L}\times G_\mathrm{R}/G_\mathrm{V}$ discussed in \refex{ex:chiralcoset}. Here we can choose the coset representative as $U=(u,u^{-1})$ where $u\in G$; this satisfies the requirement that under the automorphism $\Raut(g_\mathrm{L},g_\mathrm{R})=(g_\mathrm{R},g_\mathrm{L})$ of the chiral group $G_\mathrm{L}\times G_\mathrm{R}$, $U$ is turned into its inverse. The general transformation rule as given by the first line of~\eqref{CCWZstandard} then translates to
\begin{equation}
(u,u^{-1})\xrightarrow{(g_\mathrm{L},g_\mathrm{R})}(g_\mathrm{L},g_\mathrm{R})(u,u^{-1})(h^{-1},h^{-1})\;,
\end{equation}
where $h\in G$. The linearly transforming variable~\eqref{Sigmadef} is given by $\S=U^2=(u^2,u^{-2})$, where $u^2\equiv\U$ transforms under $G_\mathrm{L}\times G_\mathrm{R}$ as $\U\to g_\mathrm{L}\U g_\mathrm{R}^{-1}$.
\end{illustration}

It turns out that~\eqref{Sigmadef} can be further simplified in case $\Raut$ is an inner automorphism of $G$, that is, when there is an element $R\in G$ such that
\begin{equation}
\Raut(g)=R^{-1}gR\;,\qquad
g\in G\;.
\label{inneraut}
\end{equation}
Then a slight modification of~\eqref{Sigmadef} leads to a variable that transforms linearly under the adjoint action of $G$,
\begin{equation}
N(\pi)\equiv\S(\pi)R=U(\pi)RU(\pi)^{-1}\;,\quad
N(\pi)\xrightarrow{g}N(\pi'(\pi,g))=gN(\pi)g^{-1}\;.
\label{Ndef}
\end{equation}
An advantage of trading $\S(\pi)$ for $N(\pi)$ is that $N(\pi)^2$ is a constant independent of $\pi^a$, which may be convenient in concrete applications. Indeed, since $\Raut(\Raut(g))=g$ for any $g\in G$, it follows from~\eqref{inneraut} that $R^2$ belongs to the center of $G$. Consequently, $N(\pi)^2=U(\pi)R^2U(\pi)^{-1}=R^2$.

\begin{illustration}%
\label{ex:SU(2)U(1)coset}%
Consider the coset space $\gr{SU}(2)/\gr{U}(1)$, relevant for description of quantum systems with magnetic ordering such as (anti)ferromagnets. In the fundamental representation, the generators of $\gr{SU}(2)$ are $Q_A=\pau_A/2$, and the generator of the $\gr{U}(1)$ isotropy group can be taken as $\pau_3/2$. This coset space is symmetric thanks to an inner automorphism~\eqref{inneraut} with $R=\I\pau_3$. While the factor of $\I$ here is required to make $R$ an element of $\gr{SU}(2)$, we do not need it for the definition of $N(\pi)$. Let us therefore set
\begin{equation}
N(\pi)\equiv U(\pi)\pau_3U(\pi)^{-1}\;.
\end{equation}
This matrix variable is unitary and Hermitian. Moreover, it is traceless and it squares to $\un$. It can thus be mapped on a unit vector variable $\vec n(\pi)$ such that
\begin{equation}
N(\pi)=\skal\pau n(\pi)\;,
\end{equation}
which belongs to the vector representation of $\gr{SU}(2)$. The fact that we ended up describing the coset space in terms of a unit vector is not a coincidence. Thanks to the local isomorphism of $\gr{SU}(2)$ and $\gr{SO}(3)$ we also have a local equivalence of coset spaces, $\gr{SU}(2)/\gr{U(1)}\simeq\gr{SO}(3)/\gr{SO}(2)\simeq S^2$.
\end{illustration}


\section{Geometry of the Coset Space}
\label{sec:CCWZgeometry}

In this final section of the chapter, I will show that some of the structure that enters the standard realization of symmetry on a coset space can be given a neat geometric interpretation. A reader not familiar with basics of differential geometry is advised to consult Appendix~\ref{app:diffgeom} before proceeding. Further information about the geometry of homogeneous spaces at an easily accessible level can be found in~\cite{Arvanitoyeorgos2003a}.

To start, recall that each point of a coset space $G/H$ corresponds to an entire class of elements of the group $G$. The approach I have used so far was to parameterize each $x\in G/H$ with coordinates $\pi^a$ in terms of a fixed coset representative $U(\pi)$. The choice of $U(\pi)$ is however arbitrary and can be changed locally by multiplication from the right; any $\tilde U(\pi)=U(\pi)h(\pi)$ with $h(\pi)\in H_{x_0}$ is equally good. One can then promote the basic transformation rule~\eqref{CCWZstandard} for $U(\pi)$ to
\begin{equation}
U(\pi)\xrightarrow{g,h(\pi)}U(\pi'(\pi,g))=gU(\pi)h(\pi)^{-1}\;,
\end{equation}
where $h(\pi)$ is independent of $g$. This realizes an action of the product group $G\times H_\mathrm{gauge}$. The group $G$ acts on $U(\pi)$ by left multiplication as usual. The local group $H_\mathrm{gauge}$ isomorphic to $H$ acts on $U(\pi)$ by right multiplication with $h(\pi)^{-1}$ and encodes the freedom to choose locally the coset representative. This is a typical example of a gauge redundancy; any geometrically or physically well-defined quantity must be independent of the arbitrary choice of coset representative. In theoretical physics, the approach that views the left action of $G$ separately from the right action of the local group $H_\mathrm{gauge}$ is called ``hidden local symmetry''~\cite{Bando1988a}.

\begin{figure}[t]
\sidecaption[t]
\includegraphics[width=2.9in]{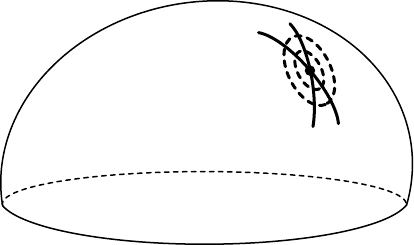}
\caption{Illustration of the action of $G$ and $H_x$ around a fixed point $x$ (\emph{black dot}) on the coset space. The thick \emph{solid lines} indicate orbits of one-parameter subgroups of $G$ passing through $x$; these correspond to generators from $\lie g/\lie h$ and act as translations on the coset space. The thick \emph{dashed lines} indicate some orbits of $H_x$ in the vicinity of $x$. The action of $H_x$ naturally induces a set of linear transformations on the tangent space at $x$}
\label{fig:orbitsofGandH}
\end{figure}

The different geometric roles of $G$ and its isotropy subgroup can be intuitively understood by looking at Fig.~\ref{fig:orbitsofGandH}. One-parameter subgroups of $G$ define a set of flows on $G/H$ that in general translate a given point $x$ to some other point. The isotropy group $H_x$ maps $x$ to itself. It does, however, act nontrivially in the vicinity of $x$. Based on the figure, we expect that the action of $H_x$ projects to a set of linear maps on the tangent space $\T_xG/H$ at $x$. By induction, such linear $H_x$-transformations should exist for any tensor at $x$.

The local group $H_\mathrm{gauge}$ gives us the freedom to do independent $H$-transformations of tangent vectors (and generally tensors) point by point. In order to be able to analyze the properties of vector (tensor) fields on $G/H$, we therefore need a way to relate tangent vectors (tensors) at different points. This resembles closely the motivation behind the construction of a local frame, or vielbein, and the corresponding connection on a manifold. The only difference to the formalism reviewed in Appendix~\ref{appsec:affine} is that here the structure group is not $\gr{GL}(\dim G/H)$ but $H$ itself. This restriction makes it possible to realize changes in the local frame in terms of a fixed matrix representation of $H$.

It turns out that we already have both the local frame and the connection: they are granted to us by the MC form~\eqref{MCformdef}. Specifically, the components $\mc^a$ of $\mcb$ define a coframe on $G/H$. By the second line of~\eqref{MCtransfo}, these indeed transform linearly under the local action of $H_\mathrm{gauge}$ as they should. Likewise, the $\lie h$ part of the MC form, $\mcu$, provides the necessary $H$-connection. This is confirmed by the transformation rule on the first line of~\eqref{MCtransfo}.

This observation stresses the significance of the MC form as a differential 1-form, taking values in the Lie algebra $\lie g$. Its exterior derivative reflects the structure of $G$,
\begin{equation}
\begin{split}
\D\mc&=-\I\D U^{-1}\w\D U=\I U^{-1}\D U\w U^{-1}\D U=-\I\mc\w\mc\\
&=-\frac\I2\mc^B\w\mc^C[Q_B,Q_C]=\frac12f^A_{BC}\mc^B\w\mc^CQ_A\;.
\end{split}
\end{equation}
This can be split into a pair of equations, one for $\mcu$ and one for $\mcb$,
\begin{equation}
\begin{split}
\D\mc^\a&=\frac12f^\a_{\b\g}\mc^\b\w\mc^\g+\frac12f^\a_{bc}\mc^b\w\mc^c\;,\\
\D\mc^a&=f^a_{\b c}\mc^\b\w\mc^c+\frac12f^a_{bc}\mc^b\w\mc^c\;,
\end{split}
\label{MCequation}
\end{equation}
where I used the facts that $f^a_{\b\g}=0$ (the Lie algebra $\lie h$ closes) and that $f^\a_{\b c}=0$ (the coset space is reductive). These so-called \emph{Maurer--Cartan equations} provide a link between the algebraic and geometric properties of the coset space.


\subsection{Canonical and Torsion-Free Connection}
\label{subsec:geometry_connection}

The local basis of 1-forms $\mc^a$ transforms under the action of the structure group $H$ as a tangent vector. It follows from~\eqref{MCtransfo} that upon an infinitesimal transformation by $h\approx e+\I\eps^\a Q_\a$, $\udelta\mc^a=-\eps^\a f^a_{\a b}\mc^b$. This determines the matrix elements of the action of the $H$-connection $\mcu$ on tangent vectors, in the notation of Appendix~\ref{appsec:affine},
\begin{equation}
\con^a_{\phantom ab}=-f^a_{\a b}\mc^\a\;.
\label{connection_canonical}
\end{equation}
The MC equations~\eqref{MCequation} encode information about both the torsion and the curvature of this so-called \emph{canonical connection} on $G/H$. The torsion 2-form follows from the second line of~\eqref{MCequation},
\begin{equation}
T^a\equiv\D\mc^a+\con^a_{\phantom ab}\w\mc^b=\frac12f^a_{bc}\mc^b\w\mc^c\;.
\label{torsion_canonical}
\end{equation}
Similarly, the curvature 2-form follows from the first line of~\eqref{MCequation},
\begin{equation}
R^a_{\phantom ab}\equiv\D\con^a_{\phantom ab}+\con^a_{\phantom ac}\w\con^c_{\phantom cb}=-\frac12f^a_{\a b}f^\a_{cd}\mc^c\w\mc^d\;,
\label{curvature_canonical}
\end{equation}
where I used the Jacobi identity~\eqref{jacobi} to simplify the result. In the language of field theory, this is nothing but the field-strength 2-form of $\mcu$.

The definition~\eqref{connection_canonical} of the canonical connection arises naturally from the splitting~\eqref{MCsplit} of the MC form into the $\lie h$ and $\lie g/\lie h$ subspaces. This is not the only possible connection on $G/H$ though. In fact, there is an infinite class of them. Let us set
\begin{equation}
\prescript{\l}{}{\con}^a_{\phantom ab}\equiv-f^a_{\a b}\mc^\a-\l f^a_{cb}\mc^c\;,\qquad
\l\in\R\;.
\label{connectionlambda}
\end{equation}
This time, it takes some manipulation to derive the corresponding torsion and curvature 2-forms. The final result is
\begin{equation}
\begin{split}
\prescript{\l}{}{T}^a&=\left(\frac12-\l\right)f^a_{bc}\mc^b\w\mc^c\;,\\
\prescript{\l}{}{R}^a_{\phantom ab}&=-\frac12\bigl(f^a_{\a b}f^\a_{cd}+\l f^a_{eb}f^e_{cd}-2\l^2f^a_{ec}f^e_{bd}\bigr)\mc^c\w\mc^d\;,
\end{split}
\end{equation}
which generalizes~\eqref{torsion_canonical} and~\eqref{curvature_canonical} to any nonzero $\l$.

Within the class of  connections~\eqref{connectionlambda}, there is one with vanishing torsion, corresponding to $\l=1/2$. One might expect that it should be possible to recover this connection from a suitable Riemannian metric on the coset space. This is indeed the case under some further technical assumptions, as I will explain below. For symmetric coset spaces, the whole class of connections~\eqref{connectionlambda} becomes degenerate, and is automatically torsion-free.


\subsection{Riemannian Metric}
\label{subsec:geometry_metric}

Every Lie algebra $\lie g$ possesses a bilinear form invariant under the adjoint action of the corresponding Lie group $G$. To see this, just take any faithful matrix representation of the generators $Q_A$ and set
\begin{equation}
g_{AB}\equiv\tr(Q_AQ_B)\;.
\label{gAB}
\end{equation}
The invariance under $Q_A\to gQ_Ag^{-1}$ for any $g\in G$ is manifest. The infinitesimal version of the invariance condition follows by setting $g\approx e+\I\eps^CQ_C$ and expanding to linear order in $\eps^C$,
\begin{equation}
f^D_{CA}g_{DB}+f^D_{CB}g_{AD}=0\;.
\label{ginvariance}
\end{equation}
Since the components $\mc^a$ of the MC form define a basis of the space of 1-forms on $G/H$, any rank-2 covariant tensor can be constructed as a linear combination of $\mc^a\otimes\mc^b$. Given the invariant bilinear form $g_{AB}$, it is then natural to introduce the following metric on $G/H$,
\begin{equation}
g_{G/H}\equiv g_{ab}\mc^a\otimes\mc^b\;,
\label{gGH}
\end{equation}
where $g_{ab}$ is the restriction of $g_{AB}$ to the $\lie g/\lie h$ subspace.

This construction is not guaranteed to work without further assumptions. A Riemannian metric should be positive-definite. A sufficient, though not necessary, condition for this is that $g_{AB}$ itself is positive-definite, which is generally only true for compact semisimple Lie algebras. For what follows, I will only need the weaker assumption that $g_{AB}$ is nondegenerate and that $g_{a\b}=0$, i.e.~that the subspace $\lie g/\lie h$ can be chosen to be ``orthogonal'' to $\lie h$. Interestingly, this alone already ensures that $G/H$ is a reductive coset space. Namely, a short manipulation using~\eqref{ginvariance} gives
\begin{equation}
0=f^\d_{\a\b}g_{\d c}=f^D_{\a\b}g_{Dc}=-f^D_{\a c}g_{\b D}=-f^\d_{\a c}g_{\b\d}\;,
\end{equation}
which implies that $f^\d_{\a c}=0$ thanks to the fact that $g_{\a\b}$ is nondegenerate. With the assumption that $g_{a\b}=0$, the invariance condition~\eqref{ginvariance} also splits into two separate conditions on $g_{ab}$,
\begin{equation}
f^d_{\g a}g_{db}+f^d_{\g b}g_{ad}=0\;,\qquad
f^d_{ca}g_{db}+f^d_{cb}g_{ad}=0\;.
\label{ginvariance2}
\end{equation}

The metric~\eqref{gGH} is invariant under the left action of $G$, since the MC form itself is. It is however also invariant under the right action of $H_\mathrm{gauge}$ as defined by~\eqref{MCtransfo}. This follows from the first condition in~\eqref{ginvariance2}. The $G$-invariance of the metric guarantees the existence of a set of Killing vector fields that realize infinitesimal group motions on the coset space. In the local coordinates $\pi^a$, these are nothing but the functions $\kil^a_A(\pi)$ introduced in~\eqref{killingdef}. According to~\eqref{cosettransfoexact}, we have $\kil^b_A(0)\mc^a_b(0)=\d^a_A$ at the origin. This means that the Killing vectors $\vec\kil_a(0)$ corresponding to generators from $\lie g/\lie h$ define a local frame dual to $\mc^a(0)$, which further illuminates the geometric nature of the MC form. Away from the origin, the duality between $\mcb$ and the subset of Killing vectors realizing infinitesimal translations on $G/H$ is still expressed by the second line of~\eqref{cosettransfoexact}. One just has to recall that the isotropy group of $x=T_{U(\pi)}x_0$ is $H_x=U(\pi)H_{x_0}U(\pi)^{-1}$. This conjugation is supplied by the matrix $\nu^B_A(\pi)$ on the left-hand side of~\eqref{cosettransfoexact}. The local basis of 1-forms $\mc^a(\pi)$ is then dual to the local frame consisting of Killing vectors of the generators $U(\pi)Q_aU(\pi)^{-1}$.

\begin{illustration}%
The metric~\eqref{gGH} is particularly easy to evaluate explicitly on symmetric coset spaces. Using the expression~\eqref{mcb_symmetric} for $\mcb$ in terms of the linearly transforming variable $\S$, we find at once that
\begin{equation}
g_{G/H}=\frac14\tr(\D\S\otimes\D\S^{-1})=\frac14\tr(\de_a\S\de_b\S^{-1})\D\pi^a\otimes\D\pi^b\;,
\end{equation}
where I used the abbreviation $\de_a\equiv\Pd{}{\pi^a}$. As an illustration, consider the coset space $\gr{SU}(2)/\gr{U}(1)\simeq\gr{SO}(3)/\gr{SO}(2)\simeq S^2$ discussed in \refex{ex:SU(2)U(1)coset}. Here we find that up to an overall factor,
\begin{equation}
g_{S^2}\propto\D\vec n\otimes\D\vec n=\D\t\otimes\D\t+\sin^2\t\,\D\vp\otimes\D\vp\;,
\end{equation}
where I used standard spherical coordinates to parameterize the unit vector $\vec n$. This is just an elaborate way to show that the unique $\gr{SO}(3)$-invariant Riemannian metric on $S^2$ is given by projecting (pulling back) the Euclidean metric on $\R^3$ to the sphere.
\end{illustration}

It remains to clarify the relationship between the metric~\eqref{gGH} and the connections~\eqref{connectionlambda}. The covariant derivative of the metric with respect to these connections, in the direction of an arbitrary vector field $\vec v$, is easily calculated in the local frame,
\begin{equation}
(\cd_{\vec v}g_{G/H})_{ab}=\vec v[g_{ab}]-g_{cb}\prescript{\l}{}\con^c_{\phantom ca}(\vec v)-g_{ac}\prescript{\l}{}\con^c_{\phantom cb}(\vec v)\;.
\end{equation}
The first term vanishes since $g_{ab}$ is merely a set of constants. The sum of the second and the third term vanishes for any $\l$ as a consequence of the combination of~\eqref{connectionlambda} and~\eqref{ginvariance2}. Thus, the whole class of connections~\eqref{connectionlambda} is compatible with the metric~\eqref{gGH}. In case the metric is (pseudo-)Riemannian, it is known that there is a unique metric connection without torsion, called the \emph{Levi-Civita connection}. As shown in the previous subsection, this corresponds to the choice $\lambda=1/2$.

\begin{watchout}%
The metric on the coset space is not uniquely fixed by the requirement of invariance under $G\times H_\mathrm{gauge}$. Although it was natural to start the construction with the Cartan--Killing form~\eqref{gAB} on the whole Lie algebra $\lie g$ and then restrict it to the $\lie g/\lie h$ subspace, we could have as well started from the latter. One then finds that possible metrics on $G/H$ invariant under $G\times H_\mathrm{gauge}$ are classified by constant symmetric tensors $g_{ab}$ invariant under the adjoint action of $H_\mathrm{gauge}$, that is $g_{ab}$ satisfying the first relation in~\eqref{ginvariance2}. For $g_{ab}$ that does not satisfy the second relation in~\eqref{ginvariance2}, the connections~\eqref{connectionlambda} with $\l\neq0$ are not metric-compatible. In particular $\prescript{1/2}{}{\con}^a_{\phantom ab}$ is not the Levi-Civita connection despite being torsion-free.
\end{watchout}


\bibliographystyle{spphys}
\bibliography{references}
\chapter{Low-Energy Effective Field Theory}
\label{chap:effLagrangian}

\abstract*{This chapter constitutes the core of the part of the book devoted to spontaneously broken internal symmetries. It develops two methods for construction of effective actions for Nambu--Goldstone bosons. The first method is based on identification of various contributions to the effective Lagrangian with tensor fields on the coset space of broken symmetry. The requirement of invariance of the action is imposed and solved directly using the tools of differential geometry. However, to understand and utilize the final result requires just elementary linear algebra and group theory. The second approach to construction of effective actions introduces a set of background gauge fields, coupled to the Noether currents of the symmetry. This makes it possible to classify invariant actions using elementary field theory, without heavy use of differential geometry. A comparison of the two approaches sheds light on the conditions under which a global symmetry can be gauged. The last section of the chapter revisits the spectrum of Nambu--Goldstone bosons. It is shown, among others, that type-A and type-B Nambu--Goldstone bosons have a distinct geometric interpretation, associated with a presymplectic structure on the coset space.}


We are now finally in the position to construct a low-energy \emph{effective field theory} (EFT) for systems with a spontaneously broken internal symmetry. We do not need any specific model to describe the dynamics of \emph{spontaneous symmetry breaking} (SSB). All we need is information about the internal symmetry group $G$ and its unbroken subgroup $H$. These determine the basic degrees of freedom of the EFT: the \emph{Nambu--Goldstone} (NG) fields. With the known symmetry and field content, one must then include in the effective Lagrangian all operators allowed by the symmetry (Chap.~12 of~\cite{Weinberg1995a}). This ensures that the predictions of the EFT match, to the desired accuracy, those of any microscopic theory with the same pattern of SSB.

The NG fields can be thought of as (local) coordinates on the coset space $G/H$. In the previous chapter, I showed that the action of the symmetry on these fields can be assumed to take the ``standard form,'' cf.~Sect.~\ref{sec:CCWZstandard}. The task to construct the most general action invariant under such nonlinear realization of symmetry is nontrivial. In Sects.~\ref{sec:effLagstructure} and~\ref{sec:effLaggauged}, I will offer two different approaches to this problem. The first of these is more general but requires substantial background in differential geometry. The second approach is based on certain simplifying assumptions but has the benefit of only requiring elementary field theory. In the concluding Sect.~\ref{sec:effLagEoMspectrum}, I use the EFT to derive the \emph{equation of motion} (EoM) for the NG fields. With the help of the latter, I then reanalyze the spectrum of NG bosons.


\section{Structure of the Effective Lagrangian}
\label{sec:effLagstructure}

To keep the discussion simple, I will assume unbroken symmetry under continuous spacetime translations and continuous spatial rotations. This is just a matter of practical convenience; the same methodology can be applied, for instance, to systems whose spatial rotation symmetry spans a (possibly finite) subgroup of $\gr{SO}(d)$. I will also assume that the spatial dimension is $d\geq2$. On the practical side, this is needed for any continuous rotations to exist. On a more fundamental note, the possibility of SSB in one spatial dimension is severely restricted; see Sect.~\ref{sec:nogo}.

The action of the low-energy EFT must inherit the symmetries of the underlying microscopic theory. As a rule, the symmetry admits an infinite number of operators in the effective Lagrangian, which require an infinite number of a priori unknown coupling constants. In order that the EFT has any predictive power, we need an organizing principle to tell us which of the allowed operators are ``relevant.'' Recall that the EFT is expected to be valid at low energies such that the only active degrees of freedom are the NG bosons. The individual operators in the Lagrangian can then be sorted by the number of derivatives they contain. The more derivatives, the smaller effects the operator is expected to produce. Such ordering of operators in the effective Lagrangian is called \emph{derivative expansion}.

Since spatial and temporal derivatives are not related by any of the assumed symmetries, we must count them separately. The low-energy effective Lagrangian can then be organized as a double series,
\begin{equation}
\La_\mathrm{eff}[\pi]=\sum_{s,t\geq0}\La_\mathrm{eff}^{(s,t)}[\pi]\;,
\label{efflag}
\end{equation}
where $s,t$ denote respectively the numbers of spatial and temporal derivatives. Following loosely~\cite{Watanabe2014a}, I will focus in this section on contributions with at most two derivatives, $s+t\leq2$. These are needed to pin down the kinetic term for all the NG fields, and thus carry information about the spectrum of NG bosons. At the same time, they encode the dominant interactions of NG bosons at low energies. Operators with a higher number of derivatives are classified more easily using the approach developed in Sect.~\ref{sec:effLaggauged}.

Within the diagrammatic expansion of quantum field theory, a given observable may receive contributions from graphs with different numbers of loops. The relative importance of various contributions depends on the operators that enter the interaction vertices and the number of loops. I will work out a precise \emph{power-counting} scheme in Chap.~\ref{chap:internalexamples}, where concrete examples of EFTs are discussed at length. For the time being, I will content myself with the rule of thumb that loops lead to parametric suppression just like derivatives in the interaction vertices. This justifies the approach adopted throughout the whole book, whereby NG bosons are treated in the classical (tree-level) approximation.

The analysis of the structure of the effective Lagrangian~\eqref{efflag} is considerably simplified using the language of differential geometry. Following the presentation below will therefore require some familiarity with the contents of Appendix~\ref{app:diffgeom}. A reader uninterested in the details will find a summary of the leading part of the effective Lagrangian (that is operators with $s+t\leq2$) in Sect.~\ref{subsec:effLagoverview}.


\subsection{Reminder of the Standard Nonlinear Realization}
\label{subsec:effLagreminder}

I will start with a brief reminder of the standard nonlinear realization of symmetry. The intention is to have the basic elements of the formalism at one place for easy reference. Further details can, if needed, be found in Chap.~\ref{chap:CCWZ}.

Every point on the coset space $G/H$ is labeled with a representative $U(\pi)\in G$ of the corresponding coset, where $\pi^a$ are local coordinates on $G/H$ (NG fields). The action of the group $G$ on the coset space can then be defined via left multiplication,
\begin{equation}
U(\pi)\xrightarrow{g}U(\pi'(\pi,g))=gU(\pi)h(\pi,g)^{-1}\;,
\label{cosetaction}
\end{equation}
where $h(\pi,g)\in H$. The choice of the representative $U(\pi)$ is not unique, but can be made so that $U(0)=e$ and that any $g\in H$ acts on $U(\pi)$ by conjugation, $\smash{U(\pi)\xrightarrow{g}gU(\pi)g^{-1}}$. In other words, $h(\pi,g)=g$ for any $g\in H$ independently of $\pi^a$.

The generators of $G$ are denoted as $Q_{A,B,\dotsc}$. A subset of these, $Q_{\a,\b,\dotsc}$, spans the Lie algebra $\lie h$ of the unbroken subgroup $H$. The remaining, broken generators of $G$ are denoted by $Q_{a,b,\dotsc}$.\footnote{The basis $Q_{a,b,\dotsc}$ is to be chosen so that it spans a subspace $\lie g/\lie h$ of the Lie algebra $\lie g$ of $G$ carrying a representation of $H$, hence $\lie g\simeq\lie h\oplus\lie g/\lie h$.} The structure of the Lie algebra $\lie g$ is fixed by the structure constants $\smash{f^C_{AB}}$ through $\smash{[Q_A,Q_B]=\I f^C_{AB}Q_C}$. Under the action of a group element infinitesimally close to unity, $\smash{g\approx e+\I\eps^AQ_A}$, \eqref{cosetaction} reduces to
\begin{equation}
\udelta\pi^a(\pi,g)\approx\eps^A\kil^a_A(\pi)\;,\qquad
h(\pi,g)\approx e+\I\eps^Ak^\a_A(\pi)Q_\a\;.
\label{cosetinfty}
\end{equation}
The functions $\kil^a_A(\pi)$ define a set of vector fields on $G/H$, $\vec\kil_A(\pi)\equiv\kil^a_A(\pi)\Pd{}{\pi^a}$, whose Lie bracket reproduces the structure of $\lie g$,
\begin{equation}
[\vec\kil_A,\vec\kil_B]=f^C_{AB}\vec\kil_C\;.
\end{equation}

An object of fundamental importance is the $\lie g$-valued \emph{Maurer--Cartan} (MC) \emph{form},
\begin{equation}
\mc(\pi)\equiv-\I U(\pi)^{-1}\D U(\pi)\equiv\mc^A(\pi)Q_A\;.
\label{efflagMCform}
\end{equation}
This can be split into a part belonging to the subspaces $\lie h$ and $\lie g/\lie h$,
\begin{equation}
\mc\equiv\mcu+\mcb\;,\qquad
\mcu\equiv\mc^\a Q_\a\;,\qquad
\mcb\equiv\mc^aQ_a\;.
\end{equation}
The components $\mc^A$ of the MC form and the vector fields $\vec\kil_A$ are mutually dual in a well-defined sense. Their precise relationship is expressed by the identities
\begin{equation}
\ix{\vec\kil_A}\mc^\a=\n^\a_A-k^\a_A\;,\qquad
\ix{\vec\kil_A}\mc^a=\n^a_A\;,
\label{MCkildual}
\end{equation}
where the matrix function $\n^B_A(\pi)$ is defined by the conjugation
\begin{equation}
U(\pi)^{-1}Q_AU(\pi)\equiv\n^B_A(\pi)Q_B\;.
\label{nudef}
\end{equation}

Under $G$, the unbroken and broken parts of the MC form transform as
\begin{equation}
\begin{split}
\mcu(\pi)&\xrightarrow{g}\mcu(\pi'(\pi,g))=h(\pi,g)\mcu(\pi)h(\pi,g)^{-1}-\I h(\pi,g)\D h(\pi,g)^{-1}\;,\\
\mcb(\pi)&\xrightarrow{g}\mcb(\pi'(\pi,g))=h(\pi,g)\mcb(\pi)h(\pi,g)^{-1}\;.
\end{split}
\label{cosetmctransfo}
\end{equation}
In the language of differential geometry, infinitesimal symmetry transformations of tensor fields on $G/H$ are given by the Lie derivative along the vector fields $\vec\kil_A$. Combining~\eqref{cosetinfty} and~\eqref{cosetmctransfo} thus corresponds to
\begin{equation}
\ld{\vec\kil_A}\mc^\a=-f^\a_{\b\g}k^\b_A\mc^\g-\D k^\a_A\;,\qquad
\ld{\vec\kil_A}\mc^a=-f^a_{\b c}k^\b_A\mc^c\;.
\label{MCld}
\end{equation}
Finally, the exterior derivative of the MC 1-form is expressed by the MC equation $\D\mc^A=(1/2)f^A_{BC}\mc^B\w\mc^C$. In terms of the $\mcu$ and $\mcb$ components of the MC form, this breaks down into
\begin{equation}
\begin{split}
\D\mc^\a&=\frac12f^\a_{\b\g}\mc^\b\w\mc^\g+\frac12f^\a_{bc}\mc^b\w\mc^c\;,\\
\D\mc^a&=f^a_{\b c}\mc^\b\w\mc^c+\frac12f^a_{bc}\mc^b\w\mc^c\;.
\end{split}
\label{dMC}
\end{equation}


\subsection{Lagrangians with Two Spatial or Two Temporal Derivatives}
\label{subsec:effLag2der}

Invariance of the effective action under a symmetry group $G$ requires that the Lagrangian (density) be \emph{quasi-invariant}, that is invariant up to a surface term. Internal symmetry transformations, covered in this part of the book, do not involve any derivatives of NG fields. This implies that all the different parts $\smash{\La_\mathrm{eff}^{(s,t)}}$ of the effective Lagrangian~\eqref{efflag} must be quasi-invariant separately. The piece without any derivatives, $\smash{\La_\mathrm{eff}^{(0,0)}}$, must be strictly $G$-invariant; its variation under $G$ cannot be a derivative of any local operator.\footnote{It is in principle possible that the variation of $\smash{\La_\mathrm{eff}^{(0,0)}}$ under $G$ is a field-independent constant. This can however only happen if $\smash{\La_\mathrm{eff}^{(0,0)}}$ is a tadpole operator, linear in $\pi^a$. Such operators must be discarded for the EFT to be perturbatively well-defined.} However, requiring $\smash{\ld{\vec\kil_A}\La_\mathrm{eff}^{(0,0)}=\kil_A^a\Pd{\La_\mathrm{eff}^{(0,0)}}{\pi^a}=0}$ leaves a constant, $\pi^a$-independent $\smash{\La_\mathrm{eff}^{(0,0)}}$ as the only option. The derivative expansion of the effective Lagrangian thus starts with operators with at least one derivative. Having a sole spatial derivative is forbidden by rotational invariance, hence $\smash{\La_\mathrm{eff}^{(1,t)}=0}$ for any $t\geq0$. The only options for $(s,t)$ with $s+t\leq2$ that are left are therefore $(0,1)$, $(0,2)$ and $(2,0)$. The Lagrangian $\smash{\La_\mathrm{eff}^{(0,1)}}$ with one temporal derivative will be analyzed in Sect.~\ref{subsec:effLag1der}. Here I will focus on $\smash{\La_\mathrm{eff}^{(2,0)}}$ and $\smash{\La_\mathrm{eff}^{(0,2)}}$, which turn out to be much easier to understand and thus constitute a good starting point.

Invariance under spacetime translations and spatial rotations restricts the two-derivative Lagrangians to the generic form
\begin{equation}
\begin{split}
\La_\mathrm{eff}^{(2,0)}&=-\frac12g_{ab}(\pi)\vec\nabla\pi^a\cdot\vec\nabla\pi^b-\frac12b_{ab}(\pi)\ve^{rs}\de_r\pi^a\de_s\pi^b\;,\\
\La_\mathrm{eff}^{(0,2)}&=\frac12\bar g_{ab}(\pi)\dot\pi^a\dot\pi^b\;,
\end{split}
\label{L20L02}
\end{equation}
where $b_{ab}(\pi)$, $g_{ab}(\pi)$ and $\bar g_{ab}(\pi)$ are some functions on $G/H$. Note that the $b_{ab}$ term can only exist in $d=2$ spatial dimensions. Also, I have discarded from the outset operators containing a second derivative of $\pi^a$, since those can be brought to the form~\eqref{L20L02} by partial integration.

The $g_{ab}$ and $\bar g_{ab}$ terms must be strictly $G$-invariant. Indeed, should an infinitesimal transformation of, say, the $g_{ab}$ term produce a surface term, it would have to be of the form $\vec\nabla\cdot[f_a(\pi)\vec\nabla\pi^a]$. But that would inevitably contain a second derivative of $\pi^a$ that was not present in the original Lagrangian. This argument does not apply to the $b_{ab}$ term, which is antisymmetric in spatial derivatives. However, in that case, possible quasi-invariance can be disregarded on physical grounds. Namely, the $b_{ab}$ term contributes to the canonical Hamiltonian density, and its mere quasi-invariance would imply that the energy density of the EFT is ill-defined. All in all, both $\smash{\La_\mathrm{eff}^{(2,0)}}$ and $\smash{\La_\mathrm{eff}^{(0,2)}}$ must be strictly invariant under the symmetry group $G$.

Let us see what strict $G$-invariance tells us about the function $g_{ab}(\pi)$. The latter gives rise to a symmetric rank-2 tensor field on $G/H$,
\begin{equation}
g(\pi)\equiv g_{ab}(\pi)\D\pi^a\otimes\D\pi^b\;.
\end{equation}
We know that the broken components of the MC form, $\mc^a(\pi)$, furnish a basis of the cotangent space (coframe) to $G/H$; see Sect.~\ref{sec:CCWZgeometry} for a detailed justification. In this basis, the tensor $g(\pi)$ can be expanded as
\begin{equation}
g(\pi)=\k_{ab}(\pi)\mc^a(\pi)\otimes\mc^b(\pi)\;,
\end{equation}
where the components $\k_{ab}(\pi)$ are some as yet unknown functions on $G/H$. Using~\eqref{MCld}, we readily calculate the Lie derivative along $\vec\kil_A$,
\begin{equation}
\ld{\vec\kil_A}g=\bigl(\ld{\vec\kil_A}\k_{ab}-\k_{cb}f^c_{\b a}k^\b_A-\k_{ac}f^c_{\b b}k^\b_A\bigr)\mc^a\otimes\mc^b\;.
\label{kappaaux}
\end{equation}
The requirement of strict $G$-invariance then leads to the following condition on $\k_{ab}$,
\begin{equation}
k^\b_A(f^c_{\b a}\k_{cb}+f^c_{\b b}\k_{ac})=\kil^c_A\PD{\k_{ab}}{\pi^c}\;.
\label{kappaaux2}
\end{equation}
It is now convenient to resort to the exponential parameterization,
\begin{equation}
U(\pi)=\exp(\I\pi^aQ_a)\;,
\label{exppar}
\end{equation}
in which $\kil^a_A(\pi)=\d^a_A+\bigO(\pi)$ and $k^\a_A(\pi)=\d^\a_A+\bigO(\pi)$, cf.~\eqref{expparamresults}. By choosing the index $A$ in~\eqref{kappaaux2} in turn as an unbroken and broken index, we thus get the following initial conditions at the origin of $G/H$,
\begin{equation}
\at{f^c_{\a a}\k_{cb}+f^c_{\a b}\k_{ac}}{\pi=0}=0\;,\qquad
\at{\PD{\k_{ab}}{\pi^c}}{\pi=0}=0\;.
\end{equation}
For a fixed \emph{broken} index $A$, \eqref{kappaaux2} can be viewed as an ordinary differential equation that uniquely determines $\k_{ab}(\pi)$ along the integral curve of $\vec\kil_A$, passing through the origin of $G/H$. We conclude that $\k_{ab}$ must be constant, subject to the constraint
\begin{equation}
f^c_{\a a}\k_{cb}+f^c_{\a b}\k_{ac}=0\;.
\label{invtensorkappa}
\end{equation}
This is equivalent to requiring that $\k_{ab}$ is a constant invariant tensor under the adjoint action of the unbroken subgroup $H$.

The same reasoning applies without change to any $G$-invariant operator constructed solely out of the broken part of the MC form, $\mcb$. As a consequence, the most general form consistent with $G$-invariance that the functions $b_{ab}(\pi)$, $g_{ab}(\pi)$ and $\bar g_{ab}(\pi)$ can take is
\begin{equation}
\begin{gathered}
b_{ab}(\pi)=\l_{cd}\mc^c_a(\pi)\mc^d_b(\pi)\;,\\
g_{ab}(\pi)=\k_{cd}\mc^c_a(\pi)\mc^d_b(\pi)\;,\qquad
\bar g_{ab}(\pi)=\bar\k_{cd}\mc^c_a(\pi)\mc^d_b(\pi)\;.
\end{gathered}
\label{bggbar}
\end{equation}
The constant matrices $\l_{ab}$, $\k_{ab}$ and $\bar\k_{ab}$ are all invariant tensors under the adjoint action of $H$. In the exponential parameterization~\eqref{exppar}, $\mc^a_b(\pi)=\d^a_b+\bigO(\pi)$ and thus $\l_{ab}=b_{ab}(0)$, $\k_{ab}=g_{ab}(0)$ and $\bar\k_{ab}=\bar g_{ab}(0)$.

For the kinetic term of the NG fields to have the correct signature, the symmetric matrices $\k_{ab}$ and $\bar\k_{ab}$ should be positive-definite. Hence, both $g(\pi)$ and its cousin $\bar g(\pi)$ constitute a $G$-invariant Riemannian metric on $G/H$. These two metrics are in principle independent from each other. However, in case $G$ is compact and semisimple and all the broken generators $Q_a$ span a single irreducible multiplet of $H$, Schur's lemma requires that $\k_{ab},\bar\k_{ab}$ be proportional to each other. The metrics $g(\pi)$ and $\bar g(\pi)$ then coincide up to an overall factor.

\begin{illustration}%
Suppose that the symmetry group $G$ is completely broken. Then the matrices $\k_{ab}$, $\bar \k_{ab}$, $\l_{ab}$ remain unconstrained by symmetry, the only physical requirement being the positive-definiteness of $\k_{ab}$ and $\bar\k_{ab}$. This, however, does not mean that when constructing an EFT for NG bosons on $G/\trgr\simeq G$, we have to work with arbitrary matrices of couplings of rank $\dim G$. The number of independent parameters can be reduced by a judicious choice of coordinates $\pi^a$ on $G$. In the theory of small oscillations of mechanical systems~\cite{Goldstein2013a}, it is known that two quadratic forms, at least one of which is positive-definite, can be simultaneously diagonalized. Thus, we can always assume that, say, $\k_{ab}=\d_{ab}$ and moreover that $\bar\k_{ab}$ is diagonal and positive-definite. We just have to keep in mind that this simplicity is a consequence of a particular choice of coordinates, not of symmetry. This may restrict the otherwise arbitrary choice of parameterization of the coset space.
\end{illustration}


\subsection{Lagrangians with One Temporal Derivative}
\label{subsec:effLag1der}

The part of the effective Lagrangian with one time derivative reads, generically,
\begin{equation}
\La_\mathrm{eff}^{(0,1)}=c_a(\pi)\dot\pi^a\;.
\end{equation}
The functions $c_a(\pi)$ define locally a 1-form on $G/H$, $c(\pi)\equiv c_a(\pi)\D\pi^a$. The requirement of quasi-invariance of $\smash{\La_\mathrm{eff}^{(0,1)}}$ under the action of $G$ is then equivalent~to
\begin{equation}
\ld{\vec\kil_A}c=\D\bar c_A\;,
\end{equation}
where $\bar c_A(\pi)$ is a set of local functions (0-forms) on $G/H$. Using the fact that the Lie derivative commutes with the exterior derivative, one deduces that $\ld{\vec\kil_A}(\D c)=\D(\ld{\vec\kil_A}c)=\D(\D\bar c_A)=0$. Hence $\D c(\pi)$ is a strictly $G$-invariant closed 2-form on $G/H$. Strict $G$-invariance requires that
\begin{equation}
\D c(\pi)=\frac12\s_{ab}\mc^a(\pi)\w\mc^b(\pi)\;,
\label{dc}
\end{equation}
where $\s_{ab}$ is a constant antisymmetric matrix, invariant under the adjoint action of $H$. This follows by the same argument I used above to derive~\eqref{bggbar}.

Not every $H$-invariant matrix $\s_{ab}$ will do, however, since the 2-form $\D c(\pi)$ must also be closed. Upon using~\eqref{dMC}, closedness of $\D c(\pi)$ is seen to be equivalent to
\begin{equation}
f^c_{\a a}\s_{cb}+f^c_{\a b}\s_{ac}=0\;,\qquad
f^d_{ab}\s_{cd}+f^d_{bc}\s_{ad}+f^d_{ca}\s_{bd}=0\;.
\label{sigmacocycle}
\end{equation}
The former of these conditions just reasserts the $H$-invariance of $\s_{ab}$. The latter constitutes a new constraint that we must now deal with. I will not be able to give a fully general explicit solution to the algebraic conditions on $\s_{ab}$, and thus a general solution for $\smash{\La_\mathrm{eff}^{(0,1)}}$. We can get very far even without an explicit solution, though. First, note that~\eqref{sigmacocycle} is a set of linear equations for $\s_{ab}$. In a concrete case where the structure constants are known, the space of solutions to~\eqref{sigmacocycle} can therefore be found using elementary linear algebra. Second, we can do even better and link possible solutions for $\s_{ab}$ to the structure of the Lie groups $G$, $H$ and their Lie algebras $\lie g$, $\lie h$.

In the exponential parameterization~\eqref{exppar}, $\D c(\pi)=(1/2)[\s_{ab}+\bigO(\pi)]\D\pi^a\w\D\pi^b$. Thus, the corresponding Lagrangian can be power-expanded in the NG fields as
\begin{equation}
\La_\mathrm{eff}^{(0,1)}=\frac12\s_{ab}\pi^a\de_0\pi^b+\bigO(\pi^3)\;.
\end{equation}
Following the argument of Sect.~\ref{subsec:NGclassificationcounting}, the matrix $\s_{ab}$ can be related to the set of \emph{vacuum expectation values} (VEVs) of commutators of broken generators,
\begin{equation}
\s_{ab}=\I\lim_{V\to\infty}\frac{\vev{[Q_a,Q_b]}}{V}\;.
\label{commutatormatrixsigmaab}
\end{equation}
There is a possibility that the representation of the symmetry group $G$ on the Hilbert space of our system features a central extension. Let us write the extended commutation relation of symmetry generators as $[Q_A,Q_B]=\I f^C_{AB}Q_C+\I Vz_{AB}$, where $z_{AB}$ are the densities of the central charges. Then we find that
\begin{equation}
\s_{ab}=-f^C_{ab}\lim_{V\to\infty}\frac{\vev{Q_C}}{V}-\vev{z_{ab}}\;.
\label{sigmaVEV}
\end{equation}

We conclude that possible solutions for $\s_{ab}$ directly reflect the Lie algebra of symmetry generators, including possible central extensions. One class of solutions to~\eqref{sigmacocycle} that always exists is
\begin{equation}
\s_{ab}=-f^C_{ab}\s_C\quad\text{such that}\quad f^A_{\a B}\s_A=0\;.
\label{sigmasolution}
\end{equation}
The latter constraint expresses, once again, the invariance of the constant tensor $\s_A$ under the adjoint action of $H$. This solution corresponds to vanishing (VEVs of) central charges, and $\s_A$ can then be interpreted as the density of $Q_A$ in the ground state. The $H$-invariance of $\s_A$ descends directly from the $H$-invariance of the ground state. For this class of solutions, \eqref{dc} is readily integrated using the MC equation~\eqref{dMC}, leading to
\begin{equation}
c(\pi)=-\s_A\mc^A(\pi)\quad\text{or}\quad
\La_\mathrm{eff}^{(0,1)}=-\s_A\mc^A_a(\pi)\dot\pi^a\;,
\label{csolution}
\end{equation}
up to addition of a closed 1-form (surface term). The variation of $\smash{\La_\mathrm{eff}^{(0,1)}}$ under infinitesimal transformations from $G$ can now be computed with the help of~\eqref{MCld},
\begin{equation}
\ld{\vec\kil_A}c=-\s_B\ld{\vec\kil_A}\mc^B=\s_B(f^B_{\a C}k^\a_A\mc^C+\d^B_\a\D k^\a_A)=\D(\s_\a k^\a_A)\;,
\end{equation}
where in the last step I used that $\s_Bf^B_{\a C}=0$ by~\eqref{sigmasolution}. The operators proportional to $\s_a$ are strictly $G$-invariant. Mere quasi-invariance of the Lagrangian thus requires nonzero $\s_\a$, that is nonzero VEV of an unbroken charge in the ground state.

\begin{illustration}%
In ferromagnets, $G/H\simeq\gr{SU}(2)/\gr{U}(1)\simeq S^2$. Up to an overall factor, there is a unique closed and $G$-invariant 2-form on $S^2$, provided by the volume (area) form. In terms of the unit-vector parameterization of the sphere, this is proportional to $\ve_{ijk}n^i\D n^j\w\D n^k$, cf.~\refex{ex:embeddingmetric}. The solution~\eqref{sigmasolution} descends from the nonzero magnetization in the ferromagnetic ground state, corresponding to a nonzero VEV of the generator of the unbroken $\gr{U}(1)$ subgroup. I will give an explicit form of the Lagrangian $\La_\mathrm{eff}^{(0,1)}$ for ferromagnets in Sect.~\ref{sec:spinwaves}.
\end{illustration}

In case the symmetry group $G$ is semisimple, its Lie algebra $\lie g$ cannot have any nontrivial central charges (Sect.~2.7 of~\cite{Weinberg1995a}). Equation~\eqref{csolution} is then the only possible solution for the 1-form $c(\pi)$ and the corresponding Lagrangian $\smash{\La_\mathrm{eff}^{(0,1)}}$. For $G$ that are not semisimple, solutions with nonzero $z_{ab}$ may exist.

\begin{illustration}%
Consider the coset space $G/H\simeq\gr{U}(1)\times\gr{U}(1)/\trgr\simeq T^2$. In this case, the symmetry group is Abelian so the conditions~\eqref{sigmacocycle} are trivially satisfied. Up to an overall factor, there is a unique rank-2 antisymmetric tensor, $\s_{ab}=\ve_{ab}$. According to~\eqref{sigmaVEV}, this necessarily arises from a central extension of the Lie algebra $\lie g$ of $G$. In the exponential parameterization~\eqref{exppar}, $\mc^a(\pi)=\D\pi^a$, which leads to $c(\pi)=(1/2)\ve_{ab}\pi^a\D\pi^b$. This is merely quasi-invariant under the action of $G$ as expected. Note that the NG fields $\pi^a$ are angular variables on the torus $T^2$ that are only locally defined, and so is therefore the 1-form $c(\pi)$. What is well-defined globally on the whole coset space is only the $G$-invariant 2-form $\D c(\pi)$. 
\end{illustration}

To conclude the analysis of the Lagrangian $\smash{\La_\mathrm{eff}^{(0,1)}}$, let us see under what conditions we can expect it to actually shift by a surface term upon a $G$-transformation. Should $\smash{\La_\mathrm{eff}^{(0,1)}}$ be strictly $G$-invariant, it would necessarily assume the form
\begin{equation}
\La_\mathrm{eff}^{(0,1)}=-\s_b\mc^b_a(\pi)\dot\pi^a\;,
\end{equation}
with a constant $H$-invariant $\s_a$, that is $f^b_{\a a}\s_b=0$. This follows by imposing directly the constraint $\ld{\vec\kil_A}(\s_a\mc^a)=0$. The corresponding closed 2-form is then
\begin{equation}
\D c=-\s_a\D\mc^a=-\frac12\s_af^a_{bc}\mc^b\w\mc^c\;,
\end{equation}
that is, $\s_{ab}=-f^c_{ab}\s_c$. Quasi-invariant Lagrangians $\smash{\La_\mathrm{eff}^{(0,1)}}$ are therefore in a one-to-one correspondence with antisymmetric matrices $\s_{ab}$ satisfying~\eqref{sigmacocycle} that \emph{cannot} be written as $\s_{ab}=-f^c_{ab}\s_c$ with some $H$-invariant tensor $\s_a$. Such antisymmetric matrices span the so-called second \emph{Lie algebra cohomology} of $\lie g$ relative to the subalgebra $\lie h$. See Sect.~3 of~\cite{Goon2012a} for a mild introduction and further references.

\begin{illustration}%
In case the symmetry under $G$ is completely spontaneously broken, only the second of the conditions in~\eqref{sigmacocycle} survives. In this case, nontrivial solutions for $\s_{ab}$, that is those that cannot be written as $\s_{ab}=-f^c_{ab}\s_c$ with some $\s_c$, are in a one-to-one correspondence with the central charges $z_{ab}$. Both of these are classified by the second Lie algebra cohomology of $\lie g$.
\end{illustration}


\subsection{Overview of the Lowest-Order Effective Lagrangian}
\label{subsec:effLagoverview}

Let me summarize what we have found so far. The basic assumptions I made are that an internal symmetry group $G$ is spontaneously broken to its subgroup $H$, but the symmetry under continuous spacetime translations and spatial rotations remains intact. Then in $d\geq2$ spatial dimensions, the parts of the effective Lagrangian with up to two derivatives of NG fields take the generic form
\begin{align}
\notag
\La_\mathrm{eff}^{(0,1)}&=c_a(\pi)\dot\pi^a\;,\\
\notag
\La_\mathrm{eff}^{(2,0)}&=-\frac12\k_{cd}\mc^c_a(\pi)\mc^d_b(\pi)\vec\nabla\pi^a\cdot\vec\nabla\pi^b-\frac12\l_{cd}\mc^c_a(\pi)\mc^d_b(\pi)\ve^{rs}\de_r\pi^a\de_s\pi^b\;,\\
\La_\mathrm{eff}^{(0,2)}&=\frac12\bar\k_{cd}\mc^c_a(\pi)\mc^d_b(\pi)\dot\pi^a\dot\pi^b\;.
\label{efflagsummary}
\end{align}
Here $\k_{ab}$ and $\bar\k_{ab}$ are constant symmetric $H$-invariant matrices,
\begin{equation}
f^c_{\a a}\k_{cb}+f^c_{\a b}\k_{ac}=0\;,\qquad
f^c_{\a a}\bar\k_{cb}+f^c_{\a b}\bar\k_{ac}=0\;.
\label{invariancekappa}
\end{equation}
Likewise, $\l_{ab}$ is a constant antisymmetric matrix invariant under $H$,
\begin{equation}
f^c_{\a a}\l_{cb}+f^c_{\a b}\l_{ac}=0\;.
\label{invariancelambda}
\end{equation}
The $\l_{ab}$ term can only exist in $d=2$ spatial dimensions. In the special case of a Lorentz-invariant system, both $\smash{\La_\mathrm{eff}^{(0,1)}}$ and the $\l_{ab}$ term in $\smash{\La_\mathrm{eff}^{(2,0)}}$ are forbidden. Moreover, $\k_{ab}=\bar\k_{ab}$ under the convention that the speed of light is set to unity. The entire effective Lagrangian with up to two derivatives then shrinks to a single term,
\begin{equation}
\La_\mathrm{eff}^{(2)}=\frac12\k_{cd}\mc^c_a(\pi)\mc^d_b(\pi)\de_\m\pi^a\de^\m\pi^b\;.
\label{efflagrelativistic}
\end{equation}

Both $\smash{\La_\mathrm{eff}^{(2,0)}}$ and $\smash{\La_\mathrm{eff}^{(0,2)}}$ are strictly $G$-invariant. On the other hand, $\smash{\La_\mathrm{eff}^{(0,1)}}$ may be only quasi-invariant. The functions $c_a(\pi)$ on $G/H$ are constrained by the requirement that the 2-form $\D c\equiv\D(c_a\D\pi^a)=(\Pd{c_b}{\pi^a})\D\pi^a\w\D\pi^b$ is closed and $G$-invariant. This implies
\begin{equation}
\D c(\pi)=\frac12\s_{cd}\mc^c_a(\pi)\mc^d_b(\pi)\D\pi^a\w\D\pi^b\;,
\label{dc2}
\end{equation}
where $\s_{ab}$ is a constant antisymmetric matrix subject to the conditions
\begin{equation}
f^c_{\a a}\s_{cb}+f^c_{\a b}\s_{ac}=0\;,\qquad
f^d_{ab}\s_{cd}+f^d_{bc}\s_{ad}+f^d_{ca}\s_{bd}=0\;.
\end{equation}
Those Lagrangians $\smash{\La_\mathrm{eff}^{(0,1)}}$ that are strictly $G$-invariant assume the form
\begin{equation}
\La_\mathrm{eff}^{(0,1)}=-\s_b\mc^b_a(\pi)\dot\pi^a\;,\quad\text{where}\quad f^b_{\a a}\s_b=0\;.
\label{L01inv}
\end{equation}
In this case, $c_a(\pi)=-\s_b\mc^b_a(\pi)$ and $\s_{ab}=-f^c_{ab}\s_c$. In general, the 1-form $c(\pi)$ is determined by~\eqref{dc2}, hence by $\s_{ab}$, up to addition of a closed 1-form, $c(\pi)\to c(\pi)+\D\tilde c(\pi)$. Such a shift only changes the Lagrangian by a total time derivative,
\begin{equation}
\La_\mathrm{eff}^{(0,1)}\to \La_\mathrm{eff}^{(0,1)}+\PD{\tilde c(\pi)}{\pi^a}\dot\pi^a=\La_\mathrm{eff}^{(0,1)}+\OD{\tilde c(\pi)}{t}\;,
\end{equation}
and so does not affect the dynamics of the low-energy EFT.

Let me conclude the overview with some remarks on the topological aspects of the effective Lagrangian. First of all, the broken part of the MC form $\mcb$ is globally well-defined on the coset space $G/H$. This is because it furnishes a coframe, dual to the frame built out of the vector fields that realize the action of $G$ on $G/H$. All the parts of the effective Lagrangian~\eqref{efflagsummary} that are strictly invariant under $G$ are constructed out of products of $\mc^a(\pi)$ with constant tensor coefficients. As a consequence, all these parts are themselves globally well-defined on $G/H$, even though their explicit form may depend on the local coordinates $\pi^a$. In particular the $\k_{ab},\bar\k_{ab}$ terms correspond to $G$-invariant Riemannian metrics on $G/H$.

The global existence of $\smash{\La_\mathrm{eff}^{(0,1)}}$, on the other hand, is not guaranteed by our construction. Problems may arise only if $\smash{\La_\mathrm{eff}^{(0,1)}}$ is merely quasi-invariant, otherwise the same argument as that for $\smash{\La_\mathrm{eff}^{(2,0)}}$ and $\smash{\La_\mathrm{eff}^{(0,2)}}$ applies. Quasi-invariant Lagrangians $\smash{\La_\mathrm{eff}^{(0,1)}}$ are classified by the second Lie algebra cohomology of $\lie g$ relative to the subalgebra $\lie h$. Should the group $G$ be compact and connected and the subgroup $H$ closed and connected, this relative Lie algebra cohomology is isomorphic to the second de Rham cohomology of $G/H$ (Theorem~7.4 in~\cite{Azcarraga2001a}). Cohomologically nontrivial $\s_{ab}$ then ensures that the 2-form $\D c$ is closed but not exact. As a consequence, the 1-form $c(\pi)$ cannot be extended from the local coordinate patch to the entire coset space.

\begin{illustration}%
Let us contrast the coset spaces $G/H\simeq G/\trgr$ with respectively $G\simeq\gr{U}(1)\times\gr{U}(1)$ and $G\simeq\R\times\R$. These have identical Lie algebras $\lie g,\lie h$ that possess a single generator of the second (relative) Lie algebra cohomology, $\s_{ab}=\ve_{ab}$. The single-derivative Lagrangian $\smash{\La_\mathrm{eff}^{(0,1)}}=(1/2)\ve_{ab}\pi^a\de_0\pi^b$ is quasi-invariant under $G$ in both cases.

Now in the first case, $G/H\simeq\gr{U}(1)\times\gr{U}(1)\simeq T^2$. The 2-form $\D c(\pi)$ is proportional to the volume form on the torus, and constitutes the single generator of the second de Rham cohomology of $T^2$. This agrees with the fact that $G$ is compact and connected and thus the two cohomology groups are necessarily isomorphic. The Lagrangian $\smash{\La_\mathrm{eff}^{(0,1)}}$ is obviously not globally well-defined on $T^2$ because the fields $\pi^a$ are not.

In the second case, on the other hand, $G/H\simeq\R^2$. In the Euclidean plane, the Poincar\'e lemma holds and the second de Rham cohomology is trivial. Accordingly, $\smash{\La_\mathrm{eff}^{(0,1)}}$ is globally well-defined, since $\pi^a$ are now nothing but the two Cartesian coordinates in the plane.
\end{illustration}

\begin{watchout}%
A pedant might object that a Lagrangian that is not well-defined everywhere on the coset space renders the whole EFT ill-defined. However, in classical field theory, one is largely interested just in the EoM. The latter only involves the functions $c_a(\pi)$ through the combinations $\Pd{c_b}{\pi^a}-\Pd{c_a}{\pi^b}$, that is, only depends on $\D c(\pi)$. Unlike the Lagrangian, the classical EoM is therefore globally well-defined on $G/H$.

The situation is quite different in quantum theory where we need to be able to perform a functional integral over all field configurations on the coset space. Luckily, we do not really need the Lagrangian $\La_\mathrm{eff}[\pi]$, or even the action $S_\mathrm{eff}$, to be well-defined. What should be unambiguous is just the phase factor $\exp(\I S_\mathrm{eff})$. To ensure this, one needs to cover the coset space with a set of coordinate patches and define the action by piecewise integration. A detailed analysis is well-beyond the scope of this book. A curious reader will find more details and further references in~\cite{Davighi2018a}, which puts forward a modern classification of topological terms in the action using a homology-based approach. Here I will just state without proof that consistency and $G$-invariance require that the so-called \emph{Manton condition} be satisfied: the 1-forms $\ix{\vec\kil_A}(\D c)$ must be exact for all the vector fields $\vec\kil_A$ realizing the action of $G$ on $G/H$. This is a stronger condition than $G$-invariance of $\D c$ that I used above, which implies merely that $\ix{\vec\kil_A}(\D c)$ is closed. The Manton condition is automatically satisfied by our $\smash{\La_\mathrm{eff}^{(0,1)}}$ whenever the first de Rham cohomology of $G/H$ is trivial. This is the case for instance when $G$ is compact and simply connected and $H$ is connected; cf.~\refex{ex:2ndcohomology}.
\end{watchout}


\section{Effective Lagrangians from Background Gauge Invariance}
\label{sec:effLaggauged}

I this section I will switch gears and outline another approach to the construction of effective Lagrangians for SSB. This is based on a technical assumption that rules out some of the more exotic theories covered by the method of Sect.~\ref{sec:effLagstructure}, notably systems with a central extension of the symmetry algebra. The reward for making this sacrifice is a drastic simplification of the classification of possible contributions to the effective action. All we shall need will be elementary field theory with no recourse to differential geometry. The method presented here was pioneered by Leutwyler~\cite{Leutwyler1994b,Leutwyler1994a}, and I will largely follow the pedagogical account of~\cite{Andersen2014a}.

The generating functional formalism constitutes an arsenal of important tools in both classical and quantum field theory. In this framework, one couples a given theory to a set of classical external (or background) fields and subsequently integrates, if only formally, over the dynamical degrees of freedom. The physical properties of the theory are encoded in the ensuing generating functional of the background fields. See Chap.~16 of~\cite{Weinberg1996a} for an introduction to the formalism.

In principle, one has the freedom to choose the background at will. However, in presence of a continuous internal symmetry group $G$, it is convenient to introduce a set of background gauge fields $A^A_\m$, one for each generator $Q_A$ of $G$. These can be clustered into a matrix-valued gauge field $\smash{A_\m\equiv A^A_\m Q_A}$. The important technical assumption I am making here is that the symmetry under $G$ is \emph{gaugeable}. That is, it is possible to add the gauge fields so that the generating functional $W\{A\}$ of the system\footnote{I follow~\cite{Leutwyler1994b,Leutwyler1994a} and use braces to indicate the arguments of a functional. Square brackets are reserved for local functions of fields and their derivatives.} is invariant under the gauge transformation
\begin{equation}
A_\m\xrightarrow{g}T_gA_\m\equiv gA_\m g^{-1}+\I g\de_\m g^{-1}\;.
\label{backgroundgauge}
\end{equation}

\begin{illustration}%
Consider a theory of a set of (not necessarily scalar) fields $\p^i$, whose Lagrangian density $\La[\p]$ is strictly invariant under a linear representation of the symmetry group $G$, $\smash{\p^i\xrightarrow{g}\rep(g)^i_{\phantom ij}\p^j}$. Such a Lagrangian can always be made invariant under the simultaneous local transformation of $\p^i$ and~\eqref{backgroundgauge}. All one has to do is to replace derivatives of $\p^i$ with gauge-covariant derivatives,
\begin{equation}
\de_\m\p^i\to\bigl[\d^i_j\de_\m-\I\rep(A_\m)^i_{\phantom ij}\bigr]\p^j\;.
\label{gaugecovder}
\end{equation}
In order to assert gauge invariance of the generating functional $W\{A\}$, one still needs to check that the functional integral measure does not change upon the local transformation of $\p^i$. That would indicate an anomaly. In general, the assumption of gauge invariance of $W\{A\}$ essentially amounts to the absence of anomalies or other obstructions to gauging.
\end{illustration}

A linear realization of symmetry is typical for microscopic field theory models. There, one can usually check explicitly that the assumption of gauge invariance of the generating functional is satisfied. The low-energy EFT for the NG bosons should then, when coupled to the same external fields, reproduce the same generating functional. In other words, we expect that the effective action $\smash{S_\mathrm{eff}\{\pi\}=\int\D^D\!x\,\La_\mathrm{eff}[\pi]}$ can be replaced with a ``gauged'' action $\smash{S_\mathrm{eff}\{\pi,A\}=\int\D^D\!x\,\La_\mathrm{eff}[\pi,A]}$ such that
\begin{equation}
\E^{\I W\{A\}}=\int\Da\pi\,\exp[\I S_\mathrm{eff}\{\pi,A\}]\;.
\end{equation}
Leutwyler~\cite{Leutwyler1994b} showed that gauge invariance of $W\{A\}$ is sufficient to ensure that the effective action $S_\mathrm{eff}\{\pi,A\}$ is invariant under the simultaneous local transformations~\eqref{cosetaction} and~\eqref{backgroundgauge} of the NG and gauge fields. The locality of~\eqref{cosetaction} is implemented by allowing $g$ therein to depend arbitrarily on spacetime coordinates as in~\eqref{backgroundgauge}.

\begin{watchout}%
It is common to gloss over this detail and simply assume right away that the effective action is $G$-invariant. I did the same in Sect.~\ref{sec:effLagstructure}. It is therefore worth stressing that this is a consequence of the properties of the generating functional, established at the microscopic level. The proof is technical and I refer the reader to~\cite{Leutwyler1994b} for details.
\end{watchout}

We have replaced the problem of classifying possible $G$-invariant actions $S_\mathrm{eff}\{\pi\}$ with that of constructing locally $G$-invariant actions $S_\mathrm{eff}\{\pi,A\}$. How does that make our task easier? In short, a lot. I will now show in a series of simple steps how the problem of finding all possible gauged actions $S_\mathrm{eff}\{\pi,A\}$ can be reduced to an exercise in group theory. If desired, one can then always discard the background gauge fields and thus recover (most of) the results of Sect.~\ref{sec:effLagstructure}. But the gauged action in fact carries more information, as it tells us how the EFT responds to certain external perturbations.


\subsection{Methodology of Construction of Effective Actions}
\label{subsec:effLagmethodology}

In the first step, we exploit the gauge invariance to eliminate explicit dependence of the action on the NG fields. Namely, by setting $g=U(\pi)^{-1}$, we get
\begin{equation}
S_\mathrm{eff}\{\pi,A\}=S_\mathrm{eff}\{0,T_{U(\pi)^{-1}}A\}\;.
\label{eliminateNG}
\end{equation}
By combining~\eqref{cosetaction} and~\eqref{backgroundgauge}, we find that under the local action of $G$, the composite field $T_{U(\pi)^{-1}}A_\m$ transforms as
\begin{equation}
T_{U(\pi)^{-1}}A_\m\xrightarrow{g}T_{U(\pi'(\pi,g))^{-1}}T_gA_\m=T_{h(\pi,g)}T_{U(\pi)^{-1}}A_\m\;.
\end{equation}
This is a special case of a local transformation of $T_{U(\pi)^{-1}}A_\m$ from $H$. Thus, gauged actions $S_\mathrm{eff}\{\pi,A\}$ locally invariant under $G$ are in a one-to-one correspondence with locally $H$-invariant functionals of $T_{U(\pi)^{-1}}A_\m$.

Let us have a closer look at this composite field. By~\eqref{backgroundgauge},
\begin{equation}
\begin{split}
T_{U(\pi)^{-1}}A_\m&=U(\pi)^{-1}A_\m U(\pi)+\I U(\pi)^{-1}\de_\m U(\pi)\\
&=\I U(\pi)^{-1}(\de_\m-\I A_\m)U(\pi)\;.
\end{split}
\end{equation}
Up to an overall sign, this is just a gauged version of the MC form~\eqref{efflagMCform}. We can make the analogy explicit by turning the gauge field into a 1-form, $A\equiv A_\m\D x^\m$, and defining the gauged MC form via\footnote{This is a slight abuse of notation. The gauge field $A$ is a 1-form on the spacetime. In order for $\MC(\pi,A)$ to be a well-defined 1-form on the spacetime as well, the MC form $\mc(\pi)$ has to be pulled back by the map $\pi^a:x\to\pi^a(x)$.}
\begin{equation}
\MC(\pi,A)\equiv\MC^A_\m(\pi,A)Q_A\D x^\m\equiv-T_{U(\pi)^{-1}}A=-\I U(\pi)^{-1}(\D-\I A)U(\pi)\;.
\label{gaugedMCform}
\end{equation}
This $\lie g$-valued 1-form can again be split into unbroken and broken parts, $\MC=\MCU+\MCB$. These inherit the transformation rules~\eqref{cosetmctransfo} under the local action of $G$,
\begin{align}
\notag
\MCU(\pi,A)&\xrightarrow{g}\MCU(\pi',A')=h(\pi,g)\MCU(\pi,A)h(\pi,g)^{-1}-\I h(\pi,g)\D h(\pi,g)^{-1}\;,\\
\MCB(\pi,A)&\xrightarrow{g}\MCB(\pi',A')=h(\pi,g)\MCB(\pi,A)h(\pi,g)^{-1}\;.
\label{cosetmcgaugetransfo}
\end{align}

The second step is to realize that we can temporarily forget about the origin of $\MCU,\MCB$ in terms of the NG fields $\pi^a$ and gauge fields $A^A_\m$. For the purposes of constructing the effective action, all we need is that $-\MCU$ transforms as a gauge field of $H$, whereas $\MCB$ transforms linearly under the adjoint action of $H$. Once a locally $H$-invariant action $S_\mathrm{eff}\{\MCU,\MCB\}$ has been found, we can reconstruct the dependence on $\pi^a,A^A_\m$ using~\eqref{gaugedMCform}. Now define two new vector fields by taking a variation of the action with respect to $\MCU$ and $\MCB$,
\begin{equation}
J^\m_\a[\MCU,\MCB]\equiv\frac{\udelta S_\mathrm{eff}\{\MCU,\MCB\}}{\udelta\MC^\a_\m}\;,\qquad
\Sigma^\m_a[\MCU,\MCB]\equiv\frac{\udelta S_\mathrm{eff}\{\MCU,\MCB\}}{\udelta\MC^a_\m}\;.
\label{JSigma}
\end{equation}
In spite of the different transformation properties of $\MCU$ and $\MCB$, both $J^\m_\a$ and $\Sigma^\m_a$ transform linearly under the adjoint action of $H$. This is just an auxiliary statement, and I therefore refer the reader to Appendix C of~\cite{Andersen2014a} for a detailed proof. The main message is that to construct such covariantly transforming objects is straightforward, and once we have done so, we can reconstruct the action.

\begin{figure}[t]
\sidecaption[t]
\includegraphics[width=2.0in]{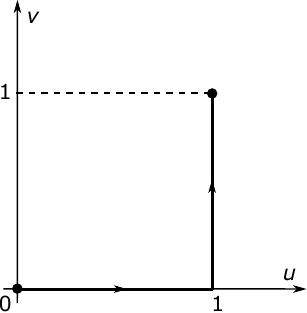}
\caption{Integration path in the space of parameters $u,v$ (shown by the oriented \emph{solid line}), used to reconstruct the effective action from its partial derivatives~\eqref{Suv}}
\label{fig:uvpath}
\end{figure}

To that end, introduce real scaling parameters $u,v$. It follows from~\eqref{JSigma} that
\begin{equation}
\begin{split}
\PD{S_\mathrm{eff}\{u\MCU,v\MCB\}}{u}&=\int\D^D\!x\,\MC^\a_\m(x)J^\m_\a[u\MCU,v\MCB](x)\;,\\
\PD{S_\mathrm{eff}\{u\MCU,v\MCB\}}{v}&=\int\D^D\!x\,\MC^a_\m(x)\Sigma^\m_a[u\MCU,v\MCB](x)\;.
\end{split}
\label{Suv}
\end{equation}
Setting without loss of generality $S_\mathrm{eff}\{0,0\}=0$, the full action is then obtained by integration along the path in the $u\,v$ space, shown in Fig.~\ref{fig:uvpath},
\begin{equation}
\begin{split}
S_\mathrm{eff}\{\MCU,\MCB\}=&\int\D^D\!x\int_0^1\D u\,\MC^\a_\m(x)J^\m_\a[u\MCU,0](x)\\
&+\int\D^D\!x\int_0^1\D v\,\MC^a_\m(x)\Sigma^\m_a[\MCU,v\MCB](x)\;.
\end{split}
\end{equation}
Equivalently, the effective Lagrangian $\La_\mathrm{eff}[\MCU,\MCB]$ can be split into two pieces, $\La_\mathrm{eff}[\MCU,\MCB]=\La_\mathrm{CS}[\MCU]+\La_\mathrm{inv}[\MCU,\MCB]$, such that
\begin{equation}
\begin{split}
\La_\mathrm{CS}[\MCU]&=\int_0^1\D u\,\MC^\a_\m J^\m_\a[u\MCU,0]\;,\\
\La_\mathrm{inv}[\MCU,\MCB]&=\int_0^1\D v\,\MC^a_\m\Sigma^\m_a[\MCU,v\MCB]\;.
\end{split}
\label{LinvCS}
\end{equation}
As a consequence of the linear transformation properties of $\MC^a_\m$ and $\Sigma^\m_a$, the part $\La_\mathrm{inv}[\MCU,\MCB]$ is strictly invariant under the local action of $G$. Possibly quasi-invariant contributions to the effective Lagrangian reside in the part $\La_\mathrm{CS}[\MCU]$, which only depends on the $\lie h$-valued gauge field $-\MCU$. I will refer to this part as \emph{Chern--Simons} (CS) due to its resemblance of the CS theory with gauge group $H$.

The strategy to reconstruct the effective Lagrangian is now as follows. The gauge-invariant part of the Lagrangian, $\La_\mathrm{inv}[\MCU,\MCB]$, can be built directly out of $\MC^a_\m$, the field-strength of $-\MC^\a_\m$, $\smash{G^\a_{\m\n}\equiv-\de_\m\MC^\a_\n+\de_\n\MC^\a_\m+f^\a_{\b\g}\MC^\b_\m\MC^\g_\n}$, and their covariant derivatives. This is common lore but if desired, a detailed proof can be found in Appendix D of~\cite{Andersen2014a}. In the same way, one can construct possible gauge-covariant ``currents'' $J^\m_\a$, except that now the available building blocks are only $G^\a_{\m\n}$ and its covariant derivatives. In both cases, products of the covariant building blocks must be contracted with tensor coefficients that ensure invariance under the residual linearly realized symmetries (spatial rotations and the unbroken subgroup $H$). Finally, the CS part of the Lagrangian is obtained from the current $J^\m_\a$ by integration over $u$ as indicated in~\eqref{LinvCS}. Here the quasi-invariance under $H$ must be checked explicitly and may impose further constraints on the tensor couplings.

Up to any desired order in the derivative expansion, only a finite number of covariant operators can contribute to $\La_\mathrm{inv}[\MCU,\MCB]$ or $J^\m_\a[\MCU]$, and they can be enumerated by inspection. See~\cite{Andersen2014a} for a full list up to order four in derivatives. The problem of constructing all gauge-invariant actions then boils down to solving the group-theoretic linear constraints on the tensor couplings of the individual operators. The full dependence on the fields $\pi^a,A^A_\m$ is already completely fixed by the structure of the operators. The practical use of this algorithmic procedure is best illustrated on the sample analysis of effective Lagrangians with up to two derivatives, worked out below.


\subsection{Lagrangians up to Order Two in Derivative Expansion}
\label{subsec:effLagorder1and2}

For the derivative expansion to be consistent, the Lorentz index on $A^A_\m$ has to count equally to the derivative $\Pd{}{x^\m}$. Each of $\MC^\a_\m,\MC^a_\m$ is then of order one in (spatial or temporal) derivatives. Up to order two, the invariant Lagrangian $\La_\mathrm{inv}[\MCU,\MCB]$ may contain the following contributions,
\begin{equation}
\MC^a_\m\quad\text{(order 1)}\;,\qquad
\MC^a_\m\MC^b_\n,D_\m\MC^a_\n,G^\a_{\m\n}\quad\text{(order 2)}\;,
\end{equation}
where $D_\m\MC^a_\n\equiv\de_\m\MC^a_\n+\I[\MC_{\parallel\m},\MC_{\perp\nu}]^a$. However, a $G$-invariant Lagrangian density built out of $D_\m\MC^a_\n$ alone would be a pure surface term. Similarly, a term of the type $\smash{c_\a G^\a_{\m\n}}$ would boil down to $\smash{c_\a f^\a_{\b\g}\MC^\b_\m\MC^\g_\n}$, the rest of $\smash{G^\a_{\m\n}}$ being a surface term. But the combination $\smash{c_\a f^\a_{\b\g}}$ would necessarily vanish due to the $H$-invariance condition imposed on the coefficient $c_\a$. Altogether, upon adding the appropriate tensor couplings, the most general gauge-invariant and rotationally invariant Lagrangian up to second order in derivatives is
\begin{equation}
\La_\mathrm{inv}[\MCU,\MCB]=-\s_a\MC^a_0+\frac12\bar\k_{ab}\MC^a_0\MC^b_0-\frac12\k_{ab}\d^{rs}\MC^a_r\MC^b_s-\frac12\l_{ab}\ve^{rs}\MC^a_r\MC^b_s\;.
\label{Linv}
\end{equation}
From~\eqref{cosetmcgaugetransfo}, the coefficients $\s_a,\k_{ab},\bar\k_{ab},\l_{ab}$ must be invariant under the adjoint action of $H$, that is satisfy the constraints
\begin{equation}
\begin{aligned}
f^b_{\a a}\s_b&=0\;,\quad
&f^c_{\a a}\l_{cb}+f^c_{\a b}\l_{ac}&=0\;,\\
f^c_{\a a}\k_{cb}+f^c_{\a b}\k_{ac}&=0\;,\quad
&f^c_{\a a}\bar\k_{cb}+f^c_{\a b}\bar\k_{ac}&=0\;.
\end{aligned}
\end{equation}

In order to pin down $\La_\mathrm{CS}[\MCU]$ up to second order in derivatives, we need $J^\m_\a[\MCU]$ up to order one. But the basic building block available, $G^\a_{\m\n}$, already has order two. The only possibility left therefore is a constant current, restricted by rotation invariance to $J^\m_\a[\MCU]=-\d^\m_0\s_\a$. By~\eqref{LinvCS}, this leads in turn to
\begin{equation}
\La_\mathrm{CS}[\MCU]=-\s_\a\MC^\a_0
\label{LCS}
\end{equation}
as the sole possibility. Quasi-invariance under $H$ imposes a constraint on $\s_\a$,
\begin{equation}
f^\g_{\a\b}\s_\g=0\;.
\end{equation}

Note how effortlessly we recovered the effective Lagrangian, obtained in Sect.~\ref{sec:effLagstructure} and summarized in~\eqref{efflagsummary}. The only difference is that out of all the possible solutions for the $c_a(\pi)$ functions, those found here correspond to vanishing VEV of the central charges $z_{ab}$ in~\eqref{sigmaVEV}. This class of solutions is selected by the requirement that the symmetry under $G$ is gaugeable.

However, our newly deduced effective Lagrangian consisting of~\eqref{Linv} and~\eqref{LCS} does not just reproduce the previously found results. It also tells us how the NG fields couple to the background gauge fields $A^A_\m$. To see how the Lagrangian depends on the latter, we combine the definition~\eqref{gaugedMCform} with~\eqref{MCkildual} and~\eqref{nudef} to write
\begin{equation}
\MCB(\pi,A)=\mcb(\pi)-A^A\n^a_A(\pi)Q_a=\mc^a_b(\pi)[\D\pi^b-A^A\kil^b_A(\pi)]Q_a\;.
\end{equation}
This makes perfect sense: the vector fields $\vec\kil_A$ define infinitesimal group motions on the coset space. As a consequence,
\begin{equation}
D_\m\pi^a\equiv\de_\m\pi^a-A^A_\m\kil^a_A(\pi)
\end{equation}
is the correct definition of a derivative of the NG field, covariant under local $G$-transformations. Altogether, the effective Lagrangian up to second order in derivatives can then be written as
\begin{align}
\notag
\La_\mathrm{eff}^{(0,1)}&=-\s_A\mc^A_a(\pi)\dot\pi^a+\s_A\n^A_B(\pi)A^B_0\;,\\
\notag
\La_\mathrm{eff}^{(2,0)}&=-\frac12\k_{cd}\mc^c_a(\pi)\mc^d_b(\pi)\vec D\pi^a\cdot\vec D\pi^b-\frac12\l_{cd}\mc^c_a(\pi)\mc^d_b(\pi)\ve^{rs}D_r\pi^aD_s\pi^b\;,\\
\La_\mathrm{eff}^{(0,2)}&=\frac12\bar\k_{cd}\mc^c_a(\pi)\mc^d_b(\pi)D_0\pi^aD_0\pi^b\;.
\label{efflaggauged}
\end{align}

\begin{illustration}%
\label{ex:mNGB}%
A chemical potential describing a statistical many-body state of a system can be introduced into its Lagrangian as a constant temporal gauge field (see Chap.~2 of~\cite{Kapusta2006a}). Replacing $\smash{A^A_\m\to\d_{\mu0}\m^A}$ in~\eqref{efflaggauged} therefore allows us to analyze the effect on the EFT of a set of chemical potentials $\m^A$. In particular, one can thus determine the ground state triggered by the chemical potentials as well as the spectrum of NG modes above it. A detailed discussion can be found in~\cite{Watanabe2013b}, where this setup was used to pin down the spectrum of massive NG bosons, introduced in Sect.~\ref{subsec:massiveNGbosons}.

In spin systems, the chemical potentials $\mu^A$ for the generators of $G\simeq\gr{SU}(2)$ can be interpreted as the components of an external magnetic field $\vec B$. This is because the magnetic field couples to the conserved charge of $\gr{SU}(2)$: spin. With the help of \refex{ex:SU(2)U(1)coset}, one can easily check that the second term in $\smash{\La_\mathrm{eff}^{(0,1)}}$ in~\eqref{efflaggauged} is proportional to $\skal Bn$. This is the Zeeman interaction of spins with the magnetic field. Its normalization is fixed by $\s_A$, hence by the magnetization of the ground state.
\end{illustration}


\subsection{Effects of Explicit Symmetry Breaking}
\label{subsec:effLagexplicit}

So far in this chapter, I have assumed the symmetry of the EFT to be perfect. There are however good reasons to consider the effects of (presumably small) perturbations breaking the symmetry \emph{explicitly}. First, I already used this idea in Sect.~\ref{sec:SSBperturbation} to isolate a unique ground state in presence of SSB. Second, symmetries of real physical systems are almost always just approximate, even if to a high precision.

I will not attempt a general analysis of explicit symmetry breaking in the EFT, but rather focus on a special case that is commonplace and admits a very simple treatment. Suppose that the microscopic Lagrangian of the system contains a contribution $m_\vr\bigO^\vr$ where $m_\vr$ are real constant parameters and $\bigO^\vr$ a set of local operators. Suppose these operators transform under some (real) linear representation $\rep$ of $G$,
\begin{equation}
\bigO^\vr\xrightarrow{g}\bigO'^\vr=\rep(g)^\vr_{\phantom\vr\s}\bigO^\s\;,\qquad g\in G\;.
\end{equation}
Such a perturbation violates the invariance of the action under $G$ unless $m_\vr=0$ for all the components $\bigO^\vr$ that belong to a nontrivial irreducible representation of $G$.

In order to understand how the effects of the perturbation propagate to the low-energy EFT, it is convenient to promote the parameters $m_\vr$ to background fields. The generating functional $W\{A,m\}$ then remains exactly invariant under the simultaneous local transformation
\begin{equation}
A_\m\xrightarrow{g}T_gA_\m=gA_\m g^{-1}+\I g\de_\m g^{-1}\;,\qquad
m_\vr\xrightarrow{g}m'_\vr=\rep(g^{-1})^\s_{\phantom\s\vr}m_\s\;.
\label{gaugetransfoextended}
\end{equation}
This translates to the invariance of the gauged effective action $S_\mathrm{eff}\{\pi,A,m\}$ under a simultaneous transformation of $\pi^a,A^A_\m,m_\vr$, generalizing~\eqref{eliminateNG} to
\begin{equation}
S_\mathrm{eff}\{\pi,A,m\}=S_\mathrm{eff}\{0,T_{U(\pi)^{-1}}A,m\rep(U(\pi))\}\;.
\label{eliminate2}
\end{equation}
The action therefore depends on $\pi^a$, $A^A_\m$ and $m_\vr$ only through two composite fields, $T_{U(\pi)^{-1}}A_\m$ and $\Xi_\vr(\pi,m)\equiv\rep(U(\pi))^\s_{\phantom\s\vr}m_\s$. The field $\Xi_\vr(\pi,m)$ transforms under $G$ through a linear representation of $H$,
\begin{align}
\notag
\Xi_\vr(\pi,m)\xrightarrow{g}\Xi_\vr(\pi',m')&=\rep(U(\pi'))^\s_{\phantom\s\vr}m'_\s=\rep(U(\pi))^\tau_{\phantom\tau\s}\rep(h(\pi,g)^{-1})^\s_{\phantom\s\vr}m_\tau\\
&=\rep(h(\pi,g)^{-1})^\s_{\phantom\s\vr}\Xi_\s(\pi,m)\;.
\end{align}
By extension of the argument in Sect.~\ref{subsec:effLagmethodology}, we can localize $\Xi_\vr$ in the strictly invariant part of the Lagrangian. The complete Lagrangian then takes the form
\begin{equation}
\La_\mathrm{eff}[\pi,A,m]=\La_\mathrm{CS}[\MCU]+\La_\mathrm{inv}[\MCU,\MCB,\Xi]\;.
\label{trickextended}
\end{equation}

Exactly which operators containing $\Xi_\vr$ should be included at a given order of the derivative expansion depends on how we decide to count $\Xi_\vr$. It does not contain any derivatives of NG fields but, being proportional to the perturbations $m_\vr$, it makes sense to treat it as small. Depending on the concrete system, only a finite number of operators including $\Xi_\vr$ is then needed at any fixed order of the derivative expansion. The leading perturbation of the effective Lagrangian is linear in $\Xi_\vr$,
\begin{equation}
\La_\mathrm{pert}[\pi,m]=\eta^\vr\Xi_\vr(\pi,m)\;.
\label{lagpert}
\end{equation}
The effective coupling $\eta^\vr$ must be $H$-invariant, which amounts to $\rep(Q_\a)^\vr_{\phantom\vr\s}\eta^\s=0$ for any generator $Q_\a$ of $H$. A more complete list of operators containing $\Xi_\vr$ and contributing to the effective Lagrangian can be found in~\cite{Andersen2014a}.

\begin{illustration}%
The choice of background fields and their transformation properties is up to us. To illustrate this freedom, note that it is possible to treat the gauge fields $A^A_\m$ themselves as linear perturbations. In this case, the operators $\bigO^\vr$ are the Noether currents $J^\m_A$ of the symmetry group $G$. These transform according to the dual of the adjoint representation of $G$,
\begin{equation}
J^\m_A\xrightarrow{g}J'^\m_A=\rep(g^{-1})^B_{\phantom BA}J^\m_B\;,
\end{equation}
where $\rep$ is the adjoint representation itself. The gauge fields $A^A_\m$ are now introduced to the microscopic Lagrangian through the linear coupling $\smash{A^A_\m J^\m_A}$. Equation~\eqref{gaugetransfoextended} assigns them the transformation rule $\smash{A^A_\m\xrightarrow{g}\rep(g)^A_{\phantom AB}A^B_\m}$, which amounts to discarding the derivative piece of the background gauge transformation~\eqref{backgroundgauge}. Following~\eqref{eliminate2} then leads to the composite field
\begin{equation}
\Xi^A_\m(\pi,A)\equiv\rep(U(\pi)^{-1})^A_{\phantom AB}A^B_\m=\n^A_B(\pi)A^B_\m\;.
\end{equation}
The leading contribution of such a perturbation to the effective Lagrangian is then $\eta_A\nu^A_B(\pi)A^B_0$. The coupling $\eta_A$ must be $H$-invariant, that is $\smash{f^A_{\a B}\eta_A=0}$.

We have successfully recovered the structure of the second term in~$\smash{\La_\mathrm{eff}^{(0,1)}}$ in~\eqref{efflaggauged}. Note that treating $A^A_\m$ as a linear perturbation leaves the effective coupling $\eta_A$ unfixed. One could in principle apply the same reasoning case by case to other operators in the effective Lagrangian containing $A^A_\m$. However, imposing full background gauge invariance is obviously much more efficient, as it fixes the dependence of the EFT on the gauge fields entirely without any free parameters.
\end{illustration}


\subsection{Coupling to Matter Fields}
\label{subsec:effLagmatter}

Throughout the whole book, I mostly assume that the NG bosons are the \emph{only} low-energy degrees of freedom present in the given system. This is justified in case any other, non-NG modes in the spectrum possess a gap. The validity of the low-energy EFT for NG bosons is then limited to energies well below this gap. However, there are physical systems where strictly gapless non-NG degrees of freedom naturally occur. One generic possibility is that the spectrum includes a Fermi sea of particles such as electrons, protons, neutrons, or quarks. Any local low-energy EFT must then necessarily include such additional gapless degrees of freedom. It is therefore worthwhile to digress and see how such modes fit into the EFT framework developed in this chapter.

The question how non-NG fields (also called \emph{matter fields}) transform under the nonlinearly realized symmetry was already resolved in Chap.~\ref{chap:CCWZ}. Namely, any set of (not necessarily scalar) matter fields $\mf^\vr$ can without loss of generality be assumed to transform under some linear representation $D$ of the unbroken subgroup $H$. The action of the whole group $G$ is then given by
\begin{equation}
\mf^\vr\xrightarrow{g}\mf'^\vr(\mf,\pi,g)=D(h(\pi,g))^\vr_{\phantom\vr\s}\mf^\s\;,
\label{matterfield}
\end{equation}
alongside~\eqref{cosetaction} which defines the matrix $h(\pi,g)\in H$.

We are now looking for the most general effective action $S_\mathrm{eff}\{\pi,A,m,\mf\}$ consistent with the background gauge invariance under $G$. Our basic trick has been to eliminate explicit dependence on the NG fields by performing a gauge transformation with $g=U(\pi)^{-1}$. For this transformation, $h(\pi,U(\pi)^{-1})=e$, hence the matter fields $\mf^\vr$ remain unaffected. The effective action can therefore be built out of the composite fields $T_{U(\pi)^{-1}}A_\m$ and $\Xi_\vr(\pi,m)$, and $\mf^\vr$. The rest is just group theory. Since the matter fields may also enter the Lagrangian with derivatives, it is useful to have at hand their covariant derivative,
\begin{equation}
D_\m\mf^\vr\equiv\de_\m\mf^\vr+\I\MC^\a_\m D(Q_\a)^\vr_{\phantom\vr\s}\mf^\s\;.
\end{equation}

\begin{watchout}%
The transformation rule~\eqref{matterfield} is not necessarily the only possible choice for the action of $G$ on matter fields. Suppose that the (presumably reducible) representation $D$ of $H$ can be extended to a linear representation of the whole group $G$ on the same set of fields $\mf^\vr$. The redefinition $\smash{\mf^\vr\to\Psi^\vr\equiv D(U(\pi))^\vr_{\phantom\vr\s}\mf^\s}$ then gives variables that transform linearly under the whole $G$, $\smash{\Psi^\vr\xrightarrow{g}D(g)^\vr_{\phantom\vr\s}\Psi^\s}$. Such fields, while superficially natural, however conceal the physical structure of the spectrum. The degenerate energy levels are still organized in multiplets of the unbroken subgroup $H$. States from different multiplets of $H$, even if formally belonging to the same multiplet of $G$, will have different dispersion relations, and different interactions. In the extreme case, it may not even be possible to form complete multiplets of $G$. The arguably more complicated nonlinear transformation rule~\eqref{matterfield} is then the only option. An explicit example will illustrate this best.
\end{watchout}

\begin{illustration}%
Most natural ferromagnets are metals, which betrays the presence of gapless, conducting electrons. While their appearance in spin doublets would seem natural, this is no longer mandatory once the $G\simeq\gr{SU}(2)$ spin symmetry is spontaneously broken down to $H\simeq\gr{U}(1)$. Let us denote the two possible polarizations of the electron as up ($\uparrow$) and down ($\downarrow$). Each of these carries a one-dimensional complex representation of $H$. The action of the full spin group $G$ is then defined in accord with~\eqref{matterfield},
\begin{equation}
\mf_\uparrow\xrightarrow{g}D_\uparrow(h(\pi,g))\mf_\uparrow\;,\qquad
\mf_\downarrow\xrightarrow{g}D_\downarrow(h(\pi,g))\mf_\downarrow\;.
\end{equation}
In the effective Lagrangian, Schr\"odinger-type terms for the up and down electrons can be added independently from each other,
\begin{equation}
\La_\text{eff}\ni\sum_{\s=\uparrow,\downarrow}c_\s\he\mf_\s\left(\I D_0+\frac{\vec D^2}{2m_\s}+\m_\s\right)\mf_\s+\dotsb\;.
\end{equation}
Here $m_\s$ and $\m_\s$ are the effective mass and chemical potential of $\mf_\s$, and the ellipsis denotes other possible operators, compatible with all the assumed symmetries. The covariant derivatives induce coupling between the electrons and the magnon (spin wave) degrees of freedom. There is nothing that prevents us from setting one of $c_\uparrow$, $c_\downarrow$ to zero: we do not need both spin polarizations to preserve the invariance under $\gr{SU}(2)$. Physically, this amounts to the possibility of magnons interacting with a fully polarized Fermi sea of electrons or other spin-$1/2$ fermions such as neutrons.
\end{illustration}

Another important application of the formalism for coupling matter fields to NG bosons is the description of interactions between baryons and pseudoscalar mesons. The latter are pseudo-NG bosons a of spontaneously broken symmetry of quantum chromodynamics. Baryons play the role of matter fields that are much heavier than the NG modes, but may nevertheless be added to the low-energy EFT if desired. An interested reader will find more details in Chap.~4 of~\cite{Scherer2012a}.


\section{Equation of Motion}
\label{sec:effLagEoMspectrum}

For many applications, it is useful to have at hand the EoM for the NG bosons. This is of obvious value in classical physics, it is however also very useful in quantum field theory. First, the linearized EoM can be used to identify the spectrum of excitations above the ground state and their perturbative propagators. Second, within a derivative expansion of the EFT, knowing the leading-order EoM helps to eliminate redundancies in the effective Lagrangian at higher orders.

Let us return to the general effective Lagrangian~\eqref{efflagsummary} without background gauge fields, this time augmented with the perturbation~\eqref{lagpert}. Deriving the EoM from the effective action amounts in principle to a mere variation with respect to the NG field $\pi^a$. However, the complicated dependence of the Lagrangian on $\pi^a$ makes this a rather odious task. I will therefore make a rare exception and leave out all details, simply displaying the final result. A reader wishing to verify it should be prepared to a repeated use of the MC equation~\eqref{dMC} and of the algebraic constraints on all the effective couplings.

I will use the shorthand notation $\mc^A_\m(\pi)\equiv\mc^A_a(\pi)\de_\m\pi^a$. Also, I will need the covariant derivative of the broken part of the MC form,
\begin{equation}
D_\m\mc^a_\n(\pi)\equiv\de_\m\mc^a_\n(\pi)-f^a_{\a b}\mc^\a_\m(\pi)\mc^b_\n(\pi)\;.
\end{equation}
With these preparations, the EoM descending from the lowest-order effective Lagrangian~\eqref{efflagsummary} can be written as
\begin{align}
\label{EoM}
&\s_{ab}\mc^b_0-(\bar\k_{ab}D_0\mc^b_0-f^d_{ab}\bar\k_{cd}\mc^b_0\mc^c_0)+(\k_{ab}\d^{rs}D_r\mc^b_s-f^d_{ab}\k_{cd}\d^{rs}\mc^b_r\mc^c_s)\\
\notag
&+\frac12(f^d_{ab}\l_{cd}+f^d_{bc}\l_{ad}+f^d_{ca}\l_{bd})\ve^{rs}\mc^b_r\mc^c_s+m_\vr\rep(U(\pi))^\vr_{\phantom\vr\s}\rep(\I Q_a)^\s_{\phantom\s\tau}\eta^\tau=0\;.
\end{align}
This does not appear particularly elegant, but it is not that bad. In concrete applications, one or more of the terms in~\eqref{EoM} are often missing. The last term obviously vanishes in the limit of exact symmetry. The term proportional to $\ve^{rs}$ is absent in $d\neq2$ spatial dimensions. Moreover, all the terms containing a structure constant with three broken indices, $f^c_{ab}$ or similar, drop for symmetric coset spaces (see Sect.~\ref{subsec:CCWZsymmetriccoset}). One last special case, in which the form of~\eqref{EoM} drastically simplifies, deserves spelling out explicitly.

\begin{illustration}%
Consider the class of EFTs for Lorentz-invariant systems. This amounts to setting $\bar\k_{ab}=\k_{ab}$ as well as to dropping the single-time-derivative term and the purely spatial two-dimensional term proportional to $\ve^{rs}$,
\begin{equation}
\k_{ab}g^{\m\n}D_\m\mc^b_\n-f^d_{ab}\k_{cd}g^{\m\n}\mc^b_\m\mc^c_\n-m_\vr\rep(U(\pi))^\vr_{\phantom\vr\s}\rep(\I Q_a)^\s_{\phantom\s\tau}\eta^\tau=0\;.
\label{EoMrelativistic}
\end{equation}
Under the additional assumption that the coset space $G/H$ is symmetric, which is the case for many relevant physical systems, the EoM takes the extremely compact form $\k_{ab}g^{\m\n}D_\m\mc^b_\n-m_\vr\rep(U(\pi))^\vr_{\phantom\vr\s}\rep(\I Q_a)^\s_{\phantom\s\tau}\eta^\tau=0$. This is essentially a nonlinear generalization of the Klein--Gordon equation to symmetric coset spaces.
\end{illustration}

A couple of remarks on the main result~\eqref{EoM} are due. First, the last term proportional to $m_\vr$ looks like it might contribute a nonzero constant in the limit $\pi^a\to0$. That would indicate an instability of the origin of the coset space under the symmetry-breaking perturbation. This is hardly surprising; in presence of the perturbation, we can no longer expect all the points of the coset space to correspond to physically equivalent vacua. In order to ensure that $\pi^a=0$ is at least a stationary point of the potential induced by the perturbation, one has to demand that such a constant term in~\eqref{EoM} is absent. That amounts to the condition
\begin{equation}
m_\vr\rep(Q_a)^\vr_{\phantom\vr\s}\eta^\s=0\;.
\end{equation}

Second, the contribution of the two-dimensional term proportional to $\ve^{rs}$ vanishes if $f^d_{ab}\l_{cd}+f^d_{bc}\l_{ad}+f^d_{ca}\l_{bd}=0$. This implies by means of~\eqref{sigmacocycle} that the 2-form $\smash{(1/2)\l_{ab}\mc^a(\pi)\w\mc^b(\pi)}$ is closed, hence contributes a mere surface term to $\smash{\La_\mathrm{eff}^{(2,0)}}$. This makes sense, since surface terms do not affect the EoM.

\begin{watchout}%
What if we want to know how the EoM depends on the external gauge fields $A^A_\m$? It might appear we could obtain a gauge-invariant EoM by replacing everywhere in~\eqref{EoM} $\mc^a_\m$ and $\mc^\a_\m$ with their gauged counterparts $\MC^a_\m,\MC^\a_\m$. Yet, there are subtleties hidden in this naive guess. First, we know from Sects.~\ref{sec:effLagstructure} and~\ref{sec:effLaggauged} that there are admissible values of $\s_{ab}$ for which the symmetry under $G$ cannot be gauged. The catch is that not every modification of~\eqref{EoM} renders it a well-defined variational equation. That is, it may not be possible to construct a Lagrangian bottom-up from the gauged EoM. It is not difficult to show that consistency of the gauged EoM and background gauge invariance of the corresponding action require that $\s_{ab}$ satisfies the constraints~\eqref{sigmasolution}.

Second, it is not obvious that even when the symmetry under $G$ can be gauged, the minimal replacement of $\mc^a_\m$ and $\mc^\a_\m$ with  $\MC^a_\m$ and $\MC^\a_\m$ gives the correct gauged EoM. Repeating the derivation of the EoM but this time starting with the gauged Lagrangian~\eqref{efflaggauged}, one finds instead of~\eqref{EoM} the following,
\begin{align}
\notag
&-f^C_{ab}\s_C\MC^b_0-(\bar\k_{ab}D_0\MC^b_0-f^d_{ab}\bar\k_{cd}\MC^b_0\MC^c_0)+(\k_{ab}\d^{rs}D_r\MC^b_s\\
&-f^d_{ab}\k_{cd}\d^{rs}\MC^b_r\MC^c_s)+\frac12(f^d_{ab}\l_{cd}+f^d_{bc}\l_{ad}+f^d_{ca}\l_{bd})\ve^{rs}\MC^b_r\MC^c_s\\
\notag
&-\frac12\l_{ab}\ve^{rs}\bigl[U(\pi)^{-1}F_{rs}U(\pi)\bigr]^b+m_\vr\rep(U(\pi))^\vr_{\phantom\vr\s}\rep(\I Q_a)^\s_{\phantom\s\tau}\eta^\tau=0\;.
\end{align}
Here $F_{\m\n}\equiv\de_\m A_\n-\de_\n A_\m-\I[A_\m,A_\n]$ is the field-strength tensor of $A_\m$. The slight modification of the first term agrees with the fact that the symmetry under $G$ can only be gauged if $\s_{ab}=-f^C_{ab}\s_C$. Most importantly, however, there is an extra term that could not have been guessed by the naive gauging of~\eqref{EoM}. This has interesting consequences. Even when the 2-form $\smash{(1/2)\l_{ab}\mc^a(\pi)\w\mc^b(\pi)}$ is closed, thus contributing a mere surface term, gauging it leads to a nontrivial modification of the Lagrangian as well as the EoM.
\end{watchout}


\subsection{Spectrum of Nambu--Goldstone Bosons Revisited}
\label{subsec:EoMspectrum}

It is common to analyze the excitation spectrum by expanding the Lagrangian to second order in fluctuations around the ground state. Here I will take an alternative route to illustrate the utility of the EoM. Namely, I will linearize~\eqref{EoM}, that is expand it to first order in the NG fields $\pi^a$. For the sake of simplicity, I will discard all external fields including the perturbations $m_\vr$. By invoking the exponential parameterization~\eqref{exppar} in which $\mc^a_b(0)=\d^a_b$, we get at once
\begin{equation}
\s_{ab}\dot\pi^b-\bar\k_{ab}\ddot\pi^b+\k_{ab}\vec\nabla^2\pi^b\approx0\;.
\label{linearizedNGEoM}
\end{equation}
The $\approx$ symbol reminds us of the linearization done. This has plane-wave solutions
\begin{equation}
\pi^a(\vec x,t)=\hat\pi^a\E^{-\I Et}\E^{\I\skal px}\;,
\end{equation}
where the amplitude $\hat\pi^a$, energy $E$ and momentum $\vec p$ satisfy
\begin{equation}
-\I E\s_{ab}\hat\pi^b+E^2\bar\k_{ab}\hat\pi^b-\vec p^2\k_{ab}\hat\pi^b=0\;.
\label{secular}
\end{equation}

We would now like to understand how the dispersion relations of the various NG modes are related to the matrices $\k_{ab}$, $\bar\k_{ab}$ and $\s_{ab}$. From Sect.~\ref{subsec:NGclassificationcounting}, we expect to find type-A and type-B NG bosons. I even declared therein that the dispersion of type-A NG modes is linear in momentum, whereas that of type-B NG modes is quadratic. We are now in the position to justify this claim. First, one can always change the basis of the variables $\hat\pi^a$ so that $\k_{ab}=\d_{ab}$. Moreover, $\s_{ab}$ can be brought by an additional orthogonal transformation to a block-diagonal form,
\begin{equation}
\s_{ab}=\left(\begin{array}{cccccc|ccc}
0 & \s_{12} & & & & & \phantom{\s_{ab}} & \phantom{\s_{ab}} & \phantom{\s_{ab}} \\
\s_{21} & 0 & & & & & & & \\
 & & 0 & \s_{34} & & & & & \\
 & & \s_{43} & 0 & & & & & \\
 & & & & \rotatebox{-8}{$\ddots$} & \phantom{\s_{56}} & & & \\
 & & & & \phantom{\s_{65}} & \rotatebox{-8}{$\ddots$} & & & \\
\hline
 & & & & & & 0 & & \\
 & & & & & & & \rotatebox{-8}{$\ddots$} & \\
 & & & & & & & & 0
\end{array}\right)\equiv
\left(\begin{array}{ccc|c}
\phantom{\s_{ab}} & \phantom{\s_{ab}} & \phantom{\s_{ab}} & \phantom{\s_{ab}} \\
 & \scalebox{1.25}{$\Sigma$} & & \scalebox{1.25}{$0$} \\
 & & & \\
\hline
\phantom{\s_{ab}} & \scalebox{1.25}{$0$} & & \scalebox{1.25}{$0$}
\end{array}\right)\;.
\label{hugematrix}
\end{equation}
The upper-left block $\Sigma$ corresponds to the type-B sector; these are the fields that enter the bilinear part of $\smash{\La_\mathrm{eff}^{(0,1)}}$. The lower-right block will analogously correspond to the type-A sector. The numbers of the different types of modes agree with our previous counting rule~\eqref{NGcounting}.

One last simplification we can make is to diagonalize the lower-right part of $\bar\k_{ab}$ by yet another orthogonal transformation without spoiling the already reduced forms of $\k_{ab}$ and $\s_{ab}$. Thus, $\bar\k_{ab}$ can be assumed to take the generic form
\begin{equation}
\bar\k_{ab}=\left(\begin{array}{ccc|c}
\phantom{\s_{ab}} & \phantom{\s_{ab}} & \phantom{\s_{ab}} & \phantom{\s_{ab}} \\
 & \scalebox{1.25}{$A$} & & \scalebox{1.25}{$B$} \\
 & & & \\
\hline
\phantom{\s_{ab}} & \raisebox{-0.2ex}{\scalebox{1.25}{$B^T$}} & & \scalebox{1.25}{$\Delta$}
\end{array}\right)\;,
\end{equation}
where $A$ is symmetric and $\Delta$ is positive-definite and diagonal. Altogether, the spectrum of NG bosons is obtained by imposing the condition that the determinant of the matrix of coefficients in~\eqref{secular} vanishes,
\begin{equation}
\det\left(\begin{array}{c|c}
-\I E\Sigma+E^2A-\vec p^2 & E^2B\\
\hline
E^2B^T & E^2\Delta-\vec p^2
\end{array}\right)=0\;.
\label{secular2}
\end{equation}
The exact solution $E(\vec p)$ for the dispersion relation of the various modes will be very complicated. However, the asymptotic behavior of the dispersion in the limit of low momentum is easy to extract from~\eqref{secular2}. Namely, in this limit, the contribution of the off-diagonal blocks can be neglected, as can be the $E^2A$ term in the upper-left block. The asymptotic dispersions in the type-A and type-B sectors therefore descend from the already diagonalized matrix equations
\begin{equation}
E^2\Delta-\vec p^2\approx0\quad\text{(type-A)}\;,\qquad
\I E\Sigma+\vec p^2\approx0\quad\text{(type-B)}\;.
\end{equation}

\begin{watchout}%
The presence of both first and second time derivatives in the type-B sector indicates that~\eqref{secular2} has solutions $E(\vec p)$ with nonzero limit as $\vec p\to\vec0$. Taken at face value, the linearized EoM~\eqref{linearizedNGEoM} thus predicts the existence of \emph{gapped} modes, accompanying type-B NG states. The required balance of operators with one and two time derivatives may however violate the derivative expansion of the EFT. In general, the presence of such gapped partners of type-B NG bosons cannot be asserted from the symmetry-breaking pattern alone. 
\end{watchout}


\subsection{More on the Geometry of the Coset Space}
\label{subsec:EoMgeometry}

We had a first look at homogeneous spaces from the point of view of differential geometry in Sect.~\ref{sec:CCWZgeometry}. I showed that any coset space $G/H$ possesses a collection of $G$-invariant (pseudo-)Riemannian metrics. In this chapter, we found a use for them: two such metrics, $g(\pi)$ and $\bar g(\pi)$, enter the part of the effective Lagrangian for NG bosons with two derivatives. In Sect.~\ref{sec:CCWZgeometry}, I also introduced a class of affine connections on coset spaces. I have not made use of the ensuing curvature and torsion so far. However, we will see in Chap.~\ref{chap:scattering} that these appear naturally in scattering amplitudes of NG bosons.

What I want to briefly discuss now is yet another geometric structure on homogeneous spaces that is intimately connected to the spectrum of NG bosons. The reader might, if needed, want to recall the contents of Sect.~\ref{subsec:symplectic} before proceeding. The exposition below follows~\cite{Watanabe2014a}, to which the reader is referred for a much more thorough discussion of geometric and topological aspects of type-B NG bosons.

Suppose that the spectrum of NG bosons were purely type-B as in ferromagnets. The matrix $\s_{ab}$ must then be nonsingular. The same applies to the 2-form $\D c(\pi)$ that defines the part of the effective Lagrangian with one time derivative, $\smash{\La_\mathrm{eff}^{(0,1)}}$. Being simultaneously closed, this 2-form therefore establishes a symplectic structure on the coset space. The 1-form $c(\pi)$ is the corresponding symplectic potential. The NG fields that block-diagonalize the bilinear part of $\smash{\La_\mathrm{eff}^{(0,1)}}$ as in~\eqref{L01} constitute a set of (local) Darboux coordinates. The existence of a symplectic structure on $G/H$ underlines that it should be treated as the phase space of the EFT. The EoM is of first order in time derivatives. The number of independent modes in the spectrum equals $(1/2)\dim G/H$.

What if the spectrum includes both type-A and type-B NG bosons? Then the 2-form $\D c(\pi)$ becomes singular, yet it still defines a new structure on $G/H$, called \emph{presymplectic}. Intuitively, the coset space becomes partially a phase space and partially a configuration space. The paired NG fields giving rise to type-B NG bosons are ``phase space coordinates.'' These are augmented with additional ``configuration space coordinates,'' corresponding to the type-A NG fields. The dynamics of the former and the latter is respectively of first and second order in time.

Let us try to be a bit more precise while remaining physically intuitive. Suppose that the commutator matrix~\eqref{commutatormatrixsigmaab} were the sole order parameter our system possesses. This would break the symmetry group $G$ to some subgroup $K$. Suppose also that $G$ is compact and that its Lie algebra $\lie g$ does not have any nontrivial central charges (for instance because $G$ is semisimple). According to the discussion in Sect.~\ref{subsec:NGclassificationcounting}, our order parameter then corresponds to the VEVs of a set of conserved charges that belong to a Cartan subalgebra of $\lie g$. These charges together generate an Abelian subgroup $T\subset G$, usually called a \emph{torus}. In group theory terminology, the subgroup $K=\{g\in G\,|\,gh=hg\ \forall h\in T\}$ is the \emph{centralizer} of the torus $T$ in $G$.

The NG modes owing their existence to the order parameter $\s_{ab}$ should be described by an EFT that lives on the coset space $G/K$. A coset space $G/K$ where $K$ is the centralizer of a torus in $G$ is called a \emph{flag manifold}; see Chap.~7 of~\cite{Arvanitoyeorgos2003a} for an introduction. Flag manifolds are known to carry a natural symplectic structure, corresponding to~\eqref{dc} with our order parameter $\s_{ab}$. In physics terms, the spectrum of the EFT on $G/K$ contains by construction only type-B NG modes. Their number is $(1/2)\dim G/K$.

\begin{illustration}%
Consider $G\simeq\gr{U}(n)$ and the torus $T$ generated by the diagonal matrices
\begin{equation}
\left(\begin{array}{c|c}
\l_1\un_{m\times m} & 0\\
\hline
0 & \l_2\un_{(n-m)\times(n-m)}
\end{array}\right)\;,
\end{equation}
where $\lambda_{1,2}\in\R$. The centralizer of this torus in $G$ consists of all block-diagonal unitary matrices, $K\simeq\gr{U}(m)\times\gr{U}(n-m)$. The flag manifold $\gr{U}(n)/[\gr{U}(m)\times\gr{U}(n-m)]$ is called the (complex) \emph{Grassmannian}. The special case of $m=1$ is the \emph{complex projective space}, $\C P^n\simeq\gr{U}(n+1)/[\gr{U}(n)\times\gr{U}(1)]$.
\end{illustration}

How do we add the type-A NG bosons then? In presence of additional order parameters, the symmetry will be broken further down from $K$ to $H$. This will lead to additional $\dim K/H$ modes, which we can identify with the type-A NG bosons. All this of course does not mean that we will have two different EFTs, one living on $K/H$ for the type-A NG bosons, and the other living on $G/K$ for the type-B NG bosons. The two coset spaces are geometrically integrated into $G/H$ through a \emph{fiber bundle} structure,\footnote{Since this is the only place in the book where the concept of a fiber bundle appears, I have decided not to include a detailed explanation in Appendix~\ref{app:diffgeom}. I hope that even a reader without the necessary mathematical background can extract some useful information from the discussion.} $\smash{K/H\to G/H\xrightarrow{\pi}G/K}$. Ignoring the type-A NG degrees of freedom amounts to the projection $\pi:G/H\to G/K$ from the \emph{total space} $G/H$ to the \emph{base space} $G/K$. The type-A NG fields span the \emph{fiber} $K/H$ above each point of the base space. The presymplectic structure on $G/H$ is obtained by pulling back the symplectic 2-form on the base space $G/K$ via the projection map $\pi$.

\newpage

\begin{illustration}%
Consider a theory with a $G\simeq\gr{U}(n)$ symmetry. Suppose that the symmetry is broken down to $H\simeq\gr{U}(n-1)$ by the expectation value of a complex scalar field $\Phi$ that transforms in the fundamental representation of $G$. Since the action of $G$ preserves the norm of $\Phi$, the coset space $\gr{U}(n)/\gr{U}(n-1)$ is obviously equivalent to $S^{2n-1}$. For convenience, we can choose the order parameter as $\vev\Phi=(1,0,\dotsc,0)^T$. The Lie algebra $\lie g$ of $G$ includes two linearly independent singlets of $H$, $Q_1\equiv\un$ and $Q_2\equiv\diag(-n+1,1,\dotsc,1)$. Their linear combination, $Q_\parallel\equiv[Q_2+(n-1)Q_1]/n=\diag(0,1,\dotsc,1)$, generates the sole $\gr{U}(1)$ factor of $H$. The generator $Q_1$ is a singlet of the whole group $G$ and its VEV therefore does not break the symmetry. On the other hand, the VEV of $Q_2$ constitutes a candidate order parameter responsible for type-B NG bosons in the spectrum. See the closely related \refex{ex:SU(n)}.

If $\vev{Q_2}=0$, or $\vev{Q_\parallel}=(n-1)/n\vev{Q_1}$, we have no order parameter, hence $K\simeq G$. In this case, the coset space $G/H$ is pure type-A; there are $2n-1$ type-A NG modes. If, however, $\vev{Q_2}\neq0$, there are $n-1$ type-B NG modes that parameterize the coset space $G/K\simeq\gr{U}(n)/[\gr{U}(1)\times\gr{U}(n-1)]\simeq\C P^{n-1}$. Above each point of this base space, there is a fiber $K/H\simeq\gr{U}(1)$, carrying one type-A NG degree of freedom.
\end{illustration}


\bibliographystyle{spphys}
\bibliography{references}
\chapter{Applications to Particle and Condensed-Matter~Physics}
\label{chap:internalexamples}

\abstract*{This chapter works out in detail two important applications of effective field theory for Nambu--Goldstone bosons. The chiral perturbation theory of mesons is used as a prototype of a relativistic effective field theory. The organization of the effective Lagrangian is explained in detail. The invariant Lagrangian is then worked out up to the next-to-leading order in the derivative expansion. The anomalous part of the effective Lagrangian is constructed in the special case of two light quark flavors. The discussion is rounded with several concrete applications to meson physics and the physics of dense nuclear matter. The next prototypical example is the effective field theory for spin waves in ferro- and antiferromagnets. The contrast between the two systems exposes the general distinction between power counting for type-A and type-B Nambu--Goldstone bosons. Realistic spin systems are often subject to perturbations breaking the ideal spin symmetry, and the text thus includes an extended discussion of such effects. The last section draws the reader's attention to some peculiar properties of ferromagnets, which can be largely traced to the topology of the coset space.}


The material of Chaps.~\ref{chap:CCWZ} and~\ref{chap:effLagrangian} is in principle sufficient to construct the low-energy \emph{effective field theory} (EFT) for any system with a spontaneously broken internal symmetry. However, the implementation of the general methodology for a concrete physical system often still requires a nontrivial amount of effort. Moreover, the actual phenomenology of the EFT may depend sensitively on the internal and spacetime symmetries present. In this chapter, I will therefore work out in detail two applications of the general formalism, one in particle physics and one in condensed-matter physics. Apart from serving as an extensive illustration, this will allow us to dive deeper into some technical details of EFTs for \emph{Nambu--Goldstone} (NG) \emph{bosons}: the consistency of the derivative expansion of the effective Lagrangian, and the topological aspects of quasi-invariant Lagrangians.


\section{Chiral Perturbation Theory of Mesons}
\label{sec:ChPT}

\begin{figure}[t]
\sidecaption[t]
\includegraphics[width=2.0in]{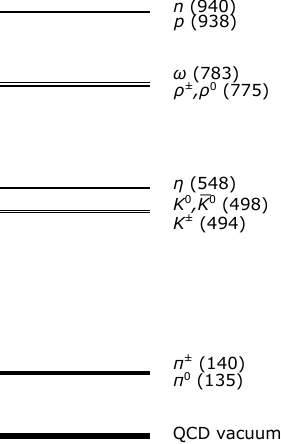}
\caption{Mass spectrum of light hadrons. All the masses in parentheses are shown in $\mathrm{MeV}$. Current numerical values as of writing this book can be found in the Review of Particle Physics~\cite{Workman2022}}
\label{fig:hadronspectrum}
\end{figure}

Historically, the development of EFT for NG bosons was largely motivated by the need for a phenomenological description of low-energy hadron physics.\footnote{This is also reflected by the widespread terminology. In the high-energy physics literature, it is still quite common to refer to NG bosons collectively as ``pions.''} On the one hand, computation of hadron properties from first principles is challenging because the strong nuclear interaction is not amenable to a perturbative treatment at energies below the intrinsic scale of \emph{quantum chromodynamics} (QCD). On the other hand, the spectrum of lightest hadrons exhibits scale separation; see Fig.~\ref{fig:hadronspectrum}. The eight lightest particles in the spectrum---the three pions, four kaons and the $\eta$-meson---are all pseudoscalars with varying quark flavor composition. Their masses are much lower that what one would expect given the constituent mass of about $300\,\mathrm{MeV}$ for the $u,d$ quarks and nearly $500\,\mathrm{MeV}$ for the $s$ quark. This suggests that the low-energy physics of the light pseudoscalar mesons should be captured by an EFT.

The key to understanding the observed scale separation in the hadron spectrum is provided by \emph{spontaneous symmetry breaking} (SSB). The gauge interaction of QCD is vector-like, that is, gluons couple equally to left- and right-handed quarks. Moreover, the interaction is insensitive to the quark flavor. As a consequence, in the limit of vanishing quark masses, QCD with $\nf$ quark flavors possesses a $G\simeq\gr{SU}(\nf)_\mathrm{L}\times\gr{SU}(\nf)_\mathrm{R}\times\gr{U}(1)_\mathrm{B}$ symmetry. Under an element $(g_\mathrm{L},g_\mathrm{R},\E^{\I\eps})\in G$, the left- and right-handed quark spinors $\Psi_{\mathrm{L},\mathrm{R}}$ transform as
\begin{equation}
\Psi_\mathrm{L}\to\E^{\I\eps/3}g_\mathrm{L}\Psi_\mathrm{L}\;,\qquad
\Psi_\mathrm{R}\to\E^{\I\eps/3}g_\mathrm{R}\Psi_\mathrm{R}\;.
\label{quarkchiraltransfo}
\end{equation}
The unitary matrices $g_{\mathrm{L},\mathrm{R}}$ act on the flavor index of the quark field. The vector-like $\gr{U}(1)_\mathrm{B}$ group corresponds to conservation of baryon number, and the factors $1/3$ in~\eqref{quarkchiraltransfo} indicate the baryon number of a single quark. In addition to this internal \emph{chiral symmetry}, QCD is invariant under the spacetime Poincar\'e group, and under the discrete symmetries of charge conjugation, parity and time reversal.

In the ground state of (still massless) QCD, the chiral symmetry is spontaneously broken down to the ``vector'' subgroup, $H\simeq\gr{SU}(\nf)_\mathrm{V}\times\gr{U}(1)_\mathrm{B}$, consisting of elements of the type $(g,g,\E^{\I\eps})$, that is, $g_\mathrm{L}=g_\mathrm{R}$. This implies the existence of $\nf^2-1$ pseudoscalar NG bosons. Restricting to $\nf=2$ accounts for the lightest mesons: the pion triplet. The additional five modes that appear with $\nf=3$ are the strange pseudoscalar mesons (kaons) and the $\eta$-meson. Of course, in reality, none of these is exactly massless. They are all pseudo-NG bosons owing to the fact that the chiral symmetry is only approximate, being explicitly broken by current quark masses. The fact that the strange mesons are heavier than the pions is a consequence of the $s$ quark being considerably heavier than the $u$ and $d$ quarks. The  masses of the $c$, $t$ and $b$ quarks are so high (above the intrinsic scale of QCD) that with $\nf\geq4$, QCD does not possess even an approximate chiral symmetry. In practice, it is therefore sufficient to focus on the $\nf=2,3$ cases. 

\begin{watchout}%
The classical Lagrangian of massless QCD is also invariant under the \emph{axial symmetry} $\gr{U}(1)_\mathrm{A}$, which amounts to $\smash{\Psi_\mathrm{L,R}\to\E^{\pm\I\eps}\Psi_\mathrm{L,R}}$. Being also spontaneously broken in the QCD vacuum, this would suggest the existence of another, flavor-singlet, pseudoscalar meson. Yet the lightest available candidate is the $\eta'$-meson with the mass of $958\,\mathrm{MeV}$. This is too high to be accounted for by explicit symmetry breaking due to quark masses. The resolution of this so-called $\gr{U}(1)_\mathrm{A}$ problem is that the axial symmetry is broken at the quantum level by the axial anomaly, arising from nonperturbative gluon dynamics. See Sect.~13.6 of~\cite{Nair2005} for further details.
\end{watchout}

The EFT for the light pseudoscalar mesons, constructed below, is called \emph{chiral perturbation theory} (ChPT).\footnote{I will adhere to this established name but stress that it is a misnomer. The derivative expansion of ChPT, worked out below, has nothing do to with ordinary perturbation theory known from quantum field theory.} The reader will find a modern graduate-level exposition of ChPT in the dedicated monograph~\cite{Scherer2012a}. However, the pioneering works of Gasser and Leutwyler~\cite{Gasser1984a,Gasser1985a} remain a valuable source of insight. The construction of ChPT will follow the general machinery developed in Sect.~\ref{sec:effLaggauged} with the appropriate coset space, $G/H\simeq[\gr{SU}(\nf)_\mathrm{L}\times\gr{SU}(\nf)_\mathrm{R}]/\gr{SU}(\nf)_\mathrm{V}$. Here I have tacitly dropped the $\gr{U}(1)_\mathrm{B}$ factor of the symmetry group of QCD. This remains unbroken and moreover leaves the meson fields intact, and so has no effect on the invariant part of the ChPT Lagrangian. I will reinstate the $\gr{U}(1)_\mathrm{B}$ symmetry in Sect.~\ref{subsec:ChPTanomaly}, where it will help us build the quasi-invariant (anomalous) part of the Lagrangian.


\subsection{Power Counting}
\label{subsec:ChPTpowercounting}

In a Lorentz-invariant EFT, the effective Lagrangian is organized by the total number, $n=s+t$, of spatial and temporal derivatives,
\begin{equation}
\La_\mathrm{eff}[\pi]=\sum_{n\geq2}\La^{(n)}_\mathrm{eff}[\pi]\;.
\label{ChPTlag}
\end{equation}
Discounting possible tadpole operators, linear in $\pi^a$, a minimum of two derivatives is enforced by the nonlinearly realized internal symmetry and Lorentz invariance. The latter is ensured by contracting Lorentz indices of spacetime derivatives with the Minkowski metric $g_{\m\n}$ or the \emph{Levi-Civita} (LC) \emph{tensor}, $\ve_{\m\n\l\dotsb}$. Hence, in an even number of spacetime dimensions $D$ (which is the case of QCD, where $D=4$), only operators with even $n$ can contribute to~\eqref{ChPTlag}.

In order to assess the relative importance of the individual parts $\smash{\La^{(n)}_\mathrm{eff}}$, consider a generic Feynman diagram $\Gamma$, contributing to a given observable. Suppose the diagram contains altogether $I$ internal propagators, $L$ loops and $V_n$ interaction vertices from each of $\smash{\La^{(n)}_\mathrm{eff}}$. Upon Fourier transform, the diagram will evaluate to a homogeneous function of energy--momenta on the external legs. Adding up all the powers of energy--momentum from the vertices, propagators and loop integrals, the degree of the diagram as a function of the external energy--momenta then reads
\begin{equation}
\deg\Gamma=DL-2I+\sum_{n\geq2}nV_n\;.
\label{graphdegreeaux}
\end{equation}
The number of propagators $I$ can be eliminated using the identity $I=L+\sum_nV_n-1$, known from graph theory to hold for any connected graph (see, for instance, Sect.~13.4 of~\cite{Matousek2009}). This leads to
\begin{equation}
\deg\Gamma=2+(D-2)L+\sum_{n\geq2}(n-2)V_n\;.
\label{graphdegree}
\end{equation}

Diagrammatic contributions to a given observable can now be organized by increasing powers of energy--momentum; higher powers imply stronger suppression at low energies. Any Feynman diagram will have $\deg\Gamma\geq2$. The \emph{leading-order} (LO) contribution to the observable corresponds to $\deg\Gamma=2$ and consists of all tree-level ($L=0$) diagrams with all vertices from $\smash{\La^{(2)}_\mathrm{eff}}$. This justifies a posteriori the focus of this book on classical Lagrangians with the lowest possible number of derivatives.

The degree $\deg\Gamma$ can be increased by adding vertices from higher-order ($n\geq3$) parts of the Lagrangian or by adding loops. In $D=4$ dimensions, the \emph{next-to-leading order} (NLO) corresponds to $\deg\Gamma=4$. This can be reached in two different ways. Either we restrict ourselves to tree-level diagrams but allow for one vertex from $\smash{\La^{(4)}_\mathrm{eff}}$, or allow one loop but keep all vertices from $\smash{\La^{(2)}_\mathrm{eff}}$. It is easy to extend this reasoning to classify possible contributions to any observable at even higher orders of the derivative expansion.

Let me conclude the discussion of power counting with several remarks. First, for pseudo-NG bosons, the propagator $\I/(p^2-m^2)$ is actually not a homogeneous function of the energy--momentum $p^\m$ because of the mass $m$. That can be fixed by assigning the mass a formal counting degree $1$ so that $m^2$ counts equally to $p^2$. This affects the classification of operators in the effective Lagrangian of ChPT that incorporate the effects of explicit breaking of chiral symmetry by the quark mass, $m_q$. We expect based on Sect.~\ref{subsec:pseudoNGbosons} that $m$ will scale as $\sqrt{m_q}$. Hence, we can continue using~\eqref{graphdegree}, as long as we count the quark mass as a small quantity of degree $\deg m_q=2$. Similarly, we will want to couple ChPT to a set of background gauge fields, $A_\m$. In order for the power counting to be consistent with background gauge invariance, we must assign these fields the degree $\deg A_\m=1$.

Second, the counting rule~\eqref{graphdegree} also tells us how to renormalize the EFT. In order to remove an overall divergence in a Feynman graph $\Gamma$, we need a counterterm among the operators in $\smash{\La^{(\deg\Gamma)}_\mathrm{eff}}$. Since all diagrams at $\deg\Gamma=2$ are tree-level, the effective couplings in $\smash{\La^{(2)}_\mathrm{eff}}$ are finite constants that do not run with the renormalization scale. The effective couplings in $\smash{\La^{(n\geq3)}_\mathrm{eff}}$ must include divergences that cancel against loop diagrams with degree $n$. Their running will be determined by the corresponding renormalization group equation. Important is that to any finite order in the derivative expansion, only a finite number of operators, and a finite number of counterterms, is needed.

Third, note that for $D=2$, all loop diagrams with the same $V_{n\geq4}$ and arbitrary $V_2$ contribute at the same order of the derivative expansion. The lack of suppression of loop effects hints that the EFT is no longer weakly coupled. Eventually, it turns out that the infrared fluctuations of NG bosons are so wild in $D=2$ dimensions that they destroy the order parameter leading to SSB. I will return to this in Sect.~\ref{sec:nogo}.

Finally, the basic counting rule~\eqref{graphdegree} and much of the above comments applies equally to any EFT with only type-A NG bosons, relativistic or not. The LO Lagrangian will still be $\smash{\La^{(2)}_\mathrm{eff}}$. The LO contribution to any observable will come from tree-level diagrams with all vertices from $\smash{\La^{(2)}_\mathrm{eff}}$. The classification of contributions at higher orders may however be modified by the presence of operators with odd $n\geq3$ regardless of the spacetime dimension $D$. The details for any particular choice of effective Lagrangian and $D$ are easy to work out using~\eqref{graphdegree}.


\subsection{Effective Lagrangian}
\label{subsec:ChPTefflag}

The coset space of QCD, $G/H\simeq[\gr{SU}(\nf)_\mathrm{L}\times\gr{SU}(\nf)_\mathrm{R}]/\gr{SU}(\nf)_\mathrm{V}$, is symmetric. Let me therefore start by recalling some general properties of symmetric coset spaces; see Sect.~\ref{subsec:CCWZsymmetriccoset} for full detail.

By definition, a coset space is symmetric if the Lie algebra $\lie g$ of $G$ possesses an automorphism $\Raut$ that acts as identity on the unbroken subalgebra $\lie h$ and as minus identity on the complementary subspace $\lie g/\lie h$ of $\lie g$. Loosely speaking, $\Raut$ changes the sign of all broken generators of $G$ while leaving all unbroken generators intact. The automorphism $\Raut$ can be, at least locally, lifted from the Lie algebra $\lie g$ to the Lie group $G$. One can then choose the coset representative $U(\pi)$ so that $\Raut(U(\pi))=U(\pi)^{-1}$. With this choice, one can build a matrix-valued NG field that transforms linearly under the entire group $G$,
\begin{equation}
\S(\pi)\equiv U(\pi)^2\;,\qquad
\S(\pi)\xrightarrow{g}\S(\pi'(\pi,g))=g\S(\pi)\Raut(g)^{-1}\;.
\end{equation}
The automorphism $\Raut$ makes it easy to project out the broken component of the gauged \emph{Maurer--Cartan} (MC) \emph{form}~\eqref{gaugedMCform},
\begin{equation}
\MCB(\pi,A)=\frac12[\MC(\pi,A)-\Raut(\MC(\pi,A))]\;,
\end{equation}
where $A\equiv A_\m\D x^\m$ is the $\lie g$-valued 1-form gauge field of $G$. Upon a brief manipulation using the definition of $\Raut$, we find that
\begin{equation}
\MCB(\pi,A)=-\frac\I2U(\pi)^{-1}D\S(\pi)U(\pi)^{-1}=\frac\I2U(\pi)D\S(\pi)^{-1}U(\pi)\;,
\label{MCBsym}
\end{equation}
where
\begin{equation}
D\S(\pi)=\D\S(\pi)-\I A\S(\pi)+\I\S(\pi)\Raut(A)
\label{DSigma}
\end{equation}
is the $G$-covariant derivative of $\S(\pi)$, and $D\S(\pi)^{-1}$ is defined analogously.

What does all this translate to in case of ChPT? Here $\Raut$ acts by swapping the transformations of left- and right-handed quarks, $\Raut(g_\mathrm{L},g_\mathrm{R})=(g_\mathrm{R},g_\mathrm{L})$. It is then natural to choose the coset representative as
\begin{equation}
U(\pi)=(u(\pi),u(\pi)^{-1})\quad\text{where }u(\pi)\in\gr{SU}(\nf)\;.
\end{equation}
The matrix variable $\S(\pi)=(u(\pi)^2,u(\pi)^{-2})$ is subject to the linear transformation $\smash{\S(\pi)\xrightarrow{(g_\mathrm{L},g_\mathrm{R})}(g_\mathrm{L},g_\mathrm{R})\S(\pi)(g_\mathrm{R},g_\mathrm{L})^{-1}}$. All information about the NG fields can then be encoded in a single $\gr{SU}(\nf)$-valued matrix variable,
\begin{equation}
\U(\pi)\equiv u(\pi)^2\;,\qquad
\U(\pi)\xrightarrow{(g_\mathrm{L},g_\mathrm{R})}\U(\pi'(\pi,g_\mathrm{L},g_\mathrm{R}))=g_\mathrm{L}\U(\pi)g_\mathrm{R}^{-1}\;.
\end{equation}
The Lie algebra of $\gr{SU}(\nf)_\mathrm{L}\times\gr{SU}(\nf)_\mathrm{R}$ is $\lie{su}(\nf)_\mathrm{L}\oplus\lie{su}(\nf)_\mathrm{R}$. In our pair notation, the matrix-valued gauge field $A_\m$ can thus be decomposed as $\smash{A_\m=(A^\mathrm{L}_\m,\un)+(\un,A^\mathrm{R}_\m)}$, where $\smash{A^{\mathrm{L},\mathrm{R}}_\m}$ are independent $\lie{su}(\nf)$-valued gauge fields acting respectively on the left- and right-handed quarks. A straightforward manipulation then leads to $D\S(\pi)=(D\U(\pi),\U(\pi)^{-1})+(\U(\pi),D\U(\pi)^{-1})$, where
\begin{equation}
\begin{split}
D_\m\U(\pi)&=\de_\m\U(\pi)-\I A_\m^\mathrm{L}\U(\pi)+\I\U(\pi)A_\m^\mathrm{R}\;,\\
D_\m\U(\pi)^{-1}&=\de_\m\U(\pi)^{-1}-\I A_\m^\mathrm{R}\U(\pi)^{-1}+\I\U(\pi)^{-1}A_\m^\mathrm{L}\;.
\end{split}
\label{DU}
\end{equation}
This eventually gives
\begin{equation}
\MCB=-\frac\I2\bigl[\bigl(u^{-1}(D\U) u^{-1},\un\bigr)+\bigl(\un,u(D\U^{-1})u\bigr)\bigr]\;.
\end{equation}

Before we can write down even the LO effective Lagrangian of ChPT, we still have to discuss the explicit breaking of chiral symmetry by the quark masses. The mass term in the microscopic Lagrangian of QCD takes the form $\adj\Psi_\mathrm{L}\MM\Psi_\mathrm{R}+\adj\Psi_\mathrm{R}\he\MM\Psi_\mathrm{L}$, where $\MM=\diag(m_u,m_d,\dotsc)$ is a real diagonal matrix, collecting the quark masses. In the spirit of Sect.~\ref{subsec:effLagexplicit}, this is promoted to a complex matrix field that transforms under the chiral symmetry as $\smash{\MM\xrightarrow{(g_\mathrm{L},g_\mathrm{R})}g_\mathrm{L}\MM g_\mathrm{R}^{-1}}$, thus restoring exact chiral invariance of the QCD Lagrangian. The basic building block for incorporating the effects of explicit symmetry breaking in ChPT is then the matrix field
\begin{equation}
\Xi(\pi,\MM)=u(\pi)^{-1}\MM u(\pi)^{-1}\;.
\end{equation}
Being linear in quark masses, this is assigned the order $\deg\Xi=2$.

The operators we now have for building the LO effective Lagrangian of ChPT are, schematically, $\MC_{\perp\m}\MCB^\m$ and $\Xi$. The only way to ensure invariance of these operators under the linearly realized unbroken subgroup, $\gr{SU}(\nf)_\mathrm{V}$, is to take a trace. Moreover, $\Xi$ has to enter the Lagrangian through $\tr(\Xi+\he\Xi)$. This is required by parity invariance of QCD and the fact that parity interchanges left- and right-handed quarks, thereby acting on both $u(\pi)$ and $\MM$ by Hermitian conjugation. At the end of the day, there are only two independent operators one can put into the LO Lagrangian,
\begin{equation}
\La_\mathrm{eff}^{(2)}=\frac{f_\pi^2}4\tr\bigl[D_\m\he{\U(\pi)}D^\m\U(\pi)\bigr]+\frac{f_\pi^2B}2\tr\bigl[\MM\he\U(\pi)+\he\MM\U(\pi)\bigr]\;.
\label{ChPTLO}
\end{equation}
Accordingly, there are two independent parameters, conventionally denoted as $f_\pi$ and $B$, both with mass dimension $1$.

\begin{illustration}%
Let us work out some immediate consequences of the LO Lagrangian~\eqref{ChPTLO}. To that end, I will drop the background gauge fields and parameterize $\U(\pi)$ in terms of a Hermitian traceless matrix $\Pi(\pi)$ as $\U(\pi)=\exp[\I\Pi(\pi)/f_\pi]$. This is a matrix version of the familiar exponential parameterization of the coset space; the factor $f_\pi$ is inserted to give $\Pi$ mass dimension $1$. Upon expansion in powers of $\Pi$, the Lagrangian becomes
\begin{equation}
\begin{split}
\La_\mathrm{eff}^{(2)}={}&\frac{f_\pi^2B}2\tr(\MM+\he\MM)\\
&+\frac14\tr\bigl[\de_\m\Pi(\pi)\de^\mu\Pi(\pi)\bigr]-\frac B4\tr\bigl[(\MM+\he\MM)\Pi(\pi)^2\bigr]+\bigO(\Pi^4)\;,
\end{split}
\label{ChPTLObilin}
\end{equation}
where I used that $\MM$ is ultimately Hermitian to drop terms linear and cubic in $\Pi(\pi)$.

The first, constant term contributes to the energy density of the chiral-symmetry-breaking vacuum. Taking the derivative of the vacuum energy density with respect to any of the quark masses in turn gives the \emph{vacuum expectation value} (VEV) of the corresponding mass operator; cf.~\eqref{vevAlambda}. The latter serves as the order parameter for spontaneous breaking of chiral symmetry. The LO prediction of ChPT, based on~\eqref{ChPTLObilin}, is that this so-called \emph{chiral condensate} is independent of the quark flavor. The total condensate, summed over all quark flavors, equals
\begin{equation}
\vev{\adj\Psi\Psi}=-\nf f_\pi^2B\;.
\end{equation}

Let us now turn attention to the bilinear part of~\eqref{ChPTLObilin}. It is convenient to further parameterize the matrix $\Pi(\pi)$ in terms of $\nf^2-1$ meson fields $\pi^a$ as $\Pi(\pi)=\pi^a\l_a$. Here $\l_a$ is a basis of traceless Hermitian $\nf\times\nf$ matrices, normalized as $\tr(\l_a\l_b)=2\d_{ab}$. This ensures canonical normalization of the kinetic term for the meson fields. In case of $\nf=2$, the $\l_a$s are just the usual Pauli matrices $\pau_a$, whereas for $\nf=3$, the so-called \emph{Gell-Mann matrices} can be used instead. The next step is to find the eigenvalues of the mass matrix for $\pi^a$. Setting $\nf=3$ and using the known flavor composition of the light pseudoscalar mesons, these eigenvalues can be identified with the masses of the individual states in Fig.~\ref{fig:hadronspectrum} as
\begin{align}
\notag
m_{\pi^0}^2&=\frac{2B}3\left(m_u+m_d+m_s-\sqrt{m_u^2+m_d^2+m_s^2-m_um_d-m_um_s-m_dm_s}\right)\;,\\
\notag
m_{\pi^\pm}^2&=B(m_u+m_d)\;,\\
\label{mesonmasses}
m_{K^\pm}^2&=B(m_u+m_s)\;,\\
\notag
m_{K^0}^2&=B(m_d+m_s)\;,\\
\notag
m_\eta^2&=\frac{2B}3\left(m_u+m_d+m_s+\sqrt{m_u^2+m_d^2+m_s^2-m_um_d-m_um_s-m_dm_s}\right)\;.
\end{align}
This determines the five different meson masses in terms of the three current quark masses, yet the latter are not uniquely fixed by~\eqref{mesonmasses}. Indeed, any overall rescaling of the quark masses can be absorbed into a redefinition of the $B$ parameter. It is however possible to use~\eqref{mesonmasses} to eliminate $B$ and the quark masses altogether and thus obtain a constraint on the meson spectrum,
\begin{equation}
2(m_{\pi^\pm}^2+m_{K^\pm}^2+m_{K^0}^2)=3(m_{\pi^0}^2+m_\eta^2)\;.
\end{equation}
This so-called \emph{Gell-Mann--Okubo formula} can be interpreted, for instance, as giving an estimate for the mass of the $\eta$-meson in terms of those of the pions and kaons. With the data shown in Fig.~\ref{fig:hadronspectrum} as input, one gets $m_\eta\approx568\,\mathrm{MeV}$, which is less than four per cent off the correct value.
\end{illustration}

The machinery developed in Sect.~\ref{sec:effLaggauged} can in principle be applied to an arbitrarily high order of the derivative expansion of ChPT. However, already at NLO ($n=4$), constructing the effective Lagrangian is a nontrivial exercise. I will therefore content myself with spelling out the final result and refer the reader to Sect.~3.2 of~\cite{Andersen2014a}, where all details are worked out  rather pedantically. For both $\nf=2$ and $\nf=3$, the invariant part of the ChPT Lagrangian at NLO can be written as
\begin{align}
\notag
\La_\mathrm{eff}^{(4)}={}&c_1\tr(D_\m\U D^\m\he\U D_\n\U D^\n\he\U)+c_2\tr(D_\m\U D^\m\he\U)\tr(D_\n\U D^\n\he\U)\\
\notag
&+c_3\tr(D_\m\U D_\n\he\U)\tr(D^\m\U D^\n\he\U)\\
\notag
&+c_4\tr(F^\mathrm{L}_{\m\n}D^\m\U D^\n\he\U+F^\mathrm{R}_{\m\n}D^\m\he\U D^\n\U)\\
\label{ChPTNLO}
&+c_5\tr(F^\mathrm{L}_{\m\n}\U F^{\mathrm{R}\m\n}\he\U)+c_6(F^\mathrm{L}_{\m\n}F^{\mathrm{L}\m\n}+F^\mathrm{R}_{\m\n}F^{\mathrm{R}\m\n})\\
\notag
&+d_1\tr(\MM\he\U)\tr(\he\MM\U)+d_2\bigl[(\tr\MM\he\U)^2+(\tr\he\MM\U)^2\bigr]\\
\notag
&+d_3\tr(\MM\he\U\MM\he\U+\he\MM\U\he\MM\U)+d_4\tr(\MM\he\MM)\\
\notag
&+d_5\tr(\MM\he\U+\he\MM\U)\tr(D_\m\U D^\m\he\U)\\
\notag
&+d_6\tr\bigl[(\MM\he\U+\U\he\MM)D_\m\U D^\m\he\U\bigr]\;.
\end{align}
Here $F^\mathrm{L}_{\m\n}$ and $F^\mathrm{R}_{\m\n}$ are the field-strength tensors of the background gauge fields. The 12 effective couplings $c_\text{1--6}$ and $d_\text{1--6}$ are mutually independent in the $\nf=3$ case. In case of $\nf=2$, special properties of $2\times2$ matrices make the $c_1$ operator redundant with $c_2$, and the $d_6$ operator redundant with $d_5$.

\begin{watchout}%
The form of the NLO Lagrangian shown in~\eqref{ChPTNLO} matches the detailed derivation offered in~\cite{Andersen2014a}. However, in the literature, a somewhat different basis of operators is often used. For the reader's convenience, I list here detailed relations between the parameters $c_\text{1--6},d_\text{1--6}$ introduced above and the more common NLO couplings of ChPT $L_\text{1--10},H_\text{1--2}$, cf.~Sect.~3.5.1 of~\cite{Scherer2012a},
\begin{equation}
\begin{gathered}
c_1=L_3\;,\qquad
c_2=L_1\;,\qquad
c_3=L_2\;,\qquad
c_4=-\I L_9\;,\\
c_5=L_{10}\;,\qquad
c_6=H_1\;,\qquad
d_1=2(L_6-L_7)\;,\qquad
d_2=L_6+L_7\;,\\
d_3=L_8\;,\qquad
d_4=H_2\;,\qquad
d_5=L_4\;,\qquad
d_6=L_5\;.
\end{gathered}
\end{equation}
\end{watchout}


\subsection{Interaction with External Fields}
\label{subsec:ChPTextfields}

For further illustrations of the use of ChPT, I will utilize its simplest version: the $\nf=2$ ChPT in the ``isospin-symmetric'' limit. In this limit, one sets $m_u=m_d=m$ so that all three pions have the same squared mass, $m^2_\pi=2Bm$. Accordingly, the LO Lagrangian~\eqref{ChPTLO} reduces to
\begin{equation}
\La_\mathrm{eff}^{(2)}=\frac{f_\pi^2}4\bigl[\tr(D_\m\he{\U}D^\m\U)+m_\pi^2\tr(\U+\he{\U})\bigr]\;.
\label{ChPTLO2}
\end{equation}
The kinetic term can be further expanded and organized by powers of the background gauge fields,
\begin{align}
\label{ChPTLO3}
\tr(D_\m\he\U D^\m\U&)=\tr\bigl[\de_\m\he\U\de^\m\U-\I A^\mathrm{L}_\m(\U\de^\m\he\U-\de^\m\U\he\U)\\
\notag
&-\I A^\mathrm{R}_\m(\he\U\de^\m\U-\de^\m\he\U\U)-2\U A^\mathrm{R}_\m\he\U A^{\mathrm{L}\m}+A^\mathrm{L}_\m A^{\mathrm{L}\m}+A^\mathrm{R}_\m A^{\mathrm{R}\m}\bigr]\;.
\end{align}
Knowing the explicit dependence on the background fields is, among others, a useful starting point for deriving the Noether currents of chiral symmetry. I will however focus on illustrating the implications of some particular choices of actual, physical background fields.

\begin{illustration}%
As explained in \refex{ex:mNGB}, a chemical potential parameterizing the statistical equilibrium of a many-body system can be introduced in the Lagrangian as a constant temporal gauge field. Thus, the effects of nonzero density of isospin, that is the diagonal generator of $\gr{SU}(2)_\mathrm{V}$, can be captured by setting
\begin{equation}
A^\mathrm{L}_\m=A^\mathrm{R}_\m=\d_{\m0}\m_\mathrm{I}\frac{\pau_3}2\;,
\end{equation}
where $\m_\mathrm{I}$ is the isospin chemical potential. The actual statistical ground state is found by minimizing the Hamiltonian of ChPT with respect to $\U$. A detailed analysis shows that the ground state can be represented by a real orthogonal matrix,
\begin{equation}
\vev\U=\begin{pmatrix}
\cos\a & \sin\a\\
-\sin\a & \cos\a
\end{pmatrix}\;.
\label{pionBEC}
\end{equation}
For $\abs{\m_\mathrm{I}}\leq m_\pi$, the value of the angle $\a$ minimizing energy is $\a=0$, implying $\vev\U=\un$. This is the usual QCD vacuum. On the other hand, for $\abs{\m_\mathrm{I}}\geq m_\pi$, the ground state corresponds to $\cos\a=m_\pi^2/\m_\mathrm{I}^2$. This state describes \emph{Bose--Einstein condensation} of charged pions. The condensate carries nonzero isospin density obtained as minus the derivative of the Hamiltonian density with respect to $\m_\mathrm{I}$,
\begin{equation}
n_\mathrm{I}=f_\pi^2\m_\mathrm{I}\sin^2\a=f_\pi^2\m_\mathrm{I}\left(1-\frac{m_\pi^4}{\m_\mathrm{I}^4}\right)\;.
\end{equation}

Next, let us look at the spectrum of excitations above the ground state~\eqref{pionBEC}. This can be extracted by expanding the Lagrangian~\eqref{ChPTLO2} to second order in fluctuations around~\eqref{pionBEC}. In the vacuum phase ($\abs{\m_\mathrm{I}}<m_\pi$), the neutral pion maintains its relativistic dispersion relation, $\smash{E_{\pi^0}(\vec p)=\sqrt{\vec p^2+m_\pi^2}}$. The energy of the charged pions is, on the other hand, trivially shifted by the chemical potential, $\smash{E_{\pi^\pm}(\vec p)=\sqrt{\vec p^2+m_\pi^2}\mp\m_\mathrm{I}}$. In the pion condensation phase ($\abs{\m_\mathrm{I}}>m_\pi$), the dispersion of the neutral pion changes to
\begin{equation}
E_{\pi^0}(\vec p)=\sqrt{\vec p^2+\m_\mathrm{I}^2}\;.
\end{equation}
This is a relativistic-looking dispersion, except that the ``mass'' equals $\abs{\m_\mathrm{I}}$. That is not a coincidence. In the pion condensation phase, the neutral pion mode can be interpreted as a massive NG boson of the isospin $\gr{SU}(2)_\mathrm{V}$ symmetry; see Sect.~\ref{subsec:massiveNGbosons} and~\cite{Watanabe2013b} for details. In the charged pion sector, it is no longer possible to distinguish isospin (or electric charge) eigenstates as a result of SSB. There are two excitation branches that are mixtures of $\pi^+$ and $\pi^-$, and their squared energies are
\begin{equation}
E_\pm(\vec p)^2=\vec p^2+\frac{\m_\mathrm{I}^2}2(1+3\cos^2\a)\pm\frac{\m_\mathrm{I}}2\sqrt{(1+3\cos^2\a)^2\m_\mathrm{I}^2+16\vec p^2\cos^2\a}\;.
\end{equation}
The lower of the two branches is gapless, $E_-(\vec0)=0$. This is the NG boson of the spontaneously broken isospin symmetry. Further discussion of meson condensates in QCD can be found for instance in the pedagogical review~\cite{Mannarelli2019a}.
\end{illustration}

QCD alone does not encompass all of particle physics. Hadrons can also interact via the weak and electromagnetic forces. ChPT makes it easy to couple pseudoscalar mesons to the electroweak sector of the Standard Model (see, for instance, Sect.~20.2 of~\cite{Peskin1995} for an overview). Indeed, we can imitate the coupling of quarks to the electroweak gauge bosons by setting
\begin{equation}
A^\mathrm{L}_\m=\frac g2\skal\pau A_\m+\frac{g'}6B_\m\;,\qquad
A^\mathrm{R}_\m=g'QB_\m\;.
\end{equation}
Here $\vec A_\m$ is a triplet of potentials of the weak isospin gauge group, $\gr{SU}(2)_\mathrm{I}$. Similarly, $B_\m$ is the potential of the hypercharge gauge group, $\gr{U}(1)_\mathrm{Y}$. The corresponding gauge couplings are $g,g'$. Finally, $Q$ is the matrix of electric charges of the $u$ and $d$ quarks,
\begin{equation}
Q=\begin{pmatrix}
2/3 & 0\\
0 & -1/3
\end{pmatrix}=\frac16\un+\frac12\pau_3\;.
\label{Qmatrix}
\end{equation}
A complete set of LO electroweak interactions of pions is then obtained by inserting the above definitions into~\eqref{ChPTLO3}.

\begin{illustration}%
A glance at~\eqref{ChPTLO3} shows that adding the electroweak gauge sector leads to nontrivial effects even in the ground state of ChPT, $\vev\U=\un$. Namely, the last three terms in~\eqref{ChPTLO3} generate a mass term for the electroweak gauge bosons,
\begin{align}
\frac{f_\pi^2}4\tr\bigl(A^\mathrm{L}_\m A^{\mathrm{L}\m}+A^\mathrm{R}_\m A^{\mathrm{R}\m}-2A^\mathrm{L}_\m A^{\mathrm{R}\m}\bigr)&=\frac{f_\pi^2}4\tr\bigl[(A^\mathrm{L}_\m-A^\mathrm{R}_\m)(A^{\mathrm{L}\m}-A^{\mathrm{R}\m})\bigr]\\
\notag
&=\frac{f_\pi^2}8\bigl[(gA^1_\m)^2+(gA^2_\m)^2+(gA^3_\m-g'B_\m)^2\bigr]\\
\notag
&=\frac{f_\pi^2}8\bigl[2g^2W_\m^+W^{-\m}+(g^2+g'^2)Z_\m Z^\m\bigr]\;.
\end{align}
In the last step, I introduced the charged gauge boson fields by $\smash{W^\pm_\m\equiv(A^1_\m\pm\I A^2_\m)/\sqrt{2}}$, and the neutral weak gauge boson via $\smash{Z_\m\equiv A^3_\m\cos\t_\mathrm{W}-B_\m\sin\t_\mathrm{W}}$. Finally, the Weinberg angle $\t_\mathrm{W}$ is related to the gauge couplings by $\smash{\cos\t_\mathrm{W}=g/\sqrt{g^2+g'^2}}$. Thus, spontaneous breaking of chiral symmetry in QCD leads to the following contributions to the masses of the electroweak gauge bosons,
\begin{equation}
m_W^2=\frac14f_\pi^2g^2\;,\qquad
m_Z^2=\frac14f_\pi^2(g^2+g'^2)\;.
\end{equation}
Given the characteristic scale of QCD, encoded in the value of $f_\pi$ (fixed precisely below), these contributions are tiny. However, the idea that the gauge boson masses might be generated by a strong dynamics that spontaneously breaks chiral symmetry is intriguing. It lies behind the ``technicolor'' scenario of dynamical electroweak symmetry breaking. In this scenario, the Higgs boson is not elementary, but rather a composite bound state of constituent ``techniquarks.'' The value of $f_\pi$ is expected to be of the order of the electroweak scale, that is a few hundreds of $\mathrm{GeV}$. See~\cite{Andersen2011} for a pedagogical introduction to technicolor models.
\end{illustration}

The physical value of $f_\pi$ can be fixed by likewise utilizing the coupling of pions to the electroweak sector of the Standard Model. All we need is a single observable that does not depend on any other as yet unknown parameter. A suitable candidate is the leptonic decay of the charged pion.

\begin{figure}[t]
\sidecaption[t]
\includegraphics[width=2.0in]{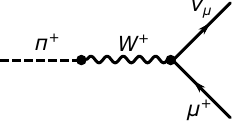}
\caption{Feynman diagram for the leading (tree-level) contribution to charged pion decay. The $\pi^+$--$W^+$ coupling is provided by ChPT whereas the interaction vertex between $W^+$ and the charged lepton current follows from the Standard Model of electroweak interactions}
\label{fig:pidecay}
\end{figure}

\begin{illustration}%
\label{ex:pidecay}%
The charged pion $\pi^+$ decays with the probability of $99.988\%$ into an antimuon $\mu^+$ and a muon neutrino $\n_\m$. The leading perturbative contribution to the amplitude for this decay is shown in Fig.~\ref{fig:pidecay}. The conversion of an on-shell pion into a virtual $W$-boson is described by our Lagrangian~\eqref{ChPTLO2}. Indeed, by expanding it to the first order in both $W^\pm_\m$ and $\smash{\pi^\pm\equiv(\pi^1\pm\I\pi^2)/\sqrt{2}}$, we find the bilinear term $\smash{-(f_\pi g/2)(W^{+\m}\de_\m\pi^-+W^{-\m}\de_\m\pi^+)}$. The subsequent decay of the virtual $W$-boson into a lepton pair is governed by the ``charged-current'' interaction of the Standard Model, specifically the operator $\smash{(g/\sqrt2)\bigl(W^+_\m\adj\m_\mathrm{L}\g^\m\n_{\m\mathrm{L}}+W^-_\m\adj\n_{\m\mathrm{L}}\g^\m\m_\mathrm{L}\bigr)}$. I used an obvious notation for the spinor fields representing the leptons. Putting all the pieces together, the invariant amplitude for the $\smash{\pi^+\to\mu^++\n_\m}$ process becomes
\begin{equation}
\Aa_{\pi^+\to\m^++\n_\m}=\I f_\pi G_\mathrm{F}\adj u(p)\slashed{k}(1-\g_5)v(q)\;.
\end{equation}
Here $k^\m,p^\m,q^\m$ are respectively the four-momenta of the pion, muon neutrino and anti\-muon. Also, $u(p)$ and $v(q)$ are the corresponding Dirac spinors; polarization indices are suppressed for clarity. Finally, $G_\mathrm{F}$ is the Fermi coupling constant,
\begin{equation}
G_\mathrm{F}=\frac{g^2}{4\sqrt2 m_W^2}\approx1.166\times10^{-5}\,\mathrm{GeV}^{-2}\;.
\end{equation}
Upon squaring the amplitude and summing over polarizations of the particles in the final state, the integrated decay rate in the rest frame of the pion is found to be
\begin{equation}
\Gamma_{\pi^+\to\mu^++\n_\m}=\frac{f_\pi^2G_\mathrm{F}^2}{4\pi}\frac{m_\m^2(m_\pi^2-m_\m^2)^2}{m_\pi^3}\;.
\label{pidecay}
\end{equation}
I have treated the neutrino as a massless particle. The masses of the pion and antimuon are, respectively, $m_\pi\approx139.6\,\mathrm{MeV}$ and $m_\m\approx105.7\,\mathrm{MeV}$. Finally, we need an input on the lifetime of the charged pion, $\tau\approx2.60\times10^{-8}\,\mathrm{s}$. This converts to the total decay rate of $\Gamma\approx2.53\times10^{-8}\,\mathrm{eV}$. At the end of the day, we get an estimate for the \emph{pion decay constant},
\begin{equation}
f_\pi\approx91\,\mathrm{MeV}\;.
\end{equation}

The result~\eqref{pidecay} tells us more than merely a good estimate for the total decay rate of the charged pion. Namely, leptons from different families have identical weak interactions. Upon replacing $m_\m$ with the electron mass $m_e$, \eqref{pidecay} therefore also gives us a decay rate for the process $\pi^+\to e^++\n_e$. This is more conveniently expressed in terms of the branching ratio,
\begin{equation}
R_{\pi^+\to e^++\n_e}\approx\frac{\Gamma_{\pi^+\to e^++\n_e}}{\Gamma_{\pi^+\to\mu^++\n_\m}}=\frac{m_e^2}{m_\m^2}\frac{(m_\pi^2-m_e^2)^2}{(m_\pi^2-m_\m^2)^2}\approx1.28\times10^{-4}\;.
\end{equation}
This is very close to the experimental value, which is about $1.23\times10^{-4}$~\cite{Workman2022}. The suppression of the electron decay channel compared to the muon one is purely kinematical. By angular momentum conservation, one of the leptons in the final state must be left-handed and one right-handed. Yet, the $W$-boson only couples to left-handed fermion fields. The combination of these two effects requires a helicity flip and is responsible for the proportionality of~\eqref{pidecay} to the lepton mass squared.
\end{illustration}


\subsection{Effects of the Chiral Anomaly}
\label{subsec:ChPTanomaly}

So far I have tacitly assumed, following Sect.~\ref{sec:effLaggauged}, that the effective action of ChPT is gauge-invariant in presence of the background fields. This allowed us to construct strictly gauge-invariant Lagrangians at LO~\eqref{ChPTLO} and NLO~\eqref{ChPTNLO} of the derivative expansion of ChPT. Are there any contributions to the ChPT Lagrangian that are merely quasi-invariant? A detailed derivation of such contributions and their coupling to background gauge fields would require a differential-geometric approach akin to that of Sect.~\ref{sec:effLagstructure}. The problem of finding all such \emph{Wess--Zumino} (WZ), or \emph{Wess--Zumino--Witten}, \emph{terms} was studied thoroughly in the 1980s and 1990s. For a discussion close in spirit to this book, I refer the reader to~\cite{DHoker1994a,DHoker1995b}. A pedagogical account of the method including explicit expressions for quasi-invariant Lagrangians for a broad class of coset spaces can be found in~\cite{Brauner2019a}.

Here I will resort to a trick, which gives the right answer in case of $\nf=2$ quark flavors. Suppose we were able to construct a current $J^\m$, conserved off-shell, that is without imposing the \emph{equation of motion} (EoM) for the NG fields in our EFT. If in addition the EFT includes an Abelian gauge field $A_\m$, then the operator $A_\m J^\m$ is quasi-invariant and can be added to the Lagrangian density. It remains to guess what $J^\m$ and $A_\m$ might be within ChPT. The current is the tricky bit. For the moment, I will simply write it down, a partial a posteriori justification will be offered below;
\begin{equation}
\begin{split}
J^\m_\mathrm{GW}=\l\ve^{\m\n\a\b}\tr\biggl\{&(\U D_\n\he\U)(\U D_\a\he\U)(\U D_\b\he\U)\\
&-\frac{3\I}2\bigl[(D_\n\U\he\U)F^\mathrm{L}_{\a\b}-(D_\n\he\U\U)F^\mathrm{R}_{\a\b}\bigr]
\biggr\}\;.
\end{split}
\label{GWcurrent}
\end{equation}
This is the  \emph{Goldstone--Wilczek} (GW) \emph{current}; the overall factor $\l$ is in principle arbitrary. In the absence of the background fields, the GW current would be manifestly conserved thanks to the antisymmetry of the LC tensor and cyclicity of trace. With the background in place, it is a nontrivial but straightforward exercise to show that\footnote{A reader willing to check this should note that the GW current is manifestly gauge-invariant. It is thus sufficient, and advantageous, to compute its gauge-covariant derivative. This maintains gauge invariance at every step and simplifies the calculation thanks to the Bianchi identity for $F^\mathrm{L,R}_{\m\n}$.}
\begin{equation}
\de_\m J^\m_\mathrm{GW}=\frac{3\l}4\ve^{\m\n\a\b}\tr\bigl(-F^\mathrm{L}_{\m\n}F^\mathrm{L}_{\a\b}+F^\mathrm{R}_{\m\n}F^\mathrm{R}_{\a\b}\bigr)\;.
\label{GWanomaly}
\end{equation}
The current is not conserved as promised, except for backgrounds that are purely vector-like, $A^\mathrm{L}_\m=A^\mathrm{R}_\m$. Luckily, this is not a problem. In fact, the nonconservation of the GW current turns out to be exactly what is needed to implement correctly the microscopic physics of QCD within ChPT.

To that end, recall that the flavor symmetry of QCD has a single $\gr{U}(1)$ factor, namely the baryon number $\gr{U}(1)_\mathrm{B}$. This can also be coupled to a background gauge field, $A^\mathrm{B}_\m$, even if such a field may not have an obvious experimental realization. The presence of a coupling $\smash{A^\mathrm{B}_\m J^\m_\mathrm{GW}}$ in the ChPT action implies that it is possible to create baryon number solely out of meson fields. This intriguing possibility was first proposed by Skyrme in the 1960s. In presence of the chiral background fields $A^\mathrm{L,R}_\m$, baryon number is not conserved due to the \emph{chiral anomaly}. An explicit calculation (see Sect.~22.3 of~\cite{Weinberg1996a}) shows that this anomaly is reproduced at the ChPT level by~\eqref{GWanomaly} if we set $\l=-1/(24\pi^2)$. This brings us to the final expression for the WZ term in the ChPT Lagrangian for $\nf=2$ quark flavors,
\begin{equation}
\begin{split}
\La_\mathrm{WZ}^{(4)}=-\frac1{24\pi^2}\ve^{\m\n\a\b}A^\mathrm{B}_\m\tr\biggl\{&(\U D_\n\he\U)(\U D_\a\he\U)(\U D_\b\he\U)\\
&-\frac{3\I}2\bigl[(D_\n\U\he\U)F^\mathrm{L}_{\a\b}-(D_\n\he\U\U)F^\mathrm{R}_{\a\b}\bigr]
\biggr\}\;.
\end{split}
\label{ChPTWZ2}
\end{equation}
The superscript ${}^{(4)}$ indicates that the WZ term enters at the NLO of the derivative expansion of ChPT. Note also that it does not come with an arbitrary coupling. The normalization of the WZ term is fixed by anomaly matching and does not receive any radiative corrections.

\begin{watchout}%
The WZ term~\eqref{ChPTWZ2} is manifestly invariant under background gauge transformations from the $\gr{SU}(2)_\mathrm{L}\times\gr{SU}(2)_\mathrm{R}$ chiral group. On the contrary, under a background $\gr{U}(1)_\mathrm{B}$ transformation, $A^\mathrm{B}_\m\to A^\mathrm{B}_\m+\de_\m\eps^\mathrm{B}$, the corresponding WZ action changes by 
\begin{equation}
\udelta S_\mathrm{WZ}=-\int\D^4x\,\eps^\mathrm{B}(x)\de_\m J^\m_\mathrm{GW}(x)\;,
\label{mixedtHooft}
\end{equation}
with the divergence of $J^\m_\mathrm{GW}$ given by~\eqref{GWanomaly}. Watch the interplay of two symmetries: the variation of the action under $\gr{U}(1)_\mathrm{B}$ is proportional to the background fields for $\gr{SU}(2)_\mathrm{L}\times\gr{SU}(2)_\mathrm{R}$. What we have here is an example of a \emph{mixed 't Hooft anomaly}. This should be contrasted to the naive $\gr{U}(1)_\mathrm{A}$ axial symmetry of QCD. The divergence of the axial current receives a  contribution from the dynamical gluon fields, giving an example of an \emph{Adler--Bell--Jackiw anomaly}. This kind of anomaly fundamentally invalidates a would-be classical symmetry of a quantum system. On the other hand, a symmetry exhibiting a 't Hooft anomaly still implies exact relations (Ward identities) for the generating functional of the theory. This makes 't Hooft anomalies a powerful tool for constraining low-energy EFTs, as I have illustrated here.
\end{watchout}

The construction of the WZ term~\eqref{ChPTWZ2} is not a mere academic exercise, as one might suspect from the presence of the ``baryon number gauge field.'' The term has measurable consequences for the electromagnetic interactions of pions. To see why, recall~\eqref{Qmatrix}, which shows that electric charge does not belong to the chiral Lie algebra $\lie{su}(2)_\mathrm{L}\times\lie{su}(2)_\mathrm{R}$ due to not being traceless. Interactions of pions with an external electromagnetic field (with all other background fields switched off) can then be generated by setting
\begin{equation}
A^\mathrm{L}_\m=A^\mathrm{R}_\m=\frac e2\pau_3A^Q_\m\;,\qquad
A^\mathrm{B}_\m=\frac e2A^Q_\m\;.
\end{equation}
Here $A^Q_\m$ is the electromagnetic gauge potential and $e$ the electromagnetic coupling. The effects of interaction with the electromagnetic field via the WZ term are most striking in case of the neutral pion.

\begin{illustration}%
Let us keep only the electromagnetic background field and the neutral pion $\pi^0$. The charged pions are discarded by using the replacement $\U\to\exp(\I\pi^0\pau_3/f_\pi)$. Upon simple integration by parts, the whole WZ term~\eqref{ChPTWZ2} then boils down to
\begin{equation}
\La_\mathrm{WZ}^{(4)}\to-\frac{e^2}{32\pi^2f_\pi}\pi^0\ve^{\m\n\a\b}F^Q_{\m\n}F^Q_{\a\b}\;.
\end{equation}
This operator governs the electromagnetic decay of $\pi^0$ into a pair of photons. Denoting the four-momentum of the pion as $k^\m$ and those of the photons as $p^\m,q^\m$, the invariant amplitude for the decay turns out to be
\begin{equation}
\Aa_{\pi^0\to\g+\g}=-\frac{e^2}{4\pi^2f_\pi}\ve^{\m\n\a\b}p_\m\eps^*_\n(p)q_\a\eps^*_\b(q)\;.
\end{equation}
Here $\eps^*_\m(p)$ and $\eps^*_\m(q)$ are the polarization vectors of the photons in the final state. It remains to take the square and sum over polarizations of the photons. The final result for the decay rate in the rest frame of the pion is
\begin{equation}
\Gamma_{\pi^0\to\g+\g}=\frac{\a^2m_\pi^3}{64\pi^3f_\pi^2}\;,
\end{equation}
where $\a\equiv e^2/(4\pi)$ is the fine structure constant. Using the numerical input $m_\pi\approx135.0\,\mathrm{MeV}$ and $\a\approx7.297\times10^{-3}$ along with the value for $f_\pi$ found in \refex{ex:pidecay}, our final result is $\smash{\Gamma_{\pi^0\to\g+\g}\approx8.0\,\mathrm{eV}}$. This is less than $3\%$ off the experimental value of~$7.81\,\mathrm{eV}$~\cite{Workman2022}.
\end{illustration}

The two-flavor WZ term~\eqref{ChPTWZ2} has remarkable physical consequences, yet vanishes by construction in the absence of external fields. A new twist in the story comes for $\nf=3$ quark flavors. Here another WZ term appears, which remains nonzero even in the absence of background fields. This governs scattering processes with an odd number of mesons, such as $K^++K^-\to\pi^++\pi^-+\pi^0$, which would otherwise be forbidden in ChPT. The mathematical origin of this WZ term parallels that of the quasi-invariant Lagrangians with one time derivative, analyzed in Sect.~\ref{sec:effLagstructure}. The Lagrangian density of the WZ term can be mapped to a 4-form $\o_\mathrm{WZ}$ such that the 5-form $\D\o_\mathrm{WZ}$ is chirally invariant and closed but not exact. Such 5-forms are classified by the fifth de Rham cohomology group of the coset space. The coset space $[\gr{SU}(3)_\mathrm{L}\times\gr{SU}(3)_\mathrm{R}]/\gr{SU}(3)_\mathrm{V}$ has a unique generator of degree-5 cohomology, 
\begin{equation}
\D\o_\mathrm{WZ}\propto\tr\bigl[(\U^{-1}\D\U)\w(\U^{-1}\D\U)\w(\U^{-1}\D\U)\w(\U^{-1}\D\U)\w(\U^{-1}\D\U)\bigr]\;.
\end{equation}
The overall normalization is again fixed by matching to the flavor anomalies of QCD. More details about the geometric nature of this WZ term and its coupling to background gauge fields can be found in the original work of Witten~\cite{Witten1983a}.

The two-flavor coset space $[\gr{SU}(2)_\mathrm{L}\times\gr{SU}(2)_\mathrm{R}]/\gr{SU}(2)_\mathrm{V}$, being three-dimensional, obviously has vanishing fifth cohomology group. However, it does have a nontrivial third de Rham cohomology with a single generator,
\begin{equation}
\o_\mathrm{GW}\propto\tr\bigl[(\U^{-1}\D\U)\w(\U^{-1}\D\U)\w(\U^{-1}\D\U)\bigr]\;.
\end{equation}
When pulled back to the four-dimensional Minkowski spacetime, $\o_\mathrm{GW}$ is just the Hodge dual of the GW current~\eqref{GWcurrent} in absence of external fields. This explains why such an identically conserved current exists in the first place. Moreover, the integral of $\smash{J^0_\mathrm{GW}}$ over $\R^3$ defines a conserved charge that is a topological invariant. Upon a suitable normalization, it coincides with the Brouwer degree of the pion fields viewed as a map $\smash{S^3\to S^3}$; cf.~\refex{ex:cohomologyinvariants}. A \emph{skyrmion} is a configuration of pion fields for which the topological charge is nonvanishing. Thanks to the coupling to $A^\mathrm{B}_\m$ in~\eqref{ChPTWZ2}, the topological charge has the interpretation as baryon number. This provides a mathematical foundation for the Skyrme model of baryons.


\section{Spin Waves in Ferro- and Antiferromagnets}
\label{sec:spinwaves}

I have already used ferromagnets repeatedly to illustrate various features of SSB, including the peculiarities of the spectrum of NG bosons in nonrelativistic systems. In order to make the present section self-contained, I will however start with a concise summary of the basic facts.

Ferro- and antiferromagnets are phases of matter that exhibit spin order. Although such order may also be induced in relativistic matter, I will have implicitly in mind its realization in ordinary crystalline solids. The advantage of this restriction is that in the nonrelativistic limit, spin can be treated as an internal degree of freedom. One can then base the construction of EFT for (anti)ferromagnets on spontaneous breakdown of the internal $G\simeq\gr{SU}(2)$ spin symmetry. In both types of systems, the unbroken subgroup is $H\simeq\gr{U}(1)$, corresponding to spin rotations about the axis of spin alignment. The coset space is therefore $G/H\simeq\gr{SU}(2)/\gr{U}(1)\simeq S^2$. From the point of view of low-energy EFT, the only difference between ferro- and antiferromagnets is a nonzero VEV of spin in the former. This is directly reflected by the spectrum of NG bosons: spin waves, or \emph{magnons}. In ferromagnets, there is a single type-B magnon, whereas antiferromagnets feature two type-A magnons.

The generators of $G\simeq\gr{SU}(2)$ can be taken as $\pau_A/2$. Without loss of generality, we may choose the spin axes so that $H\simeq\gr{U}(1)$ is generated by $\pau_3/2$. The coset space $\gr{SU}(2)/\gr{U}(1)$ is then symmetric thanks to the inner automorphism $\Raut(g)=R^{-1}gR$ with $g\in\gr{SU}(2)$ and $R=\I\pau_3$. This makes it possible to map the coset representative $U(\pi)$ on a unit-vector variable $\vec n(\pi)\in S^2$ via
\begin{equation}
\skal\pau n(\pi)\equiv N(\pi)=U(\pi)^2\pau_3=U(\pi)\pau_3 U(\pi)^{-1}\;.
\label{Nndef}
\end{equation}
The matrix field $N(\pi)$ transforms linearly in the adjoint representation of $G$. As a consequence, the $G$-covariant derivative~\eqref{DSigma} boils down to
\begin{equation}
D\vec n(\pi)=\D\vec n(\pi)+\vec A\times\vec n(\pi)\;,
\end{equation}
where $\vec A_\m$ is a triplet of background gauge fields of $\gr{SU}(2)$. The covariant derivative $D_\m\vec n(\pi)$ is the basic building block for construction of EFT for (anti)ferromagnets.

In addition to the internal spin symmetry, I will assume invariance under continuous spacetime translations and continuous spatial rotations. This is of course just a crude idealization of real materials, where such ideal symmetry may be explicitly broken by a variety of perturbations. These include especially the anisotropy induced by the underlying crystal lattice, and the effects of spin--orbit coupling. I will nevertheless initially assume the ideal, unperturbed limit. An outline of some phenomenological consequences of explicit symmetry breaking is deferred to Sect.~\ref{subsec:spinwavesperturbations}. 


\subsection{Power Counting and Effective Lagrangian}
\label{subsec:spinwavesferro}

The general philosophy of the derivative expansion of the EFT for (anti)ferromagnets copies closely that for ChPT, detailed in Sect.~\ref{subsec:ChPTpowercounting}. However, the two cases differ substantially due to the qualitatively different spectra of (anti)ferromagnetic magnons. I will start with the more nontrivial, genuinely nonrelativistic case of ferromagnets.


\subsubsection{Ferromagnets}
\label{subsubsec:ferro}

The energy of ferromagnetic magnons is quadratic in momentum, at least in the long-wavelength limit. In order to assign to a given Feynman diagram a well-defined degree, we therefore count momentum as order one and energy as order two. The Schr\"odinger-like propagator of the magnon then has overall degree $-2$. In close parallel with~\eqref{graphdegreeaux}, the degree of a Feynman diagram $\Gamma$ becomes
\begin{equation}
\deg\Gamma=(D+1)L-2I+\sum_{s,t}(s+2t)V_{s,t}\;.
\end{equation}
Here $V_{s,t}$ denotes the number of vertices from $\La_\mathrm{eff}^{(s,t)}$, the part of effective Lagrangian with $s$ spatial and $t$ temporal derivatives. As before, the number of propagators $I$ can be eliminated via $I=L+\sum_{s,t}V_{s,t}-1$, which leads to the final result
\begin{equation}
\deg\Gamma=2+(D-1)L+\sum_{s,t}(s+2t-2)V_{s,t}\;.
\label{graphdegreeB}
\end{equation}

The LO of the derivative expansion, $\deg\Gamma=2$, corresponds to tree-level diagrams ($L=0$) with all vertices satisfying $s+2t=2$. Thus, the LO effective Lagrangian consists of $\smash{\La_\mathrm{eff}^{(0,1)}}$ and $\smash{\La_\mathrm{eff}^{(2,0)}}$. Before we construct these, let us briefly consider higher orders of the derivative expansion. First, note that unlike in ChPT, here we have a well-defined derivative expansion even in $D=2$ spacetime dimensions. Indeed, ferromagnetic order that can be described by a derivatively coupled low-energy EFT exists even in one-dimensional spin chains. This is special to type-B NG bosons, as I will show in Sect.~\ref{sec:nogo}.

What exactly constitutes the NLO of the derivative expansion depends on the number of dimensions. For $D=2$ (ferromagnetic chains), the NLO corresponds to $\deg\Gamma=3$. It collects contributions from one-loop diagrams with all vertices from the LO Lagrangian, and from tree-level diagrams with one vertex from $\smash{\La_\mathrm{eff}^{(3,0)}}$, if any such operators exist. (They may be forbidden by parity.) In $D=4$ dimensions (bulk ferromagnets), on the other hand, there are two options. In case $\smash{\La_\mathrm{eff}^{(3,0)}}$ is allowed, then tree-level diagrams with one such vertex constitute the sole contribution with $\deg\Gamma=3$. Otherwise, the NLO corresponds to $\deg\Gamma=4$, and consists of tree-level diagrams with one vertex from $\smash{\La_\mathrm{eff}^{(4,0)}}$, $\smash{\La_\mathrm{eff}^{(2,1)}}$, or $\smash{\La_\mathrm{eff}^{(0,2)}}$. Up to and including NLO, there are no quantum corrections; loops only start contributing at $\deg\Gamma=5$. Perhaps the most interesting is the case of $D=3$ (thin ferromagnetic films or layers). Barring the possible existence of $\smash{\La_\mathrm{eff}^{(3,0)}}$, the NLO here is $\deg\Gamma=4$. It includes both one-loop diagrams with all vertices from the LO Lagrangian and tree-level diagrams with one vertex from the NLO Lagrangian.

Clearly, the setup of the derivative expansion depends very sensitively on the specific choice of material (which affects discrete symmetries such as parity) and sample (which controls the dimension $D$). I will therefore limit the discussion to the effective Lagrangian at LO. A detailed classification of possible operators up to order four in derivatives, including the effects of the discrete crystal and time-reversal symmetries, can be found in~\cite{Roman1999a}.

The $\smash{\La_\mathrm{eff}^{(2,0)}}$ part of the LO Lagrangian is trivial. According to the general analysis in Sect.~\ref{sec:effLaggauged}, we expect it to be of the type $\k_{ab}\d^{rs}\MC^a_r\MC^b_s$, where the coupling $\k_{ab}$ must be invariant under the adjoint action of $H\simeq\gr{U}(1)$. The broken part of the gauged MC form~\eqref{MCBsym} now takes the specific form
\begin{equation}
\MCB(\pi,A)=-\frac\I2U(\pi)^{-1}DN(\pi)U(\pi)\pau_3=\frac\I2\pau_3U(\pi)^{-1}DN(\pi)U(\pi)\;.
\label{MCferro}
\end{equation}
The $H$-invariant part of the symmetric tensor product of $\MCB$ with itself is projected out by taking the trace. This leads immediately to the Lagrangian
\begin{equation}
\La_\mathrm{eff}^{(2,0)}=-\frac{\vr_\mathrm{s}}4\tr[\vec DN(\pi)\cdot\vec DN(\pi)]=-\frac{\vr_\mathrm{s}}2\d^{rs}D_r\vec n(\pi)\cdot D_s\vec n(\pi)\;.
\label{ferro20}
\end{equation}
The parameter $\vr_\mathrm{s}$ is called \emph{spin stiffness} and controls the gradient energy arising from ``bending'' the uniform ground state magnetization.

\begin{watchout}%
In $d=2$ spatial dimensions, it is also possible to construct an invariant operator using antisymmetric tensor product, $\l_{ab}\ve^{rs}\MC^a_r\MC^b_s$. This leads to the operator $\ve^{rs}\vec n\cdot(D_r\vec n\times D_s\vec n)$. In the absence of background fields, the latter is a pure surface term; its integral over $\R^2$ gives, up to normalization, the Brouwer degree of the map $\vec n:\R^2\to S^2$. That is however no longer the case when the EFT is coupled to background gauge fields. A quick calculation shows that up to surface terms, $\ve^{rs}\vec n\cdot(D_r\vec n\times D_s\vec n)\simeq\ve^{rs}\skal nF_{rs}$, where $\vec F_{\m\n}=\de_\m\vec A_\n-\de_\n\vec A_\m+\vec A_\m\times\vec A_\n$ is the field strength of the background. When added to the Lagrangian, this operator will modify the EoM for spin waves; cf.~the discussion of EoM in Sect.~\ref{sec:effLagEoMspectrum}. In the following, I will nevertheless disregard this contribution to the EFT. First, it only exists in $d=2$ dimensions and moreover violates time reversal, under which $\vec n(\vec x,t)\to-\vec n(\vec x,-t)$. Second, it requires a specific, nontrivial background to be nonzero, and thus does not affect the propagation of free magnons.
\end{watchout}

Let us now focus on the $\smash{\La_\mathrm{eff}^{(0,1)}}$ part of the LO Lagrangian. According to the general discussion in Sect.~\ref{sec:effLaggauged}, this reads
\begin{equation}
\La_\mathrm{eff}^{(0,1)}=-M\mc^3_a(\pi)\dot\pi^a+M\n^3_A(\pi)A^A_0\;,
\label{Lagferro}
\end{equation}
where $M$ is the density of spin (magnetization) in the ferromagnetic ground state. The second term in~\eqref{Lagferro} is seen to equal $M\vec A_0\cdot\vec n(\pi)$. To evaluate the first term, we project out the third component of the MC form by taking trace with $\pau_3$. Then we apply the exponential parameterization, $U(\pi)=\exp(\I\pi^a\pau_a/2)$, and use~\eqref{AdA},
\begin{align}
\notag
-M\mc^3_a(\pi)\dot\pi^a&=\I M\tr\bigl[\pau_3U(\pi)^{-1}\de_0U(\pi)\bigr]=-\frac M2\dot\pi^a\int_0^1\D\tau\tr\bigl[\pau_3U(\tau\pi)^{-1}\pau_aU(\tau\pi)\bigr]\\
&=-M\dot\pi^a\int_0^1\D\tau\,n_a(\tau\pi)\simeq M\pi^a\int_0^1\D\tau\,\dot n_a(\tau\pi)\;,
\end{align}
where $\simeq$ indicates equality up to a total derivative. The index $a$ runs over $1,2$, we can however formally extend $\pi^a$ to a three-component vector $\vec\pi$ and write $\pi^a\dot n_a=\vec\pi\cdot\dot{\vec n}=(\vec n\times\vec\pi)\cdot(\vec n\times\dot{\vec n})$. Using the definition~\eqref{Nndef} to take the derivative $\de_\tau N(\tau\pi)$, we find that $\vec n(\tau\pi)\times\vec\pi=\de_\tau\vec n(\tau\pi)$. Putting all the pieces together, we then arrive at the final expression for the LO effective Lagrangian for ferromagnets,
\begin{equation}
\begin{split}
\La_\mathrm{eff}^\mathrm{LO}={}&M\int_0^1\D\tau\,\de_\tau\vec n(\tau\pi)\cdot[\vec n(\tau\pi)\times\dot{\vec n}(\tau\pi)]+M\vec A_0\cdot\vec n(\pi)\\
&-\frac{\vr_\mathrm{s}}2\d^{rs}D_r\vec n(\pi)\cdot D_s\vec n(\pi)\;.
\end{split}
\label{Lagferrofinal}
\end{equation}

The second term in~\eqref{Lagferrofinal} is the usual Zeeman coupling of spin to an external magnetic field, whereas the third term represents the gradient energy of the spin configuration. The first term, however, deserves a comment, since the presence of integration over the parameter $\tau$ makes it look nonlocal. To get rid of the integral, let us take a step back. Consider a one-parameter family of fields, $\vec n(\tau,\pi)$, $\tau\in[0,1]$, such that $\vec n(0,\pi)=(0,0,1)\equiv\vec n_0$ (ground state) and $\vec n(1,\pi)=\vec n(\pi)$. The explicit choice of interpolation used in~\eqref{Lagferrofinal} corresponds to $\vec n(\tau,\pi)=\vec n(\tau\pi)$. It is easy to check that upon a smooth deformation of the field, $\udelta\vec n(\tau,\pi)$, the variation of the action only depends on $\udelta\vec n(1,\pi)=\udelta\vec n(\pi)$. The concrete choice of interpolation between the $\tau=0$ and $\tau=1$ limits therefore does not matter. Assuming for simplicity that $n^3(\vec x,t)$ as a function on the spacetime is non-negative everywhere, we can change the interpolation to
\begin{equation}
\vec n(\tau,\pi)=\bigl(\tau n^1(\pi),\tau n^2(\pi),\sqrt{1-\tau^2[(n^1(\pi))^2+(n^2(\pi))^2]}\bigr)\;,
\end{equation}
which makes it possible to carry out the integral over $\tau$,
\begin{equation}
\La_\mathrm{eff}^\mathrm{LO}=-M\frac{\ve_{ab}n^a(\pi)\dot n^b(\pi)}{1+n^3(\pi)}+M\vec A_0\cdot\vec n(\pi)-\frac{\vr_\mathrm{s}}2\d^{rs}D_r\vec n(\pi)\cdot D_s\vec n(\pi)\;.
\label{Lagferrofinal2}
\end{equation}
This is as far as we can get. The Lagrangian is local and manifestly invariant under $H\simeq\gr{U}(1)$. Moreover, it only depends on the NG fields $\pi^a$ through $\vec n(\pi)$, and thus does not rely on the exponential parameterization of $U(\pi)$, originally used to derive~\eqref{Lagferrofinal}.


\subsubsection{Antiferromagnets}
\label{subsubsec:antiferro}

As far as the construction of the effective Lagrangian is concerned, antiferromagnets are much simpler than ferromagnets. The spectrum consists of two type-A NG bosons whose energy is, in the long-wavelength limit, linear in momentum. For the sake of power counting, we therefore have to treat spatial and temporal derivatives on equal footing. The resulting expression for the degree of a given Feynman diagram is a trivial generalization of~\eqref{graphdegree} we found in ChPT,
\begin{equation}
\deg\Gamma=2+(D-2)L+\sum_{s,t}(s+t-2)V_{s,t}\;.
\end{equation}
The only difference to ChPT is that spatial and temporal derivatives may enter the effective Lagrangian independently.

Owing to the same symmetry-breaking pattern, the building blocks for constructing the EFT are the same for ferro- and antiferromagnets. The main difference is that the part with a single time derivative, $\smash{\La_\mathrm{eff}^{(0,1)}}$, is missing in the latter; antiferromagnets have zero net magnetization. The LO Lagrangian then consists of $\smash{\La_\mathrm{eff}^{(2,0)}}$ and $\smash{\La_\mathrm{eff}^{(0,2)}}$, and we can write it down at once,
\begin{equation}
\La_\mathrm{eff}^\mathrm{LO}=\frac{\vr_\mathrm{s}}{2v^2}\bigl[D_0\vec n(\pi)\cdot D_0\vec n(\pi)-v^2\d^{rs}D_r\vec n(\pi)\cdot D_s\vec n(\pi)\bigr]\;.
\label{Lagantiferro}
\end{equation}
There are two independent parameters which are easy to relate to physical observables. The spin stiffness $\vr_{\mathrm{s}}$ measures the gradient energy of the order parameter, whereas $v$ turns out to be the phase velocity of antiferromagnetic magnons.

Similarly to ferromagnets, the organization of the derivative expansion beyond LO depends sensitively on $D$ and the presence of discrete symmetries such as parity or time reversal. I will therefore stop the discussion of power counting here, and turn to the consequences of the EFT at LO.


\subsection{Equation of Motion and Magnon Spectrum}
\label{subsec:spinwavesEoM}

We already know the number and type of magnons in both ferro- and antiferromagnets. However, the EFT tells us more, in particular what the corresponding fluctuations of the order parameter look like. To that end, I will drop the background gauge fields and derive the EoM corresponding to the LO effective Lagrangian.

Let us start with ferromagnets. As hinted above, taking a variation of~\eqref{Lagferrofinal} gives a surface term in $\tau$, which allows one to write the variation of the action solely in terms of $\udelta\vec n(\pi)$,
\begin{equation}
\udelta S_\mathrm{eff}^\mathrm{LO}=\int\D^D\!x\,\udelta\vec n\cdot\bigl(M\vec n\times\dot{\vec n}+\vr_\mathrm{s}\vec\nabla^2\vec n\bigr)\;.
\end{equation}
The variation $\udelta\vec n$ is not arbitrary but rather should keep $\vec n$ on the coset space, that is the unit sphere $S^2$. In other words, $\udelta\vec n$ should be a tangent vector to the sphere. The vanishing of $\smash{\udelta S_\mathrm{eff}^\mathrm{LO}}$ therefore requires that
\begin{equation}
M\vec n\times\dot{\vec n}+\vr_\mathrm{s}\vec\nabla^2\vec n=\l\vec n\;;
\end{equation}
$\l$ can be interpreted as a Lagrange multiplier for the constraint $\skal nn=1$. We can get rid of it by taking a cross product with $\vec n$, which gives the \emph{Landau--Lifshitz equation},
\begin{equation}
\dot{\vec n}=\frac{\vr_\mathrm{s}}M\vec n\times\vec\nabla^2\vec n\;.
\label{LLeq}
\end{equation}
Previously, in Sect.~\ref{subsec:symplectic}, I derived this equation using the Hamiltonian (symplectic) formulation of field theory from the postulated Poisson bracket for the spin variable $\vec n(\vec x)$; cf.~\eqref{spinbracket}. The Lagrangian and Hamiltonian descriptions of ferromagnets are of course equivalent. For instance, the symplectic 2-form~\eqref{spinsymp2form} can be recovered by noting that according to~\eqref{Lagferro}, the symplectic potential is $-M\mc^3$. The exterior derivative thereof is easily evaluated using the MC equation~\eqref{dMC}. This shows that the fundamental Poisson bracket for $\vec n(\vec x)$ is already automatically built in the low-energy Lagrangian EFT for ferromagnets.

To solve the Landau--Lifshitz equation~\eqref{LLeq} is a hard problem due to its nonlinearity. What one can do easily is to linearize the equation in small fluctuations around the ground state. Inserting $\vec n=\vec n_0+\udelta\vec n$ and keeping only terms linear in $\udelta\vec n$, we get
\begin{equation}
\udelta\dot{\vec n}=\frac{\vr_\mathrm{s}}M\vec n_0\times\vec\nabla^2\udelta\vec n\;.
\label{dndot}
\end{equation}
We can now look for plane-wave solutions by using the ansatz
\begin{equation}
\udelta\vec n(\vec x,t)=\vec A\E^{-\I Et}\E^{\I\skal px}\;,
\label{magnongplanewave}
\end{equation}
where $\vec A$ is a complex amplitude orthogonal to $\vec n_0=(0,0,1)$. Inserting the ansatz in~\eqref{dndot} shows that the energy and momentum satisfy the dispersion relation
\begin{equation}
E(\vec p)=\frac{\vr_\mathrm{s}}M\vec p^2\;,
\label{ferromagnondisp}
\end{equation}
typical for type-B NG bosons. The amplitude must satisfy the constraint $\I\vec A=\vec n_0\times\vec A$. This is solved by any $\vec A\propto(1,-\I,0)$. Ferromagnetic spin waves are circularly polarized in the plane transverse to the direction of the ground state magnetization, $\vec n_0$, regardless of the direction of momentum $\vec p$. This can be understood as Larmor precession of individual spins around the effective magnetic field generated by the spin-polarized background.

Antiferromagnets can be treated in the same way, without the complications due to the single-time-derivative operator in the Lagrangian. The EoM obtained from~\eqref{Lagantiferro} can be written as
\begin{equation}
\frac{\vr_\mathrm{s}}{v^2}\bigl(\de_0^2\vec n-v^2\vec\nabla^2\vec n\bigr)=\l\vec n\;,
\end{equation}
where $\l$ is a Lagrange multiplier. Upon linearization around the ground state, $\vec n_0$, we find plane-wave solutions of the same general form as in~\eqref{magnongplanewave}. However, the dispersion relation is now $E(\vec p)=v\abs{\vec p}$, typical for type-A NG bosons. The complex amplitude $\vec A$ must be orthogonal to $\vec n_0$. We conclude that antiferromagnetic spin waves are also polarized in the plane transverse to the direction of $\vec n_0$, regardless of the direction of momentum. However, the polarization can be both linear and circular, or in general elliptic. There are therefore two different, independent types of antiferromagnetic magnons, which can be chosen to be linearly polarized.


\subsection{Effects of Symmetry-Breaking Perturbations}
\label{subsec:spinwavesperturbations}

So far, we have assumed exact symmetry under $\gr{SU}(2)$ spin rotations, continuous spacetime translations and spatial rotations. Let us now briefly consider the effects of some phenomenologically important perturbations. These can be classified into two broad groups: perturbations controlled by external fields, and those intrinsic to the given material.

The most natural tunable perturbation that the spins in (anti)ferromagnets can be exposed to is an external magnetic field, $\vec B$. Insofar as its effect on orbital degrees of freedom can be neglected, the magnetic field couples directly to the conserved charge of $G\simeq\gr{SU}(2)$: the total spin. It can therefore be treated as a vector-valued chemical potential. In the low-energy EFT, this is implemented by setting $\vec A_\m(x)=\d_{\m0}\vec B(x)$; the magnetic moment of the spins is absorbed into the definition of $\vec B$. I will now show that treating $\vec B$ as a background gauge field allows us to make some exact statements about magnon spectrum. Importantly, we do not have to introduce any new arbitrary parameters into the Lagrangian.

\begin{illustration}%
According to~\eqref{Lagferrofinal}, the effect of an external magnetic field on ferromagnets is taken into account by adding the Zeeman term, $M\vec B\cdot\vec n(\pi)$, to the Lagrangian. As long as the magnetic field is uniform (which I will from now on assume), the ground state $\vec n_0$ will remain uniform as well. However, its orientation is no longer arbitrary, but rather has to be aligned parallel to $\vec B$. The magnetic field selects a unique stable equilibrium state. The effect on the magnon spectrum is also easy to work out. The left-hand side of the Landau--Lifshitz equation~\eqref{LLeq} has to be modified by replacing $\dot{\vec n}\to D_0\vec n=\dot{\vec n}+\vec B\times\vec n$. Upon linearization, the plane-wave magnon solutions are still found to be circularly polarized in the plane transverse to $\vec n_0\parallel\vec B$. The only effect of the magnetic field is a constant shift of the dispersion relation~\eqref{ferromagnondisp},
\begin{equation}
E(\vec p)=\abs{\vec B}+\frac{\vr_\mathrm{s}}M\vec p^2\;.
\label{ferroinB}
\end{equation}
The reason why the response of ferromagnets to a uniform magnetic field is so simple is that the conserved charge that $\vec B$ couples to remains unbroken. Energy levels can therefore be labeled by the projection of spin into the direction of $\vec B$. The excitation energy of a state with spin $S$ (relative to the ground state) will be shifted by the magnetic field by $-S\abs{\vec B}$. The ferromagnetic ground state is maximally polarized, and a single magnon carries a unit of spin less than the ground state. This explains why the magnon receives a gap equal to $\abs{\vec B}$. This is an exact result valid to all orders of the derivative expansion; the ferromagnetic magnon is an example of a massive NG boson in the sense of Sect.~\ref{subsec:massiveNGbosons}.
\end{illustration}

\begin{illustration}%
The effect of magnetic fields on antiferromagnets is somewhat less trivial. According to~\eqref{Lagantiferro}, their LO Lagrangian (in absence of other perturbations than $\vec B$) becomes
\begin{equation}
\La_\mathrm{eff}^\mathrm{LO}=\frac{\vr_\mathrm{s}}{2v^2}\bigl\{[\de_0\vec n(\pi)+\vec B\times\vec n(\pi)]^2-v^2\d^{rs}\de_r\vec n(\pi)\cdot \de_s\vec n(\pi)\bigr\}\;.
\end{equation}
The corresponding Hamiltonian density is
\begin{equation}
\Ha_\mathrm{eff}^\mathrm{LO}=\frac{\vr_\mathrm{s}}{2v^2}\bigl\{[\de_0\vec n(\pi)]^2-[\vec B\times\vec n(\pi)]^2+v^2\d^{rs}\de_r\vec n(\pi)\cdot \de_s\vec n(\pi)\bigr\}\;.
\end{equation}
The response of the ground state is quite different from ferromagnets: the energy is minimized by any $\vec n_0\perp\vec B$. Hence the $\gr{U}(1)$ group of spin rotations around the direction of $\vec B$ is spontaneously broken. We expect the spectrum to contain one true NG boson, whereas the other magnon should receive a gap from the magnetic field.

To see this explicitly, let us choose $\vec B=(0,0,\abs{\vec B})$ and $\vec n_0=(1,0,0)$. We can use $n^2$ and $n^3$ as two independent fluctuations of the order parameter, and parameterize the latter as
\begin{equation}
\vec n(x)=\bigl(\sqrt{1-[(n^2(x))^2+(n^3(x))^2]},n^2(x),n^3(x)\bigr)\;.
\label{easyplane}
\end{equation}
To the second order in the fluctuations and up to a total time derivative, the Lagrangian then reads
\begin{equation}
\La_\mathrm{eff}^\mathrm{LO}\simeq\frac{\vr_\mathrm{s}}{2v^2}\biggl\{\sum_{i=2,3}\bigl[(\de_0n^i)^2-v^2\vec\nabla n^i\cdot\vec\nabla n^i\bigr]+\vec B^2\bigl[1-(n^3)^2\bigr]\biggr\}+\dotsb\;.
\end{equation}
This makes it clear that $n^2(x)$ remains gapless, as expected. On the other hand, the $n^3(x)$ mode receives a gap, its full dispersion relation being $\smash{E(\vec p)=\sqrt{v^2\vec p^2+\vec B^2}}$. The conclusion that $E(\vec0)=\abs{\vec B}$ is exact. The spectrum of an ideal antiferromagnet in a uniform magnetic field contains one true NG boson and one massive NG boson.
\end{illustration}

As opposed to the effect of external fields, perturbations induced by the underlying crystal lattice are intrinsic to the given material and therefore cannot be ``switched off.'' A prominent position among these is occupied by anisotropy in either spatial or spin structure of the microscopic interactions. I will content myself with the simplest illustrative example of such a crystal anisotropy, whereby one spin axis is distinguished from the other two,
\begin{equation}
\La_\mathrm{pert}=\eps(n^3)^2\;.
\end{equation}
For $\eps>0$, the perturbation favors spin alignment along the third axis; this kind of anisotropy is called \emph{easy-axis}. In the opposite case of $\eps<0$, spin alignment along the first or second axis is preferred. This is an \emph{easy-plane} anisotropy. The effects of easy-axis and easy-plane anisotropy on ferro- and antiferromagnets are very different. It is therefore best to discuss them separately. In each individual case, I will proceed by first identifying the ground state and then expanding the Lagrangian to second order in fluctuations.

\begin{illustration}%
Let us start with ferromagnets. In the absence of background gauge fields but presence of the anisotropy, the Lagrangian~\eqref{Lagferrofinal2} becomes
\begin{equation}
\La_\mathrm{eff}^\mathrm{LO}=-M\frac{\ve_{ab}n^a\dot n^b}{1+n^3}-\frac{\vr_\mathrm{s}}2\d^{rs}\de_r\vec n\cdot\de_s\vec n+\eps(n^3)^2\;.
\end{equation}
In easy-axis ferromagnets, the ground state is unique up to overall sign, $\vec n_0=(0,0,1)$. The two independent fluctuations can be taken as $n^1$ and $n^2$. Both the anisotropy and the ground state preserve the $\gr{U}(1)$ group of spin rotations around the third axis. We can thus look for normal modes as eigenstates of this symmetry. This motivates the introduction of a complex field, $\psi\equiv(n^1+\I n^2)/\sqrt2$. Dropping the energy density of the ground state and expanding the Lagrangian to second order in $\psi$, we get
\begin{equation}
\La_\mathrm{eff}^\mathrm{LO}\simeq\I M\he\psi\de_0\psi-\vr_{\mathrm{s}}\vec\nabla\he\psi\cdot\vec\nabla\psi-2\eps\he\psi\psi+\dotsb\;.
\end{equation}
This leads to a Schr\"odinger-like equation, describing circularly polarized spin waves with dispersion relation
\begin{equation}
E(\vec p)=\frac{2\eps}M+\frac{\vr_{\mathrm{s}}}M\vec p^2\;.
\label{easyaxisferro}
\end{equation}
The anisotropy gives the magnon a gap since both generators of $\gr{SU}(2)$, spontaneously broken in the preferred ground state, are also broken explicitly. In contrast to~\eqref{ferroinB}, the gap predicted by~\eqref{easyaxisferro}, $E(\vec0)=2\eps/M$, is only a LO result and will receive corrections at higher orders of the derivative expansion. The same is true for all the other magnon dispersion relations, derived below. The magnon has become a pseudo-NG boson.

In the easy-plane case, any uniform state with $\vev{n^3}=0$ minimizes the energy. I will choose the ground state as $\vec n_0=(1,0,0)$ and the independent fluctuations as $n^2$ and $n^3$. Upon series expansion in the latter using the parameterization~\eqref{easyplane}, the Lagrangian becomes
\begin{equation}
\La_\mathrm{eff}^\mathrm{LO}\simeq Mn^3\de_0n^2-\frac{\vr_{\mathrm{s}}}2\sum_{i=2,3}\vec\nabla n^i\cdot\vec\nabla n^i-\abs\eps(n^3)^2+\dotsb\;.
\end{equation}
We already found the spectrum of this type of Lagrangian back in Sect.~\ref{subsec:firstmodelNRSchrodinger}, I can therefore just write down the final result,
\begin{equation}
E(\vec p)=\frac{\vr_{\mathrm{s}}}M\abs{\vec p}\sqrt{\vec p^2+\frac{2\abs\eps}{\vr_{\mathrm{s}}}}\;.
\end{equation}
The dispersion relation remains gapless but has become linear. This is because the $\gr{U}(1)$ group of spin rotations, left intact by the anisotropy, is now spontaneously broken. The anisotropy has turned the magnon into a type-A NG boson.
\end{illustration}

\begin{illustration}%
Next we turn to antiferromagnets, which are now for a change much easier to analyze. In the absence of background gauge fields but upon adding the anisotropy term, the Lagrangian~\eqref{Lagantiferro} turns into
\begin{equation}
\La_\mathrm{eff}^\mathrm{LO}=\frac{\vr_{\mathrm{s}}}{2v^2}\bigl[(\de_0\vec n)^2-v^2\d^{rs}\de_r\vec n\cdot\de_s\vec n\bigr]+\eps(n^3)^2\;.
\end{equation}
This Lagrangian is diagonal in the spin index of $n^i$. We can therefore read off the spectrum immediately upon identification of the ground state and its independent fluctuations. In easy-axis antiferromagnets, the ground state is $\vec n_0=(0,0,1)$ up to a sign, and its fluctuations are $n^1$ and $n^2$. These excite two gapped magnons,
\begin{equation}
E_{1,2}(\vec p)=\sqrt{v^2\vec p^2+\frac{2v^2\eps}{\vr_{\mathrm{s}}}}\;.
\end{equation}
We can choose the basis of independent spin waves freely, either as linear, circular, or generally elliptic. In the easy-plane case, on the other hand, we can choose the ground state as $\vec n_0=(1,0,0)$ and its fluctuations as $n^2$ and $n^3$. Here we find two different excitation branches with different dispersion relations, corresponding to linearly polarized spin waves,
\begin{equation}
E_2(\vec p)=v\abs{\vec p}\;,\qquad
E_3(\vec p)=\sqrt{v^2\vec p^2+\frac{2v^2\abs\eps}{\vr_{\mathrm{s}}}}\;.
\end{equation}
The reason why one of the excitations remains gapless is that there is an exact $\gr{U}(1)$ symmetry, left intact by the perturbation, that is spontaneously broken.
\end{illustration}

I will conclude the discussion of perturbations in spin systems with a very peculiar example that leads to fascinating phenomenology. Until now, I have ruled out the existence of $\smash{\La_\mathrm{eff}^{(1,0)}}$ based on invariance under spatial rotations. However, one can, in fact, construct an operator with a single spatial derivative that does not break rotations, as long as two conditions are satisfied. First, parity must be broken, typically by the structure of the underlying crystal lattice. Second, spin--orbit coupling must be taken into account. This breaks the separate symmetries under spatial (orbital) and spin rotations to a single $\gr{SU}(2)$, under which spatial coordinates $\vec x$ and the spin vector $\vec n$ transform simultaneously. One can then add to the Lagrangian of (anti)ferromagnets the \emph{Dzyaloshinskii--Moriya} (DM) term,
\begin{equation}
\La_\mathrm{eff}^{(1,0)}=-\frac{2\pi\vr_\mathrm{s}}{\l_\mathrm{DM}}\vec n\cdot(\vec\nabla\times\vec n)\;,
\label{DM}
\end{equation}
where $\l_\mathrm{DM}$ is a new parameter with the dimension of length. This length scale is fixed by the choice of concrete material, and is usually much larger than the scale of the underlying crystal lattice. For instance, in MnSi one finds $\l_\mathrm{DM}\approx18\,\mathrm{nm}$ and in FeGe $\l_\mathrm{DM}\approx70\,\mathrm{nm}$~\cite{Togawa2016a}. This justifies treating the perturbation coupling $2\pi\vr_\mathrm{s}/\l_\mathrm{DM}$ as a small parameter of degree one in the power counting. The Lagrangian $\smash{\La_\mathrm{eff}^{(1,0)}}$ can then be consistently included in the LO of the derivative expansion.

\begin{illustration}%
\label{ex:helimagnetgroundstate}%
The ground state induced by the DM interaction can be discussed jointly for ferro- and antiferromagnets if one restricts to time-independent spin configurations. In the absence of external gauge fields and anisotropy, the LO Hamiltonian then reduces to
\begin{equation}
\Ha_\mathrm{eff}^\mathrm{LO}=\frac{\vr_\mathrm{s}}2\d^{rs}\de_r\vec n\cdot\de_s\vec n+\frac{2\pi\vr_\mathrm{s}}{\l_\mathrm{DM}}\vec n\cdot(\vec\nabla\times\vec n)\;.
\end{equation}
The condition for minimum energy follows by ``completing the square,''
\begin{equation}
\begin{split}
\Ha_\mathrm{eff}^\mathrm{LO}&\simeq\frac{\vr_\mathrm{s}}2\bigl[(\skal\nabla n)^2+(\vec\nabla\times\vec n)^2\bigr]+\frac{2\pi\vr_\mathrm{s}}{\l_\mathrm{DM}}\vec n\cdot(\vec\nabla\times\vec n)\\
&=\frac{\vr_\mathrm{s}}2(\skal\nabla n)^2+\frac{\vr_\mathrm{s}}2\left(\vec\nabla\times\vec n+\frac{2\pi}{\l_\mathrm{DM}}\vec n\right)^2-\frac{2\pi^2\vr_\mathrm{s}}{\l_\mathrm{DM}^2}\;.
\end{split}
\end{equation}
The second term will be minimized by any spin configuration satisfying the first-order differential equation $\vec\nabla\times\vec n=-(2\pi/\l_\mathrm{DM})\vec n$. This guarantees without further conditions minimization of the first term, $(\skal\nabla n)^2$, and thus of the whole Hamiltonian.

Let us now temporarily forget that $\vec n(\vec x)$ should be a unit vector at any $\vec x$, and Fourier-transform it to momentum space. The complex amplitude $\vec n_{\vec p}$ of the plane wave with momentum $\vec p$ is then subject to the condition $\I\vec p\times\vec n_{\vec p}=-(2\pi/\l_\mathrm{DM})\vec n_{\vec p}$. This requires that $\vec p\perp\vec n_{\vec p}$ and $\abs{\vec p}=2\pi/\l_\mathrm{DM}$ for any $\vec p$ such that $n_{\vec p}$ is nonzero. This gives the parameter $\l_\mathrm{DM}$ the interpretation as the wavelength of a spatially inhomogeneous order, induced by the DM interaction. An arbitrary linear combination of plane waves with fixed $\abs{\vec p}$ is, however, not allowed by the constraint $\skal nn=1$. Without going into further technical details, let me spell out the final result. The ground state corresponds to a real helix that is right-handed for $\l_\mathrm{DM}>0$ and left-handed for $\l_\mathrm{DM}<0$. The direction of the axis of the helix is arbitrary and spontaneously breaks the symmetry under spatial and spin rotations. Choosing it for illustration along the $z$-axis, and $\vec n$ to point along the $x$-axis at $z=0$, the ground state becomes
\begin{equation}
\vev{\vec n(\vec x)}=(\cos2\pi z/\l_\mathrm{DM},\sin2\pi z/\l_\mathrm{DM},0)\;.
\end{equation}
In presence of the DM interaction, the spin system becomes a \emph{helimagnet}.

The magnon spectrum of helimagnets is no less fascinating than their ground state. In case the underlying spin order is ferromagnetic, there is a single magnon branch (cf.~\refex{ex:helimagnet}). It has an unusual, strongly anisotropic dispersion relation. Along the axis of the helix, the energy is linear in momentum for wavelengths much longer than $\l_\mathrm{DM}$, whereas in the two transverse directions it is quadratic~\cite{Kirkpatrick2005}. One can show in addition that in the former case, the spin wave is linearly polarized, while in the latter case it is polarized circularly as usual in ferromagnets.
\end{illustration}


\subsection{Some Topological Aspects of Ferromagnets}
\label{subsec:spinwavestopological}

The list of interesting features of spin systems does not end with the multitude of textures that can be induced by various perturbations. In particular ferromagnets exhibit a number of intriguing topological properties that can be traced to the part $\smash{\La_\mathrm{eff}^{(0,1)}}$ of the effective Lagrangian. A brief sample is presented below. A reader wishing to learn more about the topology of spin systems is encouraged to consult~\cite{Han2017}.

\begin{figure}[t]
\sidecaption[t]
\includegraphics[width=2.9in]{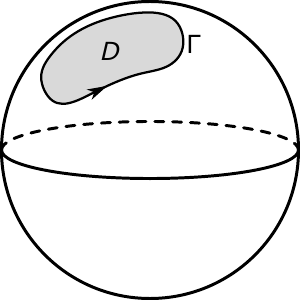}
\caption{Geometric interpretation of the single-time-derivative term in the effective Lagrangian~\eqref{Lagferrofinal} for ferromagnets. The spin configuration $\vec n(\vec x,t)$ maps the time axis to a closed curve $\Gamma$ on the coset space $S^2$. Manifest invariance of the Lagrangian under $\gr{SU}(2)$ spin rotations can be saved at the cost of extending the integration from $\Gamma$ to the disk $D$ (\emph{shaded} area) whose boundary is $\Gamma$}
\label{fig:WZferro}
\end{figure}

Let us start by rewriting the single-time-derivative part of the ferromagnet Lagrangian~\eqref{Lagferrofinal} in a way that exposes its geometric nature. Suppose the field $\vec n(\vec x,t)$ converges to a constant for $t\to\pm\infty$; this is a usual assumption when setting up the variational principle for fields. Viewed as a function of $\tau$ and $t$, $\vec n$ then maps $[0,1]\times\R$ to a ``disk'' $D$ on $S^2$ whose boundary $\Gamma$ carries the physical field ($\tau=1$). The action defined by~\eqref{Lagferrofinal} can be cast as
\begin{equation}
S_\mathrm{eff}^\mathrm{LO}\{\vec n,\vec A\}=M\int\D^d\!\vec x\int_D\Omega[\vec n](\vec x,t,\tau)+\dotsb\;,
\label{StokesS2}
\end{equation}
where $\Omega[\vec n]\equiv\vec n\cdot(\de_t\vec n\times\de_\tau\vec n)\D t\w\D\tau$; the ellipsis stands for the spacetime integral of the second and third term in~\eqref{Lagferrofinal}. Note that $\vec n\cdot(\D\vec n\times\D\vec n)$ is the area form on $S^2$. The action is therefore determined geometrically by the area of the domain on $S^2$, bounded by the curve $\Gamma$; see Fig.~\ref{fig:WZferro} for a visualization.

But how do we know which domain? I showed already that smoothly varying the interpolation $\vec n(\tau,\pi)$ between the fixed limits at $\tau=0$ and $\tau=1$ does not change the action. However, there are distinct classes of maps that cannot be smoothly deformed into each other. We could, for instance, think of the interpolation as filling the domain $D$ shown by shading in Fig.~\ref{fig:WZferro}, or its complement on the sphere. There is no a priori way to distinguish the ``inside'' and ``outside'' of the curve $\Gamma$. The interpolation $\vec n(\tau,\pi)$ could even cover the whole sphere multiple times, before converging to the curve $\Gamma$ in the limit $\tau\to1$. The only way around this intrinsic ambiguity is to ensure that it is not physically observable.

To that end, consider two domains, $D_1$ and $D_2$, swept by two different interpolations $\vec n_1(\tau,\pi)$ and $\vec n_2(\tau,\pi)$ of the same physical field, $\vec n_1(1,\pi)=\vec n_2(1,\pi)=\vec n(\pi)$. We can glue the two maps into a single one, $\tilde{\vec n}(\tau,\pi)$, $\tau\in[0,2]$, by setting
\begin{equation}
\begin{aligned}
\tilde{\vec n}(\tau,\pi)&\equiv\vec n_1(\tau,\pi)\quad&\text{for }\tau\in[0,1]\;,\\
\tilde{\vec n}(\tau,\pi)&\equiv\vec n_2(2-\tau,\pi)\quad&\text{for }\tau\in[1,2]\;.
\end{aligned}
\end{equation}
This new map satisfies $\tilde{\vec n}(0,\pi)=\tilde{\vec n}(2,\pi)=\vec n_0$. Thanks to the compactification of the time axis to the circle $S^1$, we can think of it as a map $S^2\to S^2$ with the domain spanned by the variables $\tau,t$. The points $\tau=0$ and $\tau=2$ are the poles of the domain and $\tau=1$ is the equator. As a consequence,
\begin{equation}
\int_{D_1}\Omega[\vec n_1]-\int_{D_2}\Omega[\vec n_2]=4\pi w[\tilde{\vec n}]\;,
\end{equation}
where $w[\tilde{\vec n}]$ is the integer-valued Brouwer degree~\eqref{Brouwer} of $\tilde{\vec n}$. The conclusion is that the classical action of a ferromagnet suffers from a topological ambiguity, shifting it by $4\pi MVw[\tilde{\vec n}]$ where $V$ is spatial volume. The functional integral of the EFT as a quantum theory will still be well-defined provided the action is only ambiguous up to an integer multiple of $2\pi$. This requires that the total spin $MV$ be quantized in half-integers. The low-energy EFT for ferromagnets, constructed solely based on the geometry of the coset space, ``knows'' about quantization of spin!

Now that we have established the topological nature of the $\smash{\La_\mathrm{eff}^{(0,1)}}$ Lagrangian, let us look at some of its consequences. The most immediate one is the presence of a \emph{Berry phase}; see Sect.~1.5 of~\cite{Tong2016} for a general introduction and~\cite{Watanabe2014a} for the application to EFTs for NG bosons. Let us denote the quantum-mechanical ground state of a ferromagnet, corresponding to the order parameter $\vev{\vec n}=\vec n_0$, as $\ket{\vec n_0}$. Suppose we expose the ferromagnet to a weak, uniform magnetic field $\vec B$. This forces it to align its magnetization with the field so that $\vev{\vec n}\equiv\vec n(\vec B)=\vec B/\abs{\vec B}$, with the corresponding vacuum state $\ket{\vec n(\vec B)}$.

Consider now a time-dependent magnetic field $\vec B(t)$ and arrange the ferromagnet state $\ket{\psi(0)}$ at $t=0$ to be the vacuum state $\ket{\vec n(\vec B(0))}$. If the variation of $\vec B(t)$ with time is sufficiently slow, the evolution of the state of the system $\ket{\psi(t)}$ will be adiabatic. We can set the ground state energy $E$ to zero to eliminate the trivial phase factor of $\E^{-\I Et}$. Even then, we cannot expect $\ket{\psi(t)}$ to equal $\ket{\vec n(\vec B(t))}$. Rather,
\begin{equation}
\ket{\psi(t)}=\exp[\I\g_\mathrm{B}(t)]\ket{\vec n(\vec B(t))}\;,
\label{Berry1}
\end{equation}
where $\g_\mathrm{B}(t)$ is the Berry phase. Taking the time derivative of~\eqref{Berry1} and projecting the result to $\ket{\psi(t)}$ leads to $\dot\g_\mathrm{B}(t)=\I\amplitude{\vec n(\vec B(t))}{\de_0}{\vec n(\vec B(t))}$. The total phase accumulated over an interval $[t_1,t_2]$, $\smash{\int_{t_1}^{t_2}\dot\g_\mathrm{B}(t)\,\D t}$, only depends on the path followed by $\vec B$, not on its precise time dependence. To underline the geometric nature of the Berry phase, we introduce the \emph{Berry connection},
\begin{equation}
\o_\mathrm{B}(\vec B)\equiv\I\amplitude{\vec n(\vec B)}{\D}{\vec n(\vec B)}\;,
\label{Berryconnection}
\end{equation}
which is a 1-form on the space of all quantum ground states of the ferromagnet. The gauge freedom associated with the Berry connection amounts to the arbitrary choice of phase of $\ket{\vec n(\vec B)}$ for every value of $\vec B$. Since the state $\ket{\vec n(\vec B)}$ only depends on the direction of $\vec B$, the Berry connection naturally induces a 1-form on the coset space $S^2$ of the ferromagnet. When the magnetic field varies so that it traces a closed loop in the parameter space, the total Berry phase is
\begin{equation}
\g_\mathrm{B}(\Gamma)=\int_\Gamma\o_\mathrm{B}=\int_D\D\o_\mathrm{B}\;.
\label{Berryphase}
\end{equation}
Here $\Gamma$ is the loop on $S^2$ traced by the order parameter in the process and $D$ the disk bounded by it, as in Fig.~\ref{fig:WZferro}. The 2-form $\D\o_\mathrm{B}$ is called the \emph{Berry curvature}.

To see the connection between the Berry phase and the effective Lagrangian for ferromagnets, recall the parameterization of the coset space by matrices $U(\pi)$. With a slight abuse of notation, one can write $\ket{\vec n(\vec B)}=U(\pi(\vec B))\ket{\vec n_0}$. The Berry connection then reads
\begin{equation}
\o_\mathrm{B}=\I\amplitude{\vec n_0}{\smash{U(\pi)^{-1}\D U(\pi)}}{\vec n_0}=-\amplitude{\vec n_0}{\mc(\pi)}{\vec n_0}=-MV\mc^3(\pi)\;.
\end{equation}
Comparing this to~\eqref{Lagferro}, we see that up to the factor of volume, the Berry connection is precisely the 1-form that defines the Lagrangian density $\smash{\La_\mathrm{eff}^{(0,1)}}$. On the other hand, it follows from~\eqref{StokesS2} and the Stokes theorem that $-\D\mc^3$ is the area form on $S^2$. The Berry phase~\eqref{Berryphase} therefore equals the total spin of the system times the area of $D$.

Mathematically, the area 2-form is the single generator of the second de Rham cohomology group of $S^2$. The topological features of $\smash{\La^{(0,1)}_\mathrm{eff}}$ stem from pulling this 2-form back to the space of variables $\tau,t$. However, interesting physics also arises from pulling it back to the whole spacetime. For simplicity, I will now restrict to $d=2$ spatial dimensions. The resulting closed spacetime 2-form is then Hodge-dual to a current, conventionally normalized as
\begin{equation}
J^\m[\vec n]=\frac1{8\pi}\ve^{\m\n\l}\ve_{ijk}n^i\de_\n n^j\de_\l n^k\;.
\label{topologicalcurrent}
\end{equation}
This current is conserved off-shell, similarly to the GW current introduced in Sect.~\ref{subsec:ChPTanomaly}. Following the analogy, we expect the density $J^0[\vec n]$ to give rise to a topological charge, that is a topological invariant of the field $\vec n(\vec x,t)$. Indeed, $\smash{\int\D^2\vec x\,J^0[\vec n](\vec x,t)}$ is just the Brouwer degree $w[\vec n]$ of the map $\vec n:\R^2\to S^2$. In $d=2$ dimensions, there are smooth spin configurations carrying nonzero value of $w[\vec n]$, called skyrmions, or sometimes \emph{baby skyrmions}, to distinguish them from their counterparts in QCD. See the dedicated monograph~\cite{Han2017} for further details about this fascinating subject.

Recall now the symplectic approach to ferromagnets, outlined in Sect.~\ref{subsec:symplectic}. It is easy to check explicitly that as a functional on the phase space of a ferromagnet, $w[\vec n]$ has a vanishing Poisson bracket with any other functional. This is an example of a general feature that topological charges do not generate any flow on the phase space. Let us see what happens if we deform $w[\vec n]$ by inserting an arbitrary function of spatial coordinates. Changing for convenience the overall normalization, we define a class of functionals,
\begin{equation}
Q_f[\vec n]\equiv\frac M2\int\D^2\vec x\,f(\vec x)\ve^{rs}\ve_{ijk}n^i(\vec x)\de_rn^j(\vec x)\de_sn^k(\vec x)\;.
\end{equation}
It is a simple exercise to verify that the Poisson algebra of these functionals reproduces the algebra of functions on $\R^2$,
\begin{equation}
\{Q_f,Q_g\}=Q_{\ve^{rs}\de_rf\de_sg}\equiv Q_{\{f,g\}}\;.
\label{QfQg}
\end{equation}
In addition, the functional $Q_f$ generates a flow on the phase space through
\begin{equation}
\{\vec n(\vec x),Q_f[\vec n]\}=\ve^{rs}\de_rf(\vec x)\de_s\vec n(\vec x)\;.
\end{equation}
Curiously, this makes it possible to identify $P^r\equiv Q_{\ve_{rs}x^s}$ with the generator of spatial translations, that is momentum. The real surprise comes now: according to~\eqref{QfQg},
\begin{equation}
\{P^r[\vec n],P^s[\vec n]\}=4\pi M\ve^{rs}w[\vec n]\;.
\end{equation}
For skyrmions, which carry nonzero $w[\vec n]$, the two components of momentum do not commute with each other. This has a simple physical interpretation. Namely, for nonzero $w[\vec n]$,
\begin{equation}
x_w^r[\vec n]\equiv\frac1{8\pi w[\vec n]}\int\D^2\vec x\,x^r\ve^{uv}\ve_{ijk}n^i(\vec x)\de_un^j(\vec x)\de_vn^k(\vec x)=-\frac{\ve^{rs}P_s[\vec n]}{4\pi Mw[\vec n]}
\end{equation}
represents the center of the topological charge distribution. Conservation of momentum in the absence of external forces then implies conservation of the center of charge. Skyrmions are pinned to a fixed position unless an external field is applied to them. The fact that $\{x^r_w[\vec n],x^s_w[\vec n]\}=\ve^{rs}/(4\pi Mw[\vec n])$ hints that the dynamics of skyrmions closely resembles that of a charged particle in a magnetic field.


\bibliographystyle{spphys}
\bibliography{references}
\chapter{Scattering of Nambu--Goldstone Bosons}
\label{chap:scattering}

\begin{fmffile}{feynman10}



\newcommand{\fscatJHpHp}{\parbox{15mm}{
\begin{fmfgraph}(15,15)
\fmfleftn{l}{2}
\fmfrightn{r}{2}
\fmf{dashes}{l1,v}
\fmf{phantom}{v,r1}
\fmf{plain}{l2,v,r2}
\fmfv{decor.shape=square,decor.filled=full,
decor.size=2thick}{v}
\end{fmfgraph}}}

\newcommand{\fscatJHpHps}{\parbox{20mm}{
\begin{fmfgraph}(20,15)
\fmfleftn{l}{2}
\fmfrightn{r}{2}
\fmf{dashes}{l1,vl,vr}
\fmf{phantom}{vr,r1}
\fmf{plain}{l2,vl}
\fmf{plain}{r2,vr}
\fmfdot{vl}
\fmfv{decor.shape=square,decor.filled=full,
decor.size=2thick}{vr}
\end{fmfgraph}}}

\newcommand{\fscatJHpHpt}{\parbox{15mm}{
\begin{fmfgraph}(15,15)
\fmfleftn{l}{2}
\fmfrightn{r}{2}
\fmf{plain}{vu,vd}
\fmf{plain}{l2,vu,r2}
\fmf{dashes}{l1,vd}
\fmf{phantom}{vd,r1}
\fmfdot{vu}
\fmfv{decor.shape=square,decor.filled=full,
decor.size=2thick}{vd}
\end{fmfgraph}}}

\newcommand{\fscatJHpHpu}{\parbox{15mm}{
\begin{fmfgraph}(15,15)
\fmfleftn{l}{2}
\fmfrightn{r}{2}
\fmf{dashes}{vu,vd}
\fmf{plain}{l2,vu}
\fmf{phantom}{vu,r2}
\fmf{dashes}{l1,vd}
\fmf{phantom}{vd,r1}
\fmffreeze
\fmf{phantom}{vu,r1}
\fmf{plain}{vd,r2}
\fmfdot{vd}
\fmfv{decor.shape=square,decor.filled=full,
decor.size=2thick}{vu}
\end{fmfgraph}}}

\newcommand{\ampfourpt}{\parbox{15mm}{
\begin{fmfgraph}(15,15)
\fmfleftn{l}{2}
\fmfrightn{r}{2}
\fmf{plain}{l1,v}
\fmf{plain}{l2,v}
\fmf{plain}{r1,v}
\fmf{plain}{r2,v}
\fmfv{decor.shape=circle,decor.filled=gray50,
decor.size=10thick}{v}
\end{fmfgraph}}}

\newcommand{\ampfivept}{\parbox{15mm}{
\begin{fmfgraph}(15,15)
\fmfsurroundn{v}{5}
\fmf{plain}{v1,v}
\fmf{plain}{v2,v}
\fmf{plain}{v3,v}
\fmf{plain}{v4,v}
\fmf{plain}{v5,v}
\fmfv{decor.shape=circle,decor.filled=gray50,
decor.size=10thick}{v}
\end{fmfgraph}}}


\abstract*{Interactions of Nambu--Goldstone bosons are constrained by spontaneously broken symmetry just like their spectrum. This chapter is devoted to one specific manifestation of interactions: scattering. This is of particular importance in subatomic physics. The discussion of scattering of Nambu--Goldstone bosons is therefore restricted to relativistic, Lorentz-invariant systems. Some references to literature dealing with a general setting also applicable to condensed-matter physics are provided. The first half of the chapter revisits the Adler zero phenomenon. A necessary and sufficient condition for a scattering amplitude to vanish in the soft limit for a single particle is derived. There is a small class of effective theories whose scattering amplitudes are softer than the naive Adler zero would suggest. It is shown how such theories can be constructed by systematically imposing specific scaling of the scattering amplitudes in the soft limit. The chapter concludes with a brief introduction to on-shell recursion methods. These allow one to reconstruct all tree-level scattering amplitudes of a theory from a finite number of seed, low-point amplitudes by utilizing a priori knowledge about the infrared behavior of the amplitudes.}


In our survey of \emph{spontaneous symmetry breaking} (SSB), we have so far been mostly concerned with the ground state and excitation spectrum. Yet, the \emph{effective field theory} (EFT) framework developed in Chap.~\ref{chap:effLagrangian} captures full nonlinear dependence of the action on the \emph{Nambu--Goldstone} (NG) fields. It thus also includes all interactions among NG bosons and their interactions with other excitations, if present. Microscopic interactions in quantum systems may have many different macroscopic manifestations. However, there is an important class of observables that are amenable to both systematic computation and experiment: scattering amplitudes of free asymptotic states. I will therefore conclude the part of the book devoted to spontaneous breaking of internal symmetry with a short primer on scattering of NG bosons.

I will start in Sect.~\ref{sec:Adler} by revisiting the \emph{Adler zero} principle that scattering amplitudes of NG bosons tend to vanish in the long-wavelength limit. I will offer a general justification of this phenomenon, and underline the nature of exceptions to the rule. A modern approach to scattering of NG bosons that utilizes the geometry of the coset space of broken symmetry is introduced in Sect.~\ref{sec:Adlergeom}. In Sect.~\ref{sec:beyondAdler}, I demonstrate that there are exceptional EFTs whose long-wavelength scattering amplitudes are even softer that naively expected from the Adler zero principle. These theories set a benchmark for many modern developments in the rapidly expanding field of scattering amplitudes. One important line of development concerns the recursive reconstruction of all (tree-level) scattering amplitudes from a finite number of ``seed'' amplitudes. This is the subject of Sect.~\ref{sec:softrecursion}.

Scattering amplitudes as observables are arguably more important in fundamental (subatomic) physics than in other branches such as condensed-matter physics. I will therefore restrict the discussion to Lorentz-invariant systems. Here NG bosons are massless particles and the kinematics of their scattering is well-understood. Much of the material of this chapter can however be generalized to nonrelativistic EFTs with continuous spatial rotation symmetry; see~\cite{Mojahed2022,Cheung2023} for details. Let me stress that even amplitudes in relativistic systems have grown into a vast subject, of which this chapter only exposes a small corner. A reader interested in a broader introduction to the physics of scattering amplitudes is recommended to start with the lecture notes~\cite{Cheung2017c} or the book~\cite{Elvang2015}.


\section{Adler Zero Revisited}
\label{sec:Adler}

\begin{figure}[t]
\sidecaption[t]
\includegraphics[width=2.9in]{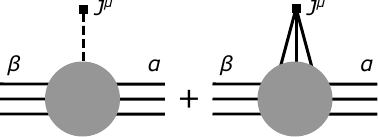}
\caption{Sketch of contributions to the matrix element $\amplitude{\b}{J^\m(x)}{\a}$. In the former, the current creates an intermediate one-particle NG state $\ket{\pi(\vec p)}$ (\emph{dashed line}) that couples to $\ket\a$ and $\ket\b$ through the amplitude $\smash{\Aa_{\a\to\b+\pi(\vec p)}}$. In the latter, the current connects to $\ket\a$, $\ket\b$ via a multiparticle intermediate state}
\label{fig:Adler}
\end{figure}

One of the hallmarks of SSB is the existence of a local conserved current $J^\m$ that couples to the NG boson state; cf.~the proof of Goldstone's theorem in Sect.~\ref{subsec:GoldstoneOpProof}. This can be used to extract nonperturbative information about scattering processes involving one or more NG bosons. We start with a generic process whose initial and final states $\ket\a$ and $\ket\b$ may include any set of particles of arbitrary mass and spin, and inspect the matrix element $\amplitude{\b}{J^\m(x)}{\a}$. This receives two qualitatively different types of contributions, schematically shown in Fig.~\ref{fig:Adler}. We shall focus on the first contribution, which amounts to the creation or annihilation of a one-particle NG boson state by the current.

The NG boson may be an elementary excitation of a field in the action. The con\-tri\-butions with an intermediate one-particle NG state then correspond to one-particle-reducible diagrams where cutting a single propagator disconnects the current from the $\a\to\b$ process. However, the NG boson may also be composite. The NG state then reveals itself through a single-particle pole in the matrix element $\amplitude{\b}{J^\m(x)}{\a}$ at energy--momentum $p^\m\equiv p^\m_\a-p^\m_\b$.  Contributions to the matrix element without such a single-particle pole are represented by the second diagram in Fig.~\ref{fig:Adler}.

We use the translation property of the current operator, $J^\m(x)=\E^{\I P\cdot x}J^\m(0)\E^{-\I P\cdot x}$, where $P^\m$ is the operator of energy--momentum. This holds whenever the current does not depend explicitly on spacetime coordinates, which is the case for internal symmetry. According to the rules of polology (Sect.~19.2 of~\cite{Weinberg1996a}), the residue of $\amplitude{\b}{J^\m(x)}{\a}$ at the one-particle NG pole is then $\E^{-\I p\cdot x}\amplitude{0}{J^\m(0)}{\pi(\vec p)}\Aa_{\a\to\b+\pi(\vec p)}$. Here $\Aa_{\a\to\b+\pi(\vec p)}$ is the invariant amplitude for the process where a NG boson of momentum $\vec p$ is added to the final state.\footnote{In contrast to Chap.~\ref{chap:NGbosons}, I use here the normalization $\braket{\pi(\vec p)}{\pi(\vec q)}=(2\pi)^d2\abs{\vec p}\d^d(\vec p-\vec q)$ of the one-particle NG states. The amplitude $\Aa_{\a\to\b}$ for a generic process $\a\to\b$ is normalized by its relation to the $S$-matrix, $S_{\a\to\b}=-(2\pi)^D\I\Aa_{\a\to\b}\d^D(p_\a-p_\b)$. Finally, adding the NG boson to $\ket\b$ rather than $\ket\a$ is a convention, amounting to the choice of orientation of $p^\m$.} With a Lorentz-invariant vacuum $\ket0$, the matrix element $\amplitude{0}{J^\m(0)}{\pi(\vec p)}$ transforms as a Lorentz vector, as one readily verifies using the properties of single-particle states (Sect.~2.5 of~\cite{Weinberg1995a}). Lorentz invariance therefore dictates that $\amplitude{0}{J^\m(0)}{\pi(\vec p)}=\I p^\m_\mathrm{on}F$, where $F$ is a constant and $p^\m_\mathrm{on}\equiv(\abs{\vec p},\vec p)$ is an on-shell energy--momentum with spatial part $\vec p$. The matrix element $\amplitude{\b}{J^\m(0)}{\a}$ can now be split into a pole part and a regular part as
\begin{equation}
\amplitude{\b}{J^\m(0)}{\a}=\amplitude{0}{J^\m(0)}{\pi(\vec p)}_\mathrm{off}\frac1{p^2}\Aa_{\a\to\b+\pi(\vec p)}+R^\m_{\b\a}(p)\;,
\label{polefactorization}
\end{equation}
where $\smash{\amplitude{0}{J^\m(0)}{\pi(\vec p)}_\mathrm{off}\equiv\I p^\m F}$. The as yet unspecified function $\smash{R^\m_{\b\a}(p)}$ collects all the nonpole contributions, and thus remains regular in the on-shell limit $p^2\to0$.

\begin{watchout}%
Polology suggests that the numerator of the pole in~\eqref{polefactorization} should be proportional to $\amplitude{0}{J^\m(0)}{\pi(\vec p)}$. Replacing this with $\amplitude{0}{J^\m(0)}{\pi(\vec p)}_\mathrm{off}$ amounts to trading $p_\mathrm{on}^\m$ for $p^\m$, and in turn to a redefinition of $\smash{R^\m_{\b\a}(p)}$ without changing the residue at the pole. The reason for doing so will become clear below when we work out an explicit example. Here let me just stress that $\amplitude{0}{J^\m(0)}{\pi(\vec p)}_\mathrm{off}$ is a mere symbolic notation for $\I p^\m F$. This object is not a well-defined matrix element but rather depends on the energy--momenta $p^\m_\a$ and $p^\m_\b$.
\end{watchout}

We now combine~\eqref{polefactorization} with current conservation expressed as $p_\m\amplitude{\b}{J^\m(0)}{\a}=0$. This leads to a master identity which allows one to reconstruct the on-shell scattering amplitude from the remainder function $\smash{R^\m_{\b\a}(p)}$,
\begin{equation}
\Aa_{\a\to\b+\pi(\vec p)}=\frac\I Fp_\m R^\m_{\b\a}(p)\;.
\label{Adlerproof}
\end{equation}

\begin{illustration}%
Recall the toy model~\eqref{toylag}, introduced in Chap.~\ref{chap:ourfirstmodel}. To simplify the notation, I will only consider the limit of exact symmetry, $\eps\to0$, and disregard the fermionic degrees of freedom. In the linear parameterization of the complex field, $\p(x)=\E^{\I\t}[v+\mf(x)+\I\pi(x)]/\sqrt2$, the axial current in~\eqref{toyconservation} becomes
\begin{equation}
J^\m_\mathrm{A}=-2\I(\p^*\de^\m\p-\de^\m\p^*\p)=2v\de^\m\pi+2(\mf\de^\m\pi-\pi\de^\m\mf)\;,
\end{equation}
whence we read off $F=-2v$. Let us have a look at the process $\mf\pi\to\mf\pi$ in which a NG boson scatters off a massive (Higgs) particle. We set $\ket\a=\ket{\mf(\vec p_1)\pi(\vec p_2)}$ and $\ket\b=\ket{\mf(\vec p_3)}$. The ``missing'' energy--momentum $p^\m_4$ is supplied by the current. Using the Feynman rules~\eqref{toyFruleslinear}, the remainder function $\smash{R^\m_{\b\a}(p_4)}$ collecting nonpole contributions to $\amplitude{\b}{\smash{J^\m_\mathrm{A}(0)}}{\a}$ turns out to be
\begin{align}
\label{Rmbaexample}
R^\m_{\b\a}(p_4)&=\fscatJHpHps+\fscatJHpHpt+\fscatJHpHpu\\
\notag
&=-4\I\l v\frac{(2p_3+p_4)^\m}s+12\I\l v\frac{(p_4-2p_2)^\m}{t-m_\mf^2}+4\I\l v\frac{(2p_1-p_4)^\m}u\;,
\end{align}
where the solid squares indicate an insertion of the axial current operator. Contracting this with $p^\m_4$ gives
\begin{equation}
\I p_{4\m}R^\m_{\b\a}(p_4)=-4\l v\biggl[1+m_\mf^2\biggl(\frac1s+\frac3{t-m_\mf^2}+\frac1u\biggr)\biggr]\;.
\end{equation}
Using finally~\eqref{Adlerproof}, this reproduces the previously calculated amplitude~\eqref{toyApsipipsipi}.

Note that when evaluating the remainder function $\smash{R^\m_{\b\a}(p_4)}$, I ignored the contributions of Feynman diagrams where an external $\pi$-type leg terminates in the current operator. These arise from the part of the current linear in $\pi$, and are included in the pole part of~\eqref{polefactorization}. Had we used therein the on-shell matrix element $\amplitude{0}{J^\m(0)}{\pi(\vec p)}$, we would have had to include in~\eqref{Rmbaexample} the contributions of the mentioned diagrams with the pole canceled by the numerator factor $\smash{p^\m-p^\m_\mathrm{on}}$. At the same time, the right-hand side of~\eqref{Adlerproof} would receive an additional factor of $2$, owing to the fact that the on-shell limit of $p\cdot p_\mathrm{on}/p^2$ is $1/2$. The conclusion that the amplitude can be reconstructed from $\smash{R^\m_{\b\a}(p_4)}$ would however remain.

It is instructive to reproduce the same result also with the other parameterization considered in Chap.~\ref{chap:ourfirstmodel}, $\p(x)=\E^{\I\t}\exp[\I\pi(x)/v][v+\mf(x)]/\sqrt2$. In this case, the axial current reads
\begin{equation}
J^\m_\mathrm{A}=2v\de^\m\pi+4\mf\de^\m\pi+\frac2v\mf^2\de^\m\pi\;,
\end{equation}
whereas the Feynman rules change to~\eqref{toyFrulesexp}. The presence of the cubic term in the current adds another Feynman diagram to the expression for the remainder function,
\begin{align}
R^\m_{\b\a}(p_4)&=\fscatJHpHps+\fscatJHpHpt+\fscatJHpHpu+\fscatJHpHp\\
\notag
&=\frac{4\I}v\frac{s-m_\mf^2}s(p_3+p_4)^\m-\frac{24\I\l vp_2^\m}{t-m_\mf^2}-\frac{4\I}v\frac{u-m_\mf^2}u(p_1-p_4)^\m-\frac{4\I p_2^\m}v\;.
\end{align}
This reproduces the same scattering amplitude, in the form shown in~\eqref{toyscatexp}.
\end{illustration}

What is most interesting about~\eqref{Adlerproof} is the extra factor of $p_\m$ on the right-hand side, arising from current conservation. This shows that
\begin{equation}
\lim_{\vec p\to\vec0}\Aa_{\a\to\b+\pi(\vec p)}=0\;,
\label{Adlerzero}
\end{equation}
provided the remainder function is nonsingular in this limit. This is the \emph{Adler zero}. The same conclusion applies to processes with a NG boson inserted in the initial state. This distinction is trivial in Lorentz-invariant systems; a pedagogical proof of~\eqref{Adlerzero} for systems with mere spatial rotation invariance can be found in Sect.~3.1 of~\cite{Mojahed2022}. I will not reproduce the details here since they have no bearing on the result.

The function $\smash{R^\m_{\b\a}(p)}$ is by construction nonsingular in the on-shell limit $p^2\to0$. The absence of singularities for $\vec p\to\vec 0$ (or equivalently $p^\m\to0$) is an additional assumption. Finding a mechanism that leads to a violation of this assumption is required for understanding the origin of exceptions to the Adler zero principle.


\subsection{Generalized Soft Theorem}
\label{subsec:generalizedsoft}

Before we embark on a more detailed analysis of the conditions under which the Adler zero~\eqref{Adlerzero} is realized, let me introduce some terminology. Taking the momentum of a massless particle (a NG boson or, for instance, a gauge boson) to zero is generally referred to as the (single) \emph{soft limit}. The statement of Adler zero is equivalent to the vanishing of the soft limit of scattering amplitudes of NG bosons. A universal statement about the soft limit of scattering amplitudes such as~\eqref{Adlerzero} is called a \emph{soft theorem}. We will see that even in theories where the soft limit of $\smash{\Aa_{\a\to\b+\pi(\vec p)}}$ is nonvanishing, one can still establish a generalized soft theorem. This relates the soft limit of $\Aa_{\a\to\b+\pi(\vec p)}$ to the amplitude $\Aa_{\a\to\b}$ with the NG boson removed.

We noticed in Chap.~\ref{chap:ourfirstmodel} that scattering the NG boson off the fermion within our toy model violates the Adler zero principle. Operationally, this could be understood as a consequence of the singularity of the fermion propagator in~\eqref{toyAfpifpigraphs} in the soft limit for the NG boson. This singularity in turn arises from the cubic coupling between the NG boson and the fermion. The origin of the singularity is however purely kinematical and not specific to coupling of NG bosons to fermions; the same obstruction to the Adler zero may occur in purely bosonic theories. It is, in fact, commonplace in EFTs where all degrees of freedom are NG bosons, as constructed in Chap.~\ref{chap:effLagrangian}. The Adler zero is therefore a much less robust consequence of SSB than, say, the very existence of gapless NG bosons.

Let us now examine the hypothesis that the violation of the Adler zero principle arises from the presence of cubic interaction vertices in the theory. To keep things simple, I will from now on restrict to relativistic EFTs of NG bosons, disregarding any non-NG degrees of freedom. The argument below closely follows~\cite{Kampf2020a}.

Consider a set of NG fields $\pi^a$ in a theory defined by the generic Lagrangian
\begin{equation}
\La_\mathrm{eff}=\frac12\d_{ab}\de_\m\pi^a\de^\m\pi^b+\frac12\l_{abc}\de_\m\pi^a\de^\m\pi^b\pi^c+\bigO(\pi^4,\de^4)\;.
\label{LagShifman}
\end{equation}
The coupling $\l_{abc}$ can be assumed to be symmetric in its $a,b$ indices. There are no terms without derivatives as guaranteed by the broken symmetry. Also, we can restrict the discussion to operators with two derivatives in line with the power counting laid out in Sect.~\ref{subsec:ChPTpowercounting}. Finally, I assume that the kinetic term is diagonal in the NG flavor indices and properly normalized. This can always be ensured by a linear transformation of the NG fields. Other than that, however, I do not implicitly assume any particular parameterization of the fields.

\begin{watchout}%
The first thing to notice is that the cubic vertex in~\eqref{LagShifman} alone cannot be the sole culprit in violating the Adler zero principle. Observable properties of scattering amplitudes must be unchanged by a field redefinition of the type
\begin{equation}
\pi^a\to\pi'^a(\pi)=\pi^a+\frac12c^a_{bc}\pi^b\pi^c+\bigO(\pi^3)\;,
\label{redefShifman}
\end{equation}
where $c^a_{bc}$ is symmetric in $b,c$. In particular, choosing $c^a_{bc}=(1/2)\d^{ad}(\l_{dbc}+\l_{dcb}-\l_{bcd})$ eliminates the cubic interaction vertex altogether. In fact, even cubic operators with more than two derivatives can be removed analogously. This redundancy of cubic couplings of NG bosons is inherent to relativistic EFTs for SSB.
\end{watchout}

For the Lagrangian~\eqref{LagShifman} to actually represent an EFT for NG bosons, it must preserve a symmetry that is spontaneously broken. We do not need to invoke the full machinery of nonlinear realizations at this stage. It is sufficient to assume the existence of a set of spontaneously broken symmetries, one for each NG field,
\begin{equation}
\udelta\pi^a\equiv \eps^b[F^a_b+G^a_{bc}\pi^c+\bigO(\pi^2)]\;.
\label{symShifman}
\end{equation}
The independence of these transformations and the fact that they are all spontaneously broken amount to the condition that $F^a_b$ is an invertible square matrix. The assumed invariance of~\eqref{LagShifman} then leads to the following set of Noether currents,
\begin{equation}
J^\m_a=\d_{bc}F^c_a\de^\m\pi^b+K_{abc}\de^\m\pi^b\pi^c+\bigO(\pi^3)\;,
\label{curShifman}
\end{equation}
where $K_{abc}\equiv\d_{bd}G^d_{ac}+F^d_a\l_{dbc}$.

\begin{figure}[t]
\sidecaption[t]
\includegraphics[width=2.0in]{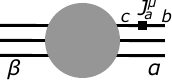}
\caption{Sketch of nonpole contributions to $\amplitude{\b}{\smash{J^\m_a(0)}}{\a}$ that are singular in the limit $\smash{p^\m\equiv p^\m_\a-p^\m_\b}\to0$. These contributions arise from inserting the bilinear part of the Noether current~\eqref{curShifman} into an external leg of the $\a\to\b$ process. The insertion is indicated by the \emph{solid square}}
\label{fig:AdlerJ}
\end{figure}

Let us now evaluate, schematically, the amplitude $\Aa_{\a\to\b+\pi^a(\vec p)}$ for a process where $\ket\a,\ket\b$ are states with an arbitrary number of NG bosons. The soft NG boson state $\ket{\smash{\pi^a(\vec p)}}$ is added to the final state $\ket\beta$. Then~\eqref{polefactorization} is replaced with
\begin{equation}
\amplitude{\b}{\smash{J^\m_a(0)}}{\a}=\sum_b\amplitude{0}{\smash{J^\m_a(0)}}{\smash{\pi^b(\vec p)}}_\mathrm{off}\frac1{p^2}\Aa_{\a\to\b+\pi^b(\vec p)}+R^\m_{\b\a a}(p)\;.
\end{equation}
From~\eqref{curShifman} we find that $\amplitude{0}{\smash{J^\m_a(0)}}{\smash{\pi^b(\vec p)}}=-\I p^\m_\mathrm{on}F^b_a$. Applying momentum conservation leads to $\smash{\sum_bF^b_a\Aa_{\a\to\b+\pi^b(\vec p)}}=-\I p_\m\smash{R^\m_{\b\a a}(p)}$. We expect the Adler zero property of $\Aa_{\a\to\b+\pi^b(\vec p)}$ to be violated if the limit $\vec p\to\vec0$ makes some of the propagators in the process on-shell. Barring fine tuning of the kinematics, this happens when the NG state $\ket{\smash{\pi^b(\vec p)}}$ is ``attached'' to an external leg of the $\alpha\to\beta$ process via a cubic interaction vertex. In terms of $\smash{R^\m_{\b\a a}(p)}$, such singular contributions map to insertions of the bilinear part of $J^\m_a$ into one of the external legs of $\a\to\b$. See Fig.~\ref{fig:AdlerJ} for a visualization and the index notation.

Near the limit of vanishing $p^\m$, inserting the current into an external leg with flavor $b$, carrying energy--momentum $p^\m_b$, gives the singular contributions
\begin{equation}
\begin{split}
R^\m_{\b\a a}(p)&\supset\sum_c(-\I\Aa^{b\to c}_{\a\to\b})\frac\I{(p_b-p)^2}\bigl[-\I K_{abc}p_b^\m+\I K_{acb}(p_b-p)^\m\bigr]\\
&\approx\frac\I2\sum_c\Aa^{b\to c}_{\a\to\b}\frac{p_b^\m}{p\cdot p_b}(K_{abc}-K_{acb})\;.
\end{split}
\end{equation}
The superscript $b\to c$ indicates a replacement of the flavor index on the leg where the insertion has been made. Upon summing over all possible insertions, we arrive at the final form of our generalized soft theorem,
\begin{equation}
\lim_{\vec p\to\vec0}\sum_bF^b_a\Aa_{\a\to\b+\pi^b(\vec p)}=\frac12\sum_{b\in\a\cup\b}\sum_c(K_{abc}-K_{acb})\Aa^{b\to c}_{\a\to\b}\;.
\label{thmShifman}
\end{equation}

Let me spell out some properties of this result explicitly. First, the soft theorem~\eqref{thmShifman} relates the soft limit of an $n$-particle amplitude to a set of $(n-1)$-particle amplitudes, in which the soft NG leg is removed and the label on one other leg is changed. The only other ingredient needed is the set of algebraic coefficients $K_{abc}$, which is independent of the concrete scattering process considered. In this sense, the soft theorem~\eqref{thmShifman} is universal.

Second, as is evident from~\eqref{curShifman}, the coefficients $K_{abc}$ receive two types of contributions: one from the cubic coupling $\l_{abc}$, the other from the part $G^a_{bc}$ of the symmetry transformation~\eqref{symShifman}, linear in NG fields. Importantly, the combination $K_{abc}-K_{acb}$, antisymmetric under the exchange $b\leftrightarrow c$, is manifestly invariant under the field redefinition~\eqref{redefShifman}. This ensures that the soft theorem~\eqref{thmShifman} has a physical meaning independent of the choice of field parameterization.


\subsection{Application to Coset Effective Theories}
\label{subsec:generalizedsoftcoset}

The above argument is fairly general and in principle also applies to scattering amplitudes in theories where some of the particles in $\ket\a$ and $\ket\b$ are not NG bosons. The only essential constraint is that there are no cubic operators without derivatives in the Lagrangian. The sum in~\eqref{thmShifman} is then to be restricted to $c$ such that $\pi^c$ has the same mass as $\pi^b$. However, once the Lagrangian~\eqref{LagShifman} does represent an EFT with only NG boson degrees of freedom, we can use the wealth of information accumulated in Chaps.~\ref{chap:CCWZ} and~\ref{chap:effLagrangian}.

Let us recall some of the basic relations that we are going to use here. First, the two-derivative part of the effective Lagrangian for the coset space $G/H$ reads
\begin{equation}
\La^{(2)}_\mathrm{eff}=\frac12\k_{cd}\mc^c_a(\pi)\mc^d_b(\pi)\de_\m\pi^a\de^\m\pi^b\;,
\label{LeffShifman}
\end{equation}
where the constant symmetric matrix $\k_{ab}$ satisfies the constraint $f^c_{\a a}\k_{cb}+f^c_{\a b}\k_{ac}=0$. The index $\a$ labels generators of the unbroken subgroup $H$. With the exponential parameterization of the coset space, $U(\pi)=\exp(\I\pi^a Q_a)$, the \emph{Maurer--Cartan} (MC) \emph{form} can be computed explicitly as a power series in the NG fields,
\begin{equation}
\mc^A_a(\pi)=\d^A_a-\frac12f^A_{ab}\pi^b+\bigO(\pi^2)\;,
\label{mcShifman}
\end{equation}
where the index $A$ runs over all generators of $G$. Likewise, infinitesimal transformations induced on the coset space by $G$, $\udelta\pi=\eps^A\kil^a_A(\pi)$, are determined by a set of Killing vectors,
\begin{equation}
\kil^a_A(\pi)=\d^a_A-\left(f^a_{Ab}-\frac12\d^e_Af^a_{eb}\right)\pi^b+\bigO(\pi^2)\;.
\label{kilShifman}
\end{equation}
To make the bilinear (kinetic) part of the Lagrangian~\eqref{LeffShifman} canonically normalized, one can set $\k_{ab}=\d_{ab}$ by a suitable linear transformation of the NG fields. To preserve the form of the exponential parameterization and hence of~\eqref{mcShifman} and~\eqref{kilShifman}, this has to be compensated by a linear transformation of the basis of broken generators $Q_a$. That in turn affects the values of the structure constants. Below, I will assume that such a choice of basis of broken generators has already been made.

With all these pieces at hand, it is easy to construct the Noether currents for the spontaneously broken generators of $G$,
\begin{equation}
\begin{split}
J^\m_a&=\PD{\La_\mathrm{eff}^{(2)}}{(\de_\m\pi^b)}\kil^b_a(\pi)\\
&=\d_{ab}\de^\m\pi^b-\left(\d_{bd}f^d_{ac}+\frac12\d_{ad}f^d_{bc}\right)\de^\m\pi^b\pi^c+\bigO(\pi^3)\;.
\end{split}
\end{equation}
By matching this to~\eqref{curShifman}, we can read off the algebraic coefficients needed in the soft theorem~\eqref{thmShifman}, namely $F^b_a=\d^b_a$ and
\begin{equation}
K_{abc}-K_{acb}=\d_{cd}f^d_{ab}-\d_{bd}f^d_{ac}-\d_{ad}f^d_{bc}\;.
\label{Kabc}
\end{equation}

\begin{illustration}%
\label{ex:semisimpleG}%
Let $G$ be compact and semisimple. Using a faithful matrix representation of $G$, one can define the Cartan--Killing form on the Lie algebra $\lie g$ as $\Delta_{AB}\equiv\tr(Q_AQ_B)$. The cyclicity of trace, $\tr([Q_A,Q_B]Q_C)=\tr(Q_A[Q_B,Q_C])$, implies the identity
\begin{equation}
f^D_{CA}\Delta_{DB}+f^D_{CB}\Delta_{AD}=0\;.
\label{cartankilling}
\end{equation}
This encodes the invariance of the Cartan--Killing form under the adjoint action of $G$ on $\lie g$. At the same time, \eqref{cartankilling} ensures that the covariant structure constant $f_{ABC}\equiv\Delta_{AD}f^D_{BC}$ is fully antisymmetric in its three indices.

It is convenient to choose the basis of broken generators $Q_a$ so that the space $\lie g/\lie h$ is ``orthogonal'' to $\lie h$, that is, $\Delta_{a\b}=0$. The unbroken and broken indices can then be raised and lowered independently using $\Delta_{\a\b}$ and $\Delta_{ab}$, respectively. The constraint $f^c_{\a a}\k_{cb}+f^c_{\a b}\k_{ac}=0$ can be rewritten as $\k^a_{\phantom ac}f^c_{\a b}-f^a_{\a c}\k^c_{\phantom cb}=0$. Thus, the matrix $\k^a_{\phantom ab}$ commutes with the representation of $H$ on the space $\lie g/\lie h$. Suppose in particular that this representation is irreducible, that is, all the NG modes of $G/H$ span a single irreducible multiplet of $H$. This is the case for a number of physically important coset spaces such as $\gr{SO}(n+1)/\gr{SO}(n)\simeq S^n$, $\gr{ISO}(n)/\gr{SO}(n)\simeq\R^n$, or $G_\mathrm{L}\times G_\mathrm{R}/G_\mathrm{V}$ with simple $G$. Then by Schur's lemma, $\k^a_{\phantom ab}$ must be proportional to $\d^a_b$, and thus $\k_{ab}$ to $\Delta_{ab}$. The choice $\k_{ab}=\d_{ab}$ here corresponds to the common normalization of symmetry generators such that $\tr(Q_AQ_B)\propto\d_{AB}$. Finally, \eqref{Kabc} reduces to
\begin{equation}
K_{abc}-K_{acb}\propto f_{cab}-f_{bac}-f_{abc}=f_{abc}\;.
\end{equation}
In this special case, the right-hand side of~\eqref{thmShifman} becomes $(1/2)\smash{\sum\limits_{b\in\a\cup\b}\sum\limits_cf_{abc}\Aa^{b\to c}_{\a\to\b}}$ up to an overall factor.
\end{illustration}

It is obvious from~\eqref{thmShifman} and~\eqref{Kabc} that a sufficient condition for the presence of Adler zero is that the coset space $G/H$ is symmetric so that $f^a_{bc}=0$. In fact, for symmetric coset spaces, the reason for vanishing of the right-hand side of~\eqref{thmShifman} is twofold. Namely, in the exponential parameterization, the automorphism $\Raut(U(\pi))=U(\pi)^{-1}$ is equivalent to the inversion $\pi^a\to-\pi^a$. As a consequence, the components of the broken part of the MC form, $\mc^a_b(\pi)$, are manifestly even functions of the NG fields. Hence, for symmetric coset spaces, the effective Lagrangian~\eqref{LeffShifman} can only give scattering amplitudes with an even number of particles. For theories with only even interaction vertices, the soft theorem~\eqref{thmShifman} applied to an amplitude with an even number of particles yields Adler zero without any further requirements on the coefficients $K_{abc}$.


\section{Geometric Framework for Scattering Amplitudes}
\label{sec:Adlergeom}

In the analysis in Sect.~\ref{subsec:generalizedsoft}, the freedom to redefine the field variables $\pi^a$ was something of a nuisance. Indeed, we had to demonstrate explicitly that our soft theorem~\eqref{thmShifman} is invariant under field reparameterization and thus physically meaningful. Here we will take control and make the freedom to redefine fields work for us. The discussion in this section loosely follows~\cite{Cheung2022}.

Suppose we are given a set of fields $\pi^a$ taking values from some manifold $\M$. For the moment, we do not even have to assume that these are NG fields and $\M$ is a coset space, $\M\simeq G/H$. All we need is that the low-energy physics of the system is captured by the Lagrangian
\begin{equation}
\La_\mathrm{eff}=\frac12g_{ab}(\pi)\de_\m\pi^a\de^\m\pi^b\;.
\label{laggeom}
\end{equation}
In particular, this requires that there are no interactions without derivatives, and any higher-derivative operators can be treated as subleading. Under a point transformation, $\pi^a\to\pi'^a(\pi)$, the symmetric matrix function $g_{ab}(\pi)$ on $\M$ behaves as a rank-2 covariant tensor. Moreover, it must be positive-definite in some neighborhood of the origin, $\pi^a=0$, to give a kinetic term with the correct signature. It can therefore be interpreted as a Riemannian metric on the target manifold $\M$.

Perturbative interactions among the fields $\pi^a$ are generated by expanding the metric in powers of the fields. Individual interaction vertices are thus determined by derivatives of the metric at the origin. However, physical observables must at the same time be invariant under an arbitrary redefinition of the fields. We can use this fact and switch to the Riemannian (geodesic) normal coordinates on $\M$, in which (see Appendix~\ref{appsubsec:normalcoord})
\begin{equation}
g_{ab}(\pi)=g_{ab}(0)-\frac13\pi^c\pi^dR_{acbd}(0)-\frac16\pi^c\pi^d\pi^e(\hat\cd_eR)_{acbd}(0)+\bigO(\pi^4)\;.
\end{equation}
Moreover, we can always assume (and I will henceforth do so) that $g_{ab}(0)=\d_{ab}$ so that the kinetic term in~\eqref{laggeom} is canonically normalized. This shows that any observable such as an on-shell scattering amplitude depends on $g_{ab}(\pi)$ solely through the Riemann curvature tensor and its covariant derivatives at the origin.

To see how the curvature tensor enters the amplitudes, let me spell out explicitly the simplest cases of the 4-particle and 5-particle amplitude,
\begin{align}
\notag
\Aa_{a_1a_2a_3a_4}(p_1,p_2,p_3,p_4)={}&\I\ampfourpt=-s_{24}R_{a_1a_2a_3a_4}-s_{34}R_{a_1a_3a_2a_4}\;,\\
\label{A5}
\Aa_{a_1a_2a_3a_4a_5}(p_1,p_2,p_3,p_4,p_5)={}&\I\ampfivept\\
\notag
={}&-s_{45}\hat\cd_{a_3}R_{a_1a_4a_2a_5}-s_{25}\hat\cd_{a_4}R_{a_1a_2a_3a_5}\\
\notag
&-s_{35}\hat\cd_{a_4}R_{a_1a_3a_2a_5}-s_{34}\hat\cd_{a_5}R_{a_1a_3a_2a_4}\\
\notag
&-(s_{24}+s_{45})\hat\cd_{a_5}R_{a_1a_2a_3a_4}\;.
\end{align}
Here I have introduced a compact notation for a scattering amplitude with particles of energy--momentum $p^\m_i$ and flavor $a_i$. All the energy--momenta are (now and henceforth) oriented inwards. The variables $\smash{s_{ij}\equiv(p_i+p_j)^2}$ generalize the standard Mandelstam variables to processes with an arbitrary number of particles. Finally, I have dropped the argument $(0)$ of the curvature tensor for notation simplicity.

\begin{watchout}%
The presentation of the amplitudes is highly ambiguous due to the redundancies among the Mandelstam variables and the components of the curvature tensor. To arrive at the expressions in~\eqref{A5}, I have adopted the following conventions. First, the variables $s_{1i}$ with any $i\geq2$ can be eliminated through $s_{1i}=-\sum_{j\geq2}s_{ji}$, which follows from overall energy--momentum conservation for massless particles. Second, one of the remaining Mandelstam variables can be eliminated by means of $0=\smash{p_1^2}=\smash{(p_2+p_3+\dotsb)^2}$. I use this to get rid of $s_{23}$. Similarly, all components $R_{a_ia_ja_ka_l}$ can be expressed solely in terms of those for which $i\leq\min(j,k,l)$ simultaneously with $k<l$ and $j\leq\max(k,l)$. This is guaranteed by the symmetries of the curvature tensor and the algebraic Bianchi identity (see Appendix~\ref{appsubsec:symofR}). Finally, the differential Bianchi identity~\eqref{bianchidiff} can be utilized to remove $\hat\cd_{a_m}R_{a_ia_ja_ka_l}$ where $m<\min(k,l)$.
\end{watchout}

With the above remarks out of the way, let us have a look at the result. First, the 4-particle amplitude vanishes in the limit where any of the energy--momenta $p^\m_i$ is taken to zero. This boils down to the very special kinematics of relativistic 4-particle scattering processes. The 5-particle amplitude is more interesting. Our expression is designed to make the soft limit for the last particle transparent,\footnote{The amplitude is of course invariant under any simultaneous permutation of the energy--momenta and the flavor labels. This permutation invariance is just hidden in the economic expression~\eqref{A5}.}
\begin{align}
\notag
\lim_{p_5\to0}\Aa_{a_1a_2a_3a_4a_5}(p_1,p_2,p_3,p_4,p_5)&=-s_{24}\hat\cd_{a_5}R_{a_1a_2a_3a_4}-s_{34}\hat\cd_{a_5}R_{a_1a_3a_2a_4}\\
&=\hat\cd_{a_5}\Aa_{a_1a_2a_3a_4}(p_1,p_2,p_3,p_4)\;.
\end{align}
This is a special case of a geometric soft theorem,
\begin{equation}
\lim_{p_n\to0}\Aa_{a_1\dotsb a_n}(p_1,\dotsc,p_n)=\hat\cd_{a_n}\Aa_{a_1\dotsb a_{n-1}}(p_1,\dotsc,p_{n-1})\;,
\label{softthm}
\end{equation}
valid for any scalar theory that does not contain nonderivative interactions. There may be other, higher-order derivative couplings additional to~\eqref{laggeom}. The covariant derivative of the amplitude in~\eqref{softthm} is taken in the target space $\M$ with respect to the \emph{Levi-Civita} (LC) connection defined by the metric $g_{ab}(\pi)$.

The reader is referred to~\cite{Cheung2022} for a general nonperturbative proof of~\eqref{softthm}. Here we are mostly interested in EFTs for NG bosons, defined on a coset space of spontaneously broken symmetry. For this class of theories, I will derive below a dedicated geometric soft theorem using the more elementary approach of Sect.~\ref{sec:Adler}.


\subsection{Geometric Soft Theorem for Nambu--Goldstone Bosons}
\label{subsec:geomsoftNG}

Inspired by the above discussion, we shall now reinterpret the results of Sect.~\ref{subsec:generalizedsoftcoset} in terms of the geometric properties of the coset space $G/H$. The Noether current for the spontaneously broken generator $Q_a$ can be written as
\begin{equation}
J^\m_a=\PD{\La_\mathrm{eff}^{(2)}}{(\de_\m\pi^b)}\kil^b_a(\pi)=g_{bc}(\pi)\kil^b_a(\pi)\de^\m\pi^c\;.
\end{equation}
Expanding this in powers of the NG fields $\pi^a$ gives
\begin{equation}
J^\m_a=g_{bc}(0)\kil^b_a(0)\de^\m\pi^c+\de_d(g_{bc}\kil^b_a)(0)\de^\m\pi^c\pi^d+\bigO(\pi^3)\;.
\label{curcoset}
\end{equation}

I have already made the assumption that the kinetic term is canonically normalized, that is $g_{ab}(0)=\d_{ab}$. Accordingly, the coupling of the current $J^\m_a(x)$ to the one-particle state $\ket{\smash{\pi^b(\vec p)}}$ is defined by the matrix element $\smash{\amplitude{0}{\smash{J^\m_a(0)}}{\smash{\pi^b(\vec p)}}}=\smash{-\I p^\m_\mathrm{on}\kil^b_a(0)}$. The bilinear part of the current~\eqref{curcoset} yields in turn the algebraic coefficient $K_{abc}=\de_c(g_{db}\kil^d_a)(0)=\de_c(\vec\kil^\flat_a)_b(0)$. We only need the part of this, antisymmetric under the exchange $b\leftrightarrow c$, which can be simplified to
\begin{equation}
K_{abc}-K_{acb}=(\hat\cd_c\vec\kil^\flat_a)_b(0)-(\hat\cd_b\vec\kil^\flat_a)_c(0)=2(\hat\cd_c\vec\kil^\flat_a)_b(0)
\end{equation}
using the Killing equation~\eqref{killing}. Putting all the pieces together, we infer from~\eqref{thmShifman} a geometric soft theorem for the amplitude $\Aa_{a_1\dotsb a_n}(p_1,\dotsc,p_n)$ for scattering of $n$ NG bosons with flavors $a_i$ and energy--momenta $p^\m_i$,
\begin{equation}
\begin{split}
\lim_{p_n\to0}\kil^b_a(0)&\Aa_{a_1\dotsb a_{n-1}b}(p_1,\dotsc,p_n)\\
&=\sum_{i=1}^{n-1}(\hat\cd^b\vec\kil^\flat_a)_{a_i}(0)\Aa_{a_1\dotsb a_{i-1}ba_{i+1}\dotsb a_{n-1}}(p_1,\dotsc,p_{n-1})\;.
\label{softfinal}
\end{split}
\end{equation}
In~\cite{Cheung2022}, the same result was obtained from~\eqref{softthm} by invoking the $G$-invariance of the scattering amplitude in the schematic form
\begin{equation}
\ld{\vec\kil_a}\Aa_{a_1\dotsb a_{n-1}}=\kil^b_a\hat\cd_b\Aa_{a_1\dotsb a_{n-1}}+\sum_{i=1}^{n-1}(\hat\cd_{a_i}\vec\kil_a)^b\Aa_{a_1\dotsb a_{i-1}ba_{i+1}\dotsb a_{n-1}}=0\;.
\end{equation}

The formulation~\eqref{softfinal} of the soft theorem is manifestly covariant under field redefinitions that preserve the normalization of the kinetic term. One can take advantage of this and compute the algebraic coefficient $\smash{(\hat\cd^b\vec\kil^\flat_a)_{a_i}(0)}$ in suitably chosen local coordinates on $G/H$. Specifically, in the exponential parameterization, we already have an expression for the Killing vector in~\eqref{kilShifman}. Likewise, the expression~\eqref{mcShifman} for the MC form can be used to recover the metric $g_{ab}(\pi)$ and in turn the Christoffel symbols, needed to evaluate the covariant derivative of the Killing vector. Putting everything together, one finds
\begin{equation}
(\hat\cd_c\vec\kil^\flat_a)_b(0)=\frac12(\d_{cd}f^d_{ab}-\d_{bd}f^d_{ac}-\d_{ad}f^d_{bc})\;,
\label{dxi}
\end{equation}
in accord with~\eqref{Kabc}. This confirms that the soft theorems~\eqref{thmShifman} and~\eqref{softfinal}, formulated respectively in terms of the algebraic and geometric properties of the coset space, are equivalent.


\subsection{Adler Zero or Not?}
\label{subsec:Adlerornot}

We have already observed that if the coset space $G/H$ is symmetric, the right-hand side of the soft theorem~\eqref{thmShifman} or~\eqref{softfinal} automatically vanishes. Is the symmetry of $G/H$ also a necessary condition for the Adler zero? We cannot exclude the possibility that the soft limit of a specific amplitude in a given theory vanishes due to fine tuning of the effective couplings. However, in order that \emph{all} amplitudes of the EFT vanish in the soft limit, the coefficients~\eqref{dxi} must vanish for any choice of $a,b,c$. Using the shorthand notation $f_{abc}\equiv\d_{ad}f^d_{bc}$, the vanishing of~\eqref{dxi} is in turn equivalent to $f_{abc}=f_{bca}+f_{cab}$. Applying this twice gives
\begin{equation}
f_{abc}=f_{bca}+f_{cab}=(f_{cab}+f_{abc})+(f_{abc}+f_{bca})=2f_{abc}+f_{bca}+f_{cab}\;.
\end{equation}
This is only possible if $f_{abc}=0$. Thus, the set of all scattering amplitudes of the EFT satisfies the Adler zero principle if and only if the coset space $G/H$ is symmetric.

Incidentally, there is a neat geometric way to understand the one-way implication, ensuring Adler zero for symmetric coset spaces. Namely, the latter have the property that all covariant derivatives of the Riemann curvature tensor identically vanish (Theorem 10.19 in~\cite{Lee2018}). The Adler zero can then be seen as a direct consequence of~\eqref{softthm}. On coset spaces that are not symmetric, the nontrivial soft limit of scattering amplitudes is governed by the structure constants $f^a_{bc}$. According to Sect.~\ref{subsec:geometry_connection}, these are in a one-to-one correspondence with the (frame components of the) torsion 2-form of the canonical connection on $G/H$. Thus, the nonvanishing soft limit can be said to arise geometrically from the torsion of the coset space.

\begin{illustration}%
One of the simplest symmetry groups leading to a nonvanishing soft limit of some scattering amplitudes is the Heisenberg group $\gr{H}_3$. This is a three-dimensional nilpotent Lie group. One can choose a basis of its Lie algebra $\lie h_3$, $Q_a$ with $a=1,2$ and $Q$, so that the only nontrivial commutation relation among the generators is
\begin{equation}
[Q_a,Q_b]=\I\ve_{ab}Q\;.
\label{Heisenbergcomm}
\end{equation}
Suppose the symmetry of a system under the Heisenberg group is completely broken. Denoting the NG fields associated with $Q_a$ and $Q$ respectively as $\pi^a$ and $\t$, the coset space $\gr{H}_3/\trgr\simeq\gr{H}_3$ can be parameterized by
\begin{equation}
U(\pi,\t)=\exp\biggl(\frac\I v\pi^aQ_a\biggr)\exp\biggl(\frac\I v\t Q\biggr)\;,
\label{Heisenbergparam}
\end{equation}
where $v$ is a positive dimensionful constant. The corresponding components of the MC form are $\mc^a=(1/v)\D\pi^a$ and $\mc^\t=(1/v)\D\t+1/(2v^2)\ve_{ab}\pi^a\D\pi^b$. The most general two-derivative effective Lagrangian would now be given by a generic rank-2 symmetric tensor built out of the MC form. For illustration, it is however sufficient to consider a particularly simple special case,
\begin{equation}
\La^{(2)}_\mathrm{eff}=\frac12\d_{ab}\de_\m\pi^a\de^\m\pi^b+\frac12\left(\de_\m\t+\frac1{2v}\ve_{ab}\pi^a\de_\m\pi^b\right)^2\;.
\label{HeisenbergLag}
\end{equation}
This makes the calculation of various scattering amplitudes easy. Namely, the sole cubic interaction operator equals, up to a surface term, $-1/(2v)\t\ve_{ab}\pi^a\Box\pi^b$. As a consequence, any Feynman diagram where a $\t$-propagator is attached to two external $\pi$-type legs will vanish on-shell.

Any on-shell 3-particle amplitude in a derivatively coupled relativistic theory of massless scalars automatically vanishes. Hence we need to consider a process with at least five particles to have a hope for a nonzero soft limit. Leaving out straightforward details, a simple example is
\begin{equation}
\Aa_{\t1222}(p_1,p_2,p_3,p_4,p_5)=\frac{s_{12}}{2v^3}\;.
\label{At1222}
\end{equation}
In order to compare this to the prediction of our soft theorem~\eqref{softfinal}, we have to keep in mind the factors of $v$ in~\eqref{Heisenbergparam}. These make the only nonzero structure constant effectively $\smash{f^\t_{ab}=\ve_{ab}/v}$. Together with $\kil^b_a(0)=\d^b_a$, \eqref{softfinal} and~\eqref{dxi} then predict
\begin{equation}
\begin{split}
\lim_{p_5\to0}\Aa_{\t1222}(p_1,\dotsc,p_5)=\frac1{2v}[&\Aa_{1122}(p_1,\dotsc,p_4)\\
&-\Aa_{\t\t22}(p_1,\dotsc,p_4)]\;.
\end{split}
\end{equation}
This is easily verified by calculating the two 4-particle amplitudes and finding that
\begin{equation}
\Aa_{1122}(p_1,p_2,p_3,p_4)=\frac{3s_{12}}{4v^2}\;,\qquad
\Aa_{\t\t22}(p_1,p_2,p_3,p_4)=-\frac{s_{12}}{4v^2}\;.
\end{equation}
By permutation invariance, the same kind of result applies to the soft limit in which either $\smash{p^\m_3}$ or $\smash{p^\m_4}$ is taken to zero. On the other hand, the soft limit of $\Aa_{\t1222}(p_1,\dotsc,p_5)$ for $\smash{p^\m_1}$ or $\smash{p^\m_2}$ manifestly vanishes, as is clear from~\eqref{At1222}.

The cubic interaction operator in~\eqref{HeisenbergLag} can be removed by the field redefinition $\pi^a=\p^a+c\t\ve^a_{\phantom ab}\p^b$ with $c\equiv1/(2v)$. This leads to the Lagrangian
\begin{equation}
\begin{split}
\La^{(2)}_\mathrm{eff}={}&\frac12(1+c^2\t^2)\de_\m\vec\p\cdot\de^\m\vec\p+\frac12(1-c^2\vec\p^2)(\de_\m\t)^2+\frac{c^2}4\de_\m(\t^2)\de^\m(\vec\p^2)\\
&+\frac{c^3}3\de_\m(\t^3)\vec\p\times\de^\m\vec\p+\frac{c^2}2\bigl[(1+c^2\t^2)\vec\p\times\de_\m\vec\p-c\vec\p^2\de_\m\t\bigr]^2\;,
\end{split}
\label{heisenbergredef}
\end{equation}
where I used the notation $\vec\p\equiv(\p^1,\p^2)$ and $\vec\p\times\de_\m\vec\p\equiv\ve_{ab}\p^a\de_\m\p^b$ to avoid proliferation of indices. The first term on the second line of~\eqref{heisenbergredef} is equivalent to $-(c^3/3)\t^3\vec\p\times\Box\vec\p$ by partial integration, and thus does not contribute to tree-level 5-particle amplitudes of the theory. All such amplitudes are therefore encoded in a single operator, $-c^3\de^\m\t\vec\p^2(\vec\p\times\de_\m\vec\p)$. This shows that $\Aa_{\t1222}$ and $\Aa_{\t2111}$ are the only nonzero 5-particle amplitudes, and makes it easy to reproduce~\eqref{At1222}.
\end{illustration}


\subsection{Symmetric Coset Spaces}
\label{subsec:geomsymcoset}

Expressing the scattering amplitudes in terms of the curvature tensor and its covariant derivatives is elegant and gives us useful insight. However, calculating the covariant derivatives of the curvature tensor from the metric $g_{ab}(\pi)$ in practice may be a tiresome task. The situation is much better for symmetric coset spaces where all the covariant derivatives of the curvature tensor vanish. Here we have a closed expression for the Riemannian metric in the normal coordinates (see Appendix~\ref{appsubsec:normalcoord}),
\begin{equation}
g_{ab}(\pi)=g_{ac}(0)\biggl[\frac{\sin^2\!\sqrt{\hat R(\pi)}}{\hat R(\pi)}\biggr]^c_{\phantom cb}=g_{ac}(0)\sum_{k=0}^\infty\frac{(-1)^k2^{2k+1}}{(2k+2)!}[\hat R(\pi)^k]^c_{\phantom cb}\;,
\label{gsymcoset}
\end{equation}
where $\hat R^a_{\phantom ab}(\pi)\equiv R^a_{\phantom acbd}(0)\pi^c\pi^d$. All we need to know are the constants $R^a_{\phantom acbd}(0)$. These can be extracted from our discussion of the geometry of coset spaces in Sect.~\ref{sec:CCWZgeometry}. First, however, a word of caution is in place.

Throughout Appendix~\ref{app:diffgeom}, I dutifully distinguish components of tensors in a local frame from those in a local coordinate basis by using different types of indices. This has not been necessary in the main text of the book so far. Here, to avoid confusion, I will use lowercase indices $a,b,\dotsc$ to indicate a local coordinate system, and underlined indices $\fr a,\fr b,\dotsc$ for components in the frame defined by the MC form. Thus, for instance, a comparison of~\eqref{LeffShifman} and~\eqref{laggeom} shows that the $G$-invariant Riemannian metric on the coset space is $g(\pi)=\k_{\fr a\fr b}\mc^{\fr a}(\pi)\otimes\mc^{\fr b}(\pi)$. The components of the metric in the MC form basis are constant, as shown explicitly in Sect.~\ref{subsec:effLag2der}.

According to Sect.~\ref{subsec:geometry_connection}, the same is true for the curvature 2-form of the LC connection on symmetric coset spaces,
\begin{equation}
R^{\fr a}_{\phantom{\fr a}\fr b}(\pi)=-\frac12f^{\fr a}_{\a\fr b}f^\a_{\fr c\fr d}\mc^{\fr c}(\pi)\w\mc^{\fr d}(\pi)\;.
\end{equation}
Combining this with~\eqref{gsymcoset} gives upon a brief manipulation a practically more useful expression for the metric,
\begin{equation}
g(\pi)=\k_{\fr a\fr c}\biggl[\frac{\sin^2\!\sqrt{\hat R(\pi)}}{\hat R(\pi)}\biggr]^{\fr c}_{\phantom{\fr c}\fr b}\D\pi^{\fr a}\otimes\D\pi^{\fr b}\;,\qquad
\hat R^{\fr a}_{\phantom{\fr a}\fr b}(\pi)\equiv-f^{\fr a}_{\a\fr c}f^\a_{\fr b\fr d}\pi^{\fr c}\pi^{\fr d}\;,
\label{gsymcosetfinal}
\end{equation}
where $\pi^{\fr a}\equiv\mc^{\fr a}_a(0)\pi^a$. Here $\mc^{\fr a}_a(0)$ is a constant matrix that can easily be evaluated from the definition of the MC form, $\mcb(\pi)=\mc^{\fr a}(\pi)Q_{\fr a}=\mc^{\fr a}_a(\pi)Q_{\fr a}\D\pi^a$. Combining~\eqref{laggeom} with~\eqref{gsymcosetfinal} determines the complete two-derivative effective Lagrangian on symmetric coset spaces in a closed form.

The structure constant $f^a_{\a c}$ in~\eqref{gsymcosetfinal} defines the adjoint action of the unbroken subgroup $H$ on the broken generators, and thus also the NG fields. For compact and semisimple groups $G$, the other type of structure constant in~\eqref{gsymcosetfinal}, $f^\a_{bd}$, can be related to the former using the Cartan--Killing form on $\lie g$; cf.~\refex{ex:semisimpleG}. This leads to the intriguing conclusion that the effective Lagrangian~\eqref{LeffShifman} is completely fixed by the set of low-energy constants $\k_{ab}$, spanning a symmetric invariant tensor of $H$, and certain linear representation of $H$. No specific information about the full symmetry group $G$ is needed, as was first emphasized in~\cite{Low2015a}.

\begin{illustration}%
The requirement that $G$ be compact and semisimple is essential. Contrast the coset spaces $\gr{SO}(n+1)/\gr{SO}(n)\simeq S^n$ and $\gr{ISO}(n)/\gr{SO}(n)\simeq\R^n$. Both of these are symmetric, share the same $H$, and in both the set of broken generators transforms as a vector of $\gr{SO}(n)$. Yet, the corresponding EFTs are different; $\gr{SO}(n+1)$ is compact but $\gr{ISO}(n)$ is not. In case of $\gr{SO}(n+1)/\gr{SO}(n)$, the low-energy EFT is most easily presented as a nonlinear sigma model for a linearly-transforming unit vector field, $\vec n\in S^n$. It implies nontrivial interactions among the $n$ NG bosons. On the other hand, in case of $\gr{ISO}(n)/\gr{SO}(n)$, the leading-order EFT~\eqref{LeffShifman} reduces to a noninteracting theory, $\La_\mathrm{eff}=(1/2)\d_{ab}\de_\m\pi^a\de^\m\pi^b$. This follows from the fact that the broken generators of $\gr{ISO}(n)$ commute with each other so that $f^\a_{bc}=0$. The curvature tensor is trivial. After all, $\gr{ISO}(n)/\gr{SO}(n)$ is just the flat Euclidean space.
\end{illustration}


\section{Beyond Adler Zero}
\label{sec:beyondAdler}

Until now, we have been solely preoccupied with the question whether or not a given scattering amplitude vanishes in the soft limit for a chosen particle. This amounts to restricting the amplitude to a special hypersurface in the space of all allowed (on-shell and energy--momentum-conserving) kinematical configurations. There are however other interesting kinematical regimes that also probe the algebraic and geometric structure of the underlying EFT. One simple possibility is to take a consecutive soft limit for two particles. Here the geometric soft theorem~\eqref{softthm} proves invaluable, giving immediately
\begin{equation}
\lim_{p_{n-1}\to0}\lim_{p_n\to0}\Aa_{a_1\dotsb a_n}(p_1,\dotsc,p_n)=\hat\cd_{a_n}\hat\cd_{a_{n-1}}\Aa_{a_1\dotsb a_{n-2}}(p_1,\dotsc,p_{n-2})\;.
\end{equation}
Of particular interest is the commutator of the two limits, probing the extent to which the limits interfere with each other,
\begin{equation}
\begin{split}
\bigl[\lim_{p_{n-1}\to0},\lim_{p_n\to0}\bigr]&\Aa_{a_1\dotsb a_n}(p_1,\dotsc,p_n)\\
&=\sum_{i=1}^{n-2}R^b_{\phantom ba_ia_{n-1}a_n}\Aa_{a_1\dotsb a_{i-1}ba_{i+1}\dotsb a_{n-2}}(p_1,\dotsc,p_{n-2})\;,
\end{split}
\end{equation}
where I used~\eqref{Rcommtenscoord}. Another possibility is to take a simultaneous soft limit for two or even more particles. This may lead to a kinematic singularity even in the absence of cubic interaction vertices. See~\cite{Cheung2022,Low2016a} for a precise formulation of such a simultaneous double-soft theorem and further details.

Here I will remain within the confines of the single soft limit, but look more closely at the asymptotic behavior of the scattering amplitude near the limit. I will promote the energy--momenta $p^\m_i$ in an $n$-particle process to functions $\smash{\hat p^\m_i(z)}$ of a scaling parameter $z\in\C$ such that $\smash{\hat p_i^\m(1)=p^\m_i}$ for all $i$ and moreover $\smash{\hat p^\m_n(z)=zp^\m_n}$. The other functions $\smash{\hat p^\m_i(z)}$ with $i=1,\dotsc,n-1$ must be chosen so as to preserve overall energy--momentum conservation, $\smash{\sum_{i=1}^n\hat p^\m_i(z)=0}$, and remain on-shell, $\smash{[\hat p_i(z)]^2=0}$, for any $z\in\C$. Also, it is required that $\smash{\hat p^\m_i(z)}$ with $i=1,\dotsc,n-1$ have a nonzero limit for $z\to0$, that is, only the $n$-th particle becomes soft for small $z$.

For a generic kinematical configuration, the complexified tree-level amplitude $\smash{\hat\Aa_{a_1\dotsb a_n}(\hat p_1(z),\dotsc,\hat p_n(z))}$ will presumably be analytic in $z$ in some neighborhood of the origin. It can thus be expanded in a series with a nonzero radius of convergence,
\begin{equation}
\hat\Aa_{a_1\dotsb a_n}(\hat p_1(z),\dotsc,\hat p_{n-1}(z),zp_n)=\sum_{k=0}^\infty z^kc^{(k)}_{a_1\dotsb a_n}\;.
\label{ts}
\end{equation}
The soft theorems derived in Sects.~\ref{sec:Adler} and~\ref{sec:Adlergeom} determine the leading term, $\smash{c^{(0)}_{a_1\dotsb a_n}}$, in terms of $(n-1)$-particle amplitudes with energy--momenta $\hat p^\m_1(0),\dotsc,\hat p^\m_{n-1}(0)$. The asymptotic behavior of the amplitude at small $z$ is dominated by the first term in~\eqref{ts} for which $\smash{c^{(k)}_{a_1\dotsb a_n}}$ is nonzero. In the following, I will refer to the corresponding value of $k$ as the \emph{soft scaling parameter} and use the symbol $\s$;
\begin{equation}
\hat\Aa_{a_1\dotsb a_n}(\hat p_1(z),\dotsc,\hat p_{n-1}(z),zp_n)=z^\s c^{(\s)}_{a_1\dotsb a_n}+\bigO(z^{\s+1})\;.
\label{scalingparameterdef}
\end{equation}

\begin{watchout}%
The notation used in~\eqref{ts} may suggest that the coefficients $\smash{c^{(k)}_{a_1\dotsb a_n}}$ are independent of the detailed choice of functions $\hat p_i(z)$ beyond the basic requirements laid out above~\eqref{ts}. I am not aware of any proof of this implicit assumption, or even a study addressing the issue. Still, it is plausible to assume that at least the leading term in~\eqref{scalingparameterdef} is independent of such arbitrary choices.
\end{watchout}

Clearly, $\s=0$ if the soft limit of the amplitude is nontrivial (nonvanishing). The Adler zero amounts to $\s\geq1$. Barring accidental cancellations, we expect a generic EFT satisfying the Adler zero principle to have $\s=1$.

\begin{illustration}%
Recall that for $n=4$, all the Mandelstam variables $s,t,u$ vanish in the soft limit for any of the four particles. This automatically guarantees Adler zero for the 4-particle amplitude in any derivatively coupled EFT of massless scalars. Moreover, in EFTs of a single species of massless scalar, permutation invariance allows only one linear function of Mandelstam variables, $s+t+u$, which is identically zero. Hence, in EFTs of a single NG boson, the 4-particle amplitude automatically satisfies $\s\geq2$.
\end{illustration}

In case $\s\geq2$ for \emph{all} (tree-level) scattering amplitudes in a given EFT, we say that the soft limit of the amplitudes is \emph{enhanced}. The rest of the present section is devoted to two immediate questions. Are there, in fact, any theories where scattering amplitudes feature an enhanced soft limit? If yes, is there any generic mechanism that makes the soft limit enhanced beyond the simple Adler zero?


\subsection{Dirac--Born--Infeld Theory}
\label{subsubsec:DBI}

The cleanest way to answer the first question is to find an explicit example. We will make an educated guess, loosely following~\cite{Cheung2015a}. Consider the class of EFTs of a single real NG field, $\pi$. We already know that $\s\geq2$ for the 4-particle amplitude. In fact, the constraint $s+t+u=0$ leaves us with a single candidate, algebraically independent permutation-invariant 4-particle amplitude with $\s=2$, namely $\Aa(p_1,p_2,p_3,p_4)=(s^2+t^2+u^2)/(4\Lambda_D)$. Here $\Lambda_D$ is a nonzero constant of mass dimension $D$, the dimension of spacetime; the factor of 4 is a mere convention. This amplitude can be reproduced by a local quartic interaction operator, $-[(\de_\m\pi)^2]^2/(8\Lambda_D)$. This interaction in turn gives nontrivial amplitudes also for any even number of particles higher than four.

The problem is that already the next, $n=6$ amplitude fails to satisfy the desired scaling with $\s=2$. By dimensional analysis, the $n=6$ amplitude is a homogeneous function of degree $6$ in energy--momenta. A detailed computation shows that the leading, $\smash{c^{(1)}}$ term in its expansion~\eqref{ts} can be canceled by adding a new interaction operator carrying six derivatives, $[(\de_\m\pi)^2]^3/(16\Lambda_D)$. This pattern extends inductively to all higher orders. For any $n$, the $\s=2$ scaling of the (tree-level) $(2n)$-particle amplitude can be rescued by adding an interaction operator $[(\de_\m\pi)^2]^n$ and tuning its coefficient to a unique value. This ensures cancellation of the $\smash{c^{(1)}}$ contribution to the $(2n)$-particle amplitude, generated by exchange diagrams with lower-point interaction vertices. Altogether, we end up with a unique theory modulo the choice of $\Lambda_D$; all the iteratively constructed interaction operators fold neatly into a closed-form Lagrangian,
\begin{equation}
\La_\mathrm{DBI}=\Lambda_D\sqrt{1+(\de_\m\pi)^2/\Lambda_D}\;.
\label{DBI}
\end{equation}
This is the so-called \emph{Dirac--Born--Infeld} (DBI) \emph{theory}.

A combination of~\eqref{DBI} with the volume element, $\smash{\D^D\!x\sqrt{1+(\de_\m\pi)^2/\Lambda_D}}$, admits a remarkable geometric interpretation. This is the induced volume measure on a $D$-dimensional hypersurface (``brane''), embedded in a flat $(D+1)$-dimensional spacetime and parameterized by the Minkowski coordinates $x^\m$. The NG field $\pi(x)$ arises from spontaneous breaking of the symmetry under translations in the extra dimension by the brane. Properly normalized, the displacement of the brane along the extra dimension is $\pi(x)/\sqrt{\abs{\Lambda_D}}$. The sign of $\Lambda_D$ determines the signature of the metric in the extra dimension. For positive $\Lambda_D$, the spacetime Lorentz group $\gr{SO}(d,1)$ is thus extended to $\gr{SO}(d,2)$, whereas for negative $\Lambda_D$ it is extended to $\gr{SO}(d+1,1)$. Up to an arbitrary choice of scale, we end up with two distinct DBI theories, differing by the geometry in the extra dimension.

The fact that the DBI Lagrangian~\eqref{DBI} does not contain any fields without derivatives automatically guarantees the Adler zero. The realization of the $\s=2$ enhancement is however highly nontrivial. At first sight, it is not even obvious why the order-by-order cancellations required to make the soft limit enhanced should be possible at all. In Chap.~\ref{chap:differences}, I will show that this is a consequence of the extended symmetry of the DBI theory. Let us therefore have a closer look at this symmetry. Under the translation in the extra dimension, the NG field shifts simply as $\pi(x)\to\pi'(x)=\pi(x)+\eps$. What is more interesting are the rotations connecting the $D$ physical dimensions to the extra dimension. These can be parameterized by a Lorentz vector $\eps^\m$ and their infinitesimal form reads
\begin{equation}
\udelta\pi(x)=\sqrt{\abs{\Lambda_D}}\eps_\m x^\m\;,\qquad
\udelta x^\m=-\frac{\sgn\Lambda_D}{\sqrt{\abs{\Lambda_D}}}\eps^\m\pi(x)\;.
\label{symDBI}
\end{equation}
Note that this is not a spacetime symmetry by the definition given in Sect.~\ref{sec:whatissym}, since $\udelta x^\m$ depends on the field $\pi(x)$. Rather, \eqref{symDBI} is an example of a generalized local transformation. Its evolutionary form is
\begin{equation}
\pi'(x)-\pi(x)=\sqrt{\abs{\Lambda_D}}\eps_\m x^\m+\frac{\sgn\Lambda_D}{\sqrt{\abs{\Lambda_D}}}\eps^\m\pi(x)\de_\m\pi(x)\;,
\label{DBIsym}
\end{equation}
which depends on both $\pi(x)$ and its derivative. It is easy to verify that~\eqref{DBI} is quasi-invariant under~\eqref{DBIsym}.


\subsection{Galileon and Special Galileon Theory}
\label{subsubsec:Galileon}

Operationally, the form of the DBI theory is so strongly constrained because its Lagrangian only depends on the first derivatives of the NG field. One might hope to find more examples of theories with scattering amplitudes enhanced beyond the plain Adler zero by allowing for higher derivatives. Indeed, one can realize $\s=2$ trivially by setting
\begin{equation}
\La_\mathrm{eff}=\frac12(\de_\m\pi)^2+\La_\mathrm{int}(\de\de\pi)\;,
\label{Lddpi}
\end{equation}
where $\La_\mathrm{int}$ is an arbitrary function of second derivatives of $\pi(x)$ whose Taylor expansion in $\pi$ starts at the fourth or higher order. This, while not very interesting per se, points in the right direction. Namely, the class of theories~\eqref{Lddpi} is manifestly invariant under the so-called \emph{Galileon} symmetry,
\begin{equation}
\udelta\pi(x)=\eps+\eps_\m x^\m\;.
\label{symgal}
\end{equation}
The interaction Lagrangian is strictly invariant, whereas the kinetic term is quasi-invariant. Mere invariance under the constant shift with scalar parameter $\eps$, together with the absence of cubic interaction vertices, already guarantees Adler zero. However, it is ultimately the full Galileon symmetry~\eqref{symgal} that makes the soft limit of the scattering amplitudes enhanced; I will again return to this in Chap.~\ref{chap:differences}. It is now natural to ask whether there might be any interactions that, similarly to the kinetic term, are merely quasi-invariant under~\eqref{symgal}. If present, such interactions would necessarily contain less than two derivatives per field, and thus realize a soft limit with $\s=2$ in a nontrivial manner. Moreover, by power counting, they would dominate over the interactions of type $\La_\mathrm{int}(\de\de\pi)$ in~\eqref{Lddpi}.

Finding all quasi-invariant Lagrangians requires an inspection of the relative Lie algebra cohomology of the spontaneously broken symmetry, as observed in a special case in Sect.~\ref{sec:effLagstructure}. In case of the Galileon symmetry, a detailed analysis shows~\cite{Goon2012a} that in $D$ spacetime dimensions, there are $D+1$ algebraically independent quasi-invariant Galileon Lagrangians,
\begin{equation}
\La^{(n)}_\mathrm{Gal}\equiv\ve^{\m_1\dotsb\m_n\l_{n+1}\dotsb\l_D}\ve^{\n_1\dotsb\n_n}_{\phantom{\n_1\dotsb\n_n}\l_{n+1}\dotsb\l_D}\pi(\de_{\m_1}\de_{\n_1}\pi)\dotsb(\de_{\m_n}\de_{\n_n}\pi)\;,
\label{galWZ}
\end{equation}
where $n=0,\dotsc,D$. The quasi-invariance of all $\smash{\La^{(n)}_\mathrm{Gal}}$ under the Galileon symmetry~\eqref{symgal} is manifest. The first of these, $\smash{\La^{(0)}_\mathrm{Gal}=(-1)^{D-1}D!\pi}$, is a mere tadpole. This should be discarded in order for the EFT to be perturbatively well-defined. The next Galileon Lagrangian, $\smash{\La^{(1)}_\mathrm{Gal}=(-1)^{D-1}(D-1)!\pi\Box\pi}$, is just the kinetic term up to a rescaling and integration by parts. Each of the interaction Lagrangians $\smash{\La^{(n)}_\mathrm{Gal}}$ with $n\geq2$, or any linear combination thereof,  realizes the $\s=2$ enhanced soft limit of scattering amplitudes. One might be concerned about the presence of the cubic operator, $\smash{\La^{(2)}_\mathrm{Gal}}$. This is however harmless. Namely, within the class of Lagrangians
\begin{equation}
\La_\mathrm{Gal}=\sum_{n=0}^Dc_n\La^{(n)}_\mathrm{Gal}\;,
\label{galclass}
\end{equation}
the couplings $c_n$ can be recalibrated by a one-parameter group of transformations, known as the \emph{Galileon duality}~\cite{Rham2014a,Kampf2014a}. This amounts to a field redefinition, and thus does not affect the $S$-matrix of the theory. It can always be used to set $c_2$ to zero.

The 4-particle amplitude in the Galileon theory~\eqref{galclass} is a homogeneous function of degree $6$ in the particle energy--momenta. Permutation invariance together with the constraint $s+t+u=0$ implies that the amplitude equals $s^3+t^3+u^3$ up to overall normalization. Hence, in Galileon theories, the 4-particle amplitude is doubly enhanced, $\s=3$. Is it possible to adjust the values of $c_n$ in~\eqref{galclass} so as to maintain $\s=3$ for all multiparticle amplitudes? The answer to this question is positive, and the resulting one-parameter family of theories is dubbed \emph{special Galileon}.

The further enhancement of the soft limit of scattering amplitudes is associated with yet another, hidden symmetry~\cite{Hinterbichler2015a}. Consider the infinitesimal transformation
\begin{equation}
\udelta_\eps\pi(x)\equiv\eps^{\m\n}\left[x_\m x_\n+\frac1{\Lambda_{D+2}}\de_\m\pi(x)\de_\n\pi(x)\right]\;,
\label{sgalsym}
\end{equation}
where $\Lambda_{D+2}$ is a (nonzero) constant with mass dimension $D+2$ and $\eps^{\m\n}$ is a traceless symmetric tensor of parameters. This is another nontrivial example of a generalized local transformation. To check whether~\eqref{sgalsym} leaves the Lagrangian~\eqref{galclass} quasi-invariant, it is sufficient to inspect the variation of the shift of the corresponding action, $\udelta_\eps S_\mathrm{Gal}$, with respect to $\pi$. A direct calculation gives
\begin{equation}
\begin{split}
\frac{\udelta(\udelta_\eps S_\mathrm{Gal})}{\udelta\pi}&=2\sum_{n=0}^Dc_n\biggl[n(n+1)X_n+\frac1{\Lambda_{D+2}}(D-n)(D-n-1)X_{n+2}\biggr]\;,\\
X_n&\equiv\ve^{\m_1\dotsb\m_n\l_{n+1}\dotsb\l_D}\ve^{\n_1\dotsb\n_n}_{\phantom{\n_1\dotsb\n_n}\l_{n+1}\dotsb\l_D}\eps_{\m_n\n_n}\prod_{i=1}^{n-1}(\de_{\m_i}\de_{\n_i}\pi)\;;
\end{split}
\label{dSgal}
\end{equation}
a reader wishing to verify this might want to take advantage of the auxiliary identity
\begin{align}
\notag
&\ve^{\m_1\dotsb\m_{n+2}\l_{n+3}\dotsb\l_D}\ve^{\n_1\dotsb\n_{n+2}}_{\phantom{\n_1\dotsb\n_{n+2}}\l_{n+3}\dotsb\l_D}\eps_{\m_{n+2}\n_{n+2}}\prod_{i=1}^{n+1}(\de_{\m_i}\de_{\n_i}\pi)\\
&=\frac{n+1}{(D-n)(D-n-1)}\ve^{\m_1\dotsb\m_n\l_{n+1}\dotsb\l_D}\ve^{\n_1\dotsb\n_n}_{\phantom{\n_1\dotsb\n_n}\l_{n+1}\dotsb\l_D}\\
\notag
&\phantom{{}={}}\times\eps^{\a\b}\biggl[n(\de_\a\de_{\m_n}\pi)(\de_\b\de_{\n_n}\pi)\prod_{i=1}^{n-1}(\de_{\m_i}\de_{\n_i}\pi)-(\de_\a\de_\b\pi)\prod_{i=1}^n(\de_{\m_i}\de_{\n_i}\pi)\biggr]\;.
\end{align}
The quasi-invariance of the Lagrangian~\eqref{galclass} is now equivalent to the vanishing of~\eqref{dSgal}, which leads to a recurrence relation for the couplings $c_n$, solved by
\begin{equation}
c_{2n}=\frac{(-1)^n}{\Lambda_{D+2}^n}\binom D{2n}\frac{c_0}{2n+1}\;,\qquad
c_{2n+1}=\frac{(-1)^n}{\Lambda_{D+2}^n}\frac1D\binom{D}{2n+1}\frac{c_1}{n+1}\;.
\end{equation}

What we have here are in principle two independent types of theories. However, the even set $c_{2n}$ inevitably includes the tadpole operator $\smash{\La^{(0)}_\mathrm{Gal}}$ that we have discarded on physical grounds. We are thus left only with the odd terms $c_{2n+1}$, that is those Galileon operators~\eqref{galWZ} that are even in the NG field $\pi$. The parameter $c_1$ is fixed by demanding proper normalization of the kinetic term. The only genuine parameter of the special Galileon theory is therefore the scale $\Lambda_{D+2}$ that enters the symmetry transformation~\eqref{sgalsym}. Specifically in $D=4$ spacetime dimensions, we have $c_3=-c_1/(2\Lambda_6)$ as the sole effective coupling. Upon some integration by parts, the Lagrangian of the special Galileon theory can then be cast as
\begin{equation}
\La^{D=4}_\mathrm{sGal}\simeq\frac12(\de_\m\pi)^2-\frac1{12\Lambda_6}(\de_\m\pi)^2\bigl[(\Box\pi)^2-(\de_\n\de_\l\pi)^2\bigr]\;.
\label{sgallag}
\end{equation}
See~\cite{Preucil2019} for a recent overview of the many intriguing properties of the special Galileon theory.

So far we have discovered three theories that realize in a nontrivial manner an enhanced soft limit of scattering amplitudes: DBI, Galileon and special Galileon. Remarkably, this is it as far as relativistic EFTs of a single NG boson are concerned~\cite{Cheung2017a}. Mapping the landscape of theories with $\s\geq2$ and multiple NG bosons remains, as of writing this book, an open problem. Below, I outline a symmetry-based approach that makes the identification of candidate multiflavor EFTs in principle straightforward. In Sect.~\ref{sec:softrecursion}, I will then introduce an entirely different method, which utilizes solely known infrared properties of on-shell amplitudes and general quantum-field-theoretic principles. This approach is very efficient in proving that other nontrivial soft behavior than that observed in the above three theories is not possible.


\subsection{Effective Theories with Enhanced Soft Limit from Symmetry}
\label{subsec:enhancedliealgebra}

I have already hinted, so far without proof, that the enhanced soft limit of scattering amplitudes in the DBI and Galileon theories is a consequence of symmetry. The additional symmetry cannot be internal, for that would lead to mere linear constraints on scattering amplitudes or, even worse, additional NG bosons if spontaneously broken. Indeed, a glance at~\eqref{symDBI} and~\eqref{symgal} shows that we are dealing with transformations with nontrivial dependence on spacetime coordinates, parameterized by a Lorentz vector. Guided by this observation, I will now show that the presence of such symmetries is severely constrained by group-theoretical consistency requirements. The discussion in this subsection is a special case of a framework developed in~\cite{Bogers2018b}.

Consider an EFT for a single species of NG bosons. The symmetry of this EFT is generated by the operators of angular momentum $J_{\m\n}$ and energy--momentum $P_\m$, and a spontaneously broken scalar generator $Q$. The commutation relations between $J_{\m\n}$ and $P_\m$ are fixed by the Poincar\'e algebra. Furthermore, $[J_{\m\n},Q]=0$ expresses the fact that $Q$ is a scalar, whereas $[P_\m,Q]=0$ is needed for $Q$ to generate an internal symmetry. Let us now add a Lorentz-vector operator $K_\m$, generating the tentative hidden symmetry responsible for enhancement of the soft limit of scattering amplitudes. Lorentz invariance restricts the commutation relations between $K_\m$ and the other generators to\footnote{Additional terms containing the LC tensor can be added to some of the commutators in $D=3$ or $D=4$ spacetime dimensions. These however turn out to be irrelevant in the sense that they are either forbidden by Jacobi identities or can be absorbed into a redefinition of the generators.}
\begin{equation}
\begin{alignedat}{3}
[J_{\m\n},K_\l]&=\I(g_{\n\l}K_\m-g_{\m\l}K_\n)\;,&\qquad
[P_\m,K_\n]&=\I(ag_{\m\n}Q+bJ_{\m\n})\;,\\
[K_\m,K_\n]&=\I cJ_{\m\n}\;,&\qquad
[K_\m,Q]&=\I(dP_\m+eK_\m)\;.
\end{alignedat}
\label{lieK}
\end{equation}
The parameters $a,b,c,d,e$ cannot take arbitrary values, since to correspond to a well-defined Lie algebra, the commutators must satisfy a set of Jacobi identities. A straightforward calculation leads to the algebraically independent constraints $ae=0$, $b=0$ and $c=-ad$. Should $[P_\m,K_\n]$ be nonzero, which is necessary for the symmetry generated by $K_\m$ to affect the momentum dependence of scattering amplitudes, $a$ must be nonzero as well. Upon absorbing it into a redefinition of $Q$ via $Q\to Q/a$, the three commutators in~\eqref{lieK} not fixed by Lorentz invariance reduce to
\begin{equation}
[P_\m,K_\n]=\I g_{\m\n}Q\;,\qquad
[K_\m,K_\n]=-\I vJ_{\m\n}\;,\qquad
[K_\m,Q]=\I vP_\m\;,
\end{equation}
with the shorthand notation $v\equiv ad$.

The case of $v=0$ describes the Lie algebra of Galileon transformations~\eqref{symgal}. Here $Q$ generates the constant shift with parameter $\eps$, whereas $K_\m$ generates the linear shift with parameter $\eps_\m$. It remains to address the case $v\neq0$. Here the parameter can be nearly eliminated by rescaling both $Q$ and $K_\m$ by $\sqrt{\abs v}$. What is left of $v$ is just its sign, entering through $[K_\m,K_\n]=\mp\I J_{\m\n}$ and $[K_\m,Q]=\pm\I P_\m$. Finally, merge $K_\m$ with $J_{\m\n}$ and $Q$ with $P_\m$ by identifying $J_{\m,D+1}\equiv K_\m$ and $P_{D+1}\equiv Q$ together with $g_{D+1,D+1}\equiv\pm1$ and voil\`a, we find the $(D+1)$-dimensional Poincar\'e algebra. Its $v=+1$ version is based on the $\gr{SO}(d,2)$ group of spacetime rotations, whereas the $v=-1$ version is based on $\gr{SO}(d+1,1)$. These correspond precisely to the two mutations of the DBI theory.

We have successfully recovered the symmetries of the DBI and Galileon theories by a simple Lie-algebraic argument. The utility of the symmetry-based approach of course does not end here. It can be pursued further towards an explicit construction of effective actions, respecting the symmetry. This however requires a generalization of the techniques developed in Chaps.~\ref{chap:CCWZ} and~\ref{chap:effLagrangian} to coordinate-dependent symmetries.

The classification of candidate symmetries leading to enhancement of the soft limit of scattering amplitudes, outlined here, can be extended in various directions. First, we may want to see how the special Galileon theory fits in. As suggested by~\eqref{sgalsym}, this requires adding another set of generators that span a traceless symmetric tensor representation of the Lorentz group. While the derivation of all the Lie-algebraic constrains is now more laborious, one eventually recovers~\eqref{sgalsym} along with~\eqref{sgallag} as the only physically relevant solution~\cite{Bogers2018b}. A similar reasoning was used in~\cite{Roest2019} to show that in $D=4$ dimensions, there are no EFTs of a single NG boson that would realize $\s\geq4$ enhancement of the soft limit in a nontrivial manner. Finally, the Lie algebra~\eqref{lieK} can be extended to allow for multiple flavors of NG bosons. This turns out to be an efficient tool for carving out the landscape of potentially interesting EFTs~\cite{Bogers2018a}. The DBI and Galileon theories have a natural multiflavor generalization. However, the precise relation between the symmetry and the asymptotic behavior of scattering amplitudes in the soft limit that would allow us to unambiguously predict the value of $\s$, remains unknown.


\section{Soft Recursion}
\label{sec:softrecursion}

No primer on scattering of NG bosons can be complete without at least briefly mentioning on-shell methods for scattering amplitudes, developed in the last decades. Given the context of this book, I will restrict the discussion to on-shell recursion relations for EFTs of NG bosons. These emerged as an adaptation of the framework, originally designed for the Yang--Mills theory by Britto, Cachazo, Feng and Witten. An interested reader will find an introduction to the latter in Chap.~3 of~\cite{Elvang2015}.

We return to the broad class of EFTs for multiple NG boson flavors. For notation simplicity, I will however suppress flavor and momentum labels and denote the on-shell tree-level amplitude for an $n$-particle scattering process simply as $\Aa_n$. Suppose now that we are able to complexify the energy--momenta in the process, $p^\m_i\to\hat p^\m_i(z)$, so that all the $\hat p^\m_i(z)$ are linear functions of $z\in\C$, satisfying $\hat p^\m_i(0)=p^\m_i$. We of course still require that the $\hat p^\m_i(z)$ add up to zero and remain on-shell, $[\hat p_i(z)]^2=0$, for any $z\in\C$. Then the scattering amplitude $\Aa_n$ is complexified to a meromorphic function $\hat\Aa_n(z)$ in the complex plane. Furthermore, for any holomorphic function $F_n(z)$ such that $F_n(0)=1$, the function $\hat\Aa_n(z)/[zF_n(z)]$ is also meromorphic and has a simple pole at $z=0$ with residue $\hat\Aa_n(0)=\Aa_n$. The residue theorem then implies that upon integration along an infinitesimal circle enclosing the origin,
\begin{equation}
\Aa_n=\frac1{2\pi\I}\oint\D z\,\frac{\hat\Aa_n(z)}{zF_n(z)}\;.
\label{softrec}
\end{equation}
By the same residue theorem, the integral can also be expressed in terms of residues at the other poles in the complex plane, possibly including the residue at infinity. This is the central idea behind on-shell recursion techniques. To progress further, we need to be more specific about the complexification of the energy--momentum variables. We also need input on the asymptotic behavior of the amplitude at $z\to\infty$. Together, these will allow us to tailor the function $F_n(z)$ to the EFT at hand.


\subsection{Complexified Kinematics}
\label{subsec:BCFWkinematics}

Our requirements that $\hat p^\m_i(z)$ be linear in $z$ and reduce to $p^\m_i$ at $z=0$ are solved by
\begin{equation}
\hat p^\m_i(z)=p^\m_i+zq^\m_i
\label{pq}
\end{equation}
with arbitrary $q^\m_i$. However, the on-shell condition $[\hat p_i(z)]^2=0$ is most easily satisfied if $q^\m_i$ is parallel to $p^\m_i$. I will parameterize such $q^\m_i$ by a real constant $c_i$,\footnote{Within this section, I always indicate summation over particles in the scattering process explicitly. Hence, no summation is implied by the repeated index ``$i$'' in~\eqref{piai}.}
\begin{equation}
\hat p^\m_i(z)\equiv p^\m_i(1-c_iz)\;.
\label{piai}
\end{equation}
This is called the \emph{all-line soft shift} of the energy--momentum variables. The last requirement we have to take care of is overall energy--momentum conservation, which leads to
\begin{equation}
\sum_{i=1}^nc_ip^\m_i=0\;.
\label{sumap}
\end{equation}
I will always assume a generic kinematical configuration, disregarding accidental linear dependencies among the energy--momenta $p^\m_i$. This is equivalent to the assumption that $p^\m_i$ as a $D\times n$ matrix has the maximum possible rank consistent with energy--momentum conservation, that is $\min(D,n-1)$. By the rank--nullity theorem, the set of solutions to~\eqref{sumap} then spans a vector space of dimension $n-\min(D,n-1)\geq1$. The one-dimensional subspace that is always guaranteed to exist corresponds to $c_1=\dotsb=c_n$. However, it is desirable to have solutions for which all the $c_i$s are different. This will allow us to probe the single soft limit for the individual particles by tuning $z\to1/c_i$. The existence of such solutions requires that $D<n-1$, or better $n\geq D+2$.

The all-line soft shift~\eqref{piai} cannot access the single soft limit of all scattering amplitudes, and the problem gets worse with increasing the spacetime dimension $D$. This motivated~\cite{Cheung2017a} to introduce alternative prescriptions, combining~\eqref{pq} and~\eqref{piai} for disjoint subsets of the $n$ particles. To illustrate the idea, I will briefly outline the minimal modification of~\eqref{piai}, dubbed the \emph{all-but-one-line soft shift}. This applies~\eqref{piai} to the first $n-1$ particles but uses~\eqref{pq} for the last one,
\begin{equation}
\hat p^\m_i(z)\equiv p^\m_i(1-c_iz)\quad\text{for }i=1,\dotsc,n-1\;,\qquad
\hat p^\m_n(z)\equiv p^\m_n+zq^\m_n\;.
\label{piaian}
\end{equation}
Energy--momentum conservation and the on-shell condition now dictate that
\begin{equation}
q^\m_n=\sum_{i=1}^{n-1}c_ip^\m_i\;,\qquad
q_n^2=p_n\cdot q_n=0\;.
\label{pqa}
\end{equation}
The first relation in~\eqref{pqa} can be viewed as a definition of $q^\m_n$. The second relation then constitutes two homogeneous constraints on $c_1,\dotsc,c_{n-1}$, one linear and one quadratic. This defines an $(n-3)$-dimensional hypersurface in the $\R^{n-1}$ space of all $c_i$. The hypersurface always includes a one-dimensional linear subspace where $c_1=\dotsb=c_{n-1}$, as guaranteed by energy--momentum conservation. The existence of other, nontrivial solutions for $c_i$ requires that $n-3>1$, or $n\geq5$, independently of the spacetime dimension $D$. We find that the all-but-one-line shift is applicable to a broader set of amplitudes than the all-line shift for any $D\geq4$. The price to pay is that it only allows us to access the soft limit for $n-1$ particles out of~$n$.


\subsection{Recursion Relation for On-Shell Amplitudes}
\label{subsec:BCFWrecursion}

We are now ready to deal with the complexified amplitudes. Consider an EFT with soft scaling parameter $\s$. By definition, any complexified $n$-particle amplitude $\hat\Aa_n(z)$ in this theory has a zero of order $\s$ at $z=1/c_i$ for any of the particles to which~\eqref{piai} has been applied. It is then advantageous to set
\begin{equation}
F_{n,\s}(z)\equiv\prod_{i=1}^{n_\mathrm s}(1-c_iz)^\s\;,
\end{equation}
where $n_\mathrm s=n$ for the all-line shift and $n_\mathrm s=n-1$ for the all-but-one-line shift. This choice of $F_n(z)$ is optimized for maximum suppression of the integrand in~\eqref{softrec} at large $z$ without introducing any new poles in it. This suppression is sufficient to eliminate the pole at infinity provided
\begin{equation}
m<n_\mathrm s\s\;,
\label{mns}
\end{equation}
where $m$ is the degree of $\Aa_n$ as a function of energy--momenta.

\begin{illustration}%
\label{ex:msigma}%
The constraint~\eqref{mns} is satisfied for a large class of EFTs. First, any EFT of the type~\eqref{LeffShifman}, defined on a symmetric coset space, has $(m,\s)=(2,1)$ for any $n$. This follows from the relation $V=I+1$ between the numbers of interaction vertices $V$ and internal propagators $I$, valid for any connected tree-level Feynman diagram. Second, the Lagrangian of the DBI theory~\eqref{DBI} contains one derivative on each field. As a result, the negative powers of energy--momentum due to propagators are exactly canceled by the adjacent interaction vertices. What is left is in effect one derivative per each external leg of the Feynman diagram, hence $(m,\s)=(n,2)$ for the DBI theory. Finally, the special Galileon theory in $D=4$ dimensions contains a single quartic interaction vertex with six derivatives. This gives $m=6V-2I$ and $n=4V-2I$, hence $(m,\s)=(2n-2,3)$.
\end{illustration}

\begin{figure}[t]
\sidecaption[t]
\includegraphics[width=2.9in]{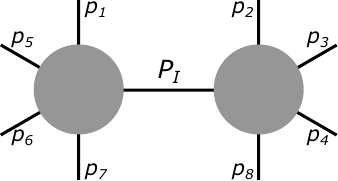}
\caption{Example of a factorization channel in an 8-particle scattering process. The two sides of the factorization channel correspond respectively to $I=\{1,5,6,7\}$ and $\tilde I=\{2,3,4,8\}$. The gray disks represent the sum of all possible Feynman diagrams with the given external legs}
\label{fig:factorization}
\end{figure}

With no pole at infinity and no new poles from $F_{n,\s}(z)$, all the $z\neq0$ poles of the integrand in~\eqref{softrec} come from the propagators inside $\hat\Aa_n(z)$. Consider a partition of all the particles in the scattering process into two disjoint subsets, $I$ and $\tilde I$. Among all diagrams contributing to the tree-level amplitude $\hat\Aa_n(z)$, there is a subset of graphs where the particles in $I$ and $\tilde I$ can be separated by cutting a single propagator. This subset of diagrams defines a \emph{factorization channel}, which I will label simply with the letter $I$; see Fig.~\ref{fig:factorization} for illustration. One can always assign the labels $I,\tilde I$ so that the $n$-th particle belongs to $\tilde I$. Then $c_i$ is well-defined for all particles in $I$ even when the all-but-one-line shift is used. With the shorthand notation
\begin{equation}
P^\m_I\equiv\sum_{i\in I}p^\m_i\;,\qquad
Q^\m_I\equiv\sum_{i\in I}c_ip^\m_i\;,
\end{equation}
the complex energy--momentum in the propagator separating $I$ and $\tilde I$ is $P^\m_I-zQ^\m_I$. This leads to two poles at $z=z^\pm_I$ where
\begin{equation}
z^\pm_I=\frac1{Q_I^2}\Bigl[P_I\cdot Q_I\pm\sqrt{(P_I\cdot Q_I)^2-P_I^2Q_I^2}\Bigr]\;.
\end{equation}
Near these poles, the complex amplitude $\hat\Aa_n(z)$ factorizes as
\begin{equation}
\hat\Aa_n(z)\xrightarrow{z\to z^\pm_I}\frac{\hat\Aa_I(z^\pm_I)\hat\Aa_{\tilde I}(z^\pm_I)}{(P_I-zQ_I)^2}=\frac{z^+_Iz^-_I}{(z-z^+_I)(z-z^-_I)}\frac{\hat\Aa_I(z^\pm_I)\hat\Aa_{\tilde I}(z^\pm_I)}{P_I^2}\;.
\end{equation}
Here $\hat\Aa_I(z^\pm_I)$ is the complexified \emph{on-shell} amplitude for scattering of particles in $I$, and likewise for $\hat\Aa_{\tilde I}(z^\pm_I)$. We arrive at the key result that the $n$-particle amplitude $\Aa_n$ can be reconstructed from on-shell amplitudes with lower numbers of particles, as long as it can be complexified using one of the prescriptions introduced in Sect.~\ref{subsec:BCFWkinematics}. Importantly, this \emph{soft recursion} applies to the entire amplitudes. This observation lies at the heart of an approach to symbolic computation of scattering amplitudes that avoids the combinatorial explosion of ordinary perturbation theory. In practice, one only needs to calculate explicitly a small number of ``seed'' amplitudes. All higher-point amplitudes can then be reconstructed recursively. 

\begin{illustration}%
In $D=4$ spacetime dimensions, the all-line shift~\eqref{piai} can be used for any amplitude with $n\geq6$. This leaves us with $\Aa_4$ and $\Aa_5$ as the seeds that need to be known a priori. For theories with just even interaction vertices, such as those from \refex{ex:msigma}, one only needs to calculate the 4-particle amplitude explicitly. This is consistent with what we learned previously about the structure of EFTs on symmetric coset spaces. Namely, it follows from~\eqref{gsymcoset} that the quartic interaction vertex already fixes uniquely the entire effective Lagrangian. The same applies to the DBI theory.
\end{illustration}

Let us now see the recursion program through to the end. Upon summation over all factorization channels, \eqref{softrec} becomes~\cite{Cheung2016a}
\begin{equation}
\Aa_n=\sum_I\frac1{P_I^2}\biggl[\frac{\hat\Aa_I(z^+_I)\hat\Aa_{\tilde I}(z^+_I)}{(1-z^+_I/z^-_I)F_{n,\s}(z^+_I)}+\frac{\hat\Aa_I(z^-_I)\hat\Aa_{\tilde I}(z^-_I)}{(1-z^-_I/z^+_I)F_{n,\s}(z^-_I)}\biggr]\;.
\label{recmaster}
\end{equation}
The sum runs over all partitions of the particles, that is each pair $I,\tilde I$ is only counted once. This is our master equation, which allows recursive construction of tree-level scattering amplitudes by symbolic computation. However, the first step of the recursion can often be performed manually. Namely, if the functions $\hat\Aa_I(z),\hat\Aa_{\tilde I}(z)$ themselves do not contain any factorization poles, the summand in~\eqref{recmaster} can be interpreted in terms of the residues of the meromorphic function $\hat\Aa_I(z)\hat\Aa_{\tilde I}(z)/[zF_{n,\s}(z)(P_I-zQ_I)^2]$. This function does have also poles at $z=1/c_i$, since unlike $\hat\Aa_n(z)$ itself, $\hat\Aa_I(z)$ and $\hat\Aa_{\tilde I}(z)$ are not on-shell and thus do not necessarily vanish at these points. We can then use the residue theorem to rewrite the sum over poles at $z^\pm_I$ in terms of a sum over poles at $z=0$ and $z=1/c_i$,
\begin{equation}
\Aa_n=\sum_I\biggl[\frac{\hat\Aa_I(0)\hat\Aa_{\tilde I}(0)}{P_I^2}+\sum_{i=1}^{n_\mathrm s}\Res_{z=1/c_i}\frac{\hat\Aa_I(z)\hat\Aa_{\tilde I}(z)}{zF_{n,\s}(z)(P_I-zQ_I)^2}\biggr]\;.
\label{recmaster2}
\end{equation}
The first term takes into account diagrammatic contributions to $\Aa_n$ with an internal propagator. The second term must therefore match the contributions to $\Aa_n$ from contact $n$-point operators in the effective Lagrangian.

\begin{illustration}%
The special Galileon theory~\eqref{sgallag} in $D=4$ dimensions has a single, quartic interaction vertex. When applied to the 6-particle amplitude, we therefore expect the second term in~\eqref{recmaster2} to vanish. To check this, rewrite the Lagrangian as
\begin{equation}
\La^{D=4}_\mathrm{sGal}\simeq\frac12(\de_\m\pi)^2-\frac1{24\Lambda_6}\ve^{\m_1\m_2\m_3\l}\ve^{\n_1\n_2\n_3}_{\phantom{\n_1\n_2\n_3}\l}\pi\prod_{i=1}^3(\de_{\m_i}\de_{\n_i}\pi)\;.
\end{equation}
The Feynman rule for the interaction vertex is $[-\I/(6\Lambda_6)]G(p_1,p_2,p_3)$, where $p^\m_1,p^\m_2,p^\m_3$ are any three of the four four-momenta in the vertex and $G$ denotes the corresponding Gram determinant. Since the latter is a homogeneous function of all its arguments of degree two, we have
\begin{equation}
\hat\Aa_I(z)\hat\Aa_{\tilde I}(z)=\frac1{(6\Lambda_6)^2}G(\{p_i\}_{i\in I})G(\{p_j\}_{j\in\tilde I})\prod_{k=1}^6(1-c_kz)^2
\label{AIGal}
\end{equation}
for any partition of the six particles into two triples $I,\tilde I$. I have implicitly used the all-line shift. Note that the criterion~\eqref{mns} is satisfied even if we take $F_{n,\s}(z)$ with $\s=2$ instead of $\s=3$. The advantage of this choice is that the denominator term $F_{n,\s}(z)$ is then completely canceled by the last factor in~\eqref{AIGal}. This indeed makes the second term in~\eqref{recmaster2} disappear since the residues at $z=1/c_i$ trivially vanish.
\end{illustration}

The derivation of~\eqref{recmaster} does not apply to theories with a nonvanishing soft limit, where $\s=0$ and~\eqref{mns} cannot hold. Here we have two alternatives: either use $F_{n,0}(z)=1$ and deal with the residue at infinity, or use $F_{n,\s}(z)$ with $\s\geq1$ and deal with the ensuing poles at $z=1/c_i$. The latter option is feasible, at least for two-derivative EFTs of the type~\eqref{LeffShifman}, where sufficient suppression of the integrand in~\eqref{softrec} is ensured by $F_{n,1}(z)$. The residues at the simple poles at $z=1/c_i$ can then be extracted from the geometric soft theorem~\eqref{softthm} or its more explicit version~\eqref{softfinal}. A minor modification of the steps leading to~\eqref{recmaster} now gives~\cite{Cheung2022}
\begin{equation}
\Aa_n=\sum_I\frac1{P_I^2}\sum_\pm\frac{\hat\Aa_I(z^\pm_I)\hat\Aa_{\tilde I}(z^\pm_I)}{(1-z^\pm_I/z^\mp_I)F_{n,1}(z^\pm_I)}+\sum_{i=1}^n\frac{\hat\cd_{a_i}\hat\Aa_{n-1}(1/c_i)}{\prod_{j\neq i}(1-c_j/c_i)}\;.
\label{recHelset}
\end{equation}
Here I used the shorthand notation $\hat\Aa_{n-1}$ for the amplitude where the $i$-th particle with flavor index $a_i$ has been removed. It is of course still possible to rewrite the first term in~\eqref{recHelset} as~\eqref{recmaster2}.


\subsection{Soft Bootstrap}
\label{subsec:BCFWbootstrap}

The modern scattering amplitude program strives to construct amplitudes, or even define what a quantum field theory is, without recourse to a Lagrangian or the symmetry thereof. It is therefore of great interest to understand what kind of soft behavior of scattering amplitudes is allowed on general grounds. Recursion relations such as~\eqref{recmaster} or its generalization~\eqref{recHelset} provide a versatile tool for constraining the landscape of possible EFTs.

In~\cite{Cheung2017a}, soft recursion was used to derive a set of bounds on the soft scaling parameter $\s$. The most striking result is that in $D\geq4$ spacetime dimensions, $\s=3$ is the maximum value that can be realized in a nontrivial manner. (Any $\s$ can be realized trivially by interactions with at least $\s$ derivatives per field.) Apart from this universal bound, the maximum achievable value of $\s$ in an EFT with a fixed set of interaction operators is also constrained by the number of derivatives per field. Consider a contact operator of the schematic type $(\de\pi)^2\de^{m-2}\pi^{n-2}$, and characterize the number of derivatives per field by the parameter $\vr\equiv(m-2)/(n-2)$. We thus have $\vr=1$ for the DBI theory and $\vr=2$ for all the Galileon interactions~\eqref{galWZ}. Then, the soft scaling parameter is bounded by $\s\leq\vr+1$. This shows that the DBI and special Galileon theories are ``exceptional'' in the sense that they maximize $\s$ in their respective classes of theories with fixed $\vr$. Similar bounds can be derived for translationally and rotationally invariant nonrelativistic EFTs~\cite{Mojahed2022}.

The values of $\s$ and $\vr$ alone do not necessarily specify a unique EFT. Here one can gain further insight by a procedure known as the \emph{soft bootstrap}, see for instance~\cite{Elvang2019,Low2019}. In $D\leq4$ dimensions, the only seed amplitudes needed to apply soft recursion in combination with the all-line shift are $\Aa_4$ and $\Aa_5$. These are contact amplitudes without any exchange contributions. As such, they are given by polynomials of degree $m$ in the particle energy--momenta, or $m/2$ in the Mandelstam variables. These polynomials are restricted by energy--momentum conservation and permutation invariance, and their complete classification is usually straightforward. With a list of candidate seed amplitudes at hand, one then proceeds to construct higher-point amplitudes via~\eqref{recmaster} or~\eqref{recHelset}. The result of the recursion must be independent of the choice of parameters $c_i$ satisfying~\eqref{sumap} or~\eqref{pqa}. Failure to pass this $c_i$-independence test indicates that the seed amplitudes do not correspond to any well-defined local field theory.

\begin{watchout}%
Soft bootstrap cannot, strictly speaking, be used to prove that a theory based on a fixed set of seed amplitudes exists. That would require checking the $c_i$-independence of recursively constructed amplitudes to all orders of the recursion. In this aspect, soft bootstrap is nicely complemented by more conventional EFT approaches based on symmetry. The latter are efficient in isolating a set of candidate EFTs with desired particle composition and soft behavior, as illustrated in Sect.~\ref{subsec:enhancedliealgebra}. What soft bootstrap can do is show that other theories with given soft behavior than those known a priori do not exist.
\end{watchout}


\end{fmffile}

\bibliographystyle{spphys}
\bibliography{references}
\begin{partbacktext}
\part{Spontaneously Broken Spacetime~Symmetry}
\label{part:spacetimeSSB}
\end{partbacktext}
\chapter{Locally Equivalent Symmetries}
\label{chap:differences}

\abstract*{The first obstacle to the application of the effective field theory program to spacetime symmetries is the identification of independent fluctuations of the order parameter. The redundancy among fluctuations induced by different broken symmetry generators often arises from the equivalence of the corresponding local symmetry transformations. This is explored in the present chapter. First, it is shown that symmetries that are locally equivalent give rise to closely related Noether currents. This explains how the Ward identities for the correlation functions of Noether currents can be saturated with a lower number of Nambu--Goldstone bosons than naively expected. The redundancy of different coordinate-dependent symmetries is then addressed from the point of view of background gauge invariance. The last part of the chapter returns to scattering of Nambu--Goldstone bosons. It shows how the presence of redundant symmetries may lead to the enhancement of scattering amplitudes in the soft limit.}


Our discussion of \emph{effective field theory} (EFT) for \emph{Nambu--Goldstone} (NG) \emph{bosons} has so far been limited to internal symmetries. The application of the EFT program to symmetries that depend on or affect spacetime coordinates is associated with numerous subtleties. The first of these lies in the identification of the degrees of freedom of the low-energy EFT. We saw already in Chap.~\ref{chap:NGbosons} that the fluctuations of the order parameter induced by independent broken symmetry generators may be related to each other. Very often, this is a consequence of a redundancy between the localized actions of different symmetry transformations. This chapter explores such local equivalence of coordinate-dependent symmetries as a preparation for the more detailed discussion of broken spacetime symmetries in the following chapters.

The first two sections follow the pedagogical exposition in~\cite{Brauner2020a}. In Sect.~\ref{sec:diffcurrentrelations}, I show that local equivalence of symmetries leads to a linear relation between the corresponding Noether currents. This lifts the intuitive classical picture of redundancy among NG modes, outlined in Chap.~\ref{chap:NGbosons}, to the language of quantum field theory. In Sect.~\ref{sec:diffgauging}, I then reinterpret the local equivalence of symmetries in terms of their simultaneous gauging. The whole Sect.~\ref{sec:diffexamples} is reserved for several physically relevant examples, illustrating the general theory of locally equivalent symmetries. Finally, in Sect.~\ref{sec:diffscattering}, I return to scattering of NG bosons of internal symmetries. I show how the relations between Noether currents derived in this chapter help explain the enhancement of the soft limit of scattering amplitudes in the \emph{Dirac--Born--Infeld} (DBI) and Galileon theories.


\section{Relations Between Noether Currents}
\label{sec:diffcurrentrelations}

Consider a theory of a set of (not necessarily scalar) fields $\psi^i$, and suppose that its action is invariant under two different classes of continuous symmetries. As we know from Chap.~\ref{chap:symmetry}, the associated Noether currents can be extracted by evaluating the variation of the action under the corresponding localized transformations. The evolutionary form of these transformations can be written generally as
\begin{equation}
\begin{split}
\udelta_1\psi^i(x)&=\eps^A_1(x)\df^i_A[\psi,x](x)+\de_\m\eps^A_1(x)\dfg^{i\m}_A[\psi,x](x)+\dotsb\;,\\
\udelta_2\psi^i(x)&=\eps^\iA_2(x)\df^i_\iA[\psi,x](x)+\de_\m\eps^\iA_2(x)\dfg^{i\m}_\iA[\psi,x](x)+\dotsb\;.
\end{split}
\label{locallyequivalentsym}
\end{equation}
As before, the square bracket notation indicates local functions of the fields and their derivatives, possibly also depending explicitly on the spacetime coordinates. Furthermore, $\eps^A_1(x)$ and $\eps^\iA_2(x)$ are the localized parameters of the transformations. The presence of derivatives of the parameters in~\eqref{locallyequivalentsym} hints that I allow for a generic localization of the transformations; the ellipses represent terms with higher derivatives of $\smash{\eps^A_1(x)}$ and $\smash{\eps^\iA_2(x)}$. The original symmetries are recovered by reducing the coordinate-dependent parameters to constants, $\smash{\eps^A_1(x)\to\eps^A_1}$ and $\smash{\eps^\iA_2(x)\to\eps^\iA_2}$. I will call the two symmetries \emph{locally equivalent} if there is a set of smooth functions $\smash{f^A_\iA(x)}$ such that setting
\begin{equation}
\eps^A_1(x)=f^A_\iA(x)\eps^\iA_2(x)
\label{localequivalence}
\end{equation}
makes the two transformations in~\eqref{locallyequivalentsym} identical, $\udelta_1\psi^i(x)=\udelta_2\psi^i(x)$.

Locally equivalent symmetry transformations should give the same variation of the action as expressed in terms of the corresponding Noether currents,
\begin{equation}
\begin{split}
\udelta S&=\int\D^D\!x\,J^\m_{2\iA}[\psi,x](x)\de_\m\eps^\iA_2(x)=\int\D^D\!x\,J^\m_{1A}[\psi,x](x)\de_\m\eps^A_1(x)\\
&=\int\D^D\!x\,J^\m_{1A}[\psi,x](x)\bigl[f^A_\iA(x)\de_\m\eps^\iA_2(x)+\eps^\iA_2(x)\de_\m f^A_\iA(x)\bigr]\;.
\end{split}
\label{noethertrick12}
\end{equation}
However, the latter expression does not automatically vanish for constant $\eps^\iA_2$ as it should. In order to ensure this, we have to impose the \emph{integrability condition}
\begin{equation}
J^\m_{1A}[\psi,x]\de_\m f^A_\iA=\de_\m N^\m_\iA[\psi,x]\;,
\label{integrabilitycondition}
\end{equation}
where $\smash{N^\m_\iA[\psi,x]}$ is some local vector function of the fields and their derivatives. Combining~\eqref{integrabilitycondition} with~\eqref{noethertrick12} and integrating by parts in the latter then leads to the Noether current \emph{equivalence relation}
\begin{equation}
J^\m_{2\iA}[\psi,x]=f^A_\iA J^\m_{1A}[\psi,x]-N^\m_\iA[\psi,x]\;.
\label{currentequivalencerelation}
\end{equation}

This is the main result of the chapter that deserves a few comments. First, an alternative approach to local equivalence is to start with constant $\smash{\eps^\iA_2}$, that is the actual symmetry with generators $Q_{2\iA}$. One then demands the existence of smooth functions $\smash{f^A_\iA(x)}$ such that this symmetry is reproduced by a localized transformation of the type $\udelta_1\psi^i$ with parameter $\smash{\eps^A_1(x)=f^A_\iA(x)\eps^\iA_2}$. This is the origin of the integrability condition~\eqref{integrabilitycondition}. If needed, the localized transformation $\udelta_2\psi^i$ can then be \emph{defined} as $\udelta_1\psi^i$ with the parameter $\eps^A_1(x)$ as given in~\eqref{localequivalence}. It is therefore reasonable to think of $\udelta_1\psi^i$ as a \emph{parent} symmetry and of $\udelta_2\psi^i$ as its \emph{descendant}.

Second, I have not imposed the equation of motion. Thus, both the equivalence relation~\eqref{currentequivalencerelation} and the integrability condition~\eqref{integrabilitycondition} must hold off-shell. By taking a divergence of~\eqref{currentequivalencerelation} and using~\eqref{integrabilitycondition}, we obtain another off-shell relation,
\begin{eqnarray}
\de_\m J^\m_{2\iA}[\psi,x]=f^A_\iA\de_\m J^\m_{1A}[\psi,x]\;.
\end{eqnarray}
This makes the Ward identities for the correlation functions of $\smash{J^\m_{2\iA}}$ a consequence of those for $\smash{J^\m_{1A}}$. By the same token, the two sets of currents couple to the same one-particle NG states. This explains in a quantum-field-theoretic language the redundancy of NG modes generated by locally equivalent symmetries.

Third, should $\smash{f^A_\iA}$ be constant, the integrability condition~\eqref{integrabilitycondition} is fulfilled for $\smash{N^\m_\iA=0}$. The equivalence relation~\eqref{currentequivalencerelation} is then strictly linear in the currents. The corresponding generators $Q_{1A}$ and $Q_{2\iA}$ are also linearly related, $\smash{Q_{2\iA}=f^A_\iA Q_{1A}}$. This special case therefore amounts to a mere change of basis of symmetry generators, or a choice of a subset thereof, and is of little interest. A nontrivial current equivalence relation only arises from a set of non-constant functions $\smash{f^A_\iA(x)}$. Without loss of generality, one can drop the constant part of $\smash{f^A_\iA(x)}$ and assume that $\smash{f^A_\iA(0)=0}$. This assumption is satisfied for all the examples worked out below. It also guarantees a posteriori that the relation~\eqref{localequivalence} cannot be inverted. In other words, the distinction between a parent symmetry and its descendant is well-defined and cannot be reversed.

Finally, recall that the Noether current is determined by the symmetry transformation only up to addition of a vector field whose divergence vanishes off-shell. Suppose that we modify the parent Noether current by $\smash{J^\m_{1A}\to J^\m_{1A}+\udelta J^\m_{1A}}$ such that $\de_\m\udelta J^\m_{1A}=0$. Then the integrability condition~\eqref{integrabilitycondition} can still be satisfied if we shift $\smash{N^\m_\iA}$ appropriately, $\smash{N^\m_\iA\to N^\m_\iA+f^A_\iA\udelta J^\m_{1A}}$. It follows that the ambiguity of the parent current does not affect the descendant current $J^\m_{2\iA}$ as given by~\eqref{currentequivalencerelation}. The descendant current of course suffers from its own ambiguity under $\smash{J^\m_{2\iA}\to J^\m_{2\iA}+\udelta J^\m_{2\iA}}$ such that $\smash{\de_\m\udelta J^\m_{2\iA}=0}$ off-shell. The ambiguities of the two currents are therefore independent of each other and do not constrain the validity of the equivalence relation~\eqref{currentequivalencerelation}. 


\section{Noether Currents from Background Gauging}
\label{sec:diffgauging}

The derivation of the Noether current of a given symmetry, based on~\eqref{noethertrick}, treats the localized symmetry transformation as a mere technical trick. There is however an alternative approach to Noether's theorem that takes the local transformation seriously and promotes it to an actual symmetry of the action. This offers additional insight into the structure of Noether currents, the price to pay being a technical assumption that has to be verified case by case.

This key assumption is that it is possible to make the action of the theory exactly invariant under the localized symmetry transformation by adding a set of background gauge fields $\smash{A^A_\m}$. More concretely, we want to promote the original action $S\{\psi\}$ to a functional $\tilde S\{\psi,A\}$ such that $\smash{\at{\tilde S\{\psi,A\}}{\mathrm{vac}}=S\{\psi\}}$, where the subscript ``vac'' indicates removing the background. I allow for a generic local transformation of $\psi^i$ with parameters $\eps^A(x)$, following~\eqref{locallyequivalentsym},
\begin{equation}
\udelta\psi^i(x)=\eps^A(x)\df^i_A[\psi,x](x)+\de_\m\eps^A(x)\dfg^{i\m}_A[\psi,x](x)+\dotsb\;.
\label{udeltapsi}
\end{equation}
As for the local transformation of the gauge field, we can assume a similar generic expansion in derivatives of the parameters $\eps^A(x)$,
\begin{equation}
\udelta A^A_\m(x)=\eps^B(x)\dA^A_{B\m}[A,x](x)+\de_\n\eps^B(x)\dAg^{A\n}_{B\m}[A,x](x)+\dotsb\;;
\label{udeltaA}
\end{equation}
the ellipsis again stands for terms with higher derivatives of $\smash{\eps^A(x)}$. It is important that $\smash{\udelta A^A_\m}$ does not depend on the dynamical fields $\psi^i$ so that $\smash{A^A_\m}$ can be treated as a fixed external background.

The variation of action under a simultaneous transformation of $\psi^i$ and $\smash{A^A_\m}$ consists of contributions from varying the former and the latter, $\udelta\tilde S\equiv\udelta_\psi\tilde S+\udelta_A\tilde S$, where
\begin{equation}
\udelta_A\tilde S=\int\D^D\!x\,\frac{\udelta\tilde S}{\udelta A^A_\m}\bigl(\eps^B \dA^A_{B\m}+\de_\n\eps^B\dAg^{A\n}_{B\m}+\dotsb\bigr)\;.
\label{udeltatildeS}
\end{equation}
Next we remove the background. In this limit, $\udelta_\psi\tilde S$ turns by construction into $\smash{\int\D^D\!x\,J_A^\m\de_\m\eps^A}$. In order that $\udelta\tilde S$ vanishes at least when $\eps^A$ is constant, there must be a local function $\smash{R_A^\m[\psi,x]}$ such that
\begin{equation}
\at{\frac{\udelta\tilde S\{\psi,A\}}{\udelta A^B_\m(x)}\dA^B_{A\m}[A,x](x)}{\mathrm{vac}}=\de_\m R_A^\m[\psi,x](x)\;.
\label{integrabilityR}
\end{equation}
With this consistency condition satisfied, the vanishing of $\udelta\tilde S$ for any $\eps^A(x)$ leads to the identification
\begin{equation}
J_A^\m[\psi,x](x)=R_A^\m[\psi,x](x)-\at{\frac{\udelta\tilde S\{\psi,A\}}{\udelta A^B_\n(x)}\dAg^{B\m}_{A\n}[A,x](x)}{\mathrm{vac}}+\dotsb\;.
\label{currentR}
\end{equation}
Equations~\eqref{integrabilityR} and~\eqref{currentR} already constitute a fairly general tool for derivation of Noether currents. Yet, the approach based on background gauge invariance can be further extended to higher-rank tensor background fields. Instead of cluttering the formalism with additional indices, I will content myself with a couple of illustrative examples, worked out in Sects.~\ref{subsec:diffexamplesrotations} and~\ref{subsec:diffexamplesgalilei}.

The notation used in~\eqref{integrabilityR} and~\eqref{currentR} is tailored to what I called the parent symmetry. What if the system possesses an additional, descendant set of symmetries? By~\eqref{localequivalence}, the localized transformations of the descendant type form a subset of all localized transformations of the parent type. All we therefore have to do is to gauge the parent symmetry by adding its background gauge fields $\smash{A^A_\m}$. That we can in fact do so is the sole technical assumption we have to make. The descendant symmetry is then automatically gauged as well without the need for any additional gauge fields. This once again underlines the redundancy of the descendant symmetry. In particular, the background gauging makes it possible to recover the equivalence relation~\eqref{currentequivalencerelation} for Noether currents; see~\cite{Brauner2020a} for a detailed proof.


\section{Examples}
\label{sec:diffexamples}

I will now illustrate the general theory developed above with several examples. These are chosen mostly for their relevance to EFTs discussed elsewhere in the book. I will therefore not shy away from showing certain amount of details.


\subsection{Galileon Symmetry}
\label{subsec:diffexamplesgalileon}

Consider a class of theories of a single real relativistic scalar field $\p$, invariant under the Galileon symmetry
\begin{equation}
\udelta\p(x)=\eps_1+\eps^\m_2x_\m\;.
\label{galileonsym}
\end{equation}
This can be localized by replacing the constant parameters with arbitrary functions, $\eps_1\to\eps_1(x)$ and $\eps^\m_2\to\eps^\m_2(x)$. The shift by $\eps_1$ is the parent symmetry. The descendant symmetry with parameter $\eps^\m_2$ can be recovered locally by setting $\eps_1(x)=\eps^\m_2(x)x_\m$. This is the local equivalence condition~\eqref{localequivalence} where $f_\m(x)=x_\m$. The integrability condition~\eqref{integrabilitycondition} and the current equivalence relation~\eqref{currentequivalencerelation} then translate to
\begin{equation}
J_{1\n}[\p,x]=\de_\m N^\m_\n[\p,x]\;,\quad
J^\m_{2\n}[\p,x](x)=x_\n J^\m_1[\p,x](x)-N^\m_\n[\p,x](x)\;.
\label{currentequivalencegalileon}
\end{equation}
Interestingly, the set of functions $N^\m_\n[\p,x]$ is in this case the primary object, which by means of~\eqref{currentequivalencegalileon} determines both currents.

Let us see how this works explicitly. Take the class of Lagrangians of the type
\begin{equation}
\La=\sum_{n=0}^Dc_n\La^{(n)}_\mathrm{Gal}+\La_\mathrm{int}(\de\de\p)\;.
\label{galileonclass}
\end{equation}
Here $\smash{\La_\mathrm{int}}$ is an arbitrary function of second derivatives of $\p$, which is manifestly invariant under~\eqref{galileonsym}. Furthermore, $\smash{\La^{(n)}_\mathrm{Gal}}$ with $n=0,\dotsc,D$ is the set of quasi-invariant Galileon Lagrangians, cf.~Sect.~\ref{subsubsec:Galileon},
\begin{equation}
\La^{(n)}_\mathrm{Gal}=\ve^{\m_1\dotsb\m_n\l_{n+1}\dotsb\l_D}\ve^{\n_1\dotsb\n_n}_{\phantom{\n_1\dotsb\n_n}\l_{n+1}\dotsb\l_D}\p(\de_{\m_1}\de_{\n_1}\p)\dotsb(\de_{\m_n}\de_{\n_n}\p)\;.
\end{equation}
It is easy to extract the current $J^\m_1$ by evaluating the variation of the action under the shift $\p(x)\to\p(x)+\eps_1(x)$. With the shorthand notation for the symmetric tensor
\begin{equation}
X^{\m\n}\equiv-\sum_{n=0}^D(n+1)c_n\ve^{\m\m_2\dotsb\m_n\l_{n+1}\dotsb\l_D}\ve^{\n\n_2\dotsb\n_n}_{\phantom{\n\n_2\dotsb\n_n}\l_{n+1}\dotsb\l_D}\prod_{i=2}^n(\de_{\m_i}\de_{\n_i}\p)\;,
\label{Xmunu}
\end{equation}
the current takes the compact form
\begin{equation}
J^\m_1=X^{\m\n}\de_\n\p-\de_\n\PD{\La_\mathrm{int}}{(\de_\m\de_\n\p)}\;.
\label{J1galileon}
\end{equation}
This matches the integrability condition in~\eqref{currentequivalencegalileon} since $X^{\m\n}$ obviously has vanishing divergence and so $X^{\m\n}\de_\n\p=\de_\n(X^{\m\n}\p)$. It follows that
\begin{equation}
N^\m_\n=X^\m_{\phantom\m\n}\p-\PD{\La_\mathrm{int}}{(\de_\m\de^\n\p)}\;,
\end{equation}
and the current $J^\m_{2\n}$ is then given by the second relation in~\eqref{currentequivalencegalileon}. This result can be easily verified by evaluating the variation of the action under $\p(x)\to\p(x)+\eps^\m_2(x)x_\m$.

\begin{illustration}%
The simplest example of a theory from the class~\eqref{galileonclass} is the theory of a free massless scalar, $\La=(1/2)(\de_\m\p)^2$. This amounts to setting $c_1=(-1)^D/[2(D-1)!]$ and discarding all the other operators contributing to~\eqref{galileonclass}. By~\eqref{Xmunu}, this leads to $X^{\m\n}=2c_1g^{\m\n}(-1)^D(D-1)!=g^{\m\n}$. We then find in turn
\begin{equation}
J^\m_1=\de^\m\p\;,\qquad
N^\m_\n=\d^\m_\n\p\;,\qquad
J^\m_{2\n}=x_\n\de^\m\p-\d^\m_\n\p\;.
\end{equation}
\end{illustration}

To illustrate the background gauging formalism, we need to gauge the parent symmetry. This is most easily done by first integrating the quasi-invariant Galileon terms by parts so that each factor of $\phi$ carries at least one derivative. All one then has to do is to replace every $\de_\m\p$ with $D_\m\p\equiv\de_\m\p-A_\m$. This gives the gauged action
\begin{align}
\tilde S\{\p,A\}=&-\sum_{n=0}^D c_n\ve^{\m_1\dotsb\m_n\l_{n+1}\dotsb\l_D}\ve^{\n_1\dotsb\n_n}_{\phantom{\n_1\dotsb\n_n}\l_{n+1}\dotsb\l_D}\\
\notag
&\times\int\D^D\!x\, D_{\m_1}\p(x)D_{\n_1}\p(x)\prod_{i=2}^n[\de_{\m_i}D_{\n_i}\p(x)]+\int\D^D\!x\,\La_\mathrm{int}(\de D\p)(x)\;.
\end{align}
This is manifestly invariant under the simultaneous gauge transformation $\udelta\p(x)=\eps_1(x)$ and $\udelta A_\m(x)=\de_\m\eps_1(x)$. The latter corresponds to~\eqref{udeltaA} with $\dA_{1\m}=0$ and $\dAg^\n_{1\m}=\d^\n_\m$; no terms with higher derivatives of $\eps_1(x)$ are present. The consistency condition~\eqref{integrabilityR} is trivially satisfied with $R^\m=0$, and~\eqref{currentR} then gives
\begin{equation}
J^\m_1[\p,x](x)=-\at{\frac{\udelta\tilde S\{\p,A\}}{\udelta A_\m(x)}}{\mathrm{vac}}\;.
\label{JU(1)fromgauged}
\end{equation}
It takes little effort to convince oneself that this exactly reproduces~\eqref{J1galileon}.

To extract the descendant current $J^\m_{2\n}$ from the same gauged action, we set $\eps_1(x)=\eps^\m_2(x)x_\m$, which changes the transformation rule for $A_\m$ to
\begin{equation}
\udelta A_\m(x)=\eps_{2\m}(x)+x_\n\de_\m\eps_2^\n(x)\;.
\end{equation}
This matches~\eqref{udeltaA} with $\smash{\dA_{2\m\n}=g_{\m\n}}$ and $\smash{\dAg^\n_{2\a\m}=\d^\n_\m x_\a}$, and~\eqref{integrabilityR} then becomes
\begin{equation}
\de_\m R^\m_\n[\p,x](x)=g_{\m\n}\at{\frac{\udelta\tilde S\{\p,A\}}{\udelta A_\m(x)}}{\mathrm{vac}}=-J_{1\n}[\p,x](x)\;.
\end{equation}
With the help of the first relation in~\eqref{currentequivalencegalileon} this is seen to be solved by $\smash{R^\m_\n=-N^\m_\n}$. The master equation~\eqref{currentR} then gives $\smash{J^\m_{2\n}}$ in agreement with the current equivalence relation, that is the second identity in~\eqref{currentequivalencegalileon}.


\subsection{Spacetime Translations and Rotations}
\label{subsec:diffexamplesrotations}

In relativistic theories with only scalar fields, spacetime translations and rotations can be implemented through their action on Minkowski coordinates,
\begin{equation}
x^\m\to x^\m+\eps^\m_1+\eps^{\m\n}_2x_\n\;.
\label{translationrotation}
\end{equation}
The vector $\eps^\m_1$ parameterizes infinitesimal translations and the antisymmetric tensor $\eps^{\m\n}_2$ infinitesimal spacetime rotations. The evolutionary form~\eqref{locallyequivalentsym} of the transformations induced on the fields $\psi^i$ is obtained by setting, respectively,\footnote{The factor $1/2$ in $\udelta_2\psi^i$ is conventional and compensates for the fact that the components $\smash{\eps^{\a\b}_2}$ and $\smash{\eps^{\b\a}_2}$ with fixed $\a\neq\b$ parameterize the same transformation. The derivation of the angular momentum tensor $\smash{M^\m_{\phantom\m\a\b}}$ below assumes that the same factor $1/2$ is used in its definition via~\eqref{noethertrick}.}
\begin{equation}
\udelta_1\psi^i=\eps^\m_1\df^i_\m=-\eps^\m_1\de_\m\psi^i\;,\qquad
\udelta_2\psi^i=\frac12\eps^{\m\n}_2\df^i_{\m\n}=\eps^{\m\n}_2x_\m\de_\n\psi^i\;.
\end{equation}
Translations and rotations are locally equivalent via~\eqref{localequivalence}; the latter can be recovered from the former by setting
\begin{equation}
\eps^\m_1(x)=\frac12f^\m_{\a\b}(x)\eps^{\a\b}_2(x)\;,\qquad
f^\m_{\a\b}(x)\equiv\d^\m_\a x_\b-\d^\m_\b x_\a\;.
\label{localtranslationrotation}
\end{equation}

The translations are the parent symmetry and the rotations its descendant. The Noether current corresponding to the parent symmetry is the \emph{canonical energy--momentum} (EM) \emph{tensor} $T^\m_{\phantom\m\n}$, defined by
\begin{equation}
\udelta_1S=\int\D^D\!x\,T^\m_{\phantom\m\n}[\psi,x](x)\de_\m\eps^\n_1(x)\;.
\label{canonicalEMtensor}
\end{equation}
The left-hand side of the integrability condition~\eqref{integrabilitycondition} then becomes
\begin{equation}
T^\m_{\phantom\m\n}\de_\m f^\n_{\a\b}=T_{\b\a}-T_{\a\b}\;.
\end{equation}
It is known that in purely scalar theories, the canonical EM tensor is symmetric; I will justify this below using the background gauging approach. The integrability condition is then satisfied by $N^\m_{\a\b}=0$. The equivalence relation~\eqref{currentequivalencerelation} in turn gives the Noether current for spacetime rotations, that is the angular momentum tensor,
\begin{equation}
M^\m_{\phantom\m\a\b}=f^\n_{\a\b}T^\m_{\phantom\m\n}=x_\b T^\m_{\phantom\m\a}-x_\a T^\m_{\phantom\m\b}\;.
\label{Mab}
\end{equation}
This is just the familiar relation between momentum and angular momentum.

\begin{watchout}%
The equivalence relation~\eqref{Mab} does not apply to theories of fields with nonzero spin without further qualification. Namely, in such theories, spacetime translations and rotations are not automatically locally equivalent. The reason for this is that the components of fields with spin undergo a transformation under spacetime rotations, which cannot be reproduced by the naive local translation, $\udelta_1\psi^i(x)=-\eps_1^\m(x)\de_\m\psi^i(x)$, for any choice of $\eps^\m_1(x)$. The problem can be circumvented by a judicious choice of the localized transformation under translations. Instead of developing a general theory, let me illustrate the idea on a simple example.
\end{watchout}

\begin{illustration}%
Recall~\refex{ex:canonicalEM} where I introduced the following theory of a real scalar field $\p$ and a vector field $A_\m$,
\begin{equation}
\La=A^\m\de_\m\p-\frac12A^\m A_\m\;.
\end{equation}
We found that the canonical EM tensor of this theory is not symmetric. However, we could solve the problem by considering a modified local translation,
\begin{equation}
\udelta_1\p(x)=-\eps^\n_1(x)\de_\n\p(x)\;,\hspace{0.5em}
\udelta_1A_\m(x)=-\eps^\n_1(x)\de_\n A_\m(x)-A_\n(x)\de_\m\eps^\n_1(x)\;.
\label{deltaphiA}
\end{equation}
This is motivated by the fact that under a general local coordinate transformation, $A_\m$ should transform as a covariant vector. Noether's theorem then gives the improved, off-shell-symmetric EM tensor
\begin{equation}
\tilde T^{\m\n}=g^{\m\n}\La+A^\m A^\n-(A^\m\de^\n\p+A^\n\de^\m\p)\;.
\label{phiAlocal}
\end{equation}

The prescription~\eqref{deltaphiA} restores local equivalence of spacetime translations and rotations in spite of the presence of the vector field. Namely, inserting $\smash{\eps^\m_1(x)=(1/2)f^\m_{\a\b}(x)\eps^{\a\b}_2}$ with constant $\smash{\eps^{\a\b}_2}$ in~\eqref{deltaphiA} gives
\begin{equation}
\udelta_1A_\m(x)\xrightarrow{\eps^\m_1(x)=\eps^{\m\n}_2x_\n}\eps^{\a\n}_2x_\a\de_\n A_\m(x)+\eps_{2\m\n}A^\n(x)\;.
\end{equation}
This is exactly how $A_\m$ should transform under a spacetime rotation. We can then localize both translations and rotations by using~\eqref{deltaphiA} augmented with~\eqref{localtranslationrotation}. Since $\tilde T^{\m\n}$ is symmetric, the integrability condition is now satisfied and we reproduce the angular momentum tensor by replacing~\eqref{Mab} with
\begin{equation}
M^\m_{\phantom\m\a\b}=x_\b\tilde T^\m_{\phantom\m\a}-x_\a\tilde T^\m_{\phantom\m\b}\;.
\label{Mabtilde}
\end{equation}
\end{illustration}

Let us now address the relation between spacetime translations and rotations within the background gauging formalism. We start by gauging the parent symmetry: spacetime translations. This amounts to promoting the flat Minkowski spacetime to a possibly curved spacetime manifold. Here it proves useful to invoke the differential-geometric language developed in Appendix~\ref{app:diffgeom}, to which I refer the reader for the basic terminology. Thus, the geometry of the spacetime manifold can be characterized by a local coframe (basis of differential 1-forms), $\smash{\vec e^{*\fr\a}\equiv e^{*\fr\a}_\m\D x^\m}$.\footnote{Following the notation introduced in Sect.~\ref{subsec:geomsymcoset}, I use underlined indices to indicate components within a local (co)frame. Ordinary Lorentz indices refer to a set of local spacetime coordinates, $x^\m$.} Under an infinitesimal coordinate transformation, $x^\m\to x^\mu+\eps^\m_1(x)$, the coordinate components of the coframe change as
\begin{equation}
\udelta e^{*\fr\a}_\m(x)=-\eps^\n_1(x)\de_\n e^{*\fr\a}_\m(x)-e^{*\fr\a}_\n(x)\de_\m\eps_1^\n(x)\;.
\label{deltacoframe}
\end{equation}
This matches~\eqref{udeltaA} provided we identify $\smash{\dA^{\fr\a}_{1\n\m}=-\de_\n e^{*\fr\a}_\m}$ and $\smash{\dAg^{\fr\a\n}_{1\l\m}=-\d^\n_\m e^{*\fr\a}_\l}$. A gauge-invariant action $\tilde S\{\psi,e^*\}$ is obtained from $S\{\psi\}$ by replacing derivatives of $\psi^i$ therein with covariant derivatives, and the flat Minkowski metric $g_{\fr\m\fr\n}$ with the spacetime metric, $\smash{g_{\m\n}(x)=g_{\fr\a\fr\b}e^{*\fr\a}_\m(x)e^{*\fr\b}_\n(x)}$. An appropriate volume element is built using the determinant of the coframe $e^{*\fr\a}_\m(x)$ as a matrix.

Eventually, we want to remove the background and go back to the flat Minkowski spacetime. I assume that the latter is described by global coordinates $x^\m$ in which the components $\smash{e^{*\fr\a}_\m}$ are constant. Then the consistency condition~\eqref{integrabilityR} is solved by $R^\m_\n=0$, and~\eqref{currentR} reduces to
\begin{equation}
T^\m_{\phantom\m\n}[\psi,x](x)=e^{*\fr\a}_\n(x)\at{\frac{\udelta\tilde S\{\psi,e^*\}}{\udelta e^{*\fr\a}_\m(x)}}{\mathrm{vac}}\;.
\label{EMtensorfrome}
\end{equation}

Suppose now that the theory at hand can be coupled to the background geometry in a way that the action $\tilde S\{\psi,e^*\}$ depends on $e^{*\fr\a}_\m$ only through the spacetime metric, $g_{\m\n}$. This is an example of a higher-rank tensor background hinted at above. Then the variation of the action~\eqref{udeltatildeS} is replaced with
\begin{equation}
\udelta_g\tilde S=\int\D^D\!x\,\frac{\udelta\tilde S}{\udelta g_{\m\n}}(-\eps_1^\l\de_\l g_{\m\n}-g_{\l\n}\de_\m\eps_1^\l-g_{\m\l}\de_\n\eps_1^\l)\;.
\end{equation}
On a flat Minkowski background, this automatically vanishes for constant $\eps^\m_1$ thanks to $\smash{\at{\de_\l g_{\m\n}}{\mathrm{vac}}=0}$. The EM tensor is then identified as
\begin{equation}
T^{\m\n}[\psi,x](x)=2\at{\frac{\udelta\tilde S\{\psi,g\}}{\udelta g_{\m\n}(x)}}{\mathrm{vac}}\;.
\label{EMtensorfromg}
\end{equation}
This is the \emph{Hilbert EM tensor}. Should our theory be Lorentz-invariant and only contain scalar fields, it can always be coupled to the spacetime background using the metric $g_{\m\n}$. At the same time, the scalar fields retain the transformation rule $\udelta_1\psi^i(x)=-\eps^\m_1(x)\de_\m\psi^i(x)$. This guarantees that the canonical and Hilbert EM tensors coincide, and shows that the canonical EM tensor is necessarily symmetric.

\begin{watchout}%
Theories of fields with spin can be coupled to background geometry in different manners, depending on how the spin index of the fields is treated. Let me illustrate this in the case of a vector field $A_\m$. One already mentioned possibility is to treat this as a covariant vector that transforms under a general coordinate transformation via the second relation in~\eqref{deltaphiA}. It is then possible to construct a generally covariant action $\tilde S\{A,g\}$ that only depends on the background through the metric. The resulting EM tensor~\eqref{EMtensorfromg} is symmetric.

On the other hand, we may project the field on the local frame $\smash{e^\m_{\fr\a}}$ and treat $\smash{A_{\fr\a}\equiv e^\m_{\fr\a}A_\m}$ as a set of spacetime scalars. This corresponds to keeping the naive transformation under local translations, $\smash{\udelta A_{\fr\a}(x)=-\eps_1^\m(x)\de_\m A_{\fr\a}(x)}$. In this case, the gauged action $\tilde S\{A,e^*\}$ may depend explicitly on the local frame. We then have to use~\eqref{EMtensorfrome} instead of~\eqref{EMtensorfromg}. This gives the canonical EM tensor which is not necessarily symmetric.

The general moral is that different choices of gauging the symmetry may lead to different expressions for the Noether current. This is tantamount to different choices of the localized symmetry transformation~\eqref{udeltapsi}. The resulting alternative Noether currents however only differ by terms that vanish on-shell, modulo the inevitable ambiguity with respect to contributions whose divergence vanishes off-shell.
\end{watchout}

With the EM tensor at hand, we can extract the Noether current for any descendant symmetry that can be locally reproduced by spacetime translations (see e.g.~Sect.~14.3 of~\cite{Burgess2021a}). Consider a one-parameter group of symmetries, generated by a vector field $\vec\kil(x)$. The corresponding local transformation with parameter $\eps_2(x)$ can be written as $x^\m\to x^\m+\eps^\m_1(x)$ with $\eps^\m_1(x)=\eps_2(x)\kil^\m(x)$. This turns~\eqref{deltacoframe} to
\begin{equation}
\udelta e^{*\fr\a}_\m=-\eps_2(\kil^\n\de_\n e^{*\fr\a}_\m+e^{*\fr\a}_\n\de_\m\kil^\n)-\kil^\n e^{*\fr\a}_\n\de_\m\eps_2\;,
\end{equation}
whence we extract $\smash{\dA^{\fr\a}_{2\m}=-(\kil^\n\de_\n e^{*\fr\a}_\m+e^{*\fr\a}_\n\de_\m\kil^\n)}$ and $\smash{\dAg^{\fr\a\n}_{2\m}=-\d^\n_\m\kil^\l e^{*\fr\a}_\l}$. The consistency condition~\eqref{integrabilityR} then boils down to
\begin{equation}
\de_\m R^\m=-T^\m_{\phantom\m\n}\de_\m\kil^\n=-\frac12T^{\m\n}(\de_\m\kil_\n+\de_\n\kil_\m)\;.
\end{equation}
In the last step, I assumed that $T^{\m\n}$ is symmetric, which we now know to be guaranteed in case the action $\tilde S\{\psi,e^*\}$ only depends on the coframe through the metric. But should $\vec\kil(x)$ actually generate a symmetry of the flat Minkowski spacetime, $\de_\m\kil_\n+\de_\n\kil_\m$ must vanish by the Killing equation~\eqref{killing}. Then~\eqref{integrabilityR} is satisfied with $R^\m=0$ and~\eqref{currentR} gives immediately
\begin{equation}
J^\m[\psi,x](x)=T^{\m\n}[\psi,x](x)\kil_\n(x)\;.
\end{equation}
This confirms, among others, a general relation between the Hilbert EM tensor and the angular momentum tensor, anticipated in~\eqref{Mabtilde}.


\subsection{Galilei Invariance}
\label{subsec:diffexamplesgalilei}

The previous two examples were relativistic in spirit. Let us therefore have a look at one more example that is intrinsically nonrelativistic. Consider a theory of a complex Schr\"odinger field $\psi(\vec x,t)$ whose excitations are nonrelativistic particles with mass $m$. Suppose that the theory is invariant under spatial translations and under the internal $\gr{U}(1)$ group of phase transformations of the field. These symmetries act respectively on the spatial coordinates, $\vec x'=\vec x+\vec\eps$, and the field, $\psi'=\E^{\I\a}\psi$, and so correspond to
\begin{equation}
\udelta_1\psi(\vec x,t)=\I\a\psi(\vec x,t)-\skal\eps\nabla\psi(\vec x,t)\;.
\label{galileitranslationU(1)}
\end{equation}
Finally, it is well-known from elementary quantum mechanics that the Schr\"odinger equation for a free particle is invariant under Galilei boosts,
\begin{equation}
\psi'(\vec x',t)=\psi'(\vec x+\vec vt,t)=\exp\left[\I m\left(\skal vx+\frac12\vec v^2t\right)\right]\psi(\vec x,t)\;.
\label{galileiboost}
\end{equation}
Here $\vec v$ is the boost velocity which plays the role of the transformation parameter. Let us assume that the invariance under~\eqref{galileiboost} is inherited by our possibly interacting nonrelativistic field theory.

The evolutionary form of an infinitesimal Galilei boost is
\begin{equation}
\udelta_2\psi(\vec x,t)=\I m\skal vx\psi(\vec x,t)-t\skal v\nabla\psi(\vec x,t)\;.
\label{localizedGal}
\end{equation}
This can be locally recovered as a combination of a spatial translation and a phase transformation if we set
\begin{equation}
\a(\vec x,t)=m\skal xv(\vec x,t)\;,\qquad
\vec\eps(\vec x,t)=t\vec v(\vec x,t)\;.
\label{localgalileiboost}
\end{equation}
In this case, therefore, the combination of spatial translations and internal $\gr{U}(1)$ is the parent symmetry, whereas the Galilei boosts are its descendants. The rest follows a familiar pattern. The integrability condition~\eqref{integrabilitycondition} requires that there is a local function $N^{\m r}[\psi,\vec x,t]$ of the field and its derivatives such that
\begin{equation}
mJ^r+T^{0r}=\de_\m N^{\m r}\;,
\label{integrabilitygalilei}
\end{equation}
where $J^\m$ is the current of the internal $\gr{U}(1)$ symmetry. The general equivalence relation~\eqref{currentequivalencerelation} then gives the current $B^{\m r}$ corresponding to Galilei boosts,
\begin{equation}
B^{\m r}=mx^rJ^\m+tT^{\m r}-N^{\m r}\;.
\label{equivalencegalilei}
\end{equation}

Should $N^{\m r}$ happen to be zero, the integrability condition~\eqref{integrabilitygalilei} would boil down to $T^{0r}=-mJ^r$. Except for the sign convention in the definition of the two currents, this is the usual nonrelativistic relation between momentum and velocity. Moreover, \eqref{equivalencegalilei} would imply $B^{0r}=mx^rJ^0+tT^{0r}=m(x^rJ^0-tJ^r)$. This generalizes to field theory the well-known fact that the conserved charge for a Galilei boost of a particle of mass $m$ is $m\vec x-\vec pt$, where $\vec p$ is the particle momentum. Whether or not, or under what conditions, $N^{\m r}$ actually is zero needs to be inspected case by case.

The story of Galilei invariance becomes very interesting, and nontrivial, once we consider background gauging. What we need is to gauge simultaneously spatial translations, Galilei boosts and the internal $\gr{U}(1)$ phase transformations. Since the Galilei boosts are descendant, we want to primarily gauge spatial translations and the internal $\gr{U}(1)$. The actions of these cannot be trivially separated as one could do in a Lorentz-invariant field theory. Namely, the gauged translations must affect the phase of the Schr\"odinger field in a way that reproduces~\eqref{galileiboost}.

It turns out that the minimal setup that does the job includes a background spatial metric $g_{rs}(x)$ and a gauge field $A_\m(x)$ for the $\gr{U}(1)$ symmetry~\cite{Son2006a}. Under a combination of an infinitesimal local translation, $\vec x'=\vec x+\vec\eps(\vec x,t)$, and a $\gr{U}(1)$ transformation with parameter $\a(\vec x,t)$, the background fields vary by
\begin{equation}
\begin{split}
\udelta g_{rs}(x)&=-\eps^u(x)\de_u g_{rs}(x)-g_{us}(x)\de_r\eps^u(x)-g_{ru}(x)\de_s\eps^u(x)\;,\\
\udelta A_0(x)&=\de_0\a(x)-\eps^u(x)\de_uA_0(x)-A_u(x)\de_0\eps^u(x)\;,\\
\udelta A_r(x)&=\de_r\a(x)-\eps^u(x)\de_uA_r(x)-A_u(x)\de_r\eps^u(x)-mg_{rs}(x)\de_0\eps^s(x)\;.
\end{split}
\label{sonwingate}
\end{equation}
The first two lines take the expected form. The only nontrivial ingredient really is the last term on the third line of~\eqref{sonwingate}, which depends on the mass parameter $m$. This reflects the fact that the realization of spatial translations and Galilei boosts on a Schr\"odinger field is centrally extended, with $m$ playing the role of the central charge. The associated nonrelativistic spacetime geometry is known as \emph{Newton--Cartan}. See~\cite{Geracie2014b,Jensen2018} for more details on nonrelativistic field theory in Newton--Cartan spacetimes, and~\cite{Bergshoeff2022} for broader background on nonrelativistic geometry.

The localized transformation of the Schr\"odinger field itself is given by~\eqref{galileitranslationU(1)}. By combining this with~\eqref{sonwingate}, we then generate at once the Noether currents of all the symmetries in terms of a gauged action $\tilde S\{\psi,g,A\}$. First, the current of the $\gr{U}(1)$ symmetry is given by a trivial modification of~\eqref{JU(1)fromgauged},
\begin{equation}
J^\m[\psi,x](x)=-\at{\frac{\udelta\tilde S\{\psi,g,A\}}{\udelta A_\m(x)}}{\mathrm{vac}}\;,
\label{galileiU(1)}
\end{equation}
where ``vac'' now indicates setting $g_{rs}$ to the flat Euclidean metric $\d_{rs}$ and $A_\m$ to zero. Next comes the EM tensor. By~\eqref{canonicalEMtensor}, the only components that the invariance under (time-dependent) spatial coordinate transformations gives us access to are
\begin{equation}
T^{0r}[\psi,x](x)=m\at{\frac{\udelta\tilde S\{\psi,g,A\}}{\udelta A_r(x)}}{\mathrm{vac}}\;,\hspace{0.9em}
T^{rs}[\psi,x](x)=2\at{\frac{\udelta\tilde S\{\psi,g,A\}}{\udelta g_{rs}(x)}}{\mathrm{vac}}\;.
\label{galileiT0r}
\end{equation}

\begin{watchout}%
Note that~\eqref{galileiU(1)} and~\eqref{galileiT0r} together imply a relation between momentum density and the particle number current, $T^{0r}=-mJ^r$. This corresponds to~\eqref{integrabilitygalilei} with vanishing $N^{\m r}$. It might come as a surprise that the background gauge invariance yields a stronger constraint on the Noether currents than the original, physical symmetry. After all, background gauge invariance per se is merely a handy tool to encode the consequences of the symmetry. However, recall that the realization of Galilei transformations on the Schr\"odinger field is centrally extended. This is an obstruction that makes gauging of the symmetry nontrivial. It is ultimately the assumption that the gauging is possible that is responsible for the vanishing of $N^{\m r}$. This does not add any new physics, but rather constrains the class of theories to which the background gauging procedure can be applied.
\end{watchout}

Next, the current $B^{\m r}$ for Galilei boosts can be identified by using~\eqref{udeltatildeS} together with $\smash{\udelta_\psi\tilde S=\int\D^D\!x\,B^\m_{\phantom\m r}\de_\m v^r}$. Upon inserting~\eqref{localgalileiboost} in~\eqref{sonwingate}, we find that the consistency condition~\eqref{integrabilityR} is satisfied with $R^\m_r=0$. This leads via~\eqref{currentR} to
\begin{equation}
\begin{split}
B^{0r}[\psi,x](x)&=-mx^r\at{\frac{\udelta\tilde S\{\psi,g,A\}}{\udelta A_0(x)}}{\mathrm{vac}}+mt\at{\frac{\udelta\tilde S\{\psi,g,A\}}{\udelta A_r(x)}}{\mathrm{vac}}\;,\\
B^{rs}[\psi,x](x)&=-mx^s\at{\frac{\udelta\tilde S\{\psi,g,A\}}{\udelta A_r(x)}}{\mathrm{vac}}+2t\at{\frac{\udelta\tilde S\{\psi,g,A\}}{\udelta g_{rs}(x)}}{\mathrm{vac}}\;.
\end{split}
\end{equation}
These can be combined using~\eqref{galileiU(1)} and~\eqref{galileiT0r} into $B^{\m r}=mx^rJ^\m+tT^{\m r}$, which is just~\eqref{equivalencegalilei} without the $N^{\m r}$ term. This is consistent with our conclusion above that $N^{\m r}=0$ as a consequence of the assumption that the gauged action $\tilde S\{\psi,g,A\}$ is invariant under~\eqref{sonwingate}.


\subsection{Changing the Background: Magnetic Translations}
\label{subsec:magtrans}

Our discussion in this chapter started with declaring the existence of certain symmetry and its subsequent localization. However, it is also possible to adopt the opposite approach. Suppose we know a priori the type of background  gauge fields and the local transformations that make the action $\tilde S\{\psi,A\}$ invariant. It is then possible to identify the original symmetry as the subset of the gauge transformations that preserve a trivial background.

\begin{illustration}%
Recall the Galileon symmetry of Sect.~\ref{subsec:diffexamplesgalileon}. Out of all the gauge transformations $\smash{A_\m(x)\to A_\m(x)+\de_\m\eps(x)}$, only those with constant $\eps$ preserve the background, in this case regardless of the specific choice of $A_\m(x)$. Constant $\eps$ corresponds exactly to the shift symmetry, acting on the Galileon scalar field $\p$.
\end{illustration}

\begin{illustration}%
In case of relativistic theories of fields with integer spin, we can consider as the background a generic spacetime metric $g_{\m\n}(x)$. The set of local general coordinate transformations that preserve such a background corresponds to (the continuous part of) its isometry group (see Appendix~\ref{appsubsec:isometries} for the necessary mathematical background). In case of the flat Minkowski spacetime, this is just the Poincar\'e group, as follows from a trivial modification of~\refex{ex:Killing}. Note that choosing the background geometry is not equivalent to fixing the coframe $e^{*\fr\a}_\m$ as~\eqref{EMtensorfrome} might naively suggest. The reason is that the coframe is not uniquely determined for a given spacetime manifold. This underlines the necessity to carefully identify the geometric data, tailored to the given spacetime symmetry and uniquely representing the spacetime background.
\end{illustration}

Once we have the gauged action $\tilde S\{\psi,A\}$, we may use it to study physics on nontrivial backgrounds. Indeed, the vanishing of the variation of $\tilde S\{\psi,A\}$ under a localized symmetry transformation is an exact statement valid for any choice of $\smash{A^A_\m}$. Imposing the equation of motion on the dynamical fields $\psi^i$ amounts to setting $\smash{\udelta_\psi\tilde S=0}$, which turns~\eqref{udeltatildeS} into a generalized ``conservation law,''
\begin{equation}
\at{\int\D^D\!x\,\frac{\udelta\tilde S}{\udelta A^A_\m}\bigl(\eps^B \dA^A_{B\m}+\de_\n\eps^B\dAg^{A\n}_{B\m}+\dotsb\bigr)}{\text{on-shell}}=0\;.
\end{equation}
For this to represent an actual divergence-type conservation law, the chosen background must possess \emph{some} symmetry. This may however be different from the symmetry that we originally gauged by introducing the fields $A^A_\m$ in the first place. I will illustrate this on an interesting example starting from Galilei invariance in $d=3$ spatial dimensions.

The flat nonrelativistic spacetime corresponds to $g_{rs}=\d_{rs}$ and $A_\m=0$. The only infinitesimal transformations of type~\eqref{sonwingate} that preserve this background are a linear combination of Euclidean translations and rotations, Galilei boosts, and the internal $\gr{U}(1)$ transformations. This is not obvious but is straightforward to check.

Let us now change the background by introducing a uniform magnetic field~$\vec B$. Such a uniform background field should preserve some notion of translation invariance. This however cannot be the naive Euclidean translations due to the coordinate-dependence of the vector potential $\vec A(\vec x)$ of the magnetic field. The details turn out to depend on the choice of gauge; I will adopt the symmetric gauge, $\vec A(\vec x)=(\vec B\times\vec x)/2$. The translation invariance can now be rescued if it is accompanied by a local $\gr{U}(1)$ transformation. In terms of the Schr\"odinger field $\psi(\vec x,t)$, this so-called \emph{magnetic translation symmetry} takes the form
\begin{equation}
\psi'(\vec x',t)=\psi'(\vec x+\vec\eps,t)=\exp\left[\frac\I2\vec\eps\cdot(\vec x\times\vec B)\right]\psi(\vec x,t)\;.
\label{magtrans}
\end{equation}
One remarkable consequence of this twisted translation symmetry is that the components of its generator $P_r$ no longer commute with each other. It is easy to check that they satisfy the commutation relation $[P_r,P_s]=-\I\ve_{rsu}B^uQ$, where $Q$ is the generator of the $\gr{U}(1)$ symmetry.

The magnetic translation symmetry~\eqref{magtrans} can be localized by promoting $\vec\eps$ to a function of spacetime coordinates and setting
\begin{equation}
\psi'(\vec x+\vec\eps(x),t)=\exp[-\I\vec\eps(x)\cdot\vec A(x)]\psi(\vec x,t)\;,
\end{equation}
where $\vec A(x)$ is now the spatial part of the variable $A_\m(x)$. At the infinitesimal level, this is equivalent to~\eqref{galileitranslationU(1)} with $\a(x)=-\vec\eps(x)\cdot\vec A(x)$. Following the same reasoning as in Sect.~\ref{subsec:diffexamplesgalilei}, we then get modified relations for the EM tensor,
\begin{equation}
\begin{split}
T^{0r}[\psi,x](x)&=\at{-mJ^r[\psi,x](x)-2A^r(x)J^0[\psi,x](x)}{\mathrm{vac}}\;,\\
T^{rs}[\psi,x](x)&=\at{2\frac{\udelta\tilde S\{\psi,g,A\}}{\udelta g_{rs}(x)}}{\mathrm{vac}}-\at{2A^s(x)J^r[\psi,x](x)}{\mathrm{vac}}\;,
\end{split}
\label{magtransequivalence}
\end{equation}
which generalize~\eqref{galileiT0r} to a uniform magnetic background. Here ``vac'' refers to the background vector potential in the symmetric gauge, $\smash{\at{A_\m(x)}{\mathrm{vac}}=(1/2)\d^r_\m\ve_{rsu}B^sx^u}$. The components of $J^\m[\psi,x]$ are still given by~\eqref{galileiU(1)}.

The magnetic translations illustrate in a very nontrivial manner many of the concepts introduced in Chap.~\ref{chap:symmetry} and this chapter. On the magnetic background, the Lagrangian of the Schr\"odinger field depends explicitly on the spacetime coordinates, yet possesses a nontrivial notion of translation symmetry. The Lie algebra of infinitesimal translations is centrally extended. The translation symmetry can be gauged simultaneously with the internal $\gr{U}(1)$ symmetry by introducing a spatial metric and allowing the background fields to transform according to~\eqref{sonwingate}. This allows one to identify a nontrivial relation between the EM tensor and the Noether current of the $\gr{U}(1)$ symmetry, shown in~\eqref{magtransequivalence}.


\section{Application to Scattering of Nambu--Goldstone Bosons}
\label{sec:diffscattering}

I will now return to the promise I made in Sect.~\ref{sec:beyondAdler}.\footnote{A reader interested primarily in the development of the general theory for spontaneously broken spacetime symmetry may proceed directly to the next chapter.} Therein, I suggested that the enhanced soft limit of scattering amplitudes of NG bosons in the DBI and Galileon theories arises from the presence of extended symmetry. We now have all we need to be able to understand why. The material of this section loosely follows~\cite{Cheung2017a}.


\subsection{Galileon Theory}
\label{subsec:diffscatteringgalileon}

Let us start with the simpler Galileon theory and recall the discussion of the Galileon symmetry in Sect.~\ref{subsec:diffexamplesgalileon}. In any translationally-invariant theory endowed with the Galileon symmetry, the parent current $J^\m_1[\p]$ will not depend explicitly on the coordinates. By the integrability condition in~\eqref{currentequivalencegalileon}, the same holds for the local function $N^\m_\n[\p]$. Following the philosophy of Sect.~\ref{sec:Adler}, we now consider an arbitrary scattering process $\a\to\b$. Using the translation property $J^\m_1[\p](x)=\E^{\I P\cdot x}J^\m_1[\p](0)\E^{-\I P\cdot x}$ and analogously $\smash{N^\m_\n[\p](x)=\E^{\I P\cdot x}N^\m_\n[\p](0)\E^{-\I P\cdot x}}$, we deduce
\begin{equation}
\amplitude{\b}{J_{1\n}[\p](0)}{\a}=-\I p_\m\amplitude{\b}{\smash{N^\m_\n[\p](0)}}{\a}\;,
\label{JNrelationgalileon}
\end{equation}
where $p^\m\equiv p^\m_\a-p^\m_\b$.

The general rules of polology dictate that when $p^\m$ approaches the mass shell of a massless particle, the matrix element $\smash{\amplitude{\b}{J_{1\n}[\p](0)}{\a}}$ develops a pole. This corresponds to the emission of a NG boson with momentum $\vec p$ and allows one to extract the on-shell scattering amplitude $\smash{\Aa_{\a\to\b+\pi(\vec p)}}$; cf.~\eqref{polefactorization}. The same reasoning can however also be applied to the operator $N^\m_\n[\p]$, leading to
\begin{equation}
\amplitude{\b}{\smash{N^\m_\n[\p](0)}}{\a}=\amplitude{0}{\smash{N^\m_\n[\p](0)}}{\pi(\vec p)}_\mathrm{off}\frac1{p^2}\Aa_{\a\to\b+\pi(\vec p)}+R^\m_{\b\a\n}(p)\;,
\label{polologyN}
\end{equation}
where $\smash{R^\m_{\b\a\n}(p)}$ is by definition free of poles. By combining this with~\eqref{JNrelationgalileon}, we find a relation between $\smash{R^\m_{\b\a\n}(p)}$ and the function $\smash{R^\m_{\b\a}(p)}$ in~\eqref{polefactorization},
\begin{equation}
R_{\b\a\n}(p)=-\I p_\m R^\m_{\b\a\n}(p)\;.
\label{RRrelation}
\end{equation}
From~\eqref{Adlerproof} we then finally get an explicit expression for the on-shell amplitude,
\begin{equation}
\Aa_{\a\to\b+\pi(\vec p)}=\frac1Fp_\m p^\n R^\m_{\b\a\n}(p)\;,
\label{ARrelationgalileon}
\end{equation}
where $F$ is determined by  $\smash{\amplitude{0}{\smash{J^\m_1[\p](0)}}{\pi(\vec p)}=\I p^\m_\mathrm{on}F}$, and $p^\m_\mathrm{on}\equiv(\abs{\vec p},\vec p)$.

The general relation~\eqref{Adlerproof} allows one to assert the existence of Adler zero, provided the function $\smash{R^\m_{\b\a}(p)}$ remains regular in the soft limit $p^\m\to0$. A sufficient condition to ensure this is, as we know from Chap.~\ref{chap:scattering}, the absence of bilinear operators in the current $J^\m_1[\p]$. This automatically translates into the absence of bilinear operators in $N^\m_\n[\p]$. We then conclude from~\eqref{ARrelationgalileon} without further assumptions that in the soft limit, the scattering amplitudes in the Galileon theory vanish at least with the second power of momentum. In other words, the soft limit of scattering amplitudes in the Galileon theory is enhanced with the soft scaling parameter $\s=2$.


\subsection{Theories with Generalized Shift Symmetry}
\label{subsec:diffscatteringgenshift}

Having warmed up on the Galileon theory, we can now generalize the above argument to a much broader class of theories. For simplicity of notation, I will restrict the discussion to Lorentz-invariant theories of a single NG field $\p$ that are invariant under the constant shift $\p\to\p+\eps$~\cite{Cheung2017a}. Further generalization to theories invariant under spacetime translations but mere spatial rotations can be found in~\cite{Mojahed2022}. A generalization to theories of multiple flavors of NG bosons presumably exists but, as far as I know, has not been worked out in the literature.

Suppose that in addition to the constant shift symmetry, our theory is also invariant under a generalized shift of the type
\begin{equation}
\udelta\p(x)=\eps\bigl\{f(x)+\Phi[\p,x](x)\bigr\}\;.
\label{genshift}
\end{equation}
Here $f(x)$ is assumed to be a polynomial function of the spacetime coordinates of degree $\deg f\equiv n\geq1$. The local operator $\Phi[\p,x]$ may depend explicitly on the coordinates as well, but is required to be polynomial in $x^\m$ of degree lower than~$n$. Examples of symmetries of this type include both the symmetry of the DBI theory~\eqref{DBIsym} and the special Galileon symmetry~\eqref{sgalsym}.

Let us now localize~\eqref{genshift} and treat it as a generalized coordinate-dependent shift of the field, that is $\udelta\p(x)=\tilde\eps[\p,x](x)$ with $\tilde\eps[\p,x](x)\equiv\eps(x)\{f(x)+\Phi[\p,x](x)\}$. The corresponding variation of the action is
\begin{equation}
\begin{split}
\udelta S
=\int\D^D\!x\,J^\m_1[\p](x)\bigl\{&\eps(x)\de_\m\bigl[f(x)+\Phi[\p,x](x)\bigr]\\
&+\bigl[f(x)+\Phi[\p,x](x)\bigr]\de_\m\eps(x)\bigr\}\;,
\end{split}
\end{equation}
where $\smash{J^\m_1[\p]}$ is the parent current due to the constant shift symmetry. An in\-te\-gra\-bility condition required so that~\eqref{genshift} really is a symmetry of the action reads
\begin{equation}
J^\m_1[\p](x)\de_\m\bigl[f(x)+\Phi[\p,x](x)\bigr]=\de_\m N^\m[\p,x](x)\;.
\label{genshiftaux}
\end{equation}
Unlike the parent current $\smash{J_1^\m[\p]}$, the local function $N^\m[\p,x]$ may depend explicitly on the coordinates. For~\eqref{genshiftaux} to hold, however, the polynomial degree of $N^\m[\p,x]$ in $x^\m$ can be at most that of $\de_\m f(x)$, that is $n-1$.

The next step is to rewrite~\eqref{genshiftaux} in the form
\begin{equation}
\begin{split}
J^\m_1[\p]\de_\m f&=\Phi[\p,x]\de_\m J^\m_1[\p]-\de_\m\bigl(\Phi[\p,x]J^\m_1[\p]-N^\m[\p,x]\bigr)\\
&\equiv\Phi[\p,x]\de_\m J^\m_1[\p]+\de_\m M^\m[\p,x]\;.
\end{split}
\label{genshiftaux2}
\end{equation}
The new local function $M^\m[\p,x]$ of the field and its derivatives is again of polynomial degree at most $n-1$ in the spacetime coordinates. It satisfies the translation property $M^\m[\p,x](x)=\E^{\I P\cdot x}M^\m[\p,x](0)\E^{-\I P\cdot x}$, with the fields on the right-hand side evaluated at the origin but the explicit coordinate dependence left intact. Applying the polology rules to this operator then gives, in analogy to~\eqref{polologyN},
\begin{equation}
\begin{split}
\amplitude{\b}{M^\m[\p,x](0)}{\a}={}&\amplitude{0}{M^\m[\p,x](0)}{\pi(\vec p)}_\mathrm{off}\frac1{p^2}\Aa_{\a\to\b+\pi(\vec p)}\\
&+R^\m_{M\b\a}(x,p)\;.
\end{split}`
\label{polologyM}
\end{equation}
The coordinate dependence of the remainder function $\smash{R^\m_{M\b\a}(x,p)}$ comes entirely from that of $M^\m[\p,x](0)$. Therefore, $\smash{R^\m_{M\b\a}(x,p)}$ is a polynomial function of the coordinates with degree at most $n-1$.

The crucial observation is that due to current conservation, the on-shell matrix element of $\smash{\Phi[\p,x]\de_\m J^\m_1[\p]}$ between any initial and final states, $\ket\a$ and $\ket\b$, vanishes. By combining~\eqref{genshiftaux2} and~\eqref{polologyM}, we then get a constraint on the remainder functions, generalizing~\eqref{RRrelation},
\begin{equation}
R^\m_{\b\a}(p)\de_\m f(x)=-\I p_\m R^\m_{M\b\a}(x,p)+\de_\m R^\m_{M\b\a}(x,p)\;.
\label{RRMrelation}
\end{equation}
To understand the implications of this identity, we expand $\smash{R^\m_{\b\a}(p)}$ and $\smash{R^\m_{M\b\a}(x,p)}$ in a formal power series in $p^\m$. Let us denote their coefficients of order $k$ in $p^\m$ respectively as $a_k$ and $b_k(x)$, and write~\eqref{RRMrelation} symbolically as
\begin{equation}
a_k\cdot\de f(x)\simeq b_{k-1}(x)+\de\cdot b_k(x)\;.
\label{abconstraint}
\end{equation}
This must hold as an equality between polynomial functions of the coordinates. We now use repeatedly the fact that $\deg b_k(x)\leq n-1$ for all $k$. Setting $k=0$, we infer that $a_0=0$ and simultaneously $\de\cdot b_0(x)=0$, which implies that $b_0(x)$ is constant, $\deg b_0(x)=0$. In case $n\geq2$, we can next take~\eqref{abconstraint} with $k=1$. This can only hold if $a_1=0$ and simultaneously $\de\cdot b_1(x)\simeq-b_0(x)$, that is, $\deg b_1(x)\leq1$. By induction, we find that for any $k\leq n-1$, $a_k=0$ and simultaneously $\deg b_k(x)\leq k$.

The main conclusion is that for a generalized shift symmetry of the type~\eqref{genshift} where $\deg f(x)=n$, the series expansion of $\smash{R^\m_{\b\a}(p)}$ starts at order $n$ in energy--momentum. By~\eqref{Adlerproof}, the soft limit of scattering amplitudes of the theory is then enhanced with the soft degree $\s=n+1$. This confirms in particular that $\s=2$ for the DBI theory and $\s=3$ for the special Galileon theory.


\bibliographystyle{spphys}
\bibliography{references}
\chapter{Nonlinear Realization of Spacetime Symmetry}
\label{chap:cosetspacetime}

\abstract*{The first step towards an effective field theory for spontaneously broken spacetime symmetry is the construction of the nonlinear realization of the symmetry. This chapter develops the theory of nonlinear realizations of spacetime symmetries in close parallel with the previous discussion of internal symmetry. There are however some important differences. Notably, the spacetime coordinates and fields have to be treated together as independent variables, spanning a larger manifold on which the symmetry acts. The points of this manifold are no longer in a one-to-one correspondence with possible values of the order parameter. Likewise, the isotropy group of any point no longer corresponds to the subgroup, left unbroken in the ground state. These obstacles can be bypassed by focusing on the subgroup of symmetries that fix a given spacetime point. This leads to an unambiguous identification of the independent Nambu--Goldstone degrees of freedom, and allows for a straightforward generalization of the standard nonlinear realization of internal symmetry. The conditions under which such standard nonlinear realization of spacetime symmetries is exhaustive are discussed explicitly. Finally, several illustrative examples are worked out in detail.}


In Part~\ref{part:internalSSB} of the book, I assumed that whatever the full symmetry of the system, only internal symmetry is spontaneously broken. Moreover, I only considered two possibilities for the symmetry of the underlying spacetime. For nonrelativistic systems, I assumed invariance under spacetime translations and spatial rotations; this symmetry is sometimes called \emph{Aristotelian}. In case of relativistic systems, the Aristotelian symmetry is turned into Poincar\'e symmetry by adding invariance under Lorentz boosts. Both of these spacetime symmetries are easy to implement provided they are not spontaneously broken. Invariance under spacetime translations is ensured by demanding that the Lagrangian density of the theory does not depend explicitly on spacetime coordinates.  Invariance under spatial or spacetime rotations can be guaranteed by contracting indices appropriately using ordinary tensor calculus.

The situation becomes subtle when the spacetime symmetry is spontaneously broken, or when a different spacetime symmetry than Poincar\'e or Aristotelian is present. It is known that there are in fact multiple mathematically consistent kinematical groups including spacetime translations, spatial rotations and boosts~\cite{Bacry1968}. Here I will consider in particular the Galilei symmetry, relevant for \emph{effective field theories} (EFTs) of many nonrelativistic systems. Due to the mathematical structure of the Galilei group, the theory of its representations, and thus construction of Galilei-invariant actions, is highly nontrivial. For an overview of other possible non-Lorentzian kinematics and the associated spacetime geometry, see~\cite{Bergshoeff2022}.

The EFT for broken spacetime symmetries finds natural application in cosmology. To make the methodology developed below useful also there, I will make an exception and initially allow for theories defined on a generic spacetime manifold. Some simplifying assumptions on the spacetime manifold will be necessary, but will only be introduced when needed. Otherwise, the general philosophy of construction of EFTs for spontaneously broken spacetime symmetry is the same as in Part~\ref{part:internalSSB}. I will start in this chapter by classifying possible nonlinear realizations of spacetime symmetry. In Chaps.~\ref{chap:spacetimequantum} and~\ref{chap:spacetimeclassical}, I will then utilize the results to build effective actions. Unfortunately, there are no general explicit expressions for effective Lagrangians akin to those for internal symmetries, worked out in Chap.~\ref{chap:effLagrangian}. I will therefore have to resort to outlining the basic algorithm and working out some illustrative examples.


\section{Reminder of Nonlinear Realization of Internal Symmetry}
\label{sec:internalreminder}

The material of this chapter closely parallels Chap.~\ref{chap:CCWZ}, which the reader is advised to remind themselves of before proceeding further. In order to underline the basic logic, let me however give at least a brief summary of the steps we took in Chap.~\ref{chap:CCWZ}.

We used the fact that internal symmetries in the sense of Sect.~\ref{subsec:symtransfo} are point transformations that only act on fields and leave spacetime coordinates intact. This allows for a geometric reformulation of the problem of classifying all nonlinear realizations of internal symmetry. Namely, if the fields take values from a manifold $\M$, then realizing the internal symmetry group $G$ on them amounts to defining an \emph{action} of $G$ on $\M$. While the action of the symmetry group is assumed to exist globally on the whole manifold $\M$, its explicit expression requires a set of local coordinates. Thus, all statements depending on a specific choice of coordinates are without further qualification valid only in a local coordinate patch.

We start by picking a point $\psi_0\in\M$.\footnote{In Chap.~\ref{chap:CCWZ}, I used the symbol $x_0$ instead of $\psi_0$, which I will now reserve for a reference point on the spacetime manifold.} The subgroup $H_{\psi_0}$ of elements of $G$ that map $\psi_0$ to itself is called its \emph{isotropy group}. Whenever $H_{\psi_0}$ is compact, there are local coordinates $(\pi^a,\mf^\vr)$ on $\M$ such that $\psi_0$ corresponds to $\pi^a=\mf^\vr=0$ , and $H_{\psi_0}$ acts separately and linearly on $\pi^a$ and $\mf^\vr$. Moreover, the set of points $(\pi^a,0)$ spans a submanifold of $\M$, equivalent to the coset space $G/H_{\psi_0}$. This makes it possible to encode $\pi^a$ in a unique representative element $U(\pi)$ of the corresponding left coset of $H_{\psi_0}$ in $G$. The representative element can be chosen so that the point $\psi_0$ is mapped to $U(0)=e$. In addition, the linear representation of the isotropy group $H_{\psi_0}$ on $\pi^a$ is realized by the adjoint action, $U(\pi)\to hU(\pi)h^{-1}$, $h\in H_{\psi_0}$. Altogether, the action of $G$ on $\M$ in the standard coordinates $(\pi^a,\mf^\vr)$ is fixed by the choice of representative $U(\pi)$ and of the linear representation of $H_{\psi_0}$ on $\mf^\vr$.

In physics terms, every point $\psi_0\in\M$ defines a specific value of the order parameter for \emph{spontaneous symmetry breaking} (SSB). The isotropy group $H_{\psi_0}$ corresponds to the unbroken subgroup of $G$. Once treated as maps from the spacetime to $\M$, $\pi^a$ become the \emph{Nambu--Goldstone} (NG) fields implied by SSB, whereas $\mf^\vr$ are usually called \emph{matter fields}. The same symmetry-breaking pattern $G\to H$ may be realized by different order parameters, and thus on different target manifolds $\M$. This however only affects the number and type of matter fields present. The NG fields are identified unambiguously as coordinates on the coset space $G/H$, and are in a one-to-one correspondence with the generators of broken~symmetry.

Moving now to spacetime symmetries, we will be able to follow the mathematical construction of nonlinear realization of internal symmetry very closely. We will only have to introduce one important modification, taking into account the distinction between fields and coordinates. However, the mapping to the physical concepts of SSB and NG bosons will turn out to be much more intricate.


\section{Spacetime Symmetry as a Point Transformation}
\label{sec:spacetimepoint}

\begin{figure}[t]
\sidecaption[t]
\includegraphics[width=2.9in]{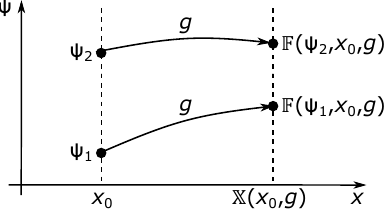}
\caption{Visualization of the action of a spacetime symmetry on the product manifold $\M\times M$. The transformation of the spacetime coordinates $x^\m$ is independent of the fields $\psi^i$. For fixed $x_0\in M$ and $g\in G$, the symmetry thus acts as a map between the slices $\M\times\{x_0\}$ and $\M\times\{\DX(x_0,g)\}$, indicated by the \emph{dashed} lines}
\label{fig:fiber}
\end{figure}

Consider a theory of a set of bosonic fields $\psi^i$ living on a $D$-dimensional spacetime manifold $M$. It is temporarily convenient to treat $\psi^i$ and the spacetime coordinates $x^\m$ on equal footing as independent variables spanning the product manifold $\M\times M$. This allows us to view the action of a spacetime symmetry (in the sense of Sect.~\ref{subsec:symtransfo}) as a special case of a point transformation on $\M\times M$. See Fig.~\ref{fig:fiber} for a visualization. Formally, the action of a spacetime symmetry group $G$ on $\M\times M$ constitutes a set of maps $T_g$, or equivalently functions $\DF^i$ and $\DX^\m$, such that
\begin{equation}
T_g:(\psi^i,x^\m)\to(\psi'^i,x'^\m)\equiv(\DF^i(\psi,x,g),\DX^\m(x,g))\;,\qquad g\in G\;.
\label{spacetimesymdef}
\end{equation}
The functions~$\DF^i$ and $\DX^\m$ cannot be chosen arbitrarily, but have to respect the structure of $G$. First, the unit element $e\in G$ must be realized by an identity transformation,
\begin{equation}
T_e=\id\quad\Leftrightarrow\quad
\DF^i(\psi,x,e)=\psi^i\;,\quad\DX^\m(x,e)=x^\m\;.
\label{actiondefunit}
\end{equation}
Second, the group multiplication law dictates that for any $g_1,g_2\in G$,
\begin{equation}
T_{g_1g_2}=T_{g_1}\circ T_{g_2}\quad\Leftrightarrow\quad\left\{
\begin{aligned}
\DF^i(\psi,x,g_1g_2)&=\DF^i(\DF(\psi,x,g_2),\DX(x,g_2),g_1)\;,\\
\DX^\m(x,g_1g_2)&=\DX^\m(\DX(x,g_2),g_1)\;.
\end{aligned}\right.
\label{actiondefcomp}
\end{equation}
Finally, for any $g\in G$ the map $T_{g^{-1}}$ must be the inverse of $T_g$,
\begin{equation}
T_{g^{-1}}=(T_g)^{-1}\quad\Leftrightarrow\quad\left\{
\begin{aligned}
\DF^i(\DF(\psi,x,g),\DX(x,g),g^{-1})&=\psi^i\;,\\
\DX^\m(\DX(x,g),g^{-1})&=x^\m\;.
\end{aligned}\right.
\label{actiondefinv}
\end{equation}

The special case of~\eqref{spacetimesymdef} where $\psi'^i=\DF^i(\psi,g)$ and $\DX^\m(x,g)=x^\m$ corresponds to an internal symmetry.\footnote{Strictly speaking, my previous definition of a spacetime symmetry included the assumption that $\DX^\m(x,g)\neq x^\m$, at least for some $x\in M$. This assumption is however immaterial in the present context. It will actually be convenient to think of the formalism developed here as a generalization rather than a modification of that in Part~\ref{part:internalSSB} of the book.} Another special case is a \emph{purely} spacetime transformation, where $\DF^i(\psi,x,g)=\psi^i$ for all $\psi\in\M$ and $x\in M$. It follows from the conditions on the functions $\DX^\m$ in~\eqref{actiondefunit}--\eqref{actiondefinv} that purely spacetime transformations form a subgroup of the symmetry group $G$. Here I will make a technical assumption that will prove essential for setting up standard coordinates on the product manifold $\M\times M$. Namely, I will assume that the subgroup of $G$ of purely spacetime transformations acts transitively on every slice $\{\psi\}\times M$ with fixed $\psi\in\M$. Loosely speaking, this requires that the spacetime manifold $M$ has sufficient symmetry that turns it into a homogeneous space. Now choose an arbitrary $x_0\in M$ and keep it fixed. By our assumption, for any $x\in M$ there is a purely spacetime transformation $\tran[x_0]{x}\in G$ that maps $x_0$ to $x$. Typically, there will be multiple such transformations; we then have to choose one $\tran[x_0]{x}$ by convention. For flat spacetimes, it is natural to use a translation connecting the two points. However, we do not have to be that specific at this stage, keeping in mind that $M$ may be curved.

Eventually, we would like to treat $\psi^i$ as functions on the spacetime. Mathematically, this amounts to replacing the action of the symmetry group $G$ on $\M\times M$ with one on maps $M\to\M$. Such an induced action is easy to write down thanks to the assumption that the transformation of the coordinates is independent of the fields. Thus, the fields $\psi^i(x)$ are mapped to $\psi'^i(x,g)$ such that
\begin{equation}
\begin{split}
\psi'^i(\DX(x,g),g)&=\DF^i(\psi(x),x,g)\;,\quad\text{or equivalently}\\
\psi'^i(x,g)&=\DF^i(\psi(\DX(x,g^{-1})),\DX(x,g^{-1}),g)\;,
\end{split}
\label{inducedaction}
\end{equation}
for any $g\in G$. We will need~\eqref{inducedaction} once we want to construct a $G$-invariant effective action for the fields. For the time being, however, I will pursue the analogy with Chap.~\ref{chap:CCWZ} and focus on the action of $G$ on the product manifold $\M\times M$.


\section{Standard Nonlinear Realization}
\label{sec:spacetimestandard}

For every point $(\psi,x)\in\M\times M$ there is an isotropy group consisting of elements of $G$ that map the point to itself,
\begin{equation}
H_{(\psi,x)}\equiv\{h\in G\,\vert\,T_h(\psi,x)=(\psi,x)\}\;.
\end{equation}
The setup of nonlinear realization of internal symmetry as reviewed in Sect.~\ref{sec:internalreminder} would naively suggest that we now choose a point $(\psi_0,x_0)\in\M\times M$. In its neighborhood, we could then establish the standard nonlinear realization of $G$ in the usual manner. There are however two problems with this naive approach. First, we eventually want to treat fields as functions, defined globally on the whole spacetime $M$, as long as global coordinates on $M$ exist. Second, the fields should be functions of the \emph{original}, physical coordinates $x^\m$, not of some new variables whose dependence on $\psi^i$ and $x^\m$ is beyond our control.

In order to reach these goals, we take an intermediate step, introducing an isotropy group of a chosen spacetime point,
\begin{equation}
H_x\equiv\{g\in G\,\vert\,\DX^\m(x,g)=x^\m\}\;.
\end{equation}
Obviously, $H_{(\psi,x)}$ is a subgroup of $H_x$ for any $\psi\in\M$. Moreover, having fixed a reference point $x_0\in M$, we can use the maps $\tran[x_0]{x}$ to show that, in analogy with~\eqref{isotropyconjugation}, isotropy groups at different spacetime points are related by conjugation,
\begin{equation}
H_x=\tran[x_0]{x}H_{x_0}\tran[x_0]{x}^{-1}\;,\qquad
H_{(\psi,x)}=\tran[x_0]{x}H_{(\psi,x_0)}\tran[x_0]{x}^{-1}\;.
\label{isotropyconjugation2}
\end{equation}

\begin{illustration}%
Let $\psi$ be a complex Schr\"odinger field so that $\M\simeq\C$, and let $M$ be the flat Galilei spacetime. The action~\eqref{galileiboost} of a Galilei boost with velocity $\vec v$ corresponds to
\begin{equation}
(\psi,\vec x,t)\xrightarrow{\vec v}\bigl(\exp[\I m(\skal vx+\vec v^2t/2)]\psi,\vec x+\vec vt,t\bigr)\;.
\label{galileiboostaction}
\end{equation}
We assume that in addition to boosts, the symmetry group $G$ also includes spacetime translations, spatial rotations, and the internal $\gr{U}(1)$ symmetry of phase transformations. For $x_0^\m=(\vec x_0,t_0)=(\vec 0,0)$, we find $H_{x_0}\simeq[\gr{SO}(d)\ltimes\R^d]\times\gr{U}(1)$, consisting of spatial rotations, Galilei boosts, and the phase transformations. In the special case of $\psi_0=0$, $H_{(\psi_0,x_0)}\simeq H_{x_0}$. Otherwise, for any $\psi_0\neq0$, $H_{(\psi_0,x_0)}\simeq\gr{SO}(d)\ltimes\R^d\subsetneq H_{x_0}$. The isotropy groups for other spacetime points are obtained by conjugation~\eqref{isotropyconjugation2} using the spacetime translation $\tran[x_0]{x}(\psi,\vec x',t')\equiv(\psi,\vec x'+\vec x-\vec x_0,t'+t-t_0)$.\footnote{In order to avoid cluttered notation, I use here and in the following the same symbol $\tran[x_0]{x}$ to denote both the element of $G$ and the corresponding map on $\M\times M$.}
\end{illustration}

For any fixed $x\in M$, the transformation of $(\psi^i,x^\m)$ under $g\in H_x$ is determined by the functions $\DF^i(\psi,x,g)$. In other words, these functions define an action of $H_x$ on the slice $\M\times\{x\}\simeq\M$. This makes it possible, following now Chap.~\ref{chap:CCWZ} verbatim, to introduce a standard set of field coordinates on $\M$. Together with the spacetime coordinates $x^\m$, this defines a parameterization of the whole manifold $\M\times M$, in which the action of $G$ takes a standard form. In order to keep the main results of this chapter together in one place, I will first give a concise overview of the ensuing standard realization of spacetime symmetry. Afterwards, I will stress some of its subtleties. A number of examples is worked out in detail in Sect.~\ref{sec:spacetimeexamples}; the reader may want to consult these alongside the formal construction developed below.


\subsection{Summary of the Construction}
\label{subsec:spacetimesummary}

Consider the action of a continuous group $G$ on the product manifold $\M\times M$ via a set of point transformations of the type~\eqref{spacetimesymdef}. It is assumed that $G$ includes a set of purely spacetime transformations that act transitively on every slice $\{\psi\}\times M$ with fixed $\psi\in\M$. I will for simplicity choose the reference point $x_0\in M$ with $x^\m_0=0$ and denote as $\tran{x}\equiv\tran[0]{x}$ the purely spacetime transformation that maps $x_0$ to $x\in M$.

The slice $\M\times\{0\}\simeq\M$ carries an action of the isotropy group $H_0$. Choose a fixed point $\psi_0\in\M$. Assuming that the isotropy subgroup $H_{(\psi_0,0)}$ is compact, it is always possible to find coordinates $(\pi^a,\mf^\vr)$ on $\M\times\{0\}$ with the following properties:
\begin{itemize}
\item The point $\psi_0$ corresponds to $(\pi^a,\mf^\vr)=(0,0)$. Also, the set of points~$\{(\pi^a,0)\}$ spans a submanifold of $\M\times\{0\}$, equivalent to the coset space $H_0/H_{(\psi_0,0)}$.
\item The coordinates $\pi^a$ parameterize uniquely a representative $U(\pi)\in H_0$ of the corresponding left coset of $H_{(\psi_0,0)}$ in $H_0$ such that $U(0)=e$.
\item The subgroup $H_{(\psi_0,0)}$ acts on $H_0/H_{(\psi_0,0)}$ by adjoint action, $U(\pi)\to hU(\pi)h^{-1}$ with $h\in H_{(\psi_0,0)}$, which induces a linear transformation of the coordinates $\pi^a$.
\item The whole group $H_0$ acts on $\M\times\{0\}$ by left multiplication as
\begin{equation}
\begin{split}
U(\pi)&\xrightarrow{g_0}U(\pi'(\pi,g_0))=g_0U(\pi)h(\pi,g_0)^{-1}\;,\\
\mf^\vr&\xrightarrow{g_0}\mf'^\vr(\mf,\pi,g_0)=D(h(\pi,g_0))^\vr_{\phantom\vr\s}\mf^\s\;,
\end{split}
\label{CCWZspacetime0}
\end{equation}
where $g_0\in H_0$ and $h(\pi,g_0)\in H_{(\psi_0,0)}$. Moreover, $D(h)$ is a matrix representation of $H_{(\psi_0,0)}$. Altogether, the action of $H_0$ on $\M\times\{0\}$ is fixed by the choice of representation $D$ of $H_{(\psi_0,0)}$ and the choice of parameterization $U(\pi)$ of the coset space $H_0/H_{(\psi_0,0)}$.
\end{itemize}

\begin{figure}[t]
\sidecaption[t]
\includegraphics[width=2.0in]{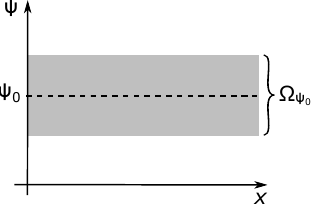}
\caption{Schematic visualization of the domain (\emph{shaded} area) on the manifold $\M\times M$ where the standard coordinates $(\pi^a,\mf^\vr,x^\m)$ are well-defined. The domain $\Omega_{\psi_0}\subset\M$ carries a local coordinate patch with the field variables $(\pi^a,\mf^\vr)$}
\label{fig:domain}
\end{figure}

This finishes the setup of coordinates and the action of $H_0$ on the domain $\Omega_{\psi_0}\times\{0\}$; $\Omega_{\psi_0}$ is a neighborhood of $\psi_0$ in $\M$ where the coordinates $(\pi^a,\mf^\vr)$ are well-defined. As the next step, we transport the coordinates $(\pi^a,\mf^\vr)$ to the whole spacetime using the maps $\tran{x}$. This leads to coordinates $(\pi^a,\mf^\vr,x^\m)$ on the domain $\Omega_{\psi_0}\times M$ (see Fig.~\ref{fig:domain}). Intuitively, we require that all points of the slice $\{\psi\}\times M$ with fixed $\psi\in\Omega_{\psi_0}$ have the same coordinates $\pi^a,\mf^\vr$. Technically, we set
\begin{equation}
(\pi^a,\mf^\vr,x^\m)\equiv\tran{x}(\pi^a,\mf^\vr,0)\;.
\label{MMparameterization}
\end{equation}
By~\eqref{isotropyconjugation2}, this fixes the action of the isotropy group $H_x$ on $\M\times\{x\}$. Indeed, for any $g_x\in H_x$ there is a unique $g_0\in H_0$ such that $g_x=\tran xg_0\tran x^{-1}$. This leads to
\begin{equation}
\begin{split}
T_{g_x}(\pi^a,\mf^\vr,x^\m)&=\tran x\circ T_{g_0}(\pi^a,\mf^\vr,0)=\tran x(\pi'^a(\pi,g_0),\mf'^\vr(\mf,\pi,g_0),0)\\
&=(\pi'^a(\pi,g_0),\mf'^\vr(\mf,\pi,g_0),x^\m)\;.
\end{split}
\label{actionofHx}
\end{equation}
Moreover, \eqref{isotropyconjugation2} guarantees that for any $g_x\in H_{(\psi_0,x)}$, $g_0\in H_{(\psi_0,0)}$. Therefore, the isotropy subgroup $H_{(\psi_0,x)}$ is realized on the slice $\M\times\{x\}$ by linear transformations of $\pi^a$ and $\mf^\vr$ for any $x\in M$.

\begin{figure}[t]
\sidecaption[t]
\includegraphics[width=2.0in]{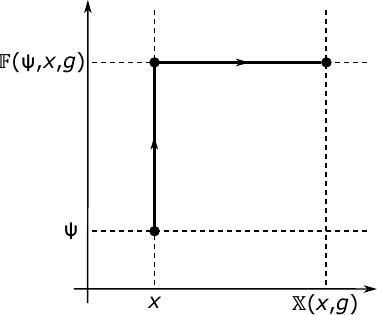}
\caption{For any fixed point $(\psi,x)\in\M\times M$, the action of a chosen element $g\in G$ can be composed from actions of an element of $H_x$ and a purely spacetime transformation}
\label{fig:composition}
\end{figure}

The structure we already have extends uniquely to an action of the entire group $G$ on the entire manifold $\M\times M$, or at least on $\Omega_{\psi_0}\times M$. Namely, note that the action of any $g\in G$ on a chosen point $(\psi,x)$ can be composed of the action of an element of $H_x$ and a purely spacetime transformation, see Fig.~\ref{fig:composition}. Indeed, decompose
\begin{equation}
g=\tran[x]{\DX(x,g)}\tran[x]{\DX(x,g)}^{-1}g\equiv\tran[x]{\DX(x,g)}g_x(x,g)\;,
\label{gdecomposition}
\end{equation}
with the shorthand notation $\smash{\tran[x]{x'}\equiv\tran[x_0]{x'}\tran[x_0]{x}^{-1}}$. The group element $g_x(x,g)$ maps $(\psi^i,x^\m)$ to $(\DF^i(\psi,x,g),x^\m)$ for any $\psi\in\M$ and thus belongs to $H_x$. It corresponds to an element $g_0(x,g)\in H_0$ by $\smash{g_0(x,g)=\tran x^{-1}g_x(x,g)\tran x=\tran{\DX(x,g)}^{-1}g\tran x}$. Together with~\eqref{actionofHx}, this leads to the final expression for the action of $G$ within $\Omega_{\psi_0}\times M$,
\begin{equation}
T_g(\pi^a,\mf^\vr,x^\m)=(\pi'^a(\pi,g_0(x,g)),\mf'^\vr(\mf,\pi,g_0(x,g)),\DX^\m(x,g))\;.
\label{CCWZspacetime}
\end{equation}
This completes the setup of the nonlinear realization of $G$. I assumed an a priori knowledge of the transformation properties of the spacetime coordinates, that is the functions $\DX^\m(x,g)$. With this provision, the action of $G$ is uniquely fixed by the structure of the isotropy groups $H_0$ and $H_{(\psi_0,0)}$, and by the representation $D(h)$ of $H_{(\psi_0,0)}$. In the process, we made some choices, including the choice of the representative $U(\pi)$ parameterized by $\pi^a$ and of the element $\tran x\in G$ representing a spacetime point $x\in M$. These are just two sides of the same coin, pertinent respectively to the manifolds $\M$ and $M$.

Complicated as it might seem, \eqref{CCWZspacetime} actually encodes in a rather simple manner the action of $G$ by left matrix multiplication. Indeed, let us represent the point $(\pi^a,\mf^\vr,x^\m)$ formally by $\tran x(U(\pi),\mf^\vr)$. Then left multiplication by any $g\in G$ gives
\begin{align}
\notag
T_g\circ\tran x(U(\pi),\mf^\vr)&=\tran{\DX(x,g)}\circ T_{g_0(x,g)}(U(\pi),\mf^\vr)\\
\label{CCWZspacetimeexp}
&=\tran{\DX(x,g)}\bigl(U(\pi'(\pi,g_0(x,g)))h(\pi,g_0(x,g)),\\
\notag
&\phantom{{}=\tran{\DX(x,g)}\bigl(}D(h(\pi,g_0(x,g)))^\vr_{\phantom\vr\s}\mf^\s\bigr)\;,
\end{align}
which exactly copies~\eqref{CCWZspacetime} in the matrix notation of~\eqref{CCWZspacetime0}. This shows that it is not actually necessary to decompose every $g\in G$ into a product of an element of $H_x$ and a purely spacetime transformation as in~\eqref{gdecomposition}. It is sufficient to observe that for any $x\in M$ and $g\in G$, $\smash{\tran{\DX(x,g)}^{-1}g\tran x}$ fixes the origin of $M$ and thus belongs to $H_0$.


\subsection{Relation to Physics of Broken Spacetime Symmetry}
\label{subsec:spacetimediscussion}

Mathematically, the generalization of the standard nonlinear realization from internal to spacetime symmetry was relatively straightforward, perhaps even deceivingly so. A number of its features thus deserve pointing out before we proceed to examples.

\runinhead{Universality of the Construction} In Chap.~\ref{chap:CCWZ} we found out that provided the isotropy group $H$ is compact, the standard nonlinear realization captures all possible actions of the symmetry group $G$. Furthermore, I remarked that even if $H$ is noncompact, the construction still goes through, but is no longer guaranteed to be exhaustive, provided the coset space $G/H$ is reductive. The same applies here with the necessary modifications. The universality of the standard nonlinear realization is ensured if $H_{(\psi_0,0)}$ is compact. Even if it is not, the construction is still consistent, yet not necessarily exhaustive, if the coset space $H_0/H_{(\psi_0,0)}$ is reductive. Reference~\cite{Joseph1970a} gives an explicit example of a group action that cannot be obtained from the standard nonlinear realization. There $G\simeq\gr{SO}(2,1)$ acts on $\M\simeq\R$, the isotropy subgroup being the noncompact affine group $\gr{Aff}(1)\simeq\R^+\ltimes\R$.

A possible point of concern might be that I assumed an a priori knowledge of the action of $G$ on spacetime coordinates. This is however not a source of any hidden ambiguity. For one thing, the action of symmetry on spacetime coordinates is usually known. It is the classification of NG fields and their transformation properties that we are after. Moreover, even if unknown a priori, the functions $\DX^\m(x,g)$ can be constrained following the same philosophy. Namely, restricting~\eqref{spacetimesymdef} to the coordinates $x^\m$ gives a group of well-defined transformations on $M$. If desired, possible forms of these transformations can be classified using the setup of Chap.~\ref{chap:CCWZ}.

\runinhead{Fate of Descendant Symmetries} Recall the discussion of locally equivalent symmetries in Sect.~\ref{sec:diffcurrentrelations}. As explained therein, one can assume without loss of generality that the functions $\smash{f^A_\iA(x)}$ vanish at the origin, $x^\m=0$. Provided the localized parent transformation in~\eqref{locallyequivalentsym} does not involve any derivatives of $\eps^A_1(x)$, the action of the descendant symmetry at the origin vanishes. In the standard nonlinear realization, such a descendant symmetry therefore automatically belongs to $H_{(\psi_0,0)}$ and does not give rise to any NG variables $\pi^a$.

\runinhead{Physical Unbroken Symmetry} In contrast to the internal symmetry case, a single point $(\psi,x)\in\M\times M$ cannot unambiguously represent the order parameter for SSB. The order parameter rather corresponds to a function $\vp^i:M\to\M$. A particular value of the order parameter then defines a submanifold, $\M_\vp\equiv\{(\vp^i(x),x^\m)\,\vert\,x\in M\}\subset\M\times M$. Accordingly, neither $H_x$ nor $H_{(\vp(x),x)}$ can be identified with the subgroup of unbroken symmetries, $H_\vp$. The latter rather consists of all elements of $G$ that preserve the manifold $\M_\vp$, that is map points on $\M_\vp$ to other points on $\M_\vp$. With the help of~\eqref{inducedaction}, this translates to the condition $\vp^i(\DX(x,g))=\DF^i(\vp(x),x,g)$ for all $x\in M$. This constraint is highly implicit; in practice it is usually much easier to check its infinitesimal form,
\begin{equation}
\df^i_A(\vp(x),x)-\dx^\m_A(x)\de_\m\vp^i(x)=0\quad\text{for all }x\in M\;.
\label{spacetimeunbroken}
\end{equation}
The functions $\df^i_A$ and $\dx^\m_A$ define the action of the generator $Q_A$ of $G$. The reconstruction of the group $H_\vp$ from the solutions to~\eqref{spacetimeunbroken} is based on the correspondence between a Lie group and its Lie algebra. An example of a possible pitfall this approach may fall into, in the context of isometries of a Riemannian manifold, is given in Appendix~\ref{appsubsec:isometries}.

\begin{illustration}%
\label{ex:constantOP}%
Suppose that the order parameter is constant, $\vp^i(x)\equiv\vp^i_0$. It follows that the ``vacuum submanifold'' is $\M_\vp=\{\vp_0\}\times M$. According to~\eqref{spacetimesymdef}, the unbroken subgroup is
\begin{equation}
H_{\vp_0}=\{g\in G\,\vert\,\DF^i(\vp_0,x,g)=\vp^i_0\text{ for all }x\in M\}\;.
\label{defHvp0}
\end{equation}
Any purely spacetime symmetry automatically belongs to $H_{\vp_0}$. In other words, purely spacetime symmetries can only be broken by a coordinate-dependent order parameter. On the other hand, for internal symmetries, the condition in~\eqref{defHvp0} reduces to $\DF^i(\vp_0,g)=\vp^i_0$. Since internal symmetries by construction do not affect the coordinates $x^\m$, the definitions of $H_{(\vp_0,x)}$ and $H_{\vp_0}$ in this special case coincide. At the same time, for internal symmetries we have trivially $H_x\simeq G$ for any $x\in M$, hence $H_x/H_{(\vp_0,x)}\simeq G/H_{\vp_0}$. This verifies that for internal symmetries, the more general formalism developed here boils down to that of Chap.~\ref{chap:CCWZ}.
\end{illustration}

\runinhead{Classification of Order Parameter Fluctuations} For internal symmetries, independent fluctuations of the order parameter are in a one-to-one correspondence with the NG fields $\pi^a$. It would therefore be tempting to conclude that in the more general case of spacetime symmetries, they are classified by the coset space $H_0/H_{(\psi_0,0)}$. The latter however only counts the NG variables $\pi^a$, realizing nonlinearly the action of $H_0$. Our parameterization~\eqref{MMparameterization} of the manifold $\M\times M$ also involves the action of the ``translations'' $\tran x$. Being purely spacetime, these may be spontaneously broken if $\de_\m\vp^i(x)\neq0$ for some $x\in M$. We will only be able to introduce the corresponding NG modes in Chap.~\ref{chap:spacetimequantum} once we treat $\psi^i$ as maps $M\to\M$. The resulting EFT can be expected to feature a nontrivial realization of spacetime translations.

It is possible to be more explicit in the special case of a constant order parameter, $\vp^i(x)=\vp^i_0$, where purely spacetime symmetries remain unbroken. Here we expect a low-energy EFT in terms of NG fields $\pi^a$ on which purely spacetime symmetries act trivially, $\pi'^a(\DX(x,g),g)=\pi^a(x)$. The number of these NG fields is identified by the dimension of $H_0/H_{(\vp_0,0)}$. This generalizes our previous counting rule for NG fields of broken internal symmetry.

\begin{illustration}%
\label{ex:galileoncoset}%
Consider a theory of a single real relativistic scalar field $\p$ that is invariant under the Galileon transformations,
\begin{equation}
(\p,x^\m)\xrightarrow{\eps,\eps_\m}(\p'(\p,x,\eps),x'^\m(x,\eps))=(\p+\eps+\eps_\m x^\m,x^\m)\;.
\label{MMgalileon}
\end{equation}
The symmetry group is $G\simeq\gr{ISO}(d,1)\ltimes\R^{D+1}$, where $\gr{ISO}(d,1)$ is the $D$-dimensional Poincar\'e group and the $\R^{D+1}$ factor collects the constant and linear shifts of $\p$ in~\eqref{MMgalileon}. Suppose the dynamics of the theory is such that the ground state has a constant \emph{vacuum expectation value} (VEV), $\vev{\p(x)}\equiv\vp$. This order parameter breaks by~\eqref{spacetimeunbroken} the entire Galileon symmetry, that is $H_\vp\simeq\gr{ISO}(d,1)$. The coset space $G/H_\vp\simeq\R^{D+1}$ obviously does not correctly identify the NG modes. After all, our theory has a Lorentz-invariant vacuum and a single scalar field. Indeed, we find that $H_0\simeq\gr{SO}(d,1)\ltimes\R^{D+1}$ where $\gr{SO}(d,1)$ is the Lorentz group. Likewise, $H_{(\vp,0)}\simeq\gr{SO}(d,1)\ltimes\R^D$. This only includes the linear shifts of $\p$ in~\eqref{MMgalileon}, which act trivially at $x^\m=0$. In the end, we thus find $H_0/H_{(\vp,0)}\simeq\R$, corresponding to the symmetry under constant shifts of $\p$. This gives the correct number of NG modes: one. The symmetry under linear shifts of $\p$ is descendant. While it constrains the effective Lagrangian, cf.~Sect.~\ref{subsec:diffexamplesgalileon}, it cannot affect the physical spectrum.
\end{illustration}

\runinhead{Domain of the Standard Nonlinear Realization} As emphasized rather stubbornly, the standard field coordinates $(\pi^a,\mf^\vr)$ are in principle only well-defined in some domain $\Omega_{\psi_0}\subset\M$. In order for our formalism to be able to capture all fluctuations of the order parameter, the manifold $\M_\vp$ had better lie entirely in $\Omega_{\psi_0}\times M$. This is obviously the case when the order parameter is constant as in~\refex{ex:constantOP}. It is however not guaranteed in general and has to be checked case by case. In fact, it is easy to imagine a situation where this condition is very nontrivial. Consider a system with an order parameter $\vp^i(x)$ where the isotropy group $H_{(\vp(x),x)}$ varies from place to place. This is in contrast to our setup where $H_{(\psi_0,x)}$ for all $x\in M$ are mutually isomorphic. It may happen e.g.~in systems where the order parameter for spontaneous breaking of internal symmetry vanishes at some spacetime points. We might then be lucky and still get away with our standard nonlinear realization. The price to pay is the presence of $\pi^a$-type variables that locally excite gapped modes in the spectrum, or $\mf^\vr$-type variables that locally excite NG modes. I am not aware of any work in the literature that would systematically address this kind of situation.

\begin{illustration}%
\label{ex:vortex}%
A two-dimensional superfluid vortex is a map $\psi:\R^2\to\C$, expressed in the polar coordinates $\vr,\t$ on $\R^2$ as $\smash{\psi(\vec x)=f(\vr)\E^{\I n\t}}$. Here $f(\vr)$ is a strictly increasing profile function such that $f(0)=0$ and the limit of $f(\vr)$ for $\vr\to\infty$ is finite. The nonzero parameter $n\in\Z$ is the so-called winding number of the vortex. The target manifold $\M\simeq\C$ carries an action of $G\simeq\gr{U}(1)$ under which $\smash{\psi\xrightarrow{\eps}\E^{\I\eps}\psi}$. This is an internal symmetry and thus $H_0\simeq G\simeq\gr{U}(1)$. For $\psi_0=0$, corresponding to the core of the vortex, we find $H_{(\psi_0,0)}\simeq G\simeq\gr{U}(1)$ as well. There are no NG variables, only two matter variables $\mf^\vr$, equivalent to the real and imaginary parts of $\psi$. These are well-defined in the entire complex plane, hence $\Omega_{\psi_0}=\C$. On the other hand, choosing $\psi_0\neq0$ as appropriate for the vortex away from the core leads to $H_{(\psi_0,0)}\simeq\trgr$. The coset space $H_0/H_{(\psi_0,0)}\simeq\gr{U(1)}$ is now parameterized by a single NG variable $\pi$. The local parameterization of $\M\simeq\C$ is completed by one matter variable $\mf$. One can view $\mf,\pi$ as polar coordinates on the target space. As such, their domain of validity $\Omega_{\psi_0}$ can be extended at most to $\C$ with a half-line starting at the origin removed.
\end{illustration}

\runinhead{Some Literature on the Subject} Nonlinear realizations of internal symmetry were classified in the pioneering work~\cite{Coleman1969a}. First attempts to generalize their results to spacetime symmetries appeared soon afterwards~\cite{Volkov1973a,Ogievetsky1974a}. These early works addressed a narrow class of relativistic systems where an extended symmetry such as the conformal symmetry is spontaneously broken to the Poincar\'e group. They followed an approach, based on an abstract coset space wherein all spontaneously broken symmetries induce relevant, independent degrees of freedom. Formally, this can be viewed as working with a set of order parameters that is sufficiently large to prevent any symmetries from being locally equivalent. Alternatively, it can be interpreted as being ``agnostic'' about the choice of order parameter. It is then necessary to include in the EFT the maximum possible set of order parameter fluctuations that can be enforced by a given symmetry-breaking pattern.

The price of this \emph{agnostic nonlinear realization} is that it introduces extra degrees of freedom that at best correspond to gapped modes in the spectrum, and at worst are outright unphysical. Writing down an EFT solely in terms of the physical NG modes requires eliminating the superfluous degrees of freedom by an operational procedure known as \emph{inverse Higgs constraint} (IHC)~\cite{Ivanov1975a}. In spite of its obvious shortcomings, this framework has influenced the narrative surrounding spontaneous breaking of spacetime symmetries for decades. Much of the subsequent work focused on the machinery of IHCs and its physical interpretation. Namely, it has been known for some time that fluctuations of the order parameter, corresponding to locally equivalent symmetries, are redundant~\cite{Low2002a,Watanabe2013a}. Eliminating the would-be NG degrees of freedom corresponding to descendant symmetries by imposing a set of IHCs can then be interpreted as gauge-fixing~\cite{Nicolis2013b}. The alternative possibility that the modes eliminated by the IHC may be physical but are necessarily gapped was recognized in~\cite{Endlich2014a,Brauner2014a}. I will elaborate on the two scenarios later when we have at hand explicit examples. Finally, much work has been done on the question to what extent imposing an IHC might interfere with the universality of the standard nonlinear realization~\cite{Creminelli2015a,Klein2017a,Finelli2020}. The moral is that a point transformation relating different parameterizations of an EFT may become nonlocal upon imposing the~IHC.

As far as I know, the fact that the would-be NG fields of descendant symmetries need not be included in the nonlinear realization at all was first pointed out in~\cite{Kharuk2018}. The approach developed here makes this explicit by identifying the relevant NG variables $\pi^a$ with coordinates on the coset space~$H_0/H_{(\psi_0,0)}$. This has a simple classical interpretation, whereby $\pi^a$ represent the deviation of the order parameter from its local value. They can be further augmented with additional NG fields, arising from the spacetime variation of the order parameter. To incorporate such additional NG fields will be the most challenging problem we will have to face in Chap.~\ref{chap:spacetimequantum}.


\section{Examples}
\label{sec:spacetimeexamples}

To illustrate the general construction of Sect.~\ref{sec:spacetimestandard}, I will now work out in detail several examples, covering a range of different symmetries. In all the examples, I assume flat (relativistic or not) spacetime where the map $\tran[x]{x'}$ is realized by the unique translation that moves the point $x\in M$ to $x'\in M$. The isotropy group $H_0$ is fixed by the transformation properties of spacetime coordinates, known a priori. The subgroup $H_{(\psi_0,0)}$, on the other hand, may depend sensitively on the choice of the reference point $\psi_0$. This is demonstrated by the examples in Sects.~\ref{subsec:spacetimexscale} to~\ref{subsec:spacetimexSchr}.


\subsection{Lorentz Scalars with Internal Symmetry}
\label{subsec:spacetimexinternal}

A good starting point is to check that our new algorithm can reproduce what we know from before about internal symmetries. Consider for simplicity a theory of a set of relativistic scalars $\psi^i$ with a symmetry group $G\simeq\gr{ISO}(d,1)\times G_\mathrm{int}$. Here $\gr{ISO}(d,1)$ is the Poincar\'e group, acting as a purely spacetime symmetry, and $G_\mathrm{int}$ is an internal symmetry. The actions of the purely spacetime and internal parts of $G$ are completely independent of each other. Mathematically speaking, $\gr{ISO}(d,1)$ and $G_\mathrm{int}$ possess a well-defined action on, respectively, $M\simeq\R^D$ and $\M$. Both actions are trivially extended to $\M\times M$ by embedding.

It follows at once that $H_0\simeq\gr{SO}(d,1)\times G_\mathrm{int}$ and $H_{(\psi_0,0)}\simeq\gr{SO}(d,1)\times H_\mathrm{int}$, where $H_\mathrm{int}\subset G_\mathrm{int}$ is the isotropy group of the point $\psi_0\in\M$. Unsurprisingly, we end up with a set of NG variables $\pi^a$, parameterizing the coset space $H_0/H_{(\psi_0,0)}\simeq G_\mathrm{int}/H_\mathrm{int}$. By~\eqref{CCWZspacetime0} and~\eqref{CCWZspacetime}, a group element $(g_\mathrm{s.t.},g_\mathrm{int})\in\gr{ISO}(d,1)\times G_\mathrm{int}$ acts on the standard coordinates $(\pi^a,\mf^\vr,x^\m)$ via
\begin{align}
\notag
U(\pi)&\xrightarrow{(g_\mathrm{s.t.},g_\mathrm{int})}U(\pi'(\pi,g_\mathrm{int}))=g_\mathrm{int}U(\pi)h_\mathrm{int}(\pi,g_\mathrm{int})^{-1}\;,\\
\label{STexamplescalars}
\mf^\vr&\xrightarrow{(g_\mathrm{s.t.},g_\mathrm{int})}\mf'^\vr(\mf,\pi,g_\mathrm{int})=D(h_\mathrm{int}(\pi,g_\mathrm{int}))^\vr_{\phantom\vr\s}\mf^\s\;,\\
\notag
x^\m&\xrightarrow{(g_\mathrm{s.t.},g_\mathrm{int})}x'^\m(x,g_\mathrm{s.t.})=(T_{g_\mathrm{s.t.}}x)^\m\;.
\end{align}
Here $h_\mathrm{int}\in H_\mathrm{int}$ and $D(h_\mathrm{int})$ is a matrix representation of $H_\mathrm{int}$. Finally, $T_{g_\mathrm{s.t.}}$ defines the action of the Poincar\'e group on the Minkowski spacetime. This is of course well-known, so the abstract notation just takes explicitly into account the possibility of using other coordinates $x^\m$ than Minkowski.

\begin{watchout}%
This might look like a mere idiosyncratic reformulation of the standard nonlinear realization of internal symmetry as laid out in Chap.~\ref{chap:CCWZ}. Indeed, should the order parameter be constant, $\vp^i(x)\equiv\vp_0^i$, the unbroken subgroup is $H_{\vp_0}\simeq\gr{ISO}(d,1)\times H_\mathrm{int}$ so that $G/H_{\vp_0}\simeq G_\mathrm{int}/H_\mathrm{int}$, in accord with~\refex{ex:constantOP}. However, the parameterization of $\M\times M$ by $(\pi^a,\mf^\vr,x^\m)$ and the corresponding group action~\eqref{STexamplescalars} are also valid for order parameters $\vp^i(x)$ with arbitrary coordinate dependence. The true unbroken subgroup $H_\vp$ can then be smaller than $\gr{ISO}(d,1)\times H_\mathrm{int}$ and in extreme cases even trivial. This is not mere pedantry. For instance, dense relativistic matter can often be described by a time-dependent condensate of scalar fields. Although such a background spontaneously breaks boosts, our construction guarantees that one can parameterize its fluctuations by the same degrees of freedom as in a Lorentz-invariant vacuum. A similar remark applies to all the other examples discussed below.
\end{watchout}

The same setup can be used even if we replace $\gr{ISO}(d,1)$ with any other purely spacetime symmetry group that contains spacetime translations. The message is fairly simple: it does not matter what the spacetime symmetry is, as long as it does not act directly on the fields.


\subsection{Lorentz Scalars with Scale Invariance}
\label{subsec:spacetimexscale}

Next, we look at a simple example of a symmetry whose actions on coordinates and fields cannot be trivially separated. Consider a set of relativistic scalar fields $\psi^i$, carrying the action of $G\simeq\R^+\ltimes\gr{ISO}(d,1)$. Here $\gr{ISO}(d,1)$ is again the Poincar\'e group. What is new is the factor $\R^+$, representing scale transformations of the spacetime coordinates. It would be possible to add an internal symmetry factor $G_\mathrm{int}$ in the same way as in Sect.~\ref{subsec:spacetimexinternal}, but I will not do so to keep the notation simple.

It is now convenient to fix the spacetime coordinates $x^\m$ as the standard Minkowski ones. The maps $\tran x$ are then realized explicitly as $\tran x=\E^{\I x\cdot P}$ where $P_\m$ is the generator of spacetime translations. The action of translations on the Minkowski coordinates is $x'^\m(x,\eps)\equiv\E^{\I\eps\cdot P}x^\m=x^\m+\eps^\m$. Likewise, an element $\E^\a\in\R^+$ with real parameter $\a$ is realized on the coordinates as $x'^\m(x,\a)\equiv\E^{\I\a D}x^\m=\E^\a x^\m$, where $D$ is the dilatation operator. Accordingly, the action of the dilatation group $\R^+$ on the Poincar\'e group is fixed by the commutation relations $[D,P_\m]=-\I P_\m$ and $[D,J_{\m\n}]=0$.

It is now clear that $H_0\simeq\R^+\times\gr{SO}(d,1)$. Since Lorentz transformations act trivially on scalar fields, we have only two options for $H_{(\psi_0,0)}$, depending on whether or not $\psi_0$ also preserves scale invariance. Let us start with $H_{(\psi_0,0)}\simeq\R^+\times\gr{SO}(d,1)$, which corresponds to a scale-invariant order parameter $\psi_0\in\M$. This implies $H_0/H_{(\psi_0,0)}\simeq\trgr$, hence there are no NG variables $\pi^a$, only matter fields. These should carry a linear representation of the dilatation symmetry. Since $\R^+$ only has one-dimensional irreducible representations, we can always choose a basis $\mf^\vr$ of coordinates on $\M\times\{0\}$ such that $\mf'^\vr(\mf,\a)\equiv\E^{\I\a D}\mf^\vr=\exp(-\a\Delta_\vr)\mf^\vr$.\footnote{From now until the end of Chap.~\ref{chap:cosetspacetime}, a repeated index $\vr$ does not imply any summation.} The parameter $\Delta_\vr$ is the scaling dimension of $\mf^\vr$. The action of dilatations on the entire manifold $\M\times M\simeq\M\times\R^D$ is then given by
\begin{equation}
(\mf^\vr,x^\m)\xrightarrow{\a}(\mf'^\vr(\mf,\a),x'^\m(x,\a))=(\exp(-\a\Delta_\vr)\mf^\vr,\E^\a x^\m)\;.
\end{equation}
It is instructive to see how this transformation rule is reproduced by~\eqref{CCWZspacetimeexp}. Namely, the commutator $[D,P_\m]=-\I P_\m$ together with the Hadamard lemma~\eqref{hadamard} gives
\begin{equation}
\E^{\I\a D}P_\m\E^{-\I\a D}=\E^\a P_\m\quad\text{or}\quad
\E^{\I\a D}\E^{\I\eps\cdot P}\E^{-\I\a D}=\exp(\I\E^\a\eps\cdot P)\;.
\label{DPconjugation}
\end{equation}
This ensures that for $g=\E^{\I\a D}$, $g_0(x,g)=\E^{\I\a D}$ for any $x\in M$. The scale transformation of $\mf^\vr$ is independent of the spacetime coordinate as it should. Mathematically, the conjugation relation~\eqref{DPconjugation} boils down to the fact that the translation generators $P_\m$ carry a representation of the dilatation group $\R^+$.

The other option for the isotropy group is $H_{(\psi_0,0)}\simeq\gr{SO}(d,1)$, in which case $H_0/H_{(\psi_0,0)}\simeq\R$. This can be thought of as arising from a Lorentz-invariant but dimensionful order parameter. We now have one NG variable, the \emph{dilaton} $\pi$, plus possibly a set of $\dim\M-1$ matter fields $\mf^\vr$. With the exponential parameterization, $U(\pi)=\E^{\I\pi D}$, \eqref{CCWZspacetime0} tells us that $\pi'(\pi,\a)=\pi+\a$ and $\mf'^\vr(\mf,\pi,\a)=\mf^\vr$. The extension of the action to other spacetime points works exactly the same as in the previous case. The final result for the action of dilatations therefore is
\begin{equation}
(\pi,\mf^\vr,x^\m)\xrightarrow{\a}(\pi'(\pi,\a),\mf'^\vr(\mf,\pi,\a),x'^\m(x,\a))=(\pi+\a,\mf^\vr,\E^\a x^\m)\;.
\end{equation}
Note that the scaling dimension of all the matter fields is now zero. This is just a matter of a choice of variables. Namely, in presence of the dilaton, any field $\Psi^\vr$ with scaling dimension $\Delta_\vr$ can be redefined to $\mf^\vr=\exp(\Delta_\vr\pi)\Psi^\vr$.

I have not spelled out explicitly the action of spacetime translations and rotations. However, these are purely spacetime transformations and only affect the spacetime coordinates, similarly to the last line of~\eqref{STexamplescalars}.


\subsection{Lorentz Vector with(out) Lorentz Scalar}
\label{subsec:spacetimexscalarvector}

Another possibility how to make the action of symmetry on coordinates and fields entangled is to keep the Poincar\'e group $G\simeq\gr{ISO}(d,1)$, but take a nonscalar field. Consider for simplicity a single Lorentz-vector field, $A^\m$, defining a $D$-dimensional target manifold $\M\simeq\R^D$. We then have $H_0\simeq\gr{SO}(d,1)$, but the isotropy group $H_{(A_0,0)}$ depends on the choice of $A^\m_0$. There are four qualitatively different options: $A^\m_0=0$, or nonvanishing $A^\m_0$ that is respectively timelike, lightlike, or spacelike.

In the simplest case of $A^\m_0=0$, we find $H_{(A_0,0)}\simeq H_0\simeq\gr{SO}(d,1)$ and consequently $H_0/H_{(A_0,0)}\simeq\trgr$. There are no NG variables and the sole degree of freedom, $A^\m$ itself, is of the matter type. It carries the vector representation of the Lorentz group, $A^\m\to A'^\m(A,g)=\Lambda(g)^\m_{\phantom\m\n}A^\n$ for any $g\in\gr{SO}(d,1)$. Since the translation generators $P_\m$ also carry the vector representation of the Lorentz group, we have a conjugation relation similar to~\eqref{DPconjugation} for dilatations. This guarantees that $g_0(x,g)=g$ for any $g\in\gr{SO}(d,1)$ and $x\in M\simeq\R^D$. From~\eqref{CCWZspacetimeexp}, we then extract the expected result for the action of Lorentz transformations,
\begin{equation}
T_g(A^\m,x^\m)=(\Lambda(g)^\m_{\phantom\m\n}A^\n,\Lambda(g)^\m_{\phantom\m\n}x^\n)\;,\qquad
g\in\gr{SO}(d,1)\;.
\label{lorentzvector}
\end{equation}
Spacetime translations only affect the coordinates, $x'^\m(x,\eps)=\E^{\I\eps\cdot P}x^\m=x^\m+\eps^\m$.

For illustration, I will work out in detail one more special case. Suppose that $\smash{A^\m_0}$ is nonzero and timelike. Without loss of generality, we can assume that $\smash{A^\m_0=a\d^{\m0}}$ with $a\neq0$. Then $H_{(A_0,0)}\simeq\gr{SO}(d)$, the group of spatial rotations. The coset space $H_0/H_{(A_0,0)}\simeq\gr{SO}(d,1)/\gr{SO}(d)$ is now a noncompact $d$-dimensional manifold, best viewed as the mass shell of a massive relativistic particle. A convenient implicit way to parameterize it is by treating it as the $d$-dimensional hyperboloid in $\M\simeq\R^D$ satisfying the constraint $A_\m A^\m=a^2$. That has the advantage of maintaining the linear transformation property~\eqref{lorentzvector} under all Lorentz transformations. This is just the noncompact version of the $\gr{SO}(d+1)/\gr{SO}(d)\simeq S^d$ coset space, conventionally parameterized by a unit vector $\vec n\in\R^{d+1}$.

If needed, it is possible to introduce explicit coordinates $\pi^r$ on $H_0/H_{(A_0,0)}$, transforming linearly as a vector under $H_{(A_0,0)}\simeq\gr{SO}(d)$. One suitable coordinatization arises from thinking of $A^\m$ as the energy--momentum of a particle of mass $\abs a$. The $d$ independent coordinates $\pi^r$ are then the components of its spatial momentum. A pure Lorentz boost can be represented by $\E^{\I\skal\eta K}$, where $\vec K$ is the boost generator and $\vec\eta$ the rapidity. The latter is parallel to the velocity $\vec v$ of the boost; its magnitude is determined implicitly by $\abs{\vec v}=\tanh\abs{\vec\eta}$. The action of the boost on $\pi^r$ in terms of the velocity $\vec v$ reads
\begin{equation}
\vec\pi'(\vec\pi,\vec v)=\vec\pi+\frac{\g_{\vec v}-1}{\vec v^2}(\skal v\pi)\vec v+\g_{\vec v}\vec v\sqrt{a^2+\vec\pi^2}\sgn a\;,\qquad
\g_{\vec v}\equiv\frac1{\sqrt{1-\vec v^2}}\;.
\end{equation}
The parameterization of the target manifold $\M\simeq\R^D$ is completed by adding to $\pi^r$ a matter field $\mf$, transforming in the singlet representation of $\gr{SO}(d)$. By~\eqref{CCWZspacetimeexp}, it then automatically transforms trivially under the whole Poincar\'e group. The transformation of coordinates $x^\m$ remains of course the same as in~\eqref{lorentzvector}.

\begin{watchout}%
From internal symmetry, we are already used to the fact that the number and type of matter fields $\mf^\vr$ present depend on the choice of order parameter. It is only the NG fields $\pi^a$ that are fixed by the symmetry-breaking pattern. Here comes the surprise: for spacetime symmetries, even the number of NG variables $\pi^a$ may depend on the order parameter. For an example, take a theory of a relativistic complex scalar field $\p$ where an internal $\gr{U}(1)$ symmetry is broken by the VEV $\vev{\p(x)}\equiv\vp(x)=\vp_0\E^{-\I\m t}$ with constant nonzero $\vp_0$ and $\m$. This state describes matter with nonzero density of the $\gr{U}(1)$ charge; $\m$ is the corresponding chemical potential. The state breaks the $G\simeq\gr{ISO}(d,1)\times\gr{U}(1)$ symmetry spontaneously down to $H_\vp\simeq\gr{ISO}(d)\times\R$, where $\gr{ISO}(d)$ now includes spatial translations and rotations and $\R$ stands for a combination of time translations and internal $\gr{U}(1)$ transformations. In line with our discussion in Sect.~\ref{subsec:spacetimexinternal}, there is a single NG variable, parameterizing the coset space $H_0/H_{(\vp_0,0)}\simeq[\gr{SO}(d,1)\times\gr{U}(1)]/\gr{SO}(d,1)$. Now add a vector $A^\m$ as a secondary order parameter. A constant background, $\smash{\vev{A^\m(x)}\equiv A^\m_0=a\d^{\m0}}$, can be interpreted for instance as the VEV of the current of the $\gr{U}(1)$ symmetry. This does not affect the unbroken subgroup $H_\vp$, yet it does add a vector $\pi^r$ of NG variables. We will deal with this puzzle in Chap.~\ref{chap:spacetimequantum}. It will turn out that the number of physical gapless NG modes in the spectrum is still uniquely fixed by the symmetry-breaking pattern $G\to H_\vp$. The new variables $\pi^r$ represent either spurious, nondynamical degrees of freedom or dynamical but gapped excitations of the order parameter.
\end{watchout}


\subsection{Schr\"odinger Scalars with Galilei Symmetry}
\label{subsec:spacetimexSchr}

To work out at least one nonrelativistic example, we now return to Galilei invariance. The action of the Galilei group on spacetime coordinates is well-known. A spatial translation by $\vec\eps$ shifts Cartesian coordinates by $\vec x\to\vec x+\vec\eps$, and likewise $t\to t+\eps$ represents a temporal translation by $\eps$. Accordingly, the translation operator $\tran x$ can be parameterized as $\tran{\vec x,t}=\E^{\I tH}\E^{\I\skal xP}$, where $H$ and $\vec P$ are the respective generators. A Galilei boost with velocity $\vec v$ acts on the coordinates as $(\vec x,t)\to(\vec x+\vec vt,t)$, and will be represented by $\E^{\I\skal vK}$ with $\vec K$ being the generator. One should also add the generator of spatial rotations, $J_{rs}$. However, since we will be mostly concerned with the boosts, I will not spell out the action of rotations explicitly.

From the transformation of the coordinates, we extract the conjugation property
\begin{equation}
\E^{\I\eps H}\vec K\E^{-\I\eps H}=\vec K-\eps\vec P\quad\text{or}\quad
[H,\vec K]=\I\vec P\;.
\label{galileiaux1}
\end{equation}
The actions of spatial translations and boosts on the coordinates commute with each other. However, it is known that representations of the Galilei symmetry admit a central extension of the commutator $[P_r,K_s]$. Let us therefore introduce a tentative central charge $Q$ so that\footnote{In $d=2$ spatial dimensions, the Galilei group admits another, exotic central extension, $[K_r,K_s]=\I\k\ve_{rs}$. The parameter $\k$ is however related to two-dimensional spin~\cite{Jackiw2000a}, and I will thus drop it.}
\begin{equation}
[P_r,K_s]=\I\d_{rs}Q\quad\text{or}\quad
\E^{\I\skal\eps P}\vec K\E^{-\I\skal\eps P}=\vec K-\vec\eps Q\;.
\label{galileiaux2}
\end{equation}
The resulting central extension of the Galilei group is known as the \emph{Bargmann group}. This has the structure $\smash{G\simeq\gr{SO}(d)\ltimes\{\R^d_K\ltimes[\R^D\times\gr{U}(1)_Q]\}}$. The first factor stands for spatial rotations, $\R^D$ for spacetime translations, and $\R^d_K$ for boosts. Finally, I use the notation $\gr{U}(1)_Q$ for the transformations generated by $Q$ although at this stage it is not clear, or even important, that the symmetry is compact. What matters is that the action of $Q$ on the coordinates is trivial. This ensures $H_0\simeq[\gr{SO}(d)\ltimes\R^d_K]\times\gr{U}(1)_Q$.

The isotropy group $H_{(\psi_0,0)}$ depends sensitively on the type of fields included. I will initially restrict to rotation scalars, $\psi^i$, which makes the rotation $\gr{SO}(d)$ into a purely spacetime symmetry. Moreover, the $\R^d_K$ group of Galilei boosts is now descendant, as we saw in Sect.~\ref{subsec:diffexamplesgalilei}. This is confirmed by an explicit classification of low-spin indecomposable representations of the Galilei group~\cite{Montigny2006a}. In the end, the only part of the symmetry whose action on $\psi^i$ at the origin, $(\vec x,t)=(\vec 0,0)$, may be nontrivial is the $\gr{U}(1)_Q$. We have a freedom to decide whether or not $\gr{U}(1)_Q$ belongs to $H_{(\psi_0,0)}$ by a suitable choice of the reference point $\psi^i_0$.

Let us first assume that $H_{(\psi_0,0)}\simeq H_0$ so that $H_0/H_{(\psi_0,0)}\simeq\trgr$. It is then possible to find a set of \emph{complex} field coordinates $\mf^\vr$ that form one-dimensional linear representations of $\gr{U}(1)_Q$, $\mf'^\vr(\mf,\a)\equiv\E^{\I\a Q}\mf^\vr=\exp(\I\a m_\vr)\mf^\vr$, where $m_\vr$ are the corresponding charges of $Q$. This simple transformation property survives at all $(\vec x,t)$ since $Q$ is a central charge and so commutes with spacetime translations. Likewise, all of the spacetime translations and spatial rotations will act solely on the coordinates. The only nontrivial piece is the action of Galilei boosts. This is determined with the help of~\eqref{CCWZspacetimeexp} and the relation
\begin{equation}
\E^{\I\skal vK}\E^{\I tH}\E^{\I\skal xP}=\E^{\I tH}\E^{\I(\vec x+\vec vt)\cdot\vec P}\exp\left[\I\left(\skal vx+\frac12\vec v^2t\right)Q\right]\E^{\I\skal vK}\;,
\label{galileiaux3}
\end{equation}
which follows from~\eqref{galileiaux1} and~\eqref{galileiaux2}. The final result is
\begin{equation}
\E^{\I\skal vK}(\mf^\vr,\vec x,t)=(\exp[\I m_\vr(\skal vx+\vec v^2t/2)]\mf^\vr,\vec x+\vec vt,t)\;,
\label{galileiaction1}
\end{equation}
which reproduces using solely the Lie algebra of the Bargmann group the transformation rule~\eqref{galileiboostaction} I simply postulated before. We can see that the central charge $Q$, introduced above ad hoc, measures the nonrelativistic (rest) mass. In systems of identical particles of a fixed mass, this is proportional to the number of particles.

The other option for the isotropy group is $H_{(\psi_0,0)}\simeq\gr{SO}(d)\ltimes\R^d_K$. This implies that $H_0/H_{(\psi_0,0)}\simeq\gr{U}(1)_Q$. The particle number symmetry $\gr{U}(1)_Q$ is going to be realized nonlinearly as for instance in superfluids. We need one NG variable, $\pi$. With the exponential parameterization of the coset space, $U(\pi)=\E^{\I\pi Q}$, this transforms under $\gr{U}(1)_Q$ as $\pi'(\pi,\a)=\pi+\a$. In addition, there will be a set of \emph{real} matter fields $\mf^\vr$, invariant under $\gr{U}(1)_Q$. Using again~\eqref{galileiaux3}, we then find quite a different result then above,
\begin{equation}
\E^{\I\skal vK}(\pi,\mf^\vr,\vec x,t)=(\pi+\skal vx+\vec v^2t/2,\mf^\vr,\vec x+\vec vt,t)\;.
\label{galileiaction2}
\end{equation}

This completes the range of options accessible with scalar fields. Let us see at least briefly what may happen when higher-spin fields are present. For simplicity, consider a single multiplet $A^\m\equiv(A^0,\vec A)$ that transforms under boosts as a Galilean vector~\cite{Montigny2006a}, $A'^\m(A,\vec v)\equiv\E^{\I\skal vK}(A^0,\vec A)=(A^0,\vec A+\vec vA^0)$. Similarly to Sect.~\ref{subsec:spacetimexscalarvector}, one can think of $\vev{A^\m(x)}\equiv A^\m_0(x)$ as the VEV of the density and current of the $\gr{U}(1)_Q$ symmetry, respectively. In this interpretation, a state with $\smash{A^\m_0(x)=(a,\vec0)}$ and constant $a\neq0$ corresponds to uniform dense matter in its rest frame. With this secondary order parameter, the isotropy group $H_{((\psi_0,A_0),0)}$ is reduced to $\gr{SO}(d)\times\gr{U}(1)_Q$ or $\gr{SO}(d)$, depending on whether or not the primary order parameter $\psi^i_0$ preserves $\gr{U}(1)_Q$. The coset space $H_0/H_{((\psi_0,A_0),0)}$ then necessarily includes a noncompact $d$-dimensional submanifold $\smash{\R^d_K}$, carrying a vector of NG variables, $\x^r$. The natural parameterization is $\smash{U(\vec\x)=\E^{\I\skal\x K}}$. Following our algorithm for the standard nonlinear realization, we find that~\eqref{galileiaction1} and~\eqref{galileiaction2} remain valid under the respective assumptions on $\gr{U}(1)_Q$. The only change is a new transformation rule for $\x^r$,
\begin{equation}
U(\vec\x'(\vec\x,\vec v))=\E^{\I\skal vK}U(\vec\x)=U(\vec\x+\vec v)\;.
\end{equation}

The modification of the realization of the Bargmann group by adding $\x^r$ is nearly trivial, so why exactly have we done this? The reason will become clear in the next chapter where we address the problem of construction of invariant actions. Namely, invariance under the linearly realized isotropy subgroup $H_{(\psi_0,0)}$ has to be ensured by brute force using representation theory. The mathematical structure of the Bargmann group makes this a difficult task. With Galilei boosts realized nonlinearly by the NG field $\x^r$, all that remains is to impose invariance under spatial rotations and possibly $\gr{U}(1)_Q$, which is straightforward.


\bibliographystyle{spphys}
\bibliography{references}
\chapter{Broken Spacetime Symmetry in~Quantum~Matter}
\label{chap:spacetimequantum}

\abstract*{This chapter constitutes the core of the part of the book devoted to spontaneously broken spacetime symmetry. It builds a framework for constructing effective actions for Nambu--Goldstone bosons, starting from the nonlinear realization of spacetime symmetry developed in the previous chapter. The framework is first applied to systems where the values of the order parameter at different spacetime points belong to the same orbit of the symmetry group. This case can be treated by a simple modification of the setup for broken internal symmetries. Next, the formalism is utilized to discuss the physical significance of so-called inverse Higgs constraints and the associated auxiliary vector fields. It is shown on concrete examples that while such fields may excite physical modes in the spectrum, these are not actual Nambu--Goldstone bosons. The rest of the chapter is devoted to genuine breaking of translation invariance by a coordinate-dependent order parameter. This requires a further improvement of the formalism to ensure proper realization of the symmetry and a one-to-one parameterization of all the fields.}


With the nonlinear realization of spacetime symmetry put forward in Chap.~\ref{chap:cosetspacetime}, we are now in the position to construct the low-energy \emph{effective field theory} (EFT). As I already occasionally stressed, the EFT machinery for spontaneously broken spacetime symmetry is less developed than that for internal symmetry. As a consequence, I will not be able to offer an  explicit general expression for the effective Lagrangian akin to what I did in Chap.~\ref{chap:effLagrangian}. Instead, I will describe the basic framework in an algorithmic fashion, and then work out some illustrative examples.

In this chapter, I only consider systems where the symmetry group $G$ is known and dictated by the microscopic dynamics. This has been the underlying assumption of the whole book so far. Yet, there are physical systems whose low-energy physics may feature additional symmetry which itself may but need not be spontaneously broken. Such symmetry is usually called \emph{emergent}, and appears among others in classical systems such as fluids or solids. I will defer a detailed discussion of such emergent symmetries and their spontaneous breaking to the next chapter.

Following some preliminary work in Sect.~\ref{sec:quantumbuildingblocks}, the survey of EFTs for spontaneously broken spacetime symmetry starts in Sect.~\ref{sec:quantumtwist}. Here I deal with systems where the values of the order parameter at all points in space belong to the same orbit of the symmetry group. This requires a minimal modification of the setup for internal symmetries, but already involves some of the subtleties associated with the spacetime ones. A major issue is that the number of \emph{Nambu--Goldstone} (NG) variables $\pi^a$ depends on the choice of order parameter realizing the given symmetry-breaking pattern. This means that some would-be NG variables are not required by the symmetry-breaking pattern per se. In Sect.~\ref{subsec:spacetimediscussion}, I suggested that such fields may, but need not, be physical. In case they are, they necessarily excite gapped modes in the spectrum, not true NG modes. I illustrate this dichotomy by several examples in Sect.~\ref{sec:quantumvector}. Then I offer some comments on the \emph{inverse Higgs constraints} (IHCs) as an operational prescription for eliminating such spurious degrees of freedom.

The generic case of systems where the order parameter may depend on coordinates in an a priori arbitrary manner is considered in Sect.~\ref{sec:quantumtranslation}. Here the methods available for the construction of EFT appear to be least developed. I will therefore have to content myself with a careful discussion of the special case of one-dimensional order parameter modulation, augmented with a couple of illustrative examples.


\section{Building Blocks for Construction of Effective Actions}
\label{sec:quantumbuildingblocks}

In Chap.~\ref{chap:effLagrangian}, I showed that the effective action for NG bosons of spontaneously broken internal symmetry can always be expressed in terms of the associated \emph{Maurer--Cartan} (MC) \emph{form}. I am not aware of any proof of the equivalent statement for spacetime symmetries. I will however make the common assumption that this is still the case.


\subsection{Maurer--Cartan Form}
\label{subsec:quantumMCform}

Recall that we deal generally with a set of bosonic fields $\psi^i$, taking values from a target manifold $\M$. The spacetime is treated as a possibly curved manifold $M$ of dimension $D$, parameterized by local coordinates $x^\m$. Together, $\psi^i$ and $x^\m$ span the product manifold $\M\times M$. In order to avoid excessive repetition, I will not review all the details of the standard nonlinear realization of spacetime symmetry. The reader is invited to consult Sect.~\ref{sec:spacetimestandard} to refresh their memory. Let me just stress the important roles of the isotropy groups $H_x$ of a spacetime point $x\in M$, and $H_{(\psi,x)}$ of the point $(\psi,x)\in\M\times M$. For chosen fixed $\psi_0\in\M$, the coset space $H_0/H_{(\psi_0,0)}$ spans a submanifold of $\M\times\{0\}\simeq\M$, parameterized by NG coordinates $\pi^a$. These are generally accompanied by a set of matter variables $\mf^\vr$. Together, $(\pi^a,\mf^\vr,x^\m)$ constitute of a complete set of local coordinates on $\M\times M$.

Let us first focus on the variables $(\pi^a,x^\m)$, \emph{required} for a successful nonlinear realization of spacetime symmetry. I will represent these jointly as
\begin{equation}
\upix(\pi,x)\equiv\tran xU(\pi)\;,
\end{equation}
where $\tran x$ is a fixed purely spacetime transformation that transports the spacetime origin to the point $x$ (cf.~Sect.~\ref{sec:spacetimepoint}). Moreover, $U(\pi)$ is a matrix representative of an element of the coset space $H_0/H_{(\psi_0,0)}$. By~\eqref{CCWZspacetimeexp}, the action of any $g\in G$ on $(\pi^a,x^\m)$ can then be expressed compactly as
\begin{equation}
\upix(\pi,x)\xrightarrow{g}\upix(\pi'(\pi,g_0(x,g)),\DX(x,g))=g\upix(\pi,x)h(\pi,g_0(x,g))^{-1}\;,
\label{Uspacetimetransfo}
\end{equation}
where $\smash{g_0(x,g)=\tran{\DX(x,g)}^{-1}g\tran x}$. Moreover, $\pi'^a(\pi,g_0(x,g))$ and $h(\pi,g_0(x,g))$ are defined by~\eqref{CCWZspacetime0}. The MC form can now be introduced through
\begin{equation}
\begin{split}
\mc(\pi,x)&\equiv-\I\upix(\pi,x)^{-1}\D\upix(\pi,x)\\
&=-\I U(\pi)^{-1}\D U(\pi)+U(\pi)^{-1}(-\I\tran x^{-1}\D\tran x)U(\pi)\;.
\end{split}
\label{MCspacetime}
\end{equation}
The first piece that depends solely on $\pi^a$ is familiar from our analysis of internal symmetries. The second piece is new and in some, as yet unclear, way reflects the action of the symmetry on the spacetime. Equation~\eqref{Uspacetimetransfo} induces the following transformation of the MC form,
\begin{equation}
\begin{split}
\mc(\pi,x)\xrightarrow{g}{}&\mc(\pi'(\pi,g_0),\DX(x,g))\\
&=h(\pi,g_0)\mc(\pi,x)h(\pi,g_0)^{-1}-\I h(\pi,g_0)\D h(\pi,g_0)^{-1}\;,
\end{split}
\label{MCspacetimetransfo}
\end{equation}
where I for brevity dropped the arguments of $g_0(x,g)$.

In order to understand better the structure of the MC form, we need to make a digression. By definition of spacetime symmetry, the transformation of coordinates $x^\m$ is independent of the field variables $\psi^i$ (or $\pi^a,\mf^\vr$). One can thus restrict the action of $G$ to the spacetime manifold, where it is defined by the maps $\smash{x^\m\xrightarrow{g}x'^\m\equiv\DX^\m(x,g)}$. Under this restricted action, the spacetime behaves as a homogeneous space, $M\simeq G/H_0$. This implies the following \emph{local} homeomorphism of manifolds,
\begin{equation}
G\simeq H_{(\psi_0,0)}\times H_0/H_{(\psi_0,0)}\times G/H_0\simeq H_{(\psi_0,0)}\times H_0/H_{(\psi_0,0)}\times M\;.
\label{isomanifold}
\end{equation}
With the obvious notation for the Lie algebras of the three groups $G$, $H_0$ and $H_{(\psi_0,0)}$, the corresponding isomorphism of tangent spaces reads
\begin{equation}
\lie g\simeq\lie h_{(\psi_0,0)}\oplus\lie h_0/\lie h_{(\psi_0,0)}\oplus\lie g/\lie h_0\;.
\label{isotangent}
\end{equation}
Here $\lie h_0/\lie h_{(\psi_0,0)}$ and $\lie g/\lie h_0$ should be viewed as shorthand notation for the respective tangent spaces; these are themselves not Lie algebras without further qualifications.

I previously assumed that at the very least, the coset space $H_0/H_{(\psi_0,0)}$ is reductive. This ensures that both $\lie h_{(\psi_0,0)}$ and $\lie h_0/\lie h_{(\psi_0,0)}$ are invariant subspaces under the adjoint action of $H_{(\psi_0,0)}$. In the following, I will need a stronger assumption, namely that all three components of the direct sum~\eqref{isotangent} are invariant under the adjoint action of $H_{(\psi_0,0)}$. This is guaranteed if $H_{(\psi_0,0)}$ is compact, but in general has to be viewed as an additional assumption.

\begin{illustration}%
Let us check the validity of this assumption for the examples of nonlinear realization worked out in Sect.~\ref{sec:spacetimeexamples}. In case of a theory of Lorentz scalars with an internal symmetry group $G_\mathrm{int}$ (Sect.~\ref{subsec:spacetimexinternal}), we have $H_{(\psi_0,0)}\simeq\gr{SO}(d,1)\times H_\mathrm{int}$. As long as the coset space $G_\mathrm{int}/H_\mathrm{int}$ itself is reductive, the tangent space $\lie h_0/\lie h_{(\psi_0,0)}\simeq\lie g_\mathrm{int}/h_\mathrm{int}$ is naturally invariant under the adjoint action of $H_{(\psi_0,0)}$. Likewise, the tangent space $\lie g/\lie h_0$ of $G/H_0\simeq M\simeq\R^D$ carries the vector representation of $\gr{SO}(d,1)$. It is easy to check that also for the examples in Sects.~\ref{subsec:spacetimexscale} and~\ref{subsec:spacetimexscalarvector}, our assumption on the decomposition~\eqref{isotangent} is satisfied.

The case of Galilei symmetry (Sect.~\ref{subsec:spacetimexSchr}) is most nontrivial. Here the commutation relation $[P_r,K_s]=\I\d_{rs}Q$ implies that the $d$-dimensional space spanned on the components of the momentum operator does not carry a representation of Galilei boosts. Hence the boosts should not be included in the isotropy subgroup $H_{(\psi_0,0)}$. The way out, as shown in Sect.~\ref{subsec:spacetimexSchr}, is to introduce a vector order parameter $A^\m$ that gives a vector of NG variables $\x^r$, carrying a nonlinear realization of the boosts. Then, $H_{((\psi_0,A_0),0)}$ is either $\gr{SO}(d)$ or $\gr{SO}(d)\times\gr{U}(1)_Q$. In both cases, the Lie algebra $\lie g$ of $G$ naturally splits into invariant subspaces of $H_{((\psi_0,A_0),0)}$ as in~\eqref{isotangent}. This observation lends further support to the mathematical realization of Galilei boosts in terms of the auxiliary vector of NG fields $\x^r$. I will return to this in Sect.~\ref{subsec:vectorunphysical}.
\end{illustration}

In order to reformulate the above basic assumption in a language more common in physics, we need a notation for the symmetry generators. In analogy with the treatment of internal symmetries in Chap.~\ref{chap:CCWZ}, I will denote generators that form a basis of $\lie g$ as $Q_{A,B,\dotsc}$. A subset of these generators that spans a basis of $\lie h_{(\psi_0,0)}$ will be $Q_{\a,\b,\dotsc}$. Analogously, the basis of $\lie h_0/\lie h_{(\psi_0,0)}$ will be $Q_{a,b,\dotsc}$. Finally, for the basis of the complementary space $\lie g/\lie h_0$, I will use the notation $P_{\fr\m,\fr\n,\dotsc}$. In the fixed basis $Q_A$, the structure of the Lie algebra $\lie g$ is defined by the structure constants $\smash{f^C_{AB}}$ via $\smash{[Q_A,Q_B]=\I f^C_{AB}Q_C}$. Then, our assumption that all the components of the direct sum~\eqref{isotangent} carry a representation of $H_{(\psi_0,0)}$ restricts the commutators of $Q_\a$ to
\begin{equation}
[Q_\a,Q_\b]=\I f^\g_{\a\b}Q_\g\;,\quad
[Q_\a,Q_b]=\I f^c_{\a b}Q_c\;,\quad
[Q_\a,P_{\fr\m}]=\I f^{\fr\n}_{\a\fr\m}P_{\fr\n}\;.
\end{equation}
The remaining commutators of the Lie algebra $\lie g$ take the generic form
\begin{align}
\notag
[Q_a,Q_b]&=\I(f^\g_{ab}Q_\g+f^c_{ab}Q_c)\;,\qquad
[Q_a,P_{\fr\m}]=\I(f^{\b}_{a\fr\m}Q_\b+f^b_{a\fr\m}Q_b+f^{\fr\n}_{a\fr\m}P_{\fr\n})\;,\\
[P_{\fr\m},P_{\fr\n}]&=\I(f^\a_{\fr\m\fr\n}Q_\a+f^a_{\fr\m\fr\n}Q_a+f^{\fr\l}_{\fr\m\fr\n}P_{\fr\l})\;.
\label{commspacetime}
\end{align}
There is no $\smash{f^{\fr\m}_{ab}P_{\fr\m}}$ term in $[Q_a,Q_b]$ since all the $Q_\a$ and $Q_a$ together span a basis of $\lie h_0$. As to $P_{\fr\m}$, at this stage we do not need to assume that they are mutually commuting translation generators.

We are now ready to return to the MC form. By its definition~\eqref{MCspacetime}, $\mc\equiv\mc^AQ_A$ is a (locally defined) 1-form on $H_0/H_{(\psi_0,0)}\times M$ that takes values in the Lie algebra $\lie g$ of $G$. It consists of three parts, corresponding to the three spaces on the right-hand side of~\eqref{isotangent},
\begin{equation}
\begin{split}
\mc&\equiv\mcu+\mcb+\mcp\;,\\
\mcu&\equiv\mc^\a Q_\a\;,\qquad
\mcb\equiv\mc^aQ_a\;,\qquad
\mcp\equiv\vec e^{*\fr\m}P_{\fr\m}\;.
\end{split}
\label{MCspacetimedecomposition}
\end{equation}
The notation for $\mcu$ and $\mcb$ follows the conventions introduced in Chap.~\ref{chap:CCWZ} for internal symmetries. The $\mcp$ component, proportional to the generators $P_{\fr\m}$, is new here. Thanks to the assumption that both spaces $\lie h_0/\lie h_{(\psi_0,0)}$ and $\lie g/\lie h_0$ are invariant under the adjoint action of $H_{(\psi_0,0)}$, the transformation rule~\eqref{MCspacetimetransfo} splits as
\begin{equation}
\begin{split}
\mcu(\pi,x)&\xrightarrow{g}h(\pi,g_0)\mcu(\pi,x)h(\pi,g_0)^{-1}-\I h(\pi,g_0)\D h(\pi,g_0)^{-1}\;,\\
\mcb(\pi,x)&\xrightarrow{g}h(\pi,g_0)\mcb(\pi,x)h(\pi,g_0)^{-1}\;,\\
\mcp(\pi,x)&\xrightarrow{g}h(\pi,g_0)\mcp(\pi,x)h(\pi,g_0)^{-1}\;.
\end{split}
\label{MCspacetimetransforcomps}
\end{equation}
Following Chap.~\ref{chap:CCWZ}, it would now be possible to work out further technical details. These include for instance an explicit expression for the action of $G$ on the coordinates $(\pi^a,x^\m)$, or the MC equations for the exterior derivative of $\mc(\pi,x)$. I will however not do so even though it would be straightforward, for we will not need these details. Let us instead turn to the physical interpretation of the various parts of the MC form.


\subsection{Covariant Derivatives of Fields}
\label{subsec:quantumcovder}

The first part of the MC form in~\eqref{MCspacetime}, $-\I U(\pi)^{-1}\D U(\pi)$, obviously takes values from the Lie algebra $\lie h_0$. It follows that $\mcp$ does not contain any contributions proportional to $\D\pi^a$. In other words, $\mcp$ is a well-defined 1-form on the slice $\{\pi\}\times M\simeq M$ for any fixed point $\pi^a$ in $H_0/H_{(\psi_0,0)}$,
\begin{equation}
\vec e^{*\fr\m}(\pi,x)\equiv e^{*\fr\m}_\n(\pi,x)\D x^\n\;.
\label{MCspacetimevielbein}
\end{equation}
By assumption, the action of $G$ on the spacetime $M$ is transitive. This implies that at $\pi^a=0$, the components of $\mcp(0,x)=(-\I\tran x^{-1}\D\tran x)^{\fr\m}P_{\fr\m}$ span the whole cotangent space to $M$. By continuity, the same must be true in some neighborhood of the origin of $H_0/H_{(\psi_0,0)}$, $\pi^a=0$. We conclude that the 1-forms $\vec e^{*\fr\m}(\pi,x)$ with fixed $\pi^a$ define a (local) coframe on the spacetime manifold $M$. We are of course free to use any (local) frame or coframe on $M$ we wish. Nevertheless, it will be convenient to stick to $\mcp$ thanks to its covariant transformation properties under $G$, as shown in~\eqref{MCspacetimetransforcomps}. This justifies a posteriori the notation $\vec e^{*\fr\m}$ for the components of $\mcp$.

Next we have a look at the $\mcb$ part of the MC form,
\begin{equation}
\mc^a(\pi,x)\equiv\mc^a_b(\pi)\D\pi^b+\mc^a_\m(\pi,x)\D x^\m\;.
\label{MCspacetimecovder}
\end{equation}
The $\smash{\mc^a_b}$ piece comes from the $-\I U(\pi)^{-1}\D U(\pi)$ term in~\eqref{MCspacetime} and is thus independent of the spacetime coordinates. The full 1-form $\mcb$ however contains components from cotangent spaces to both $H_0/H_{(\psi_0,0)}$ and $M$. Unlike in the case of internal symmetries, it therefore cannot be interpreted as giving rise to a coframe on $H_0/H_{(\psi_0,0)}$.

At this point, recall that we eventually want to treat $\pi^a$ as fields. These naturally define maps from $M$ to $H_0/H_{(\psi_0,0)}$ that take any $x\in M$ to $\pi^a(x)\in H_0/H_{(\psi_0,0)}$. Slightly more formally, one can view the fields as maps from $M$ to $H_0/H_{(\psi_0,0)}\times M$ that assign to $x\in M$ the pair $(\pi^a(x),x^\m)$. This makes it possible to pull differential forms on $H_0/H_{(\psi_0,0)}\times M$ back to $M$. Thus, the coframe~\eqref{MCspacetimevielbein} becomes $\vec e^{*\fr\m}(\pi(x),x)=e^{*\fr\m}_\n(\pi(x),x)\D x^\n$.\footnote{I take the liberty to denote 1-forms on $H_0/H_{(\psi_0,0)}\times M$ and their pull-backs to $M$ using the same symbol. It should be clear from the context which of the two is meant.} More interesting is the pull-back of~\eqref{MCspacetimecovder} to $M$,
\begin{equation}
\mc^a(\pi(x),x)=[\mc^a_b(\pi(x))\de_\m\pi^b(x)+\mc^a_\m(\pi(x),x)]\D x^\m\;.
\label{MCspacetimecovder2}
\end{equation}
This obviously carries information about the derivatives of the NG fields $\pi^a(x)$. To extract the \emph{covariant derivative} $\cd_{\fr\m}\pi^a$ of the NG field, we decompose the spacetime 1-form $\mc^a(\pi(x),x)$ in the coframe $\vec e^{*\fr\m}(\pi(x),x)$,
\begin{equation}
\mc^a\equiv(\cd_{\fr\m}\pi^a)\vec e^{*\fr\m}=(\cd_{\fr\m}\pi^a)e^{*\fr\m}_\n\D x^\n\;.
\end{equation}
Comparison with~\eqref{MCspacetimecovder2} then gives
\begin{equation}
\cd_{\fr\m}\pi^a(x)=[\mc^a_b(\pi(x))\de_\n\pi^b(x)+\mc^a_\n(\pi(x),x)]e^\n_{\fr\m}(\pi(x),x)\;,
\label{spacetimecovderpi}
\end{equation}
where $\vec e_{\fr\m}(\pi,x)\equiv e^\n_{\fr\m}(\pi,x)\de_\n$ is the local frame on $M$, dual to $\vec e^{*\fr\m}(\pi,x)$. The co\-va\-ri\-ant derivative can also be expressed compactly as $\cd_{\fr\m}\pi^a=\mc^a(\vec e_{\fr\m})$.

\begin{watchout}%
I am using here the same symbol $\cd_{\fr\m}$ for the covariant derivative as in Appendix~\ref{appsec:affine}, yet the two objects do not seem to be the same. Let me clarify the difference. First, the first term on the right-hand side of~\eqref{spacetimecovderpi} is not just the gradient of $\pi^a$ projected to the frame $\vec e_{\fr\m}$. The extra factor $\smash{\mc^a_b}$, as well as the $\mc^a_\n$ term, is needed to ensure covariance under the action of $G$. In particular the $\smash{\mc^a_b}$ factor is nontrivial even if $G$ is an internal symmetry group. Second, we are not introducing here an all-purpose connection on $M$ that would make $\cd_{\fr\m}\pi^a$ covariant under the maximal structure group $\gr{GL}(\dim M)$. As follows from~\eqref{MCspacetimetransforcomps}, we only allow changes of the local frame induced by a representation of the structure group $H_{(\psi_0,0)}$. This procedure is designed to give a $G$-invariant EFT for the NG fields. In spite of using covariant derivatives, the EFTs constructed below are not covariant under general coordinate transformations on $M$.
\end{watchout}

The remaining part of the MC form~\eqref{MCspacetimedecomposition} that we have not discussed yet is $\mcu$. This, as the transformation rule~\eqref{MCspacetimetransforcomps} suggests, plays the role of an $\lie h_{(\psi_0,0)}$-valued connection. Let us define the spacetime components of the connection after pull-back to $M$ through
\begin{equation}
\mcu(\pi(x),x)\equiv\mc^\a_\m(\pi(x),x)Q_\a\D x^\m\;.
\end{equation}
Then for a matter field $\mf^\vr$ that transforms linearly in a representation $D$ of $H_{(\psi_0,0)}$, cf.~\eqref{CCWZspacetime0}, the covariant derivative can be defined as
\begin{equation}
\cd_{\fr\m}\mf^\vr(x)\equiv\bigl[\de_\n\mf^\vr(x)+\I\mc^\a_\n(\pi(x),x)D(Q_\a)^\vr_{\phantom\vr\s}\mf^\s(x)\bigr]e^\n_{\fr\m}(\pi(x),x)\;.
\label{spacetimecovderpsi}
\end{equation}
This generalizes the covariant derivative we constructed in the context of internal symmetries in Sect.~\ref{subsec:effLagmatter}.

We conclude by checking in what precise sense the derivatives~\eqref{spacetimecovderpi} and~\eqref{spacetimecovderpsi} of the NG and matter fields are covariant. To that end, consider a transformation by an element of $g\in G$ infinitesimally close to unity, $g\approx e+\I\eps^AQ_A$, where $\eps^A$ is a set of small constant parameters. The element $h(\pi,g_0(x,g))\in H_{(\psi_0,0)}$, defined by~\eqref{CCWZspacetime0} with $\smash{g_0(x,g)=\tran{\DX(x,g)}^{-1}g\tran x}$, then reads, to linear order in $\eps^A$,
\begin{equation}
h(\pi,g_0(x,g))\approx e+\I\eps^Ak^\a_A(\pi,x)Q_\a\;.
\end{equation}
Here $k^\a_A(\pi,x)$ is a set of functions that are calculable following the guidance of Sect.~\ref{sec:CCWZstandard}, but we need not do so explicitly. It is easy to check that the last two lines of~\eqref{MCspacetimetransforcomps} now translate into the infinitesimal transformations
\begin{equation}
\begin{split}
\udelta\mc^a(\pi,x)&=-\eps^Ak^\a_A(\pi,x)f^a_{\a b}\mc^b(\pi,x)\;,\\
\udelta\vec e^{*\fr\m}(\pi,x)&=-\eps^Ak^\a_A(\pi,x)f^{\fr\m}_{\a\fr\n}\vec e^{*\fr\n}(\pi,x)\;.
\end{split}
\end{equation}
By the duality between the frame $\vec e_{\fr\m}$ and the coframe $\vec e^{*\fr\m}$, the transformation of the latter gives $\smash{\udelta\vec e_{\fr\m}=\eps^Ak^\a_Af^{\fr\n}_{\a\fr\m}\vec e_{\fr\n}}$. Equipped with these auxiliary identities, we arrive at the infinitesimal transformations of the covariant derivatives~\eqref{spacetimecovderpi} and~\eqref{spacetimecovderpsi},
\begin{equation}
\begin{split}
\udelta(\cd_{\fr\m}\pi^a)&=\eps^Ak^\a_A(-f^a_{\a b}\cd_{\fr\m}\pi^b+f^{\fr\n}_{\a\fr\m}\cd_{\fr\n}\pi^a)\;,\\
\udelta(\cd_{\fr\m}\mf^\vr)&=\eps^Ak^\a_A[\I D(Q_\a)^\vr_{\phantom\vr\s}\cd_{\fr\m}\mf^\s+f^{\fr\n}_{\a\fr\m}\cd_{\fr\n}\mf^\vr]\;.
\end{split}
\label{covdertransfo}
\end{equation}
The covariant derivatives transform linearly. This makes it possible to construct $G$-invariant Lagrangian densities out of the (possibly higher) covariant derivatives using standard methods of tensor algebra and representation theory.

\begin{illustration}%
Recall the example worked out in Sect.~\ref{subsec:spacetimexscale}. Here $G\simeq\R^+\ltimes\gr{ISO}(d,1)$ acts on a set of relativistic scalars, where $\gr{ISO}(d,1)$ is the Poincar\'e group and $\R^+$ represents scale transformations of spacetime coordinates. Suppose that the ground state of the system carries a uniform condensate, spontaneously breaking the scale invariance. We choose accordingly $\psi_0\in\M$ that breaks the dilatation group $\R^+$. Thus, $H_{(\psi_0,0)}\simeq\gr{SO}(d,1)$ whereas $H_0\simeq\R^+\times\gr{SO}(d,1)$.

Let us disregard possible matter fields and focus on the dilaton field $\pi$, parameterizing the coset space $H_0/H_{(\psi_0,0)}\simeq\R$. We fix $x^\m$ to be the standard Minkowski coordinates, and $P_{\fr\m}$ the energy--momentum operator. The manifold $H_0/H_{(\psi_0,0)}\times M\simeq\R^{d+2}$ is naturally parameterized by $\upix(\pi,x)=\E^{\I x\cdot P}\E^{\I\pi D}$.\footnote{In order to avoid notation clash with the dilatation operator $D$, I temporarily denote the dimension of spacetime as $d+1$.} With the help of the commutation relation $[D,P_{\fr\m}]=-\I P_{\fr\m}$, we compute the MC form
\begin{equation}
\mc(\pi,x)=-\I\upix(\pi,x)^{-1}\D\upix(\pi,x)=D\D\pi+\E^{-\pi}P\cdot\D x\;.
\end{equation}
From the second term, we extract the spacetime coframe $\vec e^{*\fr\m}=\d^{\fr\m}_\n\E^{-\pi}\D x^\n$ and the corresponding dual frame, $\vec e_{\fr\m}=\d_{\fr\m}^\n\E^\pi\de_\n$. The dual frame can be used as a covariant derivative operator, $\cd_{\fr\m}$, that is scale-invariant.

To produce an action, we still need a volume element. This is most easily obtained by constructing a volume form on the Minkowski spacetime using the coframe,
\begin{equation}
\vol=\vec e^{*0}\w\dotsb\w\vec e^{*d}=\E^{-(d+1)\pi}\D x^0\w\dotsb\w\D x^{d}\;.
\end{equation}
A generic Poincar\'e- and scale-invariant effective action for the dilaton then reads
\begin{equation}
S_\mathrm{eff}\{\pi\}=\int\D^{d+1}\!x\,\E^{-(d+1)\pi}\La_\mathrm{eff}(\cd\pi,\cd\cd\pi,\dotsc)\;.
\end{equation}
The Lagrangian $\La_\mathrm{eff}$ can be built by writing down the most general Lorentz-invariant operator in terms of the derivatives of $\pi$, and then replacing $\de_\m$ with $\cd_\m$ everywhere.
\end{illustration}


\section{Twisting Order Parameter for Internal Symmetry}
\label{sec:quantumtwist}

We have reached the point where we can start dealing with concrete systems featuring spontaneous breakdown of spacetime symmetry. As mentioned above, I will always assume that the effective action can be constructed solely out of the MC form. Any claim of having found the most general invariant action therefore has to be interpreted accordingly. It is not always straightforward to construct the most general effective action even in this restricted sense. To make further progress, it is convenient to narrow down the landscape of systems that we consider.

In this section, I will elaborate on the example of nonlinear realization worked out in Sect.~\ref{subsec:spacetimexinternal}. Here is a brief reminder. Suppose that the symmetry group factorizes as $G\simeq G_\mathrm{s.t.}\times G_\mathrm{int}$, where ``s.t.'' and ``int'' stand respectively for ``spacetime'' and ``internal.'' Moreover, suppose that the target manifold $\M$ is parameterized by a set of fields $\psi^i$ that are scalar in the sense that $\DF^i(\psi,x,g_\mathrm{s.t.})=\psi^i$ for all $g_\mathrm{s.t.}\in G_\mathrm{s.t.}$. This is equivalent to requiring that $G_\mathrm{s.t.}$ is a purely spacetime symmetry. Likewise, $G_\mathrm{int}$ is internal in the sense that $\DF^i(\psi,x,g_\mathrm{int})$ is independent of $x^\m$ and $\DX^\m(x,g_\mathrm{int})=x^\m$ for all $g_\mathrm{int}\in G_\mathrm{int}$. The action of the symmetry group $G$ on the standard coordinates $(\pi^a,\mf^\vr,x^\m)$ is then given by~\eqref{STexamplescalars}. The isotropy group of the spacetime origin is $H_0\simeq H_\mathrm{s.t.}\times G_\mathrm{int}$, and the spacetime manifold $M$ is equivalent to the homogeneous space $G_\mathrm{s.t.}/H_\mathrm{s.t.}$. Similarly, the isotropy group of $(\psi_0,0)\in\M\times M$ is $H_\mathrm{s.t.}\times H_\mathrm{int}$, hence $H_0/H_{(\psi_0,0)}\simeq G_\mathrm{int}/H_\mathrm{int}$. 

I will now make one additional, important assumption. Let us choose the reference point $\psi_0\in\M$ so that it represents the actual value of the order parameter at the spacetime origin, $\vev{\psi^i(0)}=\psi^i_0$. Suppose that in the resulting standard coordinates on $\M$, the \emph{vacuum expectation value} (VEV) of all the matter fields, $\vev{\mf^\vr(x)}$, is zero. This is equivalent to requiring that the order parameter is fully specified by the VEVs $\vev{\pi^a(x)}$. The values of the order parameter at different spacetime points lie on the same orbit of $G_\mathrm{int}$ in $\M$. It is then mathematically consistent to eliminate the matter fields from the EFT altogether. After all, they are expected to excite gapped modes in the spectrum, which can be ignored at sufficiently low energies.

\begin{illustration}%
\label{ex:twistedcomplexscalar}%
Consider a theory of a single real relativistic scalar field $\p$, equipped with an internal shift symmetry, $\smash{(\p,x^\m)\xrightarrow{\eps}(\p+\eps,x^\m)}$. This corresponds to $G_\mathrm{s.t.}\simeq\gr{ISO}(d,1)$, $H_\mathrm{s.t.}\simeq\gr{SO}(d,1)$, and $G_\mathrm{int}\simeq\R$. For any choice of the reference point $\p_0$, the internal isotropy group is trivial, $H_\mathrm{int}\simeq\trgr$. In this case, the only degree of freedom, that is $\p$ itself, is of the NG type. There are no matter-type variables. Therefore, our assumption that the VEV of all $\mf^\vr$ be vanishing is trivially satisfied.

A slightly less trivial example is that of a complex relativistic scalar $\p$, subject to the action of $G_\mathrm{int}\simeq\gr{U}(1)$ via $\smash{(\p,x^\m)\xrightarrow{\eps}(\E^{\I\eps}\p,x^\m)}$. For any nonzero $\p_0\in\C$, we have again $H_\mathrm{int}\simeq\trgr$. The field is then naturally parameterized by its modulus and phase, $\p=\vr\E^{\I\t}$, where $\t$ is the sole NG variable. Provided $\vev{\vr(x)}\equiv\vr_0$ is constant, $\vr$ can be traded for a matter field with vanishing VEV as desired, $\mf\equiv\vr-\vr_0$. On the other hand, the VEV of the phase, $\vev{\t(x)}$, can have arbitrary coordinate dependence. This kind of order parameter describes relativistic superfluids.
\end{illustration}

Discarding the matter fields amounts to restricting our basic setup to the manifold $H_0/H_{(\psi_0,0)}\times M\simeq G_\mathrm{int}/H_\mathrm{int}\times G_\mathrm{s.t.}/H_\mathrm{s.t.} $. Henceforth, I will only consider flat spacetimes such that $G_\mathrm{s.t.}$ possesses a set of $D$ mutually commuting translation generators $P_{\fr\m}$. With the parameterization $\upix(\pi,x)=\E^{\I x\cdot P}U(\pi)$, the MC form becomes
\begin{equation}
\mc(\pi,x)=-\I U(\pi)^{-1}\D U(\pi)+P\cdot\D x\;.
\label{MCtwisted}
\end{equation}
The spacetime coframe is trivial, $\vec e^{*\fr\m}=\d^{\fr\m}_\n\D x^\n$, and it is not necessary to distinguish frame and coordinate-basis indices. The $\mcu$ and $\mcb$ parts of the MC form~\eqref{MCtwisted} are fixed by the internal symmetry. We can therefore reuse the wealth of information accumulated in Chap.~\ref{chap:CCWZ} to evaluate them.

So far, I have not made any assumptions about the coordinate dependence of the order parameter beyond the requirement that all $\vev{\mf^\vr(x)}$ vanish. Whether or not the order parameter breaks any spacetime symmetries has no bearing on the set of degrees of freedom $\pi^a$ of the EFT, their transformation under $G$, or the MC form. Yet, the devil is in the details, as we shall now see.


\subsection{New Features of the Old Setup}
\label{subsec:twistnewfeatures}

A consistent low-energy EFT setup requires a number of ingredients to be in place. The first is of course having a set of well-defined degrees of freedom, well separated from whatever high-energy modes have been ignored. This is generally guaranteed by the broken symmetry. These low-energy degrees of freedom should map to positive-energy excitations of a stable ground state. Failure to satisfy this condition indicates that we have not identified the correct ground state. Finally, in the absence of further physical input, it is necessary to include in the effective Lagrangian all operators consistent with the given symmetry. Typically, infinitely many such operators exist. Hence, we need an organizing principle to decide which operators are relevant and which can be neglected: power counting.

\runinhead{Stability of the Ground State} Our choice of the reference point, $\psi^i_0\equiv\vev{\psi^i(0)}$, corresponds to $\vev{\pi^a(0)}=0$, or $\vev{U(\pi(0))}=e$. The actual order parameter defines a map from $M$ to the coset space $G_\mathrm{int}/H_\mathrm{int}$, $\vev{U(\pi(x))}\equiv U_0(x)$. Nontrivial coordinate dependence indicates that the order parameter is \emph{twisted}. Importantly, not every choice of $U_0(x)$ is physically consistent. A necessary condition is that $U_0(x)$ is a local minimum of the energy functional of the EFT. This is straightforward to check in two steps. First, $U_0(x)$ should be a stationary state, that is a solution of the \emph{equation of motion}  (EoM); see Sect.~\ref{sec:effLagEoMspectrum} for details on the latter. Second, the spectrum of small fluctuations about this state should only contain positive-energy excitations.

\begin{illustration}%
Recall the helimagnetic state, introduced in \refex{ex:helimagnet},
\begin{equation}
\vev{\vec n(\vec x,t)}=(\cos kz,\sin kz,0)\;,
\end{equation}
where $k$ is a positive constant. In this case, the order parameter is $\vec n\in\gr{SO}(3)/\gr{SO}(2)\simeq S^2$. It is spatially twisted so that it forms a helix oriented along the $z$-axis, with pitch $2\pi/k$. In \refex{ex:helimagnetgroundstate}, I showed that this order parameter minimizes the energy in the EFT for ferromagnets augmented with the \emph{Dzyaloshinskii--Moriya} (DM) interaction. The precise value of $k$ is fixed by the coupling of the DM term.
\end{illustration}

Ensuring the stability of the ground state is an integral part of the construction of the low-energy EFT. The requirement was already implicitly present in our analysis of internal symmetries in Part~\ref{part:internalSSB} of the book. The fact that the vacuum, $\pi^a=0$, is a solution of the EoM was guaranteed by the absence of terms linear in $\pi^a$ in the effective Lagrangian. Such terms were typically forbidden by the nonlinearly realized symmetry. Only in one exceptional example, namely the class of Galileon theories (Sect.~\ref{subsubsec:Galileon}), I enforced the absence of linear terms ad hoc by discarding the tadpole operator. Furthermore, stability under small fluctuations was ensured by fixing the signs of the parts of the effective Lagrangian bilinear in $\pi^a$, carrying respectively two temporal or two spatial derivatives.

\runinhead{Power Counting} Eventually, we would like to parameterize the EFT by NG fields $\tilde\pi^a(x)$ such that $\vev{\tilde\pi^a(x)}=0$. This makes it easy to distinguish the ground state from its fluctuations. It is often convenient to factorize the matrix variable $U(\pi)$ as
\begin{equation}
U(\pi(x))\equiv U_0(x)U(\tilde\pi(x))\;.
\label{Ufactorization}
\end{equation}
This maintains a simple transformation rule for $\tilde\pi^a$ under the action of $G_\mathrm{int}$. Indeed, a comparison with~\eqref{STexamplescalars} shows that
\begin{equation}
U(\tilde\pi)\xrightarrow{g_\mathrm{int}}U(\tilde\pi'(\tilde\pi,g_\mathrm{int}))=(U_0^{-1}g_\mathrm{int}U_0)U(\tilde\pi)h_\mathrm{int}(\tilde\pi,g_\mathrm{int})^{-1}\;.
\end{equation}
In other words, the action of $G_\mathrm{int}$ by left multiplication is simply conjugated by $U_0^{-1}$.

The challenge to face is that following blindly the machinery of Chap.~\ref{chap:effLagrangian}, we arrive at an effective Lagrangian where derivatives act on $U(\pi)$ rather than on $U(\tilde\pi)$. This is not a problem if the characteristic length scale of the symmetry breaking $G_\mathrm{int}\to H_\mathrm{int}$ is much shorter than the scale of variations of the ground state $U_0(x)$. Such a \emph{hierarchical} symmetry breaking occurs for instance in helimagnets. Indeed, the helix pitch in FeGe is about $70\,\mathrm{nm}$~\cite{Togawa2016a}, which is much longer than the scale of the crystal lattice at which the magnetic order is formed. In such systems, derivatives of both $U(\tilde\pi(x))$ and $U_0(x)$ can be considered small and counted together. Should, however, the order parameter feature short-distance variations, a different power counting is needed. The precise setup of the power counting that renders the EFT predictive is then best considered case by case.


\subsection{Case Study: Relativistic Superfluids}
\label{subsec:twistsuperfluids}

From the point of view of symmetry, a superfluid is an ordered phase of matter where an internal $\gr{U}(1)$ symmetry is spontaneously broken. We therefore need an order parameter charged under this symmetry. The simplest choice is to take a complex scalar field $\p$ carrying the action of $G_\mathrm{int}\simeq\gr{U}(1)$ via $\smash{(\p,x^\m)\xrightarrow{\eps}(\E^{\I\eps}\p,x^\m)}$. We then assume that due to microscopic dynamics, this field develops a nonzero VEV. That is however not enough. We want the vacuum to describe \emph{matter}, that is a state with nonzero density of the $\gr{U}(1)$ charge. In a Lorentz-invariant scalar theory, a state of uniform nonzero charge density corresponds to a time-dependent VEV,
\begin{equation}
\vev{\p(x)}\equiv\vp(x)=\vp_0\E^{-\I\m t}\;.
\label{OPsuperfluid}
\end{equation}
The constants $\vp_0$ and $\m$ are without loss of generality assumed to be positive. See Chap.~3 of~\cite{Schmitt2015} for an introduction to superfluidity from the field theory point of view.

\begin{watchout}%
The same superfluid state can also be represented by a constant order parameter at the cost of redefining the Lagrangian. Indeed, introduce a new scalar field, $\Phi(x)\equiv\E^{\I\m t}\p(x)$, so that our superfluid state amounts to $\vev{\Phi(x)}=\vp_0$. As a consequence of the field redefinition, $\de_0\p=(\de_0-\I\m)\Phi$. This agrees with my previous observation (see \refex{ex:mNGB}) that an equilibrium state parameterized by the chemical potential $\m$ can be described in the Lagrangian formalism by a constant background gauge field, $A_\m=(\m,\vec0)$. It is now clear that the order parameter~\eqref{OPsuperfluid} corresponds to a superfluid at rest. Using Lorentz invariance, a superfluid state in uniform motion can likewise be parameterized by a constant timelike $A_\m$, or equivalently by the VEV $\vp(x)=\vp_0\E^{-\I A\cdot x}$.
\end{watchout}

As we already saw in \refex{ex:twistedcomplexscalar}, the condensate~\eqref{OPsuperfluid} satisfies our technical assumption underlying the construction of EFT for systems with a twisted order parameter. Here we have $U(\pi(x))=\E^{\I\pi(x)Q}$, where $\pi(x)$ is the phase of $\p(x)$ and $Q$ is the generator of $G_\mathrm{int}\simeq\gr{U}(1)$. The NG field transforms under the internal symmetry as $\smash{\pi\xrightarrow{\eps}\pi+\eps}$. The MC form~\eqref{MCtwisted} reduces to $\mc(\pi,x)=Q\D\pi+P\cdot\D x$. Upon dropping the tadpole operator, $\pi$, which is itself quasi-invariant, the most general Poincar\'e- and $\gr{U}(1)$-invariant effective action can be written as
\begin{equation}
S_\mathrm{eff}\{\pi\}=\int\D^D\!x\,\La_\mathrm{eff}(\de\pi,\de\de\pi,\dotsc)\;.
\label{relsupefluidaction}
\end{equation}
The effective Lagrangian $\La_\mathrm{eff}$ is an arbitrary function of the derivatives of $\pi$ in which all Lorentz indices have been contracted in a Lorentz-invariant manner.

Let us check that the tentative ground state, $\vev{\pi(x)}=-\m t$, satisfies the EoM. This follows from the equivalence of the EoM and the conservation of the Noether current of the internal $\gr{U}(1)$ symmetry; see \refex{ex:shiftscalar} for a proof of the equivalence. The current carried by the condensate~\eqref{OPsuperfluid} is necessarily constant, and thus automatically conserved for any values of $\vp_0$ and $\m$.

Next, we have to deal with power counting. I will follow closely the logic of Sect.~\ref{subsec:ChPTpowercounting}, where a power-counting scheme for relativistic EFTs was worked out. A new complication is that we do not know a priori how large the chemical potential $\m$ is, and thus whether $\de_\m\pi$ can be considered small. We therefore cannot naively expand the effective Lagrangian in the derivatives of $\pi$. However, we can use the fact that every $\pi$ in the Lagrangian always carries at least one ``persistent'' derivative and only count \emph{additional} derivatives. This makes sense: introducing the fluctuation field $\tilde\pi$ via~\eqref{Ufactorization}, that is $\pi(x)\equiv\tilde\pi(x)-\m t$, any higher derivative kills the chemical potential and thus captures directly the variation of $\tilde\pi$.

To formalize this observation, expand the effective Lagrangian as
\begin{equation}
\La_\mathrm{eff}[\pi]=\sum_{n\geq0}\La_\mathrm{eff}^{(n)}[\pi]\;,
\end{equation}
were $\smash{\La_\mathrm{eff}^{(n)}}$ collects all Lorentz-invariant operators with $n$ additional derivatives. Now consider a generic Feynman diagram $\Gamma$, contributing to a given observable. Denote as $I$ the number of internal propagators in $\Gamma$, as $E$ the number of external legs, as $L$ the number of loops, and as $V_n$ the number of interaction vertices from each of $\smash{\La_\mathrm{eff}^{(n)}}$. Think of the Fourier representation, in which the diagram amounts to a homogeneous function of energy--momenta on the external legs. Loop integration contributes altogether $DL$ powers of energy--momenta, the internal propagators contribute $-2I$ powers. The persistent derivatives on each field give $2I+E$ powers of energy--momenta. Finally, the additional derivatives add up to $\sum_{n\geq0}nV_n$. Altogether, the naive degree of the diagram becomes
\begin{equation}
\deg\Gamma=DL+E+\sum_{n\geq0}nV_n\;.
\end{equation}

For any given observable, $E$ is fixed and $\deg\Gamma\geq E$. The \emph{leading-order} (LO) contribution comes from diagrams with $L=0$ and $V_n=0$ for any $n\geq1$. Accordingly, the LO effective Lagrangian for a relativistic superfluid is $\smash{\La_\mathrm{eff}^{(0)}}$: an arbitrary Lorentz-invariant function of $\de_\m\pi$. The \emph{next-to-leading-order} (NLO) Lagrangian requires two additional derivatives, $n=2$. For any $D\geq3$, the NLO contribution to any observable corresponds to $\deg\Gamma=E+2$ and amounts to tree-level ($L=0$) diagrams with one interaction vertex from $\smash{\La_\mathrm{eff}^{(2)}}$ and all others from $\smash{\La_\mathrm{eff}^{(0)}}$. Interestingly, the contributions from loop diagrams only start to matter beyond NLO of the derivative expansion. Most important is that at any finite degree, only a finite number of Lagrangians $\smash{\La_\mathrm{eff}^{(n)}}$ and a finite number of Feynman diagrams is required. This renders the power-counting scheme consistent and predictive.

It might still look troublesome that the LO Lagrangian can be an arbitrary Lorentz-invariant function of $\de_\m\pi$. This surely contains an infinite number of unknown parameters, so how could such an EFT be useful? To address this question, note there is only one way to make such Lagrangians Lorentz-invariant, namely by contracting the Lorentz indices on $\de_\m\pi$ pairwise. Then, upon separating the ground state from the fluctuations, the LO Lagrangian becomes
\begin{equation}
\La_\mathrm{eff}^{(0)}[\tilde\pi]=P\Bigl(\sqrt{(\de_\m\pi)^2}\Bigr)=P\Bigl(\sqrt{(\de_0\tilde\pi-\m)^2-(\vec\nabla\tilde\pi)^2}\Bigr)\;,
\label{superfluidEFT}
\end{equation}
where $P$ is some as yet unknown function. In the superfluid ground state, the Lagrangian reduces to a function of the chemical potential, $P(\m)$. This can be related to the energy density of the equilibrium state by Legendre transformation,
\begin{equation}
\at{\Ha_\mathrm{eff}^{(0)}}{\tilde\pi=0}\equiv U(\m)=\m P'(\m)-P(\m)\;.
\label{HUmuP}
\end{equation}
The prime indicates a derivative of $P$ with respect to its argument. Also, I used that $\smash{\Pd{\La_\mathrm{eff}^{(0)}}{(\de_0\pi)}=-\Pd{\La_\mathrm{eff}^{(0)}}{\m}}$ and that $\vev{\pi(x)}=-\m t$. Equation~\eqref{HUmuP} looks familiar: $P(\m)$ can be interpreted as the thermodynamic pressure of the superfluid at zero temperature, and $P'(\m)\equiv n(\m)$ as the density of the $\gr{U}(1)$ charge. Our main result therefore is that the LO EFT for superfluids at zero temperature~\eqref{superfluidEFT} is completely fixed by the thermodynamic equation of state~\cite{Son2002a}.

To get a flavor of the physical content of the EFT, let us expand~\eqref{superfluidEFT} to second order in the NG field,
\begin{equation}
\La_\mathrm{eff}^{(0)}[\tilde\pi]\simeq P(\m)+\frac{P''(\m)}2(\de_0\tilde\pi)^2-\frac{P'(\m)}{2\m}(\vec\nabla\tilde\pi)^2+\dotsb\;.
\end{equation}
This shows that the NG bosons propagate with the phase velocity
\begin{equation}
v=\sqrt{\frac{P'(\m)}{\m P''(\m)}}=\sqrt{\frac n\m\OD\m n}=\sqrt{\OD PU}\;,
\label{superfluidvelocity}
\end{equation}
where I used the thermodynamic relations $\D U=\m\D n$ and $\D P=n\D\m$. It is remarkable that the phase velocity of the superfluid NG boson at zero temperature is given by the same expression~\eqref{superfluidvelocity} as that of hydrodynamic sound.

\begin{illustration}%
For an illustration of these general results, let us have a look at dense matter consisting of relativistic (Dirac) fermions in $d=3$ dimensions~\cite{Son2002a}. Provided the interactions between the particles are sufficiently weak and their mass sufficiently small, the equation of state can be well approximated by that of a free gas of massless fermions,\footnote{This restriction is of course just a matter of convenience. Our general result for the LO EFT~\eqref{superfluidEFT} applies equally well to a strongly-interacting gas of massive fermions.}
\begin{equation}
P(\m)=\frac{\m^4}{12\pi^2}\;,\qquad
n(\m)=\frac{\m^3}{3\pi^2}\;,\qquad
U(\m)=\frac{\m^4}{4\pi^2}\;.
\end{equation}
Suppose that the interaction between the fermions is attractive so that they form Cooper pairs in the spin-singlet, $s$-wave state. The fermionic quasiparticles near the Fermi surface then become gapped and decouple at sufficiently low energies. The low-energy physics of the system reduces to the dynamics of the condensate of scalar Cooper pairs. This is an example of a fermionic superfluid. According to~\eqref{superfluidEFT}, the low-energy EFT of such a system is dominated by a LO Lagrangian that is polynomial in the NG field,
\begin{equation}
\La_\mathrm{eff}^{(0)}[\tilde\pi]=\frac1{12\pi^2}\bigl[(\de_0\tilde\pi-\m)^2-(\vec\nabla\tilde\pi)^2\bigr]^2\;.
\end{equation}
The NG bosons propagate at the speed $v=1/\sqrt3$, characteristic of a gas of weakly interacting ultrarelativistic particles. Note that in this case, the microscopic energy scale of the system is set by the chemical potential itself. The scale associated with the spontaneous breakdown of the $\gr{U}(1)$ symmetry is assumed to be much smaller, which is ensured by the weak binding of the Cooper pairs. Thus, $\m$ could not have been treated as a small parameter in any reasonable sense.
\end{illustration}


\section{Vector Modes: the Relevant, the Irrelevant and~the~Unphysical}
\label{sec:quantumvector}

As demonstrated in the previous chapter, the same symmetry-breaking pattern can be realized with various choices of order parameter. This ambiguity is innocuous for internal symmetries. On the other hand, in case of spacetime symmetries, it can lead to nonlinear realizations with apparently different numbers of NG variables. For instance, the superfluid order parameter~\eqref{OPsuperfluid} breaks the combined spacetime and internal symmetry in a rather nontrivial manner. First, it obviously breaks Lorentz boosts. Second, it breaks time translations and the internal $\gr{U}(1)$ symmetry down to a ``diagonal'' subgroup, where the phase generated by a time translation is compensated by a $\gr{U}(1)$ rotation. In the setup of Sect.~\ref{subsec:twistsuperfluids}, all this is implicitly taken into account. However, as pointed out in Sect.~\ref{subsec:spacetimexscalarvector}, it might be meaningful to add a secondary, vector order parameter, representing the VEV of the Noether current of the $\gr{U}(1)$ symmetry. This requires introducing an additional vector of NG fields. Similarly, we saw in Sect.~\ref{subsec:spacetimexSchr} that nonlinear realization of Galilei symmetry is easier to implement if one adds a vector of NG variables corresponding to Galilei boosts.

These observations raise obvious questions. Are the additional vector NG variables merely a useful mathematical tool, or do they correspond to actual modes in the spectrum? In case they are physical, what is their exact nature? To answer these questions, I will now work out several physically distinct examples, from nonlinear realization to concrete effective actions. With the additional insight, I will then in Sect.~\ref{subsec:vectorIHC} attempt to draw general conclusions.


\subsection{The Relevant: Helimagnets}
\label{subsec:vectorrelevant}

Let us start with a familiar system where a vector field turns out to excite a physical, gapless NG mode. This will set a benchmark for our discussion of the role of vector NG modes. At the same time, it will provide the first example of an EFT going beyond the class of systems covered by Sect.~\ref{sec:quantumtwist}. I already addressed spin systems in detail in Chap.~\ref{chap:internalexamples}. Moreover, I mentioned the exotic case of helimagnets on several occasions. I will therefore not spend time on reviewing the basic phenomenology, and instead start right away with a symmetry-based analysis.

We are talking about a condensed-matter system whose symmetry is captured by the Aristotelian group in three spatial dimensions, $G\simeq\gr{SO}(3)\ltimes\R^4$. Unlike in Chap.~\ref{chap:internalexamples}, I will not include a separate internal symmetry under spin rotations. This is broken by the spin-orbit coupling, which plays a key role in the formation of the helimagnetic order. The order parameter is the local magnetization density, which is a vector under spatial rotations, hence $\M\simeq\R^3$. We are of course interested in the physics of ordered states where the magnetization is nonzero. This fixes the two isotropy groups $H_0\simeq\gr{SO}(3)$ and $H_{(\psi_0,0)}\simeq\gr{SO}(2)$. The local manifold de\-com\-po\-si\-tion~\eqref{isomanifold} in this case becomes
\begin{equation}
G\simeq\gr{SO}(2)\times S^2\times\R^4\;.
\end{equation}
The submanifold relevant for the nonlinear realization of the symmetry on the NG variables is $H_0/H_{(\psi_0,0)}\times M\simeq S^2\times\R^4$. I will parameterize the spacetime using the translation operator $\tran{\vec x,t}=\E^{\I tH}\E^{\I\skal xP}$. The coset space $S^2$ is parameterized by two local coordinates $\pi^a$, encoded in an $\gr{SO}(3)$-valued matrix $U(\pi)$.

Having fixed the basic setup, we can read off the MC form from~\eqref{MCspacetime},
\begin{equation}
\begin{split}
\mc(\pi,x)&=-\I U(\pi)^{-1}\D U(\pi)+H\D t+U(\pi)^{-1}(\vec P\cdot\D\vec x)U(\pi)\\
&=-\I U(\pi)^{-1}\D U(\pi)+H\D t+P_{\fr r}[U(\pi)^{-1}]^{\fr r}_{\phantom rs}\D x^s\;.
\end{split}
\end{equation}
The second equality follows from the fact that the momentum operator carries a~vector representation of $\gr{SO}(3)$ under the adjoint action thereof.\footnote{In Chap.~\ref{chap:internalexamples}, I worked out the EFT for spin systems using the fundamental (spinor) representation of $\gr{SO}(3)$. It goes without saying that this is just a matter of convenience. The nonlinear realization of symmetry on coset spaces as detailed in Sect.~\ref{sec:CCWZstandard} only depends on the structure of the symmetry group. In the present case, the use of the vector representation of $H_0\simeq\gr{SO}(3)$ is suggested by the transformation properties of the momentum operator $\vec P$.} Using~\eqref{MCspacetimedecomposition}, we subsequently identify the elements of the spacetime coframe,
\begin{equation}
\vec e^{*\fr 0}(\pi,x)=\D t\;,\qquad
\vec e^{*\fr r}(\pi,x)=[U(\pi)^{-1}]^{\fr r}_{\phantom rs}\D x^s\;.
\label{helicoframe}
\end{equation}
The $\mcb$ part of the MC form comes entirely from $-\I U(\pi)^{-1}\D U(\pi)$, i.e.~is given by $\mc^a(\pi)=\mc^a_b(\pi)\D\pi^b$. The related covariant derivatives of the NG variables $\pi^a$ are extracted using~\eqref{spacetimecovderpi},
\begin{equation}
\cd_{\fr0}\pi^a=\mc^a_b(\pi)\de_0\pi^b\;,\qquad
\cd_{\fr r}\pi^a=\mc^a_b(\pi)U(\pi)^s_{\phantom s\fr r}\de_s\pi^b\;.
\end{equation}
Under the action of $H_{(\psi_0,0)}\simeq\gr{SO}(2)$, these split into the irreducible multiplets
\begin{equation}
\begin{gathered}
\d^{\fr a}_a\cd_{\fr a}\pi^a\;,\qquad
\ve^{ab}\d^{\fr a}_a\cd_{\fr a}\pi_b\;,\\
\cd_{\fr0}\pi^a\;,\qquad
\cd_{\fr\a}\pi^a\;,\qquad
\d^{\fr a}_a\cd_{\fr a}\pi_b+\d^{\fr a}_b\cd_{\fr a}\pi_a-\d_{ab}\d^{\fr c}_c\cd_{\fr c}\pi^c\;.
\end{gathered}
\label{heliIR}
\end{equation}
Those on the first line of~\eqref{heliIR} are singlets, while the first two items on the second line transform as vectors under $\gr{SO}(2)$. Finally, the last operator in~\eqref{heliIR} transforms as a traceless symmetric tensor of $\gr{SO}(2)$. That is also a two-dimensional real representation, but carries a double charge of $\gr{SO}(2)$ compared to the vector representation. This is important when combining the irreducible multiplets into an effective Lagrangian in a way that preserves invariance under $H_{(\psi_0,0)}\simeq\gr{SO}(2)$.

Before we can construct an effective action, we still need to consider power counting. Here I will, unlike for superfluids in Sect.~\ref{subsec:twistsuperfluids}, assume that the spatial variation of the order parameter is slow. This allows us to apply the naive scheme where each (covariant) derivative counts. The EFT will then be dominated by operators with the lowest number of derivatives. It is convenient to express the effective Lagrangian in terms of a unit-vector parameterization of the coset space, $\vec n(\pi)\in S^2$. This is related to $U(\pi)$ via $\vec n(\pi)=U(\pi)\vec n_0$, where $\vec n_0\in S^2$ is an arbitrary but fixed reference vector. The advantage of using the $\vec n$ variable is that it makes invariance under the whole $H_0\simeq\gr{SO}(3)$ manifest. Dropping surface terms, the most general invariant Lagrangian with up two derivatives consists of
\begin{align}
\notag
\La^{(1,0)}_\mathrm{eff}={}&c^{(1,0)}_1\vec n\cdot(\vec\nabla\times\vec n)\;,\\
\notag
\La^{(2,0)}_\mathrm{eff}={}&c^{(2,0)}_1\d^{rs}\de_r\vec n\cdot\de_s\vec n+c^{(2,0)}_2(\skal\nabla n)^2+c^{(2,0)}_3[\vec n\cdot(\vec\nabla\times\vec n)]^2\\
\label{helilag}
&+c^{(2,0)}_4(\skal\nabla n)[\vec n\cdot(\vec\nabla\times\vec n)]\;,\\
\notag
\La^{(1,1)}_\mathrm{eff}={}&c^{(1,1)}_1\de_0\vec n\cdot[(\skal n\nabla)\vec n]\;,\\
\notag
\La^{(0,2)}_\mathrm{eff}={}&c^{(0,2)}_1(\de_0\vec n)^2\;.
\end{align}

\begin{watchout}%
All the operators in~\eqref{helilag} are clearly invariant under spatial rotations. What is less obvious is that there are no other algebraically independent contributions to the invariant Lagrangian with up to two derivatives. This follows from the formalism based on the MC form, but would not be straightforward to check in the language using the $\vec n$ field. Let me at least clarify why some apparent candidate operators do not give anything new. Starting with $\smash{\La^{(2,0)}_\mathrm{eff}}$, there is no operator of the type $(\vec\nabla\times\vec n)^2$, since this equals $\d^{rs}\de_r\vec n\cdot\de_s\vec n-(\skal\nabla n)^2$ up to a surface term. Likewise, the operator $[\vec n\times(\vec\nabla\times\vec n)]^2$ is absent, being equal to $(\vec\nabla\times\vec n)^2-[\vec n\cdot(\vec\nabla\times\vec n)]^2$. Other candidate operators can be eliminated using the identity $(\skal n\nabla)\vec n=-\vec n\times(\vec\nabla\times\vec n)$, valid for a unit vector field $\vec n$. Finally, there is no $\de_0\vec n\cdot(\vec\nabla\times\vec n)$ in $\smash{\La^{(1,1)}_\mathrm{eff}}$, since this is itself a surface term. Similar reasoning shows that no new operators can be produced using second covariant derivatives such as $\smash{\d^{\fr a}_a\cd_{\fr0}\cd_{\fr a}\pi^a}$ or $\smash{\d^{\fr a}_a\cd_{\fr\a}\cd_{\fr a}\pi^a}$.
\end{watchout}

The presence of $\smash{\La_\mathrm{eff}^{(1,0)}}$ with a single spatial derivative is a new feature, arising from the vector order parameter. There is no invariant operator with a single time derivative. However, we know from Chap.~\ref{chap:internalexamples} that an internal $\gr{SU}(2)$ spin symmetry together with spatial rotational invariance admit a quasi-invariant Lagrangian,
\begin{equation}
\La^{(0,1)}_\mathrm{eff}=c^{(0,1)}_1\frac{\ve_{ab}n^a\de_0n^b}{1+n^3}\;,
\label{heliWZ}
\end{equation}
where $a,b\in\{1,2\}$. Being free of spatial derivatives, this is also quasi-invariant under our $H_0\simeq\gr{SO}(3)$ that acts on both $\vec n$ and spatial coordinates. Finally, note that the coframe~\eqref{helicoframe} yields the standard volume measure $\D^3\!\vec x\,\D t$ thanks to the fact that $U(\pi)$ is an $\gr{SO}(3)$-valued matrix. This justifies a posteriori dropping surface terms from~\eqref{helilag}, and makes it actually possible to include a quasi-invariant Lagrangian such as~\eqref{heliWZ}.

In ferromagnets where the coupling $c^{(0,1)}_1$ is nonzero, every time derivative counts as two spatial derivatives (see Sect.~\ref{subsec:spinwavesferro} for a detailed justification). We can then discard the $\smash{\La^{(1,1)}_\mathrm{eff}}$ and $\smash{\La^{(0,2)}_\mathrm{eff}}$ Lagrangians as subleading. Even with this simplification, there are still six independent operators in the $\smash{\La^{(1,0)}_\mathrm{eff}}$, $\smash{\La^{(0,1)}_\mathrm{eff}}$ and $\smash{\La^{(2,0)}_\mathrm{eff}}$ Lagrangians. In order to make the analysis of the physical consequences of the EFT feasible, we need to take one more step to reduce the Lagrangian. Namely, the spin-orbit interaction in real materials is typically much weaker than other interactions. It therefore makes sense to treat the couplings of all the operators that lock the spin and orbital rotations together, namely $\smash{c^{(1,0)}_1}$, $\smash{c^{(2,0)}_2}$, $\smash{c^{(2,0)}_3}$ and $\smash{c^{(2,0)}_4}$, as small. Counting all of these formally as degree one in the derivative expansion, the entire LO (degree-two) effective Lagrangian will be just
\begin{equation}
\La_\mathrm{eff}^\mathrm{LO}=-M\frac{\ve_{ab}n^a\de_0n^b}{1+n^3}-\frac{\vr_\mathrm{s}}2\d^{rs}\de_r\vec n\cdot\de_s\vec n-\frac{2\pi\vr_\mathrm{s}}{\l_\mathrm{DM}}\vec n\cdot(\vec\nabla\times\vec n)\;.
\label{helilagDM}
\end{equation}
Here I have already switched to the physical notation for the effective couplings: $M$ for the local magnetization density, $\vr_{\mathrm{s}}$ for the spin stiffness, and $\l_\mathrm{DM}$ for the length scale of the DM interaction. The latter is all that is left of the effects of the spin-orbit coupling at the LO of the derivative expansion.

I showed already in \refex{ex:helimagnetgroundstate} that the DM interaction inevitably leads to the formation of a helical structure in the ground state. One can without loss of generality choose a Cartesian system of coordinates such that
\begin{equation}
\vev{\vec n(\vec x,t)}\equiv\vec n_0(\vec x)=(\cos kz,\sin kz,0)\;,
\end{equation}
where $k\equiv2\pi/\l_\mathrm{DM}$. Let us next focus on the excitation spectrum. The first step in that regard is to reparameterize the field following the logic of~\eqref{Ufactorization},
\begin{equation}
\vec n(\vec x,t)\equiv
\begin{pmatrix}
\cos kz & -\sin kz & 0\\
\sin kz & \cos kz & 0\\
0 & 0 & 1
\end{pmatrix}
\vec N(\vec x,t)\;.
\end{equation}
The variable $\vec N(\vec x,t)$ is still a unit vector. Moreover, in the ground state it is constant, $\vev{\vec N(\vec x,t)}\equiv\vec N_0=(1,0,0)$. This allows us to identify independent fluctuations around the ground state with the $N^2$ and $N^3$ components of $\vec N$.

The task to express the Lagrangian~\eqref{helilagDM} in terms of $\vec N$ is a bit tedious, and the final result is not elegant. I will therefore skip the straightforward details and only spell out the part of the Lagrangian bilinear in $N^2,N^3$. Up to a surface term,
\begin{equation}
\setlength\arraycolsep{0.5ex}
\La^\mathrm{LO}_\mathrm{eff}\simeq\frac{\vr_\mathrm{s}}2
\bigl(\begin{matrix}
N^2 & N^3
\end{matrix}\bigr)
\raisebox{-2ex}{$\begin{pmatrix}
\vec\nabla^2 & \Delta(z)\\
-\Delta(z) & \vec\nabla^2-k^2
\end{pmatrix}
\begin{pmatrix}
N^2\\
N^3
\end{pmatrix}$}+\dotsb\;,
\label{helilagbilin}
\end{equation}
where
\begin{equation}
\Delta(z)\equiv-\frac M{\vr_\mathrm{s}}\de_0+\frac{4\pi}{\l_\mathrm{DM}}[(\cos kz)\de_1+(\sin kz)\de_2]\;.
\end{equation}
The excitation spectrum is determined by the zero modes of the matrix differential operator in~\eqref{helilagbilin}. Due to the explicit coordinate dependence, this is a complicated problem. I will therefore content myself with some simple observations that illustrate the peculiarities of helimagnets. To start with, the presence of operators with a single time derivative makes the two degrees of freedom $N^2,N^3$ canonically conjugate to each other. The spectrum should therefore consist of a single type of magnon, just like in uniform ferromagnets in the absence of the DM interaction. The magnon dispersion relation can be calculated explicitly in the special case of propagation along the helix axis. The single-particle wave function is then independent of the $x,y$ coordinates so that $\smash{\Delta\to-(M/\vr_\mathrm{s})\de_0}$. A Fourier transform in time and the $z$-coordinate then gives the energy as a function of momentum,
\begin{equation}
E(p_z)=\frac{\vr_\mathrm{s}}M\abs{p_z}\sqrt{p_z^2+k^2}\;.
\label{helidisp}
\end{equation}

The physical momentum $p_z$ should be much smaller than the inverse of the length scale associated with the formation of the ferromagnetic order, which can be thought of as $\sqrt{\vr_\mathrm{s}}$. This defines the range of validity of the EFT. Within this range, however, there are two qualitatively different regimes. At long wavelengths, $p_z\ll k$, \eqref{helidisp} is well approximated by $E(p_z)\approx(k\vr_\mathrm{s}/M)\abs{p_z}$. The dispersion relation is linear. The corresponding one-particle state is a linearly polarized spin wave with $N^3\approx0$. This amounts to spin oscillations in the $xy$ plane along the direction perpendicular locally to $\vec n_0(\vec x)$. On the other hand, at intermediate wavelengths, $k\ll p_z\ll\sqrt{\vr_\mathrm{s}}$, the dispersion relation~\eqref{helidisp} turns into $E(p_z)\approx(\vr_\mathrm{s}/M)p_z^2$. This reproduces the standard ferromagnetic magnon spectrum we found in Sect.~\ref{subsec:spinwavesEoM}. Physically, this means that such short-distance oscillations of spin are insensitive to the helical structure of the ground state. The magnon is circularly polarized in the plane perpendicular to the local magnetization $\vec n_0(\vec x)$.

This concludes our excursion to helimagnetism. We have discovered very interesting physics, but no surprises as to the number and type of NG modes in the spectrum so far. To that end, we will capitalize on the experience we have collected here to explore another fascinating physical system: liquid crystals.


\subsection{The Irrelevant: Smectic Liquid Crystals}
\label{subsec:vectorirrelevant}

Liquid crystals are an intriguing exotic state of matter; see~\cite{Gennes1993} for a thorough account of the subject. The basic ingredient underlying the liquid crystal order is a strongly anisotropic microscopic constituent. This is typically an elongated organic molecule or a polymer, with a rigid, rod-like structure. An ordered ground state arises from a geometric arrangement of the molecules that breaks invariance under spatial translations or rotations, or both.

\begin{figure}[t]
\sidecaption[t]
\includegraphics[width=2.9in]{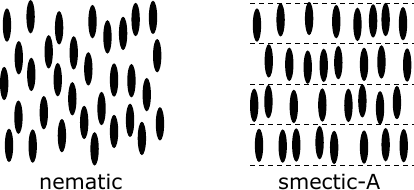}
\caption{Schematic visualization of nematic and smectic-A orders. A nematic consists of a set of mutually aligned undirected rods with random positions. In a smectic-A, the rods are in addition organized in layers. Their positions within each layer remain disordered}
\label{fig:liquidcrystal}
\end{figure}

I will be mostly concerned with the interplay of two simple liquid crystal phases: the \emph{nematic} phase and the \emph{smectic} (or more precisely \emph{smectic-A}) phase. In the nematic phase, the microscopic constituents are aligned along a common direction but their spatial positions remain disordered as in an ordinary liquid. See the left panel of Fig.~\ref{fig:liquidcrystal} for a visualization. In the smectic phase, the alignment persists but the constituents are in addition organized in parallel layers. Within each layer, however, their positions remain disordered; see the right panel of Fig.~\ref{fig:liquidcrystal}. Both of these phases feature an order parameter that defines the axis of alignment, and can be viewed as an ``unoriented vector.'' To describe smectics, one needs in addition a secondary order parameter that accounts for the layered structure of the medium.

The microscopic symmetry governing the physics of liquid crystals is the same as that of spin systems with spin-orbit coupling, $G\simeq\gr{SO}(3)\ltimes\R^4$. In the simpler nematic phase, which I will start with, the target space is $\M\simeq\R^3/\Z_2$, where any vector in $\R^3$ is identified with its opposite. This setup is almost identical to that of Sect.~\ref{subsec:vectorrelevant}, and we can therefore largely reuse the EFT developed therein. The change in the global structure of $\M$ from $\R^3$ to $\R^3/\Z_2$ affects the nature of topological defects in the medium. However, as far as the dynamics of small fluctuations of the order parameter is concerned, all we have to do is impose invariance under the reflection $\vec n\to-\vec n$. This kills, first of all, the quasi-invariant Lagrangian~\eqref{heliWZ}, for there is no equivalent of ``magnetization'' in liquid crystals that the coupling $\smash{c_1^{(0,1)}}$ could measure. Moreover, the $\smash{c_4^{(2,0)}}$ and $\smash{c_1^{(1,1)}}$ terms in~\eqref{helilag} drop out, being odd in $\vec n$. In materials that respect spatial parity, we can finally discard the $\smash{c_1^{(1,0)}}$ term.\footnote{The $\smash{c_1^{(1,0)}}$ term is relevant for \emph{cholesteric} liquid crystals whose building blocks possess intrinsic chirality. The mathematics of cholesterics is largely identical to that of helimagnets~\cite{Radzihovsky2011a}.} All in all, the LO Lagrangian for the fluctuations of the order parameter in nematics consists of the $\smash{c_1^{(2,0)}}$, $\smash{c_2^{(2,0)}}$, $\smash{c_3^{(2,0)}}$ and $\smash{c_1^{(0,2)}}$ operators. However, a somewhat different basis of operators with spatial derivatives is more common in the literature, namely
\begin{equation}
\La_\mathrm{eff}^\mathrm{LO}\supset-\frac12\left\{K_1(\skal\nabla n)^2+K_2[\vec n\cdot(\vec\nabla\times\vec n)]^2+K_3[\vec n\times(\vec\nabla\times\vec n)]^2\right\}\;.
\label{liqcrystlag}
\end{equation}
The advantage of this parameterization is that there are field configurations for which all but one of the operators in~\eqref{liqcrystlag} vanish. Global stability of the uniform state with constant $\vev{\vec n(x)}\equiv\vec n_0$ thus requires that all the couplings $K_{1,2,3}$ are positive. For completeness, let me add that they are related to the parameters in~\eqref{helilag} by $\smash{K_1=-2(c_1^{(2,0)}+c_2^{(2,0)})}$, $\smash{K_2=-2(c_1^{(2,0)}+c_3^{(2,0)})}$ and $\smash{K_3=-2c_1^{(2,0)}}$ up to integration by parts.

Unlike in the case of helimagnets, I will not use the Lagrangian~\eqref{liqcrystlag} to analyze the spectrum of excitations of a nematic. Namely, to get a physically correct picture of the dynamics, one would need to consider the interplay of the vector variable $\vec n$ with hydrodynamic degrees of freedom (see Sect.~8.5 of~\cite{Chaikin1995a}). After all, we are talking about \emph{liquids}. Our restricted setting will however be sufficient to contrast the NG degrees of freedom in the nematic and smectic-A phases.

To describe the smectic state, we need an additional order parameter accounting for the layered structure as shown in the right panel of Fig.~\ref{fig:liquidcrystal}. This can be done with a single real scalar field $\p(x)$. The layers of the medium are then defined as surfaces of constant $\p$. The translation across one layer of molecules changes $\p(x)$ by a fixed amount; I will use a normalization of $\p(x)$ where the increment equals $2\pi$. Interestingly, this picture implies the existence of a new, emergent symmetry. Namely, for the sake of identification of the individual strata of the smectic, it does not matter what the attached value of $\p$ is. None of the surfaces of constant $\p$ changes if we shift $\p(x)$ by a constant. The EFT for smectic liquid crystals should therefore be invariant under the internal symmetry $\smash{(\vec n,\p,x^\m)\xrightarrow{\eps}(\vec n,\p+\eps,x^\m)}$. This is our first example of an emergent symmetry arising from the spatial distribution of a medium; I will give a more systematic account of such symmetries in Chap.~\ref{chap:spacetimeclassical}. The complete symmetry group of a smectic is then $G\simeq[\gr{SO}(3)\ltimes\R^4]\times\R$. The corresponding isotropy groups are $H_0\simeq\gr{SO}(3)\times\R$ and $H_{(\psi_0,0)}\simeq\gr{SO}(2)$. The degrees of freedom of the EFT are determined by the coset space $H_0/H_{(\psi_0,0)}\simeq S^2\times\R$.

The values of the two order parameters in the smectic ground state are correlated. As indicated in Fig.~\ref{fig:liquidcrystal}, the axis of alignment $\vec n_0$ of the molecules is perpendicular to the entire stack of parallel layers. Moreover, the individual molecular layers are equidistantly spaced. This implies that the VEV of the gradient $\vev{\vec\nabla\p(x)}$ is a constant nonzero vector, parallel to $\vec n_0$. I will use the notation
\begin{equation}
\vev{\p(\vec x,t)}=k\vec n_0\cdot\vec x\;,
\label{smecticVEV}
\end{equation}
where the parameter $k$ defines the distance between neighboring layers, $2\pi/k$.

The shift symmetry $\R$ guarantees that $\p(x)$ can only enter the effective Lagrangian with derivatives. Being a scalar under rotations, its covariant derivatives are
\begin{equation}
\cd_{\fr0}\p=\de_0\p\;,\qquad
\cd_{\fr r}\p=U(\pi)^s_{\phantom s\fr r}\de_s\p\;,
\label{smecticcovder}
\end{equation}
where $U(\pi)$ is still the matrix related to $\vec n$ by $\vec n(\pi)=U(\pi)\vec n_0$. Also, I used the previously derived coframe~\eqref{helicoframe}, which is not affected by adding the scalar field $\p(x)$. The appearance of these covariant derivatives in the effective Lagrangian is constrained by the isotropy group $H_{(\psi_0,0)}\simeq\gr{SO}(2)$ and the reflection $\vec n\to-\vec n$. To these we should add the discrete symmetry under $\p\to-\p$ which, just like the internal shifts of $\p$, preserves the identification of the layers of the smectic. With these constraints, the contributions of $\p(x)$ to the LO effective Lagrangian are reduced to
\begin{equation}
\La_\mathrm{eff}^\mathrm{LO}\supset c_0(\de_0\p)^2+c_\parallel(\skal n\nabla\p)^2+c_\perp(\vec n\times\vec\nabla\p)^2\;.
\label{smecticbilin}
\end{equation}
In order for the alignment of $\vec\nabla\p$ and $\vec n$ to be energetically preferred, the spatial gradient couplings $c_\parallel$ and $c_\perp$ should satisfy $c_\perp<c_\parallel$.

Let us now turn the argument around and see how the background~\eqref{smecticVEV} affects the vector field $\vec n$. I will set without loss of generality $\vec n_0=(0,0,1)$ so that the two independent fluctuations of $\vec n$ correspond to $n^1$ and $n^2$. This reduces~\eqref{smecticbilin} to
\begin{equation}
\La_\mathrm{eff}^\mathrm{LO}\supset-(c_\parallel-c_\perp)k^2\bigl[(n^1)^2+(n^2)^2\bigr]\;.
\end{equation}
Thanks to the stability constraint on the couplings, $c_\parallel-c_\perp>0$. What we have here is therefore a well-defined, positive-definite mass term for the fluctuations of $\vec n$.

\begin{watchout}%
Let us pause to appreciate the implications of this result. In the nematic phase, we dealt with two NG degrees of freedom, parameterizing via the unit vector $\vec n$ the coset space $S^2$. In the smectic phase, we extended the coset space to $S^2\times\R$. The corresponding new NG variable parameterizes the fluctuations of $\p$ around its VEV~\eqref{smecticVEV}. Given that $\p$ only enters the effective Lagrangian with derivatives, it necessarily couples to a NG mode in the spectrum. However, at the same time, the fluctuations of $\vec n$ received a gap. Adding an extra NG field has \emph{reduced} the number of NG modes from two to one!

To understand what is going on, think of the order parameter~\eqref{smecticVEV} alone. Being a spatial vector, the gradient $\vec\nabla\p$ spontaneously breaks spatial rotations in addition to the internal shift symmetry of $\p$ and translations in the direction of $\vec n_0$. We may then treat $\vev{\vec n(x)}\equiv\vec n_0$ as a secondary order parameter, since it does not break any symmetries that would be left intact by $\vev{\p(x)}$. This makes the $\vec n$ mode redundant: its presence is not required to reproduce the symmetry-breaking pattern of a smectic. Consequently, the fluctuations of $\vec n$ are not protected by symmetry from acquiring a gap.
\end{watchout}

This is an explicit realization of an issue I alerted the reader to already in Chap.~\ref{chap:cosetspacetime}. The number of true, gapless NG degrees of freedom (one) is fixed by the symmetry-breaking pattern. However, the additional, would-be NG modes contained in $\vec n(x)$ are physical and may manifest themselves as low-lying excitations of the smectic ground state. The nematic and smectic phases of liquid crystals are separated by a continuous (second-order) phase transition. At the transition point, the scale $k$ vanishes. Accordingly, near the phase transition, the gap of the $\vec n$-type fluctuations is very small and they need to be included in the low-energy EFT.

Further away from the phase transition, we expect the gapped modes to have little effect on the low-energy physics. In this regime, it should be possible to construct an EFT for smectics in terms of $\p(x)$ alone. This can be arrived at in two different ways. One possibility is to start from a combination of~\eqref{liqcrystlag} and~\eqref{smecticbilin} and integrate out the vector $\vec n$. The other possibility is to construct the EFT for $\p(x)$ alone from scratch. This is arguably more straightforward but less predictive, since the effective couplings of the EFT will be unrelated to those in~\eqref{liqcrystlag} and~\eqref{smecticbilin}. For illustration, I will nevertheless choose the latter option as it highlights the physics of the NG mode in the smectic phase.

The task to find an EFT for a scalar field $\p$, subject to an internal shift symmetry $\R$, copies closely the construction of Sect.~\ref{subsec:twistsuperfluids}, with two differences. The first difference is minor. Namely, unlike in Sect.~\ref{subsec:twistsuperfluids}, we do not impose full Poincar\'e invariance, but merely the Aristotelian symmetry. The resulting effective Lagrangian is, in analogy with~\eqref{relsupefluidaction}, a function of derivatives of $\p(x)$ where spatial indices are contracted in a way preserving invariance under spatial rotations. The second difference is absolutely essential: we want our EFT to stabilize the anisotropic smectic state~\eqref{smecticVEV}. To that end, let us parameterize the fluctuations of $\p(x)$ by a NG field $\pi(x)$ such that $\p(\vec x,t)=k\vec n_0\cdot\vec x+\pi(\vec x,t)$. The basic rotationally invariant building block for the construction of the Lagrangian is $\smash{(\vec\nabla\p)^2=k^2+2k\vec n_0\cdot\vec\nabla\pi+(\vec\nabla\pi)^2}$. Dropping the constant piece, the part of the effective Lagrangian with one derivative per field will be a generic function of $\de_0\pi$ and $k\vec n_0\cdot\vec\nabla\pi+(\vec\nabla\pi)^2/2$. The static part of the effective Lagrangian, bilinear in the NG field $\pi(x)$, is thus contained in
\begin{equation}
\La_\mathrm{eff}\supset c_1\left[k\vec n_0\cdot\vec\nabla\pi+\frac{(\vec\nabla\pi)^2}2\right]+c_2\left[k\vec n_0\cdot\vec\nabla\pi+\frac{(\vec\nabla\pi)^2}2\right]^2+\dotsb\;.
\label{smecticEFT}
\end{equation}
The ellipsis includes operators that are bilinear in $\pi(x)$ but contain more than one derivative per field. The leading operator of this kind is $c_3(\vec\nabla^2\p)^2=c_3(\vec\nabla^2\pi)^2$.

Remarkably, the $c_1$ operator is forbidden. The easiest way to see this is to think of the gradient $\vec\nabla\p$ as a vector field, $\vec A\equiv\vec\nabla\p$. The part of the static Lagrangian with one derivative per $\p$ is then a mere function of $\vec A$. Should the corresponding Hamiltonian density have a minimum for nonzero $\vev{\vec A}=k\vec n_0$ as implied by~\eqref{smecticVEV}, its Hessian matrix at the minimum must have two zero modes. This is just the statement of the Goldstone theorem in terms of the eigenvalues of the mass matrix, which we proved back in Sect.~\ref{sec:firstmodelnonAbelian}. Hence, for the smectic state~\eqref{smecticVEV} to be stable within the EFT, the Lagrangian~\eqref{smecticEFT} must not contain any terms quadratic in the part of the gradient of $\pi$, perpendicular to $\vec n_0$. It is the stability criterion together with the spontaneously broken rotational invariance that forbids the $c_1$ operator.

We conclude that the leading contributions to the bilinear Lagrangian consist of the operators $(\de_0\pi)^2$, $(\vec\nabla_\parallel\pi)^2$ and $(\vec\nabla_\perp^2\pi)^2$, where $\parallel$ and $\perp$ denote projections to directions parallel and transverse to $\vec n_0$. This suggests an unusual, anisotropic power counting. For all these operators to contribute at the same order, the time derivative should have the same counting degree as $\vec\nabla_\parallel$ and $\vec\nabla_\perp^2$. The presence of a higher power of the transverse gradient has a profound impact on the stability of the smectic phase at nonzero temperature. I will relegate a more detailed discussion and further generalization of this observation to Chap.~\ref{chap:topicsnotcovered}.


\subsection{The Unphysical: Nonrelativistic Superfluids}
\label{subsec:vectorunphysical}

The final example in our exploration of the role of vector modes will be nonrelativistic superfluids. Here we do know a priori that it is possible to construct an EFT solely in terms of a scalar degree of freedom. All one has to do is to take the EFT for a relativistic superfluid, developed in Sect.~\ref{subsec:twistsuperfluids}, and perform a nonrelativistic limit. It is however instructive to carry out the construction of the nonrelativistic version of the EFT from scratch.

To get started, recall the basic properties of Galilei symmetry from Sect.~\ref{subsec:spacetimexSchr}. The Lie algebra of the Galilei group in $d$ spatial dimensions includes the generators $P_{\fr r}$ and $H$ of space and time translations and $\smash{J_{\fr r\fr s}}$ of spatial rotations. In addition, there is a vector generator $K_{\fr r}$ of Galilei boosts. I will add from the outset a central charge $Q$ whose eigenvalues measure nonrelativistic mass. This turns the Galilei group into the Bargmann group. Altogether, the nontrivial part of the local structure of the symmetry group is determined by the commutation relations
\begin{equation}
[H,K_{\fr r}]=\I P_{\fr r}\;,\qquad
[P_{\fr r},K_{\fr s}]=\I\d_{\fr r\fr s}Q\;.
\label{Galileicommutators}
\end{equation}
The commutators of $J_{\fr r\fr s}$ are fixed by rotation invariance and will not be needed. All the other commutators not listed here are vanishing. Mathematically, the Bargmann group has the structure $G\simeq\gr{SO}(d)\ltimes\{\R^d_K\ltimes[\R^D\times\gr{U}(1)_Q]\}$. All the symmetry transformations but translations leave the spacetime origin fixed, hence its isotropy group is $H_0\simeq[\gr{SO}(d)\ltimes\R^d_K]\times\gr{U}(1)_Q$. The translations themselves will be represented by the operator $\tran{\vec x,t}\equiv\E^{\I tH}\E^{\I\skal xP}$.

In order to describe the superfluid order, we need an order parameter that carries a nontrivial action of the internal $\gr{U}(1)_Q$ subgroup. Similarly to relativistic superfluids, the simplest choice is a complex scalar field $\psi$. For any $\psi_0\neq0$, the corresponding isotropy group is then $H_{(\psi_0,0)}\simeq\gr{SO}(d)\ltimes\R^d_K$. However, as pointed out before, this is not satisfactory, since it violates our requirements on the decomposition~\eqref{isotangent} of the Lie algebra $\lie g$ of $G$. Namely, the operators $P_{\fr r}$ and $H$ span a basis of $\lie g/\lie h_0$ but do not carry a representation of $H_{(\psi_0,0)}$. The way out is to add a secondary order parameter $A^\m\equiv(A^0,\vec A)$ that transforms as a vector under Galilei boosts. Choosing $A^\m_0=(a,\vec0)$ with $a\neq0$ can be interpreted as specifying the density of the $\gr{U}(1)_Q$ charge in the rest frame of the superfluid. It reduces the isotropy subgroup further to $H_{((\psi_0,A_0),0)}\simeq\gr{SO}(d)$. The coset space $\smash{H_0/H_{((\psi_0,A_0),0)}\simeq\R^d_K\times\gr{U}(1)_Q}$ can be parameterized by NG variables $\pi$ and $\x^r$ through $U(\pi,\vec\x)\equiv\E^{\I\pi Q}\E^{\I\skal\x K}$. In this parameterization, a Galilei boost with velocity $\vec v$ acts on the NG variables and the spacetime coordinates as
\begin{equation}
\E^{\I\skal vK}(\pi,\vec\x,\vec x,t)=(\pi+\skal vx+\vec v^2t/2,\vec\x+\vec v,\vec x+\vec vt,t)\;.
\label{Galileiboost}
\end{equation}

With all the pieces at hand, we can now compute the MC form~\eqref{MCspacetime},
\begin{equation}
\mc(\pi,\vec\x,\vec x,t)=Q[\D\pi-\vec\x\cdot\D\vec x+(1/2)\vec\x^2\D t]+\vec K\cdot\D\vec\x+H\D t+\vec P\cdot(\D\vec x-\vec\x\D t)\;.
\label{GalileiMC}
\end{equation}
The $\mcu$ part of the MC form is trivial. From $\mcp=H\D t+\vec P\cdot(\D\vec x-\vec\x\D t)$, we extract the spacetime coframe, $\vec e^{*\fr0}=\D t$ and $\smash{\vec e^{*\fr r}=\d^{\fr r}_s(\D x^s-\x^s\D t)}$. The $\d^{\fr r}_s\x^s\D t$ term in $\vec e^{*\fr r}$ does not affect the spacetime volume form, we can thus use the volume measure $\D^d\!\vec x\,\D t$ to turn effective Lagrangians into effective actions. To construct covariant derivatives of the NG fields, we will also need the spacetime frame, dual to the above coframe,
\begin{equation}
e^0_{\fr 0}=1\;,\qquad
e^r_{\fr 0}=\x^r\;,\qquad
e^0_{\fr r}=0\;,\qquad
e^s_{\fr r}=\d^s_{\fr r}\;.
\label{Galileiframe}
\end{equation}
The covariant derivatives of $\pi$ and $\x^r$ are then extracted from the $\mcb$ part of the MC form~\eqref{GalileiMC} with the help of~\eqref{spacetimecovderpi},
\begin{equation}
\begin{alignedat}{2}
\cd_{\fr0}\pi&=\de_0\pi+\skal\x\nabla\pi-\vec\x^2/2\;,\qquad&
\cd_{\fr r}\pi&=\d^s_{\fr r}(\de_s\pi-\x_s)\;,\\
\cd_{\fr0}\x^r&=(\de_0+\skal\x\nabla)\x^r\;,\qquad&
\cd_{\fr s}\x^r&=\d^u_{\fr s}\de_u\x^r\;.
\end{alignedat}
\label{Galileicovder}
\end{equation}
It is an easy exercise to check explicitly that all these covariant derivatives are invariant under the Galilei boost~\eqref{Galileiboost}. Thanks to their simple form, it is not necessary to dis\-tin\-guish frame and coordinate-basis indices. In the following, I will therefore happily use $\cd_r\pi=\de_r\pi-\x_r$ and $\cd_s\x^r=\de_s\x^r$.

\begin{watchout}%
In accord with the general machinery of nonlinear realizations, we could also add matter fields, organized in linear multiplets of $H_{((\psi_0,A_0),0)}\simeq\gr{SO}(d)$. By~\eqref{spacetimecovderpsi}, the covariant derivatives of a matter field $\mf^\vr$ would then be $\smash{\cd_0\mf^\vr=(\de_0+\skal\x\nabla)\mf^\vr}$ and $\smash{\cd_r\mf^\vr=\de_r\mf^\vr}$. This opens the possibility to promote (almost) any nonrelativistic theory with Aristotelian symmetry to a theory that is Galilei-invariant. Indeed, recall that in Part~\ref{part:internalSSB} of the book, we dealt largely with Aristotelian EFTs where we imposed invariance under linearly realized rotations by hand. What prevents us from introducing an auxiliary vector field $\x^r$ and replacing $\de_0\to\cd_0\equiv\de_0+\skal\x\nabla$ everywhere? This will certainly work for strictly invariant Lagrangians. (The naive replacement $\de_0\to\cd_0$ might spoil the quasi-invariance of Lagrangians that shift upon a symmetry transformation by a total time derivative.) The only fly in the ointment is that we do not have a physical interpretation for the $\x^r$ field in terms of the NG degrees of freedom of the Aristotelian EFT. Rather, $\x^r$ should be viewed as the local velocity of a medium carrying the NG modes; the combination $\de_0+\skal\x\nabla$ is often referred to as the \emph{material derivative}. The Aristotelian EFTs constructed in Chap.~\ref{chap:effLagrangian} describe the low-energy physics in the rest frame of the medium. The same remark applies to relativistic systems. There, adding a NG field for Lorentz boosts allows one to promote an Aristotelian theory to one invariant under the full Poincar\'e group.
\end{watchout}

With the basic building blocks at hand, we can proceed to the construction of an EFT for the superfluid. Following closely the relativistic counterpart in Sect.~\ref{subsec:twistsuperfluids}, we first fix a power-counting scheme. We again count both $\de_0\pi$ and $\vec\nabla\pi$ as degree zero. For the covariant derivatives in~\eqref{Galileicovder} to have a consistent counting degree, we have to assign $\x^r$ degree zero as well. Any other derivative acting on $\pi$ or $\x^r$ will then have degree one. It follows that at the LO of the derivative expansion, we can ignore $\cd_\m\x^r$. The LO Lagrangian will be some function of the rotationally invariant operators $\cd_0\pi$ and $\d^{rs}\cd_r\pi\cd_s\pi$. This underlines the fact that the field $\x^r$ is merely auxiliary and does not represent independent degrees of freedom. Indeed, it only enters the LO Lagrangian without any derivatives and can be eliminated algebraically by imposing its EoM,
\begin{equation}
(\de_r\pi-\x_r)\left[\PD{\La_\mathrm{eff}^{(0)}}{(\cd_0\pi)}-2\PD{\La_\mathrm{eff}^{(0)}}{(\d^{su}\cd_s\pi\cd_u\pi)}\right]=0\;.
\label{IHCeqEoM}
\end{equation}
The natural solution is $\vec\x=\vec\nabla\pi$, which turns the two building blocks for the LO Lagrangian to $\cd_0\pi\to\de_0\pi+(\vec\nabla\pi)^2/2$ and $\cd_r\pi\to0$. Of course, \eqref{IHCeqEoM} can also be satisfied if the expression in square brackets vanishes. This would imply that the LO Lagrangian only depends on $\cd_0\pi+(1/2)\d^{rs}\cd_r\pi\cd_s\pi=\de_0\pi+(\vec\nabla\pi)^2/2$. Either way, we end up with the same LO Lagrangian in terms of $\pi$ alone.

Finally, recall the superfluid ground state can be described by a time-dependent VEV of a complex scalar field. This amounts to $\vev{\pi(x)}=-\m t$ where $\m$ is the chemical potential. The fluctuations around this background can be parameterized by a field $\tilde\pi$ such that $\pi(x)=\tilde\pi(x)-\m t$. Following again the analogy with Sect.~\ref{subsec:twistsuperfluids}, we conclude that the LO effective Lagrangian for Galilei-invariant superfluids reads
\begin{equation}
\La_\mathrm{eff}^{(0)}[\tilde\pi]=P\left(\m-\de_0\tilde\pi-(\vec\nabla\tilde\pi)^2/2\right)\;.
\label{GalileisuperfluidLO}
\end{equation}
The function $P(\m)$ represents the thermodynamic pressure of the superfluid in equilibrium at zero temperature. 

\begin{illustration}%
As a simple application of our new EFT, expand the Lagrangian~\eqref{GalileisuperfluidLO} to second order in the NG field $\tilde\pi$. Up to a surface term, we find
\begin{equation}
\La_\mathrm{eff}^{(0)}[\tilde\pi]\simeq P(\m)+\frac{P''(\m)}2(\de_0\tilde\pi)^2-\frac{P'(\m)}2(\vec\nabla\tilde\pi)^2+\dotsb\;.
\end{equation}
This gives immediately the phase velocity of the NG boson,
\begin{equation}
v=\sqrt{\frac{P'(\m)}{P''(\m)}}=\sqrt{\OD Pn}\;,
\end{equation}
where $n(\m)\equiv P'(\m)$ should be interpreted as the mass density of the superfluid. This matches the relativistic result $v=\sqrt{\Od PU}$ in~\eqref{superfluidvelocity}, since in the nonrelativistic limit, the energy density $U(\m)$ reduces to the density of the rest mass. Note that in spite of the appearance, the EFT~\eqref{GalileisuperfluidLO} describes a type-A NG mode with a linear dispersion relation. This justifies a posteriori the use of the power-counting scheme where temporal and spatial derivatives are counted equally.
\end{illustration}

Before closing the discussion of nonrelativistic superfluids, I have a debt to pay off. Namely, when constructing the EFT, I tacitly assumed the effective Lagrangian to be strictly invariant. Could the Lagrangian also include some quasi-invariant contributions? The answer to this question is surprisingly rich, but luckily will not affect the LO Lagrangian~\eqref{GalileisuperfluidLO}.

As pointed out in Sect.~\ref{sec:effLagstructure}, quasi-invariant Lagrangians for broken internal symmetries are classified by relative Lie algebra cohomology. It is not clear whether this elegant result survives all the complications of nonlinear realization of spacetime symmetries. Let us nevertheless take a leap of faith and see where it will lead us. We thus seek $(D+1)$-forms that are closed and $H_0$-invariant. I will first split the MC form~\eqref{GalileiMC} into components corresponding to the individual generators,
\begin{equation}
\begin{alignedat}{2}
\mc_Q&=\D\pi-\vec\x\cdot\D\vec x+(1/2)\vec\x^2\D t\;,\qquad&
\mc_K^r&=\D\x^r\;,\\
\mc_H&=\D t\;,\qquad&
\mc_P^r&=\D x^r-\x^r\D t\;.
\end{alignedat}
\end{equation}
We are now to take the exterior product of $D+1$ of these 1-forms. Invariance under the whole isotropy group $H_0$ is ensured by contracting vector indices in a rotationally invariant manner. It is the closedness requirement that poses a challenge. To that end, we take note of the MC equations, reflecting the commutation relations~\eqref{Galileicommutators} between the generators,
\begin{equation}
\D\mc_Q=\d_{rs}\mc_P^r\w\mc_K^s\;,\quad
\D\mc_K^r=0\;,\quad
\D\mc_H=0\;,\quad
\D\mc_P^r=\mc_H\w\mc_K^r\;.
\end{equation}
These suggest a set of $d+1$ naturally closed, rotationally invariant $(D+1)$-forms,
\begin{align}
\MC_{D+1}^{(n)}&\equiv\frac1{d!}\ve_{r_1\dotsb r_d}\mc_H\w\mc_Q\w\mc_K^{r_1}\w\dotsb\w\mc_K^{r_n}\w\mc_P^{r_{n+1}}\w\dotsb\w\mc_P^{r_d}\\
\notag
&=\frac1{d!}\ve_{r_1\dotsb r_d}\D t\w(\D\pi-\vec\x\cdot\D\vec x)\w\D\x^{r_1}\w\dotsb\w\D\x^{r_n}\w\D x^{r_{n+1}}\w\dotsb\w\D x^{r_d}\;,
\end{align}
where $n=0,\dotsc,d$. The corresponding effective Lagrangians are equivalent to the $D$-form potentials of these $(D+1)$-forms, that is such $\smash{\MC_D^{(n)}}$ that $\smash{\D\MC_D^{(n)}=\MC_{D+1}^{(n)}}$. These $D$-form potentials are obviously not unique, since they can be shifted without consequences by the exterior derivative of any $(D-1)$-form. A suitable choice is
\begin{align}
\label{GalWZ1}
\MC_D^{(n)}=-\frac1{d!}\ve_{r_1\dotsb r_d}\D t&\w\biggl[\pi\D\x^{r_1}\w\dotsb\w\D\x^{r_n}\w\D x^{r_{n+1}}\w\dotsb\w\D x^{r_d}\\
\notag
&+\frac{n}{2(d-n+1)}\vec\x^2\D\x^{r_1}\w\dotsb\w\D\x^{r_{n-1}}\w\D x^{r_n}\w\dotsb\w\D x^{r_d}\biggr]\;.
\end{align}
A reader wishing to check this may find the following identity useful,
\begin{equation}
\begin{split}
\ve_{r_1\dotsb r_d}&\D\x^s\w\D\x^{r_1}\w\dotsb\w\D\x^{r_{n-1}}\w\D x^{r_n}\w\dotsb\w\D x^{r_d}\\
&=-\frac{d-n+1}n\ve_{r_1\dotsb r_d}\D x^s\w\D\x^{r_1}\w\dotsb\w\D\x^{r_n}\w\D x^{r_{n+1}}\w\dotsb\w\D x^{r_d}.
\end{split}
\end{equation}
Next, we eliminate the auxiliary field $\x^r$ by imposing the same constraint as above, $\vec\x=\vec\nabla\pi$. After some manipulation, the $D$-form potentials~\eqref{GalWZ1} then give a set of $d+1$ quasi-invariant Lagrangians (with adjusted overall normalization),
\begin{equation}
\La_\mathrm{quasi}^{(n)}=\ve^{r_1\dotsb r_nu_{n+1}\dotsb u_d}\ve^{s_1\dotsb s_n}_{\phantom{s_1\dotsb s_n}u_{n+1}\dotsb u_d}\pi\prod_{i=1}^n(\de_{r_i}\de_{s_i}\pi)\;.
\label{GalWZ2}
\end{equation}

The form of the Lagrangians~\eqref{GalWZ2} should ring a bell. Indeed, except for replacing spacetime indices with spatial ones, these are exactly the Galileon Lagrangians~\eqref{galWZ}. The similarity is not accidental. The generators $Q,K_\m$ of the constant and linear parts of the Galileon symmetry~\eqref{symgal} and the energy--momentum operator $P_\m$ satisfy the commutation relation $[P_\m,K_\n]=\I g_{\m\n}Q$. This is identical to the centrally extended commutator in the Bargmann algebra. We have thus killed two birds with one stone, explaining the origin of the Galileon Lagrangians as a byproduct of our discussion of Galilei-invariant superfluids. Now back to the physical content of~\eqref{GalWZ2}. The $n=0$ Lagrangian, $\smash{\La_\mathrm{quasi}^{(0)}}$, is a tadpole operator. This must be dropped to keep the EFT stable. The $n=1$ Lagrangian is equivalent to $(\vec\nabla\pi)^2$ by partial integration. This is, in spite of appearance, already included in the LO effective Lagrangian~\eqref{GalileisuperfluidLO}. Namely, it is a part of $\de_0\pi+(\vec\nabla\pi)^2/2$ where $\de_0\pi$ itself is a surface term. Finally, all the quasi-invariant Lagrangians~\eqref{GalWZ2} with $n\geq2$ contain more than one derivative per field. Consequently, they will only contribute at higher, subleading orders of the derivative expansion of the superfluid EFT.


\subsection{Inverse Higgs Constraints}
\label{subsec:vectorIHC}

The analysis of concrete examples has given us enough experience to appreciate the broader spectrum of subtleties associated with spontaneous breaking of spacetime symmetries. Let us reevaluate what we have learned in particular about the role of vector fields. I should warn the reader that this matter is not yet quite settled. Hence, the content and organization of the discussion below is necessarily more subjective than elsewhere in the book.

In case of internal symmetries, the situation is clear. There is a one-to-one mapping between broken symmetry generators and NG fields in the low-energy EFT. The construction of the EFT itself is fairly streamlined. All one needs to know is the symmetry-breaking pattern. As shown in Chap.~\ref{chap:effLagrangian}, finding the EFT then boils down to solving certain linear-algebraic constraints for the tensor couplings in the effective Lagrangian. The functional dependence of the Lagrangian on the NG fields is completely fixed by symmetry.

\begin{figure}[t]
\sidecaption[t]
\includegraphics[width=2.9in]{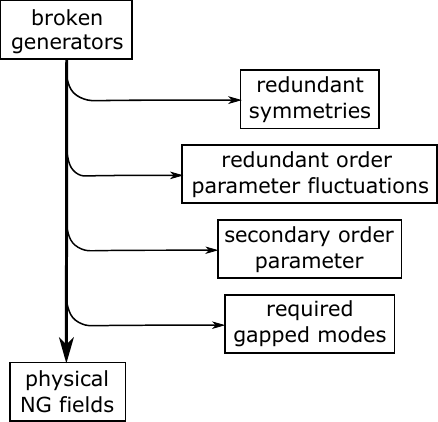}
\caption{Various mechanisms whereby the number of physical NG fields coupling to gapless, true NG modes can be lower than the number of generators parameterizing the coset space $H_0/H_{(\psi_0,0)}$. The detailed meaning of the keywords involved is explained in the main text. The physical NG fields are in turn in a one-to-one (type-A) or two-to-one (type-B) correspondence with the NG modes in the spectrum}
\label{fig:bigpictureimproved}
\end{figure}

In case of spontaneously broken spacetime symmetry, we would hope to be able to do the same. This desire lies behind the \emph{agnostic nonlinear realization} that I already mentioned in Sect.~\ref{subsec:spacetimediscussion}. In this framework, one starts with a separate NG field for each spontaneously broken symmetry generator. The actions of all the generators in an abstract group space are by construction independent of each other. This however leads to the surprising observation that the number of fields that ultimately couple to gapless excitations may be lower than that of the broken generators.\footnote{Remember that these independent NG fields may still not be in a one-to-one correspondence with NG modes in the spectrum. Some of the fields may be pairwise canonically conjugated, leading to type-B NG bosons. The canonical conjugation mechanism is the same for internal and spacetime symmetries, and thus requires no further discussion.} I will now outline the various mechanisms how the naively expected number of NG fields based on mere counting broken generators may be reduced. This will add nuance to the intuitive picture painted in Sect.~\ref{sec:NGintuitive} and summarized in Fig.~\ref{fig:bigpicture} therein. The different mechanisms are listed in Fig.~\ref{fig:bigpictureimproved}.

\runinhead{Redundancy of Symmetries} Two symmetries that are locally equivalent necessarily lead to identical fluctuations of the order parameter. This is the simplest and most important of all the mechanisms, to which I already devoted the whole Chap.~\ref{chap:differences}. Its advantage is that such a redundancy can be checked at the level of classical symmetry transformations. No specific choice of an order parameter or knowledge of the corresponding symmetry-breaking pattern is needed. See Sect.~\ref{sec:diffexamples} for examples.

\runinhead{Redundancy of Order Parameter Fluctuations} Moving down the decision tree, the next possibility is that some symmetries are locally distinct, yet lead to equivalent fluctuations of the order parameter. An example of this type is the helimagnet; see \refex{ex:helimagnet} for details. Here spatial translations and rotations are not trivially redundant due to the vector nature of the order parameter. However, thanks to the particular spatial modulation of the order parameter, three independent broken symmetry generators lead to mere two distinct types of fluctuations. This type of redundancy requires detailed knowledge of the order parameter, but can still be detected straightforwardly following the approach of Sect.~\ref{subsec:NGclassificationfluctuations}.

Both of the above types of redundancy are automatically accounted for by the standard nonlinear realization of Chap.~\ref{chap:cosetspacetime}. This ensures a one-to-one parameterization of whatever fields carrying the action of the symmetry one starts with. However, it is sometimes advantageous to add auxiliary would-be NG fields, even if these are known beforehand not to correspond to separate physical degrees of freedom. We saw an example in Sect.~\ref{subsec:vectorunphysical}. The unphysical nature of the vector field $\x^r$ therein manifested itself in the fact that we could eliminate it algebraically using its EoM. Such a procedure, while physically certainly correct, may however become exceedingly cumbersome beyond LO of the derivative expansion. There is an alternative that allows one to eliminate $\x^r$ algebraically \emph{before} one starts constructing the effective Lagrangian~\cite{Ivanov1975a}. Indeed, note that the covariant derivative $\cd_r\pi$ in~\eqref{Galileicovder} is linear in $\x^r$ without any derivatives on it. Setting it to zero gives $\vec\x=\vec\nabla\pi$, which is an algebraic condition that is covariant under all the symmetries of the EFT. This is an example of an IHC. Imposing the IHC brings us back to the nonlinear realization of Galilei symmetry solely in terms of the scalar $\pi$. Yet, invariance under Galilei boosts is still automatically implemented. Galilei-invariant Lagrangians can be constructed using the same building blocks~\eqref{Galileicovder} while imposing by hand just invariance under spatial rotations. The latter can be ensured using standard tensor methods.

\begin{watchout}%
The auxiliary field $\x^r$ in nonrelativistic superfluids is associated with Galilei boosts. In this concrete case, the corresponding would-be NG modes are clearly unphysical. However, it seems to be true generally that there are no observed single-particle NG states for which boosts would be the sole associated broken symmetry. Curiously, the Goldstone theorem for boosts can be saturated by states from a continuum spectrum rather than by stable bosonic quasiparticles, as in Fermi liquids~\cite{Rothstein2018a,Alberte2020}. A class of mathematically consistent EFTs that feature NG bosons originating in boost symmetry has been proposed in~\cite{Nicolis2015a}.
\end{watchout}

Eliminating the auxiliary vector field using its EoM or using an IHC must lead to EFTs with the same functional form. The only difference is possibly in the identification of the effective couplings, but these are usually determined subsequently by matching the predictions of the EFT to selected observables. This makes the two approaches entirely equivalent. The advantage of using the classical EoM (or generally, at loop level, integrating the auxiliary field out) is that it is physically transparent. The IHC, on the other hand, is usually technically simpler. The drawback is that there is no universal algorithm how to find a suitable IHC. As a rule, one has to follow intuition combined with insight in the concrete problem at hand. See~\cite{Finelli2020} for a discussion of some of the traps involved.

\runinhead{Gapped Modes from Secondary Order Parameter} Eliminating the redundancy of either type, we are back to the standard nonlinear realization of spacetime symmetry, developed in Sect.~\ref{sec:spacetimestandard}. Here all the fields, whether NG fields $\pi^a$ or matter fields $\mf^\vr$, describe independent, physical degrees of freedom. Barring possible NG modes of spontaneously broken translation symmetry, the NG variables are classified as coordinates on the coset space $H_0/H_{(\psi_0,0)}$. However, there may be different choices of the order parameter(s) leading to the same symmetry-breaking pattern but different coset spaces $H_0/H_{(\psi_0,0)}$, hence different numbers of NG variables. This possibility owes its existence to the distinction between the isotropy groups $H_0$, $H_{(\psi_0,0)}$ and the unbroken subgroup $H_\vp$ in case of spacetime symmetries.

A phenomenologically relevant example is the smectic liquid crystal, discussed in Sect.~\ref{subsec:vectorirrelevant}. (Another, relativistic example of this type can be found in~\cite{Kharuk2018}.) Let us inductively generalize what we observed therein. Suppose there is a secondary order parameter that can be removed from the EFT without affecting the symmetry-breaking pattern. Then the associated NG variables are not required to be present by the broken symmetry. By the same token, the corresponding modes in the spectrum are not protected by the symmetry from acquiring a gap.\footnote{A similar phenomenon exists even for internal symmetries. As stressed in Sect.~\ref{subsec:EoMgeometry}, the primary order parameter responsible for type-B NG modes in the spectrum is the commutator matrix~\eqref{commutatormatrixsigmaab}. The presence of an additional, secondary order parameter may lead to gapped ``partners'' of the type-B NG modes. See~\cite{Hayata2014b} for a detailed discussion within Hamiltonian formalism, or~Sect.~\ref{subsec:EoMspectrum} for an outline of how these gapped partner states manifest themselves in the effective Lagrangian.}

Here one has two choices how to go about the construction of the EFT. If it is possible to identify the secondary order parameter beforehand, one can drop it from the outset. The resulting \emph{minimal} nonlinear realization of the given symmetry-breaking pattern then only includes genuine NG fields. Accordingly, all modes described by the EFT are gapless, that is true NG bosons. Alternatively, one may proceed with the nonlinear realization that includes the secondary order parameter. The extra would-be NG fields are then dynamical and couple to gapped modes in the spectrum. If desired, these fields can be eliminated either by using their EoM or by imposing a suitable IHC.

\begin{illustration}%
Let us have a closer look at smectic liquid crystals (Sect.~\ref{subsec:vectorirrelevant}). Here the primary order parameter is $\vev{\p(\vec x,t)}=k\vec n_0\cdot\vec x$. This breaks rotations that do not preserve the axis of alignment defined by $\vec n_0$. Moreover, it breaks translations along $\vec n_0$ together with the emergent shift symmetry of $\p(x)$ down to the ``diagonal'' subgroup. Altogether, there are three spontaneously broken symmetry generators. One of these corresponds to the genuine NG mode described by the EFT~\eqref{smecticEFT}. In addition, there are two gapped modes that can be viewed as fluctuations of the secondary order parameter, $\vev{\vec n(x)}=\vec n_0$.

Suppose we start from an EFT including all three degrees of freedom. This might even be desirable near the nematic--smectic phase transition at which the modes excited by the vector field $\vec n(x)$ become gapless. However, should we decide to eliminate the gapped modes, we need a covariant IHC. To that end, note that the covariant derivative $\cd_{\fr r}\p$ in~\eqref{smecticcovder} splits under the action of the isotropy group $H_{(\psi_0,0)}\simeq\gr{SO}(2)$ into a singlet, $\cd_{\fr\a}\p$, and a doublet, $\cd_{\fr a}\p$. The condition $\cd_{\fr a}\p=0$ is covariant under all the symmetries of the smectic and moreover is algebraic in $\pi^a$. In a somewhat more human notation, it is equivalent to $\vec n\times\vec\nabla\p=0$. The solution to this IHC is, up to overall sign, $\vec n=\vec\nabla\p/\abs{\vec\nabla\p}$.

In fact, the IHC can do more than to eliminate unwanted degrees of freedom. According to~\eqref{covdertransfo}, covariant derivatives of NG fields $\pi^a$ extracted from the $\mcb$ part of the MC form transform exactly as matter fields with respect to some linear representation of $H_{(\psi_0,0)}$. The very fact that the constraint $\cd_{\fr a}\p=0$ can be solved for $\pi^a$ means that the Jacobian matrix $\Pd{(\cd_{\fr a}\p)}{\pi^b}$ is nonsingular, at least around the reference point $\pi^a=0$. Instead of setting it to zero, we can therefore treat $\cd_{\fr a}\p\equiv\mf_{\fr a}$ as a new field variable that replaces $\pi^a$ in the EFT. We conclude that the gapped degrees of freedom are indistinguishable from ordinary matter fields. This is the underlying reason why they can be removed from the EFT without spoiling the nonlinear realization of the broken symmetry. And for the same reason, it is no wonder that the broken symmetry does not force them to remain gapless.
\end{illustration}

The same reasoning applies to Galilei-invariant superfluids (Sect.~\ref{subsec:vectorunphysical}). Even if one does not wish to eliminate the auxiliary field $\x^r$ from the EFT, one can still trade it for $\vec\mf\equiv\vec\x-\vec\nabla\pi$. This is invariant under Galilei boosts and transforms as a matter field in the vector representation of the linearly realized isotropy group $\gr{SO}(d)$. The observation that whenever a set of would-be NG fields can be eliminated from the EFT by imposing some IHCs, they can instead be kept and traded for a set of matter fields, is general~\cite{Brauner2014a}. The change from the would-be NG fields to the matter fields is a field redefinition that is nonsingular. This is guaranteed by the fact that the IHCs are algebraically solvable. 

The distinction between the three mechanisms discussed so far crucially relies on the type of order parameter breaking the symmetry of the system. Within a naive EFT based on the agnostic nonlinear realization, there is therefore no way to distinguish these scenarios. One has to accept the possibility that some would-be NG degrees of freedom of the EFT couple to gapped states. Within the EFT itself, it is not possible to tell whether such states are physical or mere artifacts of the EFT setup. This suggests that when it comes to spacetime symmetries, the symmetry-breaking pattern itself is not restrictive enough. To get more detailed insight, one needs additional low-energy data such as the order parameter. See~\cite{Nicolis2013b,Hidaka2015a} for further discussion.

\runinhead{Gapped Modes Required by Order Parameter} So far, generators not leading to independent NG degrees of freedom could always be identified by inspecting the classical symmetry transformations and the order parameter(s). Interestingly, that is not the end of the story. There are theories featuring would-be NG fields that couple to gapped modes which cannot be discarded by dropping a secondary order parameter. See~\cite{Endlich2014a} for an example and more details. It is then still possible to eliminate such fields using either the EoM or a suitably chosen IHC. One should however expect the resulting minimal nonlinear realization of symmetry to involve generalized local transformations that depend on derivatives of the NG fields. As long as one insists on using solely point symmetries, it is not possible to recover the minimal nonlinear realization by reducing the order parameter. In this sense, the presence of the gapped modes is required for a description of the symmetry-breaking pattern based on a local order parameter.

In all the scenarios outlined above and summarized in Fig.~\ref{fig:bigpictureimproved}, the IHCs play a prominent role. It is therefore important to stress that they are just a technical tool. We always have the option to work with the full EFT based either on the agnostic nonlinear realization or the standard nonlinear realization of Chap.~\ref{chap:cosetspacetime}. We just have to live with the possibility that the EFT may include some gapped modes. These of course become irrelevant at sufficiently low energies. The set of genuine NG degrees of freedom is fixed unambiguously by the symmetry-breaking pattern.

How to choose the IHC in practice is a kind of art in its own right. The examples I presented above were all mercifully simple. Nonetheless, there is a useful rule of thumb rooted in the expression~\eqref{MCspacetime} for the MC form. Namely, if the commutator $[P_{\fr\m},Q_a]$ contains $Q_b$, then the NG field $\pi^a$ appears linearly and without derivatives in the $\mc^b$ component of the MC form. It therefore seems possible to eliminate $\pi^a$ by setting $\mc^b$ (or the irreducible multiplet of components of the MC form containing it) to zero. This naive algorithm however requires further conditions to work~\cite{Klein2017a}, and moreover depends sensitively on the exact choice of field parameterization~\cite{McArthur2010a}. It is therefore comforting to know that whether or not one chooses to impose some IHCs has no effect on the low-energy physics of the theory.


\section{Genuine Breaking of Translation Invariance}
\label{sec:quantumtranslation}

Until now I assumed that whatever the pattern of symmetry breaking, the low-energy physics is captured by an EFT that lives on the coset space $H_0/H_{(\psi_0,0)}$. As we saw, this is appropriate for systems where the value of the order parameter at all spacetime points lies on the same orbit of the symmetry group. Any matter fields $\mf^\vr$ that might be present can then be discounted. In this last section of the chapter, I will briefly outline the changes that are necessary to account for fluctuations of generic coordinate-dependent order parameters. The nonlinear realization of spacetime symmetry developed in Chap.~\ref{chap:cosetspacetime} remains valid even in this general case. This is because the construction does not depend on the order parameter as a function on spacetime, merely on the choice of the representative point $\psi_0\in\M$. By the same token, everything said in Sect.~\ref{sec:quantumbuildingblocks} about the MC form and covariant derivatives still applies.

The new complication is that we now have to explicitly include the matter fields $\mf^\vr(x)$ in the game. These may contain NG degrees of freedom that are not detected by the $H_0/H_{(\psi_0,0)}$ coset space, namely those of spontaneously broken translations. The challenge we face is to find a parameterization of the matter fields that makes the existence of the translation NG modes manifest. As of writing this book, there does not seem to be a general parameterization that would do the job in presence of other, possibly spontaneously broken, symmetries. I will thus merely sketch the basic idea in the simplest case of spacetime symmetry breaking by an otherwise featureless order parameter.


\subsection{One-Dimensional Modulation of the Order Parameter}
\label{subsec:translation1d}

To further simplify the problem, consider a single real scalar field $\p(x)$ in a theory invariant under a purely spacetime symmetry group $G$. This may be the Aristotelian group or one of its extensions including boosts such as the Galilei or Poincar\'e group. The isotropy group $H_0$ contains the transformations that act linearly on spacetime coordinates, that is rotations and boosts. For any choice of $\p_0\in\R$, the isotropy group $H_{(\p_0,0)}\simeq H_0$ makes the coset space $H_0/H_{(\p_0,0)}$ trivial. There are no NG variables and the sole degree of freedom, namely $\p$ itself, is of the matter type.

Suppose now that the VEV $\vev{\p(x)}$ varies along one particular direction in space. In order to maintain Poincar\'e (or Galilei) covariance if desired, I will characterize this preferred direction by a fixed spacelike vector $n^\m$. Our order parameter for spacetime symmetry breaking is then defined by a function $\vp$ of a single variable such that
\begin{equation}
\vev{\p(x)}\equiv\vp(n\cdot x)\;.
\label{translationOP}
\end{equation}
We would eventually like to trade $\p(x)$ for a NG field $\pi(x)$ that manifests spontaneous breaking of translations in the direction of $n^\m$. Before we proceed, let me stress that the order parameter~\eqref{translationOP} will as a rule also break rotations and (if present) Lorentz or Galilei boosts. However, these never enter our setup explicitly. For scalar fields, we already know that both spatial rotations and boosts are descendant symmetries in the sense of Chap.~\ref{chap:differences}. Their nonlinear realization will therefore be automatically taken care of by the same NG field $\pi(x)$.

\begin{figure}[t]
\sidecaption[t]
\includegraphics[width=2.0in]{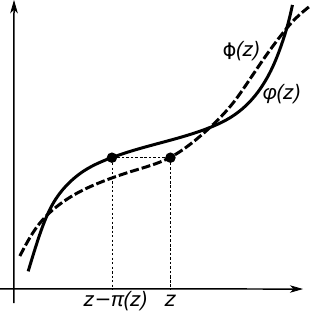}
\caption{Two parameterizations of the fluctuations of a strictly monotonic function $\vp(z)$. The first amounts to a local shift $\pi(z)$ of the graph of $\vp(z)$ in the horizontal direction. The second is given by the function $\p(z)$ corresponding to the shifted graph. Mathematically, the two parameterizations are related by $\p(z)=\vp(z-\pi(z))$}
\label{fig:parameterization}
\end{figure}

Let us now start with a purely mathematical problem on functions of a single variable. I will call the latter $z$ to distinguish it from the spacetime or spatial coordinates, $x^\m$ and $x^r$. Suppose we are given a function $\vp(z)$ that ``breaks translation invariance,'' that is, it is strictly monotonic, $\vp'(z)\neq0$. We would like to parameterize the ``fluctuations'' of $\vp(z)$ in a way that makes the breaking of translation invariance manifest. We do so by locally shifting the graph of the function by $\pi(z)$ along the $z$-axis. This defines uniquely a new function $\p(z)$ via $\p(z)\equiv\vp(z-\pi(z))$, see Fig.~\ref{fig:parameterization}. Thanks to the assumed strict monotonicity of $\vp(z)$, the mapping between $\pi(z)$ and $\p(z)$ for fixed $\vp(z)$ is one-to-one.

This little trick can be promoted to an arbitrary number $D$ of spacetime dimensions, provided $\vp$ as assumed only varies in one particular direction. The coordinates corresponding to all the other directions in spacetime play a role of fixed parameters in the mapping between $\p$ and $\pi$. Thus, I assume that $\vp(n\cdot x)$ in~\eqref{translationOP} is a strictly monotonic function of its argument. This is necessary for translations along $n^\m$ to be spontaneously broken at each spacetime point. The scalar field $\p(x)$ can then be traded for $\pi(x)$, defined implicitly by
\begin{equation}
\p(x)\equiv\vp(n\cdot x-\pi(x))\;.
\label{translationNGdef}
\end{equation}

The function $\vp$ as defined by~\eqref{translationOP} obviously depends on the choice of coordinates. For instance, $\vp(n\cdot x)$ and $\vp(n\cdot x+c)$ with any $c\in\R$ describe the same order parameter, differing only by a coordinate shift along $n^\m$. This is a hallmark of translation symmetry breaking. The introduction of the NG field $\pi(x)$ makes it possible to map a \emph{fixed} function $\vp$ to a \emph{scalar} field $\p$. This requires that under a spacetime transformation, be it a translation, rotation or boost, the NG field transforms as
\begin{equation}
\pi'(x')=\pi(x)+n'\cdot x'-n\cdot x\;.
\label{translationNGtransfo}
\end{equation}
In the special case of spatial translations, $\vec x\to\vec x+\vec\eps$, this boils down to $\pi'(\vec x+\vec\eps,t)=\pi(\vec x,t)-\skal n\eps$, where $\vec n$ is the spatial part of $n^\m$. Shifting the NG field by a constant is typical for spontaneously broken Abelian symmetries. We are on the right track.

Now that we have settled the parameterization, we would like to construct the most general effective action consistent with all the symmetries present. Since there were no NG variables to start with, the MC form~\eqref{MCspacetime} reduces to $P\cdot\D x$. This gives a trivial spacetime coframe, $\vec e^{*\fr\m}=\d^{\fr \m}_\n\D x^\n$. Also, by~\eqref{spacetimecovderpsi}, the covariant derivative of the matter field $\p$ is trivial, $\cd_{\fr\m}\p=\d_{\fr\m}^\n\de_\n\p$. Therefore, an invariant action can be built by taking any Lagrangian density that does not depend explicitly on spacetime coordinates. It can contain $\p(x)$ with an arbitrary number of derivatives (including zero). The indices on the derivatives are to be contracted in a way that preserves the linearly realized group $H_0$. In case of a relativistic, that is Poincar\'e-invariant theory, any Lorentz-invariant Lagrangian density $\La_\mathrm{eff}[\p]$ will do.

It looks like we have not made much progress. In particular, we have not learned anything so far about the possible dependence of the effective Lagrangian on the NG field $\pi(x)$. However, there are still two requirements we have not addressed yet. The first of these is that the order parameter~\eqref{translationOP} is at least a local minimum of the energy functional of the EFT. This will constrain what the effective Lagrangian may look like. The second requirement is to have a consistent power-counting scheme that would allow us to decide which operators in the Lagrangian are most important. In order to understand what is at stake, it is best to work out an illustrative example.


\subsection{Case Study: Fluctuations of a Domain Wall}
\label{subsec:translationcasestudy}

Following Chap.~5 of~\cite{Manton2004}, consider the class of relativistic scalar field theory models
\begin{equation}
\La_\mathrm{eff}=\frac12(\de_\m\p)^2-V(\p)\;,
\label{transmodellag}
\end{equation}
where $V(\p)$ is a potential function, required for stability to be bounded from below. We can assume without loss of generality that $\min V(\p)=0$. The classical EoM of the model reads
\begin{equation}
\Box\p+\OD{V(\p)}\p=0\;.
\label{transmodelEoM}
\end{equation}
Suppose the potential has (at least) two degenerate minima. Then the EoM has \emph{domain wall} (or \emph{kink}) solutions $\vp$ that interpolate between the minima. Such solutions depend on a single spatial coordinate, which I will denote as $z$, and converge to the respective minima of the potential for $z\to\pm\infty$. Upon reduction to the single variable $z$, the EoM~\eqref{transmodelEoM} has the first integral $(1/2)(\Od{\p}z)^2-V(\p)$. For the domain wall solutions, this first integral evaluates to zero since for $z\to\pm\infty$, $\vp(z)$ converges to a constant such that $\smash{\lim\limits_{z\to\pm\infty}V(\vp(z))=0}$. This reduces the EoM to the following first-order differential equation for $\vp(z)$,
\begin{equation}
\vp'(z)=\pm\sqrt{2V(\vp(z))}\;.
\label{transmodelDWeq}
\end{equation}
The prime on $\vp$ indicates a derivative with respect to its argument.

\begin{illustration}%
\label{ex:DWsol}%
One of the most common scalar potentials is the double-well potential,
\begin{equation}
V(\p)\equiv\frac\l2(\p^2-v^2)^2\;,
\label{DWpotdoublewell}
\end{equation}
where $\l$ and $v$ are positive parameters. A simple calculation gives the corresponding domain wall solution, interpolating between the minima at $\p=\pm v$,
\begin{equation}
\vp(z)=\pm v\tanh[\sqrt\l v(z-z_0)]\;.
\label{DWsolquartic}
\end{equation}
Here $z_0$ is an integration constant that determines the center of the domain wall where $\vp(z_0)=0$. The width of the domain wall is given by $1/(\sqrt\l v)$.

Another example of a phenomenologically important potential is
\begin{equation}
V(\p)\equiv m^2v^2\biggl(1-\cos\frac\p v\biggr)\;,
\label{DWpotcos}
\end{equation}
where $m$ and $v$ are positive parameters. This appears for instance in the low-energy effective theory~\eqref{ChPTLO2} of quantum chromodynamics, when restricted to neutral pions. In this realization, $m$ corresponds to the pion mass $m_\pi$ and $v$ to the pion decay constant, $f_\pi$. The potential $V(\p)$ is now periodic with minima at $\p=2n\pi v$ for any $n\in\Z$. The domain wall solution, connecting two neighboring minima, is
\begin{equation}
\vp(z)=4v\arctan\exp[\pm m(z-z_0)]\;.
\label{DWsolSineGordon}
\end{equation}
In this case, the width of the domain wall is fixed by the mass parameter to be $1/m$.
\end{illustration}

Both concrete domain wall solutions in \refex{ex:DWsol} satisfy our criteria for the translation-breaking order parameter modulated in one dimension. We would eventually like to derive a low-energy EFT for the fluctuations of the domain wall. The first step towards this goal is to use the general parameterization~\eqref{translationNGdef} with $n\cdot x=z$. The potential term in~\eqref{transmodellag} can then be eliminated by means of~\eqref{transmodelDWeq}, upon which the Lagrangian becomes
\begin{equation}
\La_\mathrm{eff}=[\vp'(z-\pi)]^2\left[\frac12(\de_\m\pi)^2-(1-\de_z\pi)\right]\;.
\label{DWfluctuationlag}
\end{equation}
The contribution of the $1-\de_z\pi$ piece to action is a mere boundary term independent of local variations of $\pi(x)$. Namely, $\smash{\int\D z\,[\vp'(z-\pi)]^2(1-\de_z\pi)=\int_{\tilde z_-}^{\tilde z_+}\D\tilde z\,[\vp'(\tilde z)]^2}$ where $\tilde z\equiv z-\pi(x)$ and $\smash{\tilde z_\pm\equiv\lim\limits_{z\to\pm\infty}\tilde z}$. Dropping this contribution, we are left with the following effective action~\cite{Hidaka2015a},
\begin{equation}
S_\mathrm{eff}\{\pi\}=\frac12\int\D^D\!x\,[\vp'(z-\pi(x))]^2[\de_\m\pi(x)]^2\;.
\label{DWfluctuationEFT}
\end{equation}

The EFT~\eqref{DWfluctuationEFT} apparently describes a derivatively coupled NG field $\pi(x)$ as we wanted. However, it is still not quite satisfactory due to the explicit appearance of the domain wall background $\vp(z)$. It is not even clear what the quasiparticle spectrum of the EFT is. To that end, we utilize the EoM, linearized in the NG field,
\begin{equation}
\Box\pi(x)-\frac{2\vp''(z)}{\vp'(z)}\de_z\pi(x)\approx0\;,
\label{DWlinEoM1}
\end{equation}
where the $\approx$ symbol indicates the linear approximation. It is convenient to switch to the linear fluctuation of the order parameter, $\mf(x)\equiv\p(x)-\vp(z)$.\footnote{While $\pi(z)$ is defined as the horizontal distance of the graphs of $\p(z)$ and $\vp(z)$ in Fig.~\ref{fig:parameterization} (the other spacetime coordinates being fixed), $\mf(z)$ can be viewed as their vertical distance.} The linearized EoM~\eqref{DWlinEoM1} is thus transformed to
\begin{equation}
\Box\mf(x)+\frac{\vp'''(z)}{\vp'(z)}\mf(x)\approx0\;,\qquad
\mf(x)\approx-\vp'(z)\pi(x)\;.
\label{DWlinEoM2}
\end{equation}
As a simple check, note that the domain wall solution satisfies $\vp'''/\vp'=\D^2V(\vp)/\D\vp^2$. Equation~\eqref{DWlinEoM2} then descends directly from the Lagrangian~\eqref{transmodellag} by expanding the latter to second order in $\mf(x)$. The form of~\eqref{DWlinEoM2} allows us to Fourier-transform in time and the transverse spatial coordinates (denoted collectively as $\vec x_\perp$). Parameterizing plane-wave solutions by energy $E$ and transverse momentum $\vec p_\perp$,
\begin{equation}
\mf(\vec x_\perp,z,t)=\hat\mf(z)\exp(-\I Et+\I\vec p_\perp\cdot\vec x_\perp)\;,
\end{equation}
the profile function $\hat\mf(z)$ solves the one-dimensional eigenvalue problem
\begin{equation}
\left[-\de_z^2+\frac{\vp'''(z)}{\vp'(z)}\right]\hat\mf(z)=(E^2-\vec p_\perp^2)\hat\mf(z)\;.
\label{DWlinEoM3}
\end{equation}
In order to appreciate the content of this equation for single-particle states, let us return to the two concrete choices of potential introduced in \refex{ex:DWsol}. In both cases, I will for simplicity consider only the solution $\vp(z)$ centered at $z_0=0$.

\newpage

\begin{illustration}%
For the double-well potential with the corresponding domain wall solution~\eqref{DWsolquartic}, $\vp'''(z)/\vp'(z)=2\l v^2[2-3\sech^2(\sqrt\l vz)]$. Introducing a dimensionless coordinate $Z\equiv\sqrt\l vz$ turns~\eqref{DWlinEoM3} into
\begin{equation}
\left(-\de_Z^2-\frac6{\cosh^2Z}\right)\hat\mf(Z)=\left(\frac{E^2-\vec p_\perp^2}{\l v^2}-4\right)\hat\mf(Z)\;.
\end{equation}
The operator on the left-hand side is a special case of the \emph{P\"oschl--Teller Hamiltonian}, $H_n\equiv-\de_Z^2-n(n+1)/\cosh^2Z$ with positive $n\in\Z$. The spectrum of this Hamiltonian is well-known; see e.g.~Chap.~11 of~\cite{Hecht2000}. There are $n$ bound states with eigenvalues $-n^2,-(n-1)^2,\dotsc,-1$. The continuous part of the spectrum covers the open interval $(0,\infty)$. In the $n=2$ case of interest to us, the ground state with eigenvalue $-4$ gives $E=\abs{\vec p_\perp}$. The corresponding profile function is $\hat\mf(Z)\propto1/\cosh^2Z$. This solution is localized in the $z$-direction to the vicinity of the domain wall. It propagates as a plane wave in the transverse directions with a massless relativistic dispersion relation. The upper bound state of $H_2$ with eigenvalue $-1$ gives $E^2=\vec p_\perp^2+3\l v^2$. Its profile function is $\hat\mf(Z)\propto\sinh Z/\cosh^2Z$. This solution is also localized to the domain wall and propagates only in the transverse directions, but it has a nonzero mass of $\sqrt{3\l}v$. Finally, the continuum part of the spectrum of the P\"oschl--Teller Hamiltonian can be parameterized by the momentum variable $p_z$ so that for $z\to\pm\infty$, $\hat\mf(z)\propto\exp(\I p_zz)$. The corresponding energy is $E^2=\vec p_\perp^2+p_z^2+4\l v^2$. This describes delocalized fluctuations of the domain wall that propagate in the entire space. They have a relativistic dispersion relation with mass $2\sqrt\l v$.

In case of the cosine potential with its domain wall solution~\eqref{DWsolSineGordon}, $\vp'''(z)/\vp'(z)=m^2[1-2\sech^2(mz)]$. With the dimensionless coordinate $Z\equiv mz$, the eigenvalue problem~\eqref{DWlinEoM3} becomes
\begin{equation}
\left(-\de_Z^2-\frac2{\cosh^2Z}\right)\hat\mf(Z)=\left(\frac{E^2-\vec p_\perp^2}{m^2}-1\right)\hat\mf(Z)\;.
\end{equation}
The left-hand side is again one of the P\"oschl--Teller Hamiltonians, $H_1$. Its sole bound state has the profile function $\hat\mf(Z)\propto1/\cosh Z$. This is a state localized to the domain wall that propagates in the transverse directions with the massless relativistic dispersion, $E=\abs{\vec p_\perp}$. The continuum part of the spectrum of this P\"oschl--Teller Hamiltonian again amounts to delocalized fluctuations of the domain wall. Their dispersion relation is $E^2=\vec p_\perp^2+p_z^2+m^2$. Far away from the domain wall, they describe particles of mass $m$ propagating in the bulk.
\end{illustration}

We are now ready to draw some general conclusions. Starting with any potential $V(\p)$ with degenerate minima, the domain wall solution will be localized in the $z$-direction. The linearized EoM~\eqref{DWlinEoM1} always admits plane-wave solutions, propagating in the transverse directions with the massless dispersion relation, $E=\abs{\vec p_\perp}$, and with a constant longitudinal profile, $\hat\pi(z)\propto1$. By~\eqref{DWlinEoM2}, this corresponds to $\hat\mf(z)\propto\vp'(z)$. The existence of this solution is guaranteed by the spontaneously broken translation invariance. Indeed, the fact that $\hat\pi(z)$ is constant indicates that this is the NG boson we are after. It does not propagate in the $z$-direction since translations in the $z$-direction are completely broken, hence there is no physical momentum $p_z$ to label one-particle states with. The NG mode propagates solely along the directions in which translations remain unbroken.

The second observation is that there is a continuum of plane-wave-like solutions that propagate in the bulk. Far away from the domain wall where $\vp(z)$ goes to one of the minima of $V(\p)$, $\vp_0$, these are simple plane waves, $\hat\mf(z)\propto\exp(\I p_zz)$. They have a relativistic dispersion relation with squared mass $V''(\vp_0)$. In fact, we could have guessed this much by expanding the Lagrangian~\eqref{transmodellag} around the uniform solution $\vev{\p(x)}=\vp_0$. In addition to the NG mode and the bulk plane-wave continuum, there may be further solutions to~\eqref{DWlinEoM2} that are localized on the domain wall and propagate only in the transverse directions. These correspond to excited bound states of the one-dimensional Hamiltonian on the left-hand side of~\eqref{DWlinEoM3}. The presence of such massive excitations propagating along the domain wall depends on the concrete choice of the potential $V(\p)$.

It is not a priori obvious why the NG mode with $\hat\pi(z)\propto1$ is necessarily the lowest-lying excitation, that is, why the domain wall solution is stable. To understand this, let us assume without loss of generality that $\p(x)$ is monotonic in $z$ and estimate its energy from below by completing the square,
\begin{align}
\notag
&\int\D^{d-1}\!\vec x_\perp\,\D z\left[\frac12(\de_0\p)^2+\frac12(\vec\nabla_\perp\p)^2+\frac12(\de_z\p)^2+V(\p)\right]\\
\notag
&=\int\D^{d-1}\!\vec x_\perp\,\D z\left\{\frac12(\de_0\p)^2+\frac12(\vec\nabla_\perp\p)^2+\frac12\bigl[\p'\mp\sqrt{2V(\p)}\bigr]^2\pm\p'\sqrt{2V(\p)}\right\}\\
&\geq\int\D^{d-1}\!\vec x_\perp\abs{\int_{\p_-}^{\p_+}\D\p\,\sqrt{2V(\p)}}\;,
\label{bogomolny}
\end{align}
where $\smash{\p_\pm\equiv\lim\limits_{z\to\pm\infty}\p(x)}$. This estimate, known as the \emph{Bogomolny bound}, shows that static solutions satisfying~\eqref{transmodelDWeq} minimize the energy per unit transverse volume on the class of field configurations with a fixed boundary condition at $z\to\pm\infty$.

\begin{watchout}%
We have discovered that our EFT~\eqref{DWfluctuationEFT} inevitably describes both the translation NG boson and specific gapped excitations. This is hardly surprising given that~\eqref{DWfluctuationEFT} is equivalent via a field redefinition to the original model~\eqref{transmodellag}. One might naively expect that an EFT for the NG mode alone could be extracted by integrating over $z$. After all, we found that the NG solution for $\pi(x)$ does not depend on $z$ at all. However, inserting a $z$-independent $\pi(\vec x_\perp,t)$ makes it possible to shift the $z$-coordinate in $\vp'(z-\pi)$, thus reducing~\eqref{DWfluctuationEFT} to
\begin{equation}
S_\mathrm{eff}\{\pi\}=\frac12\int\D z\,[\vp'(z)]^2\int\D^{d-1}\!\vec x_\perp\,\D t\,\bigl[(\de_0\pi)^2-(\vec\nabla_\perp\pi)^2]\;.
\label{freesurfacewaves}
\end{equation}
This is a noninteracting theory, describing free surface waves on the domain wall. We conclude that \emph{all} interactions among the NG modes in the EFT~\eqref{DWfluctuationEFT} are mediated by the gapped modes.
\end{watchout}

The problem of disentangling the NG mode from the gapped excitations is related to our ability to organize contributions to the EFT by a derivative expansion. Also, we have so far been forced to work with specific domain wall solutions $\vp(z)$, descending from the chosen model potential $V(\p)$. We would rather have a model-independent EFT where the stability of the ground state and the existence of a power counting serve merely as constraints. I will now address these issues jointly in a framework inspired by the background gauge invariance approach of Sect.~\ref{sec:effLaggauged}.


\subsection{Effective Action from Background Gauge Invariance}
\label{subsec:translationgauged}

The idea to use background gauging to derive an EFT for spontaneously broken translations has been utilized in multiple contexts, including inflationary cosmology~\cite{Cheung2008a} and superconductivity~\cite{Samokhin2010a}. Here I will loosely follow the exposition in~\cite{Hidaka2015a}. We are looking for an effective action $S_\mathrm{eff}\{\pi\}$ for the NG mode of spontaneously broken translations in the $z$-direction. In the spirit of Sect.~\ref{sec:effLaggauged}, we first promote the physical symmetry to a local invariance by adding a set of background gauge fields. Localizing spacetime translations leads to the group of spacetime diffeomorphisms. The precise choice of the corresponding background field is constrained by the physical symmetry and field content. In case of relativistic theories of scalar fields, one can use a spacetime metric, $g_{\m\n}(x)$, to make the gauged action $S_\mathrm{eff}\{\pi,g\}$ diffeomorphism-invariant.

The transformation of the metric under a diffeomorphism of the spacetime manifold, $x^\m\to x'^\m(x)\equiv x^\m+\eps^\m(x)$, is given by the push-forward by the inverse of the diffeomorphism; see Appendix~\ref{appsubsec:pushpull}. This must be augmented with a transformation rule for the NG field $\pi(x)$. In order to maintain the relation $\p(x)=\vp(z-\pi(x))$ between a scalar field $\p(x)$ and a fixed function $\vp(z)$, $\pi(x)$ should transform as
\begin{equation}
\pi'(x')=\pi(x)+\eps^z(x)\;.
\label{DWlocaltranslation}
\end{equation}
The NG field can now be eliminated by choosing the diffeomorphism as $\eps^\m(x)=-\d^\m_z\pi(x)$. With a slight abuse of notation, I will denote this operation as $\tran{-\pi}$. This reduces the action to $S_\mathrm{eff}\{\pi,g\}=S_\mathrm{eff}\{0,\tran{-\pi}g\}$. The dependence of the action on the composite gauge field $(\tran{-\pi}g)_{\m\n}$ is constrained by invariance under the residual group of diffeomorphisms that fix the point $\pi=0$. These are (possibly $z$-dependent) $(D-1)$-dimensional diffeomorphisms with $\eps^z(x)=0$. Once the most general admissible action $\smash{S_\mathrm{eff}\{0,\tran{-\pi}g\}}$ has been found, the EFT in flat Minkowski spacetime can be recovered by setting $g_{\m\n}(x)$ to the Minkowski metric.

To constrain the form of $\smash{S_\mathrm{eff}\{0,\tran{-\pi}g\}}$, it is convenient to separate the $z$-coordinate from the other spacetime coordinates. I will indicate those using lowercase Fraktur indices $\tri m,\tri n,\dotsc$. It will also be more practical to work with the inverse of the metric, $g^{\m\n}(x)$, rather than $g_{\m\n}(x)$ itself. The transformation of the metric under diffeomorphisms then splits as
\begin{align}
\notag
g'^{\tri{mn}}={}&g^{\tri{ab}}(\d^{\tri m}_{\tri a}+\de_{\tri a}\eps^{\tri m})(\d^{\tri n}_{\tri b}+\de_{\tri b}\eps^{\tri n})+g^{\tri az}(\d^{\tri m}_{\tri a}+\de_{\tri a}\eps^{\tri m})\de_z\eps^{\tri n}+g^{z\tri b}\de_z\eps^{\tri m}(\d^{\tri n}_{\tri b}+\de_{\tri b}\eps^{\tri n})\\
\notag
&+g^{zz}\de_z\eps^{\tri m}\de_z\eps^{\tri n}\;,\\
\notag
g'^{\tri mz}={}&g^{\tri{ab}}(\d^{\tri m}_{\tri a}+\de_{\tri a}\eps^{\tri m})\de_{\tri b}\eps^z+g^{\tri az}(\d^{\tri m}_{\tri a}+\de_{\tri a}\eps^{\tri m})(1+\de_z\eps^z)+g^{z\tri b}\de_z\eps^{\tri m}\de_{\tri b}\eps^z\\
\label{DWmetrictransfo}
&+g^{zz}\de_z\eps^{\tri m}(1+\de_z\eps^z)\;,\\
\notag
g'^{zz}={}&g^{\tri{ab}}\de_{\tri a}\eps^z\de_{\tri b}\eps^z+g^{\tri az}\de_{\tri a}\eps^z(1+\de_z\eps^z)+g^{z\tri b}(1+\de_z\eps^z)\de_{\tri b}\eps^z+g^{zz}(1+\de_z\eps^z)^2\;.
\end{align}
For the sake of brevity, I suppressed the arguments $x$ of $g^{\m\n}(x)$ and $x'$ of $g'^{\m\n}(x')$. The transformation rule for $g^{z\tri m}$ is analogous to that for $g^{\tri mz}$.

Under the restricted $(D-1)$-dimensional diffeomorphisms with $\eps^z(x)=0$, $g^{zz}(x)$ behaves as a scalar, $g'^{zz}(x')=g^{zz}(x)$. On the other hand, $g'^{\tri mz}=g^{\a z}(\d^{\tri m}_\a+\de_\a\eps^{\tri m})$ depends on both $g^{\tri mz}$ and $g^{zz}$. Finally, $\smash{g'^{\tri{mn}}=g^{\a\b}(\d^{\tri m}_\a+\de_\a\eps^{\tri m})(\d^{\tri n}_\b+\de_\b\eps^{\tri n})}$ depends on all $g^{\tri{mn}}$, $g^{\tri mz}$, $g^{z\tri n}$ and $g^{zz}$. This suggests that our main ingredient for constructing diffeomorphism-invariant actions will be
\begin{equation}
\begin{split}
(\tran{-\pi}g)^{zz}&=g^{\tri{ab}}\de_{\tri a}\pi\de_{\tri b}\pi-2g^{\tri az}\de_{\tri a}\pi(1-\de_z\pi)+g^{zz}(1-\de_z\pi)^2\\
&=g^{zz}-2g^{z\a}\de_\a\pi+g^{\a\b}\de_\a\pi\de_\b\pi\;,
\end{split}
\label{gzz}
\end{equation}
where the argument of $(\tran{-\pi}g)^{zz}$ is $\tilde x^\m\equiv x^\m-\d^\m_z\pi(x)$. The $z$-component of $\tilde x^\m$, that is $\tilde z\equiv z-\pi(x)$, is itself invariant under spacetime diffeomorphisms and can thus appear in the effective action without restrictions. Finally, the effective action may also contain $(\tran{-\pi}g)^{\tri mz}$ and $(\tran{-\pi}g)^{\tri{mn}}$. These, not being scalars, can however only enter through higher-derivative tensors such as the Riemann curvature tensor. See Appendix~A of~\cite{Cheung2008a} for a discussion of geometric structures allowed by the restricted $(D-1)$-dimensional diffeomorphism invariance. For our purposes, the main conclusion is that the part of the effective action dominant at low energies will be included in
\begin{equation}
\begin{split}
S_\mathrm{eff}\{\pi,g\}&=\int\D^D\!\tilde x\,\vol(\tran{-\pi}g)(\tilde x)\,f\bigl(\tilde z,(\tran{-\pi}g)^{zz}(\tilde x)\bigr)\\
&=\int\D^D\!x\,\vol(g)(x)\,f\bigl(z-\pi(x),(\tran{-\pi}g)^{zz}(\tilde x)\bigr)\;.
\end{split}
\label{DWactiongeneric}
\end{equation}
Here $f$ is a generic smooth function of two variables. The volume measure $\vol(g)$ for a metric of (timelike) Lorentzian signature equals $\sqrt{(-1)^d\det g}$; cf.~Appendix~\ref{appsubsec:integrationRiemannian}.

\begin{illustration}%
\label{ex:fuv}%
Using the specific choice $f(u,v)\equiv(1/2)[\vp'(u)]^2(v-1)$ and subsequently going back to the flat Minkowski spacetime gives
\begin{equation}
S_\mathrm{eff}\{\pi\}=\frac12\int\D^D\!x\,[\vp'(z-\pi)]^2[(\de_\m\pi)^2-2(1-\de_z\pi)]\;.
\end{equation}
This reproduces our previous model result~\eqref{DWfluctuationlag}.
\end{illustration}

A systematic expansion in derivatives of the NG field is easier if one uses the combination $(\tran{-\pi}g)^{zz}+1$, which goes to $2\de_z\pi+(\de_\m\pi)^2$ in the flat-spacetime limit. This suggests expanding the function $f(u,v)$ in~\eqref{DWactiongeneric} in powers of $v+1$. Up to second order in $v+1$ and thus in derivatives, we have $f(u,v)=c_0(u)+c_1(u)(v+1)+c_2(u)(v+1)^2+\dotsb$, which translates to the Minkowski-spacetime action
\begin{equation}
\begin{split}
S_\mathrm{eff}\{\pi\}=\int\D^D\!x\,\bigl\{&c_0(z-\pi)+c_1(z-\pi)[2\de_z\pi+(\de_\m\pi)^2]\\
&+c_2(z-\pi)[2\de_z\pi+(\de_\m\pi)^2]^2\bigr\}+\dotsb\;.
\end{split}
\label{DWEFTLO}
\end{equation}
This is the low-energy EFT we have been looking for. It remains to elucidate the consistency constraints on the functions $c_i(z-\pi)$. To that end, we expand the Lagrangian in~\eqref{DWEFTLO} to second order in the NG field,
\begin{align}
\label{DWEFTLOexp}
\La_\mathrm{eff}[\pi]={}&c_0(z)+[-c_0'(z)+2c_1(z)\de_z]\pi\\
\notag
&+\frac12[c_0''(z)-2c_1'(z)\de_z]\pi^2+c_1(z)(\de_\m\pi)^2+4c_2(z)(\de_z\pi)^2+\dotsb\;.
\end{align}
Bulk stability requires absence of any terms linear in $\pi$. This leads to the constraint
\begin{equation}
c_0'(z)+2c_1'(z)=0\;.
\label{bulkstability}
\end{equation}
In fact, one can demand even  $c_0(z)+2c_1(z)=0$, since any constant offset of $c_0(z)$ can be dropped from~\eqref{DWEFTLO}. This sharper relation is indeed satisfied by the choice of $f(u,v)$ in \refex{ex:fuv}, where $c_0(u)=-[\vp'(u)]^2$ and $c_1(u)=(1/2)[\vp'(u)]^2$. With the constraint~\eqref{bulkstability}, the first term on the second line of~\eqref{DWEFTLOexp} automatically drops, and the Lagrangian boils down to $c_1(z)(\de_\m\pi)^2+4c_2(z)(\de_z\pi)^2$ up to a surface term. A necessary (though not inevitably sufficient) condition for the corresponding energy functional to be bounded from below is therefore
\begin{equation}
c_1(z)\geq0\quad\text{and}\quad c_1(z)\geq4c_2(z)\quad\text{for any }z\in\R\;.
\end{equation}

So far, I have not assumed any specific choice or properties of the profile functions $c_i(z)$. The effective action~\eqref{DWEFTLO} defines a generic EFT for the NG mode of spontaneously broken translations in the $z$-direction in a relativistic system. In particular, the functions $c_i(z)$ in~\eqref{DWEFTLO} do not have to arise from a topologically nontrivial background such as a domain wall. However, if that happens to be the case, we expect them to be localized to a finite range in $z$. The NG mode will then correspond to a surface wave and propagate solely in the transverse directions. I would now like to use the general EFT~\eqref{DWactiongeneric} to shed some light on the physics of these surface waves beyond the noninteracting approximation~\eqref{freesurfacewaves}. To that end, it is convenient to switch to the $\tilde x^\m$ coordinates. Namely, $\pi(x)$ measures the local fluctuation-induced displacement of the domain wall in the $z$-direction. Geometrically, the dynamical state of the domain wall can thus be viewed as a hypersurface in $\R^D$ defined by a constant value of $\tilde z$.

With the shorthand notation $\tilde g\equiv\tran{-\pi}g$, the determinant of this $D$-dimensional metric can be decomposed as $\det\tilde g=({\tilde g}^{zz})^{-1}\det\tilde{\tri g}$. Here $\tilde{\tri g}_{\tri{mn}}$ is a projection of $\tilde g_{\m\n}$ to the $(D-1)$-dimensional space of $x^{\tri m}$, which is nothing but the metric on the domain wall induced by the bulk Lorentzian metric in $\R^D$. The factorization of $\det\tilde g$ follows from the mathematical properties of block matrices~\cite{Powell2011a}. Augmenting this observation with the constraint $c_0(z)+2c_1(z)=0$ turns the first line of~\eqref{DWactiongeneric} to
\begin{equation}
\begin{split}
S_\mathrm{eff}\{\pi,g\}=\int\D^D\!\tilde x\,&\vol(\tilde{\tri g})(\tilde x)\\
&\times\left\{c_0(\tilde z)+\left[\frac{c_0(\tilde z)}8+c_2(\tilde z)\right][\tilde g^{zz}(\tilde x)+1]^2+\dotsb\right\}\;.
\end{split}
\end{equation}
So far this is an exact rewriting of~\eqref{DWactiongeneric} except for the truncation of the power expansion in $\tilde g^{zz}+1$. Now we take the limit of flat spacetime. Simultaneously, we restrict to fields independent of the $\tilde z$ coordinate, as appropriate for the NG mode. This makes it possible to do the integral over $\tilde z$ independently of the NG field. We thus arrive at an effective $(D-1)$-dimensional theory for the surface waves on the domain wall which, up to higher-order corrections, takes the form
\begin{equation}
S_\mathrm{eff}=\int\D z\,c_0(z)\int\D^{d-1}\!\vec x_\perp\,\D t\,\vol(\tilde{\tri g})(\vec x_\perp,t)+\dotsb\;.
\label{DWDBI}
\end{equation}
In this approximation, the EFT for the domain wall fluctuations is completely geometric and fixed by the induced metric,
\begin{equation}
\tilde{\tri g}_{\tri{mn}}=g_{\tri{mn}}-\de_{\tri m}\pi\de_{\tri n}\pi\;,\qquad
\vol(\tilde{\tri g})=\sqrt{1-g^{\tri{mn}}\de_{\tri m}\pi\de_{\tri n}\pi}\;.
\end{equation}
The expression for $\vol(\tilde{\tri g})$ follows from the so-called Weinstein--Aronszajn identity for matrix determinants. The reader may recognize this as the \emph{Dirac--Born--Infeld} (DBI) \emph{theory} that we met previously in Sect.~\ref{subsubsec:DBI}. All that is left of the profile of the domain wall is the overall prefactor in~\eqref{DWDBI}.

\begin{illustration}%
The prefactor $\smash{\int\D z\,c_0(z)}$ measures, up to overall sign, the energy of the domain wall per unit transverse volume, that is its surface tension. For the class of models~\eqref{transmodellag}, it equals $\smash{-\int\D z\,[\vp'(z)]^2}$; see also~\eqref{bogomolny}. In case of the double-well potential with the domain wall solution~\eqref{DWsolquartic}, this gives $\smash{-(4/3)\sqrt\l v^3}$. For the cosine potential with the corresponding domain wall~\eqref{DWsolSineGordon}, one finds analogously $-8mv^2$.
\end{illustration}

\begin{watchout}%
The $(D-1)$-dimensional theory~\eqref{DWDBI} obviously describes only the NG mode as we wanted; the gapped modes we found in Sect.~\ref{subsec:translationcasestudy} are gone. The price to pay is that the $D$-dimensional Lorentz symmetry is now realized in a nontrivial fashion beyond our spacetime symmetry paradigm; cf.~\eqref{symDBI}.

The fact that all the modes propagating in the $D$-dimensional bulk are gapped is essential for the validity of the EFT~\eqref{DWDBI}. There are physical systems where the bulk modes are naturally gapless, for instance when two immiscible superfluids are separated by an interface. In this case, integrating the bulk modes out makes the EFT for the surface waves on the interface nonlocal. This leads to a fractional-power dispersion relation of the surface modes, $E(\vec p)\propto\abs{\vec p}^{3/2}$. See~\cite{Watanabe2014e} for a detailed discussion from an EFT perspective.
\end{watchout}


\subsection{Further Possible Applications}
\label{subsec:translationfurther}

In discussing spontaneous breaking of spatial translations, I have deliberately merely outlined the general approach and then focused on a specific type of system to illustrate it. There are however several natural, physically motivated modifications or generalizations of the setup. I will now at least briefly mention some of the possible avenues one might wish to follow.

\runinhead{Periodic Modulation of Order Parameter} Mathematically the simplest modification is one with a single, otherwise featureless real order parameter where the profile $\vp(z)$, or $c_0(z)$, is not spatially localized. The most interesting situation arises when $\vp(z)$ is a periodic function of $z$. In this case the physics changes qualitatively. The eigenvalue problem~\eqref{DWlinEoM3} for linear fluctuations will have no bound states. Instead, its continuous spectrum will have a band structure. There should still be a NG mode of the spontaneously broken translations, but this will assume the form of a sound wave (phonon) on the crystalline background $\vp(z)$. In the limit where $\vp(z)$ consists of widely separated localized kinks, the propagation of sound in the $z$-direction arises from tunneling between states localized on the individual kinks.

The detailed structure of the dispersion relation of the low-lying phonon excitations will be no less interesting. Namely, it turns out that the spontaneously broken symmetry under spatial rotations forbids the $c_1$ term in~\eqref{DWEFTLO}. As a consequence, the derivative expansion of the effective Lagrangian starts at second order in longitudinal derivatives $\de_z$ but fourth order in the transverse gradient~$\vec\nabla_\perp$~\cite{Hidaka2015a}. We already saw the same behavior in smectic liquid crystals (Sect.~\ref{subsec:vectorirrelevant}), and I will return to its physical implications in Chap.~\ref{chap:topicsnotcovered}.

\runinhead{Other Spacetime Symmetries} Another natural possibility is to consider systems with a real order parameter modulated in one spatial dimension, but with a different spacetime symmetry than Poincar\'e. This is relevant in particular for any condensed-matter system. As long as one has a concrete microscopic model, the analysis proceeds along the same steps as in Sect.~\ref{subsec:translationcasestudy}. One can however also adopt the model-independent approach of Sect.~\ref{subsec:translationgauged} based on background gauge invariance. This requires coupling the theory to an appropriate spacetime geometry. A basic discussion can be found in Appendix A of~\cite{Hidaka2015a}. For a more complete, if mathematically also more advanced, overview of non-Lorentzian geometry, see~\cite{Bergshoeff2022}.

\runinhead{Additional Degrees of Freedom} One might also want to study spontaneous breaking of translation invariance alongside other, possibly internal, broken symmetries. The field parameterization required then follows the general standard nonlinear realization of Sect.~\ref{sec:spacetimestandard} combined with the identification of the translation NG field \`a la Sect.~\ref{subsec:translation1d}. The background gauge invariance approach may again prove helpful. Its application however relies on our ability to simultaneously gauge all the relevant symmetries by adding suitable background gauge fields. This is not guaranteed a priori without further qualifications. Finally, there are also physical systems where spacetime translations are spontaneously broken in more than one direction. In this case, the idea behind Sect.~\ref{subsec:translation1d} might still be possible to apply provided one has at hand a sufficient number of fields to parameterize the order fluctuations uniquely.


\bibliographystyle{spphys}
\bibliography{references}
\chapter{Broken Spacetime Symmetry in Classical Matter}
\label{chap:spacetimeclassical}

\abstract*{The most common type of a low-energy collective mode occurring in nature is a classical matter wave, usually manifested as sound. This features many of the attributes of Nambu--Goldstone bosons, notably a vanishing energy in the long-wavelength limit. Yet, it is not obvious at first sight how thermodynamic sound fits into the general framework worked out and applied in this book so far. This chapter develops a description of classical matter that makes application of the techniques of nonlinear realization of symmetry possible. Starting from a simple toy model of a solid, it is shown that a classical medium possesses a set of emergent symmetries, reflecting its internal structure. Different choices of the emergent symmetry naturally give rise to systems that exhibit respectively solid and fluid behavior. Once all the symmetries are properly identified, the construction of their nonlinear realization proceeds as usual. By working out specific examples, the text demonstrates how well-known theories of solids (elasticity) and fluids (hydrodynamics) can be naturally recovered from the effective field theory perspective. The last section of the chapter shows how Nambu--Goldstone bosons of a spontaneously broken internal symmetry can be coupled to the collective modes of a background classical medium.}


The first example of a \emph{Nambu--Goldstone} (NG) \emph{boson} I mentioned in the introduction (Chap.~\ref{chap:intro}) was hydrodynamic sound. However, our subsequent exploration of \emph{spontaneous symmetry breaking} (SSB) took us very far away from this initial target. Indeed, I focused almost exclusively on the quantum world. This had a good reason: many conceptual subtleties of SSB are rooted in the description of quantum symmetries in terms of operators on the Hilbert space of states. We then did much work to establish the methods of nonlinear realization of symmetry and \emph{effective field theory} (EFT) for NG bosons. Eventually, we discovered that the \emph{leading order} (LO) of the derivative expansion of the EFT for quantum systems with SSB is a classical field theory. In this chapter, we shall close the circle and see how the techniques we have developed can be applied to purely classical systems.

To build the necessary intuition, I will start in Sect.~\ref{sec:classicalemergent} with a primitive toy model for elastic solids. While physically inadequate in the details, this is good enough to show that a classical medium may possess an emergent symmetry, reflecting its internal structure. A proper identification of such emergent symmetries is key for distinguishing thermodynamic phases of matter with otherwise identical microscopic dynamics, such as solids and fluids. The subsequent analysis is straightforward and follows the standard workflow from nonlinear realization to effective actions. The nonlinear realization of a purely spacetime symmetry augmented with the emergent symmetry of classical matter is detailed in Sect.~\ref{sec:classicalcoset}. In the following Sect.~\ref{sec:classicalfluidssolids}, I then work out explicit examples of EFTs for different phases of matter.


\section{Emergent Symmetry of Classical Matter}
\label{sec:classicalemergent}

Classical matter is characterized by the possibility to uniquely label its elements and individually track their evolution. While doing so, one has to carefully distinguish two different types of coordinates. The first type are the genuine spacetime coordinates, independent of whatever matter is present in the spacetime. The second type are the coordinates that label the individual elements of the medium. These are known in classical mechanics as \emph{body coordinates} (see e.g.~Sect.~8.1 of~\cite{Jose1998a}) or \emph{material coordinates} (Chap.~4 of~\cite{Soper2008}). They capture internal variations of the material properties of the medium. The difference between the two types of coordinates is best elucidated by an example.


\subsection{Introduction: Spring Model of Elasticity}
\label{subsec:classicalspring}

\begin{figure}[t]
\sidecaption[t]
\includegraphics[width=2.9in]{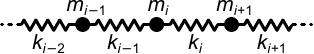}
\caption{One-dimensional chain of point masses, connected by linear springs. The springs are only allowed to vibrate in the horizontal direction. The masses and spring constants may vary along the chain as indicated}
\label{fig:spring1d}
\end{figure}

One of the most basic mechanical models, illustrating a transition between mechanics of particles and continuous field theory, is the linear spring chain, shown in Fig.~\ref{fig:spring1d}. For the sake of discussion, I will assume that the particles connected by the springs have generally unequal masses. Likewise, the springs themselves may have different spring constants, although I will for simplicity assume equal rest (unloaded) length~$a$. Denoting the position of the $i$-th particle as $x_i$, the Lagrangian of the system is
\begin{equation}
L=\sum_i\left[\frac12m_i\dot x_i^2-\frac12k_i(x_{i+1}-x_i-a)^2\right]\;.
\label{1dchainlag}
\end{equation}
The ground state of this system corresponds to a chain of point masses at rest, placed equidistantly at distance $a$ between the nearest neighbors. One can choose the origin of coordinates so that in the ground state, $\vev{x_i}=ia$. We would now like to perform the continuum limit, assuming that $a$ is very small. This requires that the displacement $x_i-ia$ from the equilibrium position only varies appreciably over distances much longer than $a$. I will replace the discrete index $i$ with a continuous body coordinate $X$ through $i\to X/a$. The individual masses $m_i$ are replaced with the linear mass density $\vr(X)$ via $m_i\to a\vr(X)$, and the spring constants $k_i$ with the local elastic (Young) modulus $E(X)$ via $k_i\to E(X)/a$. Treating the Lagrangian~\eqref{1dchainlag} as a Riemann sum then leads to the continuous approximation
\begin{equation}
L\to\int\D X\,\left\{\frac12\vr(X)[\de_0 x(X,t)]^2-\frac12E(X)[\de_Xx(X,t)-1]^2\right\}\;.
\label{1dchainlagcont}
\end{equation}
This is a field theory in one spatial dimension with the dynamical degree of freedom $x(X,t)$. Provided both $\vr(X)$ and $E(X)$ are positive for any $X\in\R$, which follows from their origin in the parameters $m_i$ and $k_i$, the ground state is $\vev{x(X,t)}=X$ up to an additive constant. This residual freedom stems from spontaneous breakdown of spatial translations. Parameterizing the displacement of the medium from equilibrium by the field $\p(X,t)\equiv x(X,t)-X$, the fluctuations around the ground state are governed by the \emph{equation of motion} (EoM),
\begin{equation}
\vr(X)\de_0^2\p(X,t)=\de_X[E(X)\de_X\p(X,t)]\;.
\label{1dchainEoM}
\end{equation}

In the limit of equal masses $m_i$ and spring constants $k_i$, the functions $\vr(X)$ and $E(X)$ become constant. The EoM~\eqref{1dchainEoM} then describes compression waves, propagating along the chain with phase velocity $v=\sqrt{E/\vr}$. In this limit, the action for $\p(X,t)$ features a set of emergent continuous symmetries that were not present in the original discrete model~\eqref{1dchainlag}. The most obvious of these is the invariance under the ``translation'' $\smash{X\xrightarrow{\eps}X+\eps}$. A straightforward application of Noether's theorem shows that the corresponding integral charge is, up to overall normalization,
\begin{equation}
P_\mathrm{pseudo}=-\int\D X\,\vr\de_0\p(X,t)\de_X\p(X,t)\;.
\label{1dchainpseudomomentum}
\end{equation}
Despite the suggestive analogy with translation invariance, this is \emph{not} the momentum carried by the oscillating masses of the original discrete chain. The latter rather equals 
\begin{equation}
P=\sum_im_i\dot x_i\to\int\D X\,\vr(X)\de_0\p(X,t)\;.
\label{1dchainmomentum}
\end{equation}

The resolution of this puzzle is that in presence of a uniform classical medium, we have two different coordinates and consequently two different translation symmetries. The position of the mass $m_i$ with respect to the laboratory inertial reference frame is given by the coordinate $x_i$. Accordingly, a genuine spatial translation amounts to the shift $\smash{x\xrightarrow{\eps}x+\eps}$. In the formulation~\eqref{1dchainlagcont} of our field theory, this acts like an ``internal symmetry.'' That is, it transforms the dependent variable $x(X,t)$ while leaving the independent variable $X$ intact. Applying Noether's theorem recovers the integral momentum~\eqref{1dchainmomentum}. This remains conserved for any choice of functions $\vr(X)$ and $E(X)$. That is because momentum conservation reflects the uniformity of the underlying space itself; it does not care about the properties of the medium. The corresponding local conservation law is equivalent to~\eqref{1dchainEoM}, which is the continuous limit of the Newtonian EoM for the point masses $m_i$.

The other integral charge, \eqref{1dchainpseudomomentum}, is usually called \emph{pseudomomentum}. This arises, as already stressed, from invariance under $\smash{X\xrightarrow{\eps}X+\eps}$. Given that $X$ was introduced as a material coordinate, conservation of $P_\mathrm{pseudo}$ should reflect the uniformity of the medium. Sure enough, it was essential for derivation of~\eqref{1dchainpseudomomentum} to assume that $\vr(X)$ and $E(X)$ are constant. The distinction between momentum and pseudomomentum has historically led to much confusion in the research on the continuum mechanics of fluids. See~\cite{Stone2000} for a pedagogical account of some of the related subtleties.

\begin{watchout}%
The comparison of momentum and pseudomomentum illustrates best the striking contrast between the fundamental spacetime symmetries and the emergent symmetries of matter. However, the field theory~\eqref{1dchainlagcont} also features other emergent symmetries than translations. Namely, for constant $\vr(X)$ and $E(X)$ its action is invariant under the whole Poincar\'e group $\gr{ISO}(1,1)$ of transformations acting on $X,t$. These include the ``translations'' of $X$, giving rise to conservation of pseudomomentum, and the usual time translations, leading to conservation of energy. In addition, the theory has an emergent $\gr{SO}(1,1)$ Lorentz invariance under ``boosts'' that mix $X$ and $t$. It is straightforward to work out the corresponding conservation law but I will not do so since we will not need it. Finally, the dynamics of our nonrelativistic spring chain should be invariant under Galilei boosts. Under the corresponding transformations, $\smash{x\xrightarrow{v}x+vt}$ with $v$ being the velocity of the boost, the Lagrangian~\eqref{1dchainlagcont} shifts by a total time derivative. Similarly to spatial translations, this is a genuine spacetime symmetry that is present for any choice of $\vr(X)$ and $E(X)$. 
\end{watchout}

\begin{figure}[t]
\sidecaption[t]
\includegraphics[width=2.0in]{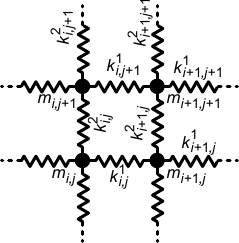}
\caption{Rectangular network of springs as a toy model of a two-dimensional elastic solid. All the point masses $m_{i,j}$ are allowed to differ, as are the spring constants $\smash{k_{i,j}^1}$ and $\smash{k_{i,j}^2}$. The rest (unloaded) length of all the springs is assumed to be the same and equal to $a$}
\label{fig:spring2d}
\end{figure}

Before closing the discussion of our toy model, let us briefly look at its generalization to a higher number of spatial dimensions $d$. For illustration, it is sufficient to take $d=2$ and consider a rectangular network of springs as in Fig.~\ref{fig:spring2d}. The position of the particle $m_{i,j}$ is now given by a two-component vector $\vec x_{i,j}$. Assuming otherwise the same dynamics as before, whereby the potential energy of each spring is a quadratic function of its extension, the Lagrangian becomes
\begin{equation}
\begin{split}
L=\sum_{i,j}\biggl[\frac12m_{i,j}\dot{\vec x}_{i,j}^2&-\frac12k_{i,j}^1\bigl(\abs{\smash{\vec x_{i+1,j}-\vec x_{i,j}}}-a\bigr)^2\\
&-\frac12k_{i,j}^2\bigl(\abs{\smash{\vec x_{i,j+1}-\vec x_{i,j}}}-a\bigr)^2\biggr]\;.
\end{split}
\label{2dchainlag}
\end{equation}
The ground state of this model corresponds to a rectangular network of equidistantly placed particles at rest. One can choose Cartesian coordinates in the plane so that $\vev{\vec x_{i,j}}=(ia,ja)$. To perform the continuum limit, we make the replacement $(i,j)\to\vec X/a$ and $m_{i,j}\to a^2\vr(\vec X)$, where $\vec X$ is a two-component vector of body coordinates and $\vr(\vec X)$ the local mass density. We also need two elastic moduli, $\smash{k_{i,j}^1\to E_1(\vec X)}$ and $\smash{k_{i,j}^2\to E_2(\vec X)}$. In terms of the displacement vector $\p^r(\vec X,t)\equiv x^r(\vec X,t)-X^r$,\footnote{I previously introduced the symbol $\dx^\m[\psi,x]$ to indicate the shift of the spacetime coordinate~$x^\m$ under an infinitesimal symmetry transformation. In this chapter, I will not use this notation in order to avoid confusion with the conventional symbol $X^r$ for body coordinates. The same applies to the symbol $\df^i[\psi,x]$, previously introduced to denote an infinitesimal transformation of a field $\psi^i$. In this chapter, I will reserve the letter $F$ for the function defining the LO effective Lagrangian.} the Lagrangian of the resulting continuous two-dimensional theory reads
\begin{equation}
\begin{split}
L\to\int\D^2\!\vec X\,\biggl\{\frac12\vr(\vec X)[\de_0\vec\p(\vec X,t)]^2&-\frac12E_1(\vec X)[\de_1\p^1(\vec X,t)]^2\\
&-\frac12E_2(\vec X)[\de_2\p^2(\vec X,t)]^2\biggr\}+\dotsb\;.
\end{split}
\label{2dchainlagcont}
\end{equation}
The ellipsis stands for terms of higher order in $\p^r$; in $d\geq2$ dimensions, the dynamics of the spring network does not resolve into a superposition of one-dimensional harmonic motions. Moreover, the model~\eqref{2dchainlagcont} does not correctly capture the physics of elastic solids even in the harmonic approximation. Namely, for purely transverse oscillations such that $\Pd{\p^r}{X^r}=0$ for each fixed $r$, there is no linear restoring force and the motion is strongly anharmonic. In spite of these flaws, the model~\eqref{2dchainlagcont} is sufficient to shed light on the fate of the emergent symmetries.

We always find exact invariance under genuine spacetime translations, spatial rotations and Galilei boosts regardless of the choice of the functions $\vr(\vec X)$ and $E_{1,2}(\vec X)$. This much is obvious from the discrete version~\eqref{2dchainlag} of the model. These symmetries reflect the properties of spacetime itself and are insensitive to the material structure of the medium. Should the medium be uniform, with constant $\vr(\vec X)$ and $E_{1,2}(\vec X)$, we will in addition have invariance under continuous ``internal translations'' of $\vec X$. This symmetry ensures conservation of pseudomomentum, and we can expect it to arise in the long-distance description of real crystalline materials. However, we cannot in general expect invariance under continuous ``internal rotations'' of $\vec X$, unless we demand $E_1=E_2$ and restrict the Lagrangian~\eqref{2dchainlagcont} to the lowest order in the power expansion in $\p^r$. This suggests that the macroscopic properties of real crystals tend to be homogeneous but anisotropic even in the continuum limit $a\to0$, which agrees with empirical observations. The dependence of the Lagrangian on $X^r$ is then constrained by the continuous internal translation invariance and the discrete group of point symmetries of the crystal lattice.


\subsection{Emergent Symmetries of Solids and Fluids}
\label{subsec:classicalemergentsym}

We shall now synthesize the above observations into a general setup for describing classical matter. Much of the discussion in this subsection is inspired by Chap.~4 of~\cite{Soper2008}. The basic assumption is that we are dealing with a thermodynamic state where quantum correlations are limited to short distances, typically due to thermal fluctuations. It is then possible to identify elements of the medium that are mutually distinguishable and can be assigned unique labels. This requirement limits the validity of the EFTs developed below to distances much longer than the scales characterizing quantum correlations and the discrete structure of matter.

The labels $X^i$ on the medium elements take values from a target space $\M$, which is typically some domain in the real space, $\R^d$. The domain would be finite for an isolated material object of a finite size. However, here we will mostly be concerned with properties of matter filling the entire space, where it is natural to take $\M\simeq\R^d$. In both cases, the labels $X^i$ represent comoving body coordinates, attached to a fixed medium element. The trajectory of the element is defined by giving its position as a function of time, $\vec x(X,t)$. This is the standard Lagrangian picture of continuum mechanics, which I used to formulate the toy models~\eqref{1dchainlagcont} and~\eqref{2dchainlagcont}. In order to highlight the genuine spacetime symmetries, it is however more convenient to invert the relation between $x^r$ and $X^i$. The time evolution of the medium is then specified by a map from the spacetime $M$ to $\M$ that assigns to every point $x^\m\in M$ the material coordinates $X^i(\vec x,t)\equiv X^i(x)$ of the element occupying this point. The uniqueness of the labeling of the medium elements is ensured by requiring that for any fixed time $t$, the map $x^r\to X^i(\vec x,t)$ is a diffeomorphism between space and $\M$. The advantage of this picture is that the labels $X^i$ take the same values in any reference frame. In other words, the functions $X^i(x)$ are scalar fields with respect to any spacetime symmetry. On the other hand, emergent symmetries acting solely on $X^i$ can then be treated as internal symmetries, that is point transformations on $\M$. 

\begin{figure}[t]
\sidecaption[t]
\includegraphics[width=2.9in]{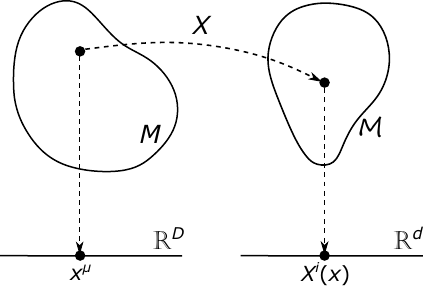}
\caption{The time evolution of the classical medium as a map from the spacetime $M$ to the target space $\M$ of body coordinates. The spacetime coordinates $x^\m$ in $M$ and the body coordinates $X^i$ in $\M$ can be chosen arbitrarily and independently of each other}
\label{fig:bodycoordinates}
\end{figure}

\begin{watchout}%
The explicit realization of the fields $X^i(x)$ depends on a choice of coordinates. Importantly, the coordinates $x^\m$ in the spacetime and the material coordinates $X^i$ can be fixed independently of each other. See Fig.~\ref{fig:bodycoordinates} for a visualization of this freedom. In the following, I will deal exclusively with matter in flat spacetimes. Accordingly, I will use standard Minkowski coordinates $x^\m$ for relativistic systems, and Cartesian coordinates $x^r$ augmented with Newtonian time $t$ for nonrelativistic systems. This still leaves us with the freedom to choose the body coordinates $X^i$ at will, which should be distinguished from any internal symmetry acting on $\M$. In Sect.~\ref{subsec:classicalspring}, I fixed this freedom tacitly by the definition of $X^r$ in terms of the discrete labels $i,j$. In order to avoid misunderstanding and to highlight the scalar nature of the body coordinates, I have now changed the notation from $X^r$ to $X^i$.
\end{watchout}

The scalar fields $X^i(\vec x,t)$ are the generalized coordinates of our continuum field-theoretic description of classical matter. The corresponding generalized velocity is $\de_0X^i(\vec x,t)$. How is this related to the actual local (Eulerian) velocity of the medium with respect to the laboratory frame? Recall that $X^i$ are comoving labels that cannot change along the trajectory of the medium element. Parameterizing the trajectory as $\vec x(t)$, the condition $\Od{X^i(\vec x(t),t)}{t}=0$ leads to $\de_0 X^i=-\dot x^r\de_rX^i$, where $\dot{\vec x}\equiv\Od{\vec x}t$ is the desired local velocity. The matrix $M^i_{\phantom ir}(\vec x,t)\equiv\de_rX^i(\vec x,t)$ is invertible thanks to the assumption that the map $x^r\to X^i(\vec x,t)$ is a diffeomorphism. It follows that
\begin{equation}
\dot x^r=-(M^{-1})^r_{\phantom ri}\de_0 X^i\;.
\label{Eulervelocity}
\end{equation}
There is also another, more elegant albeit slightly more mathematically advanced, way to express the local medium velocity in terms of derivatives of $X^i(x)$. First pick any function $f(X)$ on $\M$ and define the current
\begin{equation}
J^\m_f\equiv\frac{f(X)}{d!}\ve^{\m\n_1\dotsb\n_d}\ve_{i_1\dotsb i_d}\de_{\n_1}X^{i_1}\dotsb\de_{\n_d}X^{i_d}\;.
\label{Eulercurrent}
\end{equation}
By the antisymmetry of the \emph{Levi-Civita} (LC) symbol, $\smash{J^\m_f\de_\m X^i}=0$ for any $i$. Expan\-ding this as $\smash{J_f^0\de_0 X^i+J_f^rM^i_{\phantom ir}=0}$ gives $\smash{J_f^r=-J_f^0(M^{-1})^r_{\phantom ri}\de_0X^i}$. Using~\eqref{Eulervelocity}, we can thus rewrite the current as
\begin{equation}
J^\m_f=(J^0_f,J^0_f\dot{\vec x})\;.
\label{Jmuf}
\end{equation}

It also follows from the antisymmetry of the LC symbol that the current~\eqref{Eulercurrent} is conserved, $\smash{\de_\m J^\m_f=0}$, for any choice of the function $f(X)$. It is therefore natural to identify $\smash{J^0_f=f(X)\det M}$ with the density of a conserved charge as observed in the laboratory frame. The spatial part of the current, $\smash{J^r_f=J^0_f\dot x^r}$, then represents the flow of this charge. Finally, $f(X)$ itself is the density of the same charge in the comoving material coordinates, since $\det M$ is just the Jacobian of the transformation between the spacetime and material coordinate systems.

\begin{watchout}%
The origin of the currents $\smash{J^\m_f}$ can be traced back to the fact that the individual elements of our classical medium are distinguishable. Indeed, we could in principle attach to each element a dedicated ``charge.'' All such charges would be conserved by construction. The freedom to choose the function $f(X)$ in~\eqref{Eulercurrent} at will is just a continuous version of this observation. Operationally, the existence of infinitely many conservation laws, one for each $f(X)$, is not a problem. Namely, $\smash{J^\m_f}$ are not Noether currents in that their conservation does not require the EoM. The presence of such identically conserved currents does not constrain the local classical dynamics of the medium.

The above said does not mean that the currents $\smash{J^\m_f}$ are inevitably artifacts of the Lagrangian picture of continuum mechanics that are not macroscopically observable. Suppose our medium carries a conserved charge. This could count for instance the number of particles or, in the nonrelativistic limit, their mass. By choosing $f(X)$ as the density of the charge in the material coordinates, we get a current $\smash{J^\m_f}$ that describes macroscopic flow of this charge.
\end{watchout}

Mathematically, the currents~\eqref{Eulercurrent} descend from the $d$-forms $f(X)\D X^1\w\dotsb\w\D X^d$ on $\M$. Being top-dimensional, any such $d$-form is automatically closed. The closedness is preserved when the $d$-form is pulled back to the spacetime by the map $x^\m\to X^i(x)$. Taking the Hodge dual of the ensuing closed spacetime $d$-form then gives the current $J^\m_f$ with vanishing divergence.

Having established the basic setup, we are now finally in a position to discuss the emergent symmetries of various types of classical matter. I will focus on the two most common classical phases of matter: crystalline solids and fluids. Since this will to a certain extent merely collect and organize some observations I have already made previously, I can afford to be brief.

\runinhead{Solids} We will be interested in the quasiequilibrium dynamics of crystalline solids at distances much longer than the lattice spacing of the microscopic crystal structure. In this regime, one can expect the solid to be internally uniform, and thus be symmetric under the group $\R^d$ of continuous internal translations. These translations take a particularly simple form in ``Cartesian coordinates'' on $\M$, namely $\smash{X^i\xrightarrow{\eps}X^i+\eps^i}$. Barring possible quasi-invariant contributions, the effective Lagrangian can then only depend on $X^i(x)$ through their derivatives. In addition, the dependence on $X^i(x)$ is constrained by the point symmetry of the crystal lattice. This can be imposed order by order in powers of $X^i(x)$ using tensor methods. I will not go into detail here, and rather refer the reader to Sect.~10 of~\cite{Landau1986} for further discussion. For the sake of illustration, I will only deal with the special case of \emph{isotropic} solids, which feature a full $\gr{SO}(d)$ symmetry under continuous rotations of $X^i(x)$. Exact, full isotropy cannot really be achieved for any crystalline material. However, some polycrystalline materials are effectively isotropic at distances much longer than the typical size of a grain, thanks to the random orientation of the grains.

\runinhead{Fluids} The key difference between solids and fluids is that in the latter, there are no restoring elastic forces that would counteract shear strain. This enables macroscopic flow, whose physics is largely affected by dissipation. Unfortunately, dissipative effects are notoriously difficult to include in Lagrangian field theory. In order that the EFT for fluids we construct be meaningful, we thus need to make some simplifying assumptions. Namely, we shall restrict ourselves to physical processes that do not involve entropy production. Also, we will only consider states of the fluid that are perturbations of a uniform equilibrium where the entropy density in the material coordinates is constant. Such fluids are called \emph{barotropic}; their thermodynamic state can be described by a single variable such as pressure or density. Thus, our EFT will be able to capture processes such as sound propagation, but not, for instance, convective heat transfer due to an initial temperature gradient.

By definition, the local density of energy and entropy, and thus also of any other extensive property, of a barotropic fluid corresponds to a constant function $f$ on $\M$. At the same time, the macrostate of the fluid is completely specified by its local Eulerian velocity and density $\smash{J^0_f}$ observed in the laboratory frame. The same should be true for the Lagrangian, which is directly connected to observables such as energy density and pressure. The field-theoretic description of fluids in terms of the scalar fields $X^i(x)$ should therefore be invariant under any point transformation on $\M$ that preserves the currents~\eqref{Eulercurrent}. This is not mere freedom of choice of coordinates on $\M$. It corresponds to an actual physical reshuffling of the elements of the fluid that does not affect its macroscopic properties. Now $\smash{J^0_f=f\det M}$ requires that the point transformation on $\M$ preserves $\det M$, and so must have a unit Jacobian. We conclude that the dynamics of a barotropic fluid possesses an emergent symmetry under the group of \emph{volume-preserving diffeomorphisms} (VPDs) on $\M$, $\gr{SDiff}(\M)$. This is an infinite-dimensional symmetry group that includes the internal translations and rotations of solids as a subgroup. See~\cite{Arnold1998a} for a detailed account of the role of the diffeomorphism group in hydrodynamics, and~\cite{Jackiw2004,Andersson2021} for two somewhat complementary overviews of the variational approach to fluids.


\section{Nonlinear Realization of Emergent Symmetry}
\label{sec:classicalcoset}

We have done a serious amount of work to carefully identify the emergent symmetries of classical matter. Our effort will now pay dividends in that the next steps will be fairly straightforward. In the absence of degrees of freedom sensitive to other symmetries, the symmetry group of a classical medium is $G\simeq G_\mathrm{s.t.}\times G_\mathrm{int}$. Here $G_\mathrm{s.t.}$ is whatever spacetime symmetry is appropriate for the system at hand, such as Aristotelian, Galilei or Poincar\'e. The group $G_\mathrm{int}$ collects all the emergent symmetries that act on the scalar fields $X^i(x)$ as internal. For the reader's convenience, the choices of $G_\mathrm{int}$ in case of uniform solids and barotropic fluids are summarized in the first two columns of Table~\ref{tab:emergentsym}. The product structure, $G\simeq G_\mathrm{s.t.}\times G_\mathrm{int}$, of the symmetry group already appeared in Sect.~\ref{sec:quantumtwist}, and we will be able to largely follow the path paved therein.

\begin{table}[t]
\caption{Overview of the emergent symmetries and related subgroups in different phases of classical matter. The $G_\mathrm{cryst}$ symbol denotes the crystallographic point group of a crystal lattice. Also, $\smash{\gr{SDiff}_0(\R^d)}$ is the group of VPDs in $\R^d$ that fix the origin. The coset space $G_\mathrm{int}/H_\mathrm{int}$ is equivalent to $\R^d$ in all cases. The unbroken subgroup $H_\vp$ indicates the symmetry of a uniform, static equilibrium state. The subscript ``diag'' denotes diagonal symmetries whose actions on the spatial and body coordinates are locked to each other. The values shown in the last column assume that the subgroup of time-independent transformations in $G_\mathrm{s.t.}$, acting only on the spatial coordinates, is the Euclidean group $\gr{ISO}(d)$. The factor of $\R$ in $H_\vp$ corresponds to time translations}
\label{tab:emergentsym}
\begin{tabular}{p{2.1cm}p{3.4cm}p{2.7cm}p{2.9cm}}
\hline\noalign{\smallskip}
Material & Emergent symmetry $G_\mathrm{int}$ & Isotropy group $H_\mathrm{int}$ & Unbroken subgroup $H_\vp$ \\
\noalign{\smallskip}\svhline\noalign{\smallskip}
Solid & $G_\mathrm{cryst}\ltimes\R^d$ & $G_\mathrm{cryst}$ & $(G_\mathrm{cryst}\ltimes\R^d)_\mathrm{diag}\times\R$\\
Isotropic solid & $\gr{ISO}(d)$ & $\gr{SO}(d)$ & $[\gr{ISO}(d)]_\mathrm{diag}\times\R$ \\
Fluid & $\gr{SDiff}(\R^d)$ & $\gr{SDiff}_0(\R^d)$ & $[\gr{ISO}(d)]_\mathrm{diag}\times\R$ \\
\noalign{\smallskip}\hline\noalign{\smallskip}
\end{tabular}
\end{table}


\subsection{Field Variables and Unbroken Symmetry}
\label{subsec:classicalunbroken}

To start with, the isotropy group of the spacetime origin is $H_0\simeq H_\mathrm{s.t.}\times G_\mathrm{int}$, where $H_\mathrm{s.t.}$ collects all transformations from $G_\mathrm{s.t.}$ that fix the origin (typically rotations and boosts). I will always assume that $G_\mathrm{int}$ includes a set of mutually commuting translations that act transitively on $\M\simeq\R^d$. The isotropy subgroup $H_{(X_0,0)}$ is then the same, up to conjugation by an element of $G_\mathrm{int}$, for any choice of the reference point $X^i_0$. It is convenient to set $X^i_0=0$ so that $H_{(X_0,0)}\simeq H_\mathrm{s.t.}\times H_\mathrm{int}$ where $H_\mathrm{int}$ consists of all emergent symmetries that fix the origin in $\M$. The concrete isotropy subgroups for solids and fluids are listed in the third column of Table~\ref{tab:emergentsym}. Note that in case of fluids, both $G_\mathrm{int}\simeq\gr{SDiff}(\R^d)$ and $H_\mathrm{int}\simeq\gr{SDiff}_0(\R^d)$ are infinite-dimensional. However, the coset space $G_\mathrm{int}/H_\mathrm{int}$ is equivalent to $\R^d$ for any choice of $G_\mathrm{int}$ thanks to the assumed transitive action of $G_\mathrm{int}$ on $\M$. This reconfirms that the dynamical degrees of freedom of the EFT for classical matter will always be the $d$ NG fields $X^i(x)$. That is in contrast to the \emph{agnostic nonlinear realization} (cf.~Sects.~\ref{subsec:spacetimediscussion} and~\ref{subsec:vectorIHC}), whose application to fluids would suggest an EFT with an infinite number of would-be NG fields; see~\cite{Nicolis2014a} for details.

Before we proceed to the construction of effective actions, let us check which of the symmetries in $G$ actually are spontaneously broken. Suppose we are interested in the physics of fluctuations around a static equilibrium state, characterized by the time-independent \emph{vacuum expectation values} (VEVs)\footnote{A different choice of the reference state of our classical medium might require a modification of the discussion below. The change would however not affect the general nonlinear realization setup, which only depends on the groups $G_\mathrm{s.t.}$, $H_\mathrm{s.t.}$, $G_\mathrm{int}$ and $H_\mathrm{int}$.}
\begin{equation}
\vev{X^i(\vec x,t)}\equiv\vp^i(\vec x)\;.
\label{classicalVEV}
\end{equation}
Due to the product structure of $G$, an element $(g_\mathrm{s.t.},g_\mathrm{int})\in G$ transforms the spacetime and body coordinates as $T_g:(X^i,x^\m)\to(X'^i,x'^\m)\equiv(\DF^i(X,g_\mathrm{int}),\DX^\m(x,g_\mathrm{s.t.}))$. Here the $x^\m$-independent functions $\DF^i$ realize the action of $G_\mathrm{int}$ on $\M$, whereas the $X^i$-independent functions $\DX^\m$ realize the action of $G_\mathrm{s.t.}$ on the spacetime. The unbroken subgroup $H_\vp$ consists of transformations preserving~\eqref{classicalVEV},
\begin{equation}
\DF^i(\vp(x),h_\mathrm{int})=\vp^i(\DX(x,h_\mathrm{s.t.}))\;,\qquad
(h_\mathrm{s.t.},h_\mathrm{int})\in H_\vp\;.
\label{classicalHvp}
\end{equation}
This condition has the geometric meaning of \emph{equivariance} of $\vp^i$ as a map $M\to\M$ under the action of $H_\vp$, that is $\DF\circ\vp=\vp\circ\DX$. Since $\vp^i(\vec x)$ is a diffeomorphism between the coordinate space and $\M$, \eqref{classicalHvp} defines a one-to-one correspondence between the maps $\DF^i$ and $\DX^r$. In this sense, $H_\vp$ is the ``diagonal subgroup'' of $G_\mathrm{s.t.}\times G_\mathrm{int}$. It also includes those transformations from $G_\mathrm{s.t.}$ that act on $x^r$ trivially, that is time translations. An overview of the unbroken subgroups in solids and fluids is given in the last column of Table~\ref{tab:emergentsym}.


\subsection{Building Blocks for Construction of Effective Actions}
\label{subsec:classicalbuildingblocks}

In the absence of other degrees of freedom, we are dealing with the $d$ scalar fields $X^i(x)$.\footnote{For relativistic systems, there is an alternative formulation of the EFT where the $X^i$ are completed to a Lorentz vector of dynamical variables. See Sect.~6 of~\cite{Liu2018} for an introduction and references.} All of these are of the NG type, realizing nonlinearly the emergent internal translations on $\M\simeq\R^d$. Let us denote the generators of the translations as $\Pi_i$. I will now finally fix the freedom to choose coordinates on $\M\simeq G_\mathrm{int}/H_\mathrm{int}\simeq\R^d$ at will by parameterizing it as $U(X)\equiv\exp(\I X^i\Pi_i)$. This gives a precise definition of the previously mentioned ``Cartesian coordinates'' on $\M$, in which the translations from $G_\mathrm{int}$ act on $X^i$ by trivial shifts, $\smash{X^i\xrightarrow{\eps}X^i+\eps^i}$. According to~\eqref{MCtwisted}, the \emph{Maurer--Cartan} (MC) \emph{form} is then given by
\begin{equation}
\mc(X,x)=\Pi_i\D X^i+P\cdot\D x\;,
\label{classicalMC}
\end{equation}
where $P_{\fr\m}$ generates spacetime translations. The spacetime coframe is trivial, that is, $\smash{\vec e^{*\fr\m}=\d^{\fr\m}_\n\D x^\n}$. Accordingly, it is not necessary to use different notations for frame and coordinate-basis indices. The covariant derivatives of the NG fields are simply $\cd_\m X^i=\de_\m X^i$. This makes it possible to use the same power counting as for superfluids (Sect.~\ref{subsec:twistsuperfluids}), whereby an $n$-th partial derivative of $X^i$ is assigned the counting degree $n-1$. The LO of the EFT is thus defined by a Lagrangian density where every field $X^i$ carries exactly one derivative.

Invariance of the effective action is ensured as follows. First, invariance under the whole internal symmetry $G_\mathrm{int}$ is guaranteed by using the MC form as a building block and imposing solely the linearly realized isotropy group $H_\mathrm{int}$. Similarly, one has to impose explicitly invariance under the linearly realized spacetime isotropy group, $H_\mathrm{s.t.}$. Finally, invariance under spacetime translations requires that the Lagrangian density does not depend explicitly on the spacetime coordinates.

\begin{watchout}%
In case of fluids, the coset space $G_\mathrm{int}/H_\mathrm{int}\simeq\gr{SDiff}(\R^d)/\gr{SDiff}_0(\R^d)$ is not reductive, hence the line of reasoning using the transformation properties~\eqref{MCspacetimetransforcomps} of the MC form does not necessarily apply. In such a situation, we may have to impose by hand invariance under the entire emergent symmetry group $G_\mathrm{int}$. Luckily, we know a priori how VPDs act on the coordinates $X^i$ and thus also on the 1-forms $\D X^i$ on $\M$. This will make it possible for us to construct an EFT for fluids in Sect.~\ref{subsec:classicalperfectfluids}.
\end{watchout}


\section{Effective Field Theory of Classical Matter}
\label{sec:classicalfluidssolids}

I will now carry out the above-outlined program for several types of physical systems of interest. I will start with what in many ways is the simplest case: an isotropic relativistic solid. Having warmed up, we shall then have a look at the phenomenologically more relevant nonrelativistic solids. At the very end, we will return to the most nontrivial case of fluids with their infinite-dimensional symmetry group.


\subsection{Relativistic Solids}
\label{subsec:classicalsolids}

Consider a system with relativistic microscopic dynamics that settles to a thermodynamic equilibrium state with the symmetries of an isotropic solid. The full symmetry is $G_\mathrm{s.t.}\times G_\mathrm{int}\simeq\gr{ISO}(d,1)\times\gr{ISO}(d)$, where the first factor is the spacetime Poincar\'e group and the second one is the emergent internal symmetry. With one derivative per field, the only way to obey the linearly realized Lorentz symmetry, $H_\mathrm{s.t.}\simeq\gr{SO}(d,1)$, is to pairwise contract indices on $\de_\m X^i$. It follows that the LO effective Lagrangian is some function of the Lorentz-invariant matrix operator
\begin{equation}
\Xi^{ij}\equiv\de_\m X^i\de^\m X^j\;.
\label{Xidef}
\end{equation}
It remains to impose invariance under $H_\mathrm{int}\simeq\gr{SO}(d)$. Since $\Xi^{ij}$ is a symmetric tensor, any $H_\mathrm{int}$-invariant function of $\Xi^{ij}$ can only depend on its $d$ real eigenvalues. These can be encoded in any set of $d$ algebraically independent invariants constructed out of $\Xi^{ij}$, for instance the traces of the first $d$ powers of $\Xi^{ij}$. The LO effective action for relativistic isotropic solids then reads
\begin{equation}
S_\text{eff}\{X\}=\int\D^D\!x\,F\bigl(\tr\Xi(x),\tr\Xi(x)^2,\dotsc,\tr\Xi(x)^d\bigr)+\dotsb\;,
\label{EFTrelsolids}
\end{equation}
where $F$ is a function of the displayed arguments and the ellipsis indicates corrections of higher order in the derivative expansion. These include operators with on average more than one derivative per $X^i$, and Lorentz and internal indices contracted in a way respecting the $H_\mathrm{s.t.}\times H_\mathrm{int}\simeq\gr{SO}(d,1)\times\gr{SO}(d)$ symmetry.

With all the background we had built up, the path to~\eqref{EFTrelsolids} was very short. There are however a couple of potential loopholes to close. First, I tacitly assumed the effective Lagrangian to be strictly invariant under all the symmetries. Can there be any quasi-invariant contributions to the Lagrangian? Here I refer the reader to~\cite{Delacretaz2015a}, which showed that the symmetries of a relativistic isotropic solid do not allow any genuinely quasi-invariant Lagrangians in $d=2$ or $3$ spatial dimensions. It is plausible to assume that the same conclusion holds for any $d\geq2$.

Second, we still need to check whether the tentative equilibrium~\eqref{classicalVEV} is a stable state of the EFT~\eqref{EFTrelsolids}. To start with, note that the LO EoM of the EFT is
\begin{equation}
\de_\m\left(\PD{F}{\Xi^{ij}}\de^\m X^j\right)=0\;.
\end{equation}
In a uniform equilibrium, the VEV of the invariants~\eqref{Xidef}, $\vev{\Xi^{ij}}=-\vec\nabla\vp^i\cdot\vec\nabla\vp^j$, should be coordinate-independent. The EoM then implies that the functions $\vp^i(\vec x)$ must be harmonic, $\vec\nabla^2\vp^i=0$. Acting with the Laplace operator on $\vev{\Xi^{ij}}$, we get in turn that $(\de_r\vec\nabla\vp^i)\cdot(\de_r\vec\nabla\vp^j)=0$. Setting $i=j$ herein shows that all partial derivatives of $\vec\nabla\vp^i$ vanish. Hence, $\vp^i(\vec x)$ is a linear function of coordinates, $\vp^i(\vec x)=M^i_{\phantom ir}x^r+c^i$ with a constant invertible matrix $M^i_{\phantom ir}$ and constant $c^i$. The set of constants $c^i$ can be removed by shifting the origin of coordinates. Finally, suppose that the generators $\Pi_i$ of internal translations were chosen as a basis of the standard vector representation of $H_\mathrm{int}\simeq\gr{SO}(d)$. By using the singular value decomposition augmented with appropriate orthogonal rotations of the fields $X^i$ and coordinates $x^r$, the matrix $M^i_{\phantom ir}$ can be made diagonal and positive-semidefinite. Invariance under the unbroken diagonal $\gr{SO}(d)$ rotations of spatial and body coordinates then requires that $M^i_{\phantom ir}\propto\d^i_r$. Finally, the proportionality factor can be absorbed into a simultaneous rescaling of the body coordinates $X^i$ and the generators $\Pi_i$. We therefore conclude that, without loss of generality, the uniform equilibrium of the solid can be described by
\begin{equation}
\vev{X^i(\vec x,t)}=\d^i_rx^r\;.
\label{solidgroundstate}
\end{equation}
Below, I will implicitly assume the same ground state for both nonrelativistic solids and (relativistic or nonrelativistic) fluids.

Further constraints on the parameters of the effective Lagrangian arise from requiring that the state~\eqref{solidgroundstate} is (meta)stable with respect to small fluctuations. To that end, we parameterize the fluctuations of $X^i$ by a set of NG fields, $\pi^i(x)\equiv X^i(x)-\d^i_rx^r$. In terms of these fields, we have
\begin{equation}
\Xi^{ij}=-\d^{ij}-(\d^{ir}\de_r\pi^j+\d^{jr}\de_r\pi^i)+\de_\m\pi^i\de^\m\pi^j\;.
\label{Xiparampi}
\end{equation}
It is now a matter of straightforward algebra to expand the action~\eqref{EFTrelsolids} to second order in $\pi^i$ and compute the excitation spectrum. The reader is most welcome to do this exercise, or check~\cite{Endlich2013} where the special case of $d=3$ spatial dimensions is analyzed. I will not work out the details here, since I will address the same problem in the arguably more realistic setting of nonrelativistic solids below.


\subsection{Nonrelativistic Supersolids}
\label{subsec:classicalsupersolids}

Most naturally occurring solid materials lie safely within the nonrelativistic domain. It therefore appears more appropriate to demonstrate the utility of our EFT formalism in a setting that connects more directly to classical theory of elasticity~\cite{Landau1986}. However, in order that the discussion below is not a mere rehash of Sect.~\ref{subsec:classicalsolids} using nonrelativistic notation, I will add a new physical ingredient. Namely, some materials are known to enter at low temperatures a quantum phase called suggestively \emph{supersolid}. In this phase, matter exhibits a combination of crystalline solid order and superflow. The latter stems from the presence of an internal, spontaneously broken $\gr{U}(1)$ symmetry. Once we have found an effective action for supersolids, we will be able to recover an EFT for ordinary solids by decoupling the superfluid NG boson. The content of this subsection is heavily influenced by~\cite{Son2005a}.

The mathematical setup closely follows the discussion of nonrelativistic superfluids in Sect.~\ref{subsec:vectorunphysical}, to which I refer the reader for details. In order to avoid excessive cross-references, I will however spell out the main features of the setup explicitly. The full symmetry of the nonrelativistic supersolid is given by the Bargmann group augmented with the emergent $\gr{ISO}(d)$ symmetry of a classical isotropic solid, that is
\begin{equation}
G\simeq\gr{SO}(d)\ltimes\{\R_K^d\ltimes[\R^D\times\gr{U}(1)_Q]\}\times\gr{ISO}(d)\;.
\end{equation}
Here the $\gr{SO}(d)$ factor represents spatial rotations, $\R_K^d$ Galilei boosts, $\R^D$ spacetime translations, and $\gr{U}(1)_Q$ the internal symmetry counting the number of particles. The symmetry group does not have the simple direct product structure $G_\mathrm{s.t.}\times G_\mathrm{int}$. This is due to the dual nature of $\gr{U}(1)_Q$, which is an internal symmetry but simultaneously centrally extends the spacetime Galilei group. We will however still be able to follow the philosophy of Sect.~\ref{sec:classicalcoset} with minor modifications. In particular the spacetime isotropy group $H_0$ is in this case $\smash{H_0\simeq[\gr{SO}(d)\ltimes\R_K^d]\times\gr{U}(1)_Q\times\gr{ISO}(d)}$.

In order to be able to apply our basic EFT framework for spacetime symmetry, we need to realize the Galilei boosts nonlinearly. To that end, we need a Galilei vector order parameter $A^\m=(A^0,\vec A)$, choosing a timelike reference point, $\smash{A^\m_0=(a,\vec0)}$ with $a\neq0$. This is accompanied by two other order parameters, a complex scalar $\psi$ charged under $\gr{U}(1)_Q$, and the body coordinates $X^i$. Taking any $\psi_0\neq0$ and $\smash{X^i_0=0}$ leads to $\smash{H_{((\psi_0,A_0,X_0),0)}\simeq\gr{SO}(d)\times\gr{SO}(d)}$, where the two factors act respectively on the spatial and body coordinates. The coset space relevant for the nonlinear realization of the symmetry is $\smash{H_0/H_{((\psi_0,A_0,X_0),0)}\simeq\R_K^d\times\gr{U}(1)_Q\times\R^d}$. The last factor of $\R^d$ is new compared to Sect.~\ref{subsec:vectorunphysical} and carries the solid degrees of freedom. The whole coset space is parameterized by the NG variables $\pi$, $\x^r$ and $X^i$ through
\begin{equation}
U(\pi,\vec\x,X)\equiv\E^{\I \pi Q}\E^{\I\skal\x K}\exp(\I X^i\Pi_i)\;,
\end{equation}
where $Q$ is the generator of $\gr{U}(1)_Q$ and $K^r$ that of Galilei boosts. The MC form is calculated using the commutation relations of the Bargmann group,
\begin{equation}
\begin{split}
\mc(\pi,\vec\x,X,\vec x,t)={}&Q[\D\pi-\vec\x\cdot\D\vec x+(1/2)\vec\x^2t]+\vec K\cdot\D\vec\x+\Pi_i\D X^i\\
&+H\D t+\vec P\cdot(\D\vec x-\vec\x\D t)\;.
\end{split}
\end{equation}
The second line herein defines the spacetime coframe, $\smash{\vec e^{*\fr0}=\D t}$ and $\vec e^{*\fr r}=\d^{\fr r}_s(\D x^s-\x^s\D t)$. This allows one to extract the covariant derivatives of all the NG fields from the $\mcb$ part of the MC form,
\begin{equation}
\begin{alignedat}{2}
\cd_{0}\pi&=\de_0\pi+\skal\x\nabla\pi-\vec\x^2/2\;,\qquad&
\cd_{r}\pi&=\de_r\pi-\x_r\;,\\
\cd_{0}\x^r&=(\de_0+\skal\x\nabla)\x^r\;,\qquad&
\cd_{s}\x^r&=\de_s\x^r\;,\\
\cd_{0}X^i&=(\de_0+\skal\x\nabla)X^i\;,\qquad&
\cd_{r}X^i&=\de_rX^i\;.
\end{alignedat}
\end{equation}
For simplicity of notation, I have already dropped the underscores distinguishing spacetime and frame indices.

The would-be NG field $\x^r(x)$ is unphysical and can be eliminated by imposing the \emph{inverse Higgs constraint} (IHC) $\cd_r\pi=0$, or $\vec\x(x)=\vec\nabla\pi(x)$. This assigns $\x^r$ the counting degree zero. As a consequence, $\cd_\m\x^r$ has degree one, whereas both $\cd_\m\pi$ and $\cd_\m X^i$ are of degree zero. The building blocks we have to construct the LO effective Lagrangian are therefore
\begin{equation}
\cd_0\pi\to\de_0\pi+(\vec\nabla\pi)^2/2\;,\quad
\cd_0X^i\to(\de_0+\vec\nabla\pi\cdot\vec\nabla)X^i\;,\quad
\cd_rX^i\to\de_rX^i\;.
\label{supersolidcovder}
\end{equation}
These take automatically care of the nonlinearly realized symmetries under Galilei boosts, $\gr{U}(1)_Q$ transformations and internal translations of $X^i$. To ensure invariance under spatial rotations, note that both $\cd_0\pi$ and $\cd_0X^i$ are scalars, whereas $\cd_rX^i$ is a vector under spatial $\gr{SO}(d)$. All spatial indices thus have to be contracted into
\begin{equation}
\tilde\Xi^{ij}\equiv\d^{rs}\cd_rX^i\cd_sX^j=\vec\nabla X^i\cdot\vec\nabla X^j\;.
\label{Xitildedef}
\end{equation}
This symmetric tensor is related to the previously defined matrix $\smash{M^i_{\phantom ir}(x)\equiv\de_rX^i(x)}$ by $\tilde\Xi=MM^T$. Note that there is no additional, algebraically independent rotationally invariant operator where the spatial indices on $\cd_rX^i$ would be contracted with the LC symbol, since $\smash{\ve^{r_1\dotsb r_d}\cd_{r_1}X^{i_1}\dotsb\cd_{r_d}X^{i_d}=\ve^{i_1\dotsb i_d}\det M=\ve^{i_1\dotsb i_d}\sqrt{\det\tilde\Xi}}$. Altogether, the LO effective action for nonrelativistic supersolids is given by~\cite{Son2005a}
\begin{equation}
S_\mathrm{eff}\{\pi,X\}=\int\D^D\!x\,F\bigl(\cd_0\pi(x),\cd_0X(x),\tilde\Xi(x)\bigr)+\dotsb\;.
\label{EFTsupersolids}
\end{equation}
The indices on $\cd_0X^i$ and $\tilde\Xi^{ij}$ are to be contracted in a way that respects invariance under the internal $\gr{SO}(d)$ rotations.

Again, I have not explicitly considered the possibility of quasi-invariant contributions to the Lagrangian. There are $d+1$ such terms, constructed solely out of the superfluid mode $\pi$; cf.~Sect.~\ref{subsec:vectorunphysical}. However, these are only relevant beyond the LO of the derivative expansion. An additional quasi-invariant Lagrangian existing in $d=2$ dimensions and including the solid variable $X^i$ was found in~\cite{Delacretaz2015a}. This likewise contributes only at higher orders of the derivative expansion.


\subsection{Nonrelativistic Solids}
\label{subsec:classicalNRsolids}

Let us now see whether we can recover the physics of classical (isotropic) solids from the supersolid EFT~\eqref{EFTsupersolids}. It appears we should be able to simply erase the NG field $\pi(x)$. This certainly does not interfere with the nonlinear realization of the symmetries of solids on $X^i$, since that is entirely independent of $\pi$. The building blocks~\eqref{supersolidcovder} then boil down to $\smash{\de_0X^i}$ and $\smash{\de_rX^i}$. Invariance under spatial rotations again requires that $\smash{\de_rX^i}$ only enters the LO EFT through the combination $\tilde\Xi^{ij}$. However, the time derivative, $\smash{\de_0X^i}$, is potentially problematic since unlike $\smash{\cd_0X^i}$, it is not invariant under Galilei boosts. This actually makes sense: recall that the nonrelativistic kinetic term in~\eqref{1dchainlagcont} or~\eqref{2dchainlagcont} is only quasi-invariant under Galilei boosts. Guided by the analogy, the kinetic term for the solid should be $\smash{(1/2)\vr(\vec x,t)[\dot{\vec x}(X,t)]^2}$. Here $\vec x(X,t)$ is the Lagrangian variable we worked with in Sect.~\ref{sec:classicalemergent} and $\vr(\vec x,t)$ is the local mass density of the solid in the laboratory frame. Now $\smash{\vr(x)=\vr_0\det M(x)=\vr_0\sqrt{\det\tilde\Xi(x)}}$, where $\vr_0$ is the mass density of the solid in the body coordinates, which is constant thanks to the assumed uniformity of the solid. Using finally~\eqref{Eulervelocity}, the LO effective action for the classical nonrelativistic isotropic solid can be written as
\begin{equation}
\begin{split}
S_\mathrm{eff}\{X\}=\int\D^D\!x\,\Bigl\{&\frac{\vr_0}2\sqrt{\smash{\det\tilde\Xi(x)}\hbox to0pt{\phantom{$\sum$}}}[\tilde\Xi(x)^{-1}]_{ij}\de_0 X^i(x)\de_0X^j(x)\\
&-\Fa(\tilde\Xi(x))\Bigr\}+\dotsb\;.
\end{split}
\label{EFTsolids}
\end{equation}
The function $\smash{\Fa(\tilde\Xi)}$ defines the free energy of static elastic deformations of the solid. Our derivation of the kinetic term in~\eqref{EFTsolids} was based on an educated guess. It can however also be recovered by starting from an exactly Galilei-invariant operator in the EFT~\eqref{EFTsupersolids} for supersolids and adding a surface term, proportional to the gradient of $\pi(x)$. This explains why setting $\pi\to0$ does not destroy Galilei invariance altogether but renders~\eqref{EFTsolids} quasi-invariant. See~\cite{Son2005a} for details.

I will now review the basic elastic properties of isotropic solids, following closely~\cite{Landau1986}. As in Sect.~\ref{subsec:classicalspring}, the displacement of the solid element carrying label $X^i$ from its equilibrium position is given by $\p^i(X,t)\equiv\d^i_rx^r(X,t)-X^i$. This parameterization is however not suitable for an EFT where the scalar fields $X^i(x)$ are the dynamical variables. We need to invert the relation between $x^r$ and $\p^i$, which is not possible in a closed form. Luckily, we will only need the deviation of $X^i$ from its VEV to first order in $\p^i$,
\begin{equation}
X^i(x)=\d^i_rx^r-\p^i(x)+\bigO(\p^2)\;.
\label{Xphiparam}
\end{equation}
To the same order, the matrix variable $\tilde\Xi^{ij}$ reads
\begin{equation}
\tilde\Xi^{ij}(x)=\d^{ij}-[\d^{ir}\de_r\p^j(x)+\d^{jr}\de_r\p^i(x)]+\bigO(\p^2)\;.
\end{equation}
With the shorthand notation $\p_r(x)\equiv\d_{ri}\p^i(x)$, small deformations of the solid are therefore encoded in the \emph{strain tensor}
\begin{equation}
e_{rs}(x)\equiv\frac12[\de_r\p_s(x)+\de_s\p_r(x)]\;.
\label{straindef}
\end{equation}
By its definition, $\p^i(x)$ transforms as a vector under the unbroken diagonal $\gr{SO}(d)$ subgroup. The strain tensor therefore behaves as a rank-2 symmetric tensor.

The energy cost of a small static elastic deformation of the solid is governed by the expansion of $\Fa(\tilde\Xi)$ in powers of $\tilde\Xi^{ij}-\d^{ij}$, or directly the strain tensor. The expansion cannot contain a linear term, as that would immediately imply an instability of the equilibrium state where $\vev{\p^i(x)}=0$. The physics of small elastic deformations is therefore dominated by the part of $\Fa(\tilde\Xi)$ quadratic in the strain tensor. Invariance under unbroken diagonal rotations admits two independent quadratic operators at the lowest order in derivatives,
\begin{equation}
\Fa=\m\tr e^2+\frac\l2(\tr e)^2+\dotsb\;.
\label{FelLame}
\end{equation}
Here $\l,\m$ are known as the \emph{Lam\'e coefficients}, and the ellipsis denotes contributions of higher order in $\p^i$ or derivatives. It is convenient to rewrite~\eqref{FelLame} as
\begin{equation}
\Fa=\m\tr\left(e-\frac{\tr e}d\un\right)^2+\frac K2(\tr e)^2+\dotsb\;,
\label{Felbulk}
\end{equation}
where $K\equiv\l+(2\m/d)$. Importantly, there are configurations of the solid for which either of the two terms in~\eqref{Felbulk} vanishes whereas the other is nonzero. The $\m$-term is zero if and only if $e_{rs}$ is proportional to $\d_{rs}$, that is for uniform dilatations or contractions of the solid. Such overall volume change is detected by $\tr e=\skal\nabla\p$, which measures the deviation of the Jacobian $\det M$ from unity. For this reason, the parameter $K$ is usually called \emph{bulk modulus}. On the other hand, for deformations of shape that do not affect the volume of the solid, the $K$-term in~\eqref{Felbulk} vanishes. The energy of such deformations is measured by the \emph{shear modulus} $\m$. Since the two terms in~\eqref{Felbulk} can be set to zero independently of each other, bulk stability of the solid requires that
\begin{equation}
K>0\;,\qquad
\m>0\;.
\label{solidstabilitybound}
\end{equation}

\begin{illustration}%
The constraints~\eqref{solidstabilitybound} may look trivial, but they have very observable consequences for small oscillations of the solid. To see this, it is sufficient to consider the part of~\eqref{EFTsolids}, bilinear in the fields $\p^i(x)$,
\begin{equation}
\La_\mathrm{eff}=\frac{\vr_0}2\d^{rs}\de_0\p_r\de_0\p_s-\left[\m\tr e^2+\frac\l2(\tr e)^2\right]+\dotsb\;.
\end{equation}
The corresponding EoM reads
\begin{equation}
\vr_0\de_0^2\p_r\approx\m\vec\nabla^2\p_r+(\l+\m)\de_r\skal\nabla\p\;,
\end{equation}
where the symbol $\approx$ indicates a linear approximation. This has plane-wave solutions of the type
\begin{equation}
\p_r(\vec x,t)=\hat\p_r\E^{-\I Et}\E^{\I\skal px}\;,
\end{equation}
where $\hat\p_r$ is an amplitude. Solutions for which $\vec{\hat\p}\parallel\vec p$ and $\vec{\hat\p}\perp\vec p$ represent respec\-ti\-vely \emph{longitudinal} and \emph{transverse} sound. The corresponding phase velocities are
\begin{equation}
v_\mathrm{L}=\sqrt{\frac{\l+2\m}{\vr_0}}=\sqrt{\frac{K+2\m(1-1/d)}{\vr_0}}\;,\qquad
v_\mathrm{T}=\sqrt{\frac{\m}{\vr_0}}\;.
\end{equation}
The stability condition $K>0$ implies a specific hierarchy between the velocities of longitudinal and transverse sound,
\begin{equation}
\frac{v_\mathrm{L}}{v_\mathrm{T}}>\sqrt{2\left(1-\frac1d\right)}\;.
\end{equation}
This relation would have been completely invisible, had we focused solely on the spectrum of sound waves itself. When looking for the constraints on parameters of an EFT imposed by the stability requirement, it is therefore all-important to consider the whole configuration space of the EFT.
\end{illustration}


\subsection{Perfect Fluids}
\label{subsec:classicalperfectfluids}

As the last stop of our exploration of classical matter, we now return to fluids. I will restore full Poincar\'e invariance, partially because relativistic fluids are common in astrophysics and particle physics, and partially because it makes the analysis simpler.

As argued in Sect.~\ref{subsec:classicalsolids}, the NG fields $X^i(x)$ necessarily enter the effective Lagrangian through the combination $\smash{\Xi^{ij}\equiv\de_\m X^i\de^\m X^j}$. However, not all actions of the solid type~\eqref{EFTrelsolids} are consistent with the required invariance of the fluid action under $G_\mathrm{int}\simeq\gr{SDiff}(\R^d)$. The transition from solids to fluids can be viewed as a ``fine-tuning'' of the effective couplings of the former. There is only one algebraically independent function of $\Xi^{ij}$ that fits the bill, namely $\det\Xi\equiv\abs\Xi$. Thus, the LO effective Lagrangian of a relativistic perfect (barotropic) fluid reads
\begin{equation}
S_\mathrm{eff}\{X\}=\int\D^D\!x\,F(\abs{\Xi(x)})+\dotsb\;.
\label{EFTfluids}
\end{equation}
With the help of the identity $\Pd{\abs\Xi}{\Xi^{ij}}=\abs\Xi(\Xi^{-1})_{ji}$, we readily extract the corresponding LO EoM,
\begin{equation}
\de_\m\bigl[F'(\abs\Xi)\abs\Xi(\Xi^{-1})_{ij}\de^\m X^j\bigr]=0\;,
\label{fluidEoM}
\end{equation}
where the prime indicates a derivative of $F(\abs\Xi)$ with respect to its argument $\abs\Xi$.

We could in principle stop here, for we have accomplished our goal to construct an EFT for perfect fluids. However, to shed light on the physical content of~\eqref{EFTfluids} and~\eqref{fluidEoM} requires additional work. The first step is to derive the \emph{energy--momentum} (EM) \emph{tensor} of the EFT. This is a standard problem (see for instance \refex{ex:canonicalEM}),\footnote{I use an opposite overall sign of the EM tensor as compared to \refex{ex:canonicalEM} to ensure that the $T^{00}$ component coincides with the canonical Hamiltonian of the EFT.} the result being
\begin{equation}
T^{\m\n}=2F'(\abs\Xi)\abs\Xi(\Xi^{-1})_{ij}\de^\m X^i\de^\n X^j-g^{\m\n}F(\abs\Xi)\;.
\label{fluidEMtensor}
\end{equation}
This already tells us, upon some manipulation, that the EoM~\eqref{fluidEoM} is equivalent to the local conservation laws for energy and momentum. The next step is to trade the explicit dependence on the derivatives $\de_\m X^i$ for the local Eulerian velocity of the fluid. To that end, consider the current~\eqref{Eulercurrent} with $f(X)=1$,
\begin{equation}
J^\m\equiv J^\m_1=\frac1{d!}\ve^{\m\n_1\dotsb\n_d}\ve_{i_1\dotsb i_d}\de_{\n_1}X^{i_1}\dotsb\de_{\n_d}X^{i_d}\;.
\label{J1def}
\end{equation}
It is a matter of straightforward algebra to verify that $J^\m J_\m=(-1)^d\abs\Xi$. The extra sign compensates for our set of conventions, in which $\vev{\Xi^{ij}}=-\d^{ij}$ but $\vev{J^\m}=\d^{\m0}$. The spacetime vector
\begin{equation}
u^\m\equiv\frac{J^\m}{\sqrt{(-1)^d\abs\Xi}}
\end{equation}
is then normalized to unity, $u^2=1$. Moreover, according to~\eqref{Jmuf}, its spatial part is proportional to the local Eulerian velocity $\dot{\vec x}$ of the fluid. It follows that $u^\m$ is the velocity spacetime vector as observed in the laboratory frame.

Consider now the symmetric tensor $G_{\m\n}\equiv(\Xi^{-1})_{ij}\de_\m X^i\de_\n X^j+u_\m u_\n$. Since $\de_\m X^i$ as a spacetime vector is orthogonal to $J^\m$ and thus $u^\m$, it follows that $G_{\m\n}u^\n=u_\m$. At the same time, it is easy to see that $G_{\m\n}\de^\n X^i=\de_\m X^i$. Together, $\de^\m X^i$ for $i=1,\dotsc,d$ and $u^\m$ constitute a basis of $D$ linearly independent vectors on the spacetime, hence $G_{\m\n}A^\n=A_\m=g_{\m\n}A^\n$ for any spacetime vector $A^\m$. This implies the useful identity
\begin{equation}
(\Xi^{-1})_{ij}\de_\m X^i\de_\n X^j=g_{\m\n}-u_\m u_\n\;,
\end{equation}
which allows us to rewrite the EM tensor~\eqref{fluidEMtensor} as
\begin{equation}
T^{\m\n}=2F'(\abs\Xi)\abs\Xi(g^{\m\n}-u^\m u^\n)-g^{\m\n}F(\abs\Xi)\;.
\end{equation}
In an inertial reference frame locally comoving with the fluid (local rest frame), the EM tensor should be diagonal and isotropic, $T^{\m\n}=\diag(U,P,\dotsc,P)$. Here $U$ is the local energy density and $P$ the pressure of the fluid. This can be written in a covariant form as $T^{\m\n}=(U+P)u^\m u^\n-Pg^{\m\n}$. The \emph{scalar} functions $U$ and $P$ can be projected out of the EM tensor using $T^{\m\n}u_\m u_\n=U$ and $T^\m_{\phantom\m\m}=U-dP$. It follows that the as yet unknown function $F(\abs\Xi)$ is related to the energy and pressure by
\begin{equation}
U(\abs\Xi)=-F(\abs\Xi)\;,\qquad
P(\abs\Xi)=F(\abs\Xi)-2F'(\abs\Xi)\abs\Xi\;.
\label{fluidUP}
\end{equation}

\begin{illustration}%
To check that we have got the basic physics right, let us use our EFT to calculate the speed of sound in the fluid. For that, we need to find the bilinear part of the effective Lagrangian in the fluctuations $\pi^i(x)\equiv X^i(x)-\d^i_rx^r$. I will outline the main steps but skip straightforward details. First, the fluctuation of $\Xi^{ij}$ itself is given by~\eqref{Xiparampi},
\begin{equation}
\Xi^{ij}=-\d^{ij}+\udelta\Xi^{ij}\;,\qquad
\udelta\Xi^{ij}\equiv-(\d^{ir}\de_r\pi^j+\d^{jr}\d_r\pi^i)+\de_\m\pi^i\de^\m\pi^j\;.
\end{equation}
Using the identity $\det\Xi=\exp\tr\log\Xi$, we next obtain the expansion
\begin{equation}
(-1)^d\abs\Xi=1-\tr\udelta\Xi+\frac12\bigl[(\tr\udelta\Xi)^2-\tr(\udelta\Xi^2)\bigr]+\bigO(\udelta\Xi^3)\;.
\end{equation}
What remains to be done is just a Taylor expansion of the Lagrangian, $\La_\mathrm{eff}=F(\abs\Xi)$, in $\pi^i$. Dropping a constant and a surface term, we find
\begin{equation}
\La_\mathrm{eff}\simeq-(-1)^dF'_0(\de_0\vec\pi)^2+[(-1)^dF'_0+2F''_0](\skal\nabla\pi)^2+\dotsb\;,
\label{fluidbilin}
\end{equation}
where $F'_0$ and $F''_0$ are derivatives of $F(\abs\Xi)$ taken in the equilibrium, $\vev{\abs\Xi}=(-1)^d$. Note that~\eqref{fluidbilin} does not contain any transverse gradients of $\pi^i(x)$. This is expected, saying merely that transverse fluctuations do not propagate via harmonic oscillations due to the absence of shear elastic forces in a fluid. Fluids support only longitudinal sound waves, whose phase velocity follows as
\begin{equation}
v_\mathrm{L}=\sqrt{1+2(-1)^d\frac{F''_0}{F'_0}}\;.
\end{equation}
It is easy to see with the help of~\eqref{fluidUP} that this equals $\smash{\sqrt{P'(\abs\Xi)/U'(\abs\Xi)}=\sqrt{\Od PU}}$, as expected for thermodynamic sound.
\end{illustration}

\begin{illustration}%
Another informative check of our EFT is to take the nonrelativistic limit. To that end, we use the relation~\eqref{fluidUP} between the function $F(\abs\Xi)$ and the energy density $U(\abs\Xi)$ to decompose the former as
\begin{equation}
F(\abs\Xi)=-\vr_0\sqrt{(-1)^d\abs\Xi}-\Fa\bigl((-1)^d\abs\Xi\bigr)\;.
\label{fluidactionNRlimit}
\end{equation}
The constant $\vr_0$ is the mass density of the fluid in the body coordinates. In the local rest frame, $\vr_0\smash{\sqrt{(-1)^d\abs\Xi}}\to\vr_0\sqrt{\abs{\smash{\tilde\Xi}}}$ where $\smash{\tilde\Xi^{ij}}$ is defined by~\eqref{Xitildedef}, is the density of the rest mass of the fluid. This is expected to dominate the energy in the nonrelativistic limit. On the other hand, $\Fa\bigl((-1)^d\abs\Xi\bigr)\to\Fa(\abs{\smash{\tilde\Xi}})$ becomes the elastic free energy of the fluid. Its contribution to $F(\abs\Xi)$ is assumed to be suppressed relatively to the leading term in~\eqref{fluidactionNRlimit} by two inverse powers of the speed of light.

In the laboratory frame, we write $\Xi^{ij}=\de_0X^i\de_0X^j-\tilde\Xi^{ij}$ and again use the identity $\det\Xi=\exp\tr\log\Xi$. This gives
\begin{equation}
(-1)^d\abs\Xi=\abs{\smash{\tilde\Xi}}\bigl[1-(\tilde\Xi^{-1})_{ij}\de_0X^i\de_0X^j+\dotsb\bigr]\;,
\end{equation}
where the ellipsis stands for higher-order corrections, negligible in the nonrelativistic limit. In turn, \eqref{fluidUP} becomes
\begin{equation}
\begin{split}
U&=\vr_0\sqrt{\abs{\smash{\tilde\Xi}}}-\frac{\vr_0}2\sqrt{\abs{\smash{\tilde\Xi}}}(\tilde\Xi^{-1})_{ij}\de_0X^i\de_0X^j+\Fa(\abs{\smash{\tilde\Xi}})+\dotsb\;,\\
P&=-\Fa(\abs{\smash{\tilde\Xi}})+2\Fa'(\abs{\smash{\tilde\Xi}})\abs{\smash{\tilde\Xi}}+\dotsb\;.
\end{split}
\end{equation}
This has a simple interpretation. The three contributions to $U$ correspond, up to a sign, to the relativistic rest energy, the nonrelativistic kinetic energy and the thermodynamic (internal) energy, while $P$ is the nonrelativistic pressure. The same expansion converts the action~\eqref{EFTfluids} to
\begin{align}
\notag
S_\mathrm{eff}\{X\}=\int\D^D\!x\,\Bigl\{&-\vr_0\sqrt{\abs{\smash{\tilde\Xi(x)}}}+\frac{\vr_0}2\sqrt{\abs{\smash{\tilde\Xi(x)}}}[\tilde\Xi(x)^{-1}]_{ij}\de_0 X^i(x)\de_0X^j(x)\\
&-\Fa(\abs{\smash{\tilde\Xi(x)}})\Bigr\}+\dotsb\;.
\label{EFTfluidNR}
\end{align}
This copies the EFT~\eqref{EFTsolids} for nonrelativistic solids that we previously obtained by an educated guess. The only difference is that the gradient free energy $\Fa$ is now allowed to depend only on $\abs{\smash{\tilde\Xi}}$. (The $\smash{\vr_0\sqrt{\abs{\smash{\tilde\Xi(x)}}}}$ term gives upon integration a constant and can be dropped.) The EFT~\eqref{EFTfluidNR} for nonrelativistic fluids inherits the symmetry under VPDs of the target space $\M$ from its relativistic ancestor~\eqref{EFTfluids}. The spacetime part of its symmetry is however different, and includes spacetime translations, spatial rotations, and Galilei boosts.
\end{illustration}

The symmetry of fluids under VPDs,  $G_\mathrm{int}\simeq\gr{SDiff}(\R^d)$, implies the existence of infinitely many conserved currents. Indeed, the action of an infinitesimal VPD can be expressed as $\udelta X^i=\eps V^i(X)$, where $\eps$ is a parameter and $V^i(X)$ any smooth vector field on $\M\simeq\R^d$ with vanishing divergence, $\Pd{V^i(X)}{X^i}=0$. A straightforward application of Noether's theorem gives the current
\begin{equation}
J^\m_V=F'(\abs\Xi)\abs\Xi(\Xi^{-1})_{ij}\de^\m X^iV^j(X)\;.
\end{equation}
Its on-shell conservation is also seen as a direct consequence of the EoM~\eqref{fluidEoM}. The conservation of the corresponding integral charges generalizes the so-called Kelvin circulation theorem to relativistic fluids; see~\cite{Dubovsky2006a} for further details.


\section{Coupling Nambu--Goldstone Bosons to Classical Matter}
\label{sec:classicalNGcoupling}

Before we wrap up the discussion of EFTs for classical matter, let us step back to see how it connects to the rest of the book. We started the analysis of spontaneously broken spacetime symmetries with a fairly general construction of nonlinear realization thereof in Chap.~\ref{chap:cosetspacetime}. However, in its subsequent applications, we gradually increased the assumptions on the symmetry and the order parameter. In the present chapter, we ended up discarding altogether the possible presence of NG modes associated with long-range order in the quantum ground state. Here it is possible to tie up some loose ends with little effort. An attentive reader might have noticed two related observations I made in quite different contexts. In Sect.~\ref{subsec:vectorunphysical}, I pointed out that it might be possible to make an EFT with Aristotelian symmetry invariant under boosts, whether Galilei or Lorentz. All one needs is an auxiliary variable representing the velocity of the medium in which the EFT lives. On the other hand, we learned in Sect.~\ref{subsec:classicalemergentsym} that this velocity is naturally encoded in the identically conserved current~\eqref{Eulercurrent}. Following this link, I will now sketch concretely how an Aristotelian EFT for broken internal symmetry can be made boost-invariant by coupling the EFT to a classical medium. A general EFT framework for such hybrid systems does not seem to exist as yet. For the sake of illustration, I will restrict the discussion to systems where the classical medium is a fluid. The construction is loosely inspired by~\cite{Pavaskar2022}.

Instead of lengthy reminders, I refer the reader to Chap.~\ref{chap:effLagrangian} for the details of construction of EFTs for broken internal symmetries. (See Sect.~\ref{subsec:effLagoverview} for an executive summary.) All these EFTs were built assuming Aristotelian symmetry, that is symmetry under spacetime translations and spatial rotations. There are two details we have to attend to if we want to augment this spacetime symmetry with boosts. First, the temporal and spatial derivatives, which appear independently in the general Aristotelian effective Lagrangian~\eqref{efflagsummary}, have to be combined in a way that respects boost invariance. Second, any operator invariant under all the spacetime and internal symmetries can be multiplied with an arbitrary function of $\abs{\Xi}$ (for relativistic fluids) or $\abs{\smash{\tilde\Xi}}$ (for nonrelativistic fluids). From this point on, it is convenient to split the discussion of the two cases.

Suppose we want to make the Aristotelian EFT invariant under Galilei boosts. We need not do anything about spatial derivatives, which are already Galilei-invariant on their own. Temporal derivatives can be fixed by contraction with the current~\eqref{J1def}. In terms of the body coordinates $X^i$ and the velocity $\dot{\vec x}$ of the medium, this amounts to replacing the time derivative of the NG field $\pi^a$ of the broken internal symmetry with $J^\m\de_\m\pi^a=\det M(\de_0\pi^a+\dot{\vec x}\cdot\vec\nabla\pi^a)\equiv\det M(\cd_0\pi^a)$. The minimal modification of the two-derivative part of the effective Lagrangian~\eqref{efflagsummary} then reads
\begin{align}
\notag
\La_\mathrm{eff}^{(2,0)}={}&-\frac12\k_{cd}(\abs{\smash{\tilde\Xi}})\mc^c_a(\pi)\mc^d_b(\pi)\vec\nabla\pi^a\cdot\vec\nabla\pi^b\\
\label{efflagmediumGalilei}
&-\frac12\l_{cd}(\abs{\smash{\tilde\Xi}})\mc^c_a(\pi)\mc^d_b(\pi)\ve^{rs}\de_r\pi^a\de_s\pi^b\;,\\
\notag
\La_\mathrm{eff}^{(0,2)}={}&\frac12\bar\k_{cd}(\abs{\smash{\tilde\Xi}})\mc^c_a(\pi)\mc^d_b(\pi)\cd_0\pi^a\cd_0\pi^b\;.
\end{align}
The two factors of $\det M$ coming from the replacement $\de_0\pi^a\to J^\m\de_\m\pi^a$ are without loss of generality absorbed into $\bar\k_{ab}(\abs{\smash{\tilde\Xi}})$. To preserve the internal symmetry acting on $\pi^a$, the coupling functions $\k_{ab}(\abs{\smash{\tilde\Xi}})$, $\bar\k_{ab}(\abs{\smash{\tilde\Xi}})$ and $\l_{ab}(\abs{\smash{\tilde\Xi}})$ must satisfy the constraints~\eqref{invariancekappa} and~\eqref{invariancelambda} for all values of $X^i(x)$. The precise dependence of the coupling functions on $\abs{\smash{\tilde\Xi}}$ can be fixed by measuring the effective couplings $\k_{ab}$, $\bar\k_{ab}$ and $\l_{ab}$ as a function of the density of the underlying medium.

The only nontrivial bit of the ``Galileanization'' of the EFT~\eqref{efflagsummary} lies in the part of the Lagrangian with a single time derivative, $\smash{\La^{(0,1)}_\mathrm{eff}}$. In case this is also strictly invariant under the internal symmetry, it can be promoted to
\begin{equation}
\La_\mathrm{eff}^{(0,1)}=-\s_b(\abs{\smash{\tilde\Xi}})\mc^b_a(\pi)\cd_0\pi^a\;,
\label{J01mediuminv}
\end{equation}
where $\s_b(\abs{\smash{\tilde\Xi}})$ is subject to the condition $\smash{f^b_{\a a}\s_b(\abs{\smash{\tilde\Xi}})=0}$; cf.~\eqref{L01inv}. However, in case $\smash{\La^{(0,1)}_\mathrm{eff}}$ is merely quasi-invariant, we cannot multiply it by an arbitrary function of $\abs{\smash{\tilde\Xi}}$. The only way out then seems to be to set
\begin{equation}
\La_\mathrm{eff}^{(0,1)}=c_a(\pi)J^\m\de_\m\pi^a\;,
\label{J01medium}
\end{equation}
where the 1-form $c(\pi)\equiv c_a(\pi)\D\pi^a$ is constrained by the required strict invariance of its exterior derivative, $\D c(\pi)$. This makes invariance under Galilei boosts manifest, while quasi-invariance under the internal symmetry acting on $\pi^a$ is preserved thanks to the identical conservation of $J^\m$. Note that in this special case, imposing Galilei invariance produces a unique coupling between the NG fields $\pi^a$ and the matter variables $X^i$, without any new a priori unknown parameters. Mathematically, the Lagrangian~\eqref{J01medium} descends from the $(d+1)$-form $c(\pi)\w\D X^1\w\dotsb\w\D X^d$ on the coset space of broken internal symmetry.

\begin{illustration}%
Our main example of a system with a $\smash{\La^{(0,1)}_\mathrm{eff}}$ Lagrangian have been ferromagnets (see Sect.~\ref{sec:spinwaves}). While most natural ferromagnets are metals, there are also ferromagnetic materials that are crystalline insulators. The low-energy EFT of such materials brings together ferromagnetic magnons and the phonons of the crystal lattice. In this case, the dependence of the two-derivative Lagrangian on the tensor $\smash{\tilde\Xi^{ij}}$ is more complicated than in~\eqref{efflagmediumGalilei}. The reason for this is the reduced emergent symmetry of crystals as compared to fluids; see~\cite{Pavaskar2022} for further details. However, the one-derivative part of the Lagrangian is still~\eqref{J01medium}, generating a unique coupling between magnons and phonons.
\end{illustration}

Let us finally see how the above argument needs to be modified, should we replace the requirement of Galilei invariance with the relativistic Poincar\'e invariance. In this case, the two-derivative Lagrangian may only feature properly contracted Lorentz indices. Embedding the antisymmetric $\l_{ab}$-term into a Lorentz-invariant operator is possible by replacing $\smash{\ve^{rs}\de_r\pi^a\de_s\pi^b\to\ve^{\l\m\n}J_\l\de_\m\pi^a\de_\n\pi^a}$. Analogously, the spatial gradient operator $\vec\nabla\pi^a\cdot\vec\nabla\pi^b$ should be embedded in $\smash{-g^{\m\n}\de_\m\pi^a\de_\n\pi^b}$. In order to maintain the form of the two-derivative Lagrangian in the local rest frame of the underlying medium, it is then convenient to write it as
\begin{align}
\notag
\La_\mathrm{eff}^{(2,0)}={}&\frac12\k_{cd}(\abs\Xi)\mc^c_a(\pi)\mc^d_b(\pi)\de_\m\pi^a\de^\m\pi^b\\
\label{efflagmediumLorentz}
&-\frac12\l_{cd}(\abs\Xi)\mc^c_a(\pi)\mc^d_b(\pi)\ve^{\l\m\n}J_\l\de_\m\pi^a\de_\n\pi^b\;,\\
\notag
\La_\mathrm{eff}^{(0,2)}={}&\frac12[\bar\k_{cd}(\abs\Xi)-\k_{cd}(\abs\Xi)]\mc^c_a(\pi)\mc^d_b(\pi)(J^\m\de_\m\pi^a)(J^\n\de_\n\pi^b)\;.
\end{align}
This naturally incorporates the possibility for type-A NG bosons to propagate with a velocity unrelated to the speed of light while maintaining full Poincar\'e invariance. The coupling functions $\k_{ab}(\abs\Xi)$, $\bar\k_{ab}(\abs\Xi)$ and $\l_{ab}(\abs\Xi)$ are still subject to the constraints~\eqref{invariancekappa} and~\eqref{invariancelambda} for all values of $X^i(x)$. The one-derivative Lagrangian $\smash{\La_\mathrm{eff}^{(0,1)}}$, should it be strictly invariant, can be coupled to the medium by a slight modification of~\eqref{J01mediuminv},
\begin{equation}
\La_\mathrm{eff}^{(0,1)}=-\s_b(\abs{\Xi})\mc^b_a(\pi)J^\m\de_\m\pi^a\;.
\end{equation}
When $\smash{\La_\mathrm{eff}^{(0,1)}}$ is merely quasi-invariant, it can still be coupled to the classical medium via~\eqref{J01medium}. Namely, \eqref{J01medium} in fact respects the infinite-dimensional group of transformations preserving the spacetime volume form thanks to its origin in the $(d+1)$-form $c(\pi)\w\D X^1\w\dotsb\w\D X^d$. This group includes both Galilei and Lorentz boosts.


\bibliographystyle{spphys}
\bibliography{references}
\begin{partbacktext}
\part{Epilogue}
\label{part:epilogue}
\end{partbacktext}
\chapter{Topics Not Covered in This Book}
\label{chap:topicsnotcovered}

\abstract*{This chapter offers a survey of topics that could not have been included in the main text of the book for lack of space and time. These include some more advanced topics directly related to the material covered in the book, namely the effects of nonzero temperature and some topological aspects of broken symmetry. Moreover, the scope of the book is extended by a brief introduction to generalized symmetries and their spontaneous breaking. Finally, the text is wrapped up by a discussion of conditions under which a symmetry actually can be spontaneously broken, including an overview of no-go theorems for spontaneous symmetry breaking.}


The story of \emph{spontaneous symmetry breaking} (SSB) and its \emph{effective field theory} (EFT) description does not end here. There are several further facets of SSB that would have deserved place in the table of contents of the book. However, giving them proper credit would either increase the volume of the book beyond reasonable limits, or require a substantial amount of additional background. In this chapter, I will give a brief primer on some of these exciting advanced aspects of SSB. I will be able to go to detail where it is feasible based on the material covered elsewhere in the book. For topics that depart substantially from the main text, I will resort to a few basic comments augmented with references for further reading.


\section{Effects of Nonzero Temperature}
\label{sec:nonzeroT}

The great majority of Parts~\ref{part:internalSSB} and~\ref{part:spacetimeSSB} the book is restricted to the leading order of the derivative expansion of the EFT for \emph{Nambu--Goldstone} (NG) \emph{bosons}. This is just the classical approximation to the EFT, based on the leading-order effective Lagrangian and using only tree-level Feynman diagrams. At a few exceptional spots, such as the proof of the Goldstone theorem in Sect.~\ref{sec:GoldstoneThm} or of the Adler zero property in Sect.~\ref{sec:Adler}, the presented argument is clearly valid beyond the classical approximation. However, I have not discussed explicitly perturbative loop corrections except for an outline of their role in the derivative expansion; cf.~Sects.~\ref{subsec:ChPTpowercounting} and~\ref{subsec:spinwavesferro}.

The effects of nonzero temperature have thus fallen through the cracks together with other loop corrections. Namely, in the so-called imaginary time formalism, the temperature of a system in thermodynamic equilibrium enters through discrete sums over Matsubara frequencies in loop diagrams. This is a standard part of field theory and I will therefore not dwell on details. A reader interested in the application of thermal perturbation theory to EFT for NG bosons will find more information for instance in~\cite{Gerber1989a} (chiral perturbation theory of mesons) or~\cite{Hofmann2002a} (EFT for ferromagnetic magnons). One generic aspect of SSB worth mentioning explicitly is that the order parameter tends to be reduced by thermal fluctuations. As a rule, the phase where a symmetry is spontaneously broken will persist only up to certain critical temperature, above which the symmetry is ``restored.'' This is of course mere jargon. The symmetry of a system constrains its dynamics at all temperatures. However, above the critical temperature, the order parameter for SSB vanishes and excitations in the spectrum are organized in multiplets of the full symmetry group. See Chap.~7 of~\cite{Kapusta2006a} for a discussion of symmetry restoration within thermal field theory.

What are the thermal effects on the spectrum of NG bosons? First of all, the Goldstone theorem remains valid at nonzero temperature as long as the symmetry is spontaneously broken. See Chap.~26 of~\cite{Strocchi2021} for a proof of the existence of a stable gapless quasiparticle, the NG boson, at nonzero temperature. Likewise, the distinction between type-A and type-B NG bosons survives at nonzero temperature. A detailed investigation of the thermal spectrum of NG bosons was carried out in~\cite{Hayata2014b}. At nonzero temperature, quasiparticles manifest themselves by poles in the propagator in the lower half-plane of complex energy. The two types of NG modes differ in the way their thermal width scales with momentum in the long-wavelength limit. As a rule, the complex energy of a type-A NG boson is thus schematically $E(\vec p)\propto\abs{\vec p}-\I\vec p^2$, whereas that of a type-B NG boson is $E(\vec p)\propto \vec p^2-\I\vec p^4$. In both cases, the ratio of thermal width and the real part of the energy tends to zero in the limit $\vec p\to\vec0$. This ensures the stability of the NG modes.

The imaginary time formalism is only suitable for describing matter in thermodynamic equilibrium. In the past couple of decades, much progress has been made in the development of quantum-field-theoretic methods for thermodynamic systems out of equilibrium. The extension of these techniques to EFTs based on nonlinear realization of symmetry is fairly recent. The reader will find further details in~\cite{Landry2020,Hongo2021,Akyuz2023}.


\section{No-Go Theorems for Spontaneous Symmetry Breaking}
\label{sec:nogo}

The entire book is based on the \emph{assumption} that the symmetry of a given system is spontaneously broken. It is however equally interesting to try to understand under what circumstances SSB may in fact occur. The most striking results in this regard form a collection of ``no-go'' theorems, forbidding SSB in a specific class of theories. What follows below is, to the best of my knowledge, a representative list. The results included differ somewhat in the level of rigor at which they have been proven, and in the mathematical techniques required to establish them. Having just discussed the effect of loop corrections, it is natural to start with several related statements where the fluctuations of the order parameter play a key role.

\runinhead{Coleman Theorem} In Lorentz-invariant systems, NG bosons necessarily have a relativistic dispersion relation, $E(\vec p)=\abs{\vec p}$. It is easy to see that the corresponding free propagator in coordinate space,
\begin{equation}
\int\frac{\D^D\!p}{(2\pi)^D}\frac{\E^{-\I p\cdot x}}{p^2}\;,
\label{Colemanint}
\end{equation}
is infrared-divergent for $D=2$ spacetime dimensions. This translates into a divergence of the two-point correlation function of the order parameter at long distances. In other words, the fluctuations arising from the NG boson destroy the assumed order parameter. This is the essence of Coleman's theorem~\cite{Coleman1973a}, which forbids spontaneous breaking of a continuous symmetry in two-dimensional relativistic systems.

\begin{table}[t]
\caption{Subdivision of NG bosons into classes based on the noninteracting part of the effective Lagrangian. The latter determines the form of the (inverse) propagator, which in turn gives the dispersion relation of the NG mode. The last two columns of the table indicate spacetime dimensions, allowed by the condition that the coordinate-space propagator does not diverge at long distances}
\label{tab:CHMW}
\begin{tabular}{p{0.9cm}p{2.5cm}p{2.6cm}p{2.6cm}p{2.3cm}}
\hline\noalign{\smallskip}
Type & Inverse propagator & Dispersion relation & Dimension ($T=0$) & Dimension ($T>0$) \\
\noalign{\smallskip}\svhline\noalign{\smallskip}
$\text A_n$ & $E^2-\vec p^{2n}$ & $E(\vec p)\propto\abs{\vec p}^n$ & $D\geq n+2$ & $D\geq2n+2$ \\
$\text B_{2n}$ & $E-\vec p^{2n}$ & $E(\vec p)\propto\abs{\vec p}^{2n}$ & any $D\geq2$ & $D\geq2n+2$ \\
\noalign{\smallskip}\hline\noalign{\smallskip}
\end{tabular}
\end{table}

This argument lends itself to a broad generalization. Consider a generic effective Lagrangian whose bilinear part starts at order $2n$ in spatial derivatives. While the $n=1$ option is most natural, we also saw in Sect.~\ref{subsec:vectorirrelevant} an example of a system where $n=2$, at least in some spatial directions. More generally, the lack of low-derivative contributions to the spatial part of the kinetic term can be naturally explained by the presence of a coordinate-dependent symmetry~\cite{Griffin2015a,Griffin2015b}. As to the part of the effective Lagrangian carrying time derivatives, we expect either one or two derivatives respectively for type-A and type-B NG bosons. This leads to a refined classification of NG bosons as type-$\text A_n$ or type-$\text B_{2n}$. See the first three columns of Table~\ref{tab:CHMW} for an overview of the schematic forms of the corresponding free-particle propagators and dispersion relations. A simple modification of the argument leading to Coleman's theorem now shows that at zero temperature, type-$\text A_n$ NG bosons are forbidden in $D\leq n+1$ spacetime dimensions. Intriguingly, no such a constraint exists for type-$\text B_{2n}$ NG bosons. In order to have any spectrum of quasiparticles, at least one spatial dimension must of course be present. However, for any $D\geq2$, the propagator of a type-$\text B_{2n}$ boson at zero temperature is infrared-finite. See the fourth column of Table~\ref{tab:CHMW} for a summary.

\runinhead{Hohenberg--Mermin--Wagner Theorem} At nonzero temperature, the situation changes dramatically. Here the Minkowski-spacetime integral~\eqref{Colemanint} is replaced with an imaginary-time sum-integral of the schematic type
\begin{equation}
T\sum_{k=-\infty}^{+\infty}\int\frac{\D^d\!\vec p}{(2\pi)^d}\frac{\exp(-\I\o_k t+\I\skal px)}{(\I\o_k\ \text{or}\ \o_k^2)+\vec p^{2n}}\;,
\label{MWint}
\end{equation}
where $\o_k\equiv2k\pi T$ is a bosonic Matsubara frequency and $T$ the temperature. The type-$\text A_n$ and type-$\text B_{2n}$ cases correspond respectively to $\I\o_k$ and $\o_k^2$ in the denominator. Due to the discrete nature of the Matsubara sum, the contributions of all terms with $k\neq0$ are infrared-finite. An infrared divergence can only arise from the zero Matsubara mode, and will be absent provided $d\geq 2n+1$ or $D\geq 2n+2$~\cite{Argurio2019} (the last column of Table~\ref{tab:CHMW}). While the critical dimension is the same for type-$\text A_n$ and type-$\text B_{2n}$ NG bosons, note that $n$ here counts the number of spatial derivatives in the Lagrangian. For fixed power of momentum in the dispersion relation, the critical dimension for type-$\text A_{2n}$ NG bosons is higher than that for type-$\text B_{2n}$ ones.

\begin{watchout}
The original results on the absence of SSB at nonzero temperature are due to Hohenberg~\cite{Hohenberg1967a} (for superfluids) and Mermin and Wagner~\cite{Mermin1966a} (for isotropic lattice models of ferro- and antiferromagnets). These covered implicitly the cases of type-$\text A_1$ and type-$\text B_2$ NG bosons. However, their approach was different from the above intuitive argument and did not rely on the propagator of the would-be NG mode. Rather, they used the so-called Bogoliubov inequality to place an upper bound on the order parameter. This bound shows that the order parameter goes to zero in the limit of infinite volume and vanishing symmetry-breaking perturbation. See~\cite{Gelfert2001} for a review and further references.
\end{watchout}

\runinhead{Landau--Peierls Instability} There is another related result that applies to systems where the order parameter is spatially modulated in one direction, thus spontaneously breaking translations. This is the case for instance for liquid crystals in the smectic-A phase, as we saw in Sect.~\ref{subsec:vectorirrelevant}. Let us briefly recall the line of reasoning therein to stress its generality. Suppose that the order parameter is given by a scalar field $\p$ that develops a nonzero gradient in the ground state. We can treat $\vev{\vec\nabla\p}$ as a secondary, vector order parameter. The assumption that translations are spontaneously broken only in one dimension amounts to the condition that $\vev{\vec\nabla\p}$ points in the same direction everywhere in space. Next, we expand $\p$ around its expectation value as $\p\equiv\vev{\p}+\pi$ and focus on the part of the effective Lagrangian bilinear in the fluctuation $\pi$. The assumed (and also spontaneously broken) \emph{rotation} invariance dictates that there cannot be any term in the Lagrangian, bilinear in the part of $\vec\nabla\pi$, perpendicular to $\vev{\vec\nabla\p}$. The gradient expansion of the Lagrangian starts with terms proportional to $(\vec\nabla_\parallel\pi)^2$ and $(\vec\nabla_\perp^2\pi)^2$, where $\parallel$ and $\perp$ denote projections to subspaces parallel and perpendicular to $\vev{\vec\nabla\p}$. The denominator of the large fraction in~\eqref{MWint} should then be replaced with $(\I\o_k\ \text{or}\ \o_k^2)+\vec p_\parallel^2+\vec p_\perp^4$. Integrating the $k=0$ term over $\vec p_\parallel$ shows that the order parameter is washed out by thermal fluctuations at any nonzero temperature whenever $d\leq3$. This is known as the \emph{Landau--Peierls instability}; see Sect.~1.6 of~\cite{Gennes1993} and~\cite{Baym1982a} for a more detailed discussion within the condensed-matter and nuclear physics context, respectively.

\runinhead{Absence of Time Crystals} Speaking of spontaneously breaking translations in a single direction, it is mandatory to consider the possibility that this direction corresponds to time. The idea that time translations could be spontaneously broken was proposed in~\cite{Wilczek2012,Shapere2012}. Such systems have since been known as \emph{time crystals}. The reason why this possibility had not been noticed until the 21\textsuperscript{st} century is that time crystals are not easy to realize. It was shown soon after the original proposal that their existence is forbidden in thermodynamic equilibrium of any Hamiltonian with sufficiently short-range interactions~\cite{Bruno2013,Watanabe2015a}. There is however a nonequilibrium route towards quantum time crystals; see~\cite{Zaletel2023} for a recent review of the subject.

\runinhead{Vafa--Witten Theorem} Finally, the list of no-go theorems would not be complete without the Vafa--Witten theorem~\cite{Vafa1984a}. This is of quite a different nature than all the other results above, and applies to Lorentz-invariant gauge theories coupled to fermionic matter in a vector-like manner. An important example of such a theory is the \emph{quantum chromodynamics} (QCD). Suppose the theory possesses a vector-like symmetry, that is one that acts in the same way on left- and right-handed fermions. In QCD, this could be for instance the $\gr{U}(1)_\mathrm{B}$ baryon number symmetry or, in the limit of equal quark masses, the $\gr{SU}(2)_\mathrm{V}$ isospin symmetry. The theorem states that such vector-like symmetries cannot be spontaneously broken in the ground state. The proof is technical. See also~\cite{Nussinov2002} for a review of the relevant mathematical methods and their applications to QCD and hadron physics.


\section{Topological Aspects of Spontaneous Symmetry Breaking}
\label{sec:topologicalaspects}

We already encountered various topological aspects of EFTs for NG bosons on several occasions. Below, I will generalize some of the observations made previously, and add new interesting physics.

\begin{figure}[t]
\sidecaption[t]
\includegraphics[width=2.9in]{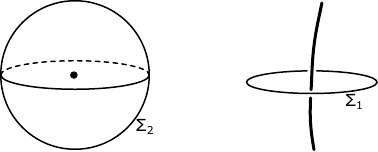}
\caption{Examples of topological defects in $d=3$ spatial dimensions together with the surfaces defining their topological charge. The left panel shows a point defect (monopole), the right panel a line defect (vortex)}
\label{fig:defect}
\end{figure}

\runinhead{Topological Defects and Solitons} The classification of \emph{topological defects} was the first major application of topology to physics; see~\cite{Mermin1979a} for an introduction to the relevant mathematics. Recall that the coset space $G/H$ can be viewed as the vacuum manifold of a theory with SSB, whose points indicate possible values of the order parameter in the ground state. A defect is a nonuniform configuration of the $G/H$-valued order parameter that is singular on some subdimensional domain in space. See Fig.~\ref{fig:defect} for the simple examples of a point and line defect. The presence of a defect can be established from the properties of the order parameter away from the singularity. Thus, a $p$-dimensional defect in $\R^d$ can be enclosed by a $(d-p-1)$-dimensional hypersurface $\S_{d-p-1}$ that is topologically a sphere, $S^{d-p-1}$. Equivalence classes of defects can then be determined by studying maps $S^{d-p-1}\to G/H$, which are classified by the \emph{homotopy group} $\pi_{d-p-1}(G/H)$. This assigns the defect a unique label: its \emph{topological charge}. An exception are domain walls, which are of codimension one in space, dividing it into two halves. A domain wall is thus characterized by the values of the order parameter on both sides. See Table~\ref{tab:defects} for an overview of basic types of topological defects.

\begin{table}[t]
\caption{Classification of basic topological defects by their codimension in space $\R^d$}
\label{tab:defects}
\begin{tabular}{p{2.5cm}p{2.5cm}p{2.8cm}}
\hline\noalign{\smallskip}
Defect & Dimension $p$ & Classifying group \\
\noalign{\smallskip}\svhline\noalign{\smallskip}
Domain wall & $d-1$ & $\pi_0(G/H)\times\pi_0(G/H)$ \\
Vortex & $d-2$ & $\pi_1(G/H)$ \\
Monopole & $d-3$ & $\pi_2(G/H)$ \\
\noalign{\smallskip}\hline\noalign{\smallskip}
\end{tabular}
\end{table}

\begin{illustration}%
We saw examples of domain walls in Sect.~\ref{sec:quantumtranslation}. The double-well potential~\eqref{DWpotdoublewell} has a discrete $G\simeq\Z_2$ symmetry under $\p\to-\p$. This is broken in its minima to $H\simeq\trgr$, hence $G/H\simeq\Z_2$ and likewise $\pi_0(G/H)\simeq\Z_2$. The cosine potential~\eqref{DWpotcos}, on the other hand, has the symmetry group $G\simeq\Z_2\ltimes\Z$. The $\Z_2$ factor is generated by the inversion $\p\to-\p$ whereas the $\Z$ factor by the shift $\p\to\p+2\pi v$. In any of the minima of the potential, the symmetry is broken to a subgroup isomorphic to $H\simeq\Z_2$. Therefore, in this case $G/H\simeq\Z$ and accordingly $\pi_0(G/H)\simeq\Z$. The elements of $\pi_0(G/H)\times\pi_0(G/H)$ shown in Table~\ref{tab:defects} correspond to domain wall solutions interpolating between different pairs of minima.
\end{illustration}

The focus on singular field configurations may be surprising; the physics should not change if we ``smooth down'' the order parameter. This is however only possible at the cost of embedding the coset space $G/H$ in a larger order parameter manifold~$\M$.

\begin{illustration}%
Consider a superfluid in $d=2$ dimensions. Here the order parameter $\vev\psi$ takes values from $\C$, while the vacuum manifold is $G/H\simeq\gr{U}(1)/\trgr\simeq S^1$ for any $\vev\psi\neq0$. A point defect (vortex) in the superfluid is described by a smooth complex field $\psi(\vec x)$. The corresponding topological charge is the \emph{winding number}, defining an element of $\pi_1(G/H)\simeq\pi_1(S^1)\simeq\Z$; see also~\refex{ex:vortex}. Importantly, nonzero winding number implies by the argument principle (Chap.~7 of~\cite{Needham1997a}) the existence of a point $\vec x\in\R^2$ where $\psi(\vec x)=0$: the core of the vortex. This is our ``singularity'' where the order parameter cannot lie on the vacuum manifold $G/H\simeq S^1$.

For another example, consider a spin system in $d=3$ dimensions with $G/H\simeq S^2$. Suppose that on the sphere $\S_2$ as shown in Fig.~\ref{fig:defect}, the order parameter represented by a unit vector $\vec n\in S^2$ points radially outwards everywhere. This spin configuration is sometimes called \emph{hedgehog}. If we now try to continue the field to the inside of the sphere, we will inevitably encounter a point singularity: a monopole. Possible values of its topological charge are classified by $\pi_2(G/H)\simeq\pi_2(S^2)\simeq\Z$. The singularity can be avoided only at the cost of deforming the spin configuration out of the coset space $G/H$. The natural way to do so is by embedding $G/H\simeq S^2$ in $\R^3$ and allowing $\vec n$ to vary its magnitude. One can then extend the hedgehog field on $\S_2$ to the inside by scaling down its magnitude so that it vanishes at the location of the monopole.
\end{illustration}

In contrast to defects are \emph{topological solitons} (Sect.~IX of \cite{Mermin1979a}). These are fields in $\R^d$ that are smooth and take values from $G/H$ everywhere. Nontrivial topology arises from imposing a specific boundary condition at spatial infinity. Thus, the field configuration of a $p$-dimensional soliton is required to converge to a constant far from a fixed $p$-dimensional hypersurface $C_p$ in $\R^d$. Now consider an open hypersurface $\S_{d-p}$ that is ``transverse'' to $C_p$. The boundary condition effectively compactifies this hypersurface to a sphere, $S^{d-p}$. Equivalence classes of $p$-dimensional solitons in $\R^d$ are therefore classified by the homotopy group $\pi_{d-p}(G/H)$.

\begin{illustration}%
The simplest type of a topological soliton is that with $p=0$. Such solitons can be viewed as quasiparticles localized in all directions in space, and typically carry a finite amount of energy. The transverse hypersurface $\S_d$ is in this case the entire space $\R^d$. We already met two examples of such solitons in Chap.~\ref{chap:internalexamples}: \emph{skyrmions} in QCD, corresponding to $d=3$ and $\pi_3(G/H)\simeq\pi_3(S^3)\simeq\Z$, and \emph{baby skyrmions} in two-dimensional ferromagnets, for which $\pi_2(G/H)\simeq\pi_2(S^2)\simeq\Z$. The latter are easily generalized to higher dimensions, simply by making the fields independent of whatever extra coordinates are present. Thus a skyrmion in a three-dimensional ferromagnet is localized to a line, $C_1$. Its topological charge is determined by the properties of a unit vector field on a two-dimensional surface $\S_2$, transverse to $C_1$.
\end{illustration}

\runinhead{Topological Terms in the Action} We saw in Sect.~\ref{sec:effLagstructure} that the part of the effective Lagrangian for NG bosons with a single time derivative may be topologically nontrivial. Similarly, it turned out in Sect.~\ref{sec:ChPT} that the part of the low-energy EFT of QCD, taking into account the chiral anomaly, is topological in nature. We shall now generalize these observations, thus uncovering a broad class of contributions to EFTs for NG bosons, arising from the topology of the coset space $G/H$.

Suppose that $G/H$ is compact and consider a closed $p$-form $\fdeg\o p$ that belongs to a nontrivial cohomology class in the $p$-th de Rham cohomology group, $H^p(G/H)$.\footnote{I follow the convention common in high-energy physics and indicate the degree of a differential form with a superscript in parentheses.} Upon pulling $\fdeg\o p$ back to a spacetime $M$ of dimension $D$ by the map $M\to G/H$ defining the NG fields, it becomes a likewise closed $p$-form on $M$. There are now three different possibilities depending on the relation between $p$ and $D$. If $p=D$, the spacetime $p$-form can be directly integrated and added to the action of the EFT. This is called a \emph{$\t$-term}. Due to its origin in a closed $D$-form, it is a mere surface term in the Lagrangian and thus does not affect the perturbative dynamics of the EFT. This is the reason why $\t$-terms have not featured in the book at all. If present, they contribute a phase factor to the generating functional though, which may affect the nonperturbative physics of the theory.

Another relevant possibility is $p=D+1$. In this case, there is a locally well-defined $D$-form $\fdeg\o D$ on $G/H$ such that $\D\fdeg\o D=\fdeg\o p$. Upon pulling back to the spacetime, $\fdeg\o D$ gives a contribution to the Lagrangian that is quasi-invariant under the action of $G$. This is a \emph{Wess--Zumino} (WZ) \emph{term}. The anomalous contribution to the low-energy effective Lagrangian of QCD is obviously of this kind; cf.~Sect.~\ref{subsec:ChPTanomaly}. However, the single-time-derivative part of the effective Lagrangian, $\smash{\La_\mathrm{eff}^{(0,1)}}$, constructed in Sect.~\ref{subsec:effLag1der}, belongs to the same category. Here only the variation of the NG fields in time matters and thus effectively $D=1$, which makes $H^2(G/H)$ the relevant de Rham cohomology group. Following the line of reasoning of Sect.~\ref{subsec:spinwavestopological}, we see moreover that such WZ terms generally give rise to a Berry phase when the ground state is driven by an external field~\cite{Watanabe2014a}. See Table~\ref{tab:WZterms} for a summary. Further technical details on the construction of topological WZ and $\t$-terms in the action can be found in the recent study~\cite{Davighi2018a}. For a somewhat more condensed-matter oriented perspective on topological terms, see Chap.~9 of~\cite{Altland2010a}.

\begin{table}[t]
\caption{Physical interpretation of generators of the de Rham cohomology group $H^p(G/H)$, depending on the dimension $D$ of spacetime they are pulled back to. For $p<D$, the pull-back leads to a closed $p$-form on the spacetime, Hodge-dual to an identically conserved tensor current of rank $D-p$. In the table, the rank of currents $J^{\m\n\dotsb}$ is indicated by the number of indices}
\label{tab:WZterms}
\begin{tabular}{p{1.7cm}p{1.7cm}p{1.7cm}p{1.7cm}p{1.7cm}p{1.1cm}}
\hline\noalign{\smallskip}
Dimension & $p=1$ & $p=2$ & $p=3$ & $p=4$ & $p=5$ \\
\noalign{\smallskip}\svhline\noalign{\smallskip}
$D=1$ & $\t$-term & WZ term & --- & --- & --- \\
$D=2$ & $J^\m$ & $\t$-term & WZ term & --- & --- \\
$D=3$ & $J^{\m\n}$ & $J^\m$ & $\t$-term & WZ term & --- \\
$D=4$ & $J^{\m\n\l}$ & $J^{\m\n}$ & $J^\m$ & $\t$-term & WZ term \\
\noalign{\smallskip}\hline\noalign{\smallskip}
\end{tabular}
\end{table}


\section{Generalized Symmetries}
\label{sec:generalizedsymmetries}

It remains to clarify the content of the lower left corner of Table~\ref{tab:WZterms}. For $p<D$, the spacetime $p$-form arising from pulling back $\fdeg\o p$ can be expressed as the Hodge dual of a $(D-p)$-form, $\ho\fdeg J{D-p}$. The form $\fdeg J{D-p}$ has a vanishing codifferential; cf.~Appendix~\ref{appsubsec:hodge}. In a flat spacetime, this can be expressed component-wise as
\begin{equation}
\de_\m J^{\m\n\dotsb}=0\;,
\label{genconslaw}
\end{equation}
where $J^{\m\n\dotsb}$ is fully antisymmetric in all its indices. Equation~\eqref{genconslaw} is nothing but a local conservation law with a tensor current of rank $D-p$. It encodes, on the classical level, a \emph{generalized symmetry}. This (and other) kind of generalization of conservation laws has attracted much attention in the last decade. An interested reader will find a comprehensive introduction for instance in~\cite{Gomes2023,Brennan2023,Bhardwaj2023}.

In the context of EFT for NG bosons, the identically conserved currents $J^{\m\n\dotsb}$ give us a tool for calculation of topological charges of defects and solitons, classified in Sect.~\ref{sec:topologicalaspects}. Mathematically, this is based on the correspondence between the homotopy and de Rham cohomology groups of $G/H$ via the Hurewicz theorem (see Appendix~\ref{appsubsec:derhamcohomology}). Physically, the conservation law~\eqref{genconslaw} suggests the existence of a conserved charge, obtained by integrating the temporal component, $J^{0rs\dotsb}$, over a spatial hypersurface. For a static defect or soliton, it is easiest to pull $\fdeg\o p$ back directly to the space $\R^d$ and express it therein as $\ho\fdeg{\mathcal J}{d-p}$. The $(d-p)$-form $\fdeg{\mathcal J}{d-p}$ is the density of our topological charge. The topological charge itself is obtained by integration over an effectively closed $p$-dimensional surface $\S_p$ in $\R^d$,
\begin{equation}
Q(\S_p)\equiv\int_{\S_p}\ho\fdeg{\mathcal J}{d-p}\;.
\label{topcharge}
\end{equation}
The fact that $\S_p$ is closed guarantees that the value of $Q(\S_p)$ does not change under smooth deformations of the surface. Depending on circumstances, \eqref{topcharge} represents the topological charge of a $(d-p-1)$-dimensional defect enclosed by $\S_p$, or that of a $(d-p)$-dimensional soliton intersecting $\S_p$.

\begin{illustration}%
The coset space $G/H\simeq\gr{U}(1)\simeq S^1$ of superfluids has nontrivial first de Rham cohomology group, $H^1(G/H)\simeq\R$. Pulling its generator back to space and integrating over a closed loop $\S_1$ defines the winding number of a $(d-2)$-dimensional defect: the vortex. The coset space $G/H\simeq\gr{SU}(2)/\gr{U}(1)\simeq S^2$ of ferro- and antiferromagnets has nontrivial second de Rham cohomology group, $H^2(G/H)\simeq\R$. Pulling its generator back to $\R^2$ and integrating over the entire plane, $\S_2=\R^2$, defines the charge of a 0-dimensional soliton: the baby skyrmion.
\end{illustration}

All the conservation laws of the type~\eqref{genconslaw} discussed until now were \emph{emergent}. Namely, their existence follows from the topology of the coset space $G/H$, which itself parameterizes the breaking of the symmetry we started with. There are however also other, important examples of generalized conservation laws. For instance, the Maxwell equations in absence of matter take the form of a rank-two conservation law, $\de_\m F^{\m\n}=0$. In this case, the charge~\eqref{topcharge} obtained by integration over a closed spatial surface is nothing but the electric flux through that surface. The ``defect'' is a localized electric charge enclosed by the surface. Intriguingly, the spontaneous breakdown of the generalized symmetry of electrodynamics is the ultimate reason why the photon is massless. This brings us back to the beginning of the book: we are constantly surrounded by a sea of NG bosons, namely sound and light. A reader wondering whether the classification of NG bosons into type-A and type-B also applies to generalized symmetries will be glad to hear that this is indeed the case. See~\cite{Hidaka2020b} for a discussion of the classification and counting of NG bosons of generalized symmetries, and~\cite{Sogabe2019a} for a nontrivial example of a system where the photon coupled to a particular type of matter behaves as a type-B NG mode.


\bibliographystyle{spphys}
\bibliography{references}
\chapter{Some Open Questions}
\label{chap:openquestions}

\abstract*{This last and very brief chapter highlights some open problems pertinent to the subject of the book. The main thrust of the book lies in the development of the effective field theory formalism for spontaneously broken symmetry. Accordingly, the list of open questions contains mostly technical issues whose resolution would further extend the validity of the results of the book. Some of the issues, related to the classification of actions of groups of manifolds, are purely mathematical. Others are more directly related to physical applications of the formalism.}


I will close the book by highlighting some technical issues that I could not have addressed properly. I label them as ``open questions,'' at the risk of revealing my own ignorance rather than a gap in the existing knowledge. Since the book revolves around developing a \emph{formalism} for spontaneously broken symmetry, some of the items are purely mathematical. The first commandment of \emph{effective field theory} (EFT) dictates that the effective Lagrangian must contain all operators that respect the symmetry of the system. Accordingly, an important ingredient of the EFT formalism is the classification of all possible Lagrangians consistent with the symmetry. This requirement lies behind most of the issues listed below.

\runinhead{Universality of the Standard Nonlinear Realization} The classification of group actions on a manifold in Chap.~\ref{chap:CCWZ} relies heavily on the assumption that the isotropy subgroup $H$ of the symmetry group $G$ is compact. Any progress towards classification of group actions such that $H$ is noncompact would be most welcome. This applies in particular to coset spaces $G/H$ that are not reductive, for which many of the nice features of our standard nonlinear realization are lost.

\runinhead{Global Existence of the Group Action} The standard nonlinear realization of symmetry developed in Chap.~\ref{chap:CCWZ} is restricted to a single local coordinate patch. Accordingly, the explicitly constructed action of the group $G$ is limited to its elements near unity. Yet, the global existence of the group action on the whole manifold $\M$, or coset space $G/H$, has been implicitly assumed throughout the book. A better understanding of possible topological obstructions to extending the group action from a local coordinate patch to the entire manifold would be desirable. This issue is particularly pressing for Lie groups consisting of several connected components. As far as I know, the first attempt to systematically deal with such cases was made in~\cite{Kharuk2018a}. However, more work is needed to establish a general, practically useful formalism.

\runinhead{Group Action via Generalized Local Transformations} Most of the book focuses on symmetries realized by point transformations on a target space $\M$ or its Cartesian product with the spacetime $M$. This allows one to use the mathematical language of group actions on a finite-dimensional manifold. However, we have also seen some physically relevant examples of generalized local symmetries, where the transformation of fields is allowed to depend on their derivatives. This suggests working directly in the infinite-dimensional space of fields as maps $M\to\M$. With the knowledge of the symmetry group and the symmetry-breaking pattern, one may then still apply the \emph{agnostic nonlinear realization}, mentioned in Chaps.~\ref{chap:cosetspacetime} and~\ref{chap:spacetimequantum}. To what extent this exhausts all possible actions of the symmetry remains unclear. Yet, a setup that makes the implementation of generalized local symmetries systematic has long been in use in the theory of differential equations~\cite{Olver1986a}. First attempts to apply this setup to EFT appeared very recently~\cite{Craig2023a,Alminawi2023}. Hopefully, they will help to place the treatment of generalized local symmetries on the same footing as that of point symmetries.

\runinhead{Interplay of Nonlinear Realization and Inverse Higgs Constraints} As stressed in Chap.~\ref{chap:spacetimequantum}, the operational way of eliminating would-be \emph{Nambu--Goldstone} (NG) fields that excite gapped modes using an \emph{inverse Higgs constraint} (IHC) is optional. The EFT for solely genuine NG degrees of freedom, obtained by applying an IHC, may carry a realization of the symmetry by generalized local transformations. On the other hand, the EFT before the IHC is imposed is ambiguous in that its field content may depend on the choice of order parameter. Either way, the universality of the nonlinear realization of the symmetry may be compromised. Here, too, more work is needed to establish that the existing formalism(s) for spontaneously broken spacetime symmetry give(s) the most general effective Lagrangian.

\runinhead{Nonlinear Realization of Translation Symmetry} Possibly the largest gap in the narrative of the book is the treatment of spontaneously broken translation symmetry. The main challenge is to establish a unique local parameterization of given fields in terms of NG variables, one for each broken translation generator. Thus, the discussion in Chap.~\ref{chap:spacetimequantum} is limited to order parameters spatially modulated just in one direction. A generalization of the formalism to ordered states of matter, modulated in several dimensions, is pending. The same obstacle hinders the application of the background gauge approach, which proved extremely efficient in case of internal symmetries (Chap.~\ref{chap:effLagrangian}). Some concrete applications of this approach beyond the single example worked out in Chap.~\ref{chap:spacetimequantum} can be found in~\cite{Hidaka2015a}. However, a fully general background gauge formalism for construction of EFTs for spontaneously broken spacetime symmetry is not available as yet.


\bibliographystyle{spphys}
\bibliography{references}
\appendix
\chapter{Elements of Differential Geometry}
\label{app:diffgeom}

\abstract*{Most of the book is phrased in an elementary language of field theory. The main mathematical tool beyond ordinary calculus is the theory of Lie groups and their representations, which the reader is expected to be familiar with. Some more advanced topics are, however, much more easily addressed using the language of differential geometry. This appendix gives an overview of the basics of differential geometry in an extent sufficient for the purposes of the book. The stress is on qualitative understanding of the various concepts rather than on rigor. The material covered includes differential and integral calculus on smooth manifolds, Riemannian geometry and de Rham cohomology, plus some adjacent topics. The appendix may also serve as a brief stand-alone introduction to differential geometry for a student of physics. References for further study are included.}


Most of the book relies on mathematical background familiar to any graduate student of theoretical physics, that is mostly advanced calculus and basic group theory. However, some topics are greatly illuminated by taking a more geometric viewpoint. For instance, differential calculus on manifolds and Riemannian geometry provide useful insight into the nature of coset spaces on which effective field theories for spontaneously broken symmetry are defined. This is discussed in Chap.~\ref{chap:CCWZ}. Likewise, the classification of effective theories for broken internal symmetries worked out in Chap.~\ref{chap:effLagrangian} is greatly simplified by using differential calculus on manifolds and elements of de Rham cohomology. Las but not least, Riemannian geometry provides a practically convenient language for the analysis of scattering of Nambu--Goldstone bosons in Chap.~\ref{chap:scattering}.

The purpose of this appendix is to give the reader the background needed to follow such more advanced parts of the book. I try to keep the discussion elementary, focusing on qualitative understanding rather than on mathematical rigor. Wherever possible, I justify the need for new concepts and motivate their formal definition. This necessarily makes the exposition biased. My approach might be natural for someone familiar with quantum field theory who seeks to develop advanced calculus methods to deal with fields that live on a manifold. The appendix can however be studied independently without any reference to the main text of the book. A more complete introduction to differential geometry at a similar level as here can be found in~\cite{Stone2009a}. A reader looking for a more thorough treatment of the subject at a level still accessible to a graduate student of theoretical physics is advised to consult~\cite{Nakahara2003a,Fecko2011a,Frankel2012a}.


\section{Smooth Manifolds}
\label{appsec:manifolds}

The basic concept of differential geometry is that of a \emph{manifold}. This is an abstraction that arose historically from the study of non-Euclidean geometry and of the geometry of surfaces. Formally, a manifold $\M$ of dimension $n$ is a set equipped with the notion of closeness (topology) that can be locally mapped to $\R^n$ in a one-to-one manner. More precisely, one assumes the existence of a collection of pairs, $\{(U_i,\vp_i)\}$, where $U_i$ is an open subset of $\M$ and $\vp_i:U_i\to\R^n$ is a \emph{homeomorphism}. (The latter is defined as a continuous map that is invertible and whose inverse is also continuous.) Borrowing terminology from cartography, each pair $(U_i,\vp_i)$ is called a \emph{chart}, and the collection of all charts is an \emph{atlas}. It is required that each point $x\in\M$ belongs to at least one of the sets $U_i$. The point $\vp_i(x)\in\R^n$ corresponds to an $n$-tuple of real numbers $x^a$, $a=1,\dotsc,n$, called the \emph{coordinates} of $x$. It is common if a bit sloppy to identify the point $x\in\M$ with its image $\vp_i(x)$ or even the coordinates $x^a$, and I will often follow this convention.

Mapping a point $x\in\M$ to its coordinates is both a blessing and a curse. On the one hand, the coordinates allow us to import to the manifold mathematical structures we are familiar with from ordinary calculus. On the other hand, it may not be possible to cover the whole manifold with a single set $U_i$. The coordinates are in general only defined locally in some neighborhood of $x$. We must therefore be able to switch from one set of coordinates to another. The first commandment of differential geometry is that whatever structure we build on the manifold, it must have an intrinsic meaning regardless of a particular choice of local coordinates.

\begin{illustration}%
The $\M=\R^n$ is itself an $n$-dimensional manifold. It can be covered with a single coordinate chart, $(U,\vp)=(\R^n,\id)$, corresponding to usual Cartesian coordinates. The coordinates can however also be chosen differently, in which case a single chart may not be sufficient to cover the whole $\R^n$. For example, let us consider $\R^2$ and use the polar coordinates $\vr,\t$ instead of the Cartesian coordinates $x,y$. The map
\begin{equation}
(x,y)\to(\vr,\t)\equiv\biggl(\sqrt{x^2+y^2},\sgn y\arccos\frac{x}{\sqrt{x^2+y^2}}\biggr)
\end{equation}
is continuous and invertible on $U=\R^2\setminus\{(x,0)\,\vert\,x\leq0\}$, making $U$ homeomorphic to the domain $(0,+\infty)\times(-\pi,+\pi)\subset\R^2$. To cover the whole plane $\M=R^2$, we would need to add at least one more chart.
\end{illustration}

The first thing we can do with the local coordinates is to introduce a differential structure on $\M$. Let us choose a chart $(U_i,\vp_i)$. We then define a \emph{smooth} real function on $U_i$ as a map $f:U_i\to\R$ such that $f\circ\vp_i^{-1}$ is an infinitely differentiable function on $\vp_i(U_i)\subset\R^n$. We would of course like the concept of smoothness to be independent of the choice of chart. This is ensured by requiring that for any two overlapping open sets $U_i,U_j$, the map $\vp_i\circ\vp_j^{-1}$ that switches between the two coordinate systems is itself infinitely differentiable. A function defined on the whole manifold, $f:\M\to\R$, is now called smooth if it is smooth when restricted to any single coordinate chart.

\begin{figure}[t]
\sidecaption[t]
\includegraphics[width=2.9in]{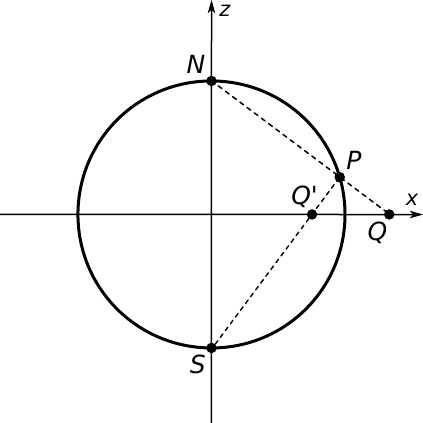}
\caption{Stereographic projection of the $n$-sphere $S^n$ to $\R^n$, here for $n=1$. First, $S^n$ is naturally embedded in $\R^{n+1}$ as a unit sphere centered at the origin. A point $P$ on $S^n$ is then projected to the point $Q$ in $\R^n$ using a straight line starting at the north pole $N$ of the sphere. An alternative projection from the south pole $S$ would map $P$ to $Q'$}
\label{fig:stereographic}
\end{figure}

\newpage

\begin{illustration}%
One way to define local coordinates on the $n$-sphere $S^n$ is through stereographic projection, see Fig.~\ref{fig:stereographic}. We imagine that $S^n$ is realized (embedded) in $\R^{n+1}$ as a unit sphere centered at the origin. The embedding is defined implicitly by $(x^1)^2+\dotsb+(x^n)^2+z^2=1$, where $(x^1,\dotsc,x^n,z)$ are Cartesian coordinates in $\R^{n+1}$. A given point $P\in S^n$ is then projected from the north pole $N$ to the hyperplane $\R^n$ defined by $z=0$. The image of $P$ is the point
\begin{equation}
Q\equiv(y^1,\dotsc,y^n)=\biggl(\frac{x^1}{1-z},\dotsc,\frac{x^n}{1-z}\biggr)\;.
\end{equation}
The set $(y^1,\dotsc,y^n)$ constitutes the desired local coordinates on $S^n$. These coordinates are well-defined on the whole sphere except for the north pole $N$ itself. To cover the entire sphere, we need another coordinate patch. This can be defined through a similar projection from the south pole $S$. The latter maps $P$ to
\begin{equation}
Q'\equiv(y'^1,\dotsc,y'^n)=\biggl(\frac{x^1}{1+z},\dotsc,\frac{x^n}{1+z}\biggr)\;.
\end{equation}
Noting that
\begin{equation}
\abs{\vec y}=\frac{\abs{\vec x}}{1-z}=\sqrt{\frac{1+z}{1-z}}\;,\qquad
\abs{\vec y'}=\frac{\abs{\vec x}}{1+z}=\sqrt{\frac{1-z}{1+z}}\;,
\end{equation}
we find that the two sets of local coordinates are related by inversion, $y'^a=y^a/\abs{\vec y}^2$. This ensures smooth transition between the two coordinate systems wherever they are both well-defined, that is on $S^n$ with both poles removed.
\end{illustration}


\section{Linear Structures on Manifolds}
\label{appsec:linearstructures}

We have now learned what a smooth function on a manifold is. Since our primary goal is to develop calculus on manifolds, we would eventually like to differentiate and integrate such functions. In order to be able to do so independently of the choice of local coordinates, we will however need additional structure on the manifold.


\subsection{Tangent Vectors and Vector Fields}
\label{appsubsec:tangentvectors}

The first step towards setting up calculus on manifolds is to define a derivative of a function. In $\R^n$ with its Cartesian coordinates, this naturally leads to the notion of a partial derivative. On a general manifold $\M$, there is however no such a preferred set of coordinates. The way out is to generalize the concept of a directional derivative in $\R^n$. Thus, for a fixed point $x\in U_i\subset\M$ and a set of coefficients $v^1,\dotsc,v^n$, we define a linear operator $\vec v$ acting on smooth functions $f$ on $U_i$ through
\begin{equation}
\vec v=v^a\PD{}{x^a}\;:\quad f\to\vec v[f]\equiv v^a\at{\PD{(f\circ\vp_i^{-1})}{x^a}}{x}\;.
\label{tangentvector}
\end{equation}
Such a linear operator is usually called a \emph{tangent vector} to $\M$ at $x$. Accordingly, the set of all tangent vectors at $x$ is called the \emph{tangent space} of $\M$ at $x$ and denoted as $\T_x\M$. Let me stress that the adjective ``tangent'' here is just a convention. There is no way to give it a literal geometric meaning unless we imagine $\M$ as a hypersurface embedded in a higher-dimensional Euclidean space.

The definition~\eqref{tangentvector} relies on the components $v^a$ in a fixed \emph{coordinate basis} of the tangent space, $\de_a\equiv\Pd{}{x^a}$. However, the concept of a tangent vector has a meaning independent of a specific choice of local coordinates. We just have to remember to recalculate the components of the vector if the local coordinates change, as the following example illustrates.

\begin{illustration}%
Suppose that we trade the Cartesian coordinates $x,y$ in $\R^2$ for the polar coordinates $\vr,\t$ through the usual prescription $(x,y)=(\vr\cos\t,\vr\sin\t)$. The coordinate basis of the tangent space is then converted by using the chain rule for partial derivatives,
\begin{equation}
\de_\vr=(\cos\t)\de_x+(\sin\t)\de_y\;,\qquad
\de_\t=-(\vr\sin\t)\de_x+(\vr\cos\t)\de_y\;.
\end{equation}
In order for a vector $\vec v$ to be well-defined independently of the choice of local coordinates, its components $(v^x,v^y)$ and $(v^\vr,v^\t)$ in the two coordinate systems must be related by
\begin{equation}
v^\vr=v^x\cos\t+v^y\sin\t\;,\qquad
v^\t=-\frac{v^x}\vr\sin\t+\frac{v^y}\vr\cos\t\;.
\end{equation}
These results agree with what elementary planar geometry would tell us if we treated $\de_\vr$ and $(1/\vr)\de_\t$ as unit vectors in the radial and azimuthal direction, respectively.
\end{illustration}

In general, if we change the local coordinates $x^a$ smoothly to $\tilde x^a(x)$, the coordinate bases and vector components in the new and old coordinate systems are related by
\begin{equation}
\tilde\de_a=\PD{x^b}{\tilde x^a}\de_b\;,\qquad
\tilde v^a=\PD{\tilde x^a}{x^b}v^b\;.
\label{coordinatechange}
\end{equation}
The first of these is just the chain rule, whereas the latter follows from the requirement that the vector $\vec v=v^a\de_a=\tilde v^a\tilde\de_a$ is independent of the choice of coordinates. Note that $\Pd{\tilde x^a}{x^b}$ is just a shorthand notation. If $x^a$ and $\tilde x^a$ correspond to coordinate maps $\vp_i,
\vp_j$, this notation represents the partial derivatives of $\vp_j\circ\vp_i^{-1}$, evaluated at $\vp_i(x)\in\R^n$. In the following, I will occasionally use such intuitive notation without further warnings.

For most applications in physics, a \emph{vector field} is of greater interest than a (tangent) vector defined at a single point $x\in\M$. Just like for real functions on $\M$, it is natural to require that a vector field varies smoothly as a function of $x$. This is most easily ensured by demanding that, within a local coordinate patch, the components $v^a$ are themselves smooth functions of $x$. It is also possible to define a vector field without referring to specific coordinates. One thus treats it as a collection of vectors $\vec v(x)\in\T_x\M$ such that $\vec v(x)[f]$ is a smooth function of $x$ for any smooth ``test function'' $f$ on $\M$.

With the notion of a vector field at hand, let us briefly return to the choice of basis of the tangent space. With local coordinates $x^a$ defined on an open set $U_i$, we get a set of coordinate basis fields $\de_a$ that are likewise well-defined only on $U_i$. It is however often advantageous to have a basis that is not limited to a particular coordinate patch, or not related to a specific set of coordinates at all. One thus defines a \emph{(local) frame} as a set of vector fields $\vec e_A$, $A=1,\dotsc,n$, defined on some open set $U\subset\M$, such that the set of vectors $\{\vec e_A(x)\}_{A=1}^n$ is a basis of $\T_x\M$ at any $x\in U$.\footnote{Within this appendix, I will carefully distinguish coordinate indices $a,b,\dotsc$ from frame indices $A,B,\dotsc$. In practice (including the main text of this book), the two notations are often blended.} Any vector field can then be expressed in terms of its components with respect to the frame, $\vec v(x)=v^A(x)\vec e_A(x)$. I will sometimes use a local frame instead of coordinate basis fields; the latter can always be viewed as a special case of the former.

\begin{watchout}%
\label{att:frameintegrability}%
Restricting the definition of a local frame to an open subset of $\M$ is not as innocuous as it might appear. The existence of a frame well-defined on the whole of $\M$ depends on the global topology of $\M$. Even on a manifold as simple as the 2-sphere $S^2$, there is no smooth vector field that would be nonzero everywhere! This is the ``hairy ball theorem'' of algebraic topology; see Sect.~13.7 of~\cite{Stone2009a} for an elementary discussion.

A second remark of caution is that a local frame is not merely a coordinate basis in disguise. By the definition of a basis, one can always relate a frame $\vec e_A$ to any coordinate basis $\de_a$ by $\vec e_A=c^a_A\de_a$, where $c^a_A$ is a matrix of coefficients with nonzero determinant. In general, there is however no set of local coordinates $y^A$, $A=1,\dotsc,n$ such that $\vec e_A=\Pd{}{y^A}$. As I will show later, this is only possible if the condition $\vec e_A[c^a_B]=\vec e_B[c^a_A]$ is satisfied for all $A,B$ and $a$.
\end{watchout}


\subsection{Tensors and Tensor Fields}
\label{appsubsec:tensors}

Once we know how to construct tangent vectors and vector fields on a manifold, we can quickly build tensors and tensor fields of an arbitrary rank. While the motivation for doing so may not be as immediate as for tangent vectors, several types of tensors are of central importance to differential geometry. To start with, the \emph{cotangent space} of $\M$ at $x$, $\T^*_x\M$, is defined as the dual of $\T_x\M$. Its elements, called \emph{cotangent vectors} or \emph{covectors}, are linear maps from $\T_x\M$ to $\R$. The counterpart of a coordinate basis $\de_a$ on $\T_x\M$ is the dual coordinate basis on $\T^*_x\M$, usually denoted as $\D x^a$. Hence $\D x^a(\de_b)=\d^a_b$. The action of any covector on any vector then follows from linearity and can be expressed component-wise,
\begin{equation}
\o=\o_a\D x^a\;,\ 
\vec v=v^a\de_a\ \Rightarrow\ 
\o(\vec v)=\o_av^a\;,\qquad
\o\in\T^*_x\M\;,\ 
\vec v\in\T_x\M\;.
\end{equation}
The choice of notation for the dual coordinate basis is not accidental. Any smooth function $f$ defined in a neighborhood of $x$ gives rise to a covector $\D f$ through
\begin{equation}
\D f(\vec v)\equiv\vec v[f]=v^a\PD{f}{x^a}\;,\qquad\vec v\in\T_x\M\;.
\label{df}
\end{equation}
I have simplified the notation by treating $f$ as a function of the local coordinates $x^a$; cf.~\eqref{tangentvector}. By the definition of the dual basis, we have $v^a=\D x^a(\vec v)$. Hence $\D f=(\Pd f{x^a})\D x^a$, generalizing the differential of a function known from ordinary calculus. It follows that upon switching the local coordinates from $x^a$ to $\tilde x^a(x)$, the dual coordinate basis and components of covectors transform to
\begin{equation}
\D\tilde x^a=\PD{\tilde x^a}{x^b}\D x^b\;,\qquad
\tilde\o_a=\PD{x^b}{\tilde x^a}\o_b\;.
\label{dualcoordinatechange}
\end{equation}

Higher-rank tensors can now be built using tensor products of $\T_x\M$ and $\T^*_x\M$. Generally, a \emph{tensor of type $(q,p)$} at $x\in\M$ is an element of $(\T_x\M)^{\otimes q}\otimes(\T^*_x\M)^{\otimes p}$. A class of tensors of special importance to both differential and integral calculus on manifolds is obtained by taking the $p$-fold antisymmetric tensor product of $\T^*_x\M$, where $p$ is a positive integer. The elements of this space, $\Omega^p_x\M$, are called \emph{$p$-forms} and define antisymmetric multilinear functions on $\T_x\M$. Note that $\Omega^1_x\M$ coincides with $\T^*_x\M$; covectors are indeed often referred to as 1-forms. An antisymmetric tensor product of $p$ covectors $\o_1,\dotsc,\o_p$ is represented compactly by the notation
\begin{equation}
\o_1\w\dotsb\w\o_p\equiv\sum_\pi\sgn\pi\,\o_{\pi(1)}\otimes\dotsb\otimes\o_{\pi(p)}\;,
\label{wedgeof1forms}
\end{equation}
where the sum is over all permutations $\pi$ of the indices $1,\dotsc,p$. Using this notation, we can write a general element $\o\in\Omega^p_x\M$ in the dual coordinate basis as
\begin{equation}
\o=\frac1{p!}\o_{a_1\dotsb a_p}\D x^{a_1}\w\dotsb\w\D x^{a_p}\;,\qquad
\o_{a_1\dotsb a_p}=\o(\de_{a_1},\dotsc,\de_{a_p})\;.
\label{pformcoordbasis}
\end{equation}
The expression for the components of the $p$-form is a direct generalization of the relation $\o_a=\o(\de_a)$ for 1-forms.

Equation~\eqref{wedgeof1forms} is more than just a useful shorthand notation. It can be extended to an operation on forms of arbitrary degree that is linear and associative: the \emph{exterior product}. With the dual coordinate basis expansion~\eqref{pformcoordbasis} at hand, linearity and associativity require that any two forms $\o\in\Omega^p_x\M$ and $\s\in\Omega^q_x\M$ are mapped to the $(p+q)$-form
\begin{equation}
\o\w\s=\frac1{p!q!}\o_{a_1\dotsb a_p}\s_{b_1\dotsb b_q}\D x^{a_1}\w\dotsb\w\D x^{a_p}\w\D x^{b_1}\w\dotsb\w\D x^{b_q}\;.
\end{equation}
The same result can be alternatively expressed in a coordinate-independent manner by the action on a set of test vectors $\vec v_1,\dotsc,\vec v_{p+q}$ from $\T_x\M$,
\begin{equation}
\begin{split}
&(\o\w\s)(\vec v_1,\dotsc,\vec v_{p+q})\\
&=\frac1{p!q!}\sum_\pi\sgn\pi\,\o(\vec v_{\pi(1)},\dotsc,\vec v_{\pi(p)})\s(\vec v_{\pi(p+1)},\dotsc,\vec v_{\pi(p+q)})\;,
\end{split}
\end{equation}
where $\pi$ now runs over permutations of $1,\dotsc,p+q$. The antisymmetry of forms as multilinear maps implies that the exterior product is graded-anticommutative,
\begin{equation}
\o\w\s=(-1)^{pq}\s\w\o\;,\qquad
\o\in\Omega^p_x\M\;,\ \s\in\Omega^q_x\M\;.
\end{equation}
The direct sum $\smash{\bigoplus_{p=0}^n\Omega^p_x\M}$ equipped with the exterior product possesses the structure of an associative algebra: the \emph{exterior (Grassmann) algebra}. Here $\Omega^0_x\M$ is identified with $\R$ and the direct sum terminates at $p=n$; by the antisymmetry property there are no $p$-forms with $p>n$ on an $n$-dimensional manifold.

Another operation on forms that will prove useful is the \emph{interior product}. This assigns to any vector $\vec v\in\T_x\M$ a linear map $\ix{\vec v}:\Omega^{p+1}_x\M\to\Omega^p_x\M$ defined by
\begin{equation}
\ix{\vec v}\o(\vec v_1,\dotsc,\vec v_p)\equiv\o(\vec v,\vec v_1,\dotsc,\vec v_p)\;,\qquad
\vec v_1,\dotsc,\vec v_p\in\T_x\M
\label{interiorproduct}
\end{equation}
for any $\o\in\Omega^{p+1}_x\M$. While this definition is coordinate-independent, the interior product takes a particularly simple form in the component notation where it amounts to contracting the vector $\vec v$ with the $(p+1)$-form $\o$, $(\ix{\vec v}\o)_{a_1\dotsb a_p}=v^b\o_{ba_1\dotsb a_p}$. Thanks to the antisymmetry of forms, the interior product is anticommutative that is $\ix{\vec u}\circ\ix{\vec v}=-\ix{\vec v}\circ\ix{\vec u}$ for any two tangent vectors $\vec u,\vec v$. Moreover, $\ix{\vec v}$ with any fixed $\vec v$ is an \emph{antiderivation} of the Grassmann algebra: for any $p$-form $\o$ and $q$-form $\s$ one has
\begin{equation}
\ix{\vec v}(\o\w\s)=(\ix{\vec v}\o)\w\s+(-1)^p\o\w(\ix{\vec v}\s)\;.
\label{antiderivationix}
\end{equation}
In fact, the properties of the interior product are already completely fixed by requiring that it is an antiderivation that reduces to $\ix{\vec v}\o=\o(\vec v)$ on 1-forms $\o$.

So far I have limited the discussion in this subsection to tensors defined at a single point $x\in\M$. However, we can build \emph{tensor fields} of arbitrary type $(q,p)$ in complete analogy with the construction of vector fields out of tangent vectors at different points.

\begin{illustration}%
\label{ex:Euclideanmetric}%
A \emph{metric} $g$	is a symmetric tensor field of type $(0,2)$ such that $g(x)\in\T^*_x\M\odot\T^*_x\M$ is a positive-definite bilinear form for any $x\in\M$. For example, $\R^n$ as a Euclidean space is equipped with the metric $g=\d_{ab}\D x^a\otimes\D x^b$ where $x^a$ are the Cartesian coordinates. The metric can of course be converted to any other coordinates by using~\eqref{dualcoordinatechange}. For instance, in $\R^3$ we may want to switch to the spherical coordinates $(r,\t,\vp)$, defined by $(x,y,z)=(r\sin\t\cos\vp,r\sin\t\sin\vp,r\cos\t)$. The metric then assumes the form
\begin{equation}
g=\D r\otimes\D r+r^2\D\t\otimes\D\t+r^2\sin^2\t\,\D\vp\otimes\D\vp\;.
\label{Euclideanmetric}
\end{equation}
An interested reader will find more about the metric tensor in Sect.~\ref{appsec:riemann}.
\end{illustration}

Promoting a $p$-form to a tensor field leads to the concept of \emph{differential $p$-form}.\footnote{For the sake of brevity, I will mostly drop the adjective ``differential'' where it is clear from the context that I speak of a tensor field rather than a tensor defined at a single point.} Just like for vector fields, we have the freedom to choose other bases for differential $p$-forms than those induced by local coordinates on $\M$. Thus, a \emph{(local) coframe} is a set of differential 1-forms $\vec e^{*A}$ defined on some open set $U\in\M$ such that the set of 1-forms $\smash{\{\vec e^{*A}(x)\}_{A=1}^n}$ spans a basis of $\Omega^1_x\M$ at any $x\in U$. Any frame $\vec e_A$ induces a unique dual coframe $\vec e^{*A}$ (and vice versa) by requiring that $\vec e^{*A}(\vec e_B)=\d^A_B$. All the properties of $p$-forms reviewed above can be immediately promoted to frame components of differential $p$-forms by the replacements $\de_a\to\vec e_A$, $\D x^a\to\vec e^{*A}$ and $\o_{a_1\dotsb a_p}\to\o_{A_1\dotsb A_p}$ wherever necessary. The same remark applies to tensor fields of an arbitrary type $(q,p)$.

\begin{illustration}%
\label{ex:spherevolumeform}%
Let us use the spherical coordinates $(\t,\vp)$ to parameterize the 2-sphere $S^2$. Following \refex{ex:Euclideanmetric}, this allows us to define a metric on $S^2$ by restricting~\eqref{Euclideanmetric} to fixed radial variable, $r=1$,
\begin{equation}
g=\D\t\otimes\D\t+\sin^2\t\,\D\vp\otimes\D\vp\;.
\label{spheremetric}
\end{equation}
The vector fields
\begin{equation}
\vec e_\t\equiv\de_\t\;,\qquad
\vec e_\vp\equiv\frac1{\sin\t}\de_\vp
\end{equation}
then constitute an orthonormal frame, $g(\vec e_A,\vec e_B)=\delta_{AB}$, wherever well-defined. The corresponding dual coframe is $\vec e^{*\t}=\D\t$ and $\vec e^{*\vp}=\sin\t\,\D\vp$. Out of these two 1-forms, we can naturally build a 2-form through
\begin{equation}
\o=\vec e^{*\t}\w\vec e^{*\vp}=\sin\t\,\D\t\w\D\vp\;.
\label{spherevolumeform}
\end{equation}
One might be tempted to think of the frame $(\vec e_\t,\vec e_\vp)$ as defining an area element on the sphere that can be subsequently used for integration. However, as I will explain in Sect.~\ref{appsec:integration}, integration on manifolds is done using differential forms. For the 2-sphere, it is exactly the differential 2-form~\eqref{spherevolumeform} that does the job of defining the proper area element.
\end{illustration}


\section{Maps Between and on Manifolds}
\label{appsec:manifoldmaps}

In Sect.~\ref{appsec:manifolds} we saw how to define a smooth function on a manifold $\M$ as a map from $\M$ to $\R$. This concept of smoothness is easily generalized to maps between two manifolds $\M$ and $\MN$. For a map $f:\M\to\MN$ that is continuous, it is always possible to choose a chart $(U,\vp)$ on $\M$ and a chart $(V,\c)$ on $\MN$ such that $f(U)\subset V$. The map is then called smooth if $\c\circ f\circ\vp^{-1}$ is infinitely differentiable. A smooth map $f:\M\to\MN$ that is a homeomorphism is called a \emph{diffeomorphism}.


\subsection{Push-Forward and Pull-Back}
\label{appsubsec:pushpull}

\begin{figure}[t]
\sidecaption[t]
\includegraphics[width=2.9in]{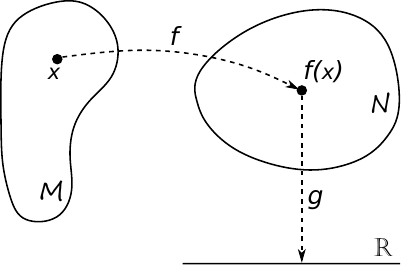}
\caption{A smooth map $f$ between manifolds $\M$ and $\MN$, defined locally in the neighborhood of the point $x\in\M$. Composing a test function $g$, defined locally around the image $f(x)\in\MN$, with $f$ allows one to lift the map $f$ to the push-forward $f_*:\T_x\M\to\T_{f(x)}\MN$}
\label{fig:pushforward}
\end{figure}

Smooth maps between manifolds naturally induce linear maps between the corresponding (co)tangent spaces. Since tangent vectors are differential operators acting on locally defined test functions, we first have to inspect how the map $f:\M\to\MN$ affects the latter. It is clear from Fig.~\ref{fig:pushforward} that by ``connecting the arrows,'' any test function $g:\MN\to\R$ can be converted into the function $g\circ f:\M\to\R$. This is enough to connect tangent vectors on the two manifolds. Formally, one starts with a tangent vector $\vec v\in\T_x\M$ and maps it to a tangent vector $f_*\vec v\in\T_{f(x)}\MN$ such that
\begin{equation}
f_*\vec v[g]\equiv\vec v[g\circ f]
\label{pushforward}
\end{equation}
for any test function $g$ defined in a neighborhood of $f(x)$ on $\MN$. The image $f_*\vec v$ is referred to as the \emph{push-forward} of $\vec v$ by $f$. Suppose that we have a set of coordinates $x^a$ in a neighborhood of $x$ on $\M$ and a set of coordinates $y^b$ in a neighborhood of $f(x)$ on $\MN$. It then follows from the chain rule that the coordinate basis vectors at $x$ are mapped to
\begin{equation}
f_*\PD{}{x^a}=\PD{f^b}{x^a}\PD{}{y^b}\;,
\label{pushforwardcoordinates}
\end{equation}
where $f^b$ is a shorthand notation for the $b$-th coordinate of $f(x)$. This looks almost like a change of variables. It is therefore worth stressing that the manifolds $\M$ and $\MN$ need not have the same dimension, hence the map $f$ need not be invertible.

\begin{illustration}%
\label{ex:embedding}%
Consider again the 2-sphere with its spherical coordinates $(\t,\vp)$ and define the map $f:S^2\to\R^3$ as the natural embedding that takes the point $(\t,\vp)\in S^2$ to $(x,y,z)=(\sin\t\cos\vp,\sin\t\sin\vp,\cos\t)\in\R^3$. Then~\eqref{pushforwardcoordinates} gives us
\begin{equation}
\begin{split}
f_*\de_\t&=(\cos\t\cos\vp)\de_x+(\cos\t\sin\vp)\de_y-(\sin\t)\de_z\;,\\
f_*\de_\vp&=-(\sin\t\sin\vp)\de_x+(\sin\t\cos\vp)\de_y\;.
\end{split}
\end{equation}
This ``realizes'' tangent vectors to the sphere as vectors in $\R^3$. The same strategy can be used to give a literal geometric interpretation to tangent vectors on any smooth manifold by embedding the latter in a higher-dimensional Euclidean space.
\end{illustration}

Once we know how to relate tangent vectors to $\M$ and $\MN$, we can do the same for covectors by using the dual nature of the cotangent space. The ``connecting the arrows'' strategy hints that the map between the cotangent spaces acts in the opposite direction than $f_*$. Thus, we assign to $\o\in\T^*_{f(x)}\MN$ a covector $f^*\o\in\T^*_x\M$ so that
\begin{equation}
f^*\o(\vec v)\equiv\o(f_*\vec v)\quad\text{for any}\quad\vec v\in\T_x\M\;.
\label{pullback}
\end{equation}
For obvious reasons, the image $f^*\o$ is referred to as the \emph{pull-back} of $\o$ by $f$. The chain rule tells us that covectors from the dual coordinate basis are mapped to
\begin{equation}
f^*\D y^b=\PD{f^b}{x^a}\D x^a\;,
\label{pullbackcoordinates}
\end{equation}
which again looks like a mere change of variables but is really much more than that.

\begin{watchout}%
By following the same line of reasoning, one can define the push-forward $f_*$ for any tensor of type $(q,0)$ and the pull-back $f^*$ for any tensor of type $(0,p)$. In general, it is however not possible to define either of the two for tensors of the mixed type $(q,p)$. In a similar vein, it is generally not possible to promote the point-wise map $f_*:\T_x\M\to\T_{f(x)}\MN$ to a push-forward of vector (or tensor) \emph{fields}. A problem arises when two points $x_1,x_2\in\M$ have the same image under $f$. Our prescription~\eqref{pushforward} would then naively try to assign to a single point $f(x_1)=f(x_2)$ two different tangent vectors. No such a problem arises for the pull-back, which can be extended from tensors of type $(0,p)$ to tensor fields for any smooth map $f$.

Both of the above problems disappear when $f$ is a diffeomorphism. In that case, we have a locally well-defined one-to-one mapping between $\M$ and $\MN$. The push-forward $f_*$ is an isomorphism between the respective tangent spaces, and the pull-back $f^*$ is an isomorphism between the respective cotangent spaces. By combining $\smash{f_*:\T_x\M\to\T_{f(x)}\MN}$ with $\smash{f^{-1})^*:\T^*_x\M\to\T^*_{f(x)}\MN}$, we can map any tensor (field) on $\M$ to a tensor (field) on $\MN$.
\end{watchout}

\begin{illustration}%
\label{ex:embeddingmetric}%
In \refex{ex:Euclideanmetric}, I introduced the Euclidean metric on $\R^n$, $g=\d_{ab}\D x^a\otimes\D x^b$, where $x^a$ are the Cartesian coordinates. This metric can be pulled back to any manifold $\M$ that can be embedded as a hypersurface in $\R^n$ by a smooth map $f$. The induced metric on $\M$ in chosen local coordinates $y^a$ on $\M$ follows from~\eqref{pullbackcoordinates},
\begin{equation}
f^*g=\d_{ab}\PD{f^a}{y^c}\PD{f^b}{y^d}\D y^c\otimes\D y^d\;.
\end{equation}
The justifies the previously used expression~\eqref{spheremetric} for the metric on the 2-sphere.

In the same manner, one can pull back any differential form defined on $\R^n$ to the embedded manifold $\M$. For instance, the area 2-form~\eqref{spherevolumeform} on the 2-sphere has an elegant representation in terms of the natural embedding of $S^2$ in $\R^3$. It is left to the reader as an exercise to check that~\eqref{spherevolumeform} is recovered by pulling back the 2-form
\begin{equation}
\frac12\vec r\cdot(\de_a\vec r\times\de_b\vec r)\D x^a\w\D x^b
\end{equation}
in $\R^3$, where $\vec r=(x,y,z)$ and the dot and cross denote respectively the conventional dot and cross products of 3-vectors. It is common to abuse the notation and represent the area 2-form on $S^2$ by simply replacing the 3-vector $\vec r$ above with a unit 3-vector $\vec n$ that defines the embedding of $S^2$ in $\R^3$. One can then write the area 2-form on $S^2$ simply as $(1/2)\vec n\cdot(\D\vec n\times\D\vec n)$.
\end{illustration}


\subsection{Flow of Vector Fields}
\label{appsubsec:flow}

In Sect.~\ref{appsubsec:tangentvectors}, I emphasized that the concept of a tangent vector does not have a literal geometric interpretation unless we embed the manifold $\M$ in a higher-dimensional Euclidean space. It is however possible to visualize a vector field $\vec v$ through the \emph{flow} it induces on $\M$. Imagine that the manifold is filled with a liquid, and think of $\vec v(x)$ as the velocity of the liquid at $x\in\M$. We can then extract information about the vector field from the trajectories of individual particles of the liquid.

Let us now formalize this concept. We need a real parameter $t\in\R$ to measure the ``time'' for the liquid motion. Next, we think of a single particle trajectory as a curve $\g:\R\to\M$ satisfying the first-order differential equation
\begin{equation}
\OD{f(\g(t))}t=\vec v[f]\Bigr\rvert_{\g(t)}
\label{flow}
\end{equation}
for any test function $f:\M\to\R$. This is just a coordinate-independent version of the relation between the particle's trajectory and its velocity, $\Od{\g^a}t=v^a(\g(t))$. By choosing the initial condition as $\g(0)=x$ and varying $x\in\M$, we get a family of trajectories that define a map $\phi_{\vec v,t}:\M\to\M$. In terms of the liquid analogy, $\phi_{\vec v,t}(x)$ is the position at time $t$ of a particle that was at $x$ at time $t=0$. For smooth vector fields $\vec v$, the flow equation~\eqref{flow} is guaranteed to have a (locally) unique solution. Hence the map $\phi_{\vec v,t}$ with fixed $\vec v$ and $t$ is a diffeomorphism on $\M$. We can thus use it to push forward tensors from one point of $\M$ to another. This is called \emph{Lie transport}.

\begin{figure}[t]
\sidecaption[t]
\includegraphics[width=2.9in]{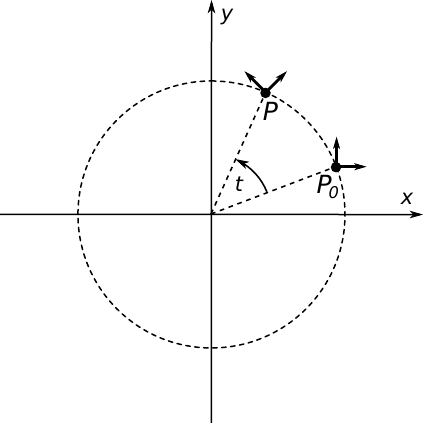}
\caption{Illustration of Lie transport of tangent vectors in $\R^2$. The Lie transport is generated by the vector field $\vec v=-y\de_x+x\de_y$, one of whose flow lines is indicated by the \emph{dashed circle}. As the point $P_0$ is transported to $P$, the attached basis of tangent space at $P_0$ (displayed as \emph{solid arrows}) is pushed forward to a basis of tangent space at $P$}
\label{fig:lietransport}
\end{figure}

\begin{illustration}%
\label{ex:R2rotation}%
Consider the vector field $\vec v=-y\de_x+x\de_y$ in $\R^2$. The flow equation~\eqref{flow} is easily solved and gives
\begin{equation}
x(t)=x_0\cos t-y_0\sin t\;,\qquad
y(t)=x_0\sin t+y_0\cos t\;,
\end{equation}
where $P_0=(x_0,y_0)$ is the starting point of the flow, see Fig.~\ref{fig:lietransport}. From the definition~\eqref{pushforward} of push-forward, or directly from~\eqref{pushforwardcoordinates}, we then see that the coordinate basis $(\de_x,\de_y)$ at $P_0$ is transported to
\begin{equation}
\phi_{\vec v,t*}\de_x=(\cos t)\de_x+(\sin t)\de_y\;,\qquad
\phi_{\vec v,t*}\de_y=-(\sin t)\de_x+(\cos t)\de_y\;
\label{lietransportR2}
\end{equation}
at $P=(x(t),y(t))$. The flow generated by $\vec v$ corresponds to a rotation of $\R^2$ around the origin by angle $t$. Indeed, switching to polar coordinates $(\vr,\t)$ through $(x,y)=(\vr\cos\t,\vr\sin\t)$ shows that $\vec v=\de_\t$. Upon changing the basis of the tangent space, \eqref{lietransportR2} is then recognized as a trivial map between the coordinate bases at $P_0$ and $P$, that is $\phi_{\vec v,t*}\de_\vr\rvert_{P_0}=\de_\vr\rvert_P$ and $\phi_{\vec v,t*}\de_\t\rvert_{P_0}=\de_\t\rvert_P$. This is a special case of a general rule that when $\vec v$ is one of the coordinate basis fields, all the coordinate basis vectors at $x$ are transported to the corresponding coordinate basis vectors at $\phi_{\vec v,t}(x)$.

As another example, we may choose $\vec v=x\de_x+y\de_y$. In this case, we find
\begin{equation}
x(t)=x_0\E^t\;,\qquad
y(t)=y_0\E^t\;,
\end{equation}
and accordingly
\begin{equation}
\phi_{\vec v,t*}\de_x=\E^t\de_x\;,\qquad
\phi_{\vec v,t*}\de_y=\E^t\de_y\;.
\label{lietransportR2alt}
\end{equation}
This is another illustration of Lie transport generated by a coordinate basis field. We just have to notice that in polar coordinates, $\vec v=\vr\de_\vr=\de_{\log\vr}$. Choosing local coordinates as $(\log\vr,\t)$ reproduces~\eqref{lietransportR2alt} as a trivial transport of $(\de_{\log\vr},\de_\t)$ at $P_0$ to $(\de_{\log\vr},\de_\t)$ at $P$.
\end{illustration}

Lie transport plays a critical role for the implementation of symmetries on manifolds. Indeed, the map $\phi_{\vec v,t}$ for fixed $\vec v$ and varying $t$ constitutes a one-parameter group of transformations on $\M$. The group structure amounts to the obvious properties
\begin{equation}
\phi_{\vec v,t}\circ\phi_{\vec v,s}=\phi_{\vec v,t+s}\;,\qquad
\phi_{\vec v,0}=\id\;,\qquad
\phi_{\vec v,t}^{-1}=\phi_{\vec v,-t}\;.
\label{1parametergroup}
\end{equation}
We can think of this as a ``representation'' of $\R$ by an additive group of transformations on $\M$. It is straightforward to generalize this concept to other (finite or Lie) groups. Formally, the \emph{action} of a group $G$ on $\M$ is defined as a set of diffeomorphisms $T_g$ on $\M$ that respect the group structure of $G$,\footnote{The concept of group action is introduced in a more pedestrian manner and used in Chap.~\ref{chap:CCWZ}.}
\begin{equation}
T_{g_1}\circ T_{g_2}=T_{g_1g_2}\;,\qquad
T_e=\id\;,\qquad
(T_g)^{-1}=T_{g^{-1}}\;,\qquad
g,g_1,g_2\in G\;.
\end{equation}
For any Lie group $G$, we can restrict to a specific one-parameter subgroup. This is mapped by~\eqref{flow} to a vector field that represents the corresponding generator of $G$ on $\M$. Combined with~\eqref{1parametergroup}, we deduce that there is a one-to-one correspondence between vector fields and one-parameter groups of transformations on $\M$.


\subsection{Lie Derivative}
\label{appsubsec:liederivative}

The practical utility of Lie transport is limited due to the necessity to solve the set of coupled first-order differential equations~\eqref{flow} for the flow. It is usually more convenient to consider the variation of tensor fields on $\M$ under infinitesimal transformations generated by $\vec v$. This variation is formally encoded in the \emph{Lie derivative}, defined for a given tensor field $\tens T$ as
\begin{equation}
\ld{\vec v}\tens T\equiv-\at{\OD{(\phi_{\vec v,t*}\tens T)}t}{t=0}=\lim_{\eps\to0}\frac{\tens T-\phi_{\vec v,\eps*}\tens T}\eps\;.
\label{liederivativedef}
\end{equation}
The overall minus sign is a convention arising from the fact that $(\phi_{\vec v,\eps*}\tens T)(x)$ corresponds to the value of $\tens T$ at $\phi_{\vec v,-\eps}(x)$, pushed forward to $x$ by the map $\phi_{\vec v,\eps}$.

Let us see how this works for vector fields. For the moment, it is practical to resort, even if only implicitly, to local coordinates. The vector field $\vec v$ then becomes a linear differential operator. The flow equation~\eqref{flow} in turn has a formal solution for $\phi_{\vec v,t}$ in terms of the (path-ordered) exponential of $\vec v$, given implicitly by
\begin{equation}
f(\phi_{\vec v,t}(x))=\E^{\vec vt}f(x)\;.
\label{pathorderedexp}
\end{equation}
The definition~\eqref{pushforward} of push-forward now tells us that for any test function $f$ and any vector field $\vec u$, $\phi_{\vec v,t*}\vec u[f]=\vec u[f\circ\phi_{\vec v,t}]=\vec u[\E^{\vec vt}f]$. If we start with $\vec u(x)$, this will give us the value of $\phi_{\vec v,t*}\vec u$ at $\phi_{\vec v,t}(x)$. To find the value of $\phi_{\vec v,t*}\vec u$ at $x$, we invert~\eqref{pathorderedexp} to $f(x)=\E^{-\vec vt}f(\phi_{\vec v,t}(x))$. This leads to
\begin{equation}
\phi_{\vec v,t*}\vec u=\E^{-\vec vt}\vec u e^{\vec vt}\,\quad\text{hence}\quad\ld{\vec v}\vec u=[\vec v,\vec u]\;,
\label{lieDvector}
\end{equation}
where I introduced the commutator of $\vec v$ and $\vec u$ as differential operators, also called the \emph{Lie bracket}. Such a commutator is only meaningful when expressed in local coordinates. However, the Lie derivative as defined by~\eqref{liederivativedef} is coordinate-independent. Hence also the Lie bracket is well-defined independently of the choice of coordinates.

Let us pause to ponder on the significance of the Lie bracket. First, as a formal commutator, this defines a Lie algebra structure on the space of vector fields on $\M$. The Jacobi identity for this Lie algebra is equivalent to the Leibniz (product) rule for the Lie derivative,
\begin{equation}
\ld{\vec u}[\vec v,\vec w]=[\ld{\vec u}\vec v,\vec w]+[\vec v,\ld{\vec u}\vec w]\;.
\end{equation}
What if we take a special set of vector fields that are the generators of the action of a Lie group $G$ on $\M$? We find, not surprisingly, that the Lie brackets of these fields reproduce the Lie algebra $\lie g$ of $G$. Instead of a formal proof of this statement, let me show a simple illustrative example.

\begin{illustration}%
In \refex{ex:R2rotation}, I introduced a vector field that generates rotations in~$\R^2$ around the origin. This is easily promoted to $\R^3$ where we have three independent rotations around the Cartesian coordinate axes, generated by
\begin{equation}
\vec v_x=y\de_z-z\de_y\;,\qquad
\vec v_y=z\de_x-x\de_z\;,\qquad
\vec v_z=x\de_y-y\de_x\;.
\end{equation}
The Lie brackets of these,
\begin{equation}
[\vec v_x,\vec v_y]=-\vec v_z\;,\qquad
[\vec v_y,\vec v_z]=-\vec v_x\;,\qquad
[\vec v_z,\vec v_x]=-\vec v_y\;,
\end{equation}
reproduce the Lie algebra of infinitesimal rotations in $\R^3$, $\lie{so}(3)$. Further generalization to $\R^n$ is straightforward and amounts to considering the set of vector fields $\vec v_{ab}\equiv x_a\de_b-x_b\de_a$ with $a,b=1,\dotsc,n$.
\end{illustration}

Second, having at hand a coordinate basis, it is easy to give a general expression for the Lie derivative of a vector field in terms of its components,
\begin{equation}
(\ld{\vec v}\vec u)^a=v^b\de_b u^a-u^b\de_b v^a=\vec v[u^a]-\vec u[v^a]\;.
\label{liebracket}
\end{equation}
The latter expression clarifies the comment on the relation between coordinate basis fields and (local) frames that I made at the end of Sect.~\ref{appsubsec:tangentvectors}. Namely, the set of coordinate basis fields $\smash{\{\de_a\}_{a=1}^n}$ always has vanishing Lie brackets. Thus the necessary condition for the existence of local coordinates $\smash{y^A}$ such that a frame $\smash{\{\vec e_A\}_{A=1}^n}$ can be represented by $\smash{\vec e_A=\Pd{}{y^A}}$ is $[\vec e_A,\vec e_B]=0$ for all $A,B$. It turns out that this condition is also sufficient; see Chap.~9 of~\cite{Lee2013a} for a proof.

Let us now return to the general discussion of the Lie derivative. This can be evaluated on other tensor fields by making use of what we already know for vector fields. For example, let us take a 1-form $\o$ and a test vector field $\vec u$. Using the definition~\eqref{pullback} of pull-back together with the inverse of~\eqref{pathorderedexp}, we get
\begin{equation}
(\phi_{\vec v,t*}\o)(\phi_{\vec v,t*}\vec u)=(\phi_{\vec v,-t}^*\o)(\phi_{\vec v,t*}\vec u)=\E^{-\vec vt}\o(\vec u)\;.
\end{equation}
Combining the general definition~\eqref{liederivativedef} of the Lie derivative with the Leibniz rule then leads immediately to
\begin{equation}
(\ld{\vec v}\o)(\vec u)=\vec v[\o(\vec u)]-\o([\vec v,\vec u])\;.
\label{lieD1form}
\end{equation}
In fact, the same reasoning can be applied without change to any differential $p$-form. For a set of test vector fields $\vec u_1,\dotsc,\vec u_p$, we find
\begin{equation}
\begin{split}
(\ld{\vec v}\o)(\vec u_1,\dotsc,\vec u_p)={}&\vec v[\o(\vec u_1,\dotsc,\vec u_p)]\\
&-\sum_{i=1}^p\o(\vec u_1,\dotsc,\vec u_{i-1},[\vec v,\vec u_i],\vec u_{i+1},\dotsc,\vec u_p)\;.
\end{split}
\label{lieDforms}
\end{equation}
For tensors of the general type $(q,p)$, it is more straightforward to work with components with respect to a coordinate basis. One can think of a tensor $\tens T$ as a multilinear map acting on a set of test vector fields and 1-forms. Combining the Leibniz rule with~\eqref{lieDvector} and~\eqref{lieD1form} then leads, after some manipulation, to
\begin{equation}
\begin{split}
(\ld{\vec v}\tens T)^{ab\dotsb}_{mn\dotsb}=v^l\de_l \tens T^{ab\dotsb}_{mn\dotsb}&+\tens T^{ab\dotsb}_{ln\dotsb}\de_m v^l+\tens T^{ab\dotsb}_{ml\dotsb}\de_n v^l+\dotsb\\
&-\tens T^{lb\dotsb}_{mn\dotsb}\de_l v^a-\tens T^{al\dotsb}_{mn\dotsb}\de_l v^b-\dotsb\;.
\end{split}
\label{Lied_formulas}
\end{equation}
It is common to extend this identity to the special case of $(q,p)=(0,0)$, corresponding to functions on $\M$. In this case, the Lie derivative coincides with the previously defined directional derivative, $\ld{\vec v}f=\vec v[f]$.

Before closing the section, let me mention a useful identity, connecting the Lie derivative of differential forms with the interior product, defined by~\eqref{interiorproduct}. For two vector fields $\vec u,\vec v$, one finds that
\begin{equation}
\ld{\vec u}\circ\ix{\vec v}-\ix{\vec v}\circ\ld{\vec u}=\ix{[\vec u,\vec v]}\;.
\end{equation}
This is another variation on the Leibniz rule, as one can easily see be rewriting it as $\ld{\vec u}(\ix{\vec v}\o)=\ix{[\vec u,\vec v]}\o+\ix{\vec v}(\ld{\vec u}\o)$ for an arbitrary test $p$-form $\o$. The proof proceeds by combining the definition~\eqref{interiorproduct} of interior product with~\eqref{lieDforms}.


\section{Exterior Derivative}
\label{appsec:exterior}

Until now, I have tried to treat on more or less equal footing tensor fields of arbitrary type $(q,p)$. However, as Sect.~\ref{appsubsec:tensors} already hinted, the (differential) $p$-forms have a special standing among all the tensors. In this section, I will introduce a new differential structure on the Grassmann algebra of differential forms. The central concept thereof is the \emph{exterior derivative}, which is a derivative-like operation that converts a $p$-form into a $(p+1)$-form. This is a tool of great importance for much of both differential and integral calculus on manifolds. There are several ways of introducing the exterior derivative with different balance of abstraction, elegance and technical simplicity. While there does not seem to be a unique, intuitively simple definition, I will at least try to make a brief comparison of the different approaches.

Consider a differential $p$-form $\o$ and a set of $p+1$ test vector fields, $\vec v_i$. We can define the exterior derivative of $\o$ as the unique $(p+1)$-form $\D\o$ such that
\begin{equation}
\begin{split}
\D\o(\vec v_1,\dotsc,\vec v_{p+1})={}&\sum_{i=1}^{p+1}(-1)^{i+1}\vec v_i[\o(\vec v_1,\dotsc,\hat{\vec v}_i,\dotsc,\vec v_{p+1})]\\
&+\sum_{i<j}(-1)^{i+j}\o([\vec v_i,\vec v_j],\vec v_1,\dotsc,\hat{\vec v}_i,\dotsc,\hat{\vec v}_j,\dotsc,\vec v_{p+1})
\end{split}
\label{ddef}
\end{equation}
for any choice of test fields, where a hat denotes an argument that is to be omitted. This definition is manifestly independent of the choice of local coordinates, but not exactly intuitive. Let us therefore try to unfold its content in some simple cases. First, for 0-forms, that is functions on $\M$, \eqref{ddef} gives simply $\D f(\vec v)=\vec v[f]$. This agrees with our previous definition~\eqref{df} of the differential of a function. Second, for a 1-form $\o$ and a 2-form $\Omega$ we get respectively
\begin{equation}
\begin{split}
\D\o(\vec u,\vec v)={}&\vec u[\o(\vec v)]-\vec v[\o(\vec u)]-\o([\vec u,\vec v])\;,\\
\D\Omega(\vec u,\vec v,\vec w)={}&\vec u[\Omega(\vec v,\vec w)]+\vec v[\Omega(\vec w,\vec u)]+\vec w[\Omega(\vec u,\vec v)]\\
&-\Omega([\vec u,\vec v],\vec w)-\Omega([\vec v,\vec w],\vec u)-\Omega([\vec w,\vec u],\vec v)\;,
\end{split}
\label{cartan}
\end{equation}
for arbitrary test vector fields $\vec u,\vec v,\vec w$. It is becoming clear that for any $p$-form $\o$, $\D\o$ only contains two types of contributions; the summation in~\eqref{ddef} is needed to ensure that $\D\o$ maintains full antisymmetry as a multilinear map on vector fields. It is possible to show (see Chap.~36 of~\cite{Needham2021a}) that~\eqref{ddef} generalizes the concept of a directional derivative from functions to $p$-forms. This accounts for the first line of~\eqref{ddef}. The second line thereof makes $\D\o$ local so that $\D\o(\vec v_1,\dotsc,\vec v_{p+1})\bigr\rvert_x$ only depends on the values of the test vector fields at the point $x$, not on their derivatives.

By picking the test vector fields $\vec v_i$ from a coordinate basis and using the component expansion of $p$-forms~\eqref{pformcoordbasis}, we find
\begin{equation}
\D\o=\frac1{p!}(\de_b\o_{a_1\dotsb a_p})\D x^b\w\D x^{a_1}\w\dotsb\w\D x^{a_p}\;.
\label{ddef2}
\end{equation}
This is in practice the easiest way to calculate the exterior derivative of a differential form. It makes manifest the fundamental property of the exterior derivative that
\begin{equation}
\D^2\o\equiv\D(\D\o)=0
\label{dd0}
\end{equation}
for any $p$-form $\o$. Moreover, it can be used as an alternative definition to~\eqref{ddef} that is technically simple and quite intuitive. The downside of this approach is that it is not obvious that an object defined by~\eqref{ddef2} actually exists, independently of the choice of local coordinates. To prove this requires some extra effort.

\newpage

\begin{illustration}%
\label{ex:hodgeR3}%
In case $\o$ is a 0-form, that is a function on $\M$, the components of $\D\o$ are $(\D\o)_a=\de_a\o$. This generalizes the gradient of functions in $\R^n$ known from ordinary vector calculus. If, on the other hand, $\o$ is a 1-form, it follows at once from~\eqref{ddef2} that the components of $\D\o$ are
\begin{equation}
(\D\o)_{ab}=\de_a\o_b-\de_b\o_a\;.
\label{d1form}
\end{equation}
This appears to generalize the concept of a curl of a vector field in $\R^3$. The problem with this interpretation is that neither $\o$ nor $\D\o$ is a vector field. We will see in Sect.~\ref{appsubsec:hodge} how to circumvent this issue.

Antisymmetric derivatives such as~\eqref{d1form} are common in physics. If $A=A_a\D x^a$ represents the potential of an electromagnetic field, then~\eqref{d1form} are the components of the corresponding field-strength tensor, $F=\D A$. In this case, the fundamental property~\eqref{dd0} of exterior derivative encodes the familiar Bianchi identity for the electromagnetic field,
\begin{equation}
\de_cF_{ab}+\de_aF_{bc}+\de_bF_{ca}=0\;.
\end{equation}
\end{illustration}

The exterior derivative interconnects nicely with other operations on differential forms that we have met. It is for example easy to prove using the coordinate definition~\eqref{ddef2} that the exterior derivative is an antiderivation of the Grassmann algebra of differential forms. Thus, for any $p$-form $\o$ and any $q$-form $\s$,
\begin{equation}
\D(\o\w\s)=\D\o\w\s+(-1)^p\o\w\D\s\;.
\label{antiderivationd}
\end{equation}
The coordinate-free definition~\eqref{ddef}, on the other hand, makes it easier to show that the exterior derivative ``commutes'' with pull-back and hence with Lie derivative. This means that for any smooth map $f:\M\to\MN$, differential form $\o$ on $\MN$ and vector field $\vec v$ on $\MN$ we have
\begin{equation}
\D(f^*\o)=f^*(\D\o)\;,\qquad
\D(\ld{\vec v}\o)=\ld{\vec v}(\D\o)\;.
\label{dpullback}
\end{equation}
The interior product, Lie derivative and exterior derivative constitute three derivative-like operations on differential forms. The first decreases the form degree by one, the second leaves it unchanged and the last increases it by one. There is a simple identity that brings them all together which, perhaps for its intrinsic beauty, is usually referred to as ``Cartan's magic formula,''
\begin{equation}
\ld{\vec v}=\ix{\vec v}\circ\D+\D\circ\ix{\vec v}\quad\text{or}\quad
\ld{\vec v}\o=\ix{\vec v}(\D\o)+\D(\ix{\vec v}\o)\;,
\label{Cartanmagicformula}
\end{equation}
for any $p$-form $\o$ and vector field $\vec v$. The proof is a straightforward combination of the coordinate-free definition~\eqref{ddef}, the Lie derivative of differential forms~\eqref{lieDforms} and the definition~\eqref{interiorproduct} of interior product. Note how the Cartan magic formula together with $\D\circ\D=0$ immediately implies the second of the identities in~\eqref{dpullback}.

\begin{illustration}%
\label{ex:MCform}%
Here is a somewhat more advanced example, closely related to the subject of this book. Take a Lie group $G$ and introduce the following object, called the \emph{Maurer--Cartan} (MC) \emph{form},
\begin{equation}
\mc\equiv-\I g^{-1}\D g\equiv\mc^AQ_A\;,\qquad g\in G\;,
\label{connLieG}
\end{equation}
where $Q_A$ are the generators of the Lie algebra $\lie g$ of $G$. In order to give a meaning to this definition, one should think of $g\in G$ in terms of a set of functions on $G$, returning the matrix elements of $g$ in some faithful representation of $G$. The identities~\eqref{hadamard} and~\eqref{AdA} of the main text of this book guarantee that $\mc$ takes values in the Lie algebra $\lie g$. Hence the components $\mc^A$ constitute a set of well-defined 1-forms on $G$.

Given that the dimension of $G$ as a manifold equals that of its Lie algebra, the 1-forms $\mc^A$ furnish a globally well-defined coframe on $G$. The exterior derivative $\D\mc^A$ should therefore be a linear combination of $\mc^B\w\mc^C$. Indeed, a short calculation shows that the expression of $\D\mc^A$ in terms of $\mc^B\w\mc^C$ reflects the algebraic structure of $G$ as a Lie group,
\begin{equation}
\begin{split}
\D\mc&=-\I\D g^{-1}\w\D g=\I g^{-1}\D g\w g^{-1}\D g=-\I\mc\w\mc\;,\\
\D\mc^A&=\frac12f^A_{BC}\mc^B\w\mc^C\;,
\end{split}
\label{MCequationG}
\end{equation}
where $f^A_{BC}$ are the structure constants of $G$. This is known as the MC equation.
\end{illustration}

Before closing the quick overview of exterior calculus, let me mention in passing another way to define the exterior derivative. This might please a reader who prefers a more axiomatic approach. Namely, the exterior derivative turns out to be a unique antiderivation of the Grassmann algebra that reduces to the ordinary differential on 0-forms and satisfies the fundamental property~\eqref{dd0}. For a reader that might prefer a more geometric definition of exterior derivative, there is yet another possibility. This is intimately related to integration on manifolds and I will return to it in Sect.~\ref{appsec:integration}.


\section{Affine Connection}
\label{appsec:affine}

In order to develop a full-fledged differential calculus on manifolds, we have to be able to differentiate not only functions, but also vector and tensor fields. In spite of the terminology, none of the two previously introduced ``derivatives'' is suitable for the purpose. First, the Lie derivative $\ld{\vec v}$ actually does not have the properties we would expect from a derivative in that it is not local in $\vec v$. The value of $\ld{\vec v}\tens T$ for a given tensor field $\tens T$ at $x\in\M$ depends not only on $\vec v(x)$ but also on its derivatives, as is clear from~\eqref{Lied_formulas}. Second, the exterior derivative is only defined for differential forms and not for other tensors. To have a well-defined notion of derivative for tensor fields of arbitrary type, we need additional structure on the manifold.

In order to see what the required structure might be, let us consider the simplest case of a test vector field $\vec u$. We might think of first finding its components $u^A$ in a chosen local frame $\vec e_A$ and then taking their directional derivative $\vec v[u^A]$ with respect to a fixed vector field $\vec v$. That is however problematic. The choice of frame $\vec e_A$ is arbitrary and can be changed locally. This is a smoking gun of what in field theory is called ``gauge freedom.'' The directional derivative of $\vec u$ must be a well-defined vector field independent of such a ``choice of gauge.'' This is not the case of $\vec v[u^A]\vec e_A$, so $\vec v[u^A]$ cannot be the components of a well-defined vector field. Following the analogy with physics, the problem can be fixed by introducing a gauge field.


\subsection{Covariant Derivative}
\label{appsubsec:covder}

Let us now formalize the above observation. The basis $\vec e_A(x)$ of $\T_x\M$ can be changed by any nonsingular linear transformation that varies (smoothly) with $x$. Upon such a transformation, the basis vectors and components of a tangent vector $\vec u$ change to
\begin{equation}
\tilde{\vec e}_A=\vec e_B(\transfo^{-1})^B_{\phantom BA}\;,\qquad
\tilde u^A=\transfo^A_{\phantom AB}u^B\;,\qquad
\transfo\in\gr{GL}(n)\;.
\label{frametransfo}
\end{equation}
The group $\gr{GL}(n)$ is our gauge group; in differential geometry, the term \emph{structure group} is common. Any geometrically well-defined object on the manifold $\M$ must be invariant under gauge transformations of the type~\eqref{frametransfo}. It is known from field theory that in order to have a well-defined notion of derivative of fields, one needs a gauge field, or gauge connection. In differential geometry, this object is called \emph{affine connection}. In a given frame, it constitutes a collection of 1-forms $\con^A_{\phantom AB}$. Treating these as the matrix elements of a matrix-valued 1-form $\con$, the action of the basis transformation~\eqref{frametransfo} on the affine connection can be defined compactly as
\begin{equation}
\tilde\con=\transfo\con\transfo^{-1}+\transfo\D\transfo^{-1}\;,\qquad
\transfo\in\gr{GL}(n)\;.
\label{gaugetransfo}
\end{equation}
This is analogous to a (non-Abelian) gauge transformation familiar from field theory.

\begin{watchout}%
The collection of 1-forms $\smash{\con^A_{\phantom AB}}$ carries one vector and one covector frame index. In spite of the suggestive notation, the object $\smash{\vec e_A\otimes\vec e^{*B}\otimes\con^A_{\phantom AB}}$ is not a well-defined tensor field of type $(1,2)$ on $\M$. This is a consequence of the second term in~\eqref{gaugetransfo}. The affine connection is therefore inseparably tied to the chosen local frame.
\end{watchout}

Following the analogy with field theory, we now define a directional \emph{covariant derivative} $\cd_{\vec v}\vec u$ of the vector field $\vec u$ along $\vec v$ component-wise as
\begin{equation}
(\cd_{\vec v}\vec u)^A\equiv\vec v[u^A]+\con^A_{\phantom AB}(\vec v)u^B\;.
\label{covder}
\end{equation}
It is easy to check that the components $(\cd_{\vec v}\vec u)^A$ transform under the local change of basis~\eqref{frametransfo} just like $u^A$. Hence the directional covariant derivative $\cd_{\vec v}\vec u\equiv(\cd_{\vec v}\vec u)^A\vec e_A$ is a well-defined vector field on $\M$ independent of the choice of frame. Moreover, $\cd_{\vec v}\vec u$ is manifestly linear and local in $\vec v$. We can therefore separate $\vec v$ out and only keep a tensor object that encodes the variation of $\vec u$ with respect to the local frame. This is done by setting $(\cd_{\vec v}\vec u)^A=v^B(\cd_{\vec e_B}\vec u)^A\equiv v^B(\cd_B\vec u)^A$, where
\begin{equation}
(\cd_B\vec u)^A=\vec e_B[u^A]+\Gamma^A_{BC}u^C\;,\qquad
\Gamma^A_{BC}\equiv\con^A_{\phantom AC}(\vec e_B)=(\cd_B\vec e_C)^A\;,
\label{covdercomp}
\end{equation}
and $\Gamma^A_{BC}$ are the \emph{Christoffel symbols}. These are independent of $\vec u$ and characterize the geometry of the manifold and of the local frame.

The field theory analogy allows us to write down at once the covariant derivative of a tensor field of any type $(q,p)$. Different types of tensors transform under different representations of the structure group, induced by the fundamental representation~\eqref{frametransfo} acting on vectors. We can thus reuse~\eqref{covder} by inserting the appropriate representation of the affine connection. For instance, tensor fields of type $(0,0)$, that is functions on $\M$, are blind to the change of local frame. This corresponds to the trivial representation of $\gr{GL}(n)$. The covariant derivative then reduces to $\cd_{\vec v}f=\vec v[f]$. For a generic tensor $\tens T$ of type $(q,p)$ we find
\begin{equation}
\begin{split}
(\cd_{\vec v}\tens T)^{AB\dotsb}_{KL\dotsb}=\vec v[\tens T^{AB\dotsb}_{KL\dotsb}]&+\con^A_{\phantom AI}(\vec v)\tens T^{IB\dotsb}_{KL\dotsb}+\con^B_{\phantom BI}(\vec v)\tens T^{AI\dotsb}_{KL\dotsb}+\dotsb\\
&-\tens T^{AB\dotsb}_{IL\dotsb}\con^I_{\phantom IK}(\vec v)-\tens T^{AB\dotsb}_{KI\dotsb}\con^I_{\phantom IL}(\vec v)-\dotsb\;.
\end{split}
\label{covdertensor}
\end{equation}
This can also be rewritten in a manner independent of the vector field $\vec v$, following the example of~\eqref{covdercomp},
\begin{equation}
\begin{split}
(\cd_C\tens T)^{AB\dotsb}_{KL\dotsb}=\vec e_C[\tens T^{AB\dotsb}_{KL\dotsb}]&+\Gamma^A_{CI}\tens T^{IB\dotsb}_{KL\dotsb}+\Gamma^B_{CI}\tens T^{AI\dotsb}_{KL\dotsb}+\dotsb\\
&-\tens T^{AB\dotsb}_{IL\dotsb}\Gamma^I_{CK}-\tens T^{AB\dotsb}_{KI\dotsb}\Gamma^I_{CL}-\dotsb\;.
\end{split}
\label{covdertensorcomp}
\end{equation}

\begin{illustration}%
\label{ex:connectionRn}%
The natural choice of $\vec e_A$ in $\R^n$ is the coordinate basis in the Cartesian coordinates. In this basis, we can set $\con=0$ so that $\cd_{\vec v}\vec u=\vec v[u^A]\vec e_A$. This expresses the fact known from elementary vector calculus that $\vec e_A$ is a globally well-defined, constant basis of vectors. Switching to another local frame $\tilde{\vec e}_A$ via~\eqref{frametransfo} induces a nontrivial connection, $\tilde\con=\transfo\D\transfo^{-1}$. This characterizes the ``variation'' of the new frame throughout $\R^n$. A short manipulation leads to an expression for the Christoffel symbols in terms of the derivatives of $\transfo$,
\begin{equation}
\begin{split}
\tilde\Gamma^A_{BC}&=-(\D\transfo\transfo^{-1})^A_{\phantom AC}(\tilde{\vec e}_B)=-(\transfo^{-1})^E_{\phantom EC}\tilde{\vec e}_B[\transfo^A_{\phantom AE}]\\
&=-(\transfo^{-1})^D_{\phantom DB}(\transfo^{-1})^E_{\phantom EC}\vec e_D[\transfo^A_{\phantom AE}]=-(\transfo^{-1})^D_{\phantom DB}(\transfo^{-1})^E_{\phantom EC}\de_D\transfo^A_{\phantom AE}\;.
\end{split}
\end{equation}
Equations~\eqref{covdercomp} and~\eqref{covdertensorcomp} then allow us to calculate derivatives of vectors and tensors in $\R^n$ in terms of their components in arbitrary curvilinear coordinates.
\end{illustration}


\subsection{Curvature and Torsion}
\label{appsubsec:curvtor}

We can pursue the analogy with field theory even further. Therein, one uses the gauge connection to construct the corresponding field strength which encodes information about the gauge field in a gauge-covariant fashion. In the geometric language, this leads to the \emph{curvature 2-form} on the manifold $\M$,
\begin{equation}
R^A_{\phantom AB}\equiv\D\con^A_{\phantom AB}+\con^A_{\phantom AC}\w\con^C_{\phantom CB}\;.
\label{curvature}
\end{equation}
The matrix elements $R^A_{\phantom AB}$ can be put together into a single object, $R\equiv \vec e_A\otimes\vec e^{*B}\otimes R^A_{\phantom AB}$. Unlike the matrix-valued connection 1-form $\con$, this is a well-defined tensor field on $\M$, of type $(1,3)$, independent of the choice of local frame. It therefore carries information about the intrinsic geometry of the manifold. The curvature 2-form is related to the commutator of directional covariant derivatives via
\begin{equation}
\bigl(\cd_{\vec u}\cd_{\vec v}\vec w-\cd_{\vec v}\cd_{\vec u}\vec w-\cd_{[\vec u,\vec v]}\vec w\bigr)^A=R^A_{\phantom AB}(\vec u,\vec v)w^B\;,
\label{curvaturecoordfree}
\end{equation}
where $\vec u$, $\vec v$ and $\vec w$ are vector fields. This identity follows directly from the definition~\eqref{covder} of covariant derivative upon using the first line of~\eqref{cartan}. Using~\eqref{covdertensor}, it can moreover be immediately generalized to the commutator of covariant derivatives of an arbitrary tensor field $\tens T$,
\begin{align}
\label{Rcommtens}
\bigl([\cd_{\vec u},\cd_{\vec v}]\tens T-\cd_{[\vec u,\vec v]}\tens T\bigr)^{AB\dotsb}_{KL\dotsb}={}&R^A_{\phantom AI}(\vec u,\vec v)\tens T^{IB\dotsb}_{KL\dotsb}+R^B_{\phantom BI}(\vec u,\vec v)\tens T^{AI\dotsb}_{KL\dotsb}+\dotsb\\
\notag
&-\tens T^{AB\dotsb}_{IL\dotsb}R^I_{\phantom IK}(\vec u,\vec v)-\tens T^{AB\dotsb}_{KI\dotsb}R^I_{\phantom IL}(\vec u,\vec v)-\dotsb\;.
\end{align}

The affine connection also gives rise to another well-defined tensor field that characterizes the geometry of the manifold. Namely, the elements of the dual coframe $\vec e^{*A}$ transform under~\eqref{frametransfo} identically to the components of a vector. We can then define a covariant exterior derivative of the coframe in parallel with~\eqref{covder},
\begin{equation}
T^A\equiv\D\vec e^{*A}+\con^A_{\phantom AB}\w\vec e^{*B}\;.
\label{torsion}
\end{equation}
This is dubbed the \emph{torsion 2-form}. It is a collection of 2-forms whose components transform under the structure group as a vector. Hence $T\equiv\vec e_A\otimes T^A$ is a well-defined tensor of type $(1,2)$. The torsion 2-form satisfies an identity similar to~\eqref{curvaturecoordfree},
\begin{equation}
\bigl(\cd_{\vec u}\vec v-\cd_{\vec v}\vec u-[\vec u,\vec v]\bigr)^A=T^A(\vec u,\vec v)\;.
\label{torsioncoordfree}
\end{equation}

Owing to the fact that $\D\circ\D=0$, one may expect to find constraints on the curvature and torsion 2-forms by taking another (covariant) exterior derivative. Indeed, a short manipulation based on the definitions~\eqref{curvature} and~\eqref{torsion} leads to the so-called \emph{Bianchi identities},
\begin{equation}
\begin{split}
\D R^A_{\phantom AB}+\con^A_{\phantom AC}\w R^C_{\phantom CB}-R^A_{\phantom AC}\w\con^C_{\phantom CB}&=0\;,\\
\D T^A+\con^A_{\phantom AB}\w T^B&=R^A_{\phantom AB}\w\vec e^{*B}\;.
\end{split}
\label{bianchi}
\end{equation}

\begin{illustration}%
Using the MC form introduced in \refex{ex:MCform}, we can get insight into the geometry of Lie groups. The simplest way to introduce an affine connection on a Lie group $G$ is to set $\con^A_{\phantom AB}=0$ in the coframe $\mc^A$ defined by~\eqref{connLieG}. The definition~\eqref{curvature} then tells us that the curvature 2-form vanishes. The torsion 2-form does not, though. From the definition~\eqref{torsion} and the MC equation~\eqref{MCequationG}, we get that $T^A=(1/2)f^A_{BC}\mc^B\w\mc^C$, where $f^A_{BC}$ are the structure constants of $G$.

Making $\con$ trivial is however not the only choice of connection we can make. Given the structure that we have available on the Lie group $G$, we can try for instance
\begin{equation}
\prescript{\l}{}{\con}^A_{\phantom AB}\equiv\l f^A_{BC}\mc^C\;,
\end{equation}
where $\l\in\R$ is a parameter. This class of connections includes the previous trivial case as $\l=0$. Using the MC equation~\eqref{MCequationG}, one now finds that
\begin{equation}
\begin{split}
\prescript{\l}{}{R}^A_{\phantom AB}&=\left(\frac\l2f^A_{BE}f^E_{CD}+\l^2f^A_{EC}f^E_{BD}\right)\mc^C\w\mc^D\;,\\
\prescript{\l}{}{T}^A&=\left(\frac12-\l\right)f^A_{BC}\mc^B\w\mc^C\;.
\end{split}
\end{equation}
If we do not impose any constraints on the structure constants $f^A_{BC}$, we can generally make the curvature 2-form vanish only by returning to the special case $\l=0$. On the other hand, we can make the torsion 2-form vanish by setting $\l=1/2$. We can however not make both 2-forms vanish simultaneously. A generic non-Abelian Lie group therefore possesses nontrivial geometry as a manifold.
\end{illustration}

Our original motivation for introducing the affine connection was to take account of the freedom to choose a local frame at will. This is largely responsible for the fact that I have phrased this section entirely in a coordinate-free language so far. It is however possible to reformulate the definitions and properties of covariant derivative, curvature and torsion in terms of local coordinates. This leads to minor simplification of some of the equations in this section. Thus, the action of the basis vector $\vec e_C$ in~\eqref{covdertensorcomp} becomes a mere partial derivative and~\eqref{covdertensorcomp} turns into
\begin{equation}
\begin{split}
(\cd_c\tens T)^{ab\dotsb}_{kl\dotsb}=\de_c\tens T^{ab\dotsb}_{kl\dotsb}&+\Gamma^a_{ci}\tens T^{ib\dotsb}_{kl\dotsb}+\Gamma^b_{ci}\tens T^{ai\dotsb}_{kl\dotsb}+\dotsb\\
&-\tens T^{ab\dotsb}_{il\dotsb}\Gamma^i_{ck}-\tens T^{ab\dotsb}_{ki\dotsb}\Gamma^i_{cl}-\dotsb\;.
\end{split}
\label{covdertensorcoord}
\end{equation}
The Christoffel symbol herein is defined by a coordinate-basis equivalent of~\eqref{covdercomp},
\begin{equation}
\Gamma^a_{bc}=(\cd_b\de_c)^a\quad\text{or}\quad
\cd_b\de_c=\Gamma^a_{bc}\de_a\;.
\end{equation}
Together with~\eqref{torsion}, this shows that the torsion tensor corresponds to the antisymmetric part of the Christoffel symbol in a coordinate basis,
\begin{equation}
\Gamma^a_{bc}-\Gamma^a_{cb}=(\cd_b\de_c-\cd_c\de_b)^a=T^a(\de_b,\de_c)\equiv T^a_{bc}\;.
\label{torsioncoord}
\end{equation}
Finally, the left-hand side of the identity~\eqref{curvaturecoordfree} for the curvature tensor reduces to a mere commutator of covariant derivatives, hence
\begin{equation}
(\cd_b\cd_c\vec u-\cd_c\cd_b\vec u)^a=R^a_{\phantom ad}(\de_b,\de_c)u^d\equiv R^a_{\phantom adbc}u^d\;.
\end{equation}
Likewise, \eqref{Rcommtens} gives the commutator of covariant derivatives of an arbitrary tensor field in the coordinate basis,
\begin{equation}
\begin{split}
(\cd_c\cd_d\tens T-\cd_d\cd_c\tens T)^{ab\dotsb}_{kl\dotsb}={}&R^a_{\phantom aicd}\tens T^{ib\dotsb}_{kl\dotsb}+R^b_{\phantom bicd}\tens T^{ai\dotsb}_{kl\dotsb}+\dotsb\\
&-\tens T^{ab\dotsb}_{il\dotsb}R^i_{\phantom ikcd}-\tens T^{ab\dotsb}_{ki\dotsb}R^i_{\phantom ilcd}-\dotsb\;.
\end{split}
\label{Rcommtenscoord}
\end{equation}


\section{Riemannian Geometry}
\label{appsec:riemann}

My exposition of differential geometry has so far been largely ahistorical. I made a conscious effort to only introduce the structure that was strictly necessary in order to develop the elementary concepts of calculus on manifolds. Thus, it only requires the basic differential structure on a manifold to define tensor fields, their Lie derivative, and the exterior calculus of differential forms. The first time that a new, additional structure was needed was in the previous section. There, we saw how the concept of affine connection naturally arises from the need to compute (covariant) derivatives of tensor fields.

However, as the Greek origin of the word ``geometry'' suggests, much of the early progress in the subject was driven by the practical problem of measuring distance. The concept of distance is very natural when the manifold in question represents space or spacetime in physics. Yet, even more abstract realizations of manifolds often possess a quantitative notion of closeness that is more refined than mere topology. The purpose of this section is to introduce the structure needed to measure distance on a manifold. Despite being anchored in geometry, we shall see that this will also give us further useful tools for our quest to develop calculus on manifolds.


\subsection{Riemannian Metric}
\label{appsubsec:riemannmetric}

Since I have already mentioned the concept of metric before, let me start right away with the formal definition. A \emph{Riemannian metric} $g$ on a manifold $\M$ is a tensor field of type $(0,2)$ such that for any $x\in\M$, $g(x)$ is a symmetric, positive-definite bilinear form on $\T_x\M$. See \refex{ex:Euclideanmetric} and \refex{ex:spherevolumeform}. A manifold endowed with a Riemannian metric is called a \emph{Riemannian manifold}. It is common to represent the metric by its components in a given coframe or dual coordinate basis,
\begin{equation}
g=g_{AB}\vec e^{*A}\otimes\vec e^{*B}=g_{ab}\D x^a\otimes\D x^b\;,
\label{metricdef}
\end{equation}
where $g_{AB}=g(\vec e_A,\vec e_B)$ and $g_{ab}=g(\de_a,\de_b)$. A Riemannian metric gives by construction rise to an inner product, both locally of tangent vectors and of vector fields. It is practically convenient to have a shorthand notation for this inner product,
\begin{equation}
\inner{\vec u}{\vec v}\equiv g(\vec u,\vec v)=g_{AB}u^Av^B\;,
\end{equation}
where $\vec u,\vec v$ are arbitrary vector fields.

For some applications, the axioms for Riemannian metric are too restrictive. By removing the requirement of positive-definiteness and insisting merely that $g$ as a bilinear form is nondegenerate, one obtains a generalization called \emph{pseudo-Riemannian metric}. Likewise, manifolds endowed with a pseudo-Riemannian metric are known as \emph{pseudo-Riemannian manifolds}. This broader concept finds an important application in the general theory of relativity, which assumes the physical spacetime to be a \emph{Lorentzian manifold}. The latter possesses by definition a metric with signature $(-,+,+,\dotsc)$ or $(+,-,-,\dotsc)$, depending on the convention adopted. Most of the contents of this section apply, with occasional minor adjustments, to Riemannian and pseudo-Riemannian manifolds alike. However, within the context of the book, pseudo-Riemannian manifolds play a little role. I will therefore content myself with this remark of caution and focus solely on Riemannian manifolds in the following.

Before illustrating the utility of the metric, it is worthwhile to contemplate its very definition. How is this actually related to the distance of two different points on the manifold? First, the metric by definition acts point-wise on tangent vectors. Second, the tangent vectors themselves were defined as differential operators on local test functions. To connect the abstract formalism to the down-to-earth problem of measuring distance, it helps to recall the concept of flow generated by the vector field $\vec v$, introduced in Sect.~\ref{appsubsec:flow}. One can interpret $\inner{\vec v}{\vec v}$ as the squared length of the vector tangent to the curve (trajectory) $\g(t)$ defined by~\eqref{flow}. In the liquid analogy of the flow generated by $\vec v$, $\sqrt{\inner{\vec v}{\vec v}}$ becomes the speed of the liquid particle. The distance of points $x_{1,2}=\g(t_{1,2})$ measured along the curve $\g(t)$ is then naturally defined by integration,
\begin{equation}
d_\g(x_1,x_2)\equiv\int_{t_1}^{t_2}\at{\sqrt{\inner{\vec v}{\vec v}}}{\g(t)}\D t\;.
\label{curvelength}
\end{equation}
This distance is independent of the choice of parameterization of the trajectory, that is of the speed at which the trajectory is traversed by the liquid particle. It is therefore an intrinsic property of the curve $\g$ as a set of points on $\M$.

Now that we know how to measure the length of a curve on $\M$, we define the distance of points $x_1,x_2\in\M$ as the minimum possible value of $d_\g(x_1,x_2)$. This is a problem in variational calculus that leads to a second-order differential equation for $\g(t)$ (see Sect.~11.5.1 of~\cite{Stone2009a}),
\begin{equation}
\frac{\D^2\g^a}{\D t^2}+\LC^a_{bc}\OD{\g^b}t\OD{\g^c}t=0\;.
\label{geodesic}
\end{equation}
Here the set of coefficients $\LC^a_{bc}$ is determined by the Riemannian metric,
\begin{equation}
\LC^a_{bc}\equiv\frac12g^{ad}(\de_bg_{dc}+\de_cg_{bd}-\de_dg_{bc})\;,
\label{LCconnection}
\end{equation}
where $g^{ab}$ are the matrix elements of the inverse of $g_{ab}$. Equation~\eqref{geodesic} is known as the \emph{geodesic equation}. The resemblance between its left-hand side and the covariant derivative $\cd_{\vec v}{\vec v}$ with $v^a=\Od{\g^a}t$ is not accidental. The coefficients~\eqref{LCconnection} are the Christoffel symbols of a special connection on the Riemannian manifold $\M$, called the \emph{Levi-Civita} (LC) \emph{connection}. The fact that $\LC^a_{bc}$ is symmetric in its lower indices implies by means of~\eqref{torsioncoord} that the LC connection has vanishing torsion. A given Riemannian manifold can support many different affine connections. The LC connection is however unique in that it is torsion-free and makes the metric of the manifold covariantly constant (see Sect.~7.4 of~\cite{Nakahara2003a}),
\begin{equation}
(\hat\cd_cg)_{ab}=\de_cg_{ab}-g_{db}\LC^d_{ca}-g_{ad}\LC^d_{cb}=0\;.
\end{equation}

\begin{illustration}%
The origin of the LC connection can be understood in elementary terms. Imagine that the manifold $\M$ is smoothly embedded in $\R^n$ for some $n>\dim\M$. We can then rely on the geometric intuition whereby tangent vectors to $\M$ are ``realized'' as vectors in $\R^n$ by the embedding $f:\M\to\R^n$, as in \refex{ex:embedding}. I will take the liberty to identify tangent vectors to $\M$ with their images under the push-forward $f_*$.

\begin{figure}[t]
\sidecaption[t]
\includegraphics[width=2.9in]{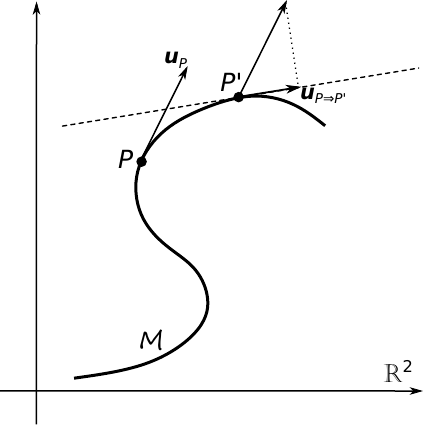}
\caption{Comparison of tangent vectors to $\M$ at different points $P$ and $P'$ is possible if the manifold is embedded in $\R^n$; here shown for simplicity for one-dimensional $\M$ embedded in $\R^2$. A tangent vector at $P\in\M$, $\vec u_P$, is transported to $P'$ as a constant vector in $\R^2$. Subsequently, it is projected to the tangent space to $\M$ at $P'$, resulting in the vector $\vec u_{P\Rightarrow P'}$. This can now be compared to any other tangent vector at $P'$. The distance of points $P$ and $P'$ is exaggerated for clarity; for the definition of covariant derivative, it should actually be infinitesimal}
\label{fig:paralleltransport}
\end{figure}

Suppose we want to define the derivative of a vector field $\vec u$ on $\M$ along the vector field $\vec v$. To do so, we need to compare $\vec u(x)$ at a given point $x\in\M$ with $\vec u(x+\eps v)$, where the shorthand notation $x+\eps v$ indicates the point on $\M$ with coordinates $x^a+\eps v^a(x)$. We cannot do this directly since the tangent spaces to $\M$ at $x$ and $x+\eps v$ are different. What we will do instead is take the vector $\vec u(x)$, transport it as a constant vector in $\R^n$ to $x+\eps v$ and then project it back to the hypersurface $\M$. See Fig.~\ref{fig:paralleltransport} for an illustration of the procedure.

It is easiest to use a frame $\vec e_A$ on $\M$ that is orthonormal with respect to the Euclidean metric in $\R^n$. The projection to $\M$ is then accomplished by taking the inner product of $\vec u(x)$ and $\vec e_A(x+\eps v)$. This leads to a definition of covariant derivative on $\M$ as
\begin{equation}
(\cd_{\vec v}\vec u)^A(x)=\lim_{\eps\to0}\frac{u^A(x+\eps v)-\inner{\vec e_A(x+\eps v)}{\vec u(x)}}\eps\;.
\end{equation}
Comparing this to~\eqref{covdercomp} allows us to identify the corresponding Christoffel symbols,
\begin{equation}
\Gamma^A_{BC}=-\inner{\de_B\vec e_A}{\vec e_C}=\inner{\vec e_A}{\de_B\vec e_C}\;,
\label{Christoffelembedding}
\end{equation}
where $\de_B\vec e_A(x)$ is defined as $\lim_{\eps\to0}[\vec e_A(x+\eps \vec e_B)-\vec e_A(x)]/\eps$. We have thus managed to define an induced affine connection on $\M$ in a way that encodes the variation of the orthonormal frame $\vec e_A$ as a set of vectors in $\R^n$. It is an easy exercise to check using~\eqref{torsioncoordfree} that this affine connection is torsion-free. Moreover, with the induced metric on $\M$ given by $g_{AB}=\inner{\vec e_A}{\vec e_B}=\d_{AB}$, one readily checks using~\eqref{covdertensorcomp} that the connection~\eqref{Christoffelembedding} makes the metric covariantly constant. Using the embedding in $\R^n$, we have therefore found the (unique) LC connection on $\M$.

As a side remark, recall that the comparison of tangent vectors at different points on $\M$ was made possible by ``transporting'' them as constant vectors in $\R^n$. This intuition lies behind an often-used geometric picture of affine connection in terms of \emph{parallel transport}. In general, a tensor field $\tens T$ is said to be parallel-transported along the vector field $\vec v$ if $\cd_{\vec v}\tens T=0$. The geodesic equation~\eqref{geodesic} then describes curves whose tangent vector is parallel-transported along the curve. The solutions of~\eqref{geodesic} thus have the dual interpretation as the shortest (or more precisely extremal) or the ``straightest'' paths on $\M$. See for example Sect.~7.2 of~\cite{Nakahara2003a} for more details. 
\end{illustration}

The metric obviously provides us with a crucial piece of information about the intrinsic geometry of a Riemannian manifold. It is nevertheless also invaluable for the calculus on manifolds, as it allows us to relate tensors of different types. It is known from elementary linear algebra that a vector space such as $\T_x\M$ for fixed $x\in\M$ and its dual, $\T^*_x\M$, have the same dimension. As such, they are necessarily isomorphic. There is, however, no natural way to set up the isomorphism without first fixing a basis in the two spaces. Here the metric comes to rescue. It allows us to define two mutually inverse \emph{musical isomorphisms}, $\flat:\T_x\M\to\T^*_x\M$ and $\sharp:\T^*_x\M\to\T_x\M$, and by extension isomorphisms between vector fields and 1-forms. Thus, any vector field $\vec v$ is mapped to its \emph{flat} $\vec v^\flat$, which is a 1-form such that
\begin{equation}
\vec v^\flat(\vec u)\equiv\inner{\vec v}{\vec u}\quad\text{for any vector field $\vec u$}\;.
\label{flat}
\end{equation}
Vice versa, any differential 1-form $\o$ is mapped to its \emph{sharp} $\o^\sharp$, which is a vector field such that
\begin{equation}
\inner{\o^\sharp}{\vec v}\equiv\o(\vec v)\quad\text{for any vector field $\vec v$}\;.
\label{sharp}
\end{equation}
Most theoretical physicists are familiar with these maps in their component forms,
\begin{equation}
v^\flat_A=g_{AB}v^B\;,\qquad
\o^{\sharp A}=g^{AB}\o_B\;.
\label{flatsharp}
\end{equation}
These correspond to the usual lowering and raising of indices by the metric and its inverse. The generalization of these operations to tensors of other types is straightforward and I will not dwell on details.

\begin{watchout}%
It is common in the physics literature to denote a tensor and its image under $\flat$ or $\sharp$ by the same symbols, and only distinguish them by the positions of their indices. In a restricted sense, it can even be meaningful to identify a vector with its image under $\flat$ or a 1-form with its image under $\sharp$. Suppose that we choose an \emph{orthonormal frame} on $\M$, that is a frame $\vec e_A$ such that $\inner{\vec e_A}{\vec e_B}=\d_{AB}$. Then it follows from~\eqref{flat} that $\flat$ maps this basis to the corresponding dual basis, $\vec e_A^\flat=\vec e^{*A}$. The same correspondence applies to the inverse isomorphism, $\smash{(\vec e^{*A})^\sharp=\vec e_A}$. By linearity, the components of any vector then remain unaffected by $\flat$, and the components of any 1-form remain unaffected by $\sharp$.
\end{watchout}


\subsection{Isometries of Riemannian Metric}
\label{appsubsec:isometries}

The whole book is devoted to symmetries and their consequences in quantum field theory. But what do we mean by a symmetry in case of a manifold? I already pointed out in Sect.~\ref{appsubsec:flow} that a group $G$ can act on a manifold $\M$ via a set of transformations (diffeomorphisms) on $\M$. However, not every transformation on $\M$ is a symmetry. For that, the transformation should in some sense respect the structure of the manifold. The most common type of symmetry one meets in differential geometry is one that preserves the Riemannian metric of the manifold.

Formally, a diffeomorphism $f:\M\to\M$ is called an \emph{isometry} if it leaves the Riemannian metric $g$ on $\M$ invariant, $f^*g=g$. Equivalently, one may require that the inner product of test vector fields $\vec u,\vec v$ is unaffected by push-forward by $f$,\footnote{This definition can at once be generalized to a \emph{local isometry}, which is a smooth but not necessarily invertible map between two different manifolds, $f:\M\to\MN$, satisfying~\eqref{isometry}. (The inner products on the two sides of the equation are then defined with the respective metrics on $\M$ and $\MN$.) In this case, it is of course not appropriate to speak of a symmetry of either $\M$ or $\MN$.}
\begin{equation}
\inner{f_*\vec u}{f_*\vec v}\bigr\rvert_{f(x)}=\inner{\vec u}{\vec v}\bigr\rvert_x\quad\text{for any }x\in\M\;.
\label{isometry}
\end{equation}
The set of isometries of a Riemannian manifold forms a group $G$ under map composition. This group may be both discrete, or even trivial, and continuous. It is however generally not easy to determine the isometry group $G$ of a given manifold $\M$ directly, unless one has some additional insight into the structure of $\M$.

The task is much easier when $G$ is a Lie group. We can then focus on symmetries of the metric under infinitesimal transformations. In line with the discussion in Sect.~\ref{appsubsec:liederivative}, this amounts to solving the condition $\ld{\vec\kil}g=0$, where $\vec\kil$ is a vector field representing a generator of the isometry group on $\M$. This is known as the \emph{Killing vector field}, or more briefly \emph{Killing vector}. The condition on $\vec\kil$ is equivalent to a set of homogeneous linear partial differential equations for the components of $\vec\kil$. These are usually much easier to solve than~\eqref{isometry}. In practice, it is often convenient to use an alternative formulation of the invariance condition, based on the LC connection instead of the Lie derivative,
\begin{equation}
\ld{\vec\kil}g=0\quad\Leftrightarrow\quad
(\hat\cd_a\vec\kil^\flat)_b+(\hat\cd_b\vec\kil^\flat)_a=0\;.
\label{killing}
\end{equation}
One then first solves for the 1-form $\vec\kil^\flat$ and subsequently recovers the Killing vector field $\vec\kil$ by acting on $\vec\kil^\flat$ with $\sharp$.

\begin{illustration}%
\label{ex:Killing}%
For illustration, let us find the continuous isometries of the Euclidean metric, $g=\d_{ab}\D x^a\otimes\D x^b$, in $\R^n$. Dropping for simplicity the superscript $\flat$ on $\vec\kil$, and using the fact that the LC connection of the Euclidean metric is trivial in Cartesian coordinates, \eqref{killing} becomes
\begin{equation}
\de_a\kil_b+\de_b\kil_a=0\;.
\label{killingeuclidean}
\end{equation}
Hence $\de_a\kil_b$ must be antisymmetric in its two indices. Using this repeatedly together with the symmetry of second partial derivatives, we infer
\begin{equation}
\de_a\de_b\kil_c=-\de_a\de_c\kil_b=\de_b\de_c\kil_a=-\de_b\de_a\kil_c=-\de_a\de_b\kil_c\;.
\end{equation}
It follows that all second partial derivatives of $\kil_a$ must vanish. This conclusion has a neat generalization to an arbitrary Riemannian manifold. Following essentially the same reasoning, one finds that
\begin{equation}
\hat\cd_a\hat\cd_b\kil_c=\kil_dR^d_{\phantom dabc}\quad\text{or}\quad
\hat\cd_a\hat\cd_b\kil^c=R^c_{\phantom cbad}\kil^d\;,
\label{killingR}
\end{equation}
where $\smash{R^d_{\phantom dabc}}$ is the curvature tensor of the LC connection. The outer covariant derivative in~\eqref{killingR} is defined by treating $\hat\cd_b\kil_c$ as a rank-2 tensor.

Together with~\eqref{killingeuclidean}, the vanishing of second partial derivatives of $\kil_a$ implies  the most general solution of the \emph{Killing equation}~\eqref{killing} for the Euclidean metric,
\begin{equation}
\kil_a(x)=\a_a+\b_{ab}x^b\;.
\label{killingeuclidean2}
\end{equation}
Here $\a_a$ is a set of constants and $\b_{ab}$ a constant antisymmetric matrix. This corresponds to a combination of infinitesimal translations and rotations, which together generate the Euclidean group $\gr{ISO}(n)$.
\end{illustration}

The Killing equation only gives us information about the Lie algebra of the isometry group $G$. The reconstruction of the whole group thus suffers from the usual ambiguities of the correspondence between Lie groups and Lie algebras. In addition, \eqref{killing} is based on local information about the metric structure of the manifold $\M$. The global topology of $\M$ may select only a subalgebra of solutions to~\eqref{killing}. The following example, adapted from Sect.~4.6 of~\cite{Fecko2011a}, illustrates this subtlety.

\begin{illustration}%
\label{ex:killingtorus}%
The set $\Gamma\equiv\{(c_1,\dotsc,c_n)\,\vert\,c_a\in\Z\}$ defines a lattice of points in $\R^n$ with integer coordinates. The quotient set $\R^n/\Gamma$ is the $n$-dimensional torus, $T^n$. In local Cartesian coordinates, the Euclidean metric on $T^n$ takes the same form as in $\R^n$, $g=\d_{ab}\D x^a\otimes\D x^b$. The Killing equation~\eqref{killing} therefore produces the same Lie algebra of local solutions as in \refex{ex:Killing}. However, not all the solutions~\eqref{killingeuclidean2} are well-defined 1-forms on $T^n$. In fact, the matrix $\b_{ab}$ has to vanish; the torus does not have continuous rotation symmetries. Moreover, the  continuous translation symmetry of $T^n$ differs from that of $\R^n$. It is $\gr{U}(1)^{\times n}$ for the torus but $\R^n$ for the Euclidean space. In addition, the torus has a discrete group of rotation isometries, which corresponds to the point group of the lattice $\Gamma$. Such discrete isometries are of course not captured by the Killing equation~\eqref{killing} at all.
\end{illustration}


\subsection{Symmetries of Curvature Tensor}
\label{appsubsec:symofR}

Before developing further the basics of Riemannian geometry, I will digress to put together some useful properties of the curvature tensor $\smash{R^A_{\phantom ABCD}}$. This is defined in any local frame by $\smash{R^A_{\phantom AB}(\vec u,\vec v)=R^A_{\phantom ABCD}u^Cv^D}$, where $\vec u,\vec v$ are arbitrary vector fields. The first property that follows immediately from the definition of $R^A_{\phantom AB}$ as a 2-form is the antisymmetry,
\begin{equation}
R^A_{\phantom AB}(\vec u,\vec v)=-R^A_{\phantom AB}(\vec v,\vec u)\quad\text{or}\quad
R^A_{\phantom ABCD}=-R^A_{\phantom ABDC}\;.
\end{equation}
This is valid for the curvature tensor based on any connection on the manifold.

Next comes a pair of relations that require the connection to be torsion-free. However, it does not have to be metric-compatible, nor does the manifold have to be endowed with a metric to start with. For vanishing torsion, the second of the Bianchi identities~\eqref{bianchi} reduces to $R^A_{\phantom AB}\w\vec e^{*B}=0$. This translates to
\begin{equation}
\begin{split}
R^A_{\phantom AB}(\vec u,\vec v)w^B+R^A_{\phantom AB}(\vec v,\vec w)u^B+R^A_{\phantom AB}(\vec w,\vec u)v^B&=0\;,\\
R^A_{\phantom ABCD}+R^A_{\phantom ACDB}+R^A_{\phantom ADBC}&=0\;,
\end{split}
\label{bianchialg}
\end{equation}
which is incidentally also known as the (first or algebraic) Bianchi identity. Likewise, the first line of~\eqref{bianchi} leads to
\begin{equation}
\begin{split}
(\cd_{\vec u}R)^A_{\phantom AB}(\vec v,\vec w)+(\cd_{\vec v}R)^A_{\phantom AB}(\vec w,\vec u)+(\cd_{\vec w}R)^A_{\phantom AB}(\vec u,\vec v)&=0\;,\\
(\cd_ER)^A_{\phantom ABCD}+(\cd_CR)^A_{\phantom ABDE}+(\cd_DR)^A_{\phantom ABEC}&=0\;,
\end{split}
\label{bianchidiff}
\end{equation}
known as the second or differential Bianchi identity. To prove this in a coordinate-free fashion requires certain amount of manipulation combining~\eqref{bianchi} with~\eqref{cartan} and the identity $\cd_{\vec u}\vec v-\cd_{\vec v}\vec u=[\vec u,\vec v]$, which follows from~\eqref{torsioncoordfree} for vanishing torsion.

In presence of a Riemannian metric, the curvature tensor of the LC connection (the \emph{Riemann curvature tensor}) has further properties that make it particularly symmetric. The starting point for uncovering these properties is the relation
\begin{equation}
\vec u[\inner{\vec v}{\vec w}]=\hat\cd_{\vec u}\inner{\vec v}{\vec w}=\inner{\hat\cd_{\vec u}\vec v}{\vec w}+\inner{\vec v}{\hat\cd_{\vec u}\vec w}\;.
\label{uvw}
\end{equation}
This follows at once from the Leibniz rule for the covariant derivative and the fact that the LC connection is metric-compatible. By a repeated application of~\eqref{uvw}, it is straightforward to show using~\eqref{curvaturecoordfree} that
\begin{equation}
\inner{\vec z}{R(\vec u,\vec v)\vec w}=-\inner{\vec w}{R(\vec u,\vec v)\vec z}\;.
\label{uvwz}
\end{equation}
Here $\vec u,\vec v,\vec w,\vec z$ are arbitrary vector fields and $R(\vec u,\vec v)\vec w$ is a shorthand notation for $\vec e_AR^A_{\phantom AB}(\vec u,\vec v)w^B$. Upon lowering the first index of the curvature tensor with the metric, \eqref{uvwz} acquires a particularly simple component form,
\begin{equation}
R_{ABCD}=-R_{BACD}\;.
\label{RABCDanti}
\end{equation}
Finally, combining~\eqref{RABCDanti} with the algebraic Bianchi identity~\eqref{bianchialg} shows that the type-$(0,4)$ tensor $R_{ABCD}$ is symmetric under swapping its two pairs of indices,
\begin{equation}
\inner{\vec z}{R(\vec u,\vec v)\vec w}=\inner{\vec v}{R(\vec w,\vec z)\vec u}\quad\text{or}\quad
R_{ABCD}=R_{CDAB}\;.
\label{RABCDswap}
\end{equation}


\subsection{Geodesic Normal Coordinates}
\label{appsubsec:normalcoord}

While the general language of differential geometry is largely coordinate-free, it is often convenient to resort to specific coordinates tailored to the problem at hand. For instance,  in Euclidean space, the global Cartesian coordinates play a privileged role. On a general Riemannian manifold, one can similarly introduce coordinates $x^a$ that are ``locally Cartesian'' (see Chap.~5 of \cite{Lee2018} for details). These are defined around an arbitrarily chosen point $x_0\in\M$ and have the following properties. First, the point $x_0$ itself is mapped to the origin, that is $x_0^a=0$. Second, the Riemannian metric at $x_0$ reduces to the standard Euclidean metric, $g_{ab}(x_0)=\d_{ab}$. (Any other choice amounts to a mere change of basis of $\T^*_{x_0}\M$.) Third, all the first derivatives $\de_cg_{ab}$ of the metric, and thus also the Christoffel symbols $\smash{\hat\Gamma^a_{bc}}$ of the LC connection, vanish at $x_0$. Finally, geodesics, that is solutions to~\eqref{geodesic}, passing through $x_0$ are straight lines $\g^a(t)=v^at$ with some fixed vector $\vec v\in\T_{x_0}\M$.

Such \emph{normal coordinates} can be constructed explicitly, if somewhat formally, in a neighborhood of $x_0$ using the last of the properties listed above. One chooses arbitrarily a basis $\vec e_a$ of the tangent space $\T_{x_0}\M$. For any vector $\vec v\equiv v^a\vec e_a\in\T_{x_0}\M$, one then denotes as $\g_{\vec v}$ the geodesic satisfying the initial conditions $\g^a(0)=0$ and $\Od{\g^a(0)}{t}=v^a$. The normal coordinates of the point $\g_{\vec v}(t)$ are defined as $x^a=v^at$.

The normal coordinates make it easy to check some differential-geometric identities involving covariant derivatives with respect to the LC connection. For instance, \eqref{Lied_formulas} is immediately seen to be equivalent to
\begin{equation}
\begin{split}
(\ld{\vec v}\tens T)^{ab\dotsb}_{mn\dotsb}=v^l(\hat\cd_l \tens T)^{ab\dotsb}_{mn\dotsb}&+\tens T^{ab\dotsb}_{ln\dotsb}(\hat\cd_m\vec v)^l+\tens T^{ab\dotsb}_{ml\dotsb}(\hat\cd_n\vec v)^l+\dotsb\\
&-\tens T^{lb\dotsb}_{mn\dotsb}(\hat\cd_l\vec v)^a-\tens T^{al\dotsb}_{mn\dotsb}(\hat\cd_l\vec v)^b-\dotsb\;.
\end{split}
\label{Lied_formulascovder}
\end{equation}
Indeed, at any chosen point $x_0$, the normal coordinates make the Christoffel symbols vanish so that~\eqref{Lied_formulascovder} agrees with~\eqref{Lied_formulas}. Since both sides of~\eqref{Lied_formulascovder} are well-defined tensors on $\M$, the identity must then hold for any choice of local coordinates.

\begin{illustration}%
For an illustration, consider the Lie derivative of a Riemannian metric,
\begin{equation}
(\ld{\vec v}g)_{ab}=g_{cb}(\hat\cd_a\vec v)^c+g_{ac}(\hat\cd_b\vec v)^c\;.
\end{equation}
Here the first term in~\eqref{Lied_formulascovder} drops owing to the fact that the metric is covariantly constant. This immediately recovers the equivalence of the two formulations of the Killing equation in~\eqref{killing}.
\end{illustration}

Perhaps even more importantly, normal coordinates allow us to systematically compute $g_{ab}(x)$ as a power series in $x^a$ and the Riemann curvature tensor and its covariant derivatives at $x_0$. Up to fourth order in the coordinates, the series reads
\begin{align}
\label{gnormalcoords}
g_{ab}(x)&=g_{ab}(0)-\frac13x^cx^dR_{acbd}(0)-\frac16x^cx^dx^e(\hat\cd_e R)_{acbd}(0)\\
\notag
&-\frac1{20}x^cx^dx^ex^f(\hat\cd_e\hat\cd_f R)_{acbd}(0)+\frac2{45}x^cx^dx^ex^fR_{acgd}(0)R^g_{\phantom gebf}(0)+\dotsb\;,
\end{align}
where I lowered the first index of $R^a_{\phantom abcd}$ using the metric. The derivation of~\eqref{gnormalcoords} is technical and I will therefore skip details. An interested reader will find a justification and a compilation of further results in an accessible form in~\cite{Brewin2009}.

A \emph{locally symmetric space} is a special type of a Riemannian manifold for which the first covariant derivative of the Riemann curvature tensor (hence also all higher covariant derivatives) vanishes. See Chap.~10, in particular Theorem 10.19, of~\cite{Lee2018}. For such manifolds, $g_{ab}(x)$ can be expressed in a closed form in terms of the curvature tensor at $x_0$. Again skipping details, the final result reads
\begin{equation}
g_{ab}(x)=g_{ac}(0)\biggl[\frac{\sin^2\!\sqrt{\hat R(x)}}{\hat R(x)}\biggr]^c_{\phantom cb}=g_{ac}(0)\sum_{k=0}^\infty\frac{(-1)^k2^{2k+1}}{(2k+2)!}[\hat R(x)^k]^c_{\phantom cb}\;,
\end{equation}
where $\hat R^a_{\phantom ab}(x)\equiv R^a_{\phantom acbd}(0)x^cx^d$.


\subsection{Hodge Star}
\label{appsubsec:hodge}

Using the $\sharp$ isomorphism, the inner product induced by the metric can be extended from vector fields to 1-forms. Thus, for any two 1-forms $\o$ and $\s$, we set
\begin{equation}
\inner\o\s\equiv\inner{\o^\sharp}{\s^\sharp}=g_{AB}\o^{\sharp A}\s^{\sharp B}=g^{AB}\o_A\s_B\;.
\end{equation}
This in turn extends to two differential forms $\o,\s$ of an arbitrary (equal) degree $p$,
\begin{equation}
\inner\o\s\equiv\frac1{p!}g^{A_1B_1}\dotsb g^{A_pB_p}\o_{A_1\dotsb A_p}\s_{B_1\dotsb B_p}\;.
\label{innerpforms}
\end{equation}

To see what this inner product of differential forms might be good for, we first pick an orthonormal frame $\vec e_A$. Then we define a new top-dimensional form ($n$-form) in terms of the corresponding dual coframe,
\begin{equation}
\vol\equiv\vec e^{*1}\w\dotsb\w\vec e^{*n}\;.
\label{volumeform}
\end{equation}
For any other local frame, $\tilde {\vec e}_A=\vec e_B(\transfo^{-1})^B_{\phantom BA}$ with $\transfo\in\gr{GL}(n)$, we get $\vol(\tilde{\vec e}_1,\dotsc,\tilde{\vec e}_n)=\det\transfo^{-1}$. This corresponds to the (oriented) volume of the parallelepiped defined by the vectors $\tilde{\vec e}_A$. The $n$-form~\eqref{volumeform} is known as the \emph{volume form}, hence the notation ``$\vol$.'' Provided that the two local frames have the same orientation, that is $\det\transfo$ is positive, the volume form can be expressed in the new dual coframe $\tilde{\vec e}^{*A}$ as
\begin{equation}
\vol=\det\transfo^{-1}\tilde{\vec e}^{*1}\w\dotsb\w\tilde{\vec e}^{*n}=\sqrt{\tilde g}\,\tilde{\vec e}^{*1}\w\dotsb\w\tilde{\vec e}^{*n}\;.
\label{volumeformtransition}
\end{equation}
Here $\tilde g$ is the determinant of the matrix representation of the metric in the new coframe, $\tilde g_{AB}=(\transfo^{-1T}\transfo^{-1})_{AB}$.

\begin{watchout}%
The frame $\vec e_A$ is in general defined only locally. We thus need to apply the definition~\eqref{volumeform} to an atlas of open sets that covers the manifold $\M$. In order that the volume form is well-defined and nonvanishing on the whole of $\M$, we have to make sure that local frames on mutually overlapping charts have the same orientation. The existence of such an atlas and a set of equally oriented local frames is the defining property of an \emph{orientable manifold}.
\end{watchout}

I now get to the main idea of this subsection. At any point $x\in\M$, the spaces $\Omega^p_x\M$ and $\Omega^{n-p}_x\M$ of $p$-forms and $(n-p)$-forms have the same dimension, $\binom{n}{p}$, and thus are isomorphic. In order to define a natural isomorphism between these spaces, and by extension between differential $p$-forms and $(n-p)$-forms, we need both the inner product~\eqref{innerpforms} and the volume form~\eqref{volumeform}. A given $p$-form $\o$ is then mapped to the unique $(n-p)$-form $\ho\o$ such that
\begin{equation}
\s\w(\ho\o)=\inner\s\o\vol
\label{hodge}
\end{equation}
for any test $p$-form $\s$. The inner product is needed to make~\eqref{hodge} linear in both $\o$ and $\s$. Likewise, the volume form serves as a reference $n$-form to compare $\s\w(\ho\o)$ to. The linear operator $\ho$ is known as the \emph{Hodge star}, and $\ho\o$ is referred to as the \emph{Hodge dual} of $\o$. In the special case of $p=0$, \eqref{hodge} gives a simple coordinate-free definition of the volume form, $\vol=\ho1$.

The definition~\eqref{hodge} is coordinate-free and elegant, but also not very transparent. For practical purposes, it is almost always more convenient to use instead the component expression for the Hodge dual,
\begin{equation}
\ho\o=\frac{\sqrt g}{p!(n-p)!}g^{A_1B_1}\dotsb g^{A_pB_p}\o_{A_1\dotsb A_p}\ve_{B_1\dotsb B_n}\vec e^{*B_{p+1}}\w\dotsb\w\vec e^{*B_n}\;.
\label{hodgecomponents}
\end{equation}
Here $\ve_{A_1\dotsb A_n}$ is the fully antisymmetric LC symbol, equal to $+1$ when $A_1,\dotsc,A_n$ is an even permutation of $1,\dotsc,n$, to $-1$ when $A_1,\dotsc,A_n$ is an odd permutation of $1,\dotsc,n$, and to $0$ otherwise. It is not to be confused with the \emph{Levi-Civita tensor}, $\sqrt g\,\ve_{A_1\dotsb A_n}$, which by~\eqref{volumeformtransition} gives the components of the volume form.

It is clear that by applying the Hodge star $\ho$ twice, we map a $p$-form $\o$ to another $p$-form, $\ho\ho\o$. What is less clear but easy to check is that the two forms are equal up to a sign,\footnote{Here and occasionally in the following, I drop the $\circ$ symbol indicating composition of operators to prevent the equations from becoming awkward.}
\begin{equation}
\ho\ho\o=(-1)^{p(n-p)}\o\;,\quad\text{or equivalently}\quad
\ho^{-1}\o=(-1)^{p(n-p)}\ho\o\;.
\label{hodgehodge}
\end{equation}
Being an isomorphism on the Grassmann algebra of differential forms, the Hodge star can be used to conjugate other linear maps defined on forms. The most important example of such a conjugate operator is the \emph{codifferential} $\udelta$. This is defined on $p$-forms by conjugating $\D$ with an extra conventional minus sign,
\begin{equation}
\udelta\o\equiv(-1)^p\ho^{-1}\D\ho\o=(-1)^{np+n+1}\ho\D\ho\o\;.
\label{codifferential}
\end{equation}
It follows from the conjugation and from~\eqref{dd0} that $\udelta\circ\udelta=0$. However, unlike $\D$ itself, the codifferential is not an antiderivation of the Grassmann algebra.

It is possible to conjugate in the same way the interior product~\eqref{interiorproduct}. This does not lead to a new interesting operator though. Indeed, combining~\eqref{interiorproduct} with~\eqref{hodgecomponents}, it is straightforward to check that for any $p$-form $\o$ and vector field $\vec v$,
\begin{equation}
(-1)^p\ho^{-1}\ix{\vec v}\ho\o=\vec v^\flat\w\o\;.
\end{equation}
Finally, one can similarly conjugate the Lie derivative $\ld{\vec v}$. Thanks to the Cartan magic formula~\eqref{Cartanmagicformula}, the result is fixed by the conjugation of $\D$ and $\ix{\vec v}$. It is however not particularly illuminating and I will therefore not show further details.

Let us now get back to the codifferential. This together with $\D$ furnishes a powerful toolkit that generalizes much of the differential structure familiar from ordinary vector calculus. The exterior derivative itself generalizes the notion of a gradient of scalar fields in $\R^n$ and curl of vector fields in $\R^3$, as noted in \refex{ex:hodgeR3}. This leads to a far-reaching generalization of integral theorems in vector calculus that I will address in Sect.~\ref{appsec:integration}. In $\R^n$ we also have the divergence operator that turns a vector field into a scalar. We can construct such a map on any orientable manifold $\M$ by combining the flat $\flat$ and the codifferential. Thus, for any vector field $\vec v$,
\begin{equation}
\divg\vec v\equiv-\udelta\vec v^\flat=\frac1{\sqrt g}\de_a(\sqrt g\,v^a)=(\hat\cd_a\vec v)^a\;,
\label{divergence}
\end{equation}
where $g$ now stands for the determinant of the matrix representation of the metric in the dual coordinate basis $\D x^a$. In the last expression in~\eqref{divergence}, the LC connection~\eqref{LCconnection} is implicitly assumed. The next-to-last expression in~\eqref{divergence} is a special case of the component form of the codifferential of a $p$-form,
\begin{equation}
(\udelta\o)_{a_2\dotsb a_p}=-\frac1{\sqrt g}\de_{b_1}\bigl[\sqrt g(\o^{\sharp})^{b_1\dotsb b_p}\bigr]g_{a_2b_2}\dotsb g_{a_pb_p}\;.
\label{codiffcomponent}
\end{equation}
This follows directly from~\eqref{hodgecomponents} and~\eqref{codifferential} upon some manipulation. Last but not least, there is another, rather elegant way of computing the divergence of a vector field through $(\divg\vec v)\vol=\ld{\vec v}\vol$. This is an infinitesimal version of the familiar fact from multivariate calculus that upon a transformation of coordinates, the volume measure in an integral changes by the Jacobian of the transformation.

\begin{illustration}%
The component expression~\eqref{codiffcomponent} makes it easy to see why the codifferential is not an antiderivation of the Grassmann algebra. Should $\udelta$ satisfy an identity analogous to~\eqref{antiderivationd}, we would have, for instance, for any 0-form $f$ and $p$-form $\o$ that
\begin{equation}
\udelta(f\w\o)=\udelta f\w\o+f\w\udelta\o=f\w\udelta\o\;.
\label{codiffaux}
\end{equation}
Here I used the fact that $\udelta f=0$ for any 0-form $f$, which is a consequence of the fact that $\D$ vanishes on any $n$-form. But~\eqref{codiffaux} cannot be correct, since~\eqref{codiffcomponent} makes it clear that $\udelta(f\w\o)$ must depend on both $f$ and its derivatives.
\end{illustration}

\begin{illustration}%
The Noether current of a continuous symmetry of a physical system is a divergence-free vector field $\vec J$, that is $\udelta\vec J^\flat=0$. The latter condition is naturally generalized to $\udelta J=0$ for any ``$p$-form current'' $J$. Such a generalized conservation law is via Noether's theorem associated with a so-called $(p-1)$-form symmetry~\cite{Gaiotto2015a}. Additional insight follows from writing the conservation condition $\udelta J=0$ equivalently as $\D\ho J=0$. Namely, if $J$ is a conserved $p$-form current and $K$ a conserved $q$-form current, then the exterior derivative of the $(2n-p-q)$-form $\ho J\w\ho K$ vanishes thanks to~\eqref{antiderivationd}. Hence $\ho(\ho J\w\ho K)$ is a $(p+q-n)$-form current that is also conserved. In this way, we can use already known generalized conservation laws to construct new ones as long as $p+q>n$~\cite{Brauner2021b}.
\end{illustration}

Finally, by combining $\D$ and $\udelta$, we can construct a ``second-order differential operator'' which maps $p$-forms to $p$-forms. While it would in principle be possible to consider separately $\D\circ\udelta$ and $\udelta\circ\D$, it is their sum that is of special interest,
\begin{equation}
\varDelta\equiv(\D+\udelta)^2=\D\circ\udelta+\udelta\circ\D\;.
\end{equation}
It follows at once from~\eqref{divergence} that on 0-forms this reduces to
\begin{equation}
\varDelta f=-\frac1{\sqrt g}\de_a\bigl(\sqrt{g}\,g^{ab}\de_bf\bigr)=-g^{ab}\hat\cd_a\hat\cd_bf\;,
\end{equation}
where the second expression again requires the LC connection. Thus, $\varDelta$ is a generalization of the Laplace operator to differential forms on orientable Riemannian manifolds. It is known as the \emph{Laplace--de Rham operator}, or alternatively as the \emph{Hodge Laplacian}.


\section{Integration on Manifolds}
\label{appsec:integration}

So far I have discussed solely the differential structure on manifolds and related concepts. It is finally time to address the other essential ingredient of calculus: integration. Following the analogy with multivariate calculus in $\R^n$, one might want to integrate functions on a given manifold $\M$. Without the privilege of having globally well-defined Cartesian coordinates, it however turns out more natural to integrate forms. I will show how to do this in the next subsection. The abstract notion of integration of forms can nevertheless be bypassed in case the manifold possesses a metric structure. It is then possible to sweep the differential form nature of integration under the rug. This results into a definition of integration in terms of a volume measure determined by the metric, which is the subject of Sect.~\ref{appsubsec:integrationRiemannian}.


\subsection{Orientable Manifolds}
\label{appsubsec:integrationorientable}

A well-defined concept of integration on a manifold must have a geometric meaning independent of the choice of coordinates. It makes no sense to just pick random coordinates $x^a$ and define an integral of a function $f:\M\to\R$ via something like
\begin{equation}
\int_\M f\equiv\int_{\vp(\M)} (f\circ\vp^{-1})(x)\,\D^nx\qquad\text{(wrong)}\;,
\end{equation}
where $\vp:\M\to\R^n$ is the (possibly only locally defined) coordinate map. This naive attempt cannot be fixed even by adding an integration measure that would correctly account for changes of coordinates. The problem is that without further structure on the manifold, there is no unique measure we could use for the purpose.

It turns out that differential forms have all the ingredients we need. Let us consider an $n$-form $\o$ on an $n$-dimensional manifold $\M$. In chosen local coordinates, it is given by~\eqref{pformcoordbasis},
\begin{equation}
\o=\frac1{n!}\o_{a_1\dotsb a_n}\D x^{a_1}\w\dotsb\w\D x^{a_n}=\o_{1\dotsb n}\D x^1\w\dotsb\w\D x^n\;.
\label{naivevolumeform}
\end{equation}
Under the change of coordinates from $x^a$ to $\tilde x^a(x)$, the single independent component $\o_{1\dotsb n}(x)$ transforms by the inverse of the Jacobian of the coordinate transformation, $\tilde\o_{1\dotsb n}(\tilde x)=|\Pd{\tilde x}{x}|^{-1}\o_{1\dotsb n}(x)$. It therefore appears that all we have to do is to strip off the factor $\D x^1\w\dotsb\w\D x^n$ in~\eqref{naivevolumeform} and integrate $\o_{1\dotsb n}(x)$ as a function on $\R^n$. Upon a change of coordinates, the transformation of $\o_{1\dotsb n}$ will exactly cancel the Jacobian generated by the Euclidean volume measure $\D^n x$.

There are two flies in the ointment though. First, a closer look reveals that while $\o_{1\dotsb n}$ indeed changes by the inverse of the Jacobian of the coordinate transformation, the Euclidean measure $\D^n x$ changes by its absolute value. Another manifestation of the same problem is that $\D^n x$ is invariant under any permutation of the coordinates, whereas $\D x^1\w\dotsb\w\D x^n$ and thus $\o_{1\dotsb n}$ changes sign under odd permutations. Yet in other words, $(\D x^1\w\dotsb\w\D x^n)(\vec v_1,\dotsc,\vec v_n)$ returns the \emph{oriented} volume of the parallelepiped defined by the test vectors $\vec v_1,\dotsc,\vec v_n$ in $\R^n$. The second, related problem is that we rarely have a set of globally defined coordinates on the manifold. It is mandatory to ensure that we can consistently sew together integrals of $\o$ over individual coordinate patches. This requires that the coordinate systems in overlapping patches have equal orientation. For that to be possible at all, the manifold $\M$ must be orientable. Even on orientable manifolds, we still have the freedom to choose global orientation. Fixing the orientation requires that we restrict coordinate transformations to those with a positive Jacobian. At the end of the day, the integral of an $n$-form will be defined up to a sign, depending on the choice of global orientation.

The problem of how to sew together integrals of $\o$ over individual coordinate patches requires more careful consideration. The idea is to ``triangulate'' the manifold in a way that each cell of the triangulation lies in a single coordinate patch. To implement this strategy, we start by defining what such an elementary cell corresponds to in $\R^n$. A \emph{$p$-simplex} $(P_0,\dotsc,P_p)$, where $0\leq p\leq n$, is the convex envelope of (that is the smallest convex set containing) the set of points $P_0,\dotsc,P_p\in\R^n$. For the triangulation approach to integration to be successful, we next need to be able to put together two or more simplexes. This is done by defining a set, $C_p$, whose elements, called \emph{$p$-chains}, are formal linear combinations of different $p$-simplexes.

\begin{watchout}%
The introduction of $p$-simplexes was motivated by the desire to triangulate a given manifold $\M$. It would therefore appear most natural to only consider linear combinations of simplexes with coefficients $\pm1$. It will however turn out convenient to allow for arbitrary real coefficients, thus making $C_p$ a real vector space. The advantage of doing so will become apparent in Sect.~\ref{appsec:cohomology} where I introduce the dual structure to $p$-chains in terms of differential forms.
\end{watchout}

For integration of differential forms, we need to define an orientation of the $p$-simplex. We can do this by identifying $(P_{\pi(0)},\dotsc,P_{\pi(p)})$ with $(\sgn\pi)(P_0,\dotsc,P_p)$ for any permutation $\pi$ on the indices $0,\dotsc,p$. With this convention, we can now also define an oriented boundary of a $p$-simplex. This is formalized by the \emph{boundary operator}, which is a linear map $\de_p:C_p\to C_{p-1}$ such that
\begin{equation}
\de_p(P_0,\dotsc,P_p)\equiv\sum_{i=0}^p(-1)^i(P_0,\dotsc,P_{i-1},P_{i+1},\dotsc,P_p)\;.
\label{boundaryop}
\end{equation}
The fundamental property of the boundary operator is that when applied twice in succession, it gives zero, $\de_{p-1}\circ\de_p=0$.

Eventually, we would like to construct a triangulation of a manifold $\M$ by lifting simplexes from $\R^n$ to $\M$. To that end, it is sufficient to consider just one, \emph{standard $p$-simplex} $s_p$ in $\R^n$. This is defined to have $P_0$ at the origin and all the other vertices on the positive Cartesian semiaxes at unit distance from the origin. The Cartesian coordinates of the vertex $P_i$ are thus $P_i^a=\d^a_i$. Now we can take any smooth map $f:s_p\to\M$. Its image, $\s_p\equiv f(s_p)$, is called a \emph{singular $p$-simplex} on $\M$. By extension, a \emph{singular $p$-chain} $c$ on $\M$ is a formal linear combination,
\begin{equation}
c=\sum_ic_i\s^i_p=\sum_ic_if^i(s_p)\quad\text{where }c_i\in\R\;,
\label{pchain}
\end{equation}
and $\s^i_p$ are singular $p$-simplexes. Finally, the boundary of a singular $p$-simplex is defined by lifting the boundary of $s_p$ with the same map $f$, $\de_p\s_p\equiv f(\de_ps_p)$.\footnote{Here I am tacitly extending the action of $f$ from the standard $p$-simplex to linear combinations of $p$-simplexes in $\R^n$ by linearity.}

After the rather long series of formal steps, we are finally ready to define integration on the manifold $\M$. In fact, we can do even more than we originally anticipated, namely integrate $p$-forms on any singular $p$-chain on $\M$ such as~\eqref{pchain}, with any $0\leq p\leq n$. The task is first reduced to individual singular $p$-simplexes by linearity. Then, the integral is pulled back to $\R^p$ using the map defining the singular $p$-simplex. At the end of the day, one sets for any $p$-form $\o$ and singular $p$-chain $c$
\begin{equation}
\int_c\o\equiv c_i\int_{\s^i_p}\o\;,\qquad
\int_{\s^i_p}\o\equiv\int_{s_p}f^{i*}\o\equiv\int_{s_p}(f^{i*}\o)_{1\dotsb p}(x)\,\D^px\;.
\label{integraldefinition}
\end{equation}
To achieve the original goal to define integration of forms on manifolds, it remains to clarify how to represent an $n$-dimensional manifold $\M$ as a singular $n$-chain. There is obviously considerable freedom in doing so, but to identify $\M$ with an $n$-chain $c=\sum_ic_i\s^i_n$, the latter must satisfy several conditions. First, the simplexes $\s^i_n$ must cover the whole of $\M$ without overlaps. Second, each $\s^i_n$ should lie in a single coordinate patch. Third, if the local coordinate system $x^a$ in which $\s^i_n$ is mapped to the standard $n$-simplex $s_n$ is compatible with the global orientation of the manifold $\M$, we should set $c_i=1$. If the orientation is opposite, we have to set $c_i=-1$. In practice, all this is easier than it might look, as the following example shows.

\begin{illustration}%
The spherical coordinates $(\t,\vp)$ on $S^2$ take values from the open rectangle $(0,\pi)\times(0,2\pi)$ in $\R^2$ and cover the whole of $S^2$ except for one selected meridian. The integral of the 2-form $\o=\sin\t\,\D\t\w\D\vp$~\eqref{spherevolumeform} then equals
\begin{equation}
\int_{S^2}\o=\int_0^{2\pi}\D\vp\int_0^\pi\sin\t\,\D\t=4\pi\;,
\end{equation}
which is the area of a unit sphere. This example shows that it is not strictly necessary to actually triangulate the integration domain. It is sufficient to dissect it into open sets, each of which is mapped by a single coordinate chart to a polytope in $\R^n$. The subsequent triangulation of the polytopes is a formal step that need not be carried out explicitly. The example also demonstrates that we can get away with a single coordinate chart even if it does not cover the whole manifold. It is sufficient to cover it up to a subset of lower dimension.
\end{illustration}

With all the necessary definitions out of the way, let me proceed directly to the most important result of integral calculus on manifolds: the \emph{Stokes theorem}. Let $c$ be a singular $(p+1)$-chain on an orientable manifold $\M$, and let $\o$ be a $p$-form. Then
\begin{equation}
\int_c\D\o=\int_{\de_{p+1}c}\o\;.
\label{stokestheorem}
\end{equation}
For a proof, see for instance Sect.~6.1 of~\cite{Nakahara2003a}. The Stokes theorem puts under the same umbrella, and further generalizes, a number of well-known results from ordinary calculus and vector calculus. These include the fundamental theorem of calculus in $\R$, the gradient theorem for line integrals in $\R^n$, the Green theorem for area integrals in $\R^2$, the eponymous Stokes theorem for surface integrals in $\R^3$, and the Gauss theorem for volume integrals in $\R^n$. 

There is an intriguing connection between the Stokes theorem and a geometric definition of the exterior derivative that I alluded to at the end of Sect.~\ref{appsec:exterior}. Namely, the exterior derivative of a $p$-form $\o$ can be defined in a coordinate-free manner through integration of $\o$ over an ``infinitesimal'' $p$-chain that is the boundary of some $(p+1)$-chain. This generalizes the usual definition of the divergence of a vector field in $\R^3$ using its flux through the surface of an infinitesimal volume element. The key problem is then to assert the consistency of such a definition; see Sect.~36 of~\cite{Arnold1989a} for details. In this sense, the exterior derivative can be defined exactly so that the Stokes theorem~\eqref{stokestheorem} holds. Rather than a nontrivial result of integral calculus, the theorem may then be viewed as arising from remarkable ingenuity in defining the right differential structure on the manifold.

\begin{illustration}%
\label{ex:stokesthmpullback}%
In~\eqref{integraldefinition}, the integral of a $p$-form $\o$ is obtained by pulling $\o$ back to $\R^p$ using the map $f:s_p\to\M$ defining the singular $p$-simplex $\s_p$. This idea can be generalized to maps between any two manifolds (see Sect.~7.8 of~\cite{Fecko2011a}). Consider two manifolds $\M$ and $\MN$ and a smooth map $f:\M\to\MN$. Then, for any $p$-form $\o$ on $\MN$ and a singular $p$-chain $c$ on $\M$, we have
\begin{equation}
\int_{f(c)}\o=\int_cf^*\o\;.
\label{stokespullback}
\end{equation}
Such integration of differential forms via pull-back is common in field theory. In this context, $f$ can be thought of as a field defined on the domain $\M$ and taking values in the target manifold $\MN$.

Suppose now that $\M$ is an $n$-dimensional manifold without boundary and $\o$ is an $n$-form on $\MN$ such that $\D\o=0$. Then $\int_\M f^*\o$ is a \emph{topological invariant}, that is, it does not change value under a smooth deformation of $f$. To see why, think of the deformation $f\to\tilde f$ as a family of functions $f_t$ where $t\in[0,1]$ is a deformation parameter and $f_0=f,f_1=\tilde f$. Formally, $f_t$ defines a smooth map from $\tilde\M\equiv[0,1]\times\M$ to $\MN$. Since $\M$ itself does not have boundary, we can view $\tilde\M$ as a cylinder whose ``bases'' at $t=0$ and $t=1$ are diffeomorphic to $\M$. Using the Stokes theorem, we then find
\begin{equation}
\begin{split}
\int_\M\tilde f^*\o-\int_\M f^*\o&=\int_\M f^*_1\o-\int_\M f^*_0\o\\
&=\int_{\de_{n+1}\tilde\M}f^*_t\o=\int_{\tilde\M}\D(f^*_t\o)=0\;.
\end{split}
\end{equation}
In the last step, I used the fact that exterior derivative commutes with pull-back~\eqref{dpullback} and the assumption that $\D\o=0$. The invariance of $\int_\M f^*\o$ under smooth deformations of $f$ can also be understood from the left-hand side of~\eqref{stokespullback}. Here the integrand $\o$ is fixed, but the integration domain $f(\M)$ is deformed. Upon tuning the parameter $t$ from 0 to 1, $f_t(\M)$ sweeps a singular $(n+1)$-chain in $\MN$ with boundaries $f_0(\M)$ and $f_1(\M)$. Invoking the Stokes theorem and the assumption $\D\o=0$ then again asserts the invariance of $\int_{f(\M)}\o$ under smooth deformations of $f$.
\end{illustration}


\subsection{Riemannian Manifolds}
\label{appsubsec:integrationRiemannian}

On a first encounter, the differential form approach to integration on manifolds may seem a bit abstract. There is however a conceptually simpler alternative in case the manifold $\M$ is endowed with a Riemannian metric. As explained in Sect.~\ref{appsubsec:hodge}, the metric induces a natural volume form~\eqref{volumeform} on the manifold. For any smooth function $f:\M\to\R$, $f\vol$ is then a well-defined $n$-form on $\M$, which can be used to define the integral of $f$ over $\M$,
\begin{equation}
\int_\M f\equiv\int_\M f\vol=\int_\M f(x)\sqrt{g(x)}\,\D^n x\;.
\label{volumemeasure}
\end{equation}
The last, coordinate expression follows from~\eqref{volumeformtransition} and is often used in physics to perform integration in curved space(time)s. The path leading to this expression can also be viewed as a natural consequence of Hodge duality, which maps the 0-form $f$ to the $n$-form $\ho f=f\ho 1=f\vol$.

\begin{watchout}%
In mathematics, it is common to define a volume form on an $n$-dimensional manifold $\M$ as any $n$-form that is nonvanishing everywhere on $\M$. One then proves that the existence of a volume form is equivalent to the requirement that $\M$ be orientable (see Sect.~6.3 of~\cite{Fecko2011a}). The definition of a volume form I gave in Sect.~\ref{appsubsec:hodge} is more restrictive in that it is tied to the metric structure on $\M$. Such volume form is normalized to unity in an orthonormal frame. It can thus be thought of as defining on $\M$ locally a generalization of the usual volume measure in $\R^n$.
\end{watchout}

Let me append two remarks of caution for more rigorously-minded readers. First, the coordinates may be defined only locally. In such a case, one has to partition the manifold and apply~\eqref{volumemeasure} separately to each coordinate patch. Second, the expression~\eqref{volumemeasure} may appear less general than~\eqref{integraldefinition}, which is well-defined for a $p$-form and a singular $p$-chain of any degree $0\leq p\leq n$. It is however easy to generalize~\eqref{volumemeasure} to integration of functions on any submanifold $\MN$ of $\M$. All one has to do is pull the metric $g$ on $\M$ back to $\MN$ using the embedding of $\MN$ in $\M$. The integral over $\MN$ is then defined as in~\eqref{volumemeasure} using the induced metric on $\MN$.

\begin{illustration}%
I will illustrate the use of~\eqref{volumemeasure} by calculating the length of a curve on a manifold. The curve is defined by a smooth map $\g:\R\to\M$ and we are interested in the length of the segment $\Gamma\equiv\{\g(t)\,|\,t\in[t_1,t_2]\}$. Treating $t$ as a coordinate on the curve, we pull the metric $g$ on $\M$ back to $\Gamma$ via
\begin{equation}
\g^*g=g_{ab}\OD{\g^a}t\OD{\g^b}t\D t\otimes\D t\;.
\end{equation}
The length of the segment of interest then follows as
\begin{equation}
\int_\Gamma1=\int_{t_1}^{t_2}\sqrt{g_{ab}\OD{\g^a}t\OD{\g^b}t}\,\D t\;.
\end{equation}
This agrees with the distance along the curve~\eqref{curvelength}, previously postulated ad hoc.
\end{illustration}


\section{Homology and Cohomology}
\label{appsec:cohomology}

Most of our discussion of differential geometry so far was restricted to local properties of manifolds. However, we have already seen how the global topology of the manifold $\M$ may place restrictions on otherwise locally defined structures. For instance, there may be no globally well-defined frame on $\M$. Also, the global topology of $\M$ may forbid some solutions of the Killing equation and thus reduce the isometry group of $\M$; see \refex{ex:killingtorus}.

Topology is a vast area of mathematics and I have no intention to cover even its very basics in a self-contained manner. Instead, I will conclude the appendix with a brief introduction to some aspects of topology, pertinent to the subject of this book. As stressed in Sect.~\ref{sec:topologicalaspects}, spontaneous symmetry breaking is intimately related to the presence of defects in the ordered medium. The classification of defects was historically the first major application of topology to physics. The classic paper by Mermin~\cite{Mermin1979a} gives a physicist-friendly introduction to the relevant branch of topology, called \emph{homotopy theory}. See also Chap.~4 of~\cite{Nakahara2003a} for a somewhat more mathematical exposition. In the present book, homotopy theory plays a marginal role. A reader that is altogether unfamiliar with the subject will get away with the basic definition of homotopy groups, found early on in the above references. What I will cover here are elements of \emph{de Rham cohomology}. This neatly connects the global topology of the manifold $\M$ to the analytic properties of differential forms on $\M$. Cohomology theory plays an important role in the construction of actions for systems with spontaneously broken symmetry. It is therefore highly relevant for Parts~\ref{part:internalSSB} and~\ref{part:spacetimeSSB} of the book.


\subsection{Singular Homology}
\label{appsubsec:singularhomology}

Let me start somewhat indirectly by recalling the concept of a singular chain, introduced in Sect.~\ref{appsubsec:integrationorientable}. The real vector spaces of singular $p$-chains on $\M$, denoted from now on more precisely as $C_p(\M,\R)$, form a sequence under the action of the boundary operator~\eqref{boundaryop},
\begin{equation}
\dotsb\xrightarrow{\de_{p+2}}C_{p+1}(\M,\R)\xrightarrow{\de_{p+1}}C_p(\M,\R)\xrightarrow{\de_p}C_{p-1}(\M,\R)\xrightarrow{\de_{p-1}}\dotsb\;.
\label{chaincomplex}
\end{equation}
The spaces $C_p(\M,\R)$ are infinite-dimensional. It is however possible to identify certain finite-dimensional spaces descending from $C_p(\M,\R)$ that carry information about the topology of $\M$. To that end, we first introduce the subspace of \emph{$p$-cycles}, $Z_p(\M,\R)\equiv\KER\de_p$, that is those $p$-chains that have no boundary. Likewise, the subspace of \emph{$p$-boundaries}, $B_p(\M,\R)\equiv\IM\de_{p+1}$, consists of those $p$-chains that are boundaries of some $(p+1)$-chains. Due to the fundamental property $\de_p\circ\de_{p+1}=0$, any $p$-boundary is a $p$-cycle. In other words, $B_p(\M,\R)$ is a subspace of $Z_p(\M,\R)$. The quotient space,
\begin{equation}
H_p(\M,\R)\equiv Z_p(\M,\R)/B_p(\M,\R)\;,\qquad
0\leq p\leq n\;,
\label{homology}
\end{equation}
is called the \emph{$p$-th homology group} of $\M$. Miraculously, this space has a finite dimension although both $Z_p(\M,\R)$ and $B_p(\M,\R)$ are infinite-dimensional. Roughly speaking, $H_p(\M,\R)$ is a formal vector space spanned on mutually inequivalent $p$-dimensional submanifolds of $\M$ that have no boundary and are themselves not the boundary of any $(p+1)$-dimensional submanifold of $\M$.

\begin{illustration}%
\label{ex:homology}%
To see how the homology groups relate to the topology of $\M$, it is good to have a look at some very simple examples. First, the homology groups of the 2-sphere are
\begin{equation}
H_0(S^2,\R)\simeq\R\;,\qquad
H_1(S^2,\R)\simeq0\;,\qquad
H_2(S^2,\R)\simeq\R\;.
\end{equation}
The $H_0(S^2,\R)$ group reflects the fact that $S^2$ is connected. Namely, 0-chains on $S^2$ are points on $S^2$ and their formal linear combinations. For any two points $P_1,P_2\in S^2$, the formal difference $P_1-P_2$ is the boundary of a 1-chain (curve) connecting the points. Hence any 0-chain equals, up to adding a 0-boundary, a multiple of a single point on $S^2$. At the same time, any 0-chain is trivially a 0-cycle. It follows that $H_0(S^2,\R)$ is one-dimensional. An obvious extension of the argument shows that in general, $H_0(\M,\R)\simeq\R^k$, where $k$ is the number of connected components of $\M$.

The triviality of $H_1(S^2,\R)$ reflects the fact that there is no 1-cycle (closed curve) on $S^2$ that would not be the boundary of some 2-chain (area) on $S^2$. Finally, as to $H_2(S^2,\R)$, there is obviously one 2-cycle on $S^2$ that is not a 2-boundary, namely the sphere itself. The fact that $H_2(S^2,\R)\simeq\R$ formalizes the observation that any 2-cycle on $S^2$ can be deformed to $S^2$ or its multiple. I will add without proof that the above results for the homology groups of the 2-sphere readily generalize to the $n$-sphere with any positive $n$,
\begin{equation}
H_0(S^n,\R)\simeq H_n(S^n,\R)\simeq\R\;,\qquad
H_p(S^n,\R)\simeq0\quad\text{for any }0<p<n\;.
\end{equation}

Let us look at another simple example, the two-dimensional torus $T^2$, for which
\begin{equation}
H_0(T^2,\R)\simeq\R\;,\qquad
H_1(T^2,\R)\simeq\R^2\;,\qquad
H_2(T^2,\R)\simeq\R\;.
\label{homologytorus}
\end{equation}
The reasoning behind the zeroth and second homology groups is the same as for the 2-sphere. The two independent generators of $H_1(T^2,\R)$ arise from the two possibilities to wind a circle around the torus so that its ``inside'' does not lie on the torus. Just like for the sphere, \eqref{homologytorus} is but a special case of a more general statement valid (without proof) for any $n$-torus with positive $n$,
\begin{equation}
H_p(T^n,\R)\simeq\R^{\binom{n}{p}}\quad
\text{for any }0\leq p\leq n\;.
\label{torushomology}
\end{equation}

As a side remark, the dimensions of the homology groups can be combined into a single quantity characterizing the topology of the manifold: the \emph{Euler characteristic},
\begin{equation}
\c(\M)\equiv\sum_{p=0}^n(-1)^p\dim H_p(\M,\R)\;.
\end{equation}
For the $n$-sphere and the $n$-torus, we have $\c(S^n)=1+(-1)^n$ and $\c(T^n)=0$.
\end{illustration}

The above example provides basic geometric intuition behind the concept of \emph{singular homology}. By no means does it explain how to actually find the homology groups for a manifold that is less trivial to visualize than a sphere. The problem how to compute homology groups belongs to algebraic topology. In physics-oriented literature, one usually encounters the older concept of \emph{simplicial homology}. This makes it possible, although not straightforward, to find the homology groups by a ``triangulation'' of the manifold. An interested reader will find more details in Chap.~3 of~\cite{Nakahara2003a}. I will instead proceed directly to the connection of homology to the Grassmann algebra of differential forms on $\M$.


\subsection{De Rham Cohomology}
\label{appsubsec:derhamcohomology}

How can the topological notion of homology be connected to differential forms? The basic idea is that any $p$-form $\o$ on $\M$ can be thought of as a linear function on singular $p$-chains on $\M$,
\begin{equation}
\o:c\to(c,\o)\equiv\int_c\o\;,\qquad
c\in C_p(\M,\R)\;.
\label{stokespairing}
\end{equation}
Thus, $\o$ can be identified with an element of the dual space $C_p^*(\M,\R)$ of $C_p(\M,\R)$. Moreover, we can set up a structure on differential forms that is ``dual'' to~\eqref{chaincomplex},
\begin{equation}
\dotsb\xleftarrow{\D}\Omega^{p+1}\M\xleftarrow{\D}\Omega^p\M\xleftarrow{\D}\Omega^{p-1}\M\xleftarrow{\D}\dotsb\;,
\label{cochaincomplex}
\end{equation}
where $\Omega^p\M$ denotes the space of differential $p$-forms on $\M$. Following the analogy with~\eqref{chaincomplex}, we then define the space of \emph{$p$-cocycles}, $Z^p(\M)\equiv\KER\at{\D}{\Omega^p\M}$, as consisting of $p$-forms with vanishing exterior derivative. Such forms are usually called \emph{closed}. Likewise, the space of \emph{$p$-coboundaries}, $B^p(\M)\equiv\IM\at{\D}{\Omega^{p-1}\M}$, consists of those $p$-forms that equal the exterior derivative of some $(p-1)$-form. Such forms are called \emph{exact}. Due to the fundamental property~\eqref{dd0}, every exact form is automatically closed, hence $B^p(\M)$ is a subspace of $Z^p(\M)$. The quotient space,
\begin{equation}
H^p(\M)\equiv Z^p(\M)/B^p(\M)\;,\qquad
0\leq p\leq n\;,
\label{cohomology}
\end{equation}
is known as the \emph{$p$-th de Rham cohomology group} of $\M$. According to the \emph{Poincar\'e lemma}, every closed $p$-form $\o$ can be locally written as the exterior derivative of a $(p-1)$-form $\s$, $\o=\D\s$; see Chap.~9 of~\cite{Fecko2011a} for a proof and numerous examples. The de Rham cohomology classifies obstructions to the global validity of the Poincar\'e lemma. It therefore gives us access to global topological information about the manifold $\M$.

To see the connection between de Rham cohomology and (singular) homology, consider a $p$-cycle $c$ and a closed $p$-form $\o$. The Stokes theorem~\eqref{stokestheorem} guarantees that the integral of $\o$ over $c$ does not change if we shift $\o$ by an exact $p$-form, or $c$ by a $p$-boundary. In other words, for any $(p-1)$-form $\s$ and any $(p+1)$-chain $d$,
\begin{equation}
\begin{split}
\int_{c+\de d}(\o+\D\s)&=\int_c\o+\int_c\D\s+\int_{\de d}\o+\int_{\de d}\D\s\\
&=\int_c\o+\int_{\de c}\s+\int_d\D\o+\int_d\D(\D\s)=\int_c\o\;,
\end{split}
\end{equation}
where I for simplicity dropped the subscript on the boundary operator. What this says is that the integral $\int_c\o$ only depends on the equivalence class of $c$ as an element of $H_p(\M,\R)$, and on the equivalence class of $\o$ as an element of $H^p(\M)$. The pairing $(c,\o)$, defined by~\eqref{stokespairing}, therefore represents a bilinear function on $H_p(\M,\R)\times H^p(\M)$. The \emph{de Rham theorem} asserts that when the manifold $\M$ is compact, this bilinear map is nondegenerate. This means that $(c,\o)$ vanishes for all closed $p$-forms $\o$ if and only if $c$ is a $p$-boundary, and for all $p$-cycles $c$ if and only if $\o$ is exact. It follows at once that the spaces $H_p(\M,\R)$ and $H^p(\M)$ are isomorphic. The bilinear map $(c,\o)$ establishes a duality between (singular) homology and (de Rham) cohomology.

\begin{watchout}%
The statement of duality is highly nontrivial. It only applies to the quotient spaces $H_p(\M,\R)$ and $H^p(\M)$ and not separately to the spaces of $p$-cycles and $p$-cocycles (closed $p$-forms), or $p$-boundaries and $p$-coboundaries (exact $p$-forms). Luckily, out of all the different classes of $p$-chains, it is the homology group $H_p(\M,\R)$ that is of greatest interest. The de Rham theorem then allows us to compute the homology groups of the manifold by means of the exterior calculus of differential forms.
\end{watchout}

\begin{illustration}%
\label{ex:toruscohomology}%
There are no $(-1)$-forms on any manifold, and hence no exact 0-forms. Thus, $H^0(\M)$ consists of all closed 0-forms. The condition $\D f=0$ for $f\in\Omega^0\M$ that should be closed is satisfied if and only if $f$ is a piecewise-constant function, that is a  constant function on every connected component of $\M$. As a consequence, $H^0(\M)$ counts the connected components of $\M$. This is in accord with our previous result for $H_0(\M,\R)$ and the de Rham theorem.

For an illustration of the cohomology groups of higher degree, let us consider the 2-sphere $S^2$. We know from \refex{ex:homology} and the de Rham theorem that $H^2(S^2)\simeq H_2(S^2,\R)\simeq\R$, which guarantees the existence of one linearly independent 2-form that has a nonzero integral over $S^2$. This can be naturally chosen as the volume form~\eqref{spherevolumeform}. On the other hand, the first de Rham cohomology group of $S^2$ is trivial since $H_1(S^2,\R)$ is. We conclude that on the 2-sphere, the Poincar\'e lemma holds globally for 1-forms; any closed 1-form is necessarily exact.

Somewhat more interesting than the 2-sphere is the 2-torus $T^2$. Here we deduce from \refex{ex:homology} that $H^1(T^2)\simeq H_1(T^2,\R)\simeq\R^2$. It is easy to guess what the two independent nontrivial closed 1-forms might be. Just think of $T^2$ as the Cartesian product $S^1\times S^1$. This gives natural angular coordinates on $T^2$, $\t^{1,2}\in(0,2\pi)$. The forms $\D\t^1,\D\t^2$ are closed but not exact since the angular coordinates are not globally well-defined. A different way to view this is to think of the basis of $H_1(T^2,\R)$, $c_a$, as two independent circles that wind around the torus. Then $\o^a\equiv\D\t^a/(2\pi)$ is the corresponding dual basis of $H^1(T^2)$ in the sense that $(c_a,\o^b)=\d^b_a$. As to $H^2(T^2)\simeq H_2(T^2,\R)\simeq\R$, it is likewise easy to guess that its single generator can be taken as $\D\t^1\w\D\t^2$, that is the volume form on the torus.

It is straightforward to generalize the above observation to the $n$-torus with any positive $n$. All we need to do is think of $T^n$ as $(S^1)^{\times n}$ and introduce a set of $n$ angular variables $\t^a$, $a=1,\dotsc,n$. We then expect that the $p$-th de Rham cohomology group of $T^n$ is generated by the set of $p$-forms of the type $\D\t^{a_1}\w\dotsb\w\D\t^{a_p}$, where $a_1,\dotsc,a_p$ is a set of different indices. This indicates that $H^p(T^n)\simeq\R^{\binom np}$, which allows us to understand  the corresponding result~\eqref{torushomology} for homology.
\end{illustration}

\begin{illustration}%
\label{ex:cohomologyinvariants}%
\refex{ex:stokesthmpullback} shows that closed forms can be used to construct topological invariants. One considers a smooth map $f:\M\to\MN$ defined on the $n$-dimensional manifold $\M$ and taking values from the target manifold $\MN$. If $\M$ has no boundary and $\o$ is a closed $n$-form on $\MN$, then the integral $\int_\M f^*\o$ does not change under smooth deformations of $f$. But if $\o$ is exact, then so is $f^*\o$ and its integral over $\M$ necessarily vanishes by the Stokes theorem. Nontrivial topological invariants can therefore only arise from nontrivial cohomology equivalence classes on $\MN$.

A simple example of a topological invariant is provided by the volume form~\eqref{spherevolumeform} on $S^2$. Let us consider a function defined on $\R^2$ and taking values on $S^2$. Following the notation of \refex{ex:embeddingmetric}, I will denote the function using a unit vector, $\vec n:\R^2\to S^2$. In physics, $\vec n$ may represent for instance a configuration of a two-dimensional (anti)ferromagnet. Suppose now that our function tends to a constant at infinity. We can then treat the infinity as a single point, which effectively turns $\R^2$ into a compact manifold without boundary. (It is common to say that we have topologically compactified $\R^2$ to $S^2$.) Using the expression for the volume form on $S^2$ in terms of $\vec n$ and pulling it back to $\R^2$, we get the topological invariant
\begin{equation}
w[\vec n]=\frac1{4\pi}\int_{\R^2}\vec n\cdot(\de_x\vec n\times\de_y\vec n)\,\D x\,\D y\;,
\label{Brouwer}
\end{equation}
where $x,y$ are the Cartesian coordinates in $\R^2$. The normalization factor $1/(4\pi)$ is conventional and ensures that $w[\vec n]$ only takes integer values. This is the \emph{Brouwer degree} of $\vec n$, treated as a map $S^2\to S^2$.
\end{illustration}

Let me conclude the appendix with a lightning summary of some further interesting properties of de Rham cohomology. A curious reader will find more details in Sect.~6.4 of~\cite{Nakahara2003a}. For an $n$-dimensional compact manifold $\M$, let us define a pairing of a $p$-form $\o$ and an $(n-p)$-form $\s$ as the bilinear map
\begin{equation}
(\o,\s)\to\int_\M\o\w\s\;.
\label{Poincareduality}
\end{equation}
It is easy to see that if $\M$ has no boundary and both $\o,\s$ are closed, then the result of the integration depends only on the respective equivalence classes of $\o,\s$ as elements of $H^p(\M)$ and $H^{n-p}(\M)$. Moreover, \eqref{Poincareduality} as a bilinear map $H^p(\M)\times H^{n-p}(\M)\to\R$ turns out to be nondegenerate. It follows that the spaces $H^p(\M)$ and $H^{n-p}(\M)$ are isomorphic. This fact is known as the \emph{Poincar\'e duality}.

\begin{illustration}%
We already saw the isomorphism $H^p(\M)\simeq H^{n-p}(\M)$ at work in \refex{ex:toruscohomology}. The Poincar\'e duality however also allows us to draw some generally valid conclusions. For instance, for any compact connected manifold $\M$ without boundary, it tells us that $H^n(\M)\simeq H^0(\M)\simeq\R$. Let us unpack the meaning of this statement in the special case that the manifold $\M$ is Riemannian. Then any $n$-form $\o$ is represented as $\o=\ho f$ in terms of some function $f$ on $\M$. The representative of the single nontrivial equivalence class of $n$-cycles can be taken as $\M$ itself. By the de Rham theorem, $\o$ is therefore exact if and only if $\int_\M f=0$. The exactness implies that $\o=\ho f=\D\s$ for some $(n-1)$-form $\s$. This leads to
\begin{equation}
f=\ho\D\s=(-1)^n\udelta\ho\s=(-1)^{n-1}\divg(\ho\s)^\sharp\;,
\end{equation}
where I used~\eqref{hodgehodge} plus the definitions of codifferential and divergence. In plain terms, the function $f$ is a divergence of some vector field if and only if the integral of $f$ over $\M$ vanishes. This generalizes a statement frequently used in physics. 
\end{illustration}

Sometimes, the manifold in question has the structure of a Cartesian product, $\M\times\MN$. Any differential form $\o$ on $\M$ can be lifted to $\M\times\MN$. Formally, this is done using the projection $\pi_\M:\M\times\MN\to\M$ which maps a point $(x,y)\in\M\times\MN$ to $x\in\M$. Then, for any $\o\in\Omega^q\M$, $\pi^*_\M\o$ is a $q$-form on $\M\times\MN$. Analogously, forms on $\MN$ are lifted to $\M\times\MN$ using the projection $\pi_\MN:\M\times\MN\to\MN$. Now for any closed $q$-form $\o$ on $\M$ and any closed $r$-form $\s$ on $\MN$, $\pi^*_\M\o\w\pi^*_\MN\s$ is a closed $(q+r)$-form on $\M\times\MN$. Moreover, the cohomology equivalence class of $\pi^*_\M\o\w\pi^*_\MN\s$ depends only on those of $\o$ and $\s$. We can thus obtain some information about de Rham cohomology groups of $\M\times\MN$ if we know those of $\M$ and $\MN$. What is less trivial is that all cohomology generators on $\M\times\MN$ can in fact be recovered in this way. This is expressed by the \emph{K\"unneth formula}
\begin{equation}
H^p(\M\times\MN)\simeq\bigoplus_{q+r=p}\bigl[H^q(\M)\otimes H^r(\MN)\bigr]\;,
\label{kunneth}
\end{equation}
where the $\otimes$ symbol denotes tensor product of vector spaces.

\begin{illustration}%
The de Rham cohomology groups of the circle, $S^1$, are $H^0(S^1)\simeq H^1(S^1)\simeq\R$. The former follows from the fact that $S^1$ is connected, and the latter most easily from Poincar\'e duality. Using the K\"unneth formula~\eqref{kunneth}, we can now recursively construct all de Rham cohomology groups of the $n$-torus. Let us see how this works for the 2-torus, $T^2\simeq S^1\times S^1$. The zeroth cohomology group is of course $\R$ since the torus is connected. The first cohomology group is
\begin{equation}
H^1(S^1\times S^1)\simeq[H^0(S^1)\otimes H^1(S^1)]\oplus[H^1(S^1)\otimes H^0(S^1)]\simeq \R\oplus\R\simeq\R^2\;.
\end{equation}
The second cohomology group is similarly
\begin{equation}
H^2(S^1\times S^1)\simeq H^1(S^1)\otimes H^1(S^1)\simeq\R\otimes\R\simeq\R\;.
\end{equation}
With the same reasoning, one then proves by induction that the de Rham cohomology groups of the $n$-torus are $\smash{H^p(T^n)\simeq H^p((S^1)^{\times n})\simeq\R^{\binom np}}$, as guessed in \refex{ex:toruscohomology}.
\end{illustration}

There is a simple relation between de Rham cohomology groups and homotopy groups that can be utilized to deduce the former from the latter. The precise statement is provided by the \emph{Hurewicz theorem} (see Sect.~22.3 of~\cite{Frankel2012a}). Let $\M$ be a compact manifold and let $\pi_p(\M)$, $p>1$ be the first nonvanishing homotopy group. Then $H^p(\M)$ is the first nonvanishing de Rham cohomology group (with nonzero degree) and $H^p(\M)\simeq\pi_p(\M)$, if the latter is treated as a real vector space.\footnote{This is a slight abuse of terminology. The Hurewicz theorem as presented in~\cite{Frankel2012a} really connects $\pi_p(\M)$ as an additive Abelian group to the homology group $H_p(\M,\Z)$ with \emph{integer} coefficients. When translating $H_p(\M,\Z)$ to $H_p(\M,\R)$, which is by the de Rham theorem in turn isomorphic to $H^p(\M)$, any finite part of $H_p(\M,\Z)$ is lost.} With the additional assumption that $\pi_1(\M)$ is Abelian, the same statement holds also for $p=1$. Let me illustrate the use of this theorem on an example of great relevance for the classification of effective actions in Chap.~\ref{chap:effLagrangian}.

\begin{illustration}%
\label{ex:2ndcohomology}%
With a bit of homotopy theory, we can find the second de Rham cohomology group of the homogeneous space $G/H$, where $G$ and $H$ are Lie groups. What we need from homotopy theory is the exact sequence of homotopy groups (see Sect.~9 of~\cite{Mermin1979a}),
\begin{equation}
\dotsb\to\pi_k(G)\to\pi_k(G/H)\to\pi_{k-1}(H)\to\pi_{k-1}(G)\to\dotsb\;.
\end{equation}
This implies that if both $\pi_k(G)$ and $\pi_{k-1}(G)$ are trivial, then $\pi_k(G/H)\simeq\pi_{k-1}(H)$. Let us assume that $G$ is compact and simply connected and $H$ is connected. Then $\pi_1(G)\simeq\pi_0(G)\simeq\trgr$, and hence $\pi_1(G/H)\simeq\pi_0(H)\simeq\trgr$. The homogeneous space is itself simply connected.

Next, we use the fact that $\pi_2(G)$ is trivial for any compact connected Lie group (Appendix~B of~\cite{Weinberg1996a}). Hence $\pi_2(G/H)\simeq\pi_1(H)$. But we have already established that $G/H$ is simply connected. The Hurewicz theorem then tells us that $H^1(G/H)\simeq0$ and $H^2(G/H)\simeq\pi_2(G/H)\simeq\pi_1(H)$. To summarize, let $G$ be a compact and simply connected Lie group, and let $H$ be its connected subgroup. Then the second de Rham cohomology group of the homogeneous space $G/H$ is isomorphic to $\pi_1(H)$ as a real vector space. Since $H$ is by assumption compact, this means that the generators of the second de Rham cohomology group of $G/H$ are in a one-to-one correspondence with $\gr{U}(1)$ factors of $H$~\cite{DHoker1995b}.
\end{illustration}


\bibliographystyle{spphys}
\bibliography{references}

\begin{thebibliography}{10}

\bibitem{Stone2009a}
M.~Stone, P.~Goldbart, \emph{Mathematics for Physics: A Guided Tour for
  Graduate Students} (Cambridge University Press, Cambridge, UK, 2009)

\bibitem{Nakahara2003a}
M.~Nakahara, \emph{Geometry, Topology and Physics} (Institute of Physics
  Publishing, Bristol, UK, 2003)

\bibitem{Fecko2011a}
M.~Fecko, \emph{{Differential Geometry and Lie Groups for Physicists}}
  (Cambridge University Press, Cambridge, UK, 2011)

\bibitem{Frankel2012a}
T.~Frankel, \emph{The Geometry of Physics: An Introduction} (Cambridge
  University Press, Cambridge, UK, 2012)

\bibitem{Lee2013a}
J.M. Lee, \emph{Introduction to Smooth Manifolds} (Springer Science+Business
  Media, New York, 2013)

\bibitem{Needham2021a}
T.~Needham, \emph{Visual Differential Geometry and Forms} (Princeton University
  Press, Princeton, US, 2021)

\bibitem{Lee2018}
J.M. Lee, \emph{Introduction to Riemannian Manifolds} (Springer International
  Publishing, Cham, Switzerland, 2018)

\bibitem{Brewin2009}
L.~Brewin, Class. Quant. Grav. \textbf{26}, 175017 (2009).
\newblock \href{https://doi.org/10.1088/0264-9381/26/17/175017}{DOI
  10.1088/0264-9381/26/17/175017}

\bibitem{Gaiotto2015a}
D.~Gaiotto, A.~Kapustin, N.~Seiberg, B.~Willett, JHEP \textbf{02}, 172 (2015).
\newblock \href{https://doi.org/10.1007/JHEP02(2015)172}{DOI
  10.1007/JHEP02(2015)172}

\bibitem{Brauner2021b}
T.~Brauner, JHEP \textbf{04}, 045 (2021).
\newblock \href{https://doi.org/10.1007/JHEP04(2021)045}{DOI
  10.1007/JHEP04(2021)045}

\bibitem{Arnold1989a}
V.I. Arnold, \emph{Mathematical Methods of Classical Mechanics}
  (Springer-Verlag, New York, 1989)

\bibitem{Mermin1979a}
N.D. Mermin, Rev. Mod. Phys. \textbf{51}, 591 (1979).
\newblock \href{https://doi.org/10.1103/RevModPhys.51.591}{DOI
  10.1103/RevModPhys.51.591}

\bibitem{Weinberg1996a}
S.~Weinberg, \emph{The Quantum Theory of Fields}, vol.~II (Cambridge University
  Press, Cambridge, UK, 1996)

\bibitem{DHoker1995b}
E.~D'Hoker, Nucl. Phys. \textbf{B451}, 725 (1995).
\newblock \href{https://doi.org/10.1016/0550-3213(95)00265-T}{DOI
  10.1016/0550-3213(95)00265-T}

\end{thebibliography}


\begin{thebibliography}{10}

\bibitem{Weinberg2021a}
S.~Weinberg, Eur. Phys. J. \textbf{H46}, 6 (2021).
\newblock \href{https://doi.org/10.1140/epjh/s13129-021-00004-x}{DOI
  10.1140/epjh/s13129-021-00004-x}

\bibitem{Georgi1999a}
H.~Georgi, \emph{Lie Algebras in Particle Physics}.
\newblock Frontiers in Physics (Perseus Books, Reading, MA, USA, 1999)

\bibitem{Stone2009a}
M.~Stone, P.~Goldbart, \emph{Mathematics for Physics: A Guided Tour for
  Graduate Students} (Cambridge University Press, Cambridge, UK, 2009)

\bibitem{Burgess2021a}
C.P. Burgess, \emph{{Introduction to Effective Field Theory}} (Cambridge
  University Press, Cambridge, UK, 2021).
\newblock \href{https://doi.org/10.1017/9781139048040}{DOI
  10.1017/9781139048040}

\bibitem{Kaplan2005a}
D.B. Kaplan,
  \href{https://arxiv.org/abs/nucl-th/0510023}{arXiv:nucl-th/0510023}

\bibitem{Manohar2018a}
A.V. Manohar, \href{https://arxiv.org/abs/1804.05863}{arXiv:1804.05863}

\bibitem{Gripaios2015a}
B.~Gripaios, \href{https://arxiv.org/abs/1506.05039}{arXiv:1506.05039}

\bibitem{Cohen2019a}
T.~Cohen, PoS \textbf{TASI2018}, 011 (2019)

\bibitem{Scherer2012a}
S.~Scherer, M.R. Schindler, \emph{{A Primer for Chiral Perturbation Theory}},
  \emph{Lecture Notes in Physics}, vol. 830 (Springer, 2012).
\newblock \href{https://doi.org/10.1007/978-3-642-19254-8}{DOI
  10.1007/978-3-642-19254-8}

\bibitem{Pich2018a}
A.~Pich, \href{https://arxiv.org/abs/1804.05664}{arXiv:1804.05664}

\bibitem{Penco2020a}
R.~Penco, \href{https://arxiv.org/abs/2006.16285}{arXiv:2006.16285}

\bibitem{Strocchi2021}
F.~Strocchi, \emph{{Symmetry Breaking}}.
\newblock Theoretical and Mathematical Physics (Springer Nature Switzerland AG,
  2021).
\newblock \href{https://doi.org/10.1007/978-3-662-62166-0}{DOI
  10.1007/978-3-662-62166-0}

\bibitem{Guralnik1968a}
G.S. Guralnik, C.R. Hagen, T.W.B. Kibble, in \emph{Advances in Particle
  Physics}, vol.~II, ed. by R.L. Cool, R.E. Marshak (Wiley, New York, NY, USA,
  1968), pp. 567--708

\bibitem{Beekman2019a}
A.J. Beekman, L.~Rademaker, J.~van Wezel, SciPost Phys. Lect. Notes \textbf{11}
  (2019).
\newblock \href{https://doi.org/10.21468/SciPostPhysLectNotes.11}{DOI
  10.21468/SciPostPhysLectNotes.11}

\bibitem{Naegels2021a}
D.~Naegels, \href{https://arxiv.org/abs/2110.14504}{arXiv:2110.14504}

\bibitem{Watanabe2020a}
H.~Watanabe, Ann. Rev. Cond. Mat. Phys. \textbf{11}, 169 (2020).
\newblock \href{https://doi.org/10.1146/annurev-conmatphys-031119-050644}{DOI
  10.1146/annurev-conmatphys-031119-050644}

\end{thebibliography}


\begin{thebibliography}{1}

\bibitem{Goldstein2013a}
H.~Goldstein, C.P. Poole, J.L. Safko, \emph{Classical Mechanics} (Pearson
  Education Limited, Harlow, UK, 2013)

\bibitem{Goldstone1961a}
J.~Goldstone, Nuovo Cim. \textbf{19}, 154 (1961).
\newblock \href{https://doi.org/10.1007/BF02812722}{DOI 10.1007/BF02812722}

\bibitem{Goldstone1962a}
J.~Goldstone, A.~Salam, S.~Weinberg, Phys. Rev. \textbf{127}, 965 (1962).
\newblock \href{https://doi.org/10.1103/PhysRev.127.965}{DOI
  10.1103/PhysRev.127.965}

\bibitem{Chisholm1961a}
J.S.R. Chisholm, Nucl. Phys. \textbf{26}, 469 (1961).
\newblock \href{https://doi.org/10.1016/0029-5582(61)90106-7}{DOI
  10.1016/0029-5582(61)90106-7}

\bibitem{Kamefuchi1961a}
S.~Kamefuchi, L.~O'Raifeartaigh, A.~Salam, Nucl. Phys. \textbf{28}, 529 (1961).
\newblock \href{https://doi.org/10.1016/0029-5582(61)90056-6}{DOI
  10.1016/0029-5582(61)90056-6}

\bibitem{Honerkamp1971}
J.~Honerkamp, K.~Meetz, Phys. Rev. \textbf{D3}, 1996 (1971).
\newblock \href{https://doi.org/10.1103/PhysRevD.3.1996}{DOI
  10.1103/PhysRevD.3.1996}

\bibitem{Burgess2021a}
C.P. Burgess, \emph{{Introduction to Effective Field Theory}} (Cambridge
  University Press, Cambridge, UK, 2021).
\newblock \href{https://doi.org/10.1017/9781139048040}{DOI
  10.1017/9781139048040}

\end{thebibliography}


\begin{thebibliography}{1}

\bibitem{Goldstone1962a}
J.~Goldstone, A.~Salam, S.~Weinberg, Phys. Rev. \textbf{127}, 965 (1962).
\newblock \href{https://doi.org/10.1103/PhysRev.127.965}{DOI
  10.1103/PhysRev.127.965}

\bibitem{Weinberg1972a}
S.~Weinberg, Phys. Rev. Lett. \textbf{29}, 1698 (1972).
\newblock \href{https://doi.org/10.1103/PhysRevLett.29.1698}{DOI
  10.1103/PhysRevLett.29.1698}

\bibitem{Coleman1973b}
S.R. Coleman, E.J. Weinberg, Phys. Rev. \textbf{D7}, 1888 (1973).
\newblock \href{https://doi.org/10.1103/PhysRevD.7.1888}{DOI
  10.1103/PhysRevD.7.1888}

\bibitem{Goldstein2013a}
H.~Goldstein, C.P. Poole, J.L. Safko, \emph{Classical Mechanics} (Pearson
  Education Limited, Harlow, UK, 2013)

\bibitem{Brauner2006b}
T.~Brauner, Phys. Rev. \textbf{D74}, 085010 (2006).
\newblock \href{https://doi.org/10.1103/PhysRevD.74.085010}{DOI
  10.1103/PhysRevD.74.085010}

\bibitem{Gongyo2016a}
S.~Gongyo, Y.~Kikuchi, T.~Hyodo, T.~Kunihiro, PTEP \textbf{2016}, 083B01
  (2016).
\newblock \href{https://doi.org/10.1093/ptep/ptw095}{DOI 10.1093/ptep/ptw095}

\bibitem{Brauner2018a}
T.~Brauner, M.F. Jakobsen, Phys. Rev. \textbf{D97}, 025021 (2018).
\newblock \href{https://doi.org/10.1103/PhysRevD.97.025021}{DOI
  10.1103/PhysRevD.97.025021}

\bibitem{Leggett2006a}
A.J. Leggett, \emph{Quantum Liquids} (Oxford University Press, Oxford, UK,
  2006)

\end{thebibliography}


\begin{thebibliography}{10}

\bibitem{Olver1986a}
P.J. Olver, \emph{Applications of Lie Groups to Differential Equations}
  (Springer-Verlag, New York, 1986)

\bibitem{Bluman2002a}
G.W. Bluman, S.C. Anco, \emph{Symmetry and Integration Methods for Differential
  Equations} (Springer-Verlag, New York, 2002)

\bibitem{Bluman2010a}
G.W. Bluman, A.F. Cheviakov, S.C. Anco, \emph{Applications of Symmetry Methods
  to Partial Differential Equations} (Springer-Verlag, New York, 2010)

\bibitem{Kosmann-Schwarzbach2011a}
Y.~Kosmann-Schwarzbach, \emph{The Noether Theorems: Invariance and Conservation
  Laws in the Twentieth Century} (Springer-Verlag, New York, 2011)

\bibitem{Gell-Mann1960a}
M.~Gell-Mann, M.~L\'evy, Nuovo Cim. \textbf{16}, 705 (1960).
\newblock \href{https://doi.org/10.1007/BF02859738}{DOI 10.1007/BF02859738}

\bibitem{Brauner2020a}
T.~Brauner, Phys. Scr. \textbf{95}, 035004 (2020).
\newblock \href{https://doi.org/10.1088/1402-4896/ab50a5}{DOI
  10.1088/1402-4896/ab50a5}

\bibitem{Kourkoulou2022a}
I.~Kourkoulou, A.~Nicolis, G.~Sun, Phys. Rev. \textbf{D106}, 125005 (2022).
\newblock \href{https://doi.org/10.1103/PhysRevD.106.125005}{DOI
  10.1103/PhysRevD.106.125005}

\bibitem{Gieres2022}
F.~Gieres, Fortschr. Phys. \textbf{70}, 2200078 (2022).
\newblock \href{https://doi.org/10.1002/prop.202200078}{DOI
  10.1002/prop.202200078}

\bibitem{Griffin2013a}
T.~Griffin, K.T. Grosvenor, P.~Ho\v{r}ava, Z.~Yan, Phys. Rev. \textbf{D88},
  101701 (2013).
\newblock \href{https://doi.org/10.1103/PhysRevD.88.101701}{DOI
  10.1103/PhysRevD.88.101701}

\bibitem{Seiberg2020a}
N.~Seiberg, SciPost Phys. \textbf{8}, 050 (2020).
\newblock \href{https://doi.org/10.21468/SciPostPhys.8.4.050}{DOI
  10.21468/SciPostPhys.8.4.050}

\bibitem{Grosvenor2022a}
K.T. Grosvenor, C.~Hoyos, F.~{Pe\~na-Ben\'{\i}tez}, P.~Sur\'owka, Front. Phys.
  \textbf{9}, 792621 (2022).
\newblock \href{https://doi.org/10.3389/fphy.2021.792621}{DOI
  10.3389/fphy.2021.792621}

\bibitem{Jose1998a}
J.V. Jos\'e, E.J. Saletan, \emph{Classical Dynamics: A Contemporary Approach}
  (Cambridge University Press, Cambridge, UK, 1998)

\bibitem{Gieres2021}
F.~Gieres, SciPost Phys. Lect. Notes \textbf{77} (2023).
\newblock \href{https://doi.org/10.21468/SciPostPhysLectNotes.77}{DOI
  10.21468/SciPostPhysLectNotes.77}

\bibitem{Arnold1989a}
V.I. Arnold, \emph{Mathematical Methods of Classical Mechanics}
  (Springer-Verlag, New York, 1989)

\bibitem{Dzyaloshinskii1980}
I.E. Dzyaloshinskii, G.E. Volovick, Ann. Phys. \textbf{125}, 67 (1980).
\newblock \href{https://doi.org/10.1016/0003-4916(80)90119-0}{DOI
  10.1016/0003-4916(80)90119-0}

\end{thebibliography}


\begin{thebibliography}{10}

\bibitem{Beekman2019a}
A.J. Beekman, L.~Rademaker, J.~van Wezel, SciPost Phys. Lect. Notes \textbf{11}
  (2019).
\newblock \href{https://doi.org/10.21468/SciPostPhysLectNotes.11}{DOI
  10.21468/SciPostPhysLectNotes.11}

\bibitem{Strocchi2021}
F.~Strocchi, \emph{{Symmetry Breaking}}.
\newblock Theoretical and Mathematical Physics (Springer Nature Switzerland AG,
  2021).
\newblock \href{https://doi.org/10.1007/978-3-662-62166-0}{DOI
  10.1007/978-3-662-62166-0}

\bibitem{Goldstein2013a}
H.~Goldstein, C.P. Poole, J.L. Safko, \emph{Classical Mechanics} (Pearson
  Education Limited, Harlow, UK, 2013)

\bibitem{Ballentine1998a}
L.E. Ballentine, \emph{Quantum Mechanics: A Modern Development} (World
  Scientific, Singapore, 1998)

\bibitem{Basdevant2002a}
J.L. Basdevant, J.~Dalibard, \emph{Quantum Mechanics} (Springer-Verlag, Berlin,
  Heidelberg, 2002)

\bibitem{Balian2007a}
R.~Balian, \emph{From Microphysics to Macrophysics}, vol.~I (Springer, Berlin,
  Heidelberg, 2007)

\bibitem{Barut1977a}
A.O. Barut, R.~Raczka, \emph{Theory of Group Representations and Applications}
  (Polish Scientific Publishers, Warszawa, Poland, 1977)

\bibitem{Chaikin1995a}
P.M. Chaikin, T.C. Lubensky, \emph{Principles of condensed matter physics}
  (Cambridge University Press, Cambridge, UK, 1995)

\bibitem{Mussardo2010}
G.~Mussardo, \emph{Statistical Field Theory} (Oxford University Press, Oxford,
  UK, 2010)

\bibitem{Koma1994}
T.~Koma, H.~Tasaki, J. Stat. Phys. \textbf{76}, 745 (1994).
\newblock \href{https://doi.org/10.1007/BF02188685}{DOI 10.1007/BF02188685}

\bibitem{Tasaki2020}
H.~Tasaki, \emph{{Physics and Mathematics of Quantum Many-Body Systems}}
  (Springer Nature Switzerland, 2020).
\newblock \href{https://doi.org/10.1007/978-3-030-41265-4}{DOI
  10.1007/978-3-030-41265-4}

\bibitem{Brauner2010a}
T.~Brauner, Symmetry \textbf{2}, 609 (2010).
\newblock \href{https://doi.org/10.3390/sym2020609}{DOI 10.3390/sym2020609}

\bibitem{Weinberg1995a}
S.~Weinberg, \emph{The Quantum Theory of Fields}, vol.~I (Cambridge University
  Press, Cambridge, UK, 1995)

\bibitem{Guralnik1968a}
G.S. Guralnik, C.R. Hagen, T.W.B. Kibble, in \emph{Advances in Particle
  Physics}, vol.~II, ed. by R.L. Cool, R.E. Marshak (Wiley, New York, NY, USA,
  1968), pp. 567--708

\bibitem{Peierls1991a}
R.~Peierls, T.A. Kaplan, P.W. Anderson, Phys. Today \textbf{44}, 13 (1991).
\newblock \href{https://doi.org/10.1063/1.2809982}{DOI 10.1063/1.2809982}

\bibitem{Weinberg1996a}
S.~Weinberg, \emph{The Quantum Theory of Fields}, vol.~II (Cambridge University
  Press, Cambridge, UK, 1996)

\end{thebibliography}


\begin{thebibliography}{10}

\bibitem{Goldstone1961a}
J.~Goldstone, Nuovo Cim. \textbf{19}, 154 (1961).
\newblock \href{https://doi.org/10.1007/BF02812722}{DOI 10.1007/BF02812722}

\bibitem{Goldstone1962a}
J.~Goldstone, A.~Salam, S.~Weinberg, Phys. Rev. \textbf{127}, 965 (1962).
\newblock \href{https://doi.org/10.1103/PhysRev.127.965}{DOI
  10.1103/PhysRev.127.965}

\bibitem{Guralnik1968a}
G.S. Guralnik, C.R. Hagen, T.W.B. Kibble, in \emph{Advances in Particle
  Physics}, vol.~II, ed. by R.L. Cool, R.E. Marshak (Wiley, New York, NY, USA,
  1968), pp. 567--708

\bibitem{Strocchi2021}
F.~Strocchi, \emph{{Symmetry Breaking}}.
\newblock Theoretical and Mathematical Physics (Springer Nature Switzerland AG,
  2021).
\newblock \href{https://doi.org/10.1007/978-3-662-62166-0}{DOI
  10.1007/978-3-662-62166-0}

\bibitem{Watanabe2012a}
H.~Watanabe, T.~Brauner, Phys. Rev. \textbf{D85}, 085010 (2012).
\newblock \href{https://doi.org/10.1103/PhysRevD.85.085010}{DOI
  10.1103/PhysRevD.85.085010}

\bibitem{Fabri1966a}
E.~Fabri, L.E. Picasso, Phys. Rev. Lett. \textbf{16}, 408 (1966).
\newblock \href{https://doi.org/10.1103/PhysRevLett.16.408.2}{DOI
  10.1103/PhysRevLett.16.408.2}

\bibitem{Watanabe2011a}
H.~Watanabe, T.~Brauner, Phys. Rev. \textbf{D84}, 125013 (2011).
\newblock \href{https://doi.org/10.1103/PhysRevD.84.125013}{DOI
  10.1103/PhysRevD.84.125013}

\bibitem{Weinberg1996a}
S.~Weinberg, \emph{The Quantum Theory of Fields}, vol.~II (Cambridge University
  Press, Cambridge, UK, 1996)

\bibitem{Low2002a}
I.~Low, A.V. Manohar, Phys. Rev. Lett. \textbf{88}, 101602 (2002).
\newblock \href{https://doi.org/10.1103/PhysRevLett.88.101602}{DOI
  10.1103/PhysRevLett.88.101602}

\bibitem{Watanabe2013a}
H.~Watanabe, H.~Murayama, Phys. Rev. Lett. \textbf{110}, 181601 (2013).
\newblock \href{https://doi.org/10.1103/PhysRevLett.110.181601}{DOI
  10.1103/PhysRevLett.110.181601}

\bibitem{Watanabe2020a}
H.~Watanabe, Ann. Rev. Cond. Mat. Phys. \textbf{11}, 169 (2020).
\newblock \href{https://doi.org/10.1146/annurev-conmatphys-031119-050644}{DOI
  10.1146/annurev-conmatphys-031119-050644}

\bibitem{Hayata2014a}
T.~Hayata, Y.~Hidaka, Phys. Lett. \textbf{B735}, 195 (2014).
\newblock \href{https://doi.org/10.1016/j.physletb.2014.06.039}{DOI
  10.1016/j.physletb.2014.06.039}

\bibitem{Kishine2015a}
J.i. Kishine, A.S. Ovchinnikov, Solid State Phys. \textbf{66}, 1 (2015).
\newblock \href{https://doi.org/10.1016/bs.ssp.2015.05.001}{DOI
  10.1016/bs.ssp.2015.05.001}

\bibitem{Togawa2016a}
Y.~Togawa, Y.~Kousaka, K.~Inoue, J.i. Kishine, J. Phys. Soc. Jpn. \textbf{85},
  112001 (2016).
\newblock \href{https://doi.org/10.7566/JPSJ.85.112001}{DOI
  10.7566/JPSJ.85.112001}

\bibitem{Watanabe2012b}
H.~Watanabe, H.~Murayama, Phys. Rev. Lett. \textbf{108}, 251602 (2012).
\newblock \href{https://doi.org/10.1103/PhysRevLett.108.251602}{DOI
  10.1103/PhysRevLett.108.251602}

\bibitem{Hidaka2013b}
Y.~Hidaka, Phys. Rev. Lett. \textbf{110}, 091601 (2013).
\newblock \href{https://doi.org/10.1103/PhysRevLett.110.091601}{DOI
  10.1103/PhysRevLett.110.091601}

\bibitem{Azcarraga1995}
J.A. de~Azc\'arraga, J.M. Izquierdo, \emph{{Lie Groups, Lie Algebras,
  Cohomology and Some Applications in Physics}} (Cambridge University Press,
  UK, 1995)

\bibitem{Kobayashi2014b}
M.~Kobayashi, M.~Nitta, Phys. Rev. Lett. \textbf{113}, 120403 (2014).
\newblock \href{https://doi.org/10.1103/PhysRevLett.113.120403}{DOI
  10.1103/PhysRevLett.113.120403}

\bibitem{Kobayashi2014a}
M.~Kobayashi, M.~Nitta, Phys. Rev. \textbf{D90}, 025010 (2014).
\newblock \href{https://doi.org/10.1103/PhysRevD.90.025010}{DOI
  10.1103/PhysRevD.90.025010}

\bibitem{Rezende2019}
S.M. Rezende, A.~Azevedo, R.L. Rodr{\'\i}guez-Su{\'a}rez, J. Appl. Phys.
  \textbf{126}, 151101 (2019).
\newblock \href{https://doi.org/10.1063/1.5109132}{DOI 10.1063/1.5109132}

\bibitem{Uchino2010a}
S.~Uchino, M.~Kobayashi, M.~Nitta, M.~Ueda, Phys. Rev. Lett. \textbf{105},
  230406 (2010).
\newblock \href{https://doi.org/10.1103/PhysRevLett.105.230406}{DOI
  10.1103/PhysRevLett.105.230406}

\bibitem{Takahashi2015a}
D.A. Takahashi, M.~Nitta, Ann. Phys. \textbf{354}, 101 (2015).
\newblock \href{https://doi.org/10.1016/j.aop.2014.12.009}{DOI
  10.1016/j.aop.2014.12.009}

\bibitem{Nitta2015a}
M.~Nitta, D.A. Takahashi, Phys. Rev. \textbf{D91}, 025018 (2015).
\newblock \href{https://doi.org/10.1103/PhysRevD.91.025018}{DOI
  10.1103/PhysRevD.91.025018}

\bibitem{Nicolis2013a}
A.~Nicolis, F.~Piazza, Phys. Rev. Lett. \textbf{110}, 011602 (2013).
\newblock \href{https://doi.org/10.1103/PhysRevLett.110.011602}{DOI
  10.1103/PhysRevLett.110.011602}

\bibitem{Watanabe2013b}
H.~Watanabe, T.~Brauner, H.~Murayama, Phys. Rev. Lett. \textbf{111}, 021601
  (2013).
\newblock \href{https://doi.org/10.1103/PhysRevLett.111.021601}{DOI
  10.1103/PhysRevLett.111.021601}

\bibitem{Brauner2018a}
T.~Brauner, M.F. Jakobsen, Phys. Rev. \textbf{D97}, 025021 (2018).
\newblock \href{https://doi.org/10.1103/PhysRevD.97.025021}{DOI
  10.1103/PhysRevD.97.025021}

\bibitem{Nicolis2013b}
A.~Nicolis, R.~Penco, F.~Piazza, R.A. Rosen, JHEP \textbf{11}, 055 (2013).
\newblock \href{https://doi.org/10.1007/JHEP11(2013)055}{DOI
  10.1007/JHEP11(2013)055}

\end{thebibliography}


\begin{thebibliography}{10}

\bibitem{Weinberg1995a}
S.~Weinberg, \emph{The Quantum Theory of Fields}, vol.~I (Cambridge University
  Press, Cambridge, UK, 1995)

\bibitem{Olver1986a}
P.J. Olver, \emph{Applications of Lie Groups to Differential Equations}
  (Springer-Verlag, New York, 1986)

\bibitem{Bluman2002a}
G.W. Bluman, S.C. Anco, \emph{Symmetry and Integration Methods for Differential
  Equations} (Springer-Verlag, New York, 2002)

\bibitem{Chisholm1961a}
J.S.R. Chisholm, Nucl. Phys. \textbf{26}, 469 (1961).
\newblock \href{https://doi.org/10.1016/0029-5582(61)90106-7}{DOI
  10.1016/0029-5582(61)90106-7}

\bibitem{Kamefuchi1961a}
S.~Kamefuchi, L.~O'Raifeartaigh, A.~Salam, Nucl. Phys. \textbf{28}, 529 (1961).
\newblock \href{https://doi.org/10.1016/0029-5582(61)90056-6}{DOI
  10.1016/0029-5582(61)90056-6}

\bibitem{Criado2019}
J.C. Criado, M.~P\'erez-Victoria, JHEP \textbf{03}, 038 (2019).
\newblock \href{https://doi.org/10.1007/JHEP03(2019)038}{DOI
  10.1007/JHEP03(2019)038}

\bibitem{Fecko2011a}
M.~Fecko, \emph{{Differential Geometry and Lie Groups for Physicists}}
  (Cambridge University Press, Cambridge, UK, 2011)

\bibitem{Coleman1969a}
S.R. Coleman, J.~Wess, B.~Zumino, Phys. Rev. \textbf{177}, 2239 (1969).
\newblock \href{https://doi.org/10.1103/PhysRev.177.2239}{DOI
  10.1103/PhysRev.177.2239}

\bibitem{Barut1977a}
A.O. Barut, R.~Raczka, \emph{Theory of Group Representations and Applications}
  (Polish Scientific Publishers, Warszawa, Poland, 1977)

\bibitem{Joseph1970a}
A.~Joseph, A.I. Solomon, J. Math. Phys. \textbf{11}, 748 (1970).
\newblock \href{https://doi.org/10.1063/1.1665205}{DOI 10.1063/1.1665205}

\bibitem{Alonso2016a}
R.~Alonso, E.E. Jenkins, A.V. Manohar, JHEP \textbf{08}, 101 (2016).
\newblock \href{https://doi.org/10.1007/JHEP08(2016)101}{DOI
  10.1007/JHEP08(2016)101}

\bibitem{Arvanitoyeorgos2003a}
A.~Arvanitoyeorgos, \emph{An Introduction to Lie Groups and the Geometry of
  Homogeneous Spaces}, \emph{Student Mathematical Library}, vol.~22 (American
  Mathematical Society, 2003)

\bibitem{Bando1988a}
M.~Bando, T.~Kugo, K.~Yamawaki, Phys. Rept. \textbf{164}, 217 (1988).
\newblock \href{https://doi.org/10.1016/0370-1573(88)90019-1}{DOI
  10.1016/0370-1573(88)90019-1}

\end{thebibliography}


\begin{thebibliography}{10}

\bibitem{Weinberg1995a}
S.~Weinberg, \emph{The Quantum Theory of Fields}, vol.~I (Cambridge University
  Press, Cambridge, UK, 1995)

\bibitem{Watanabe2014a}
H.~Watanabe, H.~Murayama, Phys. Rev. \textbf{X4}, 031057 (2014).
\newblock \href{https://doi.org/10.1103/PhysRevX.4.031057}{DOI
  10.1103/PhysRevX.4.031057}

\bibitem{Goldstein2013a}
H.~Goldstein, C.P. Poole, J.L. Safko, \emph{Classical Mechanics} (Pearson
  Education Limited, Harlow, UK, 2013)

\bibitem{Goon2012a}
G.~Goon, K.~Hinterbichler, A.~Joyce, M.~Trodden, JHEP \textbf{06}, 004 (2012).
\newblock \href{https://doi.org/10.1007/JHEP06(2012)004}{DOI
  10.1007/JHEP06(2012)004}

\bibitem{Azcarraga2001a}
J.A. de~Azc\'arraga, J.M. Izquierdo, J.C. P\'erez~Bueno, Rev. R. Acad. Cien.
  Exactas Fis. Nat. Ser. A Mat. \textbf{95}, 225 (2001)

\bibitem{Davighi2018a}
J.~Davighi, B.~Gripaios, JHEP \textbf{09}, 155 (2018).
\newblock \href{https://doi.org/10.1007/JHEP09(2018)155}{DOI
  10.1007/JHEP09(2018)155}

\bibitem{Leutwyler1994b}
H.~Leutwyler, Ann. Phys. \textbf{235}, 165 (1994).
\newblock \href{https://doi.org/10.1006/aphy.1994.1094}{DOI
  10.1006/aphy.1994.1094}

\bibitem{Leutwyler1994a}
H.~Leutwyler, Phys. Rev. \textbf{D49}, 3033 (1994).
\newblock \href{https://doi.org/10.1103/PhysRevD.49.3033}{DOI
  10.1103/PhysRevD.49.3033}

\bibitem{Andersen2014a}
J.O. Andersen, T.~Brauner, C.P. Hofmann, A.~Vuorinen, JHEP \textbf{08}, 088
  (2014).
\newblock \href{https://doi.org/10.1007/JHEP08(2014)088}{DOI
  10.1007/JHEP08(2014)088}

\bibitem{Weinberg1996a}
S.~Weinberg, \emph{The Quantum Theory of Fields}, vol.~II (Cambridge University
  Press, Cambridge, UK, 1996)

\bibitem{Kapusta2006a}
J.I. Kapusta, C.~Gale, \emph{Finite-temperature field theory: Principles and
  applications} (Cambridge University Press, Cambridge, UK, 2006)

\bibitem{Watanabe2013b}
H.~Watanabe, T.~Brauner, H.~Murayama, Phys. Rev. Lett. \textbf{111}, 021601
  (2013).
\newblock \href{https://doi.org/10.1103/PhysRevLett.111.021601}{DOI
  10.1103/PhysRevLett.111.021601}

\bibitem{Scherer2012a}
S.~Scherer, M.R. Schindler, \emph{{A Primer for Chiral Perturbation Theory}},
  \emph{Lecture Notes in Physics}, vol. 830 (Springer, 2012).
\newblock \href{https://doi.org/10.1007/978-3-642-19254-8}{DOI
  10.1007/978-3-642-19254-8}

\bibitem{Arvanitoyeorgos2003a}
A.~Arvanitoyeorgos, \emph{An Introduction to Lie Groups and the Geometry of
  Homogeneous Spaces}, \emph{Student Mathematical Library}, vol.~22 (American
  Mathematical Society, 2003)

\end{thebibliography}


\begin{thebibliography}{10}

\bibitem{Workman2022}
R.L. Workman, et~al., PTEP \textbf{2022}, 083C01 (2022).
\newblock \href{https://doi.org/10.1093/ptep/ptac097}{DOI 10.1093/ptep/ptac097}

\bibitem{Nair2005}
V.P. Nair, \emph{{Quantum Field Theory: A Modern Perspective}} (Springer
  Science$+$Business Media, New York, 2005)

\bibitem{Scherer2012a}
S.~Scherer, M.R. Schindler, \emph{{A Primer for Chiral Perturbation Theory}},
  \emph{Lecture Notes in Physics}, vol. 830 (Springer, 2012).
\newblock \href{https://doi.org/10.1007/978-3-642-19254-8}{DOI
  10.1007/978-3-642-19254-8}

\bibitem{Gasser1984a}
J.~Gasser, H.~Leutwyler, Ann. Phys. \textbf{158}, 142 (1984).
\newblock \href{https://doi.org/10.1016/0003-4916(84)90242-2}{DOI
  10.1016/0003-4916(84)90242-2}

\bibitem{Gasser1985a}
J.~Gasser, H.~Leutwyler, Nucl. Phys. \textbf{B250}, 465 (1985).
\newblock \href{https://doi.org/10.1016/0550-3213(85)90492-4}{DOI
  10.1016/0550-3213(85)90492-4}

\bibitem{Matousek2009}
J.~Matou\v{s}ek, J.~Ne\v{s}et\v{r}il, \emph{Invitation to Discrete Mathematics}
  (Oxford University Press, Oxford, UK, 2009)

\bibitem{Andersen2014a}
J.O. Andersen, T.~Brauner, C.P. Hofmann, A.~Vuorinen, JHEP \textbf{08}, 088
  (2014).
\newblock \href{https://doi.org/10.1007/JHEP08(2014)088}{DOI
  10.1007/JHEP08(2014)088}

\bibitem{Watanabe2013b}
H.~Watanabe, T.~Brauner, H.~Murayama, Phys. Rev. Lett. \textbf{111}, 021601
  (2013).
\newblock \href{https://doi.org/10.1103/PhysRevLett.111.021601}{DOI
  10.1103/PhysRevLett.111.021601}

\bibitem{Mannarelli2019a}
M.~Mannarelli, Particles \textbf{2}, 411 (2019).
\newblock \href{https://doi.org/10.3390/particles2030025}{DOI
  10.3390/particles2030025}

\bibitem{Peskin1995}
M.E. Peskin, D.V. Schroeder, \emph{{An Introduction to Quantum Field Theory}}
  (Addison-Wesley, Reading, USA, 1995)

\bibitem{Andersen2011}
J.R. Andersen, et~al., Eur. Phys. J. Plus \textbf{126}, 81 (2011).
\newblock \href{https://doi.org/10.1140/epjp/i2011-11081-1}{DOI
  10.1140/epjp/i2011-11081-1}

\bibitem{DHoker1994a}
E.~D'Hoker, S.~Weinberg, Phys. Rev. \textbf{D50}, 6050 (1994).
\newblock \href{https://doi.org/10.1103/PhysRevD.50.R6050}{DOI
  10.1103/PhysRevD.50.R6050}

\bibitem{DHoker1995b}
E.~D'Hoker, Nucl. Phys. \textbf{B451}, 725 (1995).
\newblock \href{https://doi.org/10.1016/0550-3213(95)00265-T}{DOI
  10.1016/0550-3213(95)00265-T}

\bibitem{Brauner2019a}
T.~Brauner, H.~Kole\v{s}ov\'{a}, Nucl. Phys. \textbf{B945}, 114676 (2019).
\newblock \href{https://doi.org/10.1016/j.nuclphysb.2019.114676}{DOI
  10.1016/j.nuclphysb.2019.114676}

\bibitem{Weinberg1996a}
S.~Weinberg, \emph{The Quantum Theory of Fields}, vol.~II (Cambridge University
  Press, Cambridge, UK, 1996)

\bibitem{Witten1983a}
E.~Witten, Nucl. Phys. \textbf{B223}, 422 (1983).
\newblock \href{https://doi.org/10.1016/0550-3213(83)90063-9}{DOI
  10.1016/0550-3213(83)90063-9}

\bibitem{Roman1999a}
J.M. Rom\'an, J.~Soto, Int. J. Mod. Phys. \textbf{B13}, 755 (1999).
\newblock \href{https://doi.org/10.1142/S0217979299000655}{DOI
  10.1142/S0217979299000655}

\bibitem{Togawa2016a}
Y.~Togawa, Y.~Kousaka, K.~Inoue, J.i. Kishine, J. Phys. Soc. Jpn. \textbf{85},
  112001 (2016).
\newblock \href{https://doi.org/10.7566/JPSJ.85.112001}{DOI
  10.7566/JPSJ.85.112001}

\bibitem{Kirkpatrick2005}
T.R. Kirkpatrick, D.~Belitz, Phys. Rev. \textbf{B72}, 180402 (2005).
\newblock \href{https://doi.org/10.1103/PhysRevB.72.180402}{DOI
  10.1103/PhysRevB.72.180402}

\bibitem{Han2017}
J.H. Han, \emph{{Skyrmions in Condensed Matter}}, \emph{Springer Tracts in
  Modern Physics}, vol. 278 (Springer, 2017).
\newblock \href{https://doi.org/10.1007/978-3-319-69246-3}{DOI
  10.1007/978-3-319-69246-3}

\bibitem{Tong2016}
D.~Tong, \href{https://arxiv.org/abs/1606.06687}{arXiv:1606.06687}

\bibitem{Watanabe2014a}
H.~Watanabe, H.~Murayama, Phys. Rev. \textbf{X4}, 031057 (2014).
\newblock \href{https://doi.org/10.1103/PhysRevX.4.031057}{DOI
  10.1103/PhysRevX.4.031057}

\end{thebibliography}


\begin{thebibliography}{10}

\bibitem{Mojahed2022}
M.A. Mojahed, T.~Brauner, JHEP \textbf{03}, 086 (2022).
\newblock \href{https://doi.org/10.1007/JHEP03(2022)086}{DOI
  10.1007/JHEP03(2022)086}

\bibitem{Cheung2023}
C.~Cheung, M.~Derda, A.~Helset, J.~Parra-Martinez, JHEP \textbf{08}, 103
  (2023).
\newblock \href{https://doi.org/10.1007/JHEP08(2023)103}{DOI
  10.1007/JHEP08(2023)103}

\bibitem{Cheung2017c}
C.~Cheung, \href{https://arxiv.org/abs/1708.03872}{arXiv:1708.03872}

\bibitem{Elvang2015}
H.~Elvang, Y.t. Huang, \emph{{Scattering Amplitudes in Gauge Theory and
  Gravity}} (Cambridge University Press, Cambridge, UK, 2015)

\bibitem{Weinberg1996a}
S.~Weinberg, \emph{The Quantum Theory of Fields}, vol.~II (Cambridge University
  Press, Cambridge, UK, 1996)

\bibitem{Weinberg1995a}
S.~Weinberg, \emph{The Quantum Theory of Fields}, vol.~I (Cambridge University
  Press, Cambridge, UK, 1995)

\bibitem{Kampf2020a}
K.~Kampf, J.~Novotn{\'y}, M.~Shifman, J.~Trnka, Phys. Rev. Lett. \textbf{124},
  111601 (2020).
\newblock \href{https://doi.org/10.1103/PhysRevLett.124.111601}{DOI
  10.1103/PhysRevLett.124.111601}

\bibitem{Cheung2022}
C.~Cheung, A.~Helset, J.~Parra-Martinez, JHEP \textbf{04}, 011 (2022).
\newblock \href{https://doi.org/10.1007/JHEP04(2022)011}{DOI
  10.1007/JHEP04(2022)011}

\bibitem{Lee2018}
J.M. Lee, \emph{Introduction to Riemannian Manifolds} (Springer International
  Publishing, Cham, Switzerland, 2018)

\bibitem{Low2015a}
I.~Low, Phys. Rev. \textbf{D91}, 105017 (2015).
\newblock \href{https://doi.org/10.1103/PhysRevD.91.105017}{DOI
  10.1103/PhysRevD.91.105017}

\bibitem{Low2016a}
I.~Low, Phys. Rev. \textbf{D93}, 045032 (2016).
\newblock \href{https://doi.org/10.1103/PhysRevD.93.045032}{DOI
  10.1103/PhysRevD.93.045032}

\bibitem{Cheung2015a}
C.~Cheung, K.~Kampf, J.~{Novotn\'y}, J.~Trnka, Phys. Rev. Lett. \textbf{114},
  221602 (2015).
\newblock \href{https://doi.org/10.1103/PhysRevLett.114.221602}{DOI
  10.1103/PhysRevLett.114.221602}

\bibitem{Goon2012a}
G.~Goon, K.~Hinterbichler, A.~Joyce, M.~Trodden, JHEP \textbf{06}, 004 (2012).
\newblock \href{https://doi.org/10.1007/JHEP06(2012)004}{DOI
  10.1007/JHEP06(2012)004}

\bibitem{Rham2014a}
C.~de~Rham, M.~Fasiello, A.J. Tolley, Phys. Lett. \textbf{B733}, 46 (2014).
\newblock \href{https://doi.org/10.1016/j.physletb.2014.03.061}{DOI
  10.1016/j.physletb.2014.03.061}

\bibitem{Kampf2014a}
K.~Kampf, J.~{Novotn\'y}, JHEP \textbf{10}, 006 (2014).
\newblock \href{https://doi.org/10.1007/JHEP10(2014)006}{DOI
  10.1007/JHEP10(2014)006}

\bibitem{Hinterbichler2015a}
K.~Hinterbichler, A.~Joyce, Phys. Rev. \textbf{D92}, 023503 (2015).
\newblock \href{https://doi.org/10.1103/PhysRevD.92.023503}{DOI
  10.1103/PhysRevD.92.023503}

\bibitem{Preucil2019}
F.~P\v{r}eu\v{c}il, J.~Novotn\'y, JHEP \textbf{11}, 166 (2019).
\newblock \href{https://doi.org/10.1007/JHEP11(2019)166}{DOI
  10.1007/JHEP11(2019)166}

\bibitem{Cheung2017a}
C.~Cheung, K.~Kampf, J.~Novotn{\'y}, C.H. Shen, J.~Trnka, JHEP \textbf{02}, 020
  (2017).
\newblock \href{https://doi.org/10.1007/JHEP02(2017)020}{DOI
  10.1007/JHEP02(2017)020}

\bibitem{Bogers2018b}
M.P. Bogers, T.~Brauner, JHEP \textbf{05}, 076 (2018).
\newblock \href{https://doi.org/10.1007/JHEP05(2018)076}{DOI
  10.1007/JHEP05(2018)076}

\bibitem{Roest2019}
D.~Roest, D.~Stefanyszyn, P.~Werkman, JHEP \textbf{08}, 081 (2019).
\newblock \href{https://doi.org/10.1007/JHEP08(2019)081}{DOI
  10.1007/JHEP08(2019)081}

\bibitem{Bogers2018a}
M.P. Bogers, T.~Brauner, Phys. Rev. Lett. \textbf{121}, 171602 (2018).
\newblock \href{https://doi.org/10.1103/PhysRevLett.121.171602}{DOI
  10.1103/PhysRevLett.121.171602}

\bibitem{Cheung2016a}
C.~Cheung, K.~Kampf, J.~{Novotn\'y}, C.H. Shen, J.~Trnka, Phys. Rev. Lett.
  \textbf{116}, 041601 (2016).
\newblock \href{https://doi.org/10.1103/PhysRevLett.116.041601}{DOI
  10.1103/PhysRevLett.116.041601}

\bibitem{Elvang2019}
H.~Elvang, M.~Hadjiantonis, C.R.T. Jones, S.~Paranjape, JHEP \textbf{01}, 195
  (2019).
\newblock \href{https://doi.org/10.1007/JHEP01(2019)195}{DOI
  10.1007/JHEP01(2019)195}

\bibitem{Low2019}
I.~Low, Z.~Yin, JHEP \textbf{11}, 078 (2019).
\newblock \href{https://doi.org/10.1007/JHEP11(2019)078}{DOI
  10.1007/JHEP11(2019)078}

\end{thebibliography}


\begin{thebibliography}{1}

\bibitem{Brauner2020a}
T.~Brauner, Phys. Scr. \textbf{95}, 035004 (2020).
\newblock \href{https://doi.org/10.1088/1402-4896/ab50a5}{DOI
  10.1088/1402-4896/ab50a5}

\bibitem{Burgess2021a}
C.P. Burgess, \emph{{Introduction to Effective Field Theory}} (Cambridge
  University Press, Cambridge, UK, 2021).
\newblock \href{https://doi.org/10.1017/9781139048040}{DOI
  10.1017/9781139048040}

\bibitem{Son2006a}
D.T. Son, M.~Wingate, Ann. Phys. \textbf{321}, 197 (2006).
\newblock \href{https://doi.org/10.1016/j.aop.2005.11.001}{DOI
  10.1016/j.aop.2005.11.001}

\bibitem{Geracie2014b}
M.~Geracie, D.T. Son, C.~Wu, S.F. Wu, Phys. Rev. \textbf{D91}, 045030 (2015).
\newblock \href{https://doi.org/10.1103/PhysRevD.91.045030}{DOI
  10.1103/PhysRevD.91.045030}

\bibitem{Jensen2018}
K.~Jensen, SciPost Phys. \textbf{5}, 011 (2018).
\newblock \href{https://doi.org/10.21468/SciPostPhys.5.1.011}{DOI
  10.21468/SciPostPhys.5.1.011}

\bibitem{Bergshoeff2022}
E.~Bergshoeff, J.~Figueroa-O'Farrill, J.~Gomis, SciPost Phys. Lect. Notes
  \textbf{69} (2023).
\newblock \href{https://doi.org/10.21468/SciPostPhysLectNotes.69}{DOI
  10.21468/SciPostPhysLectNotes.69}

\bibitem{Cheung2017a}
C.~Cheung, K.~Kampf, J.~Novotn{\'y}, C.H. Shen, J.~Trnka, JHEP \textbf{02}, 020
  (2017).
\newblock \href{https://doi.org/10.1007/JHEP02(2017)020}{DOI
  10.1007/JHEP02(2017)020}

\bibitem{Mojahed2022}
M.A. Mojahed, T.~Brauner, JHEP \textbf{03}, 086 (2022).
\newblock \href{https://doi.org/10.1007/JHEP03(2022)086}{DOI
  10.1007/JHEP03(2022)086}

\end{thebibliography}


\begin{thebibliography}{10}

\bibitem{Bacry1968}
H.~Bacry, J.M. L\'evy-Leblond, J. Math. Phys. \textbf{9}, 1605 (1968).
\newblock \href{https://doi.org/10.1063/1.1664490}{DOI 10.1063/1.1664490}

\bibitem{Bergshoeff2022}
E.~Bergshoeff, J.~Figueroa-O'Farrill, J.~Gomis, SciPost Phys. Lect. Notes
  \textbf{69} (2023).
\newblock \href{https://doi.org/10.21468/SciPostPhysLectNotes.69}{DOI
  10.21468/SciPostPhysLectNotes.69}

\bibitem{Joseph1970a}
A.~Joseph, A.I. Solomon, J. Math. Phys. \textbf{11}, 748 (1970).
\newblock \href{https://doi.org/10.1063/1.1665205}{DOI 10.1063/1.1665205}

\bibitem{Coleman1969a}
S.R. Coleman, J.~Wess, B.~Zumino, Phys. Rev. \textbf{177}, 2239 (1969).
\newblock \href{https://doi.org/10.1103/PhysRev.177.2239}{DOI
  10.1103/PhysRev.177.2239}

\bibitem{Volkov1973a}
D.V. Volkov, Sov. J. Part. Nucl. \textbf{4}, 1 (1973)

\bibitem{Ogievetsky1974a}
V.I. Ogievetsky, Acta Univ. Wratislav. \textbf{207}, 117 (1974)

\bibitem{Ivanov1975a}
E.~Ivanov, V.I. Ogievetsky, Theor. Math. Phys. \textbf{25}, 1050 (1975)

\bibitem{Low2002a}
I.~Low, A.V. Manohar, Phys. Rev. Lett. \textbf{88}, 101602 (2002).
\newblock \href{https://doi.org/10.1103/PhysRevLett.88.101602}{DOI
  10.1103/PhysRevLett.88.101602}

\bibitem{Watanabe2013a}
H.~Watanabe, H.~Murayama, Phys. Rev. Lett. \textbf{110}, 181601 (2013).
\newblock \href{https://doi.org/10.1103/PhysRevLett.110.181601}{DOI
  10.1103/PhysRevLett.110.181601}

\bibitem{Nicolis2013b}
A.~Nicolis, R.~Penco, F.~Piazza, R.A. Rosen, JHEP \textbf{11}, 055 (2013).
\newblock \href{https://doi.org/10.1007/JHEP11(2013)055}{DOI
  10.1007/JHEP11(2013)055}

\bibitem{Endlich2014a}
S.~Endlich, A.~Nicolis, R.~Penco, Phys. Rev. \textbf{D89}, 065006 (2014).
\newblock \href{https://doi.org/10.1103/PhysRevD.89.065006}{DOI
  10.1103/PhysRevD.89.065006}

\bibitem{Brauner2014a}
T.~Brauner, H.~Watanabe, Phys. Rev. \textbf{D89}, 085004 (2014).
\newblock \href{https://doi.org/10.1103/PhysRevD.89.085004}{DOI
  10.1103/PhysRevD.89.085004}

\bibitem{Creminelli2015a}
P.~Creminelli, M.~Serone, G.~Trevisan, E.~Trincherini, JHEP \textbf{02}, 037
  (2015).
\newblock \href{https://doi.org/10.1007/JHEP02(2015)037}{DOI
  10.1007/JHEP02(2015)037}

\bibitem{Klein2017a}
R.~Klein, D.~Roest, D.~Stefanyszyn, JHEP \textbf{10}, 051 (2017).
\newblock \href{https://doi.org/10.1007/JHEP10(2017)051}{DOI
  10.1007/JHEP10(2017)051}

\bibitem{Finelli2020}
B.~Finelli, JHEP \textbf{03}, 075 (2020).
\newblock \href{https://doi.org/10.1007/JHEP03(2020)075}{DOI
  10.1007/JHEP03(2020)075}

\bibitem{Kharuk2018}
I.~Kharuk, A.~Shkerin, Phys. Rev. \textbf{D98}, 125016 (2018).
\newblock \href{https://doi.org/10.1103/PhysRevD.98.125016}{DOI
  10.1103/PhysRevD.98.125016}

\bibitem{Jackiw2000a}
R.~Jackiw, V.P. Nair, Phys. Lett. \textbf{B480}, 237 (2000).
\newblock \href{https://doi.org/10.1016/S0370-2693(00)00379-8}{DOI
  10.1016/S0370-2693(00)00379-8}

\bibitem{Montigny2006a}
M.~de~Montigny, J.~Niederle, A.G. Nikitin, J. Phys. \textbf{A39}, 9365 (2006).
\newblock \href{https://doi.org/10.1088/0305-4470/39/29/026}{DOI
  10.1088/0305-4470/39/29/026}

\end{thebibliography}


\begin{thebibliography}{10}

\bibitem{Togawa2016a}
Y.~Togawa, Y.~Kousaka, K.~Inoue, J.i. Kishine, J. Phys. Soc. Jpn. \textbf{85},
  112001 (2016).
\newblock \href{https://doi.org/10.7566/JPSJ.85.112001}{DOI
  10.7566/JPSJ.85.112001}

\bibitem{Schmitt2015}
A.~Schmitt, \emph{{Introduction to Superfluidity}: {Field-theoretical Approach
  and Applications}}, \emph{Lecture Notes in Physics}, vol. 888 (Springer,
  2015).
\newblock \href{https://doi.org/10.1007/978-3-319-07947-9}{DOI
  10.1007/978-3-319-07947-9}

\bibitem{Son2002a}
D.T. Son, \href{https://arxiv.org/abs/hep-ph/0204199}{arXiv:hep-ph/0204199}

\bibitem{Gennes1993}
P.G. de~Gennes, J.~Prost, \emph{The Physics of Liquid Crystals} (Clarendon
  Press, Oxford, UK, 1993)

\bibitem{Radzihovsky2011a}
L.~Radzihovsky, T.C. Lubensky, Phys. Rev. \textbf{E83}, 051701 (2011).
\newblock \href{https://doi.org/10.1103/PhysRevE.83.051701}{DOI
  10.1103/PhysRevE.83.051701}

\bibitem{Chaikin1995a}
P.M. Chaikin, T.C. Lubensky, \emph{Principles of condensed matter physics}
  (Cambridge University Press, Cambridge, UK, 1995)

\bibitem{Ivanov1975a}
E.~Ivanov, V.I. Ogievetsky, Theor. Math. Phys. \textbf{25}, 1050 (1975)

\bibitem{Rothstein2018a}
I.Z. Rothstein, P.~Shrivastava, JHEP \textbf{05}, 014 (2018).
\newblock \href{https://doi.org/10.1007/JHEP05(2018)014}{DOI
  10.1007/JHEP05(2018)014}

\bibitem{Alberte2020}
L.~Alberte, A.~Nicolis, JHEP \textbf{07}, 076 (2020).
\newblock \href{https://doi.org/10.1007/JHEP07(2020)076}{DOI
  10.1007/JHEP07(2020)076}

\bibitem{Nicolis2015a}
A.~Nicolis, R.~Penco, F.~Piazza, R.~Rattazzi, JHEP \textbf{06}, 155 (2015).
\newblock \href{https://doi.org/10.1007/JHEP06(2015)155}{DOI
  10.1007/JHEP06(2015)155}

\bibitem{Finelli2020}
B.~Finelli, JHEP \textbf{03}, 075 (2020).
\newblock \href{https://doi.org/10.1007/JHEP03(2020)075}{DOI
  10.1007/JHEP03(2020)075}

\bibitem{Kharuk2018}
I.~Kharuk, A.~Shkerin, Phys. Rev. \textbf{D98}, 125016 (2018).
\newblock \href{https://doi.org/10.1103/PhysRevD.98.125016}{DOI
  10.1103/PhysRevD.98.125016}

\bibitem{Hayata2014b}
T.~Hayata, Y.~Hidaka, Phys. Rev. \textbf{D91}, 056006 (2015).
\newblock \href{https://doi.org/10.1103/PhysRevD.91.056006}{DOI
  10.1103/PhysRevD.91.056006}

\bibitem{Brauner2014a}
T.~Brauner, H.~Watanabe, Phys. Rev. \textbf{D89}, 085004 (2014).
\newblock \href{https://doi.org/10.1103/PhysRevD.89.085004}{DOI
  10.1103/PhysRevD.89.085004}

\bibitem{Nicolis2013b}
A.~Nicolis, R.~Penco, F.~Piazza, R.A. Rosen, JHEP \textbf{11}, 055 (2013).
\newblock \href{https://doi.org/10.1007/JHEP11(2013)055}{DOI
  10.1007/JHEP11(2013)055}

\bibitem{Hidaka2015a}
Y.~Hidaka, T.~Noumi, G.~Shiu, Phys. Rev. \textbf{D92}, 045020 (2015).
\newblock \href{https://doi.org/10.1103/PhysRevD.92.045020}{DOI
  10.1103/PhysRevD.92.045020}

\bibitem{Endlich2014a}
S.~Endlich, A.~Nicolis, R.~Penco, Phys. Rev. \textbf{D89}, 065006 (2014).
\newblock \href{https://doi.org/10.1103/PhysRevD.89.065006}{DOI
  10.1103/PhysRevD.89.065006}

\bibitem{Klein2017a}
R.~Klein, D.~Roest, D.~Stefanyszyn, JHEP \textbf{10}, 051 (2017).
\newblock \href{https://doi.org/10.1007/JHEP10(2017)051}{DOI
  10.1007/JHEP10(2017)051}

\bibitem{McArthur2010a}
I.N. McArthur, JHEP \textbf{11}, 140 (2010).
\newblock \href{https://doi.org/10.1007/JHEP11(2010)140}{DOI
  10.1007/JHEP11(2010)140}

\bibitem{Manton2004}
N.~Manton, P.~Sutcliffe, \emph{Topological Solitons} (Cambridge University
  Press, Cambridge, UK, 2004)

\bibitem{Hecht2000}
K.T. Hecht, \emph{Quantum Mechanics} (Springer-Verlag, New York, US, 2000)

\bibitem{Cheung2008a}
C.~Cheung, P.~Creminelli, A.L. Fitzpatrick, J.~Kaplan, L.~Senatore, JHEP
  \textbf{03}, 014 (2008).
\newblock \href{https://doi.org/10.1088/1126-6708/2008/03/014}{DOI
  10.1088/1126-6708/2008/03/014}

\bibitem{Samokhin2010a}
K.~Samokhin, Phys. Rev. \textbf{B81}, 224507 (2010).
\newblock \href{https://doi.org/10.1103/PhysRevB.81.224507}{DOI
  10.1103/PhysRevB.81.224507}

\bibitem{Powell2011a}
P.D. Powell, \href{https://arxiv.org/abs/1112.4379}{arXiv:1112.4379}

\bibitem{Watanabe2014e}
H.~Watanabe, H.~Murayama, Phys. Rev. \textbf{D89}, 101701 (2014).
\newblock \href{https://doi.org/10.1103/PhysRevD.89.101701}{DOI
  10.1103/PhysRevD.89.101701}

\bibitem{Bergshoeff2022}
E.~Bergshoeff, J.~Figueroa-O'Farrill, J.~Gomis, SciPost Phys. Lect. Notes
  \textbf{69} (2023).
\newblock \href{https://doi.org/10.21468/SciPostPhysLectNotes.69}{DOI
  10.21468/SciPostPhysLectNotes.69}

\end{thebibliography}


\begin{thebibliography}{10}

\bibitem{Jose1998a}
J.V. Jos\'e, E.J. Saletan, \emph{Classical Dynamics: A Contemporary Approach}
  (Cambridge University Press, Cambridge, UK, 1998)

\bibitem{Soper2008}
D.E. Soper, \emph{{Classical Field Theory}} (Dover Publications, 2008)

\bibitem{Stone2000}
M.~Stone, in \emph{{Artificial Black Holes}}, ed. by M.~Novello, M.~Visser,
  G.~Volovik (World Scientific, Singapore, 2002), pp. 335--364

\bibitem{Landau1986}
L.D. Landau, E.M. Lifshitz, \emph{{Theory of Elasticity}}, \emph{Course of
  Theoretical Physics}, vol.~7, 3rd edn. (Pergamon Press, Oxford, UK, 1986)

\bibitem{Arnold1998a}
V.I. Arnold, B.A. Khesin, \emph{Topological Methods in Hydrodynamics}
  (Springer-Verlag, New York, 1998)

\bibitem{Jackiw2004}
R.~Jackiw, V.P. Nair, S.Y. Pi, A.P. Polychronakos, J. Phys. \textbf{A37}, R327
  (2004).
\newblock \href{https://doi.org/10.1088/0305-4470/37/42/R01}{DOI
  10.1088/0305-4470/37/42/R01}

\bibitem{Andersson2021}
N.~Andersson, G.L. Comer, Living Rev. Rel. \textbf{24}, 3 (2021).
\newblock \href{https://doi.org/10.1007/s41114-021-00031-6}{DOI
  10.1007/s41114-021-00031-6}

\bibitem{Nicolis2014a}
A.~Nicolis, R.~Penco, R.A. Rosen, Phys. Rev. \textbf{D89}, 045002 (2014).
\newblock \href{https://doi.org/10.1103/PhysRevD.89.045002}{DOI
  10.1103/PhysRevD.89.045002}

\bibitem{Liu2018}
H.~Liu, P.~Glorioso, PoS \textbf{TASI2017}, 008 (2018).
\newblock \href{https://doi.org/10.22323/1.305.0008}{DOI 10.22323/1.305.0008}

\bibitem{Delacretaz2015a}
L.V. Delacr{\'e}taz, A.~Nicolis, R.~Penco, R.A. Rosen, Phys. Rev. Lett.
  \textbf{114}, 091601 (2015).
\newblock \href{https://doi.org/10.1103/PhysRevLett.114.091601}{DOI
  10.1103/PhysRevLett.114.091601}

\bibitem{Endlich2013}
S.~Endlich, A.~Nicolis, J.~Wang, JCAP \textbf{10}, 011 (2013).
\newblock \href{https://doi.org/10.1088/1475-7516/2013/10/011}{DOI
  10.1088/1475-7516/2013/10/011}

\bibitem{Son2005a}
D.T. Son, Phys. Rev. Lett. \textbf{94}, 175301 (2005).
\newblock \href{https://doi.org/10.1103/PhysRevLett.94.175301}{DOI
  10.1103/PhysRevLett.94.175301}

\bibitem{Dubovsky2006a}
S.~Dubovsky, T.~Gregoire, A.~Nicolis, R.~Rattazzi, JHEP \textbf{03}, 025
  (2006).
\newblock \href{https://doi.org/10.1088/1126-6708/2006/03/025}{DOI
  10.1088/1126-6708/2006/03/025}

\bibitem{Pavaskar2022}
S.~Pavaskar, R.~Penco, I.Z. Rothstein, SciPost Phys. \textbf{12}, 155 (2022).
\newblock \href{https://doi.org/10.21468/SciPostPhys.12.5.155}{DOI
  10.21468/SciPostPhys.12.5.155}

\end{thebibliography}


\begin{thebibliography}{10}

\bibitem{Gerber1989a}
P.~Gerber, H.~Leutwyler, Nucl. Phys. \textbf{B321}, 387 (1989).
\newblock \href{https://doi.org/10.1016/0550-3213(89)90349-0}{DOI
  10.1016/0550-3213(89)90349-0}

\bibitem{Hofmann2002a}
C.P. Hofmann, Phys. Rev. \textbf{B65}, 094430 (2002).
\newblock \href{https://doi.org/10.1103/PhysRevB.65.094430}{DOI
  10.1103/PhysRevB.65.094430}

\bibitem{Kapusta2006a}
J.I. Kapusta, C.~Gale, \emph{Finite-temperature field theory: Principles and
  applications} (Cambridge University Press, Cambridge, UK, 2006)

\bibitem{Strocchi2021}
F.~Strocchi, \emph{{Symmetry Breaking}}.
\newblock Theoretical and Mathematical Physics (Springer Nature Switzerland AG,
  2021).
\newblock \href{https://doi.org/10.1007/978-3-662-62166-0}{DOI
  10.1007/978-3-662-62166-0}

\bibitem{Hayata2014b}
T.~Hayata, Y.~Hidaka, Phys. Rev. \textbf{D91}, 056006 (2015).
\newblock \href{https://doi.org/10.1103/PhysRevD.91.056006}{DOI
  10.1103/PhysRevD.91.056006}

\bibitem{Landry2020}
M.J. Landry, JHEP \textbf{07}, 200 (2020).
\newblock \href{https://doi.org/10.1007/JHEP07(2020)200}{DOI
  10.1007/JHEP07(2020)200}

\bibitem{Hongo2021}
M.~Hongo, S.~Kim, T.~Noumi, A.~Ota, Phys. Rev. \textbf{D103}, 056020 (2021).
\newblock \href{https://doi.org/10.1103/PhysRevD.103.056020}{DOI
  10.1103/PhysRevD.103.056020}

\bibitem{Akyuz2023}
C.O. Akyuz, G.~Goon, R.~Penco,
  \href{https://arxiv.org/abs/2306.17232}{arXiv:2306.17232}

\bibitem{Coleman1973a}
S.R. Coleman, Commun. Math. Phys. \textbf{31}, 259 (1973).
\newblock \href{https://doi.org/10.1007/BF01646487}{DOI 10.1007/BF01646487}

\bibitem{Griffin2015a}
T.~Griffin, K.T. Grosvenor, P.~{Ho\v{r}ava}, Z.~Yan, Commun. Math. Phys.
  \textbf{340}, 985 (2015).
\newblock \href{https://doi.org/10.1007/s00220-015-2461-2}{DOI
  10.1007/s00220-015-2461-2}

\bibitem{Griffin2015b}
T.~Griffin, K.T. Grosvenor, P.~{Ho\v{r}ava}, Z.~Yan, Phys. Rev. Lett.
  \textbf{115}, 241601 (2015).
\newblock \href{https://doi.org/10.1103/PhysRevLett.115.241601}{DOI
  10.1103/PhysRevLett.115.241601}

\bibitem{Argurio2019}
R.~Argurio, D.~Naegels, A.~Pasternak, Phys. Rev. \textbf{D100}, 066002 (2019).
\newblock \href{https://doi.org/10.1103/PhysRevD.100.066002}{DOI
  10.1103/PhysRevD.100.066002}

\bibitem{Hohenberg1967a}
P.C. Hohenberg, Phys. Rev. \textbf{158}, 383 (1967)

\bibitem{Mermin1966a}
N.D. Mermin, H.~Wagner, Phys. Rev. Lett. \textbf{17}, 1133 (1966)

\bibitem{Gelfert2001}
A.~Gelfert, W.~Nolting, J. Phys.: Condens. Matter \textbf{13}, R505 (2001).
\newblock \href{https://doi.org/10.1088/0953-8984/13/27/201}{DOI
  10.1088/0953-8984/13/27/201}

\bibitem{Gennes1993}
P.G. de~Gennes, J.~Prost, \emph{The Physics of Liquid Crystals} (Clarendon
  Press, Oxford, UK, 1993)

\bibitem{Baym1982a}
G.~Baym, B.L. Friman, G.~Grinstein, Nucl. Phys. \textbf{B210}, 193 (1982).
\newblock \href{https://doi.org/10.1016/0550-3213(82)90239-5}{DOI
  10.1016/0550-3213(82)90239-5}

\bibitem{Wilczek2012}
F.~Wilczek, Phys. Rev. Lett. \textbf{109}, 160401 (2012).
\newblock \href{https://doi.org/10.1103/PhysRevLett.109.160401}{DOI
  10.1103/PhysRevLett.109.160401}

\bibitem{Shapere2012}
A.~Shapere, F.~Wilczek, Phys. Rev. Lett. \textbf{109}, 160402 (2012).
\newblock \href{https://doi.org/10.1103/PhysRevLett.109.160402}{DOI
  10.1103/PhysRevLett.109.160402}

\bibitem{Bruno2013}
P.~Bruno, Phys. Rev. Lett. \textbf{111}, 070402 (2013).
\newblock \href{https://doi.org/10.1103/PhysRevLett.111.070402}{DOI
  10.1103/PhysRevLett.111.070402}

\bibitem{Watanabe2015a}
H.~Watanabe, M.~Oshikawa, Phys. Rev. Lett. \textbf{114}, 251603 (2015).
\newblock \href{https://doi.org/10.1103/PhysRevLett.114.251603}{DOI
  10.1103/PhysRevLett.114.251603}

\bibitem{Zaletel2023}
M.P. Zaletel, M.~Lukin, C.~Monroe, C.~Nayak, F.~Wilczek, N.Y. Yao, Rev. Mod.
  Phys. \textbf{95}, 031001 (2023).
\newblock \href{https://doi.org/10.1103/RevModPhys.95.031001}{DOI
  10.1103/RevModPhys.95.031001}

\bibitem{Vafa1984a}
C.~Vafa, E.~Witten, Nucl. Phys. \textbf{B234}, 173 (1984).
\newblock \href{https://doi.org/10.1016/0550-3213(84)90230-X}{DOI
  10.1016/0550-3213(84)90230-X}

\bibitem{Nussinov2002}
S.~Nussinov, M.A. Lampert, Phys. Rept. \textbf{362}, 193 (2002).
\newblock \href{https://doi.org/10.1016/S0370-1573(01)00091-6}{DOI
  10.1016/S0370-1573(01)00091-6}

\bibitem{Mermin1979a}
N.D. Mermin, Rev. Mod. Phys. \textbf{51}, 591 (1979).
\newblock \href{https://doi.org/10.1103/RevModPhys.51.591}{DOI
  10.1103/RevModPhys.51.591}

\bibitem{Needham1997a}
T.~Needham, \emph{Visual Complex Analysis} (Oxford University Press, Oxford,
  UK, 1997)

\bibitem{Watanabe2014a}
H.~Watanabe, H.~Murayama, Phys. Rev. \textbf{X4}, 031057 (2014).
\newblock \href{https://doi.org/10.1103/PhysRevX.4.031057}{DOI
  10.1103/PhysRevX.4.031057}

\bibitem{Davighi2018a}
J.~Davighi, B.~Gripaios, JHEP \textbf{09}, 155 (2018).
\newblock \href{https://doi.org/10.1007/JHEP09(2018)155}{DOI
  10.1007/JHEP09(2018)155}

\bibitem{Altland2010a}
A.~Altland, B.~Simons, \emph{{Condensed matter field theory}} (Cambridge
  University Press, Cambridge, UK, 2010)

\bibitem{Gomes2023}
P.R.S. Gomes, SciPost Phys. Lect. Notes \textbf{74} (2023).
\newblock \href{https://doi.org/10.21468/SciPostPhysLectNotes.74}{DOI
  10.21468/SciPostPhysLectNotes.74}

\bibitem{Brennan2023}
T.D. Brennan, S.~Hong,
  \href{https://arxiv.org/abs/2306.00912}{arXiv:2306.00912}

\bibitem{Bhardwaj2023}
L.~Bhardwaj, L.E. Bottini, L.~Fraser-Taliente, L.~Gladden, D.S.W. Gould,
  A.~Platschorre, H.~Tillim, Phys. Rept. \textbf{1051}, 1 (2024).
\newblock \href{https://doi.org/10.1016/j.physrep.2023.11.002}{DOI
  10.1016/j.physrep.2023.11.002}

\bibitem{Hidaka2020b}
Y.~Hidaka, Y.~Hirono, R.~Yokokura, Phys. Rev. Lett. \textbf{126}, 071601
  (2021).
\newblock \href{https://doi.org/10.1103/PhysRevLett.126.071601}{DOI
  10.1103/PhysRevLett.126.071601}

\bibitem{Sogabe2019a}
N.~Sogabe, N.~Yamamoto, Phys. Rev. \textbf{D99}, 125003 (2019).
\newblock \href{https://doi.org/10.1103/PhysRevD.99.125003}{DOI
  10.1103/PhysRevD.99.125003}

\end{thebibliography}


\begin{thebibliography}{1}

\bibitem{Kharuk2018a}
I.~Kharuk, Phys. Rev. \textbf{D98}, 025006 (2018).
\newblock \href{https://doi.org/10.1103/PhysRevD.98.025006}{DOI
  10.1103/PhysRevD.98.025006}

\bibitem{Olver1986a}
P.J. Olver, \emph{Applications of Lie Groups to Differential Equations}
  (Springer-Verlag, New York, 1986)

\bibitem{Craig2023a}
N.~Craig, Y.T. Lee, Phys. Rev. Lett. \textbf{132}, 061602 (2024).
\newblock \href{https://doi.org/10.1103/PhysRevLett.132.061602}{DOI
  10.1103/PhysRevLett.132.061602}

\bibitem{Alminawi2023}
M.~Alminawi, I.~Brivio, J.~Davighi,
  \href{https://arxiv.org/abs/2308.00017}{arXiv:2308.00017}

\bibitem{Hidaka2015a}
Y.~Hidaka, T.~Noumi, G.~Shiu, Phys. Rev. \textbf{D92}, 045020 (2015).
\newblock \href{https://doi.org/10.1103/PhysRevD.92.045020}{DOI
  10.1103/PhysRevD.92.045020}

\end{thebibliography}


\end{document}